\documentclass[12pt]{article}
\makeatletter \@addtoreset{equation}{section} \makeatother
\renewcommand{\theequation}{\thesection.\arabic{equation}}
\newcommand{\ba}{\begin{array}}
\newcommand{\ea}{\end{array}}
\newcommand{\beq}{\begin{equation}}
\newcommand{\eeq}{\end{equation}}
\newcommand{\bea}{\begin{eqnarray}}
\newcommand{\eea}{\end{eqnarray}}
\def\bce{\begin{center}}
\def\ece{\end{center}}

\def\nonu{\nonumber}

\def\pa{\partial}

\def\be{\beta}

\def\ep{\epsilon}

\newcommand{\p}{\hat{p}}

\def\eps6{{\displaystyle \mathop{\epsilon}^{6}}{}}
\def\g6{{\displaystyle \mathop{g}^{6}}{}}

\def\nab6{{\displaystyle \mathop{\nabla}^{6}}{}}
\def\0{{\sst{(0)}}}
\def\1{{\sst{(1)}}}
\def\2{{\sst{(2)}}}
\def\3{{\sst{(3)}}}
\def\4{{\sst{(4)}}}
\def\5{{\sst{(5)}}}
\def\6{{\sst{(6)}}}
\def\7{{\sst{(7)}}}
\def\8{{\sst{(8)}}}

\def\p{\partial}

\def\ba{\begin{array}}
\def\ea{\end{array}}
\def\beq{\begin{equation}}
\def\eeq{\end{equation}}
\def\be{\begin{equation}}
\def\ee{\end{equation}}

\def\eps{\epsilon}

\def\p{\partial}

\def\ba{\begin{array}}
\def\ea{\end{array}}
\def\beq{\begin{equation}}
\def\eeq{\end{equation}}
\def\be{\begin{equation}}
\def\ee{\end{equation}}

\def\eps{\epsilon}

\def\eps6{{\displaystyle \mathop{\epsilon}^{6}}{}}

\def\nab6{{\displaystyle \mathop{\nabla}^{6}}{}}

\newcommand{\bean}{\begin{eqnarray*}}
\newcommand{\eean}{\end{eqnarray*}}

\begin{document}
\thispagestyle{empty} \addtocounter{page}{-1}
   \begin{flushright}
%PUPT-2395 \\
%CALT-68-nnnn \\
%{\tt hep-th/yymmnnn}\\
\end{flushright}

\vspace*{1.3cm}
  
\centerline{ \Large \bf   
Higher Spin Currents in Wolf Space: Part III } 
%and }
\vspace*{1.5cm}
\centerline{{\bf Changhyun Ahn } 
%and {\bf Hyunsu Kim}
%\footnote{On leave from the Department of Physics, Kyungpook National University, Taegu
%  702-701, Korea and 
%address until Aug. 31, 2011:
%Department of Physics, Princeton University, Jadwin Hall, 
%Princeton, NJ 08544, USA}
} 
\vspace*{1.0cm} 
\centerline{\it 
Department of Physics, Kyungpook National University, Taegu
702-701, Korea} 
%\centerline{\it 
%Department of Physics, Princeton University, Jadwin Hall, 
%Princeton, NJ 08544, USA}
\vspace*{0.8cm} 
\centerline{\tt ahn@knu.ac.kr 
%\qquad kimhyun@knu.ac.kr 
} 
\vskip2cm

\centerline{\bf Abstract}
\vspace*{0.5cm}

The large ${\cal N}=4$ linear superconformal algebra (generated by 
four spin-$\frac{1}{2}$ currents, seven spin-$1$ currents, four 
spin-$\frac{3}{2}$ currents and one spin-$2$ current) found by
 Sevrin, Troost and Van Proeyen (and other groups) 
was realized in the ${\cal N}=4$ superconformal
coset $\frac{SU(5)}{SU(3)}$ theory previously.
The lowest $16$ higher spin currents of spins $(1, \frac{3}{2},
\frac{3}{2}, 2)$, $(\frac{3}{2}, 2, 2, \frac{5}{2})$, $(\frac{3}{2}, 2, 2, 
\frac{5}{2})$ and $(2, \frac{5}{2}, \frac{5}{2}, 3)$ are obtained 
by starting with the operator product expansions (OPEs) between the 
four spin-$\frac{3}{2}$ currents from the above 
large ${\cal N}=4$ linear superconformal algebra  and the lowest higher 
spin-$1$ current which is the same as the one in the Wolf space coset
$\frac{SU(5)}{SU(3) \times SU(2) \times U(1)}$ theory.
These OPEs determine the four higher spin-$\frac{3}{2}$ currents 
and the next six higher spin-$2$ currents are obtained from the OPEs between 
the above four spin-$\frac{3}{2}$ currents associated with the 
${\cal N}=4$ supersymmetry  and these four higher spin-$\frac{3}{2}$ 
currents. 
The four higher spin-$\frac{5}{2}$ currents can be determined by 
calculating the OPEs between the above  four spin-$\frac{3}{2}$ currents 
and the higher spin-$2$ currents.
Similarly, the higher spin-$3$ current 
is obtained from the OPEs between 
the four spin-$\frac{3}{2}$ currents 
and the higher spin-$\frac{5}{2}$ currents.
The explicit relations between the above 
$16$ higher spin currents and the corresponding 
$16$ higher spin currents which were found in the extension of 
large ${\cal N}=4$ nonlinear superconformal algebra previously 
are given.
By examining the OPEs between the $16$ currents from 
the large ${\cal N}=4$ linear superconformal algebra and the 
$16$ higher spin currents, 
the match with the findings of 
Beccaria, Candu and Gaberdiel is also given.    
The next $16$ higher spin currents 
of spins  $(2, \frac{5}{2},
\frac{5}{2}, 3)$, $(\frac{5}{2}, 3, 3, \frac{7}{2})$, $(\frac{5}{2}, 3, 3, 
\frac{7}{2})$ and $(3, \frac{7}{2}, \frac{7}{2}, 4)$ occur
from the OPEs between the above  lowest $16$ higher spin currents. 
 
\baselineskip=18pt
\newpage
\renewcommand{\theequation}
{\arabic{section}\mbox{.}\arabic{equation}}

%%%%%%%%%%%%%%%%%%%%%%%%%%%%%%%%%%%%%%%%%%%%%%%%%%%%%%%%%%%%%%%%%%%%%
%%%%%%%%%%%%%%%%%%%%%%%%%%%%%%%%%%%%%%%%%%%%%%%%%%%%%%%%%%%%%%%%%%%%%%
\section{Introduction}
%%%%%%%%%%%%%%%%%%%%%%%%%%%%%%%%%%%%%%%%%%%%%%%%%%%%%%%%%%%%%%%%%%%%%%
%%%%%%%%%%%%%%%%%%%%%%%%%%%%%%%%%%%%%%%%%%%%%%%%%%%%%%%%%%%%%%%%%%%%%

It has been studied in \cite{GG1406} that
the better understanding for the string theory in the tensionless limit
can be described from the Vasiliev higher spin theory.  
The string theory is described by the type IIB string theory on 
$AdS_3 \times {\bf S}^3 \times {\bf T}^4$. The dual conformal field 
theory (CFT)
is believed to be on the moduli space of the free symmetric orbifold.
This CFT contains $4(N+1)$ free fermions and bosons where 
$(N+1)$ is the product of $D1$ brane and $D5$ brane charges.
On the other hand, the large ${\cal N}=4$ coset theory in two dimensional
CFT, 
which is dual to the Vasiliev higher spin theory on $AdS_3$,  has a factor 
$SO(4N+4)$ in the numerator of the coset.
This describes $(4N+4)$ real free fermions. In the large level limit of 
the large ${\cal N}=4$ coset theory, the coset factor 
$\frac{SU(N+2)}{SU(N)}$ leads to the $4(N+1)= [(N+2)^2-1]-[N^2-1]$ 
free bosons. 
Then the number of free bosons and fermions is the same both in the 
symmetric orbifold theory and in the continuous orbifold theory
(described in \cite{GG1406}).
After precise analysis of the embedding of the permutation group $S_{N+1}$ 
into the $U(N)$, they have concluded that the $U(N)$ singlet sector 
of the large ${\cal N}=4$ coset theory in the large level limit 
is a subsector of the $S_{N+1}$ singlet sector of the symmetric 
orbifold theory.
In other words, the perturbative Vasiliev theory is a subsector 
of the tensionless string theory. 
See also the recent work in \cite{GG1501} for relevant discussions. 

In the large level limit, 
the parameter $\gamma$ of large 
${\cal N}=4$ linear superconformal algebra \cite{npb1988}
becomes zero. Then the  large ${\cal N}=4$
linear superconformal algebra contracts to 
the small ${\cal N}=4$ linear superconformal algebra (generated by
spin-$2$ stress tensor, four spin-$\frac{3}{2}$ supersymmetry 
currents and three spin-$1$ currents associated with the 
affine $SU(2)$ algebra)
with four free bosons and fermions.
In the type IIB string theory on $AdS_3 \times {\bf S}^3 \times 
{\bf S}^3 \times {\bf S}^1$, the above parameter 
$\gamma$ is described in terms of the radius of $AdS_3$ space 
and the radius of one of the three spheres. 
There exists a relation between the radius of $AdS_3$ and the radii of 
two three spheres.
As $\gamma \rightarrow 0$,
one of the two three spheres 
becomes flat and decompactifies to $R^3$ (and to 
three torus ${\bf T}^3$) and then one realizes the string theory 
on $AdS_3 \times {\bf S}^3 \times {\bf T}^4$ mentioned before. 

What happens for finite level before taking the large level limit?
One should understand the dual CFT corresponding to the above type IIB
string theory on $AdS_3 \times {\bf S}^3 \times 
{\bf S}^3 \times {\bf S}^1$. 
In addition to the large ${\cal N}=4$ linear superconformal algebra,
the large ${\cal N}=4$ coset theory 
has an extended chiral algebra which contains the higher spin currents 
initiated in \cite{GG1305}. 
Therefore, one immediate thing to do  is to obtain
the lowest higher spin currents for generic $N$. Once the higher spin currents
are obtained for generic $N$, then it is straightforward to calculate 
the three point 
functions with two scalars in CFT.
Before describing the higher spin currents for general $N$, one fixes
$N=3$ in this paper because one can see most of 
the structures of the extended 
large ${\cal N}=4$ linear superconformal algebra.
One can vary $N$ and obtain the higher spin currents for several $N$ values
in near future.   
One expects that one can determine the structure constants for general $N$
from the above low $N$ values results.
The present work will give some hints for the construction of the three point
functions with two scalars for low higher spins of $s$ with arbitrary
$N$ and level $k$.
Eventually, the general spin $s$ dependence of the three point functions 
corresponding to the three point function in the $AdS_3$ bulk theory
will be obtained.   

For the $16$ currents in the large ${\cal N}=4$ linear superconformal 
algebra, the explicit results for $N=3$ in terms of the coset fields 
were present in Part I 
\cite{Ahn1311}.
The general $N$ dependence can be  
found from the earlier work in \cite{ST,Saulina}.

How does one determine the $16$ lowest higher spin currents 
in terms of coset fields?
For the lowest higher spin-$1$ current in the linear version, 
one takes the one in the nonlinear version in Part I \cite{Ahn1311}
\footnote{The fields in the linear version are defined as
the ones in an extension of large ${\cal N}=4$ `linear' superconformal 
algebra (generated by spin-$2$ current, four spin-$\frac{3}{2}$ currents, 
seven spin-$1$ currents and four spin-$\frac{1}{2}$ currents) 
where the OPEs between the 
spin-$\frac{3}{2}$ currents are linear while  the fields 
in the nonlinear version are defined as in an extension of 
large ${\cal N}=4$ `nonlinear' 
superconformal algebra  (generated by spin-$2$ current, four 
spin-$\frac{3}{2}$ currents and six spin-$1$ currents)
where the OPEs between the 
spin-$\frac{3}{2}$ currents are nonlinear.
The OPEs are nonlinear in both linear and nonlinear versions.}.    
In other words, the starting point is based on this higher spin-$1$ current 
in the linear (or nonlinear) version.
As described in the abstract, 
the next four higher spin-$\frac{3}{2}$ currents 
can be obtained from the OPEs between the four spin-$\frac{3}{2}$ currents
which are the supersymmetry generators of the large ${\cal N}=4$ linear 
superconformal algebra 
and this higher spin-$1$ current.
Each higher spin-$\frac{3}{2}$ current has the same $U(1)$ charge
(which will be discussed in next section)  
as the corresponding above spin-$\frac{3}{2}$ current because 
the $U(1)$ charge of the higher spin-$1$ current is zero.
Two of them belong to the same ${\cal N}=2$ multiplet with the higher spin-$1$
current and other two belong to two other ${\cal N}=2$ multiplets
respectively.
Then the next thing is to obtain the next six higher spin-$2$ currents.  
Since there are four spin-$\frac{3}{2}$ currents and
four higher spin-$\frac{3}{2}$ currents, 
there exist $16$ possible OPEs between them.
Among them, the only $12$ OPEs are nontrivial ones.
Then the above six higher spin-$2$ currents appear twice in these OPEs.  
One of them completes the last component of the above first 
${\cal N}=2$ multiplet containing the higher spin-$1$ current. 
Two of them belong to the above second 
${\cal N}=2$ multiplet as the second and third components. 
Two of them belong to the third ${\cal N}=2$
multiplet (second component and third component). 
The last one belongs to the new fourth ${\cal N}=2$ multiplet
as a first component.  

For the next four higher spin-$\frac{5}{2}$ currents
one continues to calculate similar OPEs.
Since there are four spin-$\frac{3}{2}$ currents and
six higher spin-$2$ currents, 
there exist $24$ possible OPEs between them.
Among them, the only $12$ OPEs are nontrivial ones.
Then the above four higher spin-$\frac{5}{2}$ currents appear three times 
in these OPEs. 
One of the four higher spin-$\frac{5}{2}$ currents plays the role of 
the last component of the above second ${\cal N}=2$ multiplet
while one of them is the last component of the third 
${\cal N}=2$ multiplet. Two of them consist of
the second and third components of the fourth ${\cal N}=2$ multiplet.
For the higher spin-$3$ current,
 there are four spin-$\frac{3}{2}$ currents and
four higher spin-$\frac{5}{2}$ currents and so there are 
$16$ OPEs between them. The nontrivial OPEs are given by four
and the higher spin-$3$ current appears four times in these OPEs.
This higher spin-$3$ current will be the last component of the fourth
${\cal N}=2$ multiplet.
One expects that the above OPEs between the spin-$\frac{3}{2}$ currents 
and the higher spin currents of $s=1, \frac{3}{2}, 2$ and $\frac{5}{2}$ 
will be generalized to the general $N$ case
without any difficulty because the structure constants will have very simple 
functions of the level $k$ and several $N$ examples will fix them 
completely.  

In \cite{GG1305}, the large ${\cal N}=4$ holography is proposed 
and some of the currents and higher spin currents are constructed using the
oscillator formalism in the context of asymptotic symmetry algebra in 
$AdS_3$ bulk theory.
One expects that the extension of large ${\cal N}=4$ linear superconformal 
algebra should match with the corresponding higher spin algebra defined in
the $AdS_3$ bulk theory.
Therefore, it is necessary to calculate the OPEs between the $16$ lowest
higher spin currents, as a first step, 
to obtain the corresponding algebra in dual CFT
associated with the above asymptotic symmetry algebra in $AdS_3$.    
One should calculate $16 \times 16=256$ OPEs but the reversed OPE 
between any two higher spin 
currents can be obtained from the original OPE between them.
Therefore, one should calculate $\sum_{n=16}^{1} n = \frac{1}{2} \times 16 \times (16+1) = 136$ OPEs which are nontrivial in the sense that there are singular 
terms. For the corresponding OPEs in the nonlinear version, the number of
OPEs is less than this number because there exist some trivial OPEs.  
The OPEs are written in terms of coset fields
because the higher spin currents are given in terms of 
those coset fields.

Therefore, in order to rewrite them in terms of known $16$ currents 
and $16$ higher spin currents (and their derivatives),
one should introduce the possible candidates with arbitrary coefficients
in the right hand side of the OPEs. 
Since there is $U(1)$ charge constraint in the given OPE, 
the number of arbitrary coefficients is reduced.
If one cannot write any OPE in terms of the known $16$ currents and $16$
higher spin currents (and their derivatives), then 
one should expect that there should exist a new primary field for given pole 
in the OPE. 
Since the extended algebra contains infinite number of higher spin currents,
one expects that there should be new primary currents in the $136$ OPEs
and actually the next $16$ higher spin currents occur.  
So it is better to obtain those $16$ next lowest higher spin currents
instead of calculating the above $136$ OPEs.
From the experience in the construction of $16$ lowest higher spin currents
one can apply the prescriptions used in there to the $16$ next lowest 
higher spin currents.    

Now then one can calculate the $136$ OPEs explicitly and each pole of each OPE 
is written in terms of multiple products of various coset fields.
One puts each pole of the OPE to be equal to be multiple product 
of known $16$ currents, $16$ higher spin currents and the next $16$
higher spin currents (and their derivatives) satisfying the $U(1)$ charge 
constraint.
Compared to the corresponding OPEs in the nonlinear version of Part II 
\cite{Ahn1408}, the present $136$ OPEs do not have any multiplet products
between the $16$ currents of large ${\cal N}=4$ linear superconformal algebra    and the $16$ lowest higher spin currents (and their derivatives) except 
a few OPEs. 
Furthermore, if one goes to different basis found in \cite{BCG}
where it is not clear how the coset fields arise explicitly,
those nonlinear terms disappear. 
One of the reasons why the current issue on the construction of $136$ OPEs
is interested in
is that if one sees the corresponding OPEs in the nonlinear version from 
Part II,
it is hard to extract the simple structure behind these OPEs 
because there are too many nonlinear terms between the $16$ currents and 
the $16$ lowest higher spin currents  in the OPEs.    
In the linear version of present paper, the right hand side of the OPEs 
consists of linear terms having either $16$ currents, $16$ lowest higher 
spin currents, or $16$ next lowest higher spin currents 
(and their derivatives) and nonlinear terms
having $16$ currents (and their derivatives) 
from the large ${\cal N}=4$ linear superconformal algebra up to the poles 
of the OPEs considered in this paper.     
This indicates that if one goes to ${\cal N}=4$ superspace description,
then the above $136$ OPEs can fit in one single OPE in ${\cal N}=4$
superspace and the right hand side of this OPE contains linear terms
as well as nonlinear terms consisting of multiple product of ${\cal N}=4$
stress tensor (and their derivatives) up to the poles considered in this 
paper.

In this paper,
the analysis done in Part I \cite{Ahn1311} and Part II \cite{Ahn1408}
for the higher spin currents is  extended to
the large ${\cal N}=4$ linear superconformal algebra.
The extra two ${\cal N}=4$ multiplets ($16$ lowest higher spin currents
and the next $16$ higher spin currents) are obtained.
In particular, the complete OPEs between the first ${\cal N}=4$ multiplet
in component approach are described.

In section $2$, 
the review of large ${\cal N}=4$ linear superconformal algebra
is described.

In section $3$, the $16$ lowest higher spin currents 
are constructed.
Let us describe the notations one uses in this paper.
In order to emphasize the extra higher spin currents, one uses 
a boldface current. In other words, the boldface currents are 
the higher spin currents.
Furthermore, in order to characterize the higher spin currents 
in the nonlinear version, the word of 'non' is added in the subscript
of the boldface higher spin currents.  
The boldface higher spin currents without 'non' are the higher spin currents
in the linear version. For example, the $16$ lowest higher spin currents  
have the following spins \footnote{In Part I \cite{Ahn1311}, 
the higher spin currents 
in the nonlinear version 
were written without `non' and there were no boldfaces. 
In part II \cite{Ahn1408}, all the higher spin currents 
were presented without `non' and the boldface higher spin currents 
were the second ${\cal N}=4$ multiplet (next $16$ lowest higher spin 
currents).}
\bea
\left(1, \frac{3}{2}, \frac{3}{2}, 2 \right)
& : & ({\bf T_{non}^{(1)}}, {\bf T_{+,non}^{(\frac{3}{2})}}, 
{\bf T_{-,non}^{(\frac{3}{2})}}, 
{\bf T_{non}^{(2)}}) \leftrightarrow
 ({\bf T^{(1)}}, {\bf T_{+}^{(\frac{3}{2})}}, 
{\bf T_{-}^{(\frac{3}{2})}}, 
{\bf T^{(2)}}),
\nonu \\
 \left(\frac{3}{2}, 2, 2, \frac{5}{2} \right) & : & 
({\bf U_{non}^{(\frac{3}{2})}}, {\bf U_{+,non}^{(2)}}, {\bf U_{-,non}^{(2)}}, 
{\bf U_{non}^{(\frac{5}{2})}} ) 
 \leftrightarrow
({\bf U^{(\frac{3}{2})}}, {\bf U_{+}^{(2)}}, {\bf U_{-}^{(2)}}, 
{\bf U^{(\frac{5}{2})}} ), 
\nonu \\
\left(\frac{3}{2}, 2, 2, \frac{5}{2} \right) & : & 
({\bf V_{non}^{(\frac{3}{2})}}, {\bf V^{(2)}_{+,non}}, {\bf V^{(2)}_{-,non}}, 
{\bf V_{non}^{(\frac{5}{2})}})
 \leftrightarrow
({\bf V^{(\frac{3}{2})}}, {\bf V^{(2)}_{+}}, {\bf V^{(2)}_{-}}, 
{\bf V^{(\frac{5}{2})}}),
  \nonu \\
\left(2, \frac{5}{2}, \frac{5}{2}, 3 \right) & : &
 ({\bf W_{non}^{(2)}}, {\bf W_{+,non}^{(\frac{5}{2})}}, {\bf W_{-,non}^{(\frac{5}{2})}}, 
{\bf W_{non}^{(3)}})
 \leftrightarrow
({\bf W^{(2)}}, {\bf W_{+}^{(\frac{5}{2})}}, {\bf W_{-}^{(\frac{5}{2})}}, 
{\bf W^{(3)}}).
\label{lowest}
\eea
These are primary fields under each stress tensor in the nonlinear or
linear version.
In section $2$, the $U(1)$ charges of these higher spin currents 
will be described. 
The precise relations between the higher spin currents 
in the nonlinear and linear versions (\ref{lowest}) are described.

In section $4$, the OPEs between the $16$ currents from the 
large ${\cal N}=4$ linear superconformal algebra and the 
the $16$ lowest higher spin currents found in section $3$
are given.
The comparison with the 
field contents in different basis \cite{BCG} is given.

In section $5$,
the $16$ next lowest higher spin currents 
are constructed.
One can specify these $16$ higher spin currents according to their spins and 
${\cal N}=2$ multiplets as follows:
\bea
\left(2, \frac{5}{2}, \frac{5}{2}, 3 \right) & : & 
( {\bf P_{non}^{(2)}}, {\bf P_{+,non}^{(\frac{5}{2})}}, {\bf P_{-,non}^{(\frac{5}{2})}}, 
{\bf P_{non}^{(3)}})
 \leftrightarrow
( {\bf P^{(2)}}, {\bf P_{+}^{(\frac{5}{2})}}, {\bf P_{-}^{(\frac{5}{2})}}, 
{\bf P^{(3)}}),
\nonu \\
\left( \frac{5}{2}, 3, 3, \frac{7}{2} \right) & : & 
( {\bf Q_{non}^{(\frac{5}{2})}}, {\bf Q_{+,non}^{(3)}}, {\bf Q_{-,non}^{(3)}}, 
{\bf Q_{non}^{(\frac{7}{2})}})
 \leftrightarrow
( {\bf Q^{(\frac{5}{2})}}, {\bf Q_{+}^{(3)}}, {\bf Q_{-}^{(3)}}, 
{\bf Q^{(\frac{7}{2})}}),
\nonu \\
\left( \frac{5}{2}, 3, 3, \frac{7}{2}  \right) & : & 
( {\bf R_{non}^{(\frac{5}{2})}}, {\bf R_{+,non}^{(3)}}, {\bf R_{-,non}^{(3)}}, 
{\bf R_{non}^{(\frac{7}{2})}})
 \leftrightarrow
( {\bf R^{(\frac{5}{2})}}, {\bf R_{+}^{(3)}}, {\bf R_{-}^{(3)}}, 
{\bf R^{(\frac{7}{2})}}),
\nonu \\
\left( 3, \frac{7}{2}, \frac{7}{2}, 4 \right)  & : & 
( {\bf S_{non}^{(3)}}, {\bf S_{+,non}^{(\frac{7}{2})}}, {\bf S_{-,non}^{(\frac{7}{2})}}, 
{\bf S_{non}^{(4)}})
 \leftrightarrow
( {\bf S^{(3)}}, {\bf S_{+}^{(\frac{7}{2})}}, {\bf S_{-}^{(\frac{7}{2})}}, 
{\bf S^{(4)}}).
\label{nextlowest}
\eea

In section $6$,
 the OPEs between the $16$ currents from the 
large ${\cal N}=4$ linear superconformal algebra and  
the $16$ next lowest higher spin currents found in section $5$
are described.

In section $7$,
the OPEs between  
the $16$ lowest higher spin currents found in section $3$
and themselves are given.

In section $8$,
the summary of this paper is given and the future directions 
are also described.

In Appendices $A$-$F$, some details appearing in previous sections 
are presented.

The mathematica package by Thielemans \cite{Thielemans}
is used.

%%%%%%%%%%%%%%%%%%%%%%%%%%%%%%%%%%%%%%%%%%%%%%%%%%%%%%%%%%%%%%%%%%%%%
%%%%%%%%%%%%%%%%%%%%%%%%%%%%%%%%%%%%%%%%%%%%%%%%%%%%%%%%%%%%%%%%%%%%%%
\section{ The large $\mathcal N = 4$ linear superconformal algebra in 
the coset: review }
%2%%%%%%%%%%%%%%%%%%%%%%%%%%%%%%%%%%%%%%%%%%%%%%%%%%%%%%%%%%%%%%%%%%%%%
%%%%%%%%%%%%%%%%%%%%%%%%%%%%%%%%%%%%%%%%%%%%%%%%%%%%%%%%%%%%%%%%%%%%%

The large ${\cal N}=4$ linear superconformal algebra is generated by
spin-$2$ stress tensor $T(z)$, four spin-$\frac{3}{2}$ supersymmetry 
generators $G_a(z)$,
seven spin-$1$ currents $A_i(z), B_i(z)$ and $U(z)$ and 
four spin-$\frac{1}{2}$ currents $F_a(z)$.
Six spin-$1$ currents are the generators of two $SU(2)$ affine algebras
where the levels are denoted by 
$(k^{+}, k^{-})$ and one spin-$1$ current is the $U(1)$ current.
The central charge appearing in the OPE between the spin-$2$ current 
and itself is given by $c = \frac{6 k^{+} k^{-}}{(k^{+} + k^{-})}$. The 
parameter $\gamma = \frac{k^{-}}{(k^{+}+k^{-})}$ appears in the large
${\cal N}=4$ linear superconformal algebra.
One presents the large ${\cal N}=4$ linear superconformal algebra in Appendix 
$A$.

The $16$ currents can be rearranged into the following 
four ${\cal N}=2$ multiplets as follows: 
\bea
\left(0, \frac{1}{2}, \frac{1}{2}, 1 \right)
& : & ( \int U dz,  F_{21}, F_{12}, A_3 \, \mbox{and} \, B_3),
\nonu \\
 \left(\frac{1}{2}, 1, 1, \frac{3}{2} \right) & : & 
(F_{11}, B_{-}, A_{+}, G_{11} \, \mbox{and} \, \pa F_{11}), 
\nonu \\
\left(\frac{1}{2}, 1, 1, \frac{3}{2} \right) & : & 
(F_{22}, A_{-}, B_{+}, G_{22} \, \mbox{and} \, \pa F_{22}),
  \nonu \\
\left(1, \frac{3}{2}, \frac{3}{2}, 2 \right) & : &
 (A_3 \, \mbox{and} \, B_3, G_{21}, G_{12}, T).
\label{scacurrents}
\eea
The last multiplet contains the currents in the 
usual ${\cal N}=2$ superconformal algebra.
The $U(1)$ charges of (\ref{scacurrents}) have the same behavior of
those of (\ref{lowest}) and $A_{\pm}(z) \equiv A_1(z) \pm i A_2(z)$
and  $B_{\pm}(z) \equiv B_1(z) \pm i B_2(z)$.

By factoring out the spin-$1$ current and four spin-$\frac{1}{2}$ currents
from the above large ${\cal N}=4$ linear superconformal algebra,
the large ${\cal N}=4$ nonlinear superconformal algebra (where the 
OPEs between the spin-$\frac{3}{2}$ currents are nonlinear) is obtained. 
For example, the spin-$\frac{3}{2}$ currents of large ${\cal N}=4$
nonlinear  superconformal algebra are written in terms of the fields 
in the linear version as follows:
\bea
\hat{G}_{11}(z)  & = & 
\left[ G_{11} +\frac{2}{(5+k)} U  F_{11} -
\frac{4}{(5+k)^2} F_{12} F_{21} F_{11} +
\frac{2i}{(5+k)} F_{11}  A_3 \right.
\nonu \\
&+ & \left. \frac{2i}{(5+k)} F_{21}  A_{+} +
\frac{2i}{(5+k)} F_{12}  B_{-} +
\frac{2i}{(5+k)} F_{11}  B_3 \right](z), 
\nonu \\
\hat{G}_{12}(z)  & = & 
\left[ G_{12} +\frac{2}{(5+k)} U  F_{12} -
\frac{4}{(5+k)^2} F_{12} F_{11} F_{22} +
\frac{2i}{(5+k)} F_{12}  A_3 \right.
\nonu \\
&- & \left. \frac{2i}{(5+k)} F_{22}  A_{+} +
\frac{2i}{(5+k)} F_{11}  B_{+} -
\frac{2i}{(5+k)} F_{12}  B_3 \right](z), 
\nonu \\
\hat{G}_{21}(z)  & = & 
\left[ G_{21} +\frac{2}{(5+k)} U  F_{21} -
\frac{4}{(5+k)^2} F_{21} F_{11} F_{22} -
\frac{2i}{(5+k)} F_{21}  A_3 \right.
\nonu \\
&+& \left. \frac{2i}{(5+k)} F_{11}  A_{-} -
\frac{2i}{(5+k)} F_{22}  B_{-} +
\frac{2i}{(5+k)} F_{21}  B_3 \right](z), 
\nonu \\
\hat{G}_{22}(z)  & = & 
 \left[ G_{22} +\frac{2}{(5+k)} U  F_{22} -
\frac{4}{(5+k)^2} F_{21} F_{12} F_{22} -
\frac{2i}{(5+k)} F_{22}  A_3 \right.
\nonu \\
&- & \left. \frac{2i}{(5+k)} F_{12}  A_{-} -
\frac{2i}{(5+k)} F_{21}  B_{+} -
\frac{2i}{(5+k)} F_{22}  B_3 \right](z). 
\label{ghatinlinear}
\eea
One can also write down the spin-$\frac{3}{2}$ currents in the 
linear version in terms of the fields in the nonlinear version
using (\ref{ghatinlinear}) and the equations $(3.6)$ and $(3.10)$
of Part I (by writing down the spin-$1$ currents in the linear version
in terms of the spin-$1$ currents in the nonlinear version plus 
spin-$\frac{1}{2}$ current terms)
\footnote{The remaining seven currents in the 
nonlinear version have the following relations with those in the linear 
version:
\bea
\hat{T}(z) & = & 
 T(z)
+ \frac{1}{(5+k)} \left( U U 
+ \pa F^{a} F_{a} \right)(z),
\nonu \\
\hat{A}_i(z) & = &
A_i(z)-\frac{1}{(5+k)} \,
\alpha_{ab}^{+i} \, F^{a} F^{b}(z), 
\nonu \\
\hat{B}_i(z) & = & 
B_i(z)-\frac{1}{(5+k)} \, \alpha_{ab}^{-i}\, F^{a} F^{b}(z).
\nonu
\eea
Also using these relations, one can express the fields in the linear version
in terms of those in the nonlinear version. The fields in the
nonlinear version have trivial OPEs between the spin-$1$ current $U(w)$
and the four spin-$\frac{1}{2}$ currents $F_a(w)$.
}. 

One has the following explicit relations between 
the spin-$\frac{3}{2}$ currents with $SO(4)$ indices 
and those with $SU(2) \times SU(2)$ indices
\bea
G^0(z) & = & 
-\frac{i}{\sqrt{2}} (G_{12}-G_{21})(z),
\qquad
G^1(z)  =  
\frac{1}{\sqrt{2}} (G_{11}+G_{22})(z),
\nonu \\
G^2(z) & = & 
\frac{i}{\sqrt{2}} (G_{11}-G_{22})(z),
\qquad
G^3(z)  =  
-\frac{1}{\sqrt{2}} (G_{12}+G_{21})(z).
\label{gsingledouble}
\eea
Similarly, one has the following relations for the spin-$\frac{1}{2}$
currents
\bea
F^0(z) & = & 
-\frac{i}{\sqrt{2}} (F_{12}-F_{21})(z),
\qquad
F^1(z)  =  
\frac{1}{\sqrt{2}} (F_{11}+F_{22})(z),
\nonu \\
F^2(z) & = & 
\frac{i}{\sqrt{2}} (F_{11}-F_{22})(z),
\qquad
F^3(z)  =  
-\frac{1}{\sqrt{2}} (F_{12}+F_{21})(z).
\label{fsingledouble}
\eea
These relations (\ref{gsingledouble}) and (\ref{fsingledouble})
will be used later to describe the OPEs in Appendix $D$.

The $U(1)$ charge can be obtained from the coefficient of the first order pole
in the OPE 
between the $U(1)$ current 
$(-2 i \gamma A_3 -2 i (1-\gamma) B_3)(z)$
and each primary field $\phi(w)$.
This $U(1)$ current appears in the half of the second order pole of
$G_{21}(z) \, G_{12}(w)$ in Appendix $A$. 
One describes the whole $U(1)$ charges for 
$16$ currents from the large ${\cal N}=4$ linear superconformal algebra
and two $16$ higher spin currents in Table $1$.
One can construct any composite fields  with definite $U(1)$ charge 
based on this Table $1$. Although the $48$ currents have different 
$U(1)$ charges compared to the ones in the nonlinear version, 
one realizes that their behaviors appearing in 
all the OPEs in this paper   are the same. In other words, 
one can consider the extra composite fields due to the 
four spin-$\frac{1}{2}$ currents and one spin-$1$ current
in addition to the Tables (with the presence of
the corresponding composite fields in the linear 
version) in Part I. 

%%%%%%%%%%%%%%%%%%%%%%%%%%%%%%%%%%%%%%%%%%%%%%%%%%%%%%%%%%%%%%%%%%%
\begin{table}[ht]
\centering % used for centering table
\begin{tabular}{|c||c| } % centered columns (4 columns)
\hline %inserts double horizontal lines
$U(1)$ charge & $16$ currents and $32$ higher spin currents   
\\ [0.5ex] % inserts table
%heading
\hline \hline % inserts single horizontal line
$\frac{(2+2k)}{(5+k)}$  & ${B}_{-}; \quad {\bf U_{+}^{(2)}};
 \quad {\bf Q_{+}^{(3)}} $ 
\\ % inserting body of the table
\hline
$1$ &  $G_{21}, \quad F_{21}; \quad {\bf T_{+}^{(\frac{3}{2})}}, \quad
{\bf W_{+}^{(\frac{5}{2})}}; \quad {\bf P_{+}^{(\frac{5}{2})}}, 
\quad {\bf S_{+}^{(\frac{7}{2})}} $ \\
\hline
$\frac{(-3+k)}{(5+k)}$ & $G_{11}, \quad F_{11}; \quad {\bf U^{(\frac{3}{2})}},
\quad {\bf U^{(\frac{5}{2})}}; \quad {\bf Q^{(\frac{5}{2})}}, \quad 
{\bf Q^{(\frac{7}{2})}} $  \\
\hline
$\frac{8}{(5+k)} $ & ${A}_{-}; \quad {\bf V_{+}^{(2)}};
\quad {\bf R_{+}^{(3)}}$  \\
\hline
$0$ & ${A}_3, \,\, {B}_3, \,\, U, \,\, T; \,\, {\bf T^{(1)}}, \,\,
{\bf T^{(2)}}, \,\, {\bf W^{(2)}}, \,\, {\bf W^{(3)}};
 \,\, {\bf P^{(2)}}, \,\,
{\bf P^{(3)}}, \,\, {\bf S^{(3)}}, \,\, {\bf S^{(4)}}
$  \\ 
\hline
$-\frac{8}{(5+k)}$ & $ {A}_{+}; \quad {\bf U_{-}^{(2)} };
\quad  {\bf Q_{-}^{(3)} } $  \\ 
\hline
$-\frac{(-3+k)}{(5+k)}$ & ${G}_{22}, \quad F_{22}; \quad
{\bf V^{(\frac{3}{2})} }, \quad {\bf V^{(\frac{5}{2})}};
\quad {\bf R^{(\frac{5}{2})} }, \quad {\bf R^{(\frac{7}{2})} }$  \\ 
\hline
$-1 $ & ${G}_{12}, \quad F_{12}; \quad {\bf T_{-}^{(\frac{3}{2})}}, 
\quad {\bf W_{-}^{(\frac{5}{2})}}; \quad
{\bf P_{-}^{(\frac{5}{2})}}, \quad {\bf S_{-}^{(\frac{7}{2})}}$  \\
\hline
$-\frac{(2+2k)}{(5+k)}$ & ${B}_{+}; \quad {\bf V_{-}^{(2)}};
\quad {\bf R_{-}^{(3)}} $  \\ 
[1ex] % [1ex] adds vertical space
\hline %inserts single line
\end{tabular}
%\label{tableone} % is used to refer this table in the text
\caption{The $U(1)$ charges for the $16$ currents from 
the large ${\cal N}=4$ linear superconformal algebra and 
two $16$ higher spin currents. 
The first $18$ upper currents with positive 
($ k > 3 $) 
$U(1)$ charges have their conjugated  currents with each 
opposite (negative) $U(1)$ charge.  
  } % title of Table
\end{table}
%%%%%%%%%%%%%%%%%%%%%%%%%%%%%%%%%%%%%%%%%%%%%%%%%%%%%%%%%%%%%%%%%%%%%%

%%%%%%%%%%%%%%%%%%%%%%%%%%%%%%%%%%%%%%%%%%%%%%%%%%%%%%%%%%%%%%%%%%%%%
%%%%%%%%%%%%%%%%%%%%%%%%%%%%%%%%%%%%%%%%%%%%%%%%%%%%%%%%%%%%%%%%%%%%%%
\section{ The higher spin currents where the lowest spin is $1$ in 
the coset }
%3%%%%%%%%%%%%%%%%%%%%%%%%%%%%%%%%%%%%%%%%%%%%%%%%%%%%%%%%%%%%%%%%%%%%%
%%%%%%%%%%%%%%%%%%%%%%%%%%%%%%%%%%%%%%%%%%%%%%%%%%%%%%%%%%%%%%%%%%%%%

Let us construct the $16$ lowest higher spin currents.

One has the two results given in $(4.9)$ and $(4.13)$ of \cite{Ahn1311}
where the two higher spin-$\frac{3}{2}$ currents in the nonlinear version 
are generated.
Then 
one considers the spin-$\frac{3}{2}$ currents $G_{21}(z)$ and $G_{12}(z)$
in the large ${\cal N}=4$ linear superconformal algebra and they act on 
the lowest spin-$1$ current ${\bf T^{(1)}}(w)$. 
From the explicit 
expressions in terms of WZW currents living in $SU(5)$, 
$(2.21)$ corresponding to $G_{21}(z)$, 
$(2.24)$ corresponding to $G_{12}(z)$ and $(4.7)$ corresponding to 
${\bf T^{(1)}}(w)$ of \cite{Ahn1311}, 
the OPEs between them can be obtained  
and it turns out that
\bea
\left(
\begin{array}{c}
G_{12}  \\
G_{21}  \end{array}
\right)(z) \, {\bf T^{(1)}}(w)
& = & \frac{1}{(z-w)} \left[ \mp \left(
\begin{array}{c}
G_{12}  \\
G_{21}  \end{array}
\right) + 2 
{\bf T_{\mp}^{(\frac{3}{2})}}
\right](w) +\cdots.
\label{g1221t1} 
\eea
This is the same as its nonlinear version
and 
each higher spin-$\frac{3}{2}$ current in (\ref{g1221t1}) 
contains $31$ terms consisting of multiple product of 
spin-$\frac{1}{2}$ current $Q^a(w)$ and the spin-$1$ current 
$V^b(w)$ (the indices $a, b$ run as $a, b = 1, 2, \cdots, 24$) 
introduced in \cite{AK1411}. 
One expects that as one generalizes the spin-$\frac{3}{2}$ currents 
and higher spin-$1$ current for $N=3$ to satisfy for general $N$,
the OPEs for general $N$ have the same form as (\ref{g1221t1}) with 
the corresponding higher spin-$\frac{3}{2}$ currents for general $N$.  
There will be no $(N, k)$ dependence in the right hand side of the OPE.

What about other higher spin-$\frac{3}{2}$ currents?
Once again $(4.20)$ and $(4.34)$ of \cite{Ahn1311} 
describe the corresponding higher spin currents in the nonlinear version.
Starting from $(2.11)$  of \cite{Ahn1311} 
associated with $G_{11}(z)$, $(2.12)$ of \cite{Ahn1311} 
associated with $G_{22}(z)$ and 
 $(4.7)$ of \cite{Ahn1311} corresponding to 
${\bf T^{(1)}}(w)$,  
the following OPEs can be obtained
%%%%%%%%%%%%%%%%%%%%%%%%%%%%%%%%%%%%%%%%%%%%%%%%%%%%%%
\bea
\left(
\begin{array}{c}
G_{11}  \\
G_{22}  \end{array}
\right)(z) \, {\bf T^{(1)}}(w)
& = & \frac{1}{(z-w)} \left[ \pm \left(
\begin{array}{c}
G_{11}  \\
G_{22}  \end{array}
\right) + 2 
\left(
\begin{array}{c}
{\bf U^{(\frac{3}{2})}}  \\
{\bf V^{(\frac{3}{2})}}  \end{array}
\right)
\right](w) +\cdots.
\label{g1122t1}
\eea
These OPES are exactly the same forms as the ones in the nonlinear version.
Moreover, each higher spin-$\frac{3}{2}$ current contains 
$26$ terms written in terms of the above 
$Q^a(w)$ and $V^b(w)$ (and their derivatives).
The relative coefficients in the right hand side of (\ref{g1122t1})
are numerical constants and one expects that the OPEs of (\ref{g1122t1})
for general $N$ remain unchanged discussed as before.

Let us move on the next higher spin currents.
The equation $(4.16)$ in \cite{Ahn1311}
determines the higher spin-$2$ current.
Now one considers the similar OPEs in the linear version.  
The spin-$\frac{3}{2}$ currents of large ${\cal N}=4$ linear superconformal 
algebra are given in \cite{Ahn1311} as before and 
the higher spin-$\frac{3}{2}$ currents are determined via
(\ref{g1221t1}). From these explicit expressions,
it turns out that 
%%%%%%%%%%%%%%%%%%%%%%%%%%%%%%%%%%%%%%%%%%%%%%%%%%%%%%%%%%
\bea
 \left(
\begin{array}{c}
G_{12} \nonu \\
G_{21}
 \end{array}
\right)(z) \, {\bf T_{\pm}^{(\frac{3}{2})}}(w)
& = & 
\mp \frac{1}{(z-w)^3} \frac{8(1+k)}{(5+k)} 
\nonu \\
& + &
\frac{1}{(z-w)^2} \left[ -\frac{8i}{(5+k)} A_3 
-\frac{2i(1+k)}{(5+k)} B_3 + {\bf T^{(1)}}
\right](w) \nonu \\
& + & 
\frac{1}{(z-w)} \left[  
%-\frac{4i}{(5+k)} \pa A_3 
%-\frac{i(1+k)}{(5+k)} \pa B_3 + \frac{1}{2} \pa {\bf T^{(1)}} 
%\mp T  
\frac{1}{2} \pa \, \mbox{(pole-2)}
\mp {\bf T^{(2)}}  \mp T \right](w) +
\cdots.
\label{g1221t3half}
\eea
Recall that the equations $(2.16)$, $(2.33)$ and $(2.39)$ of Part I
provide the currents $A_3(w)$, $B_3(w)$ and $T(w)$ living in 
the large ${\cal N}=4$ linear superconformal algebra  respectively.
In the first order pole of (\ref{g1221t3half}), 
the first term denotes the derivative of second order pole of 
(\ref{g1221t3half}).
Compared to the corresponding OPEs in the nonlinear version, 
the field contents are the same but the structure constants 
appearing in the right hand side are different. The higher 
spin-$2$ current in (\ref{g1221t3half}) contains 
$64$ terms written in terms of $Q^a(w)$ and $V^b(w)$ 
(and their derivatives).  
One expects that the corresponding OPEs for general $N$
have $(N,k)$ dependence in three places of (\ref{g1221t3half}) 
explicitly.

The footnotes $48$ and $49$ of \cite{Ahn1311}
determine other two higher spin-$2$ currents 
and the corresponding OPEs from the higher spin-$\frac{3}{2}$ currents
obtained in (\ref{g1122t1}) 
can be summarized as
%%%%%%%%%%%%%%%%%%%%%%%%%%%%%%%%%%%%%%%%%%%%%%%%%%%%%%%%%%%%%%%
\bea
\left( 
\begin{array}{c}
G_{12} \nonu \\
G_{21}
 \end{array}
\right)(z)  \left(
\begin{array}{c}
{\bf U^{(\frac{3}{2})}} \nonu \\
{\bf V^{(\frac{3}{2})}} 
\end{array}
\right)(w)
& = & 
\frac{1}{(z-w)^2}   \frac{8i}{(5+k)} A_{\pm} (w)
\nonu \\
&+& \frac{1}{(z-w)} \left[ 
% \frac{4i}{(5+k)} \pa A_{\pm} 
\frac{1}{2} \pa \, \mbox{(pole-2)}
+  \left(
\begin{array}{c}
{\bf U_{-}^{(2)}}  \\
{\bf V_{+}^{(2)}}
 \end{array}
\right)
\right](w) +\cdots.
\label{g1221uv3half}
\eea
Recall that the $A_{\pm}(w)$ appearing in the right hand side 
are obtained from the equation $(2.13)$ of \cite{Ahn1311}.
The second order pole of (\ref{g1221uv3half}) has different 
structure constant compared to the ones in the nonlinear version
and depends on $(N,k)$ dependence for general $N$.
Each higher spin-$2$ current contains $28$ terms written in terms of
$Q^a(w)$ and $V^b(w)$ (and their derivatives).

Similarly the footnotes $47$ and $50$ of \cite{Ahn1311}
determine other two higher spin-$2$ currents 
and the corresponding OPEs from the higher spin-$\frac{3}{2}$ currents
obtained in (\ref{g1122t1}) 
can be described as
%%%%%%%%%%%%%%%%%%%%%%%%%%%%%%%%%%%%%%%%%%%%%%%%%%%%%%%%%%%%%%%%%
\bea
\left( 
\begin{array}{c}
G_{12} \nonu \\
G_{21}
 \end{array}
\right)(z)  \left(
\begin{array}{c}
{\bf V^{(\frac{3}{2})}} \nonu \\
{\bf U^{(\frac{3}{2})}} 
\end{array}
\right)(w)
& = & 
-\frac{1}{(z-w)^2}   \frac{2i(1+k)}{(5+k)} B_{\pm} (w)
\nonu \\
&+& \frac{1}{(z-w)} \left[ 
%- \frac{i(1+k)}{(5+k)} \pa B_{\pm} 
\frac{1}{2} \pa \, \mbox{(pole-2)}
+  \left(
\begin{array}{c}
{\bf V_{-}^{(2)}}  \\
{\bf U_{+}^{(2)}}
 \end{array}
\right)
\right](w) +\cdots.
\label{g1221vu3half}
\eea
Recall that the $B_{\pm}(w)$ appearing in the right hand side 
are obtained from the equation $(2.31)$ of \cite{Ahn1311}.
Each higher spin-$2$ current in (\ref{g1221vu3half}) 
contains $42$ terms written in terms of
$Q^a(w)$ and $V^b(w)$ (and their derivatives).
The second order pole of (\ref{g1221vu3half}) has different 
structure constant compared to the ones in the nonlinear version
and depends on $(N,k)$ dependence for general $N$ as before.

The equation $(4.48)$ of \cite{Ahn1311}
determines the remaining higher spin-$2$ current.
From the higher spin-$\frac{3}{2}$ currents determined in (\ref{g1122t1}),
one has the following OPEs
%%%%%%%%%%%%%%%%%%%%%%%%%%%%%%%%%%%%%%%%%%%%%%%%%%%%%%%%%%%%%%%%%%%%%%%%
\bea
\left( 
\begin{array}{c}
G_{11} \nonu \\
G_{22}
 \end{array}
\right)(z)  \left(
\begin{array}{c}
{\bf V^{(\frac{3}{2})}} \nonu \\
{\bf U^{(\frac{3}{2})}} 
\end{array}
\right)(w)
& = & 
\pm \frac{1}{(z-w)^3} \frac{8(1+k)}{(5+k)}
\nonu \\
& + & \frac{1}{(z-w)^2} \left[  \frac{8i}{(5+k)} A_3 -
\frac{2i(1+k)}{(5+k)} B_3 + {\bf T^{(1)}} \right](w)
\nonu \\
&+& \frac{1}{(z-w)} \left[  
%\frac{4i}{(5+k)} \pa A_3 -
%\frac{i(1+k)}{(5+k)} \pa B_3 + \frac{1}{2} \pa {\bf T^{(1)}} 
\frac{1}{2} \pa \, \mbox{(pole-2)}
\pm  {\bf W^{(2)}} \pm  T \right](w) +\cdots.
\label{g1122vu3half}
\eea
Compared to the OPEs in the nonlinear version, the fields contents 
in (\ref{g1122vu3half}) 
are the same form but the structure constants have different forms.
Note that the footnote $13$ of \cite{AK1411} describes 
the primary current $\hat{W}^{(2)}(w)$ while the higher spin-$2$ current 
$W^{(2)}(w)$ is a quasi primary field.
The higher spin-$2$ current in (\ref{g1122vu3half}) 
contains $76$ terms written in terms of
$Q^a(w)$ and $V^b(w)$ (and their derivatives).

The equation $(4.31)$ of \cite{Ahn1311} determines 
the higher spin-$\frac{5}{2}$ current. 
From the explicit expressions appearing in the OPEs in (\ref{g1221uv3half}),
one can construct the following OPEs
%%%%%%%%%%%%%%%%%%%%%%%%%%%%%%%%%%%%%%%%%%%%%%%%%%%%%%%%%%%%%%%%%%
\bea
\left(
 \begin{array}{c}
G_{12} \nonu \\
G_{21}
 \end{array}
\right)(z) \, 
\left(
\begin{array}{c}
{\bf V_{+}^{(2)}} \nonu \\
{\bf U_{-}^{(2)}}
 \end{array}
\right)(w) & = &
\frac{1}{(z-w)^2} \frac{2(3+k)}{(5+k)} \left[ 
\mp \left(
 \begin{array}{c}
G_{22} \nonu \\
G_{11}
 \end{array}
\right) + 2   \left(
\begin{array}{c}
{\bf V^{(\frac{3}{2})}} \nonu \\
{\bf U^{(\frac{3}{2})}} 
\end{array}
\right)
\right](w) 
\nonu \\
& + & \frac{1}{(z-w)} \left[  
%- \frac{2(3+k)}{3(5+k)} \pa \left(
% \begin{array}{c}
%G_{22} \nonu \\
%-G_{11}
% \end{array}
%\right) +   
% \frac{4(3+k)}{3(5+k)} \pa 
% \left(
%\begin{array}{c}
%{\bf V^{(\frac{3}{2})}} \nonu \\
%{\bf U^{(\frac{3}{2})}} 
%\end{array}
%\right)
\frac{1}{3} \pa \, \mbox{(pole-2)}
 \mp     \left(
\begin{array}{c}
{\bf V^{(\frac{5}{2})}}  \\
{\bf U^{(\frac{5}{2})}} 
\end{array}
\right) \right](w) + \cdots. 
\label{g1221vu2}
\eea
Compared to the corresponding OPEs in the nonlinear version, 
the field contents are the same but the structure constant in the second order
pole behaves differently. 
Furthermore, the second order pole contains 
the overall $k$ dependent factor. Note that the corresponding 
nonlinear version does not have this kind of behavior. There is no common 
factor in the second order pole in $(4.31)$ of \cite{Ahn1311}. 
We will see this feature in next section.
Definitely this structure constant depends on 
$(N, k)$ explicitly for general $N$.
The higher spin-$\frac{5}{2}$ currents in (\ref{g1221vu2}) 
contain $283$ and $279$ terms
written in terms of 
$Q^a(w)$ and $V^b(w)$ (and their derivatives).

Again the equations $(4.52)$ and $(4.55)$ of Part I
determine the  remaining higher spin-$\frac{5}{2}$
currents.
With the help of (\ref{g1122vu3half}), one can calculate 
the following OPEs
%%%%%%%%%%%%%%%%%%%%%%%%%%%%%%%%%%%%%%%%%%%%%%%%%%%%%%%%%%%%%%%%%%%%
\bea
\left(
\begin{array}{c}
G_{12} \nonu \\
G_{21}
 \end{array}
\right)(z) \, {\bf W^{(2)}}(w) & = & 
\frac{1}{(z-w)^2} \frac{(-3+k)}{2(5+k)} \left[  \left(
\begin{array}{c}
G_{12} \nonu \\
G_{21}
 \end{array}
\right) \mp 2 {\bf T_{\mp}^{(\frac{3}{2})}} \right](w)
\nonu \\
&+ &  \frac{1}{(z-w)} \left[ 
%\frac{ 
%(-3+k)}{6(5+k)} \pa  \left(
%\begin{array}{c}
%G_{12} \nonu \\
%G_{21}
% \end{array}
%\right) \mp \frac{(-3+k)}{3(5+k)}  {\bf T_{\mp}^{(\frac{3}{2})}}
\frac{1}{3} \pa \, \mbox{(pole-2)}  
+  {\bf W_{\mp}^{(\frac{5}{2})}} \right](w) +\cdots.
\label{g1221w2}
\eea
The higher spin-$\frac{5}{2}$ currents in (\ref{g1221w2}) 
contain $279$ and $287$ terms
written in terms of 
$Q^a(w)$ and $V^b(w)$ (and their derivatives).
One cannot compare the above OPEs and the equation $(4.52)$ of 
\cite{Ahn1311} directly because the higher spin-$2$ current 
in the nonlinear version
is a quasi primary current.
As before,  the second order pole contains 
the overall $k$ dependent factor.
Recall that the $(k-3)$ factor can be written in terms of
$(k^{+} - k^{-})=(k+1)-(N+1)=(k-N)$ for general $N$ \cite{AK1411}.
Therefore, for $k^{+} =k^{-}$, there is no second order pole in (\ref{g1221w2}).
The combination of spin-$\frac{3}{2}$ currents and higher 
spin-$\frac{3}{2}$ currents will appear in next section.

Let us describe the final higher spin-$3$ current.
The nonlinear version appears in the equation $(4.59)$ of \cite{Ahn1311}.
With the help of (\ref{g1221w2}), one can construct the following OPEs 
%%%%%%%%%%%%%%%%%%%%%%%%%%%%%%%%%%%%%%%%%%%%%%%%%%%%%%%%%%%%%%
\bea
 \left(
\begin{array}{c}
G_{12} \nonu \\
G_{21}
 \end{array}
\right)(z) \, {\bf W_{\pm}^{(\frac{5}{2})}}(w) & = & 
\mp
\frac{1}{(z-w)^3} \frac{8{\bf (-3+k)}}{3(5+k)} {\bf T^{(1)}}(w) 
\nonu \\
& + &  
\frac{1}{(z-w)^2} \left[ \frac{4(-3+k)}{3(5+k)} {\bf T^{(2)}} +
4 {\bf W^{(2)}} \right](w) \nonu \\
& + & \frac{1}{(z-w)} \left[ \frac{1}{4} \pa \, \mbox{(pole-2)} 
\mp {\bf W^{(3)}} 
% \frac{(-3+k)}{3(5+k)} \pa {\bf T^{(2)}} +
% \pa {\bf W^{(2)}}  
\right. \nonu \\
& \mp & \left. \frac{4{\bf (-3+k)}}{(17+13k)} \left(
T \, {\bf T^{(1)}} -\frac{1}{2} \pa^2 {\bf T^{(1)}} 
\right) \right](w)  +\cdots.
\label{g1221w5half}
\eea
The field contents are different compared to the nonlinear version
but the right hand side is simpler than the one in the nonlinear version.
There are no nonlinear terms in the second order pole of (\ref{g1221w5half}).
Furthermore, one regards the first order pole except the derivative term
as a quasi primary field because the last two terms is a 
quasi primary field. 
We will describe this issue further in next section.
The higher spin-$3$ current in (\ref{g1221w5half}) 
contain $1130$ terms
written in terms of 
$Q^a(w)$ and $V^b(w)$ (and their derivatives).
As explained before, the $(k-3)$ factor, emphasized by a boldface,  
appears in the 
higher spin-$1$ current dependent terms in the third order and the first order
poles.
In particular, for $k^{+}=k^{-}$, there is no nonlinear term in the OPE of
(\ref{g1221w5half}) and the coefficient of ${\bf T^{(2)}}(w)$ vanishes.

In Appendix $B$, the precise relations of the higher spin currents 
in the nonlinear version and those in the linear version are presented.

%%%%%%%%%%%%%%%%%%%%%%%%%%%%%%%%%%%%%%%%%%%%%%%%%%%%%%%%%%%%%%%%%%%%%%%%%
%%%%%%%%%%%%%%%%%%%%%%%%%%%%%%%%%%%%%%%%%%%%%%%%%%%%%%%%%%%%%%%%%%%%%%%%%%
\section{The OPEs between the $16$ currents  in the large 
${\cal N}=4$ linear superconformal algebra 
and the $16$ lowest higher spin currents}
%4%%%%%%%%%%%%%%%%%%%%%%%%%%%%%%%%%%%%%%%%%%%%%%%%%%%%%%%%%%%%%%%%%%%%%%%%%%%%%%
%%%%%%%%%%%%%%%%%%%%%%%%%%%%%%%%%%%%%%%%%%%%%%%%%%%%%%%%%%%%%%%%%%%%%%%%%%%%%%%%

In order to compare the results in \cite{BCG}, 
one should calculate the OPEs between the $16$ currents 
in the large ${\cal N}=4$ linear superconformal algebra and 
the $16$ higher spin currents found in previous section.
The complete OPEs between them are presented in Appendix $C$. 
In Appendix $D$, the fourth equation in Appendix $D.1$ 
is the definition of their 
higher spin-$\frac{3}{2}$ currents for given higher spin-$1$ current and 
the four spin-$\frac{3}{2}$ currents. 
The four spin-$\frac{3}{2}$ currents have explicit relations in 
(\ref{gsingledouble}).
From the results of Part I \cite{Ahn1311} (or (\ref{g1122t1}) and 
(\ref{g1221t1})), one has the following OPEs
\bea
\left(
\begin{array}{c}
G_{11} \nonu \\
G_{22}  \end{array}
\right)(z) \, {\bf T^{(1)}}(w)
& = & \frac{1}{(z-w)} \left[ \pm \left(
\begin{array}{c}
G_{11} \nonu \\
G_{22}  \end{array}
\right) + 2 
\left(
\begin{array}{c}
{\bf U^{(\frac{3}{2})}} \nonu \\
{\bf V^{(\frac{3}{2})}}  \end{array}
\right)
\right](w) +\cdots,
\nonu \\
\left(
\begin{array}{c}
G_{12}  \\
G_{21}  \end{array}
\right)(z) \, {\bf T^{(1)}}(w)
& = &  \frac{1}{(z-w)} \left[ \mp \left(
\begin{array}{c}
G_{12}  \\
G_{21}  \end{array}
\right) + 2 
{\bf T_{\mp}^{(\frac{3}{2})}}
\right](w) +\cdots.
\label{g11221221t1} 
\eea
With the identification of 
$V_0^{(1)}(w) = {\bf T^{(1)}}(w)$ \footnote{Strictly speaking,
the exact relation is given by $V_0^{(1)}(w) = \pm i \, {\bf T^{(1)}}(w)$.},
one can make the linear combinations from (\ref{g11221221t1})
with the help of (\ref{gsingledouble}). 
By reading off the first order poles 
in (\ref{g11221221t1}) and in the fourth equation of Appendix $D.1$,
one obtains the higher spin-$\frac{3}{2}$ currents 
in \cite{BCG} in terms of the spin-$\frac{3}{2}$ currents 
and the higher spin-$\frac{3}{2}$ currents as follows:
%%%%%%%%%%%%%%%%%%%%%%%%%%%%%%%%%%%%%%%%%%%%%%%%%%%%%%%%%%%%%%%
\bea
V_{\frac{1}{2}}^{(1),0}(z) & = & 
-\frac{i}{\sqrt{2}} \left( - G_{12} - G_{21}
- 2 {\bf T_{+}^{(\frac{3}{2})}} + 2 
{\bf T_{-}^{(\frac{3}{2})}} \right)(z), 
\nonu \\
V_{\frac{1}{2}}^{(1),1}(z) & = & 
\frac{1}{\sqrt{2}} \left(  G_{11} - G_{22}
+ 2 {\bf U^{(\frac{3}{2})}} + 2 
{\bf V^{(\frac{3}{2})}} \right)(z),
\nonu \\
V_{\frac{1}{2}}^{(1),2}(z) & = & 
 \frac{i}{\sqrt{2}} \left( G_{11} + G_{22}
+  2 {\bf U^{(\frac{3}{2})}} - 2 
{\bf V^{(\frac{3}{2})}} \right)(z), 
\nonu \\
V_{\frac{1}{2}}^{(1),3}(z) & = & - \frac{1}{\sqrt{2}}
\left(  - G_{12} + G_{21}
+ 2 {\bf T_{+}^{(\frac{3}{2})}} +  2
{\bf T_{-}^{(\frac{3}{2})}} \right)(z). 
\label{v3half}
\eea

Let us identify other higher spin currents.
The eighth equation of Appendix $D.1$ implies that
one should calculate the OPEs 
between the spin-$\frac{3}{2}$ currents and 
the higher spin-$\frac{3}{2}$ currents in (\ref{v3half}).
%%%%%%%%%%%%%%%%%%%%%%%%%%%%%%%%%%%%%%%%%%%%%%%%%%%%%%%%%%%%%%%
%\bea
%V_{\frac{1}{2}}^{(1),0}(z) & = & 
%\left( \frac{i}{\sqrt{2}} G_{12} +\frac{i}{\sqrt{2}} G_{21}
%+i \sqrt{2} {\bf T_{+}^{(\frac{3}{2})}} - i \sqrt{2} 
%{\bf T_{-}^{(\frac{3}{2})}} \right)(z) 
%\nonu \\
%V_{\frac{1}{2}}^{(1),1}(z) & = & 
%\left( \frac{1}{\sqrt{2}} G_{11} -\frac{1}{\sqrt{2}} G_{22}
%+ \sqrt{2} {\bf U^{(\frac{3}{2})}} + \sqrt{2} 
%{\bf V^{(\frac{3}{2})}} \right)(z) 
%\nonu \\
%V_{\frac{1}{2}}^{(1),2}(z) & = & 
%\left( \frac{i}{\sqrt{2}} G_{11} +\frac{i}{\sqrt{2}} G_{22}
%+ i \sqrt{2} {\bf U^{(\frac{3}{2})}} - i \sqrt{2} 
%{\bf V^{(\frac{3}{2})}} \right)(z) 
%\nonu \\
%V_{\frac{1}{2}}^{(1),3}(z) & = & 
%\left( \frac{1}{\sqrt{2}} G_{12} -\frac{1}{\sqrt{2}} G_{21}
%- \sqrt{2} {\bf T_{+}^{(\frac{3}{2})}} -  \sqrt{2} 
%{\bf T_{-}^{(\frac{3}{2})}} \right)(z). 
%\nonu 
%\eea
When the indices $(a,b)$ of the eighth equation of Appendix $D.1$
become $(a,b)=(0,1)$, then the right hand side of the 
first order pole is given by 
$-\frac{1}{2} (V_1^{(1),1} -V_1^{(1),-1})(w)$.
Furthermore the indices are given by $(a,b)=(2,3)$, then 
the first order pole is given by 
$\frac{1}{2} (V_1^{(1),1} +V_1^{(1),-1})(w)$.
On the other hand, one can express the relevant OPEs 
from the relations (\ref{v3half})
and the defining OPEs between the spin-$\frac{3}{2}$ currents 
in the large ${\cal N}=4$ linear superconformal algebra.
For the former, one has the relevant OPEs
given by (\ref{g1221uv3half}) and (\ref{g1221vu3half}).
For the latter, one has the following OPEs
\bea
\left(
\begin{array}{c}
G_{11} \nonu \\
G_{22}
 \end{array}
\right)(z) \, {\bf T_{\pm}^{(\frac{3}{2})}}(w)
& = & \frac{1}{(z-w)^2} \frac{2i(1+k)}{(5+k)} B_{\mp}(w) 
\nonu \\
& + &
\frac{1}{(z-w)} \left[ 
%\frac{i(1+k)}{(5+k)} \pa B_{\mp}
\frac{1}{2} \pa \, \mbox{(pole-2)}
-  \left(
\begin{array}{c}
{\bf U_{+}^{(2)}} \nonu \\
{\bf V_{-}^{(2)}}
 \end{array}
\right) \right](w) +
\cdots, 
\nonu \\
 \left(
\begin{array}{c}
G_{11} \nonu \\
G_{22}
 \end{array}
\right)(z) \, {\bf T_{\mp}^{(\frac{3}{2})}}(w)
& = & \frac{1}{(z-w)^2} \frac{8i}{(5+k)} A_{\pm}(w) 
\nonu \\
& + &
\frac{1}{(z-w)} \left[ 
%\frac{4i}{(5+k)} \pa A_{\pm}
\frac{1}{2} \pa \, \mbox{(pole-2)}
-  \left(
\begin{array}{c}
{\bf U_{-}^{(2)}}  \\
{\bf V_{+}^{(2)}}
 \end{array}
\right) \right](w) +
\cdots. 
\label{g1122tptm}
\eea
These are obtained from the relation (\ref{g1221t1})
and the result on the spin-$\frac{3}{2}$ currents from Part I.
Eventually, one can write down $V_1^{(1),\pm1}(z)$ as
a linear combinations of ${\bf U_{\pm}^{(2)}}(z)$ and 
${\bf V_{\pm}^{(2)}}(z)$ which will be shown later.

Similarly, 
when the indices $(a,b)$ of the eighth equation of Appendix $D.1$
become $(a,b)=(0,2)$, then the right hand side of the 
first order pole is given by 
$-\frac{1}{2} (V_1^{(1),2} -V_1^{(1),-2})(w)$.
For the indices $(a,b)=(1,3)$, then 
the first order pole is given by 
$-\frac{1}{2} (V_1^{(1),2} +V_1^{(1),-2})(w)$.
One can write down $V_1^{(1),\pm 2}(z)$ as
a linear combinations of ${\bf U_{\pm}^{(2)}}(z)$ and 
${\bf V_{\pm}^{(2)}}(z)$ as analyzed before (\ref{g1122tptm}).

Finally, 
when the indices $(a,b)$ of the eighth equation of Appendix $D.1$
become $(a,b)=(0,3)$, then the right hand side of the 
first order pole is given by 
$-\frac{1}{2} (V_1^{(1),3} -V_1^{(1),-3})(w)$.
For the indices $(a,b)=(1,2)$, then 
the first order pole is given by 
$\frac{1}{2} (V_1^{(1),3} +V_1^{(1),-3})(w)$.
For the former, the relevant OPEs are described in (\ref{g1221t3half}).
For the latter, one has the OPEs in (\ref{g1122vu3half}).
One can write down $V_1^{(1),\pm 3}(z)$ as
a linear combinations of ${\bf T^{(2)}}(z)$ and 
${\bf W^{(2)}}(z)$.
Therefore, the explicit higher spin-$2$ currents
have the following relations
%%%%%%%%%%%%%%%%%%%%%%%%%%%%%%%%%%%%%%%%%%%%%%%%%%%%%%%%%%%%%
\bea
V_1^{(1), \pm 1}(z) & = & 
2i \left( {\bf U_{\mp}^{(2)}} - 
{\bf V_{\pm}^{(2)}} \right)(z), 
\nonu \\
V_1^{(1), \pm 2}(z) & = & 
-2 \left( {\bf U_{\mp}^{(2)}} + 
{\bf V_{\pm}^{(2)}} \right)(z), 
\nonu \\
V_1^{(1), \pm 3}(z) & = & 
\pm 2i \left( {\bf T^{(2)}} \mp
{\bf W^{(2)}} \right)(z). 
\label{vspin2}
\eea

Let us move to the next higher spin-$\frac{5}{2}$ current.
When the indices $(a,i)$ of the thirteenth equation of Appendix 
$D.1$ become $(a,i )=(3,3)$, then the nonderivative term in the 
right hand side of the 
first order pole is given by 
$-\frac{1}{2} V_{\frac{3}{2}}^{(1),0}(w)$.
Then the relevant OPEs are given by (\ref{g1221w2})
and furthermore one can calculate the following OPEs
\bea
 \left(
\begin{array}{c}
G_{12} \nonu \\
G_{21}
 \end{array}
\right)(z) \, {\bf T^{(2)}}(w) & = & 
\frac{1}{(z-w)^2} \frac{3}{2} \left[  - \left(
\begin{array}{c}
G_{12} \nonu \\
G_{21}
 \end{array}
\right) \pm  2 {\bf T_{\mp}^{(\frac{3}{2})}} \right](w)
\nonu \\
&+&  \frac{1}{(z-w)} 
%\frac{1}{2} \left[ - \pa 
%\left(
% \begin{array}{c}
%G_{12} \nonu \\
%G_{21}
% \end{array}
%\right) +  2 \pa  {\bf T_{\mp}^{(\frac{3}{2})}} 
\frac{1}{3} \pa \, \mbox{(pole-2)}(w) +\cdots.
\label{g1221t2}
\eea
One realizes that one can write down $V_\frac{3}{2}^{(1),0}(z)$ as
a linear combinations of ${\bf W_{\pm}^{(\frac{5}{2})}}(z)$.

When the indices $(a,i)$ of the thirteenth equation of Appendix $D.1$
become $(a,i )=(0,3)$, then the nonderivative term in the 
right hand side of the 
first order pole is given by 
$\frac{1}{2} V_{\frac{3}{2}}^{(1),3}(w)$.
In this case also,  one can write down $V_\frac{3}{2}^{(1),3}(z)$ as
a linear combinations of ${\bf W_{\pm}^{(\frac{5}{2})}}(z)$ as analyzed before
in (\ref{g1221t2}).

When the indices $(a,i)$ of the thirteenth equation of Appendix $D.1$
become $(a,i )=(0,2)$, then the nonderivative term in the 
right hand side of the 
first order pole is given by 
$\frac{1}{2} V_{\frac{3}{2}}^{(1),2}(w)$.
%%%%%%%%%%%%%%%%%%%%%%%%%%%%%%%%%%%%%%%%%%%%%%%%%%%%%%%%%%%%%
%\bea
%V_1^{(1), +1}(z) & = & 
%2i \left( {\bf U_{-}^{(2)}} - 
%{\bf V_{+}^{(2)}} \right)(z), 
%\nonu \\
%V_1^{(1), +2}(z) & = & 
%-2 \left( {\bf U_{-}^{(2)}} + 
%{\bf V_{+}^{(2)}} \right)(z), 
%\nonu \\
%V_1^{(1), +3}(z) & = & 
%2i \left( {\bf T^{(2)}} - 
%{\bf W^{(2)}} \right)(z), 
%\nonu \\
%V_1^{(1), -1}(z) & = & 
%2i \left( {\bf U_{+}^{(2)}} - 
%{\bf V_{-}^{(2)}} \right)(z), 
%\nonu \\
%V_1^{(1), -2}(z) & = & 
%-2 \left( {\bf U_{+}^{(2)}} + 
%{\bf V_{-}^{(2)}} \right)(z), 
%\nonu \\
%V_1^{(1), -3}(z) & = & 
%-2i \left( {\bf T^{(2)}} + 
%{\bf W^{(2)}} \right)(z). 
%\nonu 
%\eea
In this case, the corresponding OPEs are given by
(\ref{g1221vu2}).
Similarly, 
 for the indices $(a,i )=(0,1)$, the nonderivative term in the 
right hand side of the 
first order pole is given by 
$\frac{1}{2} V_{\frac{3}{2}}^{(1),1}(w)$.
Then the remaining higher spin-$\frac{5}{2}$ currents
can be written in terms of the linear combinations of 
${\bf U^{(\frac{5}{2})}}(z)$ and ${\bf V^{(\frac{5}{2})}}(z)$.
Therefore one obtains the following simple relations
%%%%%%%%%%%%%%%%%%%%%%%%%%%%%%%%%%%%%%%%%%%%%%%%%%%%%%%%%%%%%%%%%%%%%
\bea
V_{\frac{3}{2}}^{(1), 0}(z) & = & 
-2 i \sqrt{2} \left( {\bf W_{+}^{(\frac{5}{2})}} + 
{\bf W_{-}^{(\frac{5}{2})}} \right)(z),  
\qquad
V_{\frac{3}{2}}^{(1), 1}(z)  =  
-2  \sqrt{2} \left( {\bf U^{(\frac{5}{2})}} - 
{\bf V^{(\frac{5}{2})}} \right)(z),  
\nonu \\
V_{\frac{3}{2}}^{(1), 2}(z) & = & 
-2 i \sqrt{2} \left( {\bf U^{(\frac{5}{2})}} + 
{\bf V^{(\frac{5}{2})}} \right)(z),  
\qquad
V_{\frac{3}{2}}^{(1), 3}(z)  =  
2  \sqrt{2} \left( {\bf W_{+}^{(\frac{5}{2})}} - 
{\bf W_{-}^{(\frac{5}{2})}} \right)(z).  
\label{v5half}
\eea

Let us consider the final higher spin-$3$ current.
According to the  OPE between the spin-$\frac{3}{2}$ currents 
and the higher spin-$\frac{5}{2}$ currents in Appendix $D$,
the first order pole with the indices $(a,b)=(0,0)$ 
provides the higher spin-$3$ current $V_2^{(1)}(z)$ 
which is a quasi primary field. 
One has the OPEs (\ref{g1221w5half}).
Then by examining the first order pole, one arrives at the final relation
%%%%%%%%%%%%%%%%%%%%%%%%%%%%%%%%%%%%%%%%%%%%%%%%%%%%%%%%%%%%%%%%%%%%
\bea
V_{2}^{(1)}(z) & = & 
4 \left[ {\bf W^{(3)}} + 
 \frac{4{\bf (-3+k)}}{(17+13k)} \left(  T \, {\bf T^{(1)}} -
\frac{1}{2} \pa^2 {\bf T^{(1)}}
\right)  \right](z).  
\label{bcgspin3}
\eea
When $k=3$, this quasi primary field becomes a primary one.
One can check that using the right hand side of (\ref{bcgspin3})
instead of using ${\bf W^{(3)}}(w)$ itself allows us to remove 
all the nonlinear terms  appearing in Appendix $C$.  
Some of nonlinear terms in Appendix $F$ will disappear 
if one uses $V_{2}^{(1)}(w)$ rather than ${\bf W^{(3)}}(w)$. 
It would be interesting to see their precise relation 
in the ${\cal N}=4$ superspace.

%%%%%%%%%%%%%%%%%%%%%%%%%%%%%%%%%%%%%%%%%%%%%%%%%%%%%%%%%%%%%%%%%%%%%%%%
%\bea
%U(z) \,  
%\frac{4{\bf (-3+k)}}{(17+13k)} \left(  T \, {\bf T^{(1)}} -\frac{1}{2} 
%\pa^2 {\bf T^{(1)}}
%\right)(w) & = & \frac{1}{(z-w)^2} \frac{4{\bf (-3+k)}}{(17+13k)}
%U \, {\bf T^{(1)}}(w) + \cdots,
%\nonu \\
%T(z) \,  
%\frac{4{\bf (-3+k)}}{(17+13k)} \left(  T \, {\bf T^{(1)}} -\frac{1}{2} \pa^2 
%{\bf T^{(1)}}
%\right)(w) & = & \frac{1}{(z-w)^4} \frac{4{\bf (-3+k)}}{(5+k)} {\bf T^{(1)}}(w)
%+\cdots.
%\nonu
%\eea

%%%%%%%%%%%%%%%%%%%%%%%%%%%%%%%%%%%%%%%%%%%%%%%%%%%%%%%%%%%%%%%%%%%%%%%%%
%%%%%%%%%%%%%%%%%%%%%%%%%%%%%%%%%%%%%%%%%%%%%%%%%%%%%%%%%%%%%%%%%%%%%%%%%%
\section{ The higher spin currents where the lowest spin is $2$ in 
the coset }
%5%%%%%%%%%%%%%%%%%%%%%%%%%%%%%%%%%%%%%%%%%%%%%%%%%%%%%%%%%%%%%%%%%%%%%%%%%%%%%%
%%%%%%%%%%%%%%%%%%%%%%%%%%%%%%%%%%%%%%%%%%%%%%%%%%%%%%%%%%%%%%%%%%%%%%%%%%%%%%%

In order to the complete OPEs between the lowest higher spin currents,
one should find out the next higher spin currents (\ref{nextlowest}) 
as in Part II \cite{Ahn1408}.

The higher spin-$1$ current has the same form as the one given in Part I 
\cite{Ahn1311} in the nonlinear
version (\ref{newspinonelinear}).
From the explicit results in (\ref{g1221vu2}), 
one can calculate the following OPEs 
%%%%%%%%%%%%%%%%%%%%%%%%%%%%%%%%%%%%%%%%%%%%%%%%%%%%%%%%%%%%%%%%%%%%
\bea
{\bf T^{(1)}}(z) \left(
\begin{array}{c}
{\bf U^{(\frac{5}{2})}} \nonu \\
{\bf V^{(\frac{5}{2})}}  \end{array}
\right)(w) 
& = & 
\mp \frac{1}{(z-w)^3} \frac{12k}{(5+k)^2}  
\left(
\begin{array}{c}
F_{11} \nonu \\
F_{22}  
 \end{array}
\right)(w) 
\nonu \\
& + & \frac{1}{(z-w)^2} \left[ 
%\frac{12k}{(5+k)^2} \pa 
%\left(
%\begin{array}{c}
%F_{11} \nonu \\
%-F_{22}  
% \end{array}
%\right) 
- \pa \, \mbox{(pole-3)} 
+ \frac{4i(6+k)}{3(5+k)^2}
\left(
\begin{array}{c}
F_{11} \nonu \\
F_{22}  
 \end{array}
\right) B_3
\right. \nonu \\
& \mp &  \frac{4(-3+k)}{3(5+k)}
 \left(
\begin{array}{c}
G_{11} \nonu \\
G_{22}  
 \end{array}
\right)
\mp \frac{8(-3+k)}{3(5+k)^2} U 
\left(
\begin{array}{c}
F_{11} \nonu \\
F_{22}  
 \end{array}
\right) \nonu \\
&+ &  \frac{4(-3+k)}{3(5+k)^3} F_{12} F_{21} 
\left(
\begin{array}{c}
F_{11} \nonu \\
F_{22}  
 \end{array}
\right) 
-\frac{4i(3+2k)}{3(5+k)^2} \left(
\begin{array}{c}
F_{11} \nonu \\
F_{22}  
 \end{array}
\right) A_3
\nonu \\
&- & \left. \frac{4i(3+2k)}{3(5+k)^2} 
\left(
\begin{array}{c}
F_{21} \nonu \\
F_{12}  
 \end{array}
\right) A_{\pm}
+\frac{4i(6+k)}{3(5+k)^2}
\left(
\begin{array}{c}
F_{12} \nonu \\
F_{21}  
 \end{array}
\right) B_{\mp} 
\right](w) \nonu \\
& + & \frac{1}{(z-w)} 
 \left[ \pm \left(
\begin{array}{c}
{\bf U^{(\frac{5}{2})}}  \\
{\bf V^{(\frac{5}{2})}}  \end{array}
\right)  + 
 \left(
\begin{array}{c}
{\bf Q^{(\frac{5}{2})}}  \\
{\bf R^{(\frac{5}{2})}}  \end{array}
\right)(w) 
\right](w)   +\cdots.
\label{t1uv5half}
\eea
Compared to the corresponding OPEs in the nonlinear version,
the second order pole does not contain the higher spin-$\frac{3}{2}$ 
currents. Of course, the extra terms due to the spin-$\frac{1}{2}$ currents 
and  the spin-$1$ current in the linear version of this paper occur in the 
above OPEs.
In the basis of (\ref{v5half}), the higher spin-$1$ current transforms 
the higher spin-$\frac{5}{2}$ current $V_{\frac{3}{2}}^{(1),1}(w)$ into
the higher spin-$\frac{5}{2}$ current $V_{\frac{3}{2}}^{(1),2}(w)$ and vice versa.

Similarly, from the explicit expression in (\ref{g1221w2})
one can calculate the following OPEs 
\bea
%%%%%%%%%%%%%%%%%%%%%%%%%%%%%%%%%%%%%%%%%%%%%%%%%%
{\bf T^{(1)}}(z) \,
{\bf W_{\pm}^{(\frac{5}{2})}}(w) 
& = & 
%%%%%%%%%%%%%%%%%%%%%%%%%%%%%%%%%%%%%%%%%%%%%%%%%%%%%%%
\mp \frac{1}{(z-w)^3} \frac{12k}{(5+k)^2}  
\left(
\begin{array}{c}
F_{21} \nonu \\
F_{12}  
 \end{array}
\right)(w) 
\nonu \\
& + & \frac{1}{(z-w)^2} \left[ 
%\frac{12k}{(5+k)^2} \pa 
%\left(
%\begin{array}{c}
%F_{21} \nonu \\
%-F_{12}  
% \end{array}
%\right) 
- \pa \, \mbox{(pole-3)} 
+  \frac{4i(6+k)}{3(5+k)^2}
\left(
\begin{array}{c}
F_{21} \nonu \\
F_{12}  
 \end{array}
\right) B_3
\right. \nonu \\
&\mp &  \frac{4(-3+k)}{3(5+k)}
 \left(
\begin{array}{c}
G_{21} \nonu \\
G_{12}  
 \end{array}
\right)
\mp \frac{8(-3+k)}{3(5+k)^2} U 
\left(
\begin{array}{c}
F_{21} \nonu \\
F_{12}  
 \end{array}
\right) \nonu \\
&- &  \frac{4(-3+k)}{3(5+k)^3} 
\left(
\begin{array}{c}
F_{21} \nonu \\
F_{12}  
 \end{array}
\right) F_{11} F_{22}
+\frac{4i(3+2k)}{3(5+k)^2} \left(
\begin{array}{c}
F_{21} \nonu \\
F_{12}  
 \end{array}
\right) A_3
\nonu \\
&- & \left. \frac{4i(3+2k)}{3(5+k)^2} 
\left(
\begin{array}{c}
F_{11} \nonu \\
F_{22}  
 \end{array}
\right) A_{\mp}
-\frac{4i(6+k)}{3(5+k)^2}
\left(
\begin{array}{c}
F_{22} \nonu \\
F_{11}  
 \end{array}
\right) B_{\mp} 
\right](w) \nonu \\
& + &  \frac{1}{(z-w)} 
 \left[ 
\pm {\bf W_{\pm}^{(\frac{5}{2})}}   +
{\bf P_{\pm}^{(\frac{5}{2})}} 
\right](w)   +\cdots.
\label{t1w5half+-}
\eea
In this case also, the higher spin-$\frac{3}{2}$ currents 
are not present in the second order pole.
In the basis of (\ref{v5half}), the higher spin-$1$ current transforms 
the higher spin-$\frac{5}{2}$ current $V_{\frac{3}{2}}^{(1),0}(w)$ into
the higher spin-$\frac{5}{2}$ current $V_{\frac{3}{2}}^{(1),3}(w)$ and vice versa.

In order to obtain the next higher spin currents, one should
calculate the spin-$\frac{3}{2}$ currents and the higher spin-$\frac{5}{2}$
currents obtained before. 
Let us take one of the spin-$\frac{3}{2}$ currents in the 
large ${\cal N}=4$ linear superconformal algebra
and the higher spin-$\frac{5}{2}$ current from (\ref{t1w5half+-}).
It turns out that
\bea
%%%%%%%%%%%%%%%%%%%%%%%%%%%%%%%%%%%%%%%%%%%%%%%%%%%%%%%%%
&& G_{21}(z) \, {\bf P_{-}^{(\frac{5}{2})}}(w) 
 =  \frac{1}{(z-w)^3} \left[  \frac{8{\bf (-3+k)}}{3(5+k)} 
{\bf T^{(1)}}-\frac{4i(18+5k)}{(5+k)^2} A_3 +
%%%%%%%%%%%%%%%%%%%%%%%%%%%%%%%%%%%%%%%%%%%%%%%%%%%%%%%%%%%%%%%%%%%%%
\frac{4ik(21+4k)}{3(5+k)^2} B_3   \right.
\nonu \\
& &-  \left. \frac{8(27+9k+2k^2)}{3(5+k)^3} F_{11} F_{22}
+ \frac{8(-3+k)(9+2k)}{3(5+k)^3} F_{12} F_{21} -\frac{12k}{(5+k)^2} U  
\right](w)
\nonu \\
& &+ \frac{1}{(z-w)^2} {\bf P^{(2)}}(w) 
+  
\frac{1}{(z-w)} \left[ \frac{1}{4} \pa {\bf P^{(2)}} +\frac{4{\bf (-3+k)}}{
(17+13k)}
\left( T {\bf T^{(1)}} -\frac{1}{2} \pa^2 {\bf T^{(1)}} \right) 
+ {\bf P^{(3)}}
\right. \nonu \\
&& -  \frac{6i(18+5k)}{(5+k)(17+13k)} \left( T A_3 -\frac{1}{2} \pa^2 
A_3 \right) + \frac{2ik(21+4k)}{(5+k)(17+13k)}  
\left( T B_3 -\frac{1}{2} \pa^2 
B_3 \right)  
\nonu \\
& & -     
\frac{18k}{(5+k)(17+13k)}  \left( T U -\frac{1}{2} \pa^2 
U \right) 
-   \frac{4(27+9k+2k^2)}{(5+k)^2(17+13k)} \left( T F_{11} F_{22} -
\frac{1}{2} \pa^2 (F_{11} F_{22}) \right)  \nonu \\
& & +  \left. 
\frac{4(-3+k)(9+2k)}{(5+k)^2(17+13k)}
\left( T F_{12} F_{21} -
\frac{1}{2} \pa^2 (F_{12} F_{21}) \right)
 \right](w) 
+  \cdots.
\label{g21p5half}
\eea
Note the appearance of the factor $(k-3)$ in the higher spin-$1$
current. The corresponding terms in the nonlinear version
appear in the above OPEs also. The last three extra expressions 
in the first order pole are quasi primary fields.
Note that the higher spin-$2$ current ${\bf P^{(2)}}(w)$
is not equal to the one in the basis of \cite{BCG}. The explicit relation 
will be described in next section.

Now one considers the following OPEs with the explicit 
results in (\ref{t1uv5half})
\bea
%%%%%%%%%%%%%%%%%%%%%%%%%%%%%%%%%%%%%%%%%%%%%%%%%%%%%%%%%%
&& \left(
\begin{array}{c}
G_{21} \nonu \\
G_{12}  
 \end{array}
\right)(z) 
 \left(
\begin{array}{c}
{\bf Q^{(\frac{5}{2})}} \nonu \\
{\bf R^{(\frac{5}{2})}}  \end{array}
\right)(w) 
 =  \frac{1}{(z-w)^3} \left[ \frac{4ik(21+4k)}{3(5+k)^2}
%%%%%%%%%%%%%%%%%%%%%%%%%%%%%%%%%%%%%%%%%%%%%%%%%%%%%%%%%%%%%%%%%%%%%%%%
B_{\mp} \mp \frac{32k(3+k)}{3(5+k)^3} \left(
\begin{array}{c}
F_{11} F_{21} \nonu \\
F_{22} F_{12} 
 \end{array}
\right)  \right](w)
\nonu \\
& &+  \frac{1}{(z-w)^2} \left[ \frac{16(3+k)}{3(5+k)} 
\left(
\begin{array}{c}
{\bf U_{+}^{(2)}} \nonu \\
{\bf V_{-}^{(2)}}
 \end{array}
\right)  \mp \frac{4(-3+k)}{3(5+k)^2} A_3 B_{\mp} 
+  \frac{4i(-3+k)}{3(5+k)^3} A_3 
\left(
\begin{array}{c}
F_{11} F_{21} \nonu \\
F_{22} F_{12} 
 \end{array}
\right) 
\right. \nonu \\
&& -  \frac{4i(33+13k)}{3(5+k)^3} B_3 
\left(
\begin{array}{c}
F_{11} F_{21} \nonu \\
F_{22} F_{12} 
 \end{array}
\right) 
\pm  \frac{8i(3+k)}{(5+k)^3} B_{\mp} F_{11} F_{22}
\mp \frac{4i(15+7k)}{3(5+k)^3} B_{\mp} F_{12} F_{21}
\nonu \\
& & \mp   \frac{2(21+5k)}{3(5+k)^2} 
\left(
\begin{array}{c}
F_{11} G_{21} \nonu \\
F_{22} G_{12} 
 \end{array}
\right) 
\mp \frac{4(9+5k)}{(5+k)^3}
\left(
\begin{array}{c}
\pa F_{11} F_{21} \nonu \\
\pa F_{22} F_{12} 
 \end{array}
\right)  
 \mp     \frac{4(39+11k)}{3(5+k)^3}
\left(
\begin{array}{c}
 F_{11} \pa F_{21} \nonu \\
 F_{22} \pa F_{12} 
 \end{array}
\right)
\nonu \\
&& \left. \pm  \frac{4(6+k)}{3(5+k)^2} 
\left(
\begin{array}{c}
F_{21} G_{11} \nonu \\
F_{12} G_{22} 
 \end{array}
\right) 
+  \frac{4ik}{(5+k)^2} U B_{\mp}
\mp \frac{4(33+13k)}{3(5+k)^3} U
\left(
\begin{array}{c}
 F_{11}  F_{21} \nonu \\
 F_{22}  F_{12} 
 \end{array}
\right) \right](w)
\nonu \\
& & +  \frac{1}{(z-w)} \left[ \frac{1}{4} \pa 
 (\mbox{pole-2}) 
+ \left(
\begin{array}{c}
 {\bf Q_{+}^{(3)}} \nonu \\
 {\bf R_{-}^{(3)}} 
 \end{array}
\right)
%\left(
%\begin{array}{c}
% \{ G_{21}  Q^{(\frac{5}{2})} \}_{-2} \nonu \\
% \{ G_{12}  R^{(\frac{5}{2})} \}_{-2} 
% \end{array}
%\right) 
 +   
\frac{2ik(21+4k)}{(5+k)(17+13k)}  \left( T B_{\mp} -\frac{1}{2} \pa^2 
B_{\mp} \right) \right. \nonu \\
& & \mp    \left.  \frac{16k(3+k)}{(5+k)^2(17+13k)} 
 \left(
\begin{array}{c}
 T F_{11} F_{21} -\frac{1}{2} \pa^2 (F_{11} F_{21})  \\
  T F_{22} F_{12} -\frac{1}{2} \pa^2 (F_{22} F_{12}) 
 \end{array}
\right) 
\right](w)   +  \cdots.
\label{g2112qr} 
\eea
Compared to the corresponding OPEs in the nonlinear version,
the nonlinear term between the higher spin current and the currents
from the large ${\cal N}=4$ linear superconformal algebra 
does not appear in the OPE.
The last expression containing the spin-$\frac{1}{2}$ currents 
in the first order pole is a quasi primary field.

Now one can consider the following OPEs where the 
spin-$\frac{3}{2}$ currents are interchanged in the left hand side of 
(\ref{g2112qr}) 
\bea
%%%%%%%%%%%%%%%%%%%%%%%%%%%%%%%%%%%%%%%%%%%%%%%%%%%%%%%%%%%%%
&& \left(
\begin{array}{c}
G_{12} \nonu \\
G_{21}  
 \end{array}
\right)(z) 
 \left(
\begin{array}{c}
{\bf Q^{(\frac{5}{2})}} \nonu \\
{\bf R^{(\frac{5}{2})}}  \end{array}
\right)(w) 
 =  \frac{1}{(z-w)^3} \left[ \frac{4i(18+5k)}{(5+k)^2}
%%%%%%%%%%%%%%%%%%%%%%%%%%%%%%%%%%%%%%%%%%%%%%%%%%%%%%%%%%%%%%%%%%%%%%%%%%%
A_{\pm} \pm \frac{16(9+k)}{(5+k)^3} \left(
\begin{array}{c}
F_{11} F_{12} \nonu \\
F_{22} F_{21} 
 \end{array}
\right)  \right](w)
\nonu \\
& & +  \frac{1}{(z-w)^2} \left[ -\frac{8(9+k)}{3(5+k)} 
\left(
\begin{array}{c}
{\bf U_{-}^{(2)}} \nonu \\
{\bf V_{+}^{(2)}}
 \end{array}
\right)  \mp \frac{4(-3+k)}{3(5+k)^2} A_{\pm} B_3 
 -  \frac{4i(57+5k)}{3(5+k)^3} A_3 
\left(
\begin{array}{c}
F_{11} F_{12} \nonu \\
F_{22} F_{21} 
 \end{array}
\right) 
\right. \nonu \\
&& -  \frac{4i(-3+k)}{3(5+k)^3} B_3 
\left(
\begin{array}{c}
F_{11} F_{12} \nonu \\
F_{22} F_{21} 
 \end{array}
\right) 
\pm  \frac{4i(9+k)}{(5+k)^3} A_{\pm} F_{11} F_{22}
\pm \frac{8i(15+k)}{3(5+k)^3} A_{\pm} F_{12} F_{21}
\nonu \\
 & & \mp   \frac{8(6+k)}{3(5+k)^2} 
\left(
\begin{array}{c}
F_{11} G_{12} \nonu \\
F_{22} G_{21} 
 \end{array}
\right) 
\pm \frac{4(21+k)}{(5+k)^3}
\left(
\begin{array}{c}
\pa F_{11} F_{12} \nonu \\
\pa F_{22} F_{21} 
 \end{array}
\right)  
 \pm     \frac{4(51+7k)}{3(5+k)^3}
\left(
\begin{array}{c}
 F_{11} \pa F_{12} \nonu \\
 F_{22} \pa F_{21} 
 \end{array}
\right)
\nonu \\
& & \left. \pm  \frac{2(15+k)}{3(5+k)^2} 
\left(
\begin{array}{c}
F_{12} G_{11} \nonu \\
F_{21} G_{22} 
 \end{array}
\right) 
-  \frac{12i}{(5+k)^2} U A_{\pm}
\mp \frac{4(57+5k)}{3(5+k)^3} U
\left(
\begin{array}{c}
 F_{11}  F_{12} \nonu \\
 F_{22}  F_{21} 
 \end{array}
\right) \right](w)
\nonu \\
 & &+  \frac{1}{(z-w)} \left[ \frac{1}{4} \pa 
(\mbox{pole-2})
+ \left(
\begin{array}{c}
 {\bf Q_{-}^{(3)}} \nonu \\
 {\bf R_{+}^{(3)}} 
 \end{array}
\right)
% \left(
%\begin{array}{c}
% \{ G_{12}  Q^{(\frac{5}{2})} \}_{-2} \nonu \\
% \{ G_{21}  R^{(\frac{5}{2})} \}_{-2} 
% \end{array}
%\right) 
 +   
\frac{6i(18+5k)}{(5+k)(17+13k)}  \left( T A_{\pm} -\frac{1}{2} \pa^2 
A_{\pm} \right) \right. \nonu \\
& & \pm   \left.  \frac{24(9+k)}{(5+k)^2(17+13k)} 
 \left(
\begin{array}{c}
 T F_{11} F_{12} -\frac{1}{2} \pa^2 (F_{11} F_{12})  \\
  T F_{22} F_{21} -\frac{1}{2} \pa^2 (F_{22} F_{21}) 
 \end{array}
\right) 
\right](w)  +  \cdots.
\label{g1221qr5half} 
\eea
As in previous OPEs, the nonlinear term in the corresponding OPEs
in the nonlinear version does not appear in the second order pole. 

For the last unknown higher spin-$3$ current,
one considers the following OPE
\bea
%%%%%%%%%%%%%%%%%%%%%%%%%%%%%%%%%%%%%%%%%%%%%%%%%%%%%%%%%%%%%%%%%
&& G_{22}(z) \, {\bf Q^{(\frac{5}{2})}}(w)  =  
%%%%%%%%%%%%%%%%%%%%%%%%%%%%%%%%%%%%%%%%%%%%%%%%%%%%%%%%%%%%%%%
\frac{1}{(z-w)^3} \left[\frac{8{\bf (-3+k)}}{3(5+k)} {\bf T^{(1)}} 
+\frac{4i(18+5k)}{(5+k)^2} A_3 +
\frac{4ik(21+4k)}{3(5+k)^2} B_3 
\right. \nonu \\
&&  +  \left. \frac{12k}{(5+k)^2} U -
\frac{8(-3+k)(9+2k)}{3(5+k)^3} F_{11} F_{22} +
\frac{8(27+9k+2k^2)}{3(5+k)^3} F_{12} F_{21} \right](w) 
\nonu \\
&& + \frac{1}{(z-w)^2} \left[ 
- {\bf P^{(2)}} -\frac{8(9+k)}{3(5+k)} {\bf T^{(2)}}
+\frac{8(9+k)}{3(5+k)} {\bf W^{(2)}}
-\frac{8(-3+k)}{3(5+k)^2} A_3 B_3
\right.
\nonu \\
& & + \frac{4i(-3+k)}{3(5+k)^3} A_3 F_{11} F_{22}
-\frac{4i(-3+k)}{3(5+k)^3} A_3 F_{12} F_{21}
-\frac{4i(-3+k)}{3(5+k)^3} B_3 F_{11} F_{22}
\nonu \\
& &- 
\frac{4i(-3+k)}{3(5+k)^3} B_3 F_{12} F_{21}
-\frac{4i(33+13k)}{3(5+k)^3} B_{-} F_{12} F_{22}
 +   \frac{4i(33+13k)}{3(5+k)^3} B_{+} F_{11} F_{21}
\nonu \\
&& -\frac{2(21+5k)}{3(5+k)^2} F_{11} G_{22}
-\frac{4(9+5k)}{(5+k)^3} \pa F_{11} F_{22}
-  \frac{4(39+11k)}{3(5+k)^3} F_{11} \pa F_{22} 
\nonu \\
&&
+\frac{2(21+5k)}{3(5+k)^2} F_{12} G_{21}
+\frac{4(9+5k)}{(5+k)^3} \pa F_{12} F_{21}
 + \frac{4(39+11k)}{3(5+k)^3} F_{12} \pa F_{21}
-\frac{4(6+k)}{3(5+k)^2} F_{21} G_{12}
\nonu \\
&& \left.
+\frac{4(6+k)}{3(5+k)^2} F_{22} G_{11}
+  \frac{8ik}{(5+k)^2} U B_3 -
\frac{4(33+13k)}{3(5+k)^3} U F_{11} F_{22}
+ \frac{4(33+13k)}{3(5+k)^3} U F_{12} F_{21}
\right](w)
\nonu \\
& & + \frac{1}{(z-w)} \left[
\frac{1}{4} \pa \{ G_{22} \, {\bf Q^{(\frac{5}{2})}} \}_{-2}
+\frac{4{\bf (-3+k)}}{(17+13k)} \left( T {\bf T^{(1)}} -
\frac{1}{2} \pa^2 {\bf T^{(1)}} \right)
+ {\bf S^{(3)}}
\right.
\nonu \\
& & +  \frac{18k}{(5+k)(17+13k)}  \left( T U -\frac{1}{2} \pa^2 U \right)
+  
\frac{6i(18+5k)}{(5+k)(17+13k)} \left( T A_3 -\frac{1}{2} \pa^2 A_3 \right)
\nonu \\
& & + \frac{2ik(21+4k)}{(5+k)(17+13k)}
\left( T B_3 -\frac{1}{2} \pa^2 B_3 \right)
 -  \frac{4(-3+k)(9+2k)}{(5+k)^2(17+13k)}
\left( T F_{11} F_{22} -\frac{1}{2} \pa^2 (F_{11} F_{22}) \right) 
\nonu \\
& &+  \left. 
\frac{4(27+9k+2k^2)}{(5+k)^2(17+13k)}
\left( T F_{12} F_{21} -\frac{1}{2} \pa^2 (F_{12} F_{21}) \right) 
\right](w) + \cdots.
\label{g22q5half}
\eea
In the first order pole, one sees the nonlinear term between the 
higher spin-$1$ current and the spin-$2$ stress tensor
from the large ${\cal N}=4$ linear superconformal algebra. 
Note that in Appendix $C$, the OPE between $G_{22}(z)$ and 
${\bf U^{(\frac{5}{2})}}(w)$ contains this kind of nonlinear term in the 
first order pole. Therefore simply adding the higher spin-$\frac{5}{2}$ current
 ${\bf U^{(\frac{5}{2})}}(w)$ to the left hand side of the OPE (\ref{g22q5half})
will eliminate the above nonlinear term because they have opposite signs.

So far, the four higher spin-$\frac{5}{2}$ currents and the six higher spin-$3$
currents are obtained.  
The remaining four higher spin-$\frac{7}{2}$ currents and one single 
higher spin-$4$ current will be present in Appendix $E$.
Contrary to the previous section, the exact relations between the
higher spin currents in (\ref{nextlowest}) and those in \cite{BCG}
are complicated because the precise relation between the 
lowest higher spin-$2$ currents is rather involved and this will further 
give rise to the complexity as 
the spin increases. 

%%%%%%%%%%%%%%%%%%%%%%%%%%%%%%%%%%%%%%%%%%%%%%%%%%%%%%%%%%%%%%%%%%%%%%%%%
%%%%%%%%%%%%%%%%%%%%%%%%%%%%%%%%%%%%%%%%%%%%%%%%%%%%%%%%%%%%%%%%%%%%%%%%%%
\section{The OPEs between the $16$ currents in the 
 large ${\cal N}=4$ linear superconformal algebra and 
the next $16$ lowest higher spin currents}
%4%%%%%%%%%%%%%%%%%%%%%%%%%%%%%%%%%%%%%%%%%%%%%%%%%%%%%%%%%%%%%%%%%%%%%%%%%%%%%%
%%%%%%%%%%%%%%%%%%%%%%%%%%%%%%%%%%%%%%%%%%%%%%%%%%%%%%%%%%%%%%%%%%%%%%%%%%%%%%%%

Because the $16$ currents of large ${\cal N}=4$ linear superconformal algebra
and the $16$ next higher spin currents are known in terms of 
the coset fields, it is straightforward to obtain 
the OPEs between them.
For example, the previous OPEs, (\ref{g21p5half}),
(\ref{g2112qr}), (\ref{g1221qr5half}), and (\ref{g22q5half})
are some of those OPEs.
The right hand side of these OPEs contain the nonlinear terms. 
On the other hands, the description of \cite{BCG}
implies that there exist linear OPEs between 
the $16$ currents  of large ${\cal N}=4$ linear superconformal algebra
and the $16$ next higher spin currents. 

Recall that the lowest higher spin-$2$ current \cite{BCG} living in the 
next $16$ higher spin currents has the same form in the linear and 
nonlinear version.  
Then it is easy to obtain their higher spin-$2$ current in terms of 
the higher spin-$2$ currents (and some composite fields) of this paper
by requiring that the higher spin-$2$ current should satisfy the first three
equations of Appendix $D.1$. 
It turns out that
one obtains 
\bea
V_0^{(2)}(z) & = & \left[ {\bf P^{(2)}} - 4 {\bf W^{(2)}} 
-\frac{4 (k-3)}{3 (k+5)} {\bf T^{(2)}}
+ \frac{(41 k^2+400 k+783)}{5 (k+5)^2} {\bf T^{(1)}} {\bf T^{(1)}}
\right. \nonu \\
& - & \frac{4 (k+3) (5 k+34)}{3 (k+5)^2} T
-\frac{4 (5 k+24)}{3 (k+5)^2} A_3 A_3
-\frac{4 i (5 k+24)}{3 (k+5)^2} \pa A_3
\nonu \\
& + & \frac{8 i (k+15)}{3 (k+5)^3} A_3 F_{11} F_{22}
+ \frac{4 i (k+9)}{(k+5)^3} A_3 F_{12} F_{21}
+ \frac{4 i (5 k+57)}{3 (k+5)^3} A_{-} F_{11} F_{12}
\nonu \\
&-& \frac{4 (5 k+24)}{3 (k+5)^2} A_{+} A_{-}
-\frac{4 \left(9 k^2+70 k+102\right)}{15 (k+5)^2} B_3 B_3
-\frac{4 i \left(9 k^2+70 k+102\right)}{15 (k+5)^2} \pa B_3
\nonu \\
& - &  \frac{8 i \left(9 k^2+80 k+177\right)}{15 (k+5)^3}
B_3 F_{11} F_{22} +
\frac{4 i \left(6 k^2+55 k+113\right)}{5 (k+5)^3} B_3 F_{12} F_{21}
\nonu \\
& - & 
\frac{8 i \left(9 k^2+65 k+132\right)}{15 (k+5)^3} B_{-} F_{12} F_{22}
-\frac{4 \left(9 k^2+70 k+102\right)}{15 (k+5)^2} B_{+} B_{-}
\nonu \\
& - & \frac{4 i \left(6 k^2+65 k+143\right)}{5 (k+5)^3} +
\frac{10 (k+3)}{3 (k+5)^2} F_{11} G_{22}
-\frac{8 \left(9 k^2+80 k+127\right)}{5 (k+5)^4} F_{11} F_{12} F_{21} F_{22}
\nonu \\
& -& \frac{2 \left(77 k^2+630 k+1401\right)}{15 (k+5)^3} \pa F_{11} F_{22}
+ \frac{2 \left(77 k^2+810 k+2301\right)}{15 (k+5)^3} F_{11} \pa F_{22}
\nonu \\
&-& \frac{4}{(k+5)^2} F_{12} G_{21}
-\frac{2 \left(77 k^2+780 k+1671\right)}{15 (k+5)^3} \pa F_{12} F_{21}
+ \frac{14 \left(11 k^2+100 k+273\right)}{15 (k+5)^3} F_{12} \pa F_{21}
\nonu \\
&+& \frac{4}{(k+5)} F_{21} G_{12} +
\frac{4 (2 k+9)}{3 (k+5)^2} F_{22} G_{11}
-\frac{12 i}{(k+5)^2} U A_3 
-\frac{4 i k}{(k+5)^2} U B_3
\nonu \\
&-& \left. \frac{4 (k+3) (5 k+34)}{3 (k+5)^3} U U 
+ \frac{16 (k-3)}{3 (k+5)^3} U F_{11} F_{22}
-\frac{12}{(k+5)^2} U F_{12} F_{21} \right](z).
\label{v02}
\eea
Therefore, one can check that the OPEs between the $16$ currents
of large ${\cal N}=4$ linear superconformal algebra and the higher
spin-$2$ current ${\bf P^{(2)}}(w)$
can be obtained from the coset fields and they will exactly produce
the OPEs between the $16$ currents and $V_0^{(2)}(w) + \cdots (= 
{\bf P^{(2)}}(w))$ in (\ref{v02}).

The four higher spin-$\frac{5}{2}$ currents 
\cite{BCG} can be written in terms of cost fields 
because the higher spin-$2$ current 
is given in (\ref{v02}) and also
further can be written in terms of other higher spin currents 
as in Appendix $E$.
Then the OPEs between the $16$ currents in the large ${\cal N}=4$
linear superconformal algebra and the four higher spin-$\frac{5}{2}$
currents, ${\bf P_{\pm}^{(\frac{5}{2})}}(w)$, ${\bf Q^{\frac{5}{2}}}(w)$
and ${\bf R^{(\frac{5}{2})}}(w)$
can be obtained from the results in \cite{BCG}
and Appendices $A$ and $C$. On the other hands,
they can be also obtained from the explicit expressions for the coset fields.  
In this case, the nontrivial thing is to identify 
the OPEs in terms of the known $16$ currents and higher spin currents 
(and their derivatives).
Again the eighth equation of Appendix $D.1$ allows us to obtain 
the six higher spin-$3$ currents, $V_1^{(2),+i}(w)$ and $V_1^{(2),-i}(w)$ 
from the previous higher 
spin-$\frac{5}{2}$ currents, $V_{\frac{1}{2}}^{(2),a}(w)$ as done in section $4$.
Then the corresponding OPEs between the above $16$ currents 
and the six higher spin-$3$ currents will be obtained by following previous 
procedure. In this way, one can  obtain the OPEs between 
the above $16$ currents 
and the single higher spin-$4$ current which is the last component of the 
$16$ next lowest higher spin currents.

%%%%%%%%%%%%%%%%%%%%%%%%%%%%%%%%%%%%%%%%%%%%%%%%%%%%%%%%%%%%%%%%%%%%%%%%%
%%%%%%%%%%%%%%%%%%%%%%%%%%%%%%%%%%%%%%%%%%%%%%%%%%%%%%%%%%%%%%%%%%%%%%%%%%
\section{The OPEs between the 
higher spin currents where the lowest spin is $1$}
%4%%%%%%%%%%%%%%%%%%%%%%%%%%%%%%%%%%%%%%%%%%%%%%%%%%%%%%%%%%%%%%%%%%%%%%%%%%%%%%
%%%%%%%%%%%%%%%%%%%%%%%%%%%%%%%%%%%%%%%%%%%%%%%%%%%%%%%%%%%%%%%%%%%%%%%%%%%%%%%%

The full OPEs for these are given in Appendix $F$.
When the OPEs 
for low spins are calculated, one follows what has been done 
in the nonlinear version. In other words, 
because the $16$ currents and the $16$ higher spin currents 
are known in terms of coset fields explicitly, 
after calculating the OPE, the right hand side of the OPE 
is written in terms of the coset fields.
Then one should rewrite them in terms of the above known 
$16$ currents and $16$ higher spin currents 
by assuming the possible composite fields with definite 
$U(1)$ charges introduced in section $2$ together with arbitrary coefficient
functions which depend on $k$. All the results in Appendix $F$ 
are done by using this procedure explicitly. 

On the other hands, one can use the results of Part I and Part II 
in the nonlinear version because the above $16$ currents and $16$ higher spin
currents in the linear version have the corresponding currents in the 
nonlinear version. 
For example, let us consider 
the OPE ${\bf T^{(1)}}(z) \, {\bf T_{\pm}^{(\frac{3}{2})}}(w)$.
The higher spin-$1$ current has the same form in the linear and nonlinear 
versions. The higher spin-$\frac{3}{2}$ current is given in the last 
expression in (\ref{t3halfnon}) where ${\bf T_{\pm}^{(\frac{3}{2})}}(w)$
can be written in terms of ${\bf T_{\pm,non}^{(\frac{3}{2})}}(w)$ plus other
six terms which depend on the spin-$\frac{1}{2}$ current, the spin-$1$ current,
the other spin-$1$ currents $\hat{A}_i(w)$ and $\hat{B}_i(w)$ 
living in the nonlinear version. Now one should calculate the OPE
between ${\bf T^{(1)}}(z)$ and  ${\bf T_{\pm,non}^{(\frac{3}{2})}}(w)$.
However, this was done in Part II already but the right hand side is written
in terms of the field  ${\bf T_{\pm,non}^{(\frac{3}{2})}}(w)$ 
in the nonlinear version.
Therefore, one should rewrite this in the linear version using 
the first 
expression in (\ref{t3halfnon}) where all the expressions are given in terms 
of the fields in the linear version.
Now one should calculate the OPE between the  ${\bf T^{(1)}}(z)$ and other
six terms mentioned above. By the definition of the higher spin-$1$ current,
the four spin-$\frac{1}{2}$ currents and the spin-$1$ current do not have any 
singular terms in the OPE with the higher spin-$1$ current. Furthermore, 
the higher spin-$1$ current does not have any singular terms with the 
above spin-$1$ currents, $\hat{A}_i(w)$ and $\hat{B}_i(w)$.
Then there is no contribution from the OPE between 
 the  ${\bf T^{(1)}}(z)$ and other
six terms. Therefore, the OPE  
 ${\bf T^{(1)}}(z) \, {\bf T_{\pm}^{(\frac{3}{2})}}(w)$
has the first order pole given in  ${\bf T_{\pm,non}^{(\frac{3}{2})}}(w)$
and this should be written in terms of the fields in the linear version
using  (\ref{t3halfnon}). Then the explicit result is given in Appendix $F$.
Some of the structures (the composite fields of spin-$\frac{7}{2}$, 
spin-$4$, spin-$\frac{9}{2}$ or spin-$5$) 
in the OPEs of Appendix $F$ are not presented but 
they can be obtained using this procedure. 

There are three cases in Appendix $F$.
The first case is that the right hand side of the OPEs do not 
have the higher spin currents but they have only the multiple product 
(including the derivatives) of the $16$ currents of large ${\cal N}=4$
linear superconformal algebra. The second case is that
the right hand side of the OPEs contains the higher spin currents
in (\ref{lowest}) or the higher spin currents (\ref{nextlowest}) as well as
the composite fields from the $16$ currents.
The third case is that the right hand side of OPEs 
possess the higher spin currents in (\ref{lowest}) and in (\ref{nextlowest})
as well as the multiple products in the above $16$ currents.

One realizes that any OPE between any higher spin current and itself 
does not contain the higher spin currents in the right hand side.
The only exception is the higher spin-$3$ current. 
Moreover, the OPEs between ${\bf U^{\frac{3}{2}}}(z) [{\bf V^{\frac{3}{2}}}(z)]$ 
and any other component
fields in the second [third] ${\cal N}=2$ multiplet do not contain any 
higher spin currents in the right hand side. 
Some the OPEs containing ${\bf T^{(1)}}(z)$ or ${\bf T_{\pm}^{\frac{3}{2}}}(z)$
do not have the higher spin currents in the right hand side from 
Appendix $F$.   
One sees that the OPEs between the second components of
${\cal N}=2$ second, third or fourth multiplets do not have the higher 
spin currents in the right hand side. 
Similarly, the OPEs between the third components of these ${\cal N}=2$ 
multiplets contain only the $16$ currents in the large ${\cal N}=4$
linear superconformal algebra. Finally, the OPEs between 
the higher spin-$1$ current and the higher spin currents of integer 
spin do not contain the higher spin currents in the right hand side.

If one use the  OPEs $V_{\frac{1}{2}}^{(1),0}(z) \, V_1^{(1),\pm 3}(w)$
and  $V_{\frac{1}{2}}^{(1),3}(z) \, V_1^{(1),\pm 3}(w)$ in different basis,
then the right hand side do not contain the higher spin currents
${\bf T_{\pm}^{(\frac{3}{2})}}(w)$
in the OPEs of ${\bf T_{\pm}^{(\frac{3}{2})}}(z) \, {\bf T^{(2)}}(w)$.
Similarly, the OPEs  $V_{\frac{1}{2}}^{(1),1}(z) \, V_1^{(1),\pm 3}(w)$
and  $V_{\frac{1}{2}}^{(1),2}(z) \, V_1^{(1),\pm 3}(w)$ remove the 
higher spin currents ${\bf U^{(\frac{3}{2})}}(w)$ and ${\bf V^{(\frac{3}{2})}}(w)$
which appear in the OPEs  ${\bf U^{(\frac{3}{2})}}(z) \, {\bf W^{(2)}}(w)$ and 
${\bf V^{(\frac{3}{2})}}(z) \, {\bf W^{(2)}}(w)$.  
Some of OPEs contain the higher spin currents 
${\bf Q_{\pm}^{(3)}}(w)$ or ${\bf R_{\pm}^{(3)}}(w)$
in the right hand side where there are no other types of higher spin 
currents. For example, the OPE between 
${\bf T^{(2)}}(z)$ and ${\bf U_{+}^{(2)}}(w)$
has only the higher spin current ${\bf Q_{+}^{(3)}}(w)$ in the first order pole.
Most of the OPEs have both the lowest higher spin currents 
and the next lowest higher spin currents as well as 
the $16$ currents in the large ${\cal N}=4$ linear superconformal algebra 
in the right hand side of the 
OPEs in general.

%%%%%%%%%%%%%%%%%%%%%%%%%%%%%%%%%%%%%%%%%%%%%%%%%%%%%%%%%%%%%%%%%%%%%%%%%%
\section{Conclusions and outlook }
%6%%%%%%%%%%%%%%%%%%%%%%%%%%%%%%%%%%%%%%%%%%%%%%%%%%%%%%%%%%%%%%%%%%%%%%%%%%%%%%
%%%%%%%%%%%%%%%%%%%%%%%%%%%%%%%%%%%%%%%%%%%%%%%%%%%%%%%%%%%%%%%%%%%%%%%%%%%%%%%%

The main result of this paper is the fact that 
the OPEs between the $16$ lowest higher spin currents and 
itself (presented in Appendix $F$)
do not have any nonlinear terms between the 
$16$ currents from the large ${\cal N}=4$ linear superconformal algebra
and the $16$ lowest higher spin currents except few cases.
For the nonlinear version, this kind of nonlinear terms occurs in the 
corresponding various OPEs. 
In particular, for the OPEs where the left hand side contains the higher 
spin-$3$ current ${\bf W^{(3)}}(w)$ there exist the nonlinear terms containing 
the higher spin-$1$ current ${\bf T^{(1)}}(w)$.
One can easily see that these nonlinear terms 
disappear once one uses the higher spin-$3$ current (quasi primary field) 
defined in 
(\ref{bcgspin3}).
Note also that the OPE $V_2^{(1)}(z) \, V_2^{(1)}(w)$ corresponding to 
the OPE  ${\bf W^{(3)}}(z) \, {\bf W^{(3)}}(w)$ does not contain the 
nonlinear term ${\bf T^{(1)} \, T^{(1)}}(w)$ in the fourth order pole because
the contribution from the extra terms in (\ref{bcgspin3})
cancel exactly.

Among the OPEs between the higher spin-$\frac{3}{2}$ currents and the 
higher spin-$\frac{5}{2}$ currents, the different kind of nonlinear terms 
occur: $T \, {\bf T^{(1)}}(w)$.
Specifically, there are four OPEs,
${\bf T_{\pm}^{(\frac{3}{2})}}(z) \, {\bf W_{\mp}^{(\frac{5}{2})}}(w)$,
${\bf U^{(\frac{3}{2})}}(z) \, {\bf V^{(\frac{5}{2})}}(w)$
and ${\bf V^{(\frac{3}{2})}}(z) \, {\bf U^{\frac{5}{2}}}(w)$ from Appendix $F$. 
Also these nonlinear terms can be removed by using the higher 
spin-$\frac{3}{2}$ currents (\ref{v3half}) and the higher spin-$\frac{5}{2}$
currents (\ref{v5half}) which are primary fields. 
In other words, the first order poles of the following OPEs do not 
contain the above nonlinear term:
$V_{\frac{1}{2}}^{(1),0}(z) \, V_{\frac{3}{2}}^{(1),0}(w)$,
$V_{\frac{1}{2}}^{(1),0}(z) \, V_{\frac{3}{2}}^{(1),3}(w)$,
$V_{\frac{1}{2}}^{(1),3}(z) \, V_{\frac{3}{2}}^{(1),0}(w)$,
$V_{\frac{1}{2}}^{(1),3}(z) \, V_{\frac{3}{2}}^{(1),3}(w)$,
$V_{\frac{1}{2}}^{(1),1}(z) \, V_{\frac{3}{2}}^{(1),1}(w)$,
$V_{\frac{1}{2}}^{(1),1}(z) \, V_{\frac{3}{2}}^{(1),2}(w)$,
$V_{\frac{1}{2}}^{(1),2}(z) \, V_{\frac{3}{2}}^{(1),1}(w)$, and
$V_{\frac{1}{2}}^{(1),2}(z) \, V_{\frac{3}{2}}^{(1),2}(w)$.
So far, the nonlinearity in the OPEs disappears if one uses 
the higher spin-$3$ current in the quasi-primary basis. It would be interesting
to obtain the precise relation between the nonlinearity and the property of
quasi-primary field (The OPE between the spin-$2$ stress tensor
and the quasi primary field has the vanishing third order pole and nontrivial 
higher order poles). 

The small ${\cal N}=4$ linear superconformal algebra can be obtained
by rescaling the four spin-$\frac{1}{2}$ currents, one spin-$1$ current and 
three spin-$1$ currents with level $k^{+}$ and taking 
$k^{+} \rightarrow \infty$ limit. For example, in the OPE
between ${\bf T^{(1)}}(z)$ and ${\bf T_{\pm}^{(\frac{3}{2})}}(w)$ given in Appendix 
$F$, the third, the sixth and the seventh terms in the right hand side 
are vanishing while 
the first, the second, the fourth and the fifth terms  
survive in this limit. 
One can analyze all the other OPEs presented in Appendix $F$ and obtains 
$136$ OPEs in an extension of small ${\cal N}=4$ linear superconformal 
algebra.
It would be interesting to see how this plays the role in \cite{GG1406}. 

$\bullet$ The higher spin currents for general $N$

In this paper, the $N$ in the coset $\frac{SU(N+2)}{SU(N)}$ 
is fixed as $N=3$. The next question one can ask is
how one can obtain the $16$ higher spin currents for generic $N$.
There are several ways to obtain the higher spin currents for general $N$.
One of them is to use the higher spin currents for generic $N$ in the 
nonlinear version. From the experience of this paper, 
there should exist the exact relations between the higher spin currents 
in the nonlinear and linear versions for generic $N$.   
Their relations for  $N=3$ are presented in Appendix $B$.
All the $k$ dependent coefficients can be generalized to the ones 
with $N$ dependence explicitly. Or one can start with the four 
spin-$\frac{3}{2}$ currents of large ${\cal N}=4$ linear superconformal algebra
and the higher spin-$1$ current fro general $N$. 
They have explicit their coset field 
realizations. Then it is straightforward to calculate the various OPEs 
and to extract the higher spin currents by hand.

$\bullet$ The ${\cal N}=4$ superspace description

For the $16$ currents of large ${\cal N}=4$ linear superconformal algebra,
one can write down them in one single ${\cal N}=4$ superfield (of superspin $0$)
and the various OPEs in Appendix $A$ can be written as a single 
OPE in ${\cal N}=4$ superspace. 
According to the observation of \cite{BCG}, the other single
${\cal N}=4$ superfield (of superspin $1$) 
representing the lowest $16$ higher spin 
currents transform as linearly under the above ${\cal N}=4$ superfield.
In other words, the Appendix $D$ can be written in terms of 
the OPE between the ${\cal N}=4$ supercurrent with superspin $0$
and the ${\cal N}=4$ supercurrent with superspin $1$ and the right hand side
of this OPE depends on the ${\cal N}=4$ supercurrent with superspin $1$ 
linearly.  
It would be an open problem to rewrite Appendix $F$ 
in terms of one single ${\cal N}=4$ super OPE.
The nontrivial thing is to express the nonlinear terms between the
above ${\cal N} =4$ super current of superspin zero which will be complicated. 
The linear terms from the ${\cal N}=4$ higher spin currents of 
superspin $1$ or $2$ will be rather simple form. 
Furthermore, 
in ${\cal N}=2$ superspace, because the ${\cal N}=4$ multiplet corresponds to
four independent ${\cal N}=2$ multiplets as in (\ref{lowest}), 
one has the ten ${\cal N}=2$
super OPEs. It would be also interesting to write down the
above single ${\cal N}=4$ super OPE in terms of
ten ${\cal N}=2$ super OPEs.  

$\bullet$ The higher spin currents in the linear version 
in the orthogonal Wolf space

In  \cite{AP1410}, the higher spin currents construction was given for the 
different type of ${\cal N}=4$ super coset theory.
It would be interesting to describe the linear version of the higher spin 
currents.

\vspace{.7cm}

%%%%%%%%%%%%%%%%%%%%%%%%%%%%%%%%%%%%%%%%%%%%%%%%%%%%%%%%%%%%%%
%%%%%%%%%%%%%%%%%%%%%%%%%%%%%%%%%%%%%%%%%%%%%%%%%%%%%%%%%%%%%%%
\centerline{\bf Acknowledgments}
%%%%%%%%%%%%%%%%%%%%%%%%%%%%%%%%%%%%%%%%%%%%%%%%%%%%%%%%%%%%%%%
%%%%%%%%%%%%%%%%%%%%%%%%%%%%%%%%%%%%%%%%%%%%%%%%%%%%%%%%%%%%%%%

We would like to thank H. Kim and M.H. Kim for discussions. 
%and R. Gopakumar and H. Kim for discussions. 
%CA would like to thank his previous collaborators K. Schoutens and 
%A. Sevrin for earlier works.
This work was supported by the Mid-career Researcher Program through
the National Research Foundation of Korea (NRF) grant 
funded by the Korean government (MEST) 
(No. 2012-045385/2013-056327/2014-051185).
%CA would like to thank the participants of the focus program of 
%Asia Pacific Center
%for Theoretical Physics (APCTP) 
%on
%``Liouville, Integrability and Branes (10) Focus Program at Asia-Pacific
%Center for Theoretical Physics'',
%Sept. 03-14, 2014 for their feedbacks.
%We thank the Galileo Galilei Institute for Theoretical Physics 
%for the hospitality and the INFN for partial support during 
%the completion of this work. 
%CA appreciates APCTP for its hospitality during completion of this work.
CA acknowledges warm hospitality from 
the School of  Liberal Arts (and Institute of Convergence Fundamental
Studies), Seoul National University of Science and Technology.

\newpage

\appendix

\renewcommand{\thesection}{\large \bf \mbox{Appendix~}\Alph{section}}
\renewcommand{\theequation}{\Alph{section}\mbox{.}\arabic{equation}}

%%%%%%%%%%%%%%%%%%%%%%%%%%%%%%%%%%%%%%%%%%%%%%%%%%%%%%%%%%%%%%%%%%%%%
%%%%%%%%%%%%%%%%%%%%%%%%%%%%%%%%%%%%%%%%%%%%%%%%%%%%%%%%%%%%%%%%%%%%%
\section{ The  large ${\cal N}=4$ linear superconformal 
algebra}
%A%%%%%%%%%%%%%%%%%%%%%%%%%%%%%%%%%%%%%%%%%%%%%%%%%%%%%%%%%%%%%%%%%%%
%%%%%%%%%%%%%%%%%%%%%%%%%%%%%%%%%%%%%%%%%%%%%%%%%%%%%%%%%%%%%%%%%%%%

The large ${\cal N}=4$ linear superconformal algebra (generated by
spin-$2$ current $T(z)$, four spin-$\frac{3}{2}$ currents $G_a(z)$, 
three spin-$1$ currents $A^{+i}(z)$, three spin-$1$ currents 
$A^{- i}(z)$, spin-$1$ current $U(z)$ and four spin-$\frac{1}{2}$
currents $\Gamma_a(z)$)
can be summarized by \cite{npb1988} 
\bea
T(z) \, T(w) & = & \frac{1}{(z-w)^4} \, \frac{c}{2} +\frac{1}{(z-w)^2}
\, 2 T(w) + \frac{1}{(z-w)} \, \pa T(w) + \cdots,
\nonu \\
T(z) \, \phi(w) & = & \frac{1}{(z-w)^2}
\, h_{\phi} \, \phi(w) + \frac{1}{(z-w)} \, \pa \phi(w) + \cdots,
\nonu \\
G_a(z) \, G_b(w)  & = & \frac{1}{(z-w)^3} \, \frac{4c}{3} \, \delta_{ab}
- \frac{1}{(z-w)^2} \, 16 \left[ \gamma \, \alpha_{ab}^{+ i} \, A_i^{+}
+ (1-\gamma) \alpha_{ab}^{-i} \, A_i^{-} \right](w)
\nonu \\
& + & 
\frac{1}{(z-w)} \, \left[ \frac{1}{2} \pa (\mbox{pole-2}) +  4 \delta_{ab} \,
 T \right](w)
+\cdots,
\nonu \\
A^{+i}(z) \, G_a(w) & = & - \frac{1}{(z-w)^2} \, 2(1-\gamma)  
\,  \alpha_{a}^{+i b} \, 
\Gamma_b(w) + \frac{1}{(z-w)} \,  \alpha_{a}^{+i b} \, 
G_b(w) + \cdots,
\nonu \\
A^{-i}(z) \, G_a(w) & = &  \frac{1}{(z-w)^2} \, 2 \gamma  
\,  \alpha_{a}^{-i b} \, 
\Gamma_b(w) + \frac{1}{(z-w)} \,  \alpha_{a}^{-i b} \, 
G_b(w) + \cdots,
\nonu \\
A^{+i}(z) \, A^{+j}(w) & = & - \frac{1}{(z-w)^2} \, \frac{1}{2} k^{+} 
\delta^{ij}  +\frac{1}{(z-w)} \, \ep^{ijk} A_{k}^{+}(w) +\cdots,
\nonu \\
A^{-i}(z) \, A^{-j}(w) & = & - \frac{1}{(z-w)^2} \, \frac{1}{2} k^{-} 
\delta^{ij}  +\frac{1}{(z-w)} \, \ep^{ijk} A_{k}^{-}(w) +\cdots,
\nonu \\
\Gamma_a(z) \, G_b(w) & = & \frac{1}{(z-w)} \, 
2 \left[ 2 ( \alpha_{ab}^{+i} A_i^{+}
- \alpha_{ab}^{-i} A_i^{-}) + \delta_{ab} \, U  \right](w) 
+\cdots,
\nonu \\
A^{+i}(z) \, \Gamma_a(w) & = & \frac{1}{(z-w)} \,  \alpha_{a}^{+i b} \, 
\Gamma_b(w)+ \cdots,
\nonu \\
A^{-i}(z) \, \Gamma_a(w) & = & \frac{1}{(z-w)} \,  \alpha_{a}^{-i b} \, 
\Gamma_b(w)+ \cdots,
\nonu \\
U(z) \, G_a(w) & = & \frac{1}{(z-w)^2} \, \Gamma_a(w) + \cdots,
\nonu \\
\Gamma_a(z) \, \Gamma_b(w) & = & -\frac{1}{(z-w)} \,
 \frac{c}{ 6 \gamma(1-\gamma)} \, \delta_{ab} + \cdots, 
\nonu \\
U(z) \, U(w)  & = & -\frac{1}{(z-w)^2} \, \frac{c}{12 \gamma(1-\gamma)} 
+ \cdots. 
\nonu
\eea
Here the conformal dimension is as follows:
$h_{G_a} = \frac{3}{2}$, $h_{A^{\pm i}}=1$, $h_{U}=1$ and $h_{\Gamma_a}=\frac{1}{2}$.
The numerical values for the quantities in the right hand side 
are given in \cite{npb1988}. The indices $a, b$ are 
bispinor notations. One has the following letters where 
$A^{+i}(z) \equiv A_i(z)$, $A^{-i}(z) \equiv B_i(z)$ and $\Gamma_a(z) \equiv 
F_a(z)$.
The coset field realization representing this algebra for $N=3$ 
is already given in Part I \cite{Ahn1311}.

%%%%%%%%%%%%%%%%%%%%%%%%%%%%%%%%%%%%%%%%%%%%%%%%%%%%%%%%%%%%%%%%%%%%%
%%%%%%%%%%%%%%%%%%%%%%%%%%%%%%%%%%%%%%%%%%%%%%%%%%%%%%%%%%%%%%%%%%%%%
\section{ The  $16$ lowest 
higher spin currents in the nonlinear version in terms of those in 
linear version}
%A%%%%%%%%%%%%%%%%%%%%%%%%%%%%%%%%%%%%%%%%%%%%%%%%%%%%%%%%%%%%%%%%%%%
%%%%%%%%%%%%%%%%%%%%%%%%%%%%%%%%%%%%%%%%%%%%%%%%%%%%%%%%%%%%%%%%%%%%

In this Appendix, the explicit relations between the $16$ lowest higher
spin currents in the nonlinear version and the linear version are given.
For the lowest higher spin-$1$ current, the nonlinear version 
in \cite{Ahn1311} and the linear version of this paper
are the same each other
\bea
{\bf T_{non}^{(1)}}(z) = {\bf T^{(1)}}(z).
\label{newspinonelinear}
\eea

In (\ref{g1221t1}), the higher spin-$\frac{3}{2}$ currents are 
determined. The corresponding currents in the 
nonlinear version can be obtained from the equations 
$(4.9)$ and $(4.13)$ of \cite{Ahn1311}.
The spin-$\frac{3}{2}$ currents of large ${\cal N}=4$ superconformal 
algebra are related to each other in equation $(3.13)$ of \cite{Ahn1311}.
Then one can substitute the right hand side of $(3.13)$ of Part I into 
 the equations 
$(4.9)$ and $(4.13)$ of Part I. The nontrivial OPEs come from 
only the spin-$\frac{3}{2}$ currents because
there are no singular terms between $(U, F_a, \hat{A}_i, \hat{B}_{i})(z)$
and ${\bf T^{(1)}}(w)$.
This implies that the left hand sides of the OPEs in 
the nonlinear version and the 
linear version are the same and one has the following relations
\bea
\mp \left(
\begin{array}{c}
\hat{G}_{12}  \\
\hat{G}_{21}  \end{array}
\right)(z) + 2 
{\bf T_{\mp,non}^{(\frac{3}{2})}}(z) =
\mp \left(
\begin{array}{c}
G_{12}  \\
G_{21}  \end{array}
\right)(z) + 2 
{\bf T_{\mp}^{(\frac{3}{2})}}(z).
\nonu
\eea
Or the higher spin-$\frac{3}{2}$ currents can be written as 
\bea
{\bf T_{\mp,non}^{(\frac{3}{2})}}(z) = {\bf T_{\mp}^{(\frac{3}{2})}}(z)
\pm \frac{1}{2} \left(
\begin{array}{c}
\hat{G}_{12}  \\
\hat{G}_{21}  \end{array}
\right)(z)
\mp \frac{1}{2} \left(
\begin{array}{c}
G_{12}  \\
G_{21}  \end{array}
\right)(z).
\nonu 
\eea
Therefore, the extra terms between the higher spin-$\frac{3}{2}$ currents 
in the nonlinear and linear versions are 
exactly those from the spin-$\frac{3}{2}$
currents.
We arrive at the following results 
%%%%%%%%%%%%%%%%%%%%%%%%%%%%%%%%%%%%%%%%%%%%%%%%%%%%%%%%%%%%%%%%%%%%%%
\bea
{\bf T_{\pm,non}^{(\frac{3}{2})}}(z) & = & 
\left[ {\bf T_{\pm}^{(\frac{3}{2})}} \mp \frac{1}{(5+k)} U  
\left(
\begin{array}{c}
F_{21} \\
F_{12}
\end{array}
\right)+
\frac{4}{(5+k)^2} 
\left(
\begin{array}{c}
F_{21} \\
F_{12}
\end{array}
\right)
 F_{11} F_{22} +
\frac{i}{(5+k)}   
\left(
\begin{array}{c}
F_{21} \\
F_{12}
\end{array}
\right)
A_3 \right. 
\nonu \\
&- & \left. \frac{i}{(5+k)} 
\left(
\begin{array}{c}
F_{11} \\
F_{22}
\end{array}
\right)
  A_{\mp} +
\frac{i}{(5+k)} 
\left(
\begin{array}{c}
F_{22} \\
F_{11}
\end{array}
\right)
  B_{\mp} -
\frac{i}{(5+k)} 
\left(
\begin{array}{c}
F_{21} \\
F_{12}
\end{array}
\right)
 B_3 \right](z) 
\nonu \\
& = & 
\left[ {\bf T_{\pm}^{(\frac{3}{2})}} \mp \frac{1}{(5+k)} U  
\left(
\begin{array}{c}
F_{21} \\
F_{12}
\end{array}
\right)-
\frac{2}{(5+k)^2} 
\left(
\begin{array}{c}
F_{21} \\
F_{12}
\end{array}
\right)
 F_{11} F_{22} +
\frac{i}{(5+k)}   
\left(
\begin{array}{c}
F_{21} \\
F_{12}
\end{array}
\right)
\hat{A}_3 \right. 
\nonu \\
&- & \left. \frac{i}{(5+k)} 
\left(
\begin{array}{c}
F_{11} \\
F_{22}
\end{array}
\right)
  \hat{A}_{\mp} +
\frac{i}{(5+k)} 
\left(
\begin{array}{c}
F_{22} \\
F_{11}
\end{array}
\right)
  \hat{B}_{\mp} -
\frac{i}{(5+k)} 
\left(
\begin{array}{c}
F_{21} \\
F_{12}
\end{array}
\right)
 \hat{B}_3 \right](z). 
\label{t3halfnon}
%& = &
% \left[ {\bf T_{+}^{(\frac{3}{2})}} -\frac{1}{(5+k)} U \, F_{21} -
%\frac{2}{(5+k)^2} F_{21}\, F_{11} \,F_{22} +
%\frac{i}{(5+k)} F_{21} \, \hat{A}_3 \right. 
%\nonu \\
%&- & \left. \frac{i}{(5+k)} F_{11} \, \hat{A}_{-} +
%\frac{i}{(5+k)} F_{22} \, \hat{B}_{-} -
%\frac{i}{(5+k)} F_{21} \, \hat{B}_3 \right](z), 
%\nonu \\
%{\bf T_{-,non}^{(\frac{3}{2})}}(z) & = & 
%\left[ {\bf T_{-}^{(\frac{3}{2})}} +\frac{1}{(5+k)} U \, F_{12} +
%\frac{4}{(5+k)^2} F_{12}\, F_{11} \,F_{22} +
%\frac{i}{(5+k)} F_{12} \, A_3 \right.
%\nonu \\
%&- & \left. \frac{i}{(5+k)} F_{22} \, A_{+} +
%\frac{i}{(5+k)} F_{11} \, B_{+} -
%\frac{i}{(5+k)} F_{12} \, B_3 \right](z) 
%\nonu \\
% & = & 
%\left[ {\bf T_{-}^{(\frac{3}{2})}} +\frac{1}{(5+k)} U \, F_{12} -
%\frac{2}{(5+k)^2} F_{12}\, F_{11} \,F_{22} +
%\frac{i}{(5+k)} F_{12} \, \hat{A}_3 \right.
%\nonu \\
%&- & \left. \frac{i}{(5+k)} F_{22} \, \hat{A}_{+} +
%\frac{i}{(5+k)} F_{11} \, \hat{B}_{+} -
%\frac{i}{(5+k)} F_{12} \, \hat{B}_3 \right](z). 
%\nonu 
\eea
In the first expression of (\ref{t3halfnon}), 
the extra term in the 
right hand side is written in terms of 
the fields in the linear version
and in the second expression of (\ref{t3halfnon}), 
the extra term in the right hand side 
is given by the fields in the nonlinear version using the equations 
$(3.5)$, $(3.6)$ and $(3.10)$ of \cite{Ahn1311}.

Similarly, one has 
the following relations from the equation (\ref{g1122t1}) together with
$(4.20)$ and $(4.34)$ of \cite{Ahn1311} 
\bea
\left(
\begin{array}{c}
{\bf U_{non}^{(\frac{3}{2})}} \\
{\bf V_{non}^{(\frac{3}{2})}}
\end{array}
\right)  
(z) = \left(
\begin{array}{c}
{\bf U^{(\frac{3}{2})}} \\
{\bf V^{(\frac{3}{2})}}
\end{array}
\right)  
(z)
\pm \frac{1}{2} \left(
\begin{array}{c}
\hat{G}_{11}  \\
\hat{G}_{22}  \end{array}
\right)(z)
\mp \frac{1}{2} \left(
\begin{array}{c}
G_{11}  \\
G_{22}  \end{array}
\right)(z).
\label{uv3halfnon}
\eea
In other words, one has, as done before,
\bea
\left(
\begin{array}{c}
{\bf U_{non}^{(\frac{3}{2})}} \\
{\bf V_{non}^{(\frac{3}{2})}} 
\end{array}
\right)
(z) & = & 
\left[ \left(
\begin{array}{c}
{\bf U^{(\frac{3}{2})}} \\
{\bf V^{(\frac{3}{2})}} 
\end{array}
\right) \mp \frac{1}{(5+k)} U  
\left(
\begin{array}{c}
F_{11} \\
F_{22}
\end{array}
\right)-
\frac{4}{(5+k)^2} 
\left(
\begin{array}{c}
F_{11} \\
F_{22}
\end{array}
\right)
 F_{12} F_{21} -
\frac{i}{(5+k)}   
\left(
\begin{array}{c}
F_{11} \\
F_{22}
\end{array}
\right)
A_3 \right. 
\nonu \\
&- & \left. \frac{i}{(5+k)} 
\left(
\begin{array}{c}
F_{21} \\
F_{12}
\end{array}
\right)
  A_{\pm} -
\frac{i}{(5+k)} 
\left(
\begin{array}{c}
F_{12} \\
F_{21}
\end{array}
\right)
  B_{\mp} -
\frac{i}{(5+k)} 
\left(
\begin{array}{c}
F_{11} \\
F_{22}
\end{array}
\right)
 B_3 \right](z) 
\nonu \\
& = & 
\left[  \left(
\begin{array}{c}
{\bf U^{(\frac{3}{2})}} \\
{\bf V^{(\frac{3}{2})}} 
\end{array}
\right) \mp \frac{1}{(5+k)} U  
\left(
\begin{array}{c}
F_{11} \\
F_{22}
\end{array}
\right)+
\frac{2}{(5+k)^2} 
\left(
\begin{array}{c}
F_{11} \\
F_{22}
\end{array}
\right)
 F_{12} F_{21} -
\frac{i}{(5+k)}   
\left(
\begin{array}{c}
F_{11} \\
F_{22}
\end{array}
\right)
\hat{A}_3 \right. 
\nonu \\
&- & \left. \frac{i}{(5+k)} 
\left(
\begin{array}{c}
F_{21} \\
F_{12}
\end{array}
\right)
  \hat{A}_{\mp} -
\frac{i}{(5+k)} 
\left(
\begin{array}{c}
F_{12} \\
F_{21}
\end{array}
\right)
  \hat{B}_{\mp} -
\frac{i}{(5+k)} 
\left(
\begin{array}{c}
F_{11} \\
F_{22}
\end{array}
\right)
 \hat{B}_3 \right](z). 
\label{uv3halfnon1}
%& = & 
%\left[ {\bf U^{(\frac{3}{2})}} -\frac{1}{(5+k)} U \, F_{11} -
%\frac{4}{(5+k)^2} F_{12}\, F_{21} \,F_{11} -
%\frac{i}{(5+k)} F_{11} \, A_3 \right.
%\nonu \\
%&- & \left. \frac{i}{(5+k)} F_{21} \, A_{+} -
%\frac{i}{(5+k)} F_{12} \, B_{-} -
%\frac{i}{(5+k)} F_{11} \, B_3 \right](z) 
%\nonu \\
% & = & 
%\left[ {\bf U^{(\frac{3}{2})}} -\frac{1}{(5+k)} U \, F_{11} +
%\frac{2}{(5+k)^2} F_{12}\, F_{21} \,F_{11} -
%\frac{i}{(5+k)} F_{11} \, \hat{A}_3 \right.
%\nonu \\
%&- & \left. \frac{i}{(5+k)} F_{21} \, \hat{A}_{+} -
%\frac{i}{(5+k)} F_{12} \, \hat{B}_{-} -
%\frac{i}{(5+k)} F_{11} \, \hat{B}_3 \right](z), 
%\nonu \\
%{\bf V_{non}^{(\frac{3}{2})}}(z) & = & 
%\left[ {\bf V^{(\frac{3}{2})}} +\frac{1}{(5+k)} U \, F_{22} +
%\frac{4}{(5+k)^2} F_{21}\, F_{12} \,F_{22} -
%\frac{i}{(5+k)} F_{22} \, A_3 \right.
%\nonu \\
%&- & \left. \frac{i}{(5+k)} F_{12} \, A_{-} -
%\frac{i}{(5+k)} F_{21} \, B_{+} -
%\frac{i}{(5+k)} F_{22} \, B_3 \right](z)
%\nonu \\
% & = & 
%\left[ {\bf V^{(\frac{3}{2})}} +\frac{1}{(5+k)} U \, F_{22} -
%\frac{2}{(5+k)^2} F_{21}\, F_{12} \,F_{22} -
%\frac{i}{(5+k)} F_{22} \, \hat{A}_3 \right.
%\nonu \\
%&- & \left. \frac{i}{(5+k)} F_{12} \, \hat{A}_{-} -
%\frac{i}{(5+k)} F_{21} \, \hat{B}_{+} -
%\frac{i}{(5+k)} F_{22} \, \hat{B}_3 \right](z). 
%\nonu
\eea
One can use the second relation of (\ref{uv3halfnon1})
to express the higher spin-$\frac{3}{2}$ currents in the linear version 
in terms of the fields in the nonlinear version.

For the higher spin-$2$ currents, one 
has the equation (\ref{g1221vu3half}) and 
the corresponding footnotes $47$ and $50$ of \cite{Ahn1311}.
Now one can substitute the spin-$\frac{3}{2}$ currents 
in the nonlinear version  using $(3.13)$ of \cite{Ahn1311}
and  the relation (\ref{uv3halfnon}) into those equations.
Then one can identify the following relations as follows:
%%%%%%%%%%%%%%%%%%%%%%%%%%%%%%%%%%%%%%%%%%%%%%%%%%%%%%%%%%%%%%%%%%%%%%%%%
\bea
\left(
\begin{array}{c}
{\bf U_{+,non}^{(2)}} \\
{\bf V_{-,non}^{(2)}}
\end{array}
\right) 
(z) & = &  \left[ \left(
\begin{array}{c}
{\bf U_{+}^{(2)}} \\
{\bf V_{-}^{(2)}}
\end{array}
\right) 
+\frac{2}{(5+k)} 
\left(
\begin{array}{c}
F_{11}  \\
F_{22} 
\end{array}
\right) {\bf T_{\pm}^{(\frac{3}{2})}}
-  \frac{2}{(5+k)} 
\left(
\begin{array}{c}
F_{21} {\bf U^{(\frac{3}{2})}}  \\
F_{12} {\bf V^{(\frac{3}{2})}} 
\end{array}
\right) 
\right.
\nonu \\
& \mp &
\frac{2}{(5+k)} A_3 \, B_{\mp}-\frac{4i}{(5+k)^2} A_3  
\left(
\begin{array}{c}
F_{11}  F_{21} \\
F_{22} F_{12}
\end{array}
\right)
\pm \frac{2i}{(5+k)^2} B_{\mp}  F_{11}  F_{22}
\nonu \\
&\pm & \frac{2i}{(5+k)^2} B_{\mp} \, F_{12} \, F_{21}
\pm \frac{1}{(5+k)} 
\left(
\begin{array}{c}
F_{11}  G_{21} \\
F_{22} G_{12}
\end{array}
\right)
\nonu \\
& \pm & \left. \frac{2}{(5+k)^2} 
\left(
\begin{array}{c}
\pa F_{11}  F_{21} \\
\pa F_{22} F_{12}
\end{array}
\right)
\mp \frac{2}{(5+k)^2} 
\left(
\begin{array}{c}
F_{11} \pa F_{21} \\
F_{22} \pa F_{12}
\end{array}
\right)
\mp
\frac{1}{(5+k)} 
\left(
\begin{array}{c}
F_{21}  G_{11} \\
F_{12} G_{22}
\end{array}
\right)
 \right](z)
\nonu \\
 & = &  \left[ 
 \left(
\begin{array}{c}
{\bf U_{+}^{(2)}} \\
{\bf V_{-}^{(2)}}
\end{array}
\right) 
+\frac{2}{(5+k)} 
\left(
\begin{array}{c}
F_{11} \\
F_{22}
\end{array}
\right)
{\bf T_{\pm,non}^{(\frac{3}{2})}}
-  \frac{2}{(5+k)} 
\left(
\begin{array}{c}
F_{21}  {\bf U_{non}^{(\frac{3}{2})}} \\
F_{12}  {\bf V_{non}^{(\frac{3}{2})}}
\end{array}
\right) \right.
\nonu \\
& \mp & \left.
\frac{2}{(5+k)} \hat{A}_3  \hat{B}_{\mp}
\pm  
\frac{1}{(5+k)} 
\left(
\begin{array}{c}
F_{11}  \hat{G}_{21} \\
F_{22} \hat{G}_{12}
\end{array}
\right) 
\mp \frac{1}{(5+k)} 
\left(
\begin{array}{c}
F_{21}  \hat{G}_{11} \\
F_{12} \hat{G}_{22}
\end{array}
\right) \right](z).
\label{uvplusminusnon}
\eea
In the second equation of (\ref{uvplusminusnon}),
the extra terms are written in terms of the fields 
in the nonlinear version.
Note the appearance of the nonlinear terms between the higher spin currents 
and the spin-$\frac{1}{2}$ currents.

Similarly, from the equation (\ref{g1221uv3half})
and the footnotes $48$ and $49$ of \cite{Ahn1311},
one can obtain the following relations 
\bea
\left(
\begin{array}{c}
{\bf U_{-,non}^{(2)}} \\
{\bf V_{+,non}^{(2)}}
\end{array}
\right) 
(z) 
& = &  \left[ \left(
\begin{array}{c}
{\bf U_{-}^{(2)}} \\
{\bf V_{+}^{(2)}}
\end{array}
\right) 
-\frac{2}{(5+k)} 
\left(
\begin{array}{c}
F_{11}  \\
F_{22} 
\end{array}
\right) {\bf T_{\pm}^{(\frac{3}{2})}}
+  \frac{2}{(5+k)} 
\left(
\begin{array}{c}
F_{12} {\bf U^{(\frac{3}{2})}}  \\
F_{21} {\bf V^{(\frac{3}{2})}} 
\end{array}
\right) 
\right.
\nonu \\
& \mp &
\frac{2}{(5+k)} A_{\pm} \, B_{3} \mp \frac{2i}{(5+k)^2} A_{\pm}  
F_{11} F_{22}
+ \frac{2i}{(5+k)^2} A_{\pm}  F_{12}  F_{21}
\nonu \\
&+ & \frac{4i}{(5+k)^2} B_{3}  
\left(
\begin{array}{c}
F_{11}  F_{21} \\
F_{22} F_{12}
\end{array}
\right)
\pm \frac{1}{(5+k)} 
\left(
\begin{array}{c}
F_{11}  G_{12} \\
F_{22} G_{21}
\end{array}
\right)
\nonu \\
& \pm & \left. \frac{2}{(5+k)^2} 
\left(
\begin{array}{c}
\pa F_{11}  F_{12} \\
\pa F_{22} F_{21}
\end{array}
\right)
\mp \frac{2}{(5+k)^2} 
\left(
\begin{array}{c}
F_{11} \pa F_{12} \\
F_{22} \pa F_{21}
\end{array}
\right)
\pm
\frac{1}{(5+k)} 
\left(
\begin{array}{c}
F_{12}  G_{11} \\
F_{21} G_{22}
\end{array}
\right)
 \right](z)
\nonu \\
 & = &  \left[ 
 \left(
\begin{array}{c}
{\bf U_{-}^{(2)}} \\
{\bf V_{+}^{(2)}}
\end{array}
\right) 
-\frac{2}{(5+k)} 
\left(
\begin{array}{c}
F_{11} \\
F_{22}
\end{array}
\right)
{\bf T_{\mp,non}^{(\frac{3}{2})}}
+  \frac{2}{(5+k)} 
\left(
\begin{array}{c}
F_{12}  {\bf U_{non}^{(\frac{3}{2})}} \\
F_{21}  {\bf V_{non}^{(\frac{3}{2})}}
\end{array}
\right) \right.
\nonu \\
& \mp & \left.
\frac{2}{(5+k)} \hat{A}_{\pm}  \hat{B}_{3}
\pm  
\frac{1}{(5+k)} 
\left(
\begin{array}{c}
F_{11}  \hat{G}_{12} \\
F_{22} \hat{G}_{21}
\end{array}
\right) 
\pm \frac{1}{(5+k)} 
\left(
\begin{array}{c}
F_{12}  \hat{G}_{11} \\
F_{21} \hat{G}_{22}
\end{array}
\right) \right](z).
\label{uvminusplusnon}
%{\bf U_{-,non}^{(2)}}(z) & = & \left[ {\bf U_{-}^{(2)}}
%-\frac{2}{(5+k)} F_{11} {\bf T_{-}^{(\frac{3}{2})}}
%+  \frac{2}{(5+k)} F_{12} \, {\bf U^{(\frac{3}{2})}} \right.
%\nonu \\
%& - & \left.
%\frac{2}{(5+k)} \hat{A}_{+} \, \hat{B}_{3}
%+\frac{1}{(5+k)} F_{11} \, \hat{G}_{12}
%+\frac{1}{(5+k)} F_{12} \, \hat{G}_{11} \right](z)
%\nonu \\
% & = & \left[ {\bf U_{-}^{(2)}}
%-\frac{2}{(5+k)} F_{11} {\bf T_{-,non}^{(\frac{3}{2})}}
%+  \frac{2}{(5+k)} F_{12} \, {\bf U_{non}^{(\frac{3}{2})}} \right.
%\nonu \\
%& - & \left.
%\frac{2}{(5+k)} \hat{A}_{+} \, \hat{B}_{3}
%+  
%\frac{1}{(5+k)} F_{11} \, \hat{G}_{12}
%+\frac{1}{(5+k)} F_{12} \, \hat{G}_{11} \right](z),
%\nonu \\
%{\bf V_{+,non}^{(2)}}(z) & = & \left[ {\bf V_{+}^{(2)}}
%-\frac{2}{(5+k)} F_{22} {\bf T_{+}^{(\frac{3}{2})}}
%+  \frac{2}{(5+k)} F_{21} \, {\bf V^{(\frac{3}{2})}} \right.
%\nonu \\
%& + &
%\frac{2}{(5+k)} A_{-} \, B_{3}-\frac{2i}{(5+k)^2} A_{-} \, F_{22} \, F_{11}
%+\frac{2i}{(5+k)^2} A_{-} \, F_{21} \, F_{12}
%\nonu \\
%&+ & \frac{4i}{(5+k)^2} B_{3} \, F_{22} \, F_{21} 
%-\frac{1}{(5+k)} F_{22} \, G_{21}
%\nonu \\
%& - & \left. \frac{2}{(5+k)^2} \pa F_{22} \, F_{21}
%+\frac{2}{(5+k)^2} F_{22} \pa F_{21}
%-\frac{1}{(5+k)} F_{21} \, G_{22} \right](z)
%\nonu \\
% & = & \left[ {\bf V_{+}^{(2)}}
%-\frac{2}{(5+k)} F_{22} {\bf T_{+,non}^{(\frac{3}{2})}}
%+  \frac{2}{(5+k)} F_{21} \, {\bf V_{non}^{(\frac{3}{2})}} \right.
%\nonu \\
%& + & \left.
%\frac{2}{(5+k)} \hat{A}_{-} \, \hat{B}_{3}
%-\frac{1}{(5+k)} F_{22} \, \hat{G}_{21}
%-\frac{1}{(5+k)} F_{21} \, \hat{G}_{22} \right](z),
%\nonu 
\eea
In this case, the second equation of (\ref{uvminusplusnon})
contains the extra terms written in terms of the fields in the 
nonlinear version.
As before, one sees 
the appearance of the nonlinear terms between the higher spin currents 
and the spin-$\frac{1}{2}$ currents.

The equation $(4.16)$ of \cite{Ahn1311}
and the equation (\ref{g1221t3half})
determine the explicit relation between the higher spin-$2$ current
in the nonlinear version and in the linear version 
\bea
{\bf T_{non}^{(2)}}(z) & = & 
\left[ {\bf T^{(2)}} 
-\frac{2}{(5+k)} F_{11}  {\bf V^{(\frac{3}{2})}}
+\frac{2}{(5+k)} F_{22}  {\bf U^{(\frac{3}{2})}} 
+\frac{(1+k)(21+k)}{(5+k)^2(3+7k)} \pa F_{12}  F_{21}
\right.
\nonu \\
& + & \frac{(3+k)}{(3+7k)} T +\frac{1}{(5+k)} A_3  A_3
-\frac{2}{(5+k)} A_3  B_3 +\frac{i}{(5+k)} \pa A_3
-  \frac{4i}{(5+k)^2} A_3  F_{11}  F_{22}
\nonu \\
& - & \frac{2i}{(5+k)^2} A_{-}  F_{11}  F_{12}
+\frac{1}{(5+k)} A_{+}  A_{-}
-  \frac{2i}{(5+k)^2} A_{+}  F_{21}  F_{22}
+\frac{1}{(5+k)} B_3  B_3 +\frac{i}{(5+k)} \pa B_3
\nonu \\
&+& \frac{2i}{(5+k)^2} B_{-}  F_{12}  F_{22}
+\frac{1}{(5+k)} B_{+}  B_{-}
+\frac{2i}{(5+k)^2} B_{+}  F_{11}  F_{21}
+\frac{1}{(5+k)} F_{22}  G_{11}
\nonu \\
&+& \frac{1}{(5+k)} F_{11} G_{22}
+\frac{(27+36k+k^2)}{(5+k)^2(3+7k)} \pa F_{11}  F_{22}
+\frac{4i}{(5+k)^2} B_3 F_{11} F_{22}
\nonu \\
&- & \left.
\frac{(27+36k+k^2)}{(5+k)^2(3+7k)} F_{11}  \pa F_{22}
-
\frac{(1+k)(21+k)}{(5+k)^2(3+7k)}  F_{12}  \pa F_{21}
+\frac{(3+k)}{(5+k)(3+7k)} U U \right](z)
\nonu \\
 & = & 
\left[ {\bf T^{(2)}} 
-\frac{2}{(5+k)} F_{11}  {\bf V_{non}^{(\frac{3}{2})}}
+\frac{2}{(5+k)} F_{22}  {\bf U_{non}^{(\frac{3}{2})}} 
+\frac{i}{(5+k)} \pa \hat{B}_3
+ \frac{1}{(5+k)} \hat{B}_{+}  \hat{B}_{-}
\right.
\nonu \\
& + & \frac{(3+k)}{(3+7k)} \hat{T} +\frac{1}{(5+k)} \hat{A}_3  \hat{A}_3
-\frac{2}{(5+k)} \hat{A}_3  \hat{B}_3 +\frac{i}{(5+k)} \pa \hat{A}_3
\nonu \\
& + & \left.
\frac{1}{(5+k)} \hat{A}_{+}  \hat{A}_{-} 
+\frac{1}{(5+k)} \hat{B}_3  \hat{B}_3 
+ \frac{1}{(5+k)} F_{11} \hat{G}_{22}
+\frac{1}{(5+k)} F_{22}  \hat{G}_{11}
\right](z).
\label{t2non}
\eea
As explained before, one 
can express the higher spin-$2$ current in the linear version in terms of
the fields in the nonlinear version using the second equation of 
(\ref{t2non}).
The nonlinear terms between the higher spin currents 
and the spin-$\frac{1}{2}$ currents occur in the right hand side.

Let us continue to consider the last higher spin-$2$ current.
With the help of (\ref{g1122vu3half}) and the equation $(4.48)$ of 
\cite{Ahn1311}, one can express the following relation between the 
higher spin-$2$ current in the nonlinear version and in the linear version 
\bea
{\bf W_{non}^{(2)}}(z) & = & 
\left[ {\bf W^{(2)}} 
+\frac{2}{(5+k)} F_{12}  {\bf T_{+}^{(\frac{3}{2})}}
-\frac{2}{(5+k)} F_{21}  {\bf T_{-}^{(\frac{3}{2})}} 
+\frac{1}{(5+k)} B_3  B_3 +\frac{i}{(5+k)} \pa B_3
\right.
\nonu \\
& + &  T +\frac{1}{(5+k)} A_3  A_3
+\frac{2}{(5+k)} A_3  B_3 +\frac{i}{(5+k)} \pa A_3
-  \frac{4i}{(5+k)^2} A_3  F_{12}  F_{21}
\nonu \\
& - & \frac{2i}{(5+k)^2} A_{-}  F_{11}  F_{12}
+\frac{1}{(5+k)} A_{+}  A_{-}
-  \frac{2i}{(5+k)^2} A_{+}  F_{21}  F_{22}
\nonu \\
&+& \frac{2i}{(5+k)^2} B_{-}  F_{12}  F_{22}
+\frac{1}{(5+k)} B_{+}  B_{-}
+\frac{2i}{(5+k)^2} B_{+}  F_{11}  F_{21}
+ \frac{1}{(5+k)} F_{12} G_{21}
\nonu \\
& + & \frac{(7+k)}{(5+k)^2} \pa F_{11}  F_{22}
-\frac{4i}{(5+k)^2} B_3  F_{12} F_{21}
-  
\frac{(7+k)}{(5+k)^2} F_{11}  \pa F_{22}
+\frac{(9+k)}{(5+k)^2} \pa F_{12}  F_{21}
\nonu \\
& + & \left. \frac{1}{(5+k)} F_{21}  G_{12}
- 
\frac{(9+k)}{(5+k)^2}  F_{12}  \pa F_{21}
+\frac{1}{(5+k)} U U \right](z)
\nonu \\
 & = & 
\left[ {\bf W^{(2)}} 
+\frac{2}{(5+k)} F_{12}  {\bf T_{+,non}^{(\frac{3}{2})}}
-\frac{2}{(5+k)} F_{21}  {\bf T_{-,non}^{(\frac{3}{2})}}
+\frac{1}{(5+k)} F_{21}  \hat{G}_{12}
 \right.
\nonu \\
& + &  \hat{T} +\frac{1}{(5+k)} \hat{A}_3  \hat{A}_3
+\frac{2}{(5+k)} \hat{A}_3  \hat{B}_3 +\frac{i}{(5+k)} \pa \hat{A}_3
+ \frac{1}{(5+k)} F_{12} \hat{G}_{21}
\nonu \\
& + & \left.
\frac{1}{(5+k)} \hat{A}_{+}  \hat{A}_{-}
+\frac{1}{(5+k)} \hat{B}_3  \hat{B}_3 +\frac{i}{(5+k)} \pa \hat{B}_3
+\frac{1}{(5+k)} \hat{B}_{+}  \hat{B}_{-}
 \right](z).
\label{w2non}
\eea
Furthermore, the second equation of (\ref{w2non}) 
allows us to write down the higher spin-$2$ current in the linear version 
in terms of the fields 
in the nonlinear version. 
One observes 
the appearance of the nonlinear terms between the higher spin currents 
and the spin-$\frac{1}{2}$ currents in the right hand side.

From the relation in (\ref{g1221vu2}) and the explicit result for the 
corresponding higher spin currents in Part I, the following relation holds 
%%%%%%%%%%%%%%%%%%%%%%%%%%%%%%%%%%%%%%%%%%%%%%%%%%%%%%%%%%%%%%%
\bea
{\bf U_{non}^{(\frac{5}{2})}}(z) & = & 
\left[ {\bf U^{(\frac{5}{2})}} 
-\frac{2i}{(5+k)} A_3  {\bf U^{(\frac{3}{2})}} 
+ \frac{2i}{(5+k)} B_3  {\bf U^{(\frac{3}{2})}}
+ \frac{2i}{(5+k)} B_{-}  {\bf T_{-}^{(\frac{3}{2})}}
 +
\frac{4}{(5+k)^2} F_{11}  F_{12}  {\bf T_{+}^{(\frac{3}{2})}}
 \right.
\nonu \\
& + & \frac{2}{(5+k)} F_{11}  {\bf W^{(2)}}
-  \frac{8}{(5+k)^2} F_{11}  F_{21}  {\bf T_{-}^{(\frac{3}{2})}}
-\frac{4}{(5+k)^2} F_{11}  F_{22}  {\bf U^{(\frac{3}{2})}}
-\frac{2}{(5+k)} \pa F_{11}  {\bf T^{(1)}}
\nonu \\
&+ & \frac{1}{(5+k)} F_{11}  \pa {\bf T^{(1)}}
-\frac{2}{(5+k)} F_{12}  {\bf U_{+}^{(2)}}
+\frac{4}{(5+k)^2} F_{12}  F_{21}   {\bf U^{(\frac{3}{2})}}
+ \frac{2}{(5+k)} U   {\bf U^{(\frac{3}{2})}}
\nonu \\
& - & \frac{2}{(5+k)} F_{21}   {\bf U_{-}^{(2)}}
+\frac{4}{3(5+k)} \pa  {\bf U^{(\frac{3}{2})}}
+
\frac{4}{(5+k)^2} F_{11}  F_{21}  G_{12}
\nonu \\
&- & \frac{i}{(5+k)} A_3  G_{11}
+
\frac{4}{(5+k)^2} A_3  B_3  F_{11}
+  \frac{4}{(5+k)^2} B_3  B_3  F_{11}
-\frac{8i}{3(5+k)^2} \pa B_3  F_{11}
\nonu \\
& - & \frac{8i}{3(5+k)^2} \pa A_3  F_{11} 
-\frac{4i}{3(5+k)^2} A_3  \pa F_{11}(z)
-\frac{8i}{(5+k)^3} A_3  F_{11}  F_{12}  F_{21}
+  \frac{i}{(5+k)} A_{+}  G_{21}
\nonu \\
& + & \frac{2}{(5+k)^2} A_{+}  A_3  F_{21}
-\frac{2}{(5+k)^2} A_{+}  A_{-}  F_{11}
+ \frac{2}{(5+k)^2} A_{+}  B_3  F_{21}
+\frac{2}{(5+k)^2} A_{+}  B_{-}  F_{22}
\nonu \\
& + & \frac{4i}{(5+k)^3} A_{+}  F_{11} F_{21}  F_{22}
-  \frac{4i}{3(5+k)^2} \pa A_{+}  F_{21}
-\frac{4i}{3(5+k)^2} A_{+}  \pa F_{21}
-\frac{i}{(5+k)} B_3  G_{11}
\nonu \\
&- & \frac{4i}{3(5+k)^2} B_3  \pa F_{11}
-\frac{16i}{(5+k)^3} B_3  F_{11} F_{12} F_{21}
-\frac{i}{(5+k)} B_{-}  G_{12}
+ 
\frac{4}{(5+k)^2} B_{-}  B_3  F_{12}
\nonu \\
& - & \frac{8i}{(5+k)^3} B_{-}  F_{11}  F_{12}  F_{22}
+  \frac{8i}{3(5+k)^2} \pa B_{-}  F_{12}
-  \frac{4i}{3(5+k)^2} B_{-}  \pa F_{12}
+ \frac{4}{(5+k)^2} F_{11}  F_{12}  G_{21}
\nonu \\
& - & \frac{4}{3(5+k)^3} \pa F_{11}  F_{12}  F_{21}
+  \frac{8}{3(5+k)^3} F_{11} \pa F_{12} F_{21}
-\frac{4}{3(5+k)^3} F_{11}  F_{12}  \pa F_{21}
\nonu \\
& + & \frac{1}{(5+k)} U  G_{11}
+ \frac{2i}{(5+k)^2} U  A_{+}  F_{21}
-\frac{4i}{(5+k)^2} U  B_3  F_{11}
\nonu \\
& - & \left. \frac{4}{3(5+k)^2} U  \pa F_{11}
-\frac{4}{3(5+k)^2} F_{11}  G_{22}  F_{11}
-\frac{4}{(5+k)^3} F_{11}  \pa F_{11}  F_{22} \right](z)
\nonu \\
 & = & 
\left[ {\bf U^{(\frac{5}{2})}} 
-\frac{2i}{(5+k)} \hat{A}_3  {\bf U_{non}^{(\frac{3}{2})}} 
+ \frac{2i}{(5+k)} \hat{B}_3  {\bf U_{non}^{(\frac{3}{2})}}
+ \frac{2i}{(5+k)} \hat{B}_{-}  {\bf T_{-,non}^{(\frac{3}{2})}} 
-\frac{2}{(5+k)} \pa F_{11}  {\bf T^{(1)}}
\right.
\nonu \\
& + & \frac{2}{(5+k)} F_{11}  {\bf W_{non}^{(2)}}
 -
\frac{4}{(5+k)^2} F_{11}  F_{12}  {\bf T_{+,non}^{(\frac{3}{2})}}
+  \frac{4}{(5+k)^2} F_{11}  F_{21}  {\bf T_{-,non}^{(\frac{3}{2})}}
+ \frac{2}{(5+k)} U   {\bf U_{non}^{(\frac{3}{2})}}
\nonu \\
&+ & \frac{1}{(5+k)} F_{11}  \pa {\bf T^{(1)}}
-\frac{2}{(5+k)} F_{12}  {\bf U_{+,non}^{(2)}}
-\frac{4}{(5+k)^2} F_{12}  F_{21}   {\bf U_{non}^{(\frac{3}{2})}}
+\frac{4}{3(5+k)} \pa  {\bf U_{non}^{(\frac{3}{2})}}
\nonu \\
& - & \frac{2}{(5+k)} F_{21}   {\bf U_{-,non}^{(2)}}
- 
\frac{i}{(5+k)} \hat{B}_3  \hat{G}_{11}
-  \frac{2}{(5+k)^2} \hat{B}_3  \hat{B}_3  F_{11}
-\frac{2i}{(5+k)^2} \pa \hat{B}_3  F_{11}
\nonu \\
&- & \frac{i}{(5+k)} \hat{A}_3  \hat{G}_{11}
-
\frac{2}{(5+k)^2} \hat{A}_3  \hat{A}_3  F_{11}
-
\frac{4}{(5+k)^2} \hat{A}_3 \hat{B}_3  F_{11}
-
\frac{4}{(5+k)^2} \hat{A}_3  \hat{B}_{-}  F_{12}
\nonu \\
& - & \frac{2i}{(5+k)^2} \pa \hat{A}_3  F_{11} 
+  \frac{i}{(5+k)} \hat{A}_{+}  \hat{G}_{21}
-\frac{2}{(5+k)^2} \hat{A}_{+}  \hat{A}_{-}  F_{11}
- \frac{4}{(5+k)^2} \hat{A}_{+}  \hat{B}_3  F_{21}
\nonu \\
& - & \frac{2}{(5+k)^2} \hat{B}_{+} \, \hat{B}_{-} \, F_{11}
-
\frac{i}{(5+k)} \hat{B}_{-}  \hat{G}_{12}
- \frac{2}{(5+k)^2} F_{11}  F_{12}  \hat{G}_{21}
-\frac{2}{(5+k)} \hat{T}  F_{11}
\nonu \\
&- & \left.
\frac{2}{(5+k)^2} F_{11}  F_{21}  \hat{G}_{12}
+
\frac{4}{(5+k)^2} F_{12}  F_{21}  \hat{G}_{11}
+  \frac{1}{(5+k)} U  \hat{G}_{11}
 \right](z), \label{u5halfnon}
 \\
{\bf V_{non}^{(\frac{5}{2})}}(z) & = & 
\left[ {\bf V^{(\frac{5}{2})}} 
+\frac{2i}{(5+k)} A_3  {\bf V^{(\frac{3}{2})}} 
+\frac{2i}{(5+k)} A_{-}  {\bf T_{-}^{(\frac{3}{2})}}
- \frac{2i}{(5+k)} B_3  {\bf V^{(\frac{3}{2})}} \right.
+  \frac{2}{(5+k)} F_{22}  {\bf W^{(2)}}
\nonu \\
& - & \frac{8}{(5+k)^2} F_{22}  F_{21}  {\bf T_{-}^{(\frac{3}{2})}}
-\frac{2}{(5+k)} U  {\bf V^{(\frac{3}{2})}}
+ 
\frac{4}{(5+k)^2} F_{22}  F_{12}   {\bf T_{+}^{(\frac{3}{2})}}
-
\frac{4}{(5+k)^2} F_{22}  F_{11}   {\bf V^{(\frac{3}{2})}}
\nonu \\
&+ & \frac{2}{(5+k)} \pa F_{22}  {\bf T^{(1)}}
-\frac{1}{(5+k)} F_{22}  \pa {\bf T^{(1)}}
+\frac{2}{(5+k)}  F_{21}  {\bf V_{-}^{(2)}}
+\frac{4}{3(5+k)} \pa  {\bf V^{(\frac{3}{2})}}
\nonu \\
&- & \frac{4}{(5+k)^2} F_{21}  F_{12}  {\bf V^{(\frac{3}{2})}}
+\frac{2}{(5+k)} F_{12}  {\bf V_{+}^{(2)}}
\nonu \\
& + & \frac{i}{(5+k)} A_3  G_{22}
+
\frac{4}{(5+k)^2} A_3  A_3  F_{22}
+  \frac{16i}{(5+k)^3} A_{3}  F_{22}  F_{21}   F_{12}
-\frac{i}{(5+k)} A_{-}   G_{12}
\nonu \\
& + & \frac{4}{(5+k)^2}  A_3  B_3 F_{22} 
+\frac{2}{(5+k)^2} A_3  B_{+}  F_{21}
-\frac{4i}{3(5+k)^2} A_3   \pa F_{22}
+ \frac{4}{(5+k)^2} A_{-}  A_3  F_{12}
\nonu \\
& + & \frac{2}{(5+k)^2} A_{-}  B_{+}  F_{11}
-\frac{8i}{(5+k)^3} A_{-}  F_{22} F_{12}  F_{11}
+  \frac{8i}{3(5+k)^2} \pa A_{-}  F_{12}
-\frac{4i}{3(5+k)^2} A_{-}  \pa F_{12}
\nonu \\
& + & 
\frac{4i}{(5+k)^2} \pa B_3   F_{22}
-\frac{4i}{3(5+k)^2} B_3  \pa F_{22}
-  \frac{2}{(5+k)^2}  B_{-}  B_{+}  F_{22}
+\frac{4}{(5+k)^2} F_{22}  F_{21}  G_{12}
\nonu \\
&+ & \frac{8i}{(5+k)^3} B_3   F_{22}  F_{21}  F_{12}
-\frac{i}{(5+k)} B_{+}  G_{21}
+\frac{2}{(5+k)^2} B_{+}  B_3  F_{21}
+  \frac{i}{(5+k)} B_3  G_{22}
\nonu \\
&+& \frac{4i}{(5+k)^3} B_{+}  F_{22}  F_{21}  F_{11}
-\frac{4i}{3(5+k)^2} \pa B_{+}   F_{21}
-\frac{4i}{3(5+k)^2} B_{+}  \pa  F_{21}
+\frac{4}{3(5+k)^2} U  \pa F_{22}
\nonu \\
& + &
\frac{4}{3(5+k)^3} \pa F_{22}  F_{21}  F_{12}
+\frac{4}{3(5+k)^3}  F_{22}  \pa F_{21}  F_{12}
-  \frac{8}{3(5+k)^3} F_{22}  F_{21} \pa F_{12}
\nonu \\
& + &  \frac{1}{(5+k)} U   G_{22}
+\frac{4i}{(5+k)^2} U  A_3  F_{22}
-\frac{2i}{(5+k)^2} U   B_{+}  F_{21}
+\frac{8}{3(5+k)^2} \pa U   F_{22}
\nonu \\
& - & \left. \frac{4}{3(5+k)^2} F_{22}   G_{11}  F_{22}
-\frac{4}{(5+k)^3} F_{22}  \pa F_{22}  F_{11} \right](z)
\nonu \\
 & = & 
\left[ {\bf V^{(\frac{5}{2})}} 
+\frac{2i}{(5+k)} \hat{A}_3  {\bf V_{non}^{(\frac{3}{2})}} 
+\frac{2i}{(5+k)} \hat{A}_{-}  {\bf T_{-,non}^{(\frac{3}{2})}}
- \frac{2i}{(5+k)} \hat{B}_3  {\bf V_{non}^{(\frac{3}{2})}} 
+\frac{4}{3(5+k)} \pa  {\bf V_{non}^{(\frac{3}{2})}}
\right.
\nonu \\
&+ & \frac{2}{(5+k)} F_{22}  {\bf W_{non}^{(2)}}
+\frac{4}{(5+k)^2} F_{22}  F_{21}  {\bf T_{-,non}^{(\frac{3}{2})}}
-\frac{2}{(5+k)} U  {\bf V_{non}^{(\frac{3}{2})}}
+\frac{2}{(5+k)} F_{12}  {\bf V_{+,non}^{(2)}}
\nonu \\
& - & 
\frac{4}{(5+k)^2} F_{22}  F_{12}   {\bf T_{+,non}^{(\frac{3}{2})}}
+  \frac{2}{(5+k)} \pa F_{22}  {\bf T^{(1)}}
-\frac{1}{(5+k)} F_{22}  \pa {\bf T^{(1)}}
+\frac{2}{(5+k)}  F_{21}  {\bf V_{-,non}^{(2)}}
\nonu \\
&+ & \frac{4}{(5+k)^2} F_{21}  F_{12}  {\bf V_{non}^{(\frac{3}{2})}}
-  
\frac{2}{(5+k)^2}  \hat{B}_3  \hat{B}_3   F_{22}+
\frac{2i}{(5+k)^2} \pa \hat{B}_3   F_{22}
-  
\frac{i}{(5+k)} \hat{B}_{+}  \hat{G}_{21}
\nonu \\
& + & \frac{i}{(5+k)} \hat{A}_3  \hat{G}_{22}
-
\frac{2}{(5+k)^2} \hat{A}_3  \hat{A}_3  F_{22}
-  \frac{4}{(5+k)^2}  \hat{A}_3  \hat{B}_3 F_{22} 
-\frac{2}{(5+k)^2} F_{22}  F_{12}  \hat{G}_{21}
\nonu \\
& - & \frac{4}{(5+k)^2} \hat{A}_3  \hat{B}_{+}  F_{21}
+\frac{2i}{(5+k)^2} \pa \hat{A}_3    F_{22}
- 
\frac{i}{(5+k)} \hat{A}_{-}   \hat{G}_{12}
-\frac{2}{(5+k)^2} F_{22}  F_{21}  \hat{G}_{12}
\nonu \\
&-&
\frac{2}{(5+k)^2} \hat{A}_{-}  \hat{A}_{+}  F_{22}
-\frac{4}{(5+k)^3} \hat{A}_{-}  \hat{B}_{3}  F_{12}
+ 
\frac{i}{(5+k)} \hat{B}_3  \hat{G}_{22}
-  \frac{2}{(5+k)^2}  \hat{B}_{-}  \hat{B}_{+}  F_{22}
\nonu \\
& - &  \left. \frac{2}{(5+k)^2} F_{21}  F_{12}  \hat{G}_{22} 
-  \frac{2}{(5+k)} \hat{T} F_{22}
+  \frac{1}{(5+k)} U   \hat{G}_{22}
 \right](z).
\label{v5halfnon} 
\eea
As before, the extra terms except the higher spin currents in the 
linear version are written in terms of composite fields living in the 
nonlinear version.
Note the appearance of the nonlinear terms (containing the higher spin
currents) between the higher spin currents 
and the spin-$\frac{1}{2}$ currents (the spin-$1$ currents).

For the remaining higher spin-$\frac{5}{2}$ currents, 
one has the following relations which obtained from the relation 
(\ref{g1221w2}) and some results from Part I 
\bea
{\bf W_{+,non}^{(\frac{5}{2})}}(z) & = &
\left[ {\bf W_{+}^{(\frac{5}{2})}}
-\frac{2i}{(5+k)} A_3  {\bf T_{+}^{(\frac{3}{2})}}  
+ \frac{2i}{(5+k)} B_3  {\bf T_{+}^{(\frac{3}{2})}}
+\frac{2}{(5+k)} F_{11}  {\bf V_{+}^{(2)}} 
+\frac{2}{(5+k)} U  {\bf T_{+}^{(\frac{3}{2})}}
\right.
\nonu \\
& + & 
\frac{4}{(5+k)^2} F_{11}  F_{21}  {\bf V^{(\frac{3}{2})}}
- \frac{8}{(5+k)^2} F_{11}  F_{22}  {\bf T_{+}^{(\frac{3}{2})}}
+\frac{2}{(5+k)} F_{21}  {\bf T^{(2)}}
+  \frac{2}{(5+k)} F_{22} {\bf U_{+}^{(2)}}
\nonu \\
& + & \frac{4}{(5+k)^2} F_{21}  F_{22}  {\bf U^{(\frac{3}{2})}}
-\frac{2}{(5+k)} \pa F_{21}  {\bf T^{(1)}}
+ \frac{1}{(5+k)} F_{21}  \pa {\bf T^{(1)}}
\nonu \\
& - & \frac{3i}{(5+k)} A_3  G_{21}
-\frac{4}{(5+k)^2} A_3  A_3  F_{21}
+ \frac{2}{(5+k)^2} B_{+}  B_{-}  F_{21}
- \frac{2}{(5+k)^2} A_3  B_{-}  F_{22}
\nonu \\
& - & \frac{8i}{(5+k)^3} A_3  F_{11}  F_{21}  F_{22}
-\frac{4i}{3(5+k)^2} \pa A_3  F_{21}
+ \frac{8i}{3(5+k)^2} A_3  \pa F_{21}
+\frac{i}{(5+k)} A_{-}  G_{11} 
\nonu \\
& + &
\frac{2}{(5+k)^2} A_{-}  A_3  F_{11}
+ \frac{2}{(5+k)^2} A_{-}  B_3  F_{11}
+\frac{4i}{3(5+k)^2} \pa A_{-}  F_{11}
-\frac{8i}{3(5+k)^2} A_{-} \pa F_{11}
\nonu \\
&-& \frac{4i}{(5+k)^3} A_{-}  F_{11}  F_{12}  F_{21}
-\frac{2}{(5+k)^2} A_{+}  A_{-}  F_{21}
-\frac{i}{(5+k)} B_3  G_{21}
+  \frac{4}{(5+k)^2} B_3  B_3  F_{21}
\nonu \\
& - & \frac{8i}{(5+k)^3} B_3  F_{11}  F_{21}  F_{22}
+ \frac{8i}{3(5+k)^2} B_{-}  \pa F_{22}
+  \frac{4i}{3(5+k)^2} \pa B_3  F_{21}
-\frac{8i}{3(5+k)^2} B_3  \pa F_{21}
\nonu \\
& + & \frac{i}{(5+k)} B_{-}  G_{22}
-  \frac{2}{(5+k)^2} B_{-}  B_{3}  F_{22}
+\frac{4i}{(5+k)^3} B_{-}  F_{12}  F_{21}  F_{22}
-\frac{4i}{3(5+k)^2} \pa B_{-}  F_{22}
\nonu \\
&- & \frac{4}{(5+k)^2} F_{11}  F_{21}  G_{22}
+\frac{4}{3(5+k)^3} \pa F_{11}  F_{21}  F_{22}
-\frac{4i}{(5+k)^2} U  A_3  F_{21}
+\frac{2i}{(5+k)^2} U  B_{-}  F_{22} 
\nonu \\
&- & \frac{8}{3(5+k)^3} F_{11}  \pa F_{21}  F_{22}
+\frac{4}{3(5+k)^3} F_{11}  F_{21}  \pa F_{22}
-\frac{4}{(5+k)^2} F_{11}  F_{22}  G_{21}
\nonu \\
& - & \frac{4}{(5+k)^2} F_{12}  F_{21}  G_{21}
+ \frac{4}{(5+k)^2} F_{21}  F_{22}  G_{11}
-  \frac{4}{3(5+k)} \pa G_{21}
-\frac{4i}{(5+k)^2} U  B_3  F_{21}
\nonu \\
&+& \left. \frac{2i}{(5+k)^2} U  A_{-}  F_{11}
- \frac{8}{3(5+k)^2} \pa U  F_{21}
-\frac{8}{3(5+k)^2} U  \pa F_{21} 
+ \frac{1}{(5+k)} U  G_{21}
\right](z)
\nonu \\
 & = &
\left[ {\bf W_{+}^{(\frac{5}{2})}}
-\frac{2i}{(5+k)} \hat{A}_3  {\bf T_{+,non}^{(\frac{3}{2})}}  
+ \frac{2i}{(5+k)} \hat{B}_3  {\bf T_{+,non}^{(\frac{3}{2})}}
+\frac{2}{(5+k)} F_{11}  {\bf V_{+,non}^{(2)}} 
\right. \nonu \\
& + &  \frac{1}{(5+k)} F_{21}  \pa {\bf T^{(1)}}
- 
\frac{4}{(5+k)^2} F_{11}  F_{21}  {\bf V_{non}^{(\frac{3}{2})}}
+ \frac{4}{(5+k)^2} F_{11}  F_{22}  {\bf T_{+,non}^{(\frac{3}{2})}}
\nonu \\
& + & \frac{2}{(5+k)} F_{21}  {\bf T_{non}^{(2)}}
-\frac{2}{(5+k)} \pa F_{21}  {\bf T^{(1)}}
-  \frac{4}{(5+k)^2} F_{21}  F_{22}  {\bf U_{non}^{(\frac{3}{2})}}
\nonu \\
& + &  \frac{2}{(5+k)} F_{22} {\bf U_{+,non}^{(2)}}
+\frac{2}{(5+k)} U  {\bf T_{+,non}^{(\frac{3}{2})}}
+  \frac{2}{(5+k)^2} F_{11}  F_{21}  \hat{G}_{22}
\nonu \\
& - & \frac{3i}{(5+k)} \hat{A}_3  \hat{G}_{21}
-\frac{2}{(5+k)^2} \hat{A}_3  \hat{A}_3  F_{21}
+\frac{4}{(5+k)^2} \hat{A}_3  \hat{B}_3  F_{21}
+ \frac{4}{(5+k)^2} \hat{A}_3  \hat{B}_{-}  F_{22}
\nonu \\
& + & 
\frac{i}{(5+k)} \hat{A}_{-}  \hat{G}_{11} 
- \frac{4}{(5+k)^2} \hat{A}_{-}  \hat{B}_3  F_{11}
- 
\frac{2}{(5+k)^2} \hat{A}_{+}  \hat{A}_{-}  F_{21}
-\frac{i}{(5+k)} \hat{B}_3  \hat{G}_{21}
\nonu \\
&- & \frac{2i}{(5+k)^2} \pa \hat{A}_3  F_{21}-
 \frac{2i}{(5+k)^2} \pa \hat{B}_3  F_{21}
+\frac{i}{(5+k)} \hat{B}_{-}  \hat{G}_{22}
- 
\frac{2}{(5+k)^2} \hat{B}_{+}  \hat{B}_{-}  F_{21}
\nonu \\
&+ & 
\frac{2}{(5+k)^2} F_{11}  F_{22}  \hat{G}_{21}
-  
\frac{2}{(5+k)^2} F_{21}  F_{22}  \hat{G}_{11}
-  \frac{4}{3(5+k)} \pa \hat{G}_{21}
-   \frac{2}{(5+k)^2} \hat{B}_3  \hat{B}_3  F_{21}
\nonu \\
&+& \left. \frac{1}{(5+k)} U  \hat{G}_{21}
-\frac{2(k+3)}{(5+k)(7k+3)} \hat{T}  F_{21}
 \right](z),
\label{w+5halfnon} \\
{\bf W_{-,non}^{(\frac{5}{2})}}(z) & = & \left[ {\bf W_{-}^{(\frac{5}{2})}}
-\frac{2i}{(5+k)} A_3  {\bf T_{-}^{(\frac{3}{2})}}
+ \frac{2i}{(5+k)} B_3   {\bf T_{-}^{(\frac{3}{2})}}
-\frac{2}{(5+k)} F_{11}  {\bf V_{-}^{(2)}} 
-\frac{2}{(5+k)} U   {\bf T_{-}^{(\frac{3}{2})}}
\right.
\nonu \\
& + & \frac{4}{(5+k)^2} F_{11}  F_{12}  {\bf V^{(\frac{3}{2})}}
-  \frac{8}{(5+k)^2} F_{11}  F_{22}   {\bf T_{-}^{(\frac{3}{2})}}
+\frac{2}{(5+k)} F_{12}  {\bf T^{(2)}}
-  \frac{2}{(5+k)} F_{22}  {\bf U_{-}^{(2)}}
\nonu \\
& + & \frac{4}{(5+k)^2} F_{12}  F_{22}  {\bf U^{(\frac{3}{2})}}
+\frac{2}{(5+k)} \pa F_{12}  {\bf T^{(1)}}
-  \frac{1}{(5+k)} F_{12}  \pa {\bf T^{(1)}}
\nonu \\
& + & \frac{3i}{(5+k)} A_3  G_{12}
-\frac{4}{(5+k)^2} A_3  A_3  F_{12}
-\frac{8i}{(5+k)^3} B_3  F_{11}  F_{12}  F_{22}
-  \frac{2}{(5+k)^2} A_3  B_{+}  F_{11}
\nonu \\
& - & \frac{8i}{(5+k)^3} A_3  F_{11}  F_{12}  F_{22}
-\frac{8i}{3(5+k)^2} \pa A_3  F_{12}
-\frac{8i}{3(5+k)^2} A_3  \pa F_{12}
-\frac{i}{(5+k)} A_{+}  G_{22}
\nonu \\ 
& + & \frac{2}{(5+k)^2} A_{+}  A_3 F_{22}
- \frac{2}{(5+k)^2} A_{+}  A_{-}  F_{12}
+\frac{2}{(5+k)^2} A_{+}  B_3  F_{22}
-\frac{4i}{(5+k)^3} A_{+}  F_{12}  F_{21}  F_{22}
\nonu \\
& - & \frac{4i}{3(5+k)^2} \pa A_{+}  F_{22} +
\frac{8i}{3(5+k)^2} A_{+}  \pa F_{22}
+\frac{i}{(5+k)} B_3  G_{12}
+ \frac{8i}{3(5+k)^2} \pa B_3  F_{12}
\nonu \\
& + & \frac{8i}{3(5+k)^2} B_3  \pa F_{12}
-\frac{i}{(5+k)} B_{+}  G_{11}
- \frac{2}{(5+k)^2} B_{+}  B_3  F_{11}
+\frac{2}{(5+k)^2} B_{+}  B_{-}  F_{12}
\nonu \\
& + & \frac{4i}{3(5+k)^2} \pa B_{+}  F_{11}
-  \frac{8i}{3(5+k)^2} B_{+} \pa F_{11}
+\frac{4i}{(5+k)^3} B_{+}  F_{11}  F_{12}  F_{21}
+
\frac{4}{(5+k)^2} B_3  B_3  F_{12}
\nonu \\
&-& \frac{4}{(5+k)^2} F_{11}  F_{12}  G_{22}
-\frac{4}{3(5+k)^3} \pa F_{11}  F_{12}  F_{22}
+\frac{4i}{(5+k)^2} U  A_3  F_{12}
-\frac{2i}{(5+k)^2} U  B_{+}  F_{11}
\nonu \\
&+& \frac{8}{3(5+k)^3} F_{11}  \pa F_{12}  F_{22}
-\frac{4}{3(5+k)^3} F_{11}  F_{12}  \pa F_{22}
+\frac{4}{(5+k)^2} F_{11}  F_{22}  G_{12}
\nonu \\
& + & \frac{4}{(5+k)^2} F_{12}  F_{21}  G_{12}
+ \frac{4}{(5+k)^2} F_{12}  F_{22}  G_{11}
-\frac{4}{3(5+k)} \pa G_{12}
+\frac{4i}{(5+k)^2} U  B_3  F_{12}
\nonu \\
&- & \left. \frac{2i}{(5+k)^2} U  A_{+}  F_{22}
-  \frac{8}{3(5+k)^2} \pa U  F_{12} -\frac{8}{3(5+k)^2}
U  \pa F_{12} 
+ \frac{1}{(5+k)} U  G_{12} 
\right](z)
\nonu \\
 & = & \left[ {\bf W_{-}^{(\frac{5}{2})}}
-\frac{2i}{(5+k)} A_3  {\bf T_{-,non}^{(\frac{3}{2})}}
+ \frac{2i}{(5+k)} B_3   {\bf T_{-,non}^{(\frac{3}{2})}}
-\frac{2}{(5+k)} F_{11}  {\bf V_{-,non}^{(2)}} 
\right.
\nonu \\
& - & \frac{4}{(5+k)^2} F_{11}  F_{12}  {\bf V_{non}^{(\frac{3}{2})}}
+  \frac{4}{(5+k)^2} F_{11}  F_{22}   {\bf T_{-,non}^{(\frac{3}{2})}}
+\frac{2}{(5+k)} F_{12}  {\bf T_{non}^{(2)}}
\nonu \\
& + & \frac{2}{(5+k)} \pa F_{12}  {\bf T^{(1)}}
-  \frac{4}{(5+k)^2} F_{12}  F_{22}  {\bf U_{non}^{(\frac{3}{2})}}
  -  \frac{2}{(5+k)} F_{22}  {\bf U_{-,non}^{(2)}}
\nonu \\
& - & \frac{2}{(5+k)} U   {\bf T_{-,non}^{(\frac{3}{2})}}
-  \frac{1}{(5+k)} F_{12}  \pa {\bf T^{(1)}}
+ \frac{2}{(5+k)^2} F_{11}  F_{12}  \hat{G}_{22}
\nonu \\
& + & \frac{3i}{(5+k)} \hat{A}_3  \hat{G}_{12}
-\frac{2}{(5+k)^2} \hat{A}_3  \hat{A}_3  F_{12}
+ \frac{2}{(5+k)^2} \hat{A}_3  \hat{B}_3  F_{12}
-  
\frac{2i}{(5+k)^2} \pa \hat{A}_3  F_{12}
\nonu \\
&-&
\frac{i}{(5+k)} \hat{A}_{+}  \hat{G}_{22}
- \frac{2}{(5+k)^2} \hat{A}_{+}  \hat{A}_{-}  F_{12}
-\frac{2}{(5+k)^2} \hat{A}_{+}  \hat{B}_3  F_{22}
+ \frac{4}{(5+k)^2} \hat{A}_3  \hat{B}_{+}  F_{11}
\nonu \\
& + & 
\frac{i}{(5+k)} \hat{B}_3  \hat{G}_{12}
+
\frac{4i}{(5+k)^3} \hat{B}_3  F_{11}  F_{12}  F_{22}
- \frac{2i}{(5+k)^2} \pa \hat{B}_3  F_{12}
-\frac{i}{(5+k)} \hat{B}_{+}  \hat{G}_{11}
\nonu \\
&-& \frac{2}{(5+k)^2} \hat{B}_{+}  \hat{B}_3  F_{11}
-\frac{2}{(5+k)^2} \hat{B}_{+}  \hat{B}_{-}  F_{12}
+\frac{2i}{(5+k)^2} \pa \hat{B}_{+}  F_{11}
\nonu \\
&-&
\frac{2}{(5+k)^2} F_{11}  F_{22}  \hat{G}_{12}
 -  
 \frac{2}{(5+k)^2} F_{12}  F_{22}  \hat{G}_{11}
-\frac{4}{3(5+k)} \pa \hat{G}_{12}
\nonu \\
&+& \left. \frac{1}{(5+k)} U  \hat{G}_{12}
 -\frac{2 (k+3)}{(k+5) (7 k+3)} \hat{T} F_{12}
+ 
\frac{2i}{(5+k)^2} U  \hat{B}_3  F_{12}
 \right](z).
\label{w-5halfnon}
\eea
The nonlinear terms between the higher spin currents 
and the spin-$\frac{1}{2}$ currents (and the spin-$1$ currents) are present.
Due to these nonlinear terms, some of the nonlinear terms in the 
nonlinear version will disappear in the context of Appendix $F$. 

Similarly, the final higher spin-$3$ current, from (\ref{g1221w5half})
and the corresponding equation in Part I has the following relation
%%%%%%%%%%%%%%%%%%%%%%%%%%%%%%%%%%%%%%%%%%%%%%%%%%%%%%%%%%%%%%%%%%%
\bea
{\bf W^{(3)}_{non}}(z)  & = &
\left[ {\bf W^{(3)}} -\frac{4i}{(5+k)} A_3  {\bf T^{(2)}}
+ \frac{8i}{(5+k)^2} A_3  F_{11}  {\bf V^{(\frac{3}{2})}}
-  \frac{8i}{(5+k)^2} A_3  F_{22}  {\bf U^{(\frac{3}{2})}} \right.
\nonu \\
& - & \frac{i}{(5+k)} \pa A_3  {\bf T^{(1)}}
+\frac{i}{(5+k)} A_3  \pa {\bf T^{(1)}}
-  \frac{2i}{(5+k)} A_{-}  {\bf U_{-}^{(2)}}
+ \frac{4i}{(5+k)^2} A_{-}  F_{11}  {\bf T_{-}^{(\frac{3}{2})}}
\nonu \\
& - & \frac{4i}{(5+k)^2} A_{-}  F_{12}  {\bf U^{(\frac{3}{2})}}
+\frac{4i}{(5+k)} B_3  {\bf T^{(2)}}
-  \frac{8i}{(5+k)^2} B_3  F_{11}  {\bf V^{(\frac{3}{2})}}
+ \frac{8i}{(5+k)^2} B_3  F_{22}  {\bf U^{(\frac{3}{2})}}
\nonu \\
& + & \frac{i}{(5+k)} \pa B_3 {\bf T^{(1)}}
-  \frac{i}{(5+k)} B_3  \pa {\bf T^{(1)}}
+ \frac{2i}{(5+k)} B_{-}  {\bf V_{-}^{(2)}} 
+\frac{4i}{(5+k)^2} B_{-}  F_{22}  {\bf T_{-}^{(\frac{3}{2})}}
\nonu \\
& - & \frac{4i}{(5+k)^2} B_{-}  F_{12}  {\bf V^{(\frac{3}{2})}}
-  \frac{8}{(5+k)^3} F_{11}  F_{12}  F_{21}  {\bf V^{(\frac{3}{2})}}
-  \frac{4}{(5+k)^2} F_{11}  F_{21}  {\bf V_{-}^{(2)}}
\nonu \\
& - & \frac{8}{(5+k)^2} F_{11}  F_{22}  {\bf T^{(2)}}
-\frac{2(-41+31k+4k^2)}{(5+k)^2(19+23k)} \pa F_{11}  F_{22}  {\bf T^{(1)}}
-\frac{2}{(5+k)} U  \pa {\bf T^{(1)}} 
\nonu \\
& + & \frac{2(-79-15k+4k^2)}{(5+k)^2(19+23k)} F_{11}  \pa F_{22}  
{\bf T^{(1)}}
+ 
\frac{8(-3+k)}{(5+k)(19+23k)}  F_{12}  \pa F_{21}  {\bf T^{(1)}}
\nonu \\
& + & \frac{2}{(5+k)^2} F_{11}  F_{22}  \pa {\bf T^{(1)}}
-\frac{6}{(5+k)} \pa F_{11}  {\bf V^{(\frac{3}{2})}}
+ \frac{2}{(5+k)} F_{11}  \pa {\bf V^{(\frac{3}{2})}}
+\frac{2}{(5+k)} \pa U {\bf T^{(1)}}
\nonu \\
&-& \frac{8}{(5+k)^3} F_{12}  F_{21}  F_{22}  {\bf U^{(\frac{3}{2})}}
-\frac{8(-3+k)}{(5+k)(19+23k)} \pa F_{12}  F_{21}  {\bf T^{(1)}}
- \frac{2(14+3k)}{(5+k)^2} \pa F_{12}  {\bf T_{+}^{(\frac{3}{2})}}
\nonu \\
& + & \frac{2(6+k)}{(5+k)^2} F_{12}  \pa  {\bf T_{+}^{(\frac{3}{2})}}
-\frac{4}{(5+k)^2} F_{21}  F_{22}  {\bf U_{-}^{(2)}}
-  \frac{12(-3+k)(5+k)}{(17+13k)(19+23k)} T  {\bf T^{(1)}}
\nonu \\
& - & \frac{2(16+3k)}{(5+k)^2} \pa F_{21}  {\bf T_{-}^{(\frac{3}{2})}}
+ 
\frac{2(4+k)}{(5+k)^2} F_{21}  \pa  {\bf T_{-}^{(\frac{3}{2})}}
- \frac{6}{(5+k)} \pa F_{22}  {\bf U^{(\frac{3}{2})}}
\nonu \\
& + & \frac{2}{(5+k)} F_{22}  \pa  {\bf U^{(\frac{3}{2})}}
+ \frac{1}{(5+k)} \pa {\bf W^{(2)}}
+ \frac{6(-3+k)(5+k)}{(17+13k)(19+23k)} \pa^2 {\bf T^{(1)}}
\nonu \\
& - & 
\frac{8(-3+k)}{(5+k)(19+23k)} U  U {\bf T^{(1)}}
\nonu \\
& + & \frac{4i}{(5+k)^2} A_3  A_3  A_3
-\frac{4i}{(5+k)^2} A_3  A_3  B_3
 -  \frac{10i}{(5+k)^3} \pa A_{-}  F_{11}  F_{12}
+\frac{6i}{(5+k)^3} A_{-} \pa F_{11}  F_{12}
\nonu \\
& - &  \frac{4}{(5+k)^2} \pa A_3  A_3
+  \frac{16}{(5+k)^3} A_3  A_3  F_{11}  F_{22}
-\frac{4i}{(5+k)^2} A_3  B_3  B_3
-  \frac{2(k-3)}{(5+k)^2} \pa A_3  B_3
\nonu \\
& + & \frac{2(-3+k)}{(5+k)^2} A_3  \pa B_3
-\frac{8}{(5+k)^3} A_3  B_{-}  F_{12}  F_{22}
-  \frac{8}{(5+k)^3} A_3  B_{+}  F_{11}  F_{21}
\nonu \\
& - & \frac{(-3-85k+10k^2)}{(5+k)^2(19+23k)} i \pa^2 A_3
-\frac{4i}{(5+k)^2} A_3  F_{11}  G_{22}
+  \frac{16i}{(5+k)^4} A_3  F_{11}  F_{12}  F_{21}  F_{22}
\nonu \\
& + & \frac{24i}{(5+k)^3} \pa A_3  F_{11}  F_{22}
 -  \frac{4i}{(5+k)^2} A_3  F_{12}  G_{21}
-\frac{2i(-225-172k+13k^2)}{(5+k)^3(19+23k)} A_3  \pa F_{12}  F_{21} 
\nonu \\
&+& \frac{2i(263+256k+33k^2)}{(5+k)^3(19+23k)} A_3  \pa F_{11}  F_{22}
+\frac{2i(-377-356k+13k^2)}{(5+k)^3(19+23k)} A_3  F_{11}  \pa F_{22}
\nonu \\
&- & \frac{6i(37+24k+11k^2)}{(5+k)^3(19+23k)} A_3  F_{12}  \pa F_{21}
-\frac{4i}{(5+k)^2} A_3  F_{21}  G_{12}
+  \frac{8}{(5+k)^3} A_{-}  B_3  F_{11}  F_{12}
\nonu \\
& + & \frac{2i}{(5+k)^2} A_{-}  F_{11}  G_{12}
-\frac{4i}{(5+k)^2} A_3  F_{22}  G_{11}
-  \frac{2i}{(5+k)^3} A_{-}  F_{11}  \pa F_{12}
\nonu \\
& - & \frac{2i}{(5+k)^2} A_{-}  F_{12}  G_{11}
+\frac{4}{(5+k)^3} A_{+}  A_{-}  F_{12}  F_{21}
-\frac{1}{(5+k)^2} \pa A_{+}  A_{-}
+\frac{1}{(5+k)^2} A_{+}  \pa A_{-}
\nonu \\
&+& \frac{12}{(5+k)^3} A_{+}  A_{-}  F_{11} F_{22}
+ \frac{4i}{(5+k)^2} B_3  F_{11}  G_{22}
-\frac{16i}{(5+k)^4} B_3  F_{11}  F_{12}  F_{21}  F_{22}
\nonu \\
&+& \frac{10i}{(5+k)^3} \pa A_{+}  F_{21}  F_{22}
+\frac{2i}{(5+k)^3} A_{+}  \pa F_{21}  F_{22}
-\frac{6i}{(5+k)^3} A_{+}  F_{21}  \pa F_{22}
+ \frac{4i}{(5+k)^2} A_{+}  A_{-}  A_3
\nonu \\
& + & \frac{4i}{(5+k)^2} A_{+}  F_{22}  G_{21}
+\frac{4i}{(5+k)^2} B_3  B_3  B_3
+\frac{8}{(5+k)^3} A_{+}  B_3  F_{21}  F_{22}
+\frac{4i}{(5+k)^2} B_3  F_{21}  G_{12}
\nonu \\
&- & 
\frac{4}{(5+k)^2} \pa B_3  B_3
-\frac{16}{(5+k)^3} B_3  B_3  F_{11}  F_{22}
+\frac{i(-456+29k+97k^2)}{3(5+k)^2(19+23k)} \pa^2 B_3
\nonu \\
&+& \frac{2i(323+248k+29k^2)}{(5+k)^3(19+23k)} B_3  \pa F_{11}  F_{22}
+\frac{2i(-437-348k+17k^2)}{(5+k)^3(19+23k)} B_3  F_{11}  \pa F_{22}
\nonu \\
& +& \frac{4i}{(5+k)^2} B_3  F_{12}  G_{21}
+\frac{8i}{(5+k)^3} \pa B_3  F_{12}  F_{21}
+\frac{4i}{(5+k)^2} B_3  F_{22}  G_{11}
-\frac{2 i k}{(5+k)^3} B_{-}  F_{12}  \pa F_{22}
\nonu \\
&+ & \frac{2i(-285-164k+17k^2)}{(5+k)^3(19+23k)} B_3  F_{12}  
\pa F_{21}
+\frac{2i(171+64k+29k^2)}{(5+k)^3(19+23k)} B_3  \pa F_{12}  F_{21}
\nonu \\
& + & \frac{2i}{(5+k)^2} B_{-}  F_{12}  G_{22}
+ \frac{2i(2+k)}{(5+k)^3} \pa B_{-}  F_{12}  F_{22}
-\frac{2i(-4+k)}{(5+k)^3} B_{-} \pa F_{12}  F_{22}
\nonu \\
&-& \frac{6i}{(5+k)^2} B_{-}  F_{22}  G_{12}
+\frac{4i}{(5+k)^2} B_{+}  B_{-}  B_3
+
\frac{4i}{(5+k)^2} B_3  F_{22}  G_{11}
- \frac{(-2+k)}{(5+k)^2} \pa B_{+}  B_{-}
\nonu \\
& + & \frac{(-2+k)}{(5+k)^2} B_{+}  \pa B_{-}
-\frac{12}{(5+k)^3} B_{+}  B_{-}  F_{11}  F_{22}
+ \frac{(-3+k)(-71+5k)}{3(5+k)^3(19+23k)} F_{11}  \pa^2 F_{22}
\nonu \\
&+& \frac{4}{(5+k)^3} B_{+}  B_{-}  F_{12}  F_{21}
-\frac{4i}{(5+k)^2} B_{+}  F_{11}  G_{21}
-\frac{2i(2+k)}{(5+k)^3} \pa B_{+}  F_{11}  F_{21}
\nonu \\
&+& \frac{2 i k}{(5+k)^3} B_{+}  \pa F_{11}  F_{21}
+\frac{2i(-4+k)}{(5+k)^3} B_{+}  F_{11}  \pa F_{21}
+\frac{(6+k)}{(5+k)^2} F_{12}  \pa G_{21}
\nonu \\
& + & \frac{24(33+26k+9k^2)}{(5+k)^4(19+23k)} \pa F_{11}  F_{12}  F_{21}
 F_{22}
+\frac{4}{(5+k)^3} F_{11}  F_{12}  F_{21}  G_{22}
-\frac{4}{(5+k)^3} F_{12}  F_{21}  F_{22}  G_{11}
\nonu \\
& - & \frac{16(-3+k)(7+11k)}{(5+k)^4(19+23k)} F_{11}  \pa F_{12} 
F_{21}  F_{22}
-\frac{8(-3+k)}{(5+k)^3(19+23k)} F_{11}  F_{12}  \pa F_{21}  F_{22}
\nonu \\
&-& \frac{32(39+32k+k^2)}{(5+k)^4(19+23k)} F_{11}  F_{12}  F_{21}  \pa
F_{22} +\frac{16}{(5+k)^3} F_{11}  F_{12}  F_{22}  G_{21}
+  \frac{16}{(5+k)^3} F_{11}  F_{21}  F_{22}  G_{12}
\nonu \\
&+& \frac{(-3+k)(-71+5k)}{3(5+k)^3(19+23k)} \pa^2 F_{11}  F_{22}
-\frac{16(-3+k)}{3(5+k)^3} \pa F_{11}  \pa F_{22}
-\frac{(11+3k)}{(5+k)^2} \pa F_{22}  G_{11}
\nonu \\
&+ & \frac{(11+3k)}{(5+k)^2} \pa F_{11}  G_{22}
-\frac{(1+k)}{(5+k)^2} F_{11} \pa G_{22}
 -  \frac{(14+3k)}{(5+k)^2} \pa F_{12}  G_{21}
+\frac{8i}{(5+k)^3} U  A_{-}  F_{11}  F_{12}
\nonu \\
& - & 
\frac{(1611+874k+95k^2)}{3(5+k)^3(19+23k)} \pa^2 F_{12}  F_{21}
+\frac{8(3+2k)}{3(5+k)^3} \pa F_{12}  \pa F_{21}
-\frac{4i}{(5+k)^2} U  \pa A_3
\nonu \\
&-& \frac{(1611+874k+95k^2)}{3(5+k)^3(19+23k)} F_{12} \pa^2 F_{21}
+ \frac{(16+3k)}{(5+k)^2} \pa F_{21}  G_{12}
-\frac{(4+k)}{(5+k)^2} F_{21}  \pa G_{12}
\nonu \\
& + & \frac{(1+k)}{(5+k)^2} F_{22} \pa G_{11}
-\frac{2}{(5+k)} G_{12}  G_{21}
+  \frac{4 i(13+5k)}{(5+k)(19+23k)} T  A_3 
\nonu \\
&+& \frac{4i(19+3k)}{(5+k)(19+23k)} T  B_3
+\frac{2}{(5+k)} \pa T
+  \frac{8(-3+k)}{(5+k)^3(19+23k)} U  U F_{11}  F_{22}
\nonu \\
&+& \frac{8(-3+k)}{(5+k)^2(19+23k)} T  F_{11}  F_{22}
+\frac{32(4+k)}{(5+k)^2(19+23k)} T  F_{12}  F_{21}
+  \frac{4i}{(5+k)^2} \pa U  A_3
\nonu \\
&-& \frac{8i}{(5+k)^3} U  A_{+}  F_{21}  F_{22}
-\frac{4i}{(5+k)^2} \pa U  B_3
+\frac{4i}{(5+k)^2} U \pa B_3
+ \frac{8i}{(5+k)^3} U  B_{-}  F_{12}  F_{22}
\nonu \\
& - & \frac{8i}{(5+k)^3} U  B_{+}  F_{11}  F_{21}
-\frac{4}{(5+k)^2} U  F_{12}  G_{21}
+ \frac{4}{(5+k)^2} U  F_{21}  G_{12}
+\frac{(-1+k)}{(5+k)^3} F_{21}  G_{12} F_{21}  F_{12}
\nonu \\
&+ &  \frac{4i(13+5k)}{(5+k)^2(19+23k)} U  U A_3
+\frac{4i(19+3k)}{(5+k)^2(19+23k)} U  U  B_3
-  
\frac{8(3+19k)}{(5+k)^3(19+23k)} U  U F_{12}  F_{21}
\nonu \\
& - & \left. \frac{(-5+k)}{(5+k)^3} F_{22}  G_{11}  F_{22}  F_{11}
-  
\frac{(11+k)}{(5+k)^3} F_{11}  G_{22}  F_{11}  F_{22}
+\frac{(-1+k)}{(5+k)^3} F_{12}  G_{21} F_{12} F_{21} \right](z)
\nonu \\
  & = &
\left[ {\bf W^{(3)}} -\frac{4i}{(5+k)} A_3  {\bf T_{non}^{(2)}}
-  \frac{i}{(5+k)} \pa A_3  {\bf T^{(1)}}
+\frac{i}{(5+k)} A_3  \pa {\bf T^{(1)}}
+ \frac{2}{(5+k)} F_{11}  \pa {\bf V_{non}^{(\frac{3}{2})}}
\right. \nonu \\
& - &  \frac{2i}{(5+k)} A_{-}  {\bf U_{-,non}^{(2)}}
+ 
\frac{4i}{(5+k)} B_3  {\bf T_{non}^{(2)}}
+  
\frac{i}{(5+k)} \pa B_3 {\bf T^{(1)}}
- 
\frac{6}{(5+k)} \pa F_{11}  {\bf V_{non}^{(\frac{3}{2})}}
\nonu \\
& - & \frac{i}{(5+k)} B_3  \pa {\bf T^{(1)}}
+ \frac{2i}{(5+k)} B_{-}  {\bf V_{-,non}^{(2)}} 
-  
\frac{4(k-3)}{(5+k)(17+13k)} \pa F_{11}  F_{22}  {\bf T^{(1)}}
\nonu \\
& + & \frac{4(k-3)}{(5+k)(17+13k)} F_{11}  \pa F_{22}  
{\bf T^{(1)}}
+ 
\frac{4(-3+k)}{(5+k)(17+13k)}  F_{12}  \pa F_{21}  {\bf T^{(1)}}
+\frac{2}{(5+k)} F_{12}  \pa  {\bf T_{+,non}^{(\frac{3}{2})}}
\nonu \\
&-& 
\frac{4(-3+k)}{(5+k)(17+13k)} \pa F_{12}  F_{21}  {\bf T^{(1)}}
 -  
\frac{4(-3+k)}{(5+k)(17+13k)} U  U {\bf T^{(1)}}
- \frac{6}{(5+k)} \pa F_{12}  {\bf T_{+,non}^{(\frac{3}{2})}}
\nonu \\
& - & \frac{6}{(5+k)} \pa F_{21}  {\bf T_{-,non}^{(\frac{3}{2})}}
+ 
\frac{2}{(5+k)} F_{21}  \pa  {\bf T_{-,non}^{(\frac{3}{2})}}
- \frac{6}{(5+k)} \pa F_{22}  {\bf U_{non}^{(\frac{3}{2})}}
\nonu \\
& + & \frac{2}{(5+k)} F_{22}  \pa  {\bf U_{non}^{(\frac{3}{2})}}
+ \frac{1}{(5+k)} \pa {\bf W_{non}^{(2)}}
+ \frac{6(-3+k)(5+k)}{(17+13k)(19+23k)} \pa^2 {\bf T^{(1)}}
\nonu \\
& - & \frac{12(-3+k)(5+k)}{(17+13k)(19+23k)} \hat{T}  {\bf T^{(1)}}
+\frac{2}{(5+k)} \pa U {\bf T^{(1)}}
-\frac{2}{(5+k)} U  \pa {\bf T^{(1)}} 
\nonu \\
& + & \frac{8i}{(5+k)^2} \hat{A}_3  \hat{A}_3  \hat{A}_3
-\frac{16i}{(5+k)^2} \hat{A}_3  \hat{A}_3  \hat{B}_3
- \frac{6}{(5+k)^2} \pa \hat{A}_3  \hat{A}_3
+\frac{8i}{(5+k)^2} \hat{A}_3  \hat{B}_3  \hat{B}_3
\nonu \\
& - & \frac{2(k-6)}{(5+k)^2} \pa \hat{A}_3  \hat{B}_3
+\frac{2(-3+k)}{(5+k)^2} \hat{A}_3  \pa \hat{B}_3
-  
\frac{2 i \left(58 k^3+k^2+460 k+309\right)}{(k+5)^2 (7 k+3) (23 k+19)} 
\pa^2 \hat{A}_3
\nonu \\
& - &  
\frac{8i}{(5+k)^2} \hat{A}_{+}  \hat{A}_{-}  \hat{B}_3
+ \frac{8i}{(5+k)^2} \hat{A}_{+}  \hat{A}_{-}  \hat{A}_3
+\frac{6}{(5+k)^2} \pa \hat{A}_{+}  \hat{A}_{-}
-\frac{3}{(5+k)} \pa F_{22}  \hat{G}_{11}
\nonu \\
& - & \frac{4}{(5+k)^2} \hat{A}_{+}  \pa \hat{A}_{-}
- \frac{(-1+k)}{(5+k)^2} \pa \hat{B}_{+}  \hat{B}_{-}
+\frac{(1+k)}{(5+k)^2} \hat{B}_{+}  \pa \hat{B}_{-}
-\frac{1}{(5+k)} F_{21}  \pa \hat{G}_{12}
\nonu \\
&+ & 
\frac{2}{(5+k)^2} \pa \hat{B}_3  \hat{B}_3
+ \frac{i \left(817 k^3+1229 k^2-729 k+171\right)}
{3 (k+5)^2 (7 k+3) (23 k+19)}    \pa^2 \hat{B}_3
+ \frac{3}{(5+k)} \pa F_{21}  \hat{G}_{12}
\nonu \\
&+ & \frac{3}{(5+k)} \pa F_{11}  \hat{G}_{22}
-\frac{(1+k)}{(5+k)} F_{11} \pa \hat{G}_{22}
-  \frac{3}{(5+k)} \pa F_{12}  \hat{G}_{21}
+\frac{1}{(5+k)} F_{12}  \pa \hat{G}_{21}
\nonu \\
& + & \frac{1}{(5+k)} F_{22} \pa \hat{G}_{11}
-\frac{2}{(5+k)} \hat{G}_{12}  \hat{G}_{21}
+ \frac{8 i \left(29 k^2+97 k+48\right)}{(k+5) (7 k+3) (23 k+19)}  
\hat{T}  \hat{A}_3 
\nonu \\
&-& \left.
\frac{8 i (k-27) k}{(k+5) (7 k+3) (23 k+19)}  \hat{T}  \hat{B}_3
+\frac{1}{(5+k)} \pa \hat{T}
 \right](z).
\label{w3non}
\eea

Now the higher spin currents in the linear version can be written 
in terms of those in the nonlinear version and this implies that 
using the previous results in Part I and Part II one can 
write $136$ OPEs in the linear version (although 
one should rewrite all the fields written in the nonlinear version in the 
basis of linear version and make the mathematica file which defines the results 
of Part I and Part II inside Thielemans package).   

Due to these nonlinear terms between the boldface 
higher spin currents and the currents from the large ${\cal N}=4$ linear 
superconformal algebra, some of the nonlinear terms in the 
nonlinear version will disappear in the context of Appendix $F$.

%%%%%%%%%%%%%%%%%%%%%%%%%%%%%%%%%%%%%%%%%%%%%%%%%%%%%%%%%%%%%%%%%%%%%
%%%%%%%%%%%%%%%%%%%%%%%%%%%%%%%%%%%%%%%%%%%%%%%%%%%%%%%%%%%%%%%%%%%%%
\section{ The complete
OPEs between the $16$ currents  of large ${\cal N}=4$
 linear superconformal algebra and the $16$ lowest 
higher spin currents }
%B%%%%%%%%%%%%%%%%%%%%%%%%%%%%%%%%%%%%%%%%%%%%%%%%%%%%%%%%%%%%%%%%%%%
%%%%%%%%%%%%%%%%%%%%%%%%%%%%%%%%%%%%%%%%%%%%%%%%%%%%%%%%%%%%%%%%%%%%

In this Appendix, one describes 
the complete
OPEs between the $16$ currents  of large ${\cal N}=4$
 linear superconformal algebra and the $16$ lowest 
higher spin currents.
Except the few cases, these are linear.  

%%%%%%%%%%%%%%%%%%%%%%%%%%%%%%%%%%%%%%%%%%%%%%%%%%%%%%%%%%%%%%%%%%%%%
\subsection{ The OPEs between the spin-$\frac{1}{2}$ currents and 
the $16$ lowest higher spin currents }
%C%%%%%%%%%%%%%%%%%%%%%%%%%%%%%%%%%%%%%%%%%%%%%%%%%%%%%%%%%%%%%%%%%%%

One can perform the various OPEs between the four spin-$\frac{1}{2}$ currents
found in Part I
and $16$ higher spin currents obtained in section $3$
as follows:
\bea
\left(
\begin{array}{c}
F_{11} \nonu \\
F_{22}
\end{array}
\right)(z) \, {\bf T_{\pm}^{(\frac{3}{2})}}(w) & = & 
\frac{1}{(z-w)} \frac{i}{2} 
B_{\mp}
(w) + \cdots,
\nonu \\
\left(
\begin{array}{c}
F_{11} \nonu \\
F_{22}
\end{array}
\right)(z) \, {\bf T_{\mp}^{(\frac{3}{2})}}(w) & = & 
-\frac{1}{(z-w)} \frac{i}{2} 
A_{\pm}
(w) + \cdots,
\nonu \\
\left(
\begin{array}{c}
F_{11} \nonu \\
F_{22} \end{array}
\right)(z) \,  {\bf T^{(2)}}(w) & = & \frac{1}{(z-w)} \left[ 
\frac{1}{2} \left(
\begin{array}{c} 
G_{11} \nonu \\
G_{22} 
\end{array}
\right)  \pm
\left(
\begin{array}{c}
{\bf U^{(\frac{3}{2})}} \nonu \\
{\bf V^{(\frac{3}{2})}} 
\end{array}
\right) 
\right](w) +\cdots,
\nonu \\
\left(
\begin{array}{c}
F_{11} \nonu \\
F_{22}
\end{array}
\right)(z) \,
\left(
\begin{array}{c}
{\bf V^{(\frac{3}{2})}}
\nonu \\
{\bf U^{(\frac{3}{2})}}
\end{array}
\right)(w) & = & 
\frac{1}{(z-w)} \frac{1}{2} \left[ - i A_3 - i B_3 \pm U 
\right](w) +\cdots,
\nonu \\
\left(
\begin{array}{c}
F_{11} \nonu \\
F_{22}
\end{array}
\right)(z) \,
\left(
\begin{array}{c}
{\bf V_{+}^{(2)}}
\nonu \\
{\bf U_{-}^{(2)}}
\end{array}
\right)(w) & = & 
-\frac{1}{(z-w)}  \left[  \pm \frac{1}{2} \left(
\begin{array}{c} 
G_{21} \nonu \\
G_{12} 
\end{array}
\right) + {\bf T_{\pm}^{(\frac{3}{2})}}
\right](w) +\cdots,
\nonu \\
\left(
\begin{array}{c}
F_{11} \nonu \\
F_{22}
\end{array}
\right)(z) \,
\left(
\begin{array}{c}
{\bf V_{-}^{(2)}}
\nonu \\
{\bf U_{+}^{(2)}}
\end{array}
\right)(w) & = & 
\frac{1}{(z-w)}  \left[ \mp  \frac{1}{2} \left(
\begin{array}{c} 
G_{12} \nonu \\
G_{21} 
\end{array}
\right) + {\bf T_{\mp}^{(\frac{3}{2})}}
\right](w) +\cdots,
\nonu \\
\left(
\begin{array}{c}
F_{11} \nonu \\
F_{22}
\end{array}
\right)(z) \,
\left(
\begin{array}{c}
{\bf V^{(\frac{5}{2})}}
\nonu \\
{\bf U^{(\frac{5}{2})}}
\end{array}
\right)(w) & = & 
\pm \frac{1}{(z-w)^2} {\bf T^{(1)}}(w) 
+\frac{1}{(z-w)} \left[ \mp \frac{1}{2} \pa {\bf T^{(1)}} + {\bf W^{(2)}}
\right](w) + \cdots, 
\nonu \\
\left(
\begin{array}{c}
F_{11} \nonu \\
F_{22}
\end{array}
\right)(z) \,
{\bf W_{\pm}^{(\frac{5}{2})}}(w) & = &  \pm 
\frac{1}{(z-w)} \left(
\begin{array}{c}
{\bf U_{+}^{(2)}} \nonu \\
{\bf V_{-}^{(2)}}
\end{array}
\right)(w) + \cdots, 
\nonu \\
\left(
\begin{array}{c}
F_{11} \nonu \\
F_{22}
\end{array}
\right)(z) \,
{\bf W_{\mp}^{(\frac{5}{2})}}(w) & = &  
\mp \frac{1}{(z-w)} \left(
\begin{array}{c}
{\bf U_{-}^{(2)}} \nonu \\
{\bf V_{+}^{(2)}}
\end{array}
\right)(w) + \cdots, 
\nonu \\
\left(
\begin{array}{c}
F_{11} \nonu \\
F_{22}
\end{array}
\right)(z) 
{\bf W^{(3)}}(w) & = &
\frac{1}{(z-w)^2} \left[ \mp  \frac{3}{2} 
\left( 
\begin{array}{c} 
G_{11} \nonu \\
G_{22} \end{array}
\right) - 3 
\left(
\begin{array}{c}
{\bf U^{(\frac{3}{2})}}
\nonu \\
{\bf V^{(\frac{3}{2})}}
\end{array}
\right)
- \frac{2{\bf (-3+k)}}{(17+13k)} {\bf T^{(1)}} \left(
\begin{array}{c} 
F_{11} \nonu \\
F_{22} \end{array}
\right)
 \right](w) 
\nonu \\
& + & \frac{1}{(z-w)} \left[  
-\frac{1}{3} \pa 
 (\mbox{pole-2})
%\left(
%\begin{array}{c}
%G_{11} \nonu \\
%-G_{22} 
%\end{array}
%\right)
+ 
%\pa 
%\left(
%\begin{array}{c}
%{\bf U^{(\frac{3}{2})}} 
%\nonu \\
%{\bf V^{(\frac{3}{2})}} 
%\end{array}
%\right)
\frac{4{\bf (-3+k)}}{3(17+13k)} \pa 
\left(
\begin{array}{c}
F_{11} \nonu \\
F_{22} \end{array} 
\right) {\bf T^{(1)}}  
\right. \nonu \\
& - & \left. \frac{2{\bf (-3+k)}}{3(17+13k)} 
\left(
\begin{array}{c}
F_{11} \nonu \\
F_{22} \end{array} 
\right) \pa {\bf T^{(1)}} 
\right](w) +\cdots,
\nonu \\
\left(
\begin{array}{c}
F_{12} \nonu \\
F_{21}
\end{array}
\right)(z) \,
{\bf T_{\pm}^{(\frac{3}{2})}}(w)  & = & 
\frac{1}{(z-w)} \frac{1}{2} \left[  i A_3 - i B_3 \mp  U 
\right](w) +\cdots,
\nonu \\
\left(
\begin{array}{c}
F_{12} \nonu \\
F_{21}
\end{array}
\right)(z) \,
\left(
\begin{array}{c}
{\bf U^{(\frac{3}{2})}}
\nonu \\
{\bf V^{(\frac{3}{2})}}
\end{array}
\right)(w) & = & 
-\frac{1}{(z-w)} \frac{i}{2} A_{\pm}(w) +\cdots,
\nonu \\
\left(
\begin{array}{c}
F_{12} \nonu \\
F_{21}
\end{array}
\right)(z) \,
\left(
\begin{array}{c}
{\bf U_{+}^{(2)}}
\nonu \\
{\bf V_{-}^{(2)}}
\end{array}
\right)(w) & = & 
-\frac{1}{(z-w)}  \left[  \pm \frac{1}{2} \left(
\begin{array}{c} 
G_{11} \nonu \\
G_{22} 
\end{array}
\right) + \left(
\begin{array}{c}
{\bf U^{(\frac{3}{2})}}
\nonu \\
{\bf V^{(\frac{3}{2})}}
\end{array}
\right)
\right](w) +\cdots,
\nonu \\
\left(
\begin{array}{c}
F_{12} \nonu \\
F_{21}
\end{array}
\right)(z) \,
\left(
\begin{array}{c}
{\bf U^{(\frac{5}{2})}}
\nonu \\
{\bf V^{(\frac{5}{2})}}
\end{array}
\right)(w) & = & 
\mp \frac{1}{(z-w)} 
\left(
\begin{array}{c}
{\bf U_{-}^{(2)}}
\nonu \\
{\bf V_{+}^{(2)}}
\end{array}
\right)
(w) + \cdots, 
\nonu \\
\left(
\begin{array}{c}
F_{12} \nonu \\
F_{21}
\end{array}
\right)(z) \,
\left(
\begin{array}{c}
{\bf V^{(\frac{3}{2})}}
\nonu \\
{\bf U^{(\frac{3}{2})}}
\end{array}
\right)(w) & = & 
-\frac{1}{(z-w)} \frac{i}{2} B_{\pm}(w) +\cdots,
\nonu \\
\left(
\begin{array}{c}
F_{12} \nonu \\
F_{21}
\end{array}
\right)(z) \,
\left(
\begin{array}{c}
{\bf V_{+}^{(2)}}
\nonu \\
{\bf U_{-}^{(2)}}
\end{array}
\right)(w) & = & 
-\frac{1}{(z-w)}  \left[ \pm  \frac{1}{2} \left(
\begin{array}{c} 
G_{22} \nonu \\
G_{11} 
\end{array}
\right) - \left(
\begin{array}{c}
{\bf V^{(\frac{3}{2})}}
\nonu \\
{\bf U^{(\frac{3}{2})}}
\end{array}
\right)
\right](w) +\cdots,
\nonu \\
\left(
\begin{array}{c}
F_{12} \nonu \\
F_{21}
\end{array}
\right)(z) \,
\left(
\begin{array}{c}
{\bf V^{(\frac{5}{2})}}
\nonu \\
{\bf U^{(\frac{5}{2})}}
\end{array}
\right)(w) & = & 
\pm \frac{1}{(z-w)} 
\left(
\begin{array}{c}
{\bf V_{-}^{(2)}}
\nonu \\
{\bf U_{+}^{(2)}}
\end{array}
\right)
(w) + \cdots, 
\nonu \\
\left(
\begin{array}{c}
F_{12} \nonu \\
F_{21} \end{array}
\right)(z) \,  {\bf W^{(2)}}(w) & = & \frac{1}{(z-w)} \left[ 
\frac{1}{2} \left(
\begin{array}{c} 
G_{12} \nonu \\
G_{21} 
\end{array}
\right)  \mp
{\bf T_{\mp}^{(\frac{3}{2})}} \right](w) +\cdots,
\nonu \\
\left(
\begin{array}{c}
F_{12} \nonu \\
F_{21}
\end{array}
\right)(z) \,
{\bf W_{\pm}^{(\frac{5}{2})}}(w) & = &  
\mp \frac{1}{(z-w)^2} {\bf T^{(1)}}(w)  
+  \frac{1}{(z-w)} \left[ \pm \frac{1}{2} \pa {\bf T^{(1)}} + 
{\bf T^{(2)}}  \right](w) + \cdots, 
\nonu \\
\left(
\begin{array}{c}
F_{12} \nonu \\
F_{21}
\end{array}
\right)(z) \, 
{\bf W^{(3)}}(w) & = &
\frac{1}{(z-w)^2} \left[  \pm \frac{3}{2} 
\left( 
\begin{array}{c} 
G_{12} \nonu \\
G_{21} \end{array}
\right) - 3 
{\bf T_{\mp}^{(\frac{3}{2})}}
- \frac{2{\bf (-3+k)}}{(17+13k)} {\bf T^{(1)}} \left(
\begin{array}{c} 
F_{12} \nonu \\
F_{21} \end{array}
\right)
 \right](w) 
\nonu \\
& + & \frac{1}{(z-w)} \left[  
-\frac{1}{3} \pa (\mbox{pole-2})
%\left(
%\begin{array}{c}
%-G_{12} \nonu \\
%G_{21} 
%\end{array}
%\right)
%+ \pa 
%{\bf T_{\mp}^{(\frac{3}{2})}} 
+\frac{4{\bf (-3+k)}}{3(17+13k)} \pa 
\left(
\begin{array}{c}
F_{12} \nonu \\
F_{21} \end{array} 
\right) {\bf T^{(1)}}  
\right. \nonu \\
& - & \left. \frac{2{\bf (-3+k)}}{3(17+13k)}  
\left(
\begin{array}{c}
F_{12} \nonu \\
F_{21} \end{array} 
\right) \pa {\bf T^{(1)}} 
\right](w) +\cdots.
\nonu
\eea
It is easy to see that the OPEs containing the higher spin-$3$ current
(which is a primary field)
do have the nonlinear terms in the right hand side.
Again, by considering the quasi primary field containing 
the higher spin-$1$ current (that is 
(\ref{bcgspin3})) one can remove  these nonlinear terms with the relations 
in (\ref{fsingledouble}).
For the corresponding OPEs in the nonlinear version,
there are trivial OPEs because the OPEs between the higher spin currents 
and the spin-$\frac{1}{2}$ currents are regular by construction.
One also checks that the above OPEs are equivalent to the corresponding
OPEs in \cite{BCG} via the field identifications in section $4$.

%%%%%%%%%%%%%%%%%%%%%%%%%%%%%%%%%%%%%%%%%%%%%%%%%%%%%%%%%%%%%%%%%%%%%
\subsection{ The OPEs between the spin-$1$ current and 
the $16$ lowest higher spin currents  }
%C%%%%%%%%%%%%%%%%%%%%%%%%%%%%%%%%%%%%%%%%%%%%%%%%%%%%%%%%%%%%%%%%%%%

Let us perform the various OPEs between the spin-$1$ current
found in Part I
and $16$ higher spin currents obtained in section $3$
as follows:
\bea
U(z) \, {\bf T_{\pm}^{(\frac{3}{2})}}(w) & = & \mp \frac{1}{(z-w)^2} 
\frac{1}{2} 
\left(
\begin{array}{c}
F_{21} \\
 F_{12} 
\end{array}
\right)(w)
+\cdots,
\nonu \\
U(z) \, 
\left(
\begin{array}{c}
{\bf U^{(\frac{3}{2})}}
\nonu \\
{\bf V^{(\frac{3}{2})}}
\end{array}
\right)(w) & = & \mp \frac{1}{(z-w)^2} 
\frac{1}{2} 
\left(
\begin{array}{c}
F_{11} \\
 F_{22} 
\end{array}
\right)(w)
+\cdots,
\nonu \\
U(z) \left(
\begin{array}{c} 
{\bf U^{(\frac{5}{2})}} \nonu \\
{\bf V^{(\frac{5}{2})}} \end{array} 
\right)(w) & = & 
\frac{1}{(z-w)^2} \left[ \frac{1}{2} 
\left(
\begin{array}{c}
G_{11} \nonu \\
G_{22}  \end{array}
\right) \pm   \left( 
\begin{array}{c} 
{\bf U^{(\frac{3}{2})}} \nonu \\
{\bf V^{(\frac{3}{2})}}
\end{array}
\right)
\right](w) +\cdots,
\nonu \\
U(z) \left(
\begin{array}{c} 
{\bf W_{+}^{(\frac{5}{2})}} \nonu \\
{\bf W_{-}^{(\frac{5}{2})}} \end{array} 
\right)(w) & = & 
\frac{1}{(z-w)^2} \left[ \frac{1}{2} 
\left(
\begin{array}{c}
G_{21} \nonu \\
G_{12}  \end{array}
\right) \pm
{\bf T_{\pm}^{(\frac{3}{2})}} 
\right](w) +\cdots,
\nonu \\
U(z) \, {\bf W^{(3)}}(w) & = &
\frac{1}{(z-w)^3} 2 {\bf T^{(1)}}(w) 
\nonu \\
& - & \frac{1}{(z-w)^2}  \left[ \frac{1}{2} \pa (\mbox{pole-3}) 
%{\bf T^{(1)}}
 + \frac{4{\bf (-3+k)}}{(17+13k)} 
U \, {\bf T^{(1)}} \right](w) 
+\cdots.
\nonu
\eea
It is easy to see that the last OPE containing the higher spin-$3$ current
has the nonlinear terms in the right hand side.
This will disappear by using the quasi primary field in (\ref{bcgspin3}).
For the corresponding OPEs in the nonlinear version,
there are trivial OPEs because the OPEs between the higher spin currents 
and the spin-$1$ current are regular by construction.
One also checks that the above OPEs are equivalent to the corresponding
OPEs in \cite{BCG} (or in Appendix $D$) 
via the field identifications in section $4$.

%%%%%%%%%%%%%%%%%%%%%%%%%%%%%%%%%%%%%%%%%%%%%%%%%%%%%%%%%%%%%%%%%%%%%
\subsection{ The OPEs between the spin-$1$ currents and 
the $16$ lowest higher spin currents  }
%C%%%%%%%%%%%%%%%%%%%%%%%%%%%%%%%%%%%%%%%%%%%%%%%%%%%%%%%%%%%%%%%%%%%

The OPEs between the three spin-$1$ currents found in Part I
and the $16$ higher spin currents obtained in section $3$
are
\bea
A_{\pm}(z) \, {\bf T_{\pm}^{(\frac{3}{2})}}(w) & = & 
-\frac{1}{(z-w)^2} \frac{i(1+k)}{(5+k)} \left(
\begin{array}{c}
F_{11} \nonu \\
F_{22} 
\end{array}
\right)(w)  \mp  \frac{1}{(z-w)} i  
\left(
\begin{array}{c}
{\bf U^{(\frac{3}{2})}} \nonu \\
 {\bf V^{(\frac{3}{2})}} 
\end{array}
\right)(w)+\cdots,
\nonu \\
A_{\pm}(z) \, {\bf T^{(2)}}(w) & = & 
\frac{1}{(z-w)} i 
\left(
\begin{array}{c}
{\bf U_{-}^{(2)}} \nonu \\
 {\bf V_{+}^{(2)}} 
\end{array}
\right)(w)+\cdots,
\nonu \\
A_{\pm}(z)  
\left(
\begin{array}{c}
{\bf V^{(\frac{3}{2})}} \nonu \\
 {\bf U^{(\frac{3}{2})}} 
\end{array}
\right)(w) & = & -
\frac{1}{(z-w)^2} \frac{i(1+k)}{(5+k)}
\left(
\begin{array}{c}
F_{12} \nonu \\
F_{21} 
\end{array}
\right)(w) \pm 
\frac{1}{(z-w)} i  {\bf T_{\mp}^{(\frac{3}{2})}}(w)
+ \cdots,
\nonu \\
A_{\pm}(z) \left(
\begin{array}{c}
{\bf V_{+}^{(2)}} \nonu \\
 {\bf U_{-}^{(2)}} 
\end{array}
\right)(w) & = & 
\mp \frac{1}{(z-w)^2} i {\bf T^{(1)}}(w) 
+  \frac{1}{(z-w)} i \left[ {\bf T^{(2)}} - {\bf W^{(2)}} \right](w) +
\cdots, 
\nonu \\
A_{\pm}(z) \left(
\begin{array}{c}
{\bf V^{(\frac{5}{2})}} \nonu \\
{\bf U^{(\frac{5}{2})}} 
\end{array}
\right)(w) & = & 
\frac{1}{(z-w)^2} \frac{i(21+5k)}{3(5+k)} \left[ 2 
{\bf T_{\mp}^{(\frac{3}{2})}}  \mp   \left(
\begin{array}{c}
G_{12} \nonu \\
G_{21}  \end{array}
\right) \right](w) 
\nonu \\
& \pm & \frac{1}{(z-w)} i {\bf W_{\mp}^{(\frac{5}{2})}}(w) +\cdots,
\nonu \\
A_{\pm}(z) \, {\bf W^{(2)}}(w) & = & 
-\frac{1}{(z-w)} i
 \left(
\begin{array}{c}
{\bf U_{-}^{(2)}} \nonu \\
 {\bf V_{+}^{(2)}} 
\end{array}
\right)(w) + \cdots,
\nonu \\
A_{\pm}(z) \,
{\bf W_{\pm}^{(\frac{5}{2})}}(w) & = & 
\frac{1}{(z-w)^2} \frac{i(21+5k)}{3(5+k)} \left[ 2 
\left(
\begin{array}{c}
{\bf U^{(\frac{3}{2})}} \nonu \\
 {\bf V^{(\frac{3}{2})}} 
\end{array}
\right) 
\pm   \left(
\begin{array}{c}
G_{11} \nonu \\
G_{22}  \end{array}
\right) \right](w) 
\nonu \\
& \mp & \frac{1}{(z-w)} i \left(
\begin{array}{c}
{\bf U^{(\frac{5}{2})}} \nonu \\
 {\bf V^{(\frac{5}{2})}} 
\end{array}
\right)(w) +\cdots,
\nonu \\
A_{\pm}(z) \, {\bf W^{(3)}}(w) & = & 
\frac{1}{(z-w)^2} \left[  \mp 4 i  \left(
\begin{array}{c}
{\bf U_{-}^{(2)}} \nonu \\
 {\bf V_{+}^{(2)}} 
\end{array}
\right)  -\frac{4{\bf (-3+k)}}{(17+13k)} A_{\pm } {\bf T^{(1)}} 
\right](w) +\cdots,
\nonu \\
A_3 (z) \, {\bf T_{\pm}^{(\frac{3}{2})}}(w) & = &
\frac{1}{(z-w)^2} \frac{i(1+k)}{2(5+k)} 
\left(
\begin{array}{c}
F_{21} \nonu \\
F_{12} 
\end{array}
\right)(w) \pm  
\frac{1}{(z-w)} \frac{i}{2}   {\bf T_{\pm}^{(\frac{3}{2})}}(w)+
\cdots,
\nonu \\
A_3(z) \, 
\left(
\begin{array}{c}
{\bf T^{(2)}} \nonu \\
{\bf W^{(2)}} 
\end{array}
\right)(w) & = & 
\pm \frac{1}{(z-w)^2} \frac{i}{2} {\bf T^{(1)}}(w) +\cdots,
\nonu \\
A_3(z) \left(
\begin{array}{c}
{\bf U^{(\frac{3}{2})}} \nonu \\
 {\bf V^{(\frac{3}{2})}} 
\end{array}
\right)(w) & = & 
-\frac{1}{(z-w)^2} \frac{i(1+k)}{2(5+k)} 
\left(
\begin{array}{c}
F_{11} \nonu \\
F_{22} 
\end{array}
\right)(w) \mp   \frac{1}{(z-w)} \frac{i}{2}
 \left(
\begin{array}{c}
{\bf U^{(\frac{3}{2})}} \nonu \\
{\bf V^{(\frac{3}{2})}} 
\end{array}
\right)(w)
 +\cdots,
\nonu \\
A_3(z)   \left(
\begin{array}{c}
{\bf U_{-}^{(2)}} \nonu \\
 {\bf V_{+}^{(2)}} 
\end{array}
\right)(w) & = & \mp \frac{1}{(z-w)} i  
\left(
\begin{array}{c}
{\bf U_{-}^{(2)}} \nonu \\
 {\bf V_{+}^{(2)}} 
\end{array}
\right)(w) + \cdots,
\nonu \\
A_3(z) \left(
\begin{array}{c}
{\bf U^{(\frac{5}{2})}} \nonu \\
 {\bf V^{(\frac{5}{2})}} 
\end{array}
\right)(w) & = & 
\frac{1}{(z-w)^2} \frac{i(21+5k)}{6(5+k)} \left[ 2 
 \left(
\begin{array}{c}
{\bf U^{(\frac{3}{2})}} \nonu \\
 {\bf V^{(\frac{3}{2})}} 
\end{array}
\right) \pm  \left(
\begin{array}{c}
G_{11} \nonu \\
G_{22}  \end{array}
\right) \right](w) 
\nonu \\
& \mp & \frac{1}{(z-w)} \frac{i}{2}  \left(
\begin{array}{c}
{\bf U^{(\frac{5}{2})}} \nonu \\
 {\bf V^{(\frac{5}{2})}} 
\end{array}
\right) (w) +\cdots,
\nonu \\
A_3(z) \, {\bf W_{\pm}^{(\frac{5}{2})}}(w)
& = &-
\frac{1}{(z-w)^2} \frac{i(21+5k)}{6(5+k)} \left[ 
2 {\bf T_{\pm}^{(\frac{3}{2})}} \pm
 \left(
\begin{array}{c}
G_{21} \nonu \\
G_{12}  \end{array}
\right) 
\right](w) 
 \pm 
\frac{1}{(z-w)} \frac{i}{2}  {\bf W_{\pm}^{(\frac{5}{2})}}(w) + 
\cdots,
\nonu \\
A_3(z) \, {\bf W^{(3)}}(w) & = & 
\frac{1}{(z-w)^2} \left[  -2 i {\bf T^{(2)}}  
+ 2i {\bf W^{(2)}} -\frac{4{\bf (-3+k)}}{
(17+13k)} A_3 {\bf T^{(1)}} \right](w) +\cdots.
\nonu 
\eea
It is easy to see that the OPEs containing the higher spin-$3$ current
do have the nonlinear terms in the right hand side.
This will be eliminated if one uses (\ref{bcgspin3}).
One can also compare the expressions in the nonlinear version in Part I 
and the above expressions. Some of the OPEs in Part I were present but
the corresponding OPEs in this paper do not appear. The nonlinear terms 
in Part I disappear in the above OPEs.  
It is easy to see that the above OPEs are equivalent to the corresponding
OPEs in \cite{BCG} via the field identifications in section $4$.

%%%%%%%%%%%%%%%%%%%%%%%%%%%%%%%%%%%%%%%%%%%%%%%%%%%%%%%%%%%%%%%%%%%%%
\subsection{ The OPEs between the spin-$1$ currents and 
the $16$ lowest higher spin currents }
%C%%%%%%%%%%%%%%%%%%%%%%%%%%%%%%%%%%%%%%%%%%%%%%%%%%%%%%%%%%%%%%%%%%%

The OPEs between the other three spin-$1$ currents found in Part I
and the $16$ higher spin currents obtained in section $3$
are
\bea
B_{\pm}(z) \, {\bf T_{\pm}^{(\frac{3}{2})}}(w) & = & 
\frac{1}{(z-w)^2} \frac{4i}{(5+k)} \left(
\begin{array}{c}
F_{22} \nonu \\
F_{11} 
\end{array}
\right)(w)  +  \frac{1}{(z-w)} i  
\left(
\begin{array}{c}
-{\bf V^{(\frac{3}{2})}} + G_{22} \nonu \\
 {\bf U^{(\frac{3}{2})}} + G_{11} 
\end{array}
\right)(w)+\cdots,
\nonu \\
B_{\pm}(z) \, {\bf T^{(2)}}(w) & = & 
\frac{1}{(z-w)} i 
\left(
\begin{array}{c}
{\bf V_{-}^{(2)}} \nonu \\
 { \bf U_{+}^{(2)}} 
\end{array}
\right)(w)+\cdots,
\nonu \\
B_{\pm}(z)  
\left(
\begin{array}{c}
{\bf U^{(\frac{3}{2})}} \nonu \\
 {\bf V^{(\frac{3}{2})}} 
\end{array}
\right)(w) & = & -
\frac{1}{(z-w)^2} \frac{4i}{(5+k)}
\left(
\begin{array}{c}
F_{12} \nonu \\
F_{21} 
\end{array}
\right)(w) \nonu \\
& + & 
\frac{1}{(z-w)} i \left[   \pm {\bf T_{\mp}^{(\frac{3}{2})}}  - 
\left(
\begin{array}{c}
 G_{12} \nonu \\
 G_{21} 
\end{array}
\right)\right](w)
+ \cdots,
\nonu \\
B_{\pm}(z) \left(
\begin{array}{c}
{\bf U_{+}^{(2)}} \nonu \\
 {\bf V_{-}^{(2)}} 
\end{array}
\right)(w) & = & 
\mp \frac{1}{(z-w)^2} i {\bf T^{(1)}}(w) 
+ 
\frac{1}{(z-w)} i \left[ {\bf T^{(2)}} + {\bf W^{(2)}} \right](w) +
\cdots, 
\nonu \\
B_{\pm}(z) \left(
\begin{array}{c}
{\bf U^{(\frac{5}{2})}} \nonu \\
 {\bf V^{(\frac{5}{2})}} 
\end{array}
\right)(w) & = & 
\frac{1}{(z-w)^2} \frac{4i(6+k)}{3(5+k)} \left[ 2 
{\bf T_{\mp}^{(\frac{3}{2})}} \mp  \left(
\begin{array}{c}
G_{12} \nonu \\
G_{21}  \end{array}
\right) \right](w) 
\mp  \frac{1}{(z-w)} i {\bf W_{\mp}^{(\frac{5}{2})}}(w) +\cdots,
\nonu \\
B_{\pm}(z) \, {\bf W^{(2)}}(w) & = & 
\frac{1}{(z-w)} i
 \left(
\begin{array}{c}
{\bf V_{-}^{(2)}} \nonu \\
 {\bf U_{+}^{(2)}} 
\end{array}
\right)(w) + \cdots,
\nonu \\
B_{\pm}(z) \,
{\bf W_{\pm}^{(\frac{5}{2})}}(w) & = & 
\frac{1}{(z-w)^2} \frac{4i(6+k)}{3(5+k)} \left[ -2 
\left(
\begin{array}{c}
{\bf V^{(\frac{3}{2})}} \nonu \\
 {\bf U^{(\frac{3}{2})}} 
\end{array}
\right) 
\pm   \left(
\begin{array}{c}
G_{22} \nonu \\
G_{11}  \end{array}
\right) \right](w) 
\nonu \\
& \pm & \frac{1}{(z-w)} i \left(
\begin{array}{c}
{\bf V^{(\frac{5}{2})}} \nonu \\
 { \bf U^{(\frac{5}{2})}} 
\end{array}
\right)(w) +\cdots,
\nonu \\
B_{\pm}(z) \, {\bf W^{(3)}}(w) & = & 
\frac{1}{(z-w)^2} \left[  \pm 4 i  \left(
\begin{array}{c}
{\bf V_{-}^{(2)}} \nonu \\
{\bf U_{+}^{(2)}} 
\end{array}
\right)  -\frac{4{\bf (-3+k)}}{(17+13k)} B_{\pm } {\bf T^{(1)}} 
\right](w) +\cdots,
\nonu \\
B_3 (z) \, {\bf T_{\pm}^{(\frac{3}{2})}}(w) & = &
-\frac{1}{(z-w)^2} \frac{2i}{(5+k)} 
\left(
\begin{array}{c}
F_{21} \nonu \\
F_{12} 
\end{array}
\right)(w) \pm 
\frac{1}{(z-w)} \frac{i}{2}   {\bf T_{\pm}^{(\frac{3}{2})}}(w)+
\cdots,
\nonu \\
B_3(z) \, 
\left(
\begin{array}{c}
{\bf T^{(2)}} \nonu \\
{\bf W^{(2)}} 
\end{array}
\right)(w) & = & 
 \frac{1}{(z-w)^2} \frac{i}{2} {\bf T^{(1)}}(w) +\cdots,
\nonu \\
B_3(z) \left(
\begin{array}{c}
{\bf U^{(\frac{3}{2})}} \nonu \\
 {\bf V^{(\frac{3}{2})}} 
\end{array}
\right)(w) & = & 
-\frac{1}{(z-w)^2} \frac{2i}{(5+k)} 
\left(
\begin{array}{c}
F_{11} \nonu \\
F_{22} 
\end{array}
\right)(w)  \pm   \frac{1}{(z-w)} \frac{i}{2}
 \left(
\begin{array}{c}
{\bf U^{(\frac{3}{2})}} \nonu \\
 {\bf V^{(\frac{3}{2})}} 
\end{array}
\right)(w)
 +\cdots,
\nonu \\
B_3(z)   \left(
\begin{array}{c}
{\bf U_{+}^{(2)}} \nonu \\
 {\bf V_{-}^{(2)}} 
\end{array}
\right)(w) & = & \pm \frac{1}{(z-w)} i  
\left(
\begin{array}{c}
{\bf U_{+}^{(2)}} \nonu \\
 {\bf V_{-}^{(2)}} 
\end{array}
\right)(w) + \cdots,
\nonu \\
B_3(z) \left(
\begin{array}{c}
{\bf U^{(\frac{5}{2})}} \nonu \\
 {\bf V^{(\frac{5}{2})}} 
\end{array}
\right)(w) & = & 
\frac{1}{(z-w)^2} \frac{2i(6+k)}{3(5+k)} \left[ 2 
 \left(
\begin{array}{c}
{\bf U^{(\frac{3}{2})}} \nonu \\
{\bf  V^{(\frac{3}{2})}} 
\end{array}
\right) \pm  \left(
\begin{array}{c}
G_{11} \nonu \\
G_{22}  \end{array}
\right) \right](w) 
\nonu \\
& \pm  & \frac{1}{(z-w)} \frac{i}{2}  \left(
\begin{array}{c}
{\bf U^{(\frac{5}{2})}} \nonu \\
 {\bf V^{(\frac{5}{2})}} 
\end{array}
\right) (w) +\cdots,
\nonu \\
B_3(z) \, {\bf W_{\pm}^{(\frac{5}{2})}}(w)
& = &
\frac{1}{(z-w)^2} \frac{2i(6+k)}{3(5+k)} \left[ 
2 {\bf T_{\pm}^{(\frac{3}{2})}} \pm
 \left(
\begin{array}{c}
G_{21} \nonu \\
G_{12}  \end{array}
\right) 
\right](w) \pm 
\frac{1}{(z-w)} \frac{i}{2}  {\bf W_{\pm}^{(\frac{5}{2})}}(w) + 
\cdots,
\nonu \\
B_3(z) \, {\bf W^{(3)}}(w) & = & 
\frac{1}{(z-w)^2} \left[  2 i {\bf T^{(2)}}  
+ 2i {\bf W^{(2)}} -\frac{4{\bf (-3+k)}}{(17+13k)} B_3 {\bf T^{(1)}} 
\right](w) +\cdots.
\nonu
\eea
It is easy to see that the OPEs containing the higher spin-$3$ current
do have the nonlinear terms in the right hand side.
This can be removed if one uses (\ref{bcgspin3}).
One can compare the expressions in the nonlinear version in Part I 
and the above expressions. Some of the OPEs in Part I were present but
the corresponding OPEs in this paper do not appear. The nonlinear terms 
in Part I disappear in the above OPEs.  
Due to the symmetry in the structure constants one can present the above 
OPEs by combining separate OPEs together.
One also checks that the above OPEs are equivalent to the corresponding
OPEs in \cite{BCG} (or in Appendix $D$) 
via the field identifications in section $4$. 

%%%%%%%%%%%%%%%%%%%%%%%%%%%%%%%%%%%%%%%%%%%%%%%%%%%%%%%%%%%%%%%%%%%%%
\subsection{ The OPEs between the spin-$\frac{3}{2}$ currents and 
the $16$ lowest higher spin currents }
%C%%%%%%%%%%%%%%%%%%%%%%%%%%%%%%%%%%%%%%%%%%%%%%%%%%%%%%%%%%%%%%%%%%%

The OPEs between the four spin-$\frac{3}{2}$ currents found in Part I
and the $16$ higher spin currents obtained in section $3$
are
\bea
\left(
\begin{array}{c}
G_{11} \nonu \\
G_{22}  \end{array}
\right)(z) \, {\bf T^{(1)}}(w)
& = & \frac{1}{(z-w)} \left[ \pm \left(
\begin{array}{c}
G_{11} \nonu \\
G_{22}  \end{array}
\right) + 2 
\left(
\begin{array}{c}
{\bf U^{(\frac{3}{2})}} \nonu \\
{\bf V^{(\frac{3}{2})}}  \end{array}
\right)
\right](w) +\cdots,
\nonu \\
 \left(
\begin{array}{c}
G_{11} \nonu \\
G_{22}
 \end{array}
\right)(z) \, {\bf T_{\pm}^{(\frac{3}{2})}}(w)
& = & \frac{1}{(z-w)^2} \frac{2i(1+k)}{(5+k)} B_{\mp}(w) 
\nonu \\
& + &
\frac{1}{(z-w)} \left[ 
%\frac{i(1+k)}{(5+k)} \pa B_{\mp}
\frac{1}{2} \pa \, \mbox{(pole-2)}
-  \left(
\begin{array}{c}
{\bf U_{+}^{(2)}} \nonu \\
{\bf V_{-}^{(2)}}
 \end{array}
\right) \right](w) +
\cdots, 
\nonu \\
 \left(
\begin{array}{c}
G_{11} \nonu \\
G_{22}
 \end{array}
\right)(z) \, {\bf T_{\mp}^{(\frac{3}{2})}}(w)
& = & \frac{1}{(z-w)^2} \frac{8i}{(5+k)} A_{\pm}(w) 
\nonu \\
& + &
\frac{1}{(z-w)} \left[ 
%\frac{4i}{(5+k)} \pa A_{\pm}
\frac{1}{2} \pa \, \mbox{(pole-2)}
-  \left(
\begin{array}{c}
{\bf U_{-}^{(2)}} \nonu \\
{\bf V_{+}^{(2)}}
 \end{array}
\right) \right](w) +
\cdots, 
\nonu \\
 \left(
\begin{array}{c}
G_{11} \nonu \\
G_{22}
 \end{array}
\right)(z) \, {\bf T^{(2)}}(w) & = & 
\frac{1}{(z-w)^2} \frac{(-3+k)}{2(5+k)} \left[  \left(
\begin{array}{c}
G_{11} \nonu \\
G_{22}
 \end{array}
\right) \pm 2 \left(
\begin{array}{c}
{\bf U^{(\frac{3}{2})}} \nonu \\
{\bf V^{(\frac{3}{2})}}  \end{array}
\right)  \right](w)
\nonu \\
&+&  \frac{1}{(z-w)} \left[  
\frac{1}{3} \pa \, \mbox{(pole-2)} +
\left(
\begin{array}{c}
{\bf U^{(\frac{5}{2})}} \nonu \\
{\bf V^{(\frac{5}{2})}}  \end{array}
\right)  
%+\frac{(-3+k)}{6(5+k)} \pa 
%\left(
% \begin{array}{c}
%G_{11} \nonu \\
%G_{22}
% \end{array}
%\right) + \frac{(-3+k)}{3(5+k)} \pa  \left(
%\begin{array}{c}
%{\bf U^{(\frac{3}{2})}} \nonu \\
%-{\bf V^{(\frac{3}{2})}}  \end{array}
%\right) 
\right](w) +\cdots, 
\nonu \\
\left( 
\begin{array}{c}
G_{11} \nonu \\
G_{22}
 \end{array}
\right)(z)  \left(
\begin{array}{c}
{\bf V^{(\frac{3}{2})}} \nonu \\
{\bf U^{(\frac{3}{2})}} 
\end{array}
\right)(w)
& = & 
\pm \frac{1}{(z-w)^3} \frac{8(1+k)}{(5+k)}
\nonu \\
& + & \frac{1}{(z-w)^2} \left[  \frac{8i}{(5+k)} A_3 -
\frac{2i(1+k)}{(5+k)} B_3 + {\bf T^{(1)}} \right](w)
\nonu \\
&+& \frac{1}{(z-w)} \left[  
%\frac{4i}{(5+k)} \pa A_3 -
%\frac{i(1+k)}{(5+k)} \pa B_3 + \frac{1}{2} \pa {\bf T^{(1)}} 
\frac{1}{2} \pa \, \mbox{(pole-2)} 
\pm  {\bf W^{(2)}} \pm  T \right](w) +\cdots,
\nonu \\
\left(
 \begin{array}{c}
G_{11} \nonu \\
G_{22}
 \end{array}
\right)(z) \, 
\left(
\begin{array}{c}
{\bf V_{+}^{(2)}} \nonu \\
{\bf U_{-}^{(2)}}
 \end{array}
\right)(w) & = &
-\frac{1}{(z-w)^2} \frac{2(3+k)}{(5+k)} \left[ 
\pm \left(
 \begin{array}{c}
G_{21} \nonu \\
G_{12}
 \end{array}
\right) + 2  {\bf T_{\pm}^{(\frac{3}{2})}}
\right](w) 
\nonu \\
& + & \frac{1}{(z-w)} \left[   
%\frac{2(3+k)}{3(5+k)} \pa \left(
% \begin{array}{c}
%G_{21} \nonu \\
%-G_{12}
% \end{array}
%\right) +   
% \frac{4(3+k)}{3(5+k)} \pa 
%{\bf T_{\pm}^{(\frac{3}{2})}} 
\frac{1}{3} \pa \, \mbox{(pole-2)} 
\mp  {\bf W_{\pm}^{(\frac{5}{2})}} \right](w) + \cdots, 
\nonu \\
\left(
 \begin{array}{c}
G_{11} \nonu \\
G_{22}
 \end{array}
\right)(z) \, 
\left(
\begin{array}{c}
{\bf V_{-}^{(2)}} \nonu \\
{\bf U_{+}^{(2)}}
 \end{array}
\right)(w) & = &
\frac{1}{(z-w)^2} \frac{(9+k)}{(5+k)} \left[ 
\pm \left(
 \begin{array}{c}
G_{12} \nonu \\
G_{21}
 \end{array}
\right) - 2  {\bf T_{\mp}^{(\frac{3}{2})}}
\right](w) 
\nonu \\
& + & \frac{1}{(z-w)} \left[   
%\frac{(9+k)}{3(5+k)} \pa \left(
% \begin{array}{c}
%G_{12} \nonu \\
%-G_{21}
% \end{array}
%\right) -  
% \frac{2(9+k)}{3(5+k)} \pa 
%{\bf T_{\mp}^{(\frac{3}{2})}} 
\frac{1}{3} \pa \, \mbox{(pole-2)} 
\mp  {\bf W_{\mp}^{(\frac{5}{2})}} \right](w) + \cdots, 
\nonu \\
\left(
 \begin{array}{c}
G_{11} \nonu \\
G_{22}
 \end{array}
\right)(z) \, 
 \left(
\begin{array}{c}
 {\bf V^{(\frac{5}{2})}} \nonu \\
{\bf U^{(\frac{5}{2})}} 
\end{array}
\right)(w)
& = &
\pm \frac{1}{(z-w)^3} \frac{8{\bf (-3+k)}}{3(5+k)} {\bf T^{(1)}}(w)
\nonu \\
& + & \frac{1}{(z-w)^2} \left[ 4 {\bf T^{(2)}} +
\frac{4(-3+k)}{3(5+k)} {\bf W^{(2)}} \right](w)
\nonu \\
&+& \frac{1}{(z-w)} \left[ 
\frac{1}{4} \pa \, \mbox{(pole-2)} 
\pm {\bf W^{(3)}}  
%+ \pa {\bf T^{(2)}} +
%\frac{(-3+k)}{3(5+k)} \pa {\bf W^{(2)}} 
\right. \nonu \\
& \pm & \left. 
\frac{4{\bf (-3+k)}}{(17+13k)} 
\left( T {\bf T^{(1)}} -\frac{1}{2} \pa^2 {\bf T^{(1)}} 
\right)
\right](w) +\cdots,
\nonu \\
 \left(
\begin{array}{c}
G_{11} \nonu \\
G_{22}
 \end{array}
\right)(z) \, {\bf W^{(2)}}(w) & = & 
-\frac{1}{(z-w)^2} \frac{3}{2} \left[  \left(
\begin{array}{c}
G_{11} \nonu \\
G_{22}
 \end{array}
\right) \pm 2 \left(
\begin{array}{c}
{\bf U^{(\frac{3}{2})}} \nonu \\
{\bf V^{(\frac{3}{2})}}  \end{array}
\right)  \right](w)
\nonu \\
& + &  \frac{1}{(z-w)} 
\frac{1}{3} \pa (\mbox{pole-2})(w)
%\frac{1}{2} \left[  \pa 
%\left(
% \begin{array}{c}
%G_{11} \nonu \\
%G_{22}
% \end{array}
%\right) + 2 \pa  \left(
%\begin{array}{c}
%{\bf U^{(\frac{3}{2})}} \nonu \\
%-{\bf V^{(\frac{3}{2})}}  \end{array}
%\right) \right](w) 
+\cdots, 
\nonu \\
 \left(
\begin{array}{c}
G_{11} \nonu \\
G_{22}
 \end{array}
\right)(z) \, {\bf W_{\pm}^{(\frac{5}{2})}}(w) & = & \pm
\frac{1}{(z-w)^2} \frac{16(3+k)}{3(5+k)} 
\left(
\begin{array}{c}
{\bf U_{+}^{(2)}} \nonu \\
{\bf V_{-}^{(2)}}
 \end{array}
\right)(w) \nonu \\
& + & \frac{1}{(z-w)} 
%\frac{4(3+k)}{3(5+k)} 
%\pa \left(
%\begin{array}{c}
%{\bf U_{+}^{(2)}} \nonu \\
%-{\bf V_{-}^{(2)}}
% \end{array}
\frac{1}{4} \pa \, \mbox{(pole-2)}(w)  +\cdots,
\nonu \\
 \left(
\begin{array}{c}
G_{11} \nonu \\
G_{22}
 \end{array}
\right)(z) \, {\bf W_{\mp}^{(\frac{5}{2})}}(w) & = & 
\pm \frac{1}{(z-w)^2} \frac{8(9+k)}{3(5+k)} 
\left(
\begin{array}{c}
{\bf U_{-}^{(2)}} \nonu \\
{\bf V_{+}^{(2)}}
 \end{array}
\right)(w)  +  \frac{1}{(z-w)} 
\frac{1}{4} \pa (\mbox{pole-2})(w) +\cdots,
\nonu \\
\left(
 \begin{array}{c}
G_{11} \nonu \\
G_{22}
 \end{array}
\right)(z) \, {\bf W^{(3)}}(w) & = & -
\frac{1}{(z-w)^3} \frac{2(-3+k)(113+61k)}{
3(5+k)(17+13k)} \left[\pm  \left(
 \begin{array}{c}
G_{11} \nonu \\
G_{22}  
 \end{array}
\right) +2 
 \left(
\begin{array}{c}
{\bf U^{(\frac{3}{2})}} \nonu \\
{\bf V^{(\frac{3}{2})}}  \end{array}
\right) \right](w) 
\nonu \\
&+& \frac{1}{(z-w)^2} \left[ \mp 5  \left(
\begin{array}{c}
{\bf U^{(\frac{5}{2})}} \nonu \\
{\bf V^{(\frac{5}{2})}}  \end{array}
\right)
\pm  \frac{4 (-3+k)}{(17+13k)} \pa \left(
 \begin{array}{c}
G_{11} \nonu \\
G_{22}  
 \end{array}
\right) \right. \nonu \\
& + & \left.  \frac{8(-3+k)}{(17+13k)}
 \pa \left(
\begin{array}{c}
{\bf U^{(\frac{3}{2})}} \nonu \\
{\bf V^{(\frac{3}{2})}}  \end{array}
\right) -\frac{6{\bf (-3+k)}}{(17+13k)} \left(
\begin{array}{c} G_{11}
\nonu \\
G_{22}
\end{array}
\right) {\bf T^{(1)}}
\right](w)
\nonu \\
& + & \frac{1}{(z-w)} \left[ \frac{1}{5} \pa (\mbox{pole-2})  
%\pa \left( \begin{array}{c}
%-{\bf U^{(\frac{5}{2})}} \nonu \\
%{\bf V^{(\frac{5}{2})}}  \end{array}
%\right) 
\pm \frac{6(-3+k)}{5(17+13k)} \pa^2 
 \left(
 \begin{array}{c}
G_{11} \nonu \\
G_{22}  
 \end{array}
\right)
 \right. \nonu \\
& -&  \frac{4{\bf (-3+k)}}{5(17+13k)} \pa   \left(
 \begin{array}{c}
G_{11} \nonu \\
G_{22}  
 \end{array}
\right) {\bf T^{(1)}} +\frac{12(-3+k)}{5(17+13k)} \pa^2   
 \left(
\begin{array}{c}
{\bf U^{(\frac{3}{2})}} \nonu \\
{\bf V^{(\frac{3}{2})}}  \end{array}
\right) \nonu \\
& \mp & \left. \frac{4 (-3+k)}{(17+13k)} T 
 \left(
 \begin{array}{c}
G_{11} \nonu \\
G_{22}  
 \end{array}
\right) -\frac{8{\bf (-3+k)}}{(17+13k)} T 
\left(
\begin{array}{c}
{\bf U^{(\frac{3}{2})}} \nonu \\
{\bf V^{(\frac{3}{2})}}  \end{array}
\right)
\right. \nonu \\
&+ & \left.  \frac{6{\bf (-3+k)}}{5(17+13k)}    \left(
 \begin{array}{c}
G_{11} \nonu \\
G_{22}  
 \end{array}
\right) \pa {\bf T^{(1)}}
  \right](w) +\cdots,
\nonu \\
\left(
\begin{array}{c}
G_{12} \nonu \\
G_{21}  \end{array}
\right)(z) \, {\bf T^{(1)}}(w)
& = &  \frac{1}{(z-w)} \left[ \mp \left(
\begin{array}{c}
G_{12} \nonu \\
G_{21}  \end{array}
\right) + 2 
{\bf T_{\mp}^{(\frac{3}{2})}}
\right](w) +\cdots,
\nonu \\
 \left(
\begin{array}{c}
G_{12} \nonu \\
G_{21}
 \end{array}
\right)(z) \, {\bf T_{\pm}^{(\frac{3}{2})}}(w)
& = & 
\mp
\frac{1}{(z-w)^3} \frac{8(1+k)}{(5+k)} 
\nonu \\
& + &
\frac{1}{(z-w)^2} \left[ -\frac{8i}{(5+k)} A_3 
-\frac{2i(1+k)}{(5+k)} B_3 + {\bf T^{(1)}}
\right](w) \nonu \\
& + & 
\frac{1}{(z-w)} \left[  
%-\frac{4i}{(5+k)} \pa A_3 
%-\frac{i(1+k)}{(5+k)} \pa B_3 + \frac{1}{2} \pa {\bf T^{(1)}} 
%\mp T  
\frac{1}{2} \pa \, \mbox{(pole-2)} 
\mp {\bf T^{(2)}} \mp T \right](w) +
\cdots, 
\nonu \\
 \left(
\begin{array}{c}
G_{12} \nonu \\
G_{21}
 \end{array}
\right)(z) \, {\bf T^{(2)}}(w) & = & 
\frac{1}{(z-w)^2} \frac{3}{2} \left[  - \left(
\begin{array}{c}
G_{12} \nonu \\
G_{21}
 \end{array}
\right) \pm  2 {\bf T_{\mp}^{(\frac{3}{2})}} \right](w)
+  \frac{1}{(z-w)} 
%\frac{1}{2} \left[ - \pa 
%\left(
% \begin{array}{c}
%G_{12} \nonu \\
%G_{21}
% \end{array}
%\right) +  2 \pa  {\bf T_{\mp}^{(\frac{3}{2})}} 
\frac{1}{3} \pa \, \mbox{(pole-2)}(w) +\cdots, 
\nonu \\
\left( 
\begin{array}{c}
G_{12} \nonu \\
G_{21}
 \end{array}
\right)(z)  \left(
\begin{array}{c}
{\bf U^{(\frac{3}{2})}} \nonu \\
{\bf V^{(\frac{3}{2})}} 
\end{array}
\right)(w)
& = & 
\frac{1}{(z-w)^2}   \frac{8i}{(5+k)} A_{\pm} (w)
\nonu \\
&+& \frac{1}{(z-w)} \left[  
%\frac{4i}{(5+k)} \pa A_{\pm} 
\frac{1}{2} \pa \, \mbox{(pole-2)} 
+  \left(
\begin{array}{c}
{\bf U_{-}^{(2)}} \nonu \\
{\bf V_{+}^{(2)}}
 \end{array}
\right)
\right](w) +\cdots,
\nonu \\
\left(
 \begin{array}{c}
G_{12} \nonu \\
G_{21}
 \end{array}
\right)(z) \, 
\left(
\begin{array}{c}
{\bf U_{+}^{(2)}} \nonu \\
{\bf V_{-}^{(2)}}
 \end{array}
\right)(w) & = &
\frac{1}{(z-w)^2} \frac{(9+k)}{(5+k)} \left[ 
\pm
\left(
 \begin{array}{c}
G_{11} \nonu \\
G_{22}
 \end{array}
\right) + 2   \left(
\begin{array}{c}
{\bf U^{(\frac{3}{2})}} \nonu \\
{\bf V^{(\frac{3}{2})}} 
\end{array}
\right)
\right](w) 
\nonu \\
& + & \frac{1}{(z-w)} \left[   
%\frac{(9+k)}{3(5+k)} \pa \left(
% \begin{array}{c}
%G_{11} \nonu \\
%-G_{22}
% \end{array}
%\right) +   
% \frac{2(9+k)}{3(5+k)} \pa 
% \left(
%\begin{array}{c}
%{\bf U^{(\frac{3}{2})}} \nonu \\
%{\bf V^{(\frac{3}{2})}} 
%\end{array}
%\right)
\frac{1}{3} \pa \, \mbox{(pole-2)} 
\mp     \left(
\begin{array}{c}
{\bf U^{(\frac{5}{2})}} \nonu \\
{\bf V^{(\frac{5}{2})}} 
\end{array}
\right) \right](w) + \cdots, 
\nonu \\
\left(
 \begin{array}{c}
G_{12} \nonu \\
G_{21}
 \end{array}
\right)(z) \, 
 \left(
\begin{array}{c}
{\bf U^{(\frac{5}{2})}} \nonu \\
{\bf V^{(\frac{5}{2})}} 
\end{array}
\right)(w)
& = & \pm
\frac{1}{(z-w)^2} \frac{8(9+k)}{3(5+k)} \left(
\begin{array}{c}
{\bf U_{-}^{(2)}} \nonu \\
{ \bf V_{+}^{(2)}}
 \end{array}
\right) (w)
+ \frac{1}{(z-w)} 
%\frac{2(9+k)}{3(5+k)} \pa \left(
%\begin{array}{c}
%{\bf U_{-}^{(2)}} \nonu \\
%-{\bf V_{+}^{(2)}}
% \end{array}
%\right)
\frac{1}{4} \pa \, \mbox{(pole-2)}(w) +\cdots,
\nonu \\
\left( 
\begin{array}{c}
G_{12} \nonu \\
G_{21}
 \end{array}
\right)(z)  \left(
\begin{array}{c}
{\bf V^{(\frac{3}{2})}} \nonu \\
{\bf U^{(\frac{3}{2})}} 
\end{array}
\right)(w)
& = & 
-\frac{1}{(z-w)^2}   \frac{2i(1+k)}{(5+k)} B_{\pm} (w)
\nonu \\
&+& \frac{1}{(z-w)} \left[ 
%- \frac{i(1+k)}{(5+k)} \pa B_{\pm} 
\frac{1}{2} \pa \, \mbox{(pole-2)} 
+  \left(
\begin{array}{c}
{\bf V_{-}^{(2)}} \nonu \\
{\bf U_{+}^{(2)}}
 \end{array}
\right)
\right](w) +\cdots,
\nonu \\
\left(
 \begin{array}{c}
G_{12} \nonu \\
G_{21}
 \end{array}
\right)(z) \, 
\left(
\begin{array}{c}
{\bf V_{+}^{(2)}} \nonu \\
{\bf U_{-}^{(2)}}
 \end{array}
\right)(w) & = &
\frac{1}{(z-w)^2} \frac{2(3+k)}{(5+k)} \left[ 
\mp
\left(
 \begin{array}{c}
G_{22} \nonu \\
G_{11}
 \end{array}
\right) + 2   \left(
\begin{array}{c}
{\bf V^{(\frac{3}{2})}} \nonu \\
{\bf U^{(\frac{3}{2})}} 
\end{array}
\right)
\right](w) 
\nonu \\
& + & \frac{1}{(z-w)} \left[  
%- \frac{2(3+k)}{3(5+k)} \pa \left(
% \begin{array}{c}
%G_{22} \nonu \\
%-G_{11}
% \end{array}
%\right) +   
% \frac{4(3+k)}{3(5+k)} \pa 
% \left(
%\begin{array}{c}
%{\bf V^{(\frac{3}{2})}} \nonu \\
%{\bf U^{(\frac{3}{2})}} 
%\end{array}
%\right)
% \right. \nonu \\
\frac{1}{3} \pa \, \mbox{(pole-2)} 
\mp     \left(
\begin{array}{c}
{\bf V^{(\frac{5}{2})}} \nonu \\
{\bf U^{(\frac{5}{2})}} 
\end{array}
\right) \right](w) + \cdots, 
\nonu \\
\left(
 \begin{array}{c}
G_{12} \nonu \\
G_{21}
 \end{array}
\right)(z) \, 
 \left(
\begin{array}{c}
{\bf V^{(\frac{5}{2})}} \nonu \\
{\bf U^{(\frac{5}{2})}} 
\end{array}
\right)(w)
& = &
\pm
\frac{1}{(z-w)^2} \frac{16(3+k)}{3(5+k)} \left(
\begin{array}{c}
{\bf V_{-}^{(2)}} \nonu \\
{\bf U_{+}^{(2)}}
 \end{array}
\right) (w)
\nonu \\
&+& \frac{1}{(z-w)} 
%\frac{4(3+k)}{3(5+k)} \pa \left(
%\begin{array}{c}
%{\bf V_{-}^{(2)}} \nonu \\
%-{\bf U_{+}^{(2)}}
% \end{array}
%\right)
\frac{1}{4} \pa \, \mbox{(pole-2)}(w) +\cdots,
\nonu \\
\left(
\begin{array}{c}
G_{12} \nonu \\
G_{21}
 \end{array}
\right)(z) \, {\bf W^{(2)}}(w) & = & 
\frac{1}{(z-w)^2} \frac{(-3+k)}{2(5+k)} \left[  \left(
\begin{array}{c}
G_{12} \nonu \\
G_{21}
 \end{array}
\right) \mp 2 { \bf T_{\mp}^{(\frac{3}{2})}} \right](w)
\nonu \\
&+ &  \frac{1}{(z-w)} \left[ 
%\frac{ 
%(-3+k)}{6(5+k)} \pa  \left(
%\begin{array}{c}
%G_{12} \nonu \\
%G_{21}
% \end{array}
%\right) \mp \frac{(-3+k)}{3(5+k)}  {\bf T_{\mp}^{(\frac{3}{2})}}
\frac{1}{3} \pa \, \mbox{(pole-2)} 
  +  
{\bf W_{\mp}^{(\frac{5}{2})}} \right](w) +\cdots, 
\nonu \\
 \left(
\begin{array}{c}
G_{12} \nonu \\
G_{21}
 \end{array}
\right)(z) \, {\bf W_{\pm}^{(\frac{5}{2})}}(w) & = & 
\mp
\frac{1}{(z-w)^3} \frac{8{\bf (-3+k)}}{3(5+k)} {\bf T^{(1)}}(w) 
\nonu \\
& + &  
\frac{1}{(z-w)^2} \left[ \frac{4(-3+k)}{3(5+k)} {\bf T^{(2)}} +
4 {\bf W^{(2)}} \right](w) \nonu \\
& + & \frac{1}{(z-w)} \left[ 
\frac{1}{4} \pa \, \mbox{(pole-2)} 
\mp {\bf W^{(3)}} 
%+
% \frac{(-3+k)}{3(5+k)} \pa {\bf T^{(2)}} +
% \pa {\bf W^{(2)}}  
\right. \nonu \\
& \mp & \left. \frac{4{\bf (-3+k)}}{(17+13k)} \left(
T \, {\bf T^{(1)}} -\frac{1}{2} \pa^2 {\bf T^{(1)}} \right) \right](w)  +\cdots,
\nonu \\
\left(
 \begin{array}{c}
G_{12} \nonu \\
G_{21}
 \end{array}
\right)(z) \, {\bf W^{(3)}}(w) & = & 
\frac{1}{(z-w)^3} \frac{2(-3+k)(113+61k)}{
3(5+k)(17+13k)} \left[ \pm \left(
 \begin{array}{c}
G_{12} \nonu \\
G_{21}  
 \end{array}
\right) -2 {\bf T_{\mp}^{(\frac{3}{2})}} \right](w) 
\nonu \\
&+& \frac{1}{(z-w)^2} \left[ \pm 5  {\bf W_{\mp}^{(\frac{5}{2})}} 
\mp  \frac{4(-3+k)}{(17+13k)} \pa \left(
 \begin{array}{c}
G_{12} \nonu \\
G_{21}  
 \end{array}
\right) \right. \nonu \\
& + & \left.  \frac{8(-3+k)}{(17+13k)}
 \pa 
 {\bf T_{\mp}^{(\frac{3}{2})}}
 -\frac{6{\bf (-3+k)}}{(17+13k)} \left(
\begin{array}{c} G_{12}
\nonu \\
G_{21}
\end{array}
\right) {\bf T^{(1)}}
\right](w)
\nonu \\
& + & \frac{1}{(z-w)} \left[ 
%\pm  \pa   {\bf W_{\mp}^{(\frac{5}{2})}} 
\frac{1}{5} \pa (\mbox{pole-2}) \mp
\frac{6(-3+k)}{5(17+13k)} \pa^2 
 \left(
 \begin{array}{c}
G_{12} \nonu \\
G_{21}  
 \end{array}
\right)
 \right. \nonu \\
& -&  \frac{4{\bf (-3+k)}}{5(17+13k)} \pa   \left(
 \begin{array}{c}
G_{12} \nonu \\
G_{21}  
 \end{array}
\right) {\bf T^{(1)}} +\frac{12(-3+k)}{5(17+13k)} \pa^2   
 {\bf T_{\mp}^{(\frac{3}{2})}} \nonu \\
& \pm  & \left. \frac{4 (-3+k)}{(17+13k)} T 
 \left(
 \begin{array}{c}
G_{12} \nonu \\
G_{21}  
 \end{array}
\right) -\frac{8 (-3+k)}{(17+13k)} T 
{\bf T_{\mp}^{(\frac{3}{2})}}
\right. \nonu \\
&+ & \left.  
 \frac{6{\bf (-3+k)}}{5(17+13k)}    \left(
 \begin{array}{c}
G_{12} \nonu \\
G_{21}  
 \end{array}
\right) \pa {\bf T^{(1)}}
 \right](w) +\cdots.
\nonu
\eea
It is easy to see that the OPEs containing the higher spin-$\frac{5}{2}$ 
or spin-$3$ current
do have the nonlinear terms in the right hand side.
These nonlinear terms will disappear in the different basis in next 
Appendix. In doing this, 
one should realize that there exist nontrivial OPEs 
(\ref{g11221221t1}) in section $4$.
In the right hand side of those OPEs, the higher spin-$\frac{3}{2}$ currents
occur and this will contribute to the corresponding 
terms in the first order pole above.
In this case, the symmetry of the structure constants allows us to write the 
OPEs together and this is new feature compared to the expressions in Part I.
The nonlinear terms in Part I do not appear in the above OPEs.
One also checks that the above OPEs are equivalent to the corresponding
OPEs in \cite{BCG} (or in Appendix $D$) 
via the field identifications in section $4$. 
In next Appendix, the above OPEs can be written in different basis 
where the nonlinear terms will disappear.

%%%%%%%%%%%%%%%%%%%%%%%%%%%%%%%%%%%%%%%%%%%%%%%%%%%%%%%%%%%%%%%%%%%%%
%%%%%%%%%%%%%%%%%%%%%%%%%%%%%%%%%%%%%%%%%%%%%%%%%%%%%%%%%%%%%%%%%%%%%
\section{ Other basis where the OPEs are linear }
%B%%%%%%%%%%%%%%%%%%%%%%%%%%%%%%%%%%%%%%%%%%%%%%%%%%%%%%%%%%%%%%%%%%%
%%%%%%%%%%%%%%%%%%%%%%%%%%%%%%%%%%%%%%%%%%%%%%%%%%%%%%%%%%%%%%%%%%%%

In order to compare the fields in \cite{BCG} with the ones in this paper,
their defining relations are given.

%%%%%%%%%%%%%%%%%%%%%%%%%%%%%%%%%%%%%%%%%%%%%%%%%%%%%%%%%%%%%%%%%%%%%
\subsection{ The OPEs in the linear version }
%C%%%%%%%%%%%%%%%%%%%%%%%%%%%%%%%%%%%%%%%%%%%%%%%%%%%%%%%%%%%%%%%%%%%

Let us present the description of \cite{BCG} as follows: 
\bea
U (z) \, V_0^{(s)}(w)  & = & +\cdots,
\nonu \\
Q^a(z) \,  V_0^{(s)}(w)  & = & +\cdots,
\nonu \\
A^{\pm, i}(z) \, V_0^{(s)}(w)  & = & +\cdots,
\nonu \\
G^a(z) \, V_0^{(s)}(w)  & = & \frac{1}{(z-w)} \, V_{\frac{1}{2}}^{(s), a}(w) 
+\cdots,  
\nonu \\
U (z) \, V_{\frac{1}{2}}^{(s), a}(w)  & = & +\cdots,
\nonu \\
Q^a(z) \,  V_{\frac{1}{2}}^{(s), b}(w)  & = & +\cdots,
\nonu \\
A^{\pm, i}(z) \, V_{\frac{1}{2}}^{(s), a}(w)  & = & \frac{1}{(z-w)} \,
\alpha_{ab}^{\pm, i} \, V_{\frac{1}{2}}^{(s), b}(w)+\cdots,
\nonu \\
G^a(z) \, V_{\frac{1}{2}}^{(s), b}(w)  & = & \frac{1}{(z-w)^2} \, 2 s 
\, \delta^{ab} \, V_0^{(s)}(w) \nonu \\
& + &
\frac{1}{(z-w)} \, \left[ \alpha_{ab}^{+, i} \, V_{1}^{(s),+, i} +
\alpha_{ab}^{-, i} \, V_{1}^{(s),-, i} +\delta^{ab} \,
\pa V_{0}^{(s)} \right](w) 
+\cdots,  
\nonu \\
U (z) \, V_{1}^{(s), \pm i}(w)  & = & +\cdots,
\nonu \\
Q^a(z) \,  V_{1}^{(s),\pm, i}(w)  & = & \pm \frac{1}{(z-w)} \,
 2 \, \alpha_{ab}^{\pm, i} \, V_{\frac{1}{2}}^{(s),b}(w) +\cdots,
\nonu \\
A^{\pm, i}(z) \, V_{1}^{(s), \pm, j}(w)  & = & \frac{1}{(z-w)^2} \,
2s \, \delta^{ij} \, V_0^{(s)}(w) + \frac{1}{(z-w)} \,
\ep^{ijk} \, V_{1}^{(s), \pm, k}(w)+\cdots,
\nonu \\
A^{\pm, i}(z) \, V_{1}^{(s), \mp, j}(w)  & = & +\cdots,
\nonu \\
G^a(z) \, V_1^{(s), \pm, i}(w)  & = & \frac{1}{(z-w)^2} \, 
4 \, (s + \gamma_{\mp})  \, \alpha_{ab}^{\pm, i} \, V_{\frac{1}{2}}^{(s), b}(w)  
\nonu \\
& + &
\frac{1}{(z-w)} \, \left[ \frac{1}{(2s+1)} \pa (\mbox{pole-2}) \mp
\alpha_{ab}^{\pm, i} \, V_{\frac{3}{2}}^{(s), b} \right](w) 
+\cdots,  
\nonu \\
U (z) \, V_{\frac{3}{2}}^{(s), a}(w)  & = & -\frac{1}{(z-w)^2} \, 
2 \, V_{\frac{1}{2}}^{(s), a}(w) +\cdots,
\nonu \\
Q^a(z) \,  V_{\frac{3}{2}}^{(s),b}(w)  & = & 
\frac{1}{(z-w)^2} \,
4 \, s \, \delta^{ab} \, V_0^{(s)}(w)
\nonu \\
& + &   \frac{1}{(z-w)} \,
2 \left[ \alpha_{ab}^{+, i} \, V_{1}^{(s),+, i} 
+ \alpha_{ab}^{-, i} \, V_{1}^{(s),-, i} -\delta^{ab} \, 
\pa V_{0}^{(s)} \right](w)  +\cdots,
\nonu \\
A^{\pm, i}(z) \, V_{\frac{3}{2}}^{(s), a}(w)  & = & \pm \frac{1}{(z-w)^2} \,
\left[ \frac{8s(s+1) + 4 \gamma_{\mp}}{(2s+1)} \right]  \,
\alpha_{ab}^{\pm, i} \,
V_{\frac{1}{2}}^{(s), b}(w) \nonu \\
& + &  \frac{1}{(z-w)} \,
\alpha_{ab}^{\pm, i} \, 
V_{\frac{3}{2}}^{(s), b}(w)+\cdots,
\nonu \\
G^a(z) \, V_{\frac{3}{2}}^{(s), b}(w)  & = & 
- \frac{1}{(z-w)^3} \, \left[ \frac{16s(s+1)(2\gamma-1)}
{(2s+1)}\right] \delta^{ab} \, V_0^{(s)}(w)  
\nonu \\
& - &  \frac{1}{(z-w)^2} \, \frac{8(s+1)}{(2s+1)} \,
\left[ (s + \gamma_{+}) \,  \alpha_{ab}^{+, i} \, V_{1}^{(s), +, i}  
- (s + \gamma_{-})  \, \alpha_{ab}^{-, i} \, V_{1}^{(s), -, i}  
\right](w)
\nonu \\
& + &
\frac{1}{(z-w)} \, \left[ \frac{1}{2(s+1)}\, \pa (\mbox{pole-2}) +
\delta^{ab} \, V_{2}^{(s)} \right](w) 
+\cdots,  
\nonu \\
U (z) \, V_{2}^{(s)}(w)  & = & \frac{1}{(z-w)^3} \,
8 \, s \, V_{0}^{(s)}(w) -\frac{1}{(z-w)^2} \, 
4 \, \pa V_{0}^{(s)}(w) +\cdots,
\nonu \\
Q^a(z) \,  V_{2}^{(s)}(w)  & = & 
- \frac{1}{(z-w)^2} \,
2  \, (2s+1)  \, V_{\frac{1}{2}}^{(s), a}(w)
+    \frac{1}{(z-w)} \,
2 \,
\pa V_{\frac{1}{2}}^{(s), a}(w)  +\cdots,
\nonu \\
A^{\pm, i}(z) \, V_{2}^{(s)}(w)  & = & \pm \frac{1}{(z-w)^2} \,
2(s+1) \, V_{1}^{(s), \pm, i}(w) +\cdots,
\nonu \\
G^a(z) \, V_{2}^{(s)}(w)  & = & 
\frac{1}{(z-w)^3} \, \left[ \frac{16s(s+1)(2\gamma-1)}
{(2s+1)}\right]  \, V_{\frac{1}{2}}^{(s), a}(w)  
\nonu \\
& + &  \frac{1}{(z-w)^2} \, (2s+3) \,
V_{\frac{3}{2}}^{(s), a}(w)
+ 
\frac{1}{(z-w)} \, \pa  V_{\frac{3}{2}}^{(s), a}(w) 
+\cdots,  
\nonu \\
T(z) \, V_{2}^{(s)}(w)  & = & 
-\frac{1}{(z-w)^4} \, \left[ \frac{24s(s+1)(2\gamma-1)}
{(2s+1)}\right]  \, V_{0}^{(s)}(w)  
\nonu \\
& + &  \frac{1}{(z-w)^2} \, (s+2) \,
V_{2}^{(s)}(w)
+ 
\frac{1}{(z-w)} \, \pa  V_{2}^{(s)}(w) 
+\cdots,  
\nonu \\
T(z) \, V_{0}^{(s)}(w)  & = & 
  \frac{1}{(z-w)^2} \, s \,
V_{0}^{(s)}(w)
+ 
\frac{1}{(z-w)} \, \pa  V_{0}^{(s)}(w) 
+\cdots,  
\nonu \\
T(z) \, V_{\frac{1}{2}}^{(s), a}(w)  & = & 
  \frac{1}{(z-w)^2} \, (s+\frac{1}{2}) \,
V_{\frac{1}{2}}^{(s), a}(w)
+ 
\frac{1}{(z-w)} \, \pa  V_{\frac{1}{2}}^{(s), a}(w) 
+\cdots,  
\nonu \\
T(z) \, V_{1}^{(s), \pm, i}(w)  & = & 
  \frac{1}{(z-w)^2} \, (s+1) \,
V_{1}^{(s), \pm, i}(w)
+ 
\frac{1}{(z-w)} \, \pa  V_{1}^{(s), \pm, i}(w) 
+\cdots,  
\nonu \\
T(z) \, V_{\frac{3}{2}}^{(s), a}(w)  & = & 
  \frac{1}{(z-w)^2} \, (s+\frac{3}{2}) \,
V_{\frac{3}{2}}^{(s), a}(w)
+ 
\frac{1}{(z-w)} \, \pa  V_{\frac{3}{2}}^{(s), a}(w) 
+\cdots.  
\nonu 
\eea
Here the spin-$\frac{1}{2}$ currents $Q^a(z)$ play the role of previous 
one $\Gamma^a(z)$ in Appendix $A$. In the present notation, 
this is equivalent to the spin-$\frac{1}{2}$ current $F_a(z)$.
One has $\gamma_{+} \equiv \gamma$ and $\gamma_{-} \equiv 1-\gamma$.
Some of the typos appearing in \cite{BCG} are corrected. 
In particular, note that the first order pole of 
the OPE $G^a(z) \, V_{\frac{3}{2}}^{(s),b}(w)$ should 
possess the $\delta^{ab} \, V_2^{(s)}$ term in the right hand side.  
These OPEs hold for any spin $s=1, 2, \cdots$.
The higher spin current $V_2^{(s)}(w)$ does not contain 
the third order pole in the OPE with $T(z)$ and has nontrivial 
fourth order pole.
The spin-$\frac{3}{2}$ currents which are the supersymmetry
generators determine all the higher spin currents via the OPEs 
containing $G_a(z)$ in the above OPEs.  

%%%%%%%%%%%%%%%%%%%%%%%%%%%%%%%%%%%%%%%%%%%%%%%%%%%%%%%%%%%%%%%%%%%%%
\subsection{ The higher spin currents in the nonlinear version }
%C%%%%%%%%%%%%%%%%%%%%%%%%%%%%%%%%%%%%%%%%%%%%%%%%%%%%%%%%%%%%%%%%%%%

One has the explicit relations between the higher spin currents 
in the nonlinear version \cite{BCG} and the 
corresponding higher spin currents in this paper.
For the higher spin-$2$ currents, one has
%%%%%%%%%%%%%%%%%%%%%%%%%%%%%%%%%%%%%%%%%%%%%%%%%%%%%%%%%%%%
\bea
\widetilde{V}_1^{(1), +1}(z) & = & 
\left[ 2i \left( {\bf U_{-,non}^{(2)}} - 
{\bf V_{+,non}^{(2)}} \right) + \frac{8i}{(5+k)} \hat{A}_1 \hat{B}_3 \right](z),
\nonu \\
\widetilde{V}_1^{(1), +2}(z) & = & 
\left[ -2 \left( {\bf U_{-,non}^{(2)}} + 
{\bf V_{+,non}^{(2)}} \right) - \frac{8i}{(5+k)} \hat{A}_2 \hat{B}_3 \right](z),
\nonu \\
\widetilde{V}_1^{(1), +3}(z) & = & 
\left[ 2i \left( {\bf T_{non}^{(2)}} - 
{\bf W_{non}^{(2)}} \right) + \frac{8i}{(5+k)} \hat{A}_3 \hat{B}_3
+\frac{12 i k}{(3+7k)} \hat{T} \right](z),
\nonu \\
\widetilde{V}_1^{(1), -1}(z) & = & 
\left[ 2i \left( {\bf U_{+,non}^{(2)}} - 
{\bf V_{-,non}^{(2)}} \right) + \frac{8i}{(5+k)} \hat{A}_3 \hat{B}_1 \right](z),
\nonu \\
\widetilde{V}_1^{(1), -2}(z) & = & 
\left[ -2 \left( {\bf U_{+,non}^{(2)}} + 
{\bf V_{-,non}^{(2)}} \right) + \frac{8i}{(5+k)} \hat{A}_3 \hat{B}_2 \right](z),
\nonu \\
\widetilde{V}_1^{(1), +3}(z) & = & 
\left[ -2i \left( {\bf T_{non}^{(2)}} + 
{\bf W_{non}^{(2)}} \right) + \frac{4i}{(5+k)} (
\hat{A}_i \hat{A}_i + \hat{B}_i \hat{B}_i)
+\frac{4 i (3+4k)}{(3+7k)} \hat{T} \right](z).
\label{tildehigherspin2}
\eea
From the relations $(4.23)$-$(4.25)$ of \cite{BCG}, one can rewrite 
the higher spin currents in the nonlinear version in terms of those in the
linear version. Then using the explicit relations (\ref{v3half}), 
(\ref{v5half}) and (\ref{bcgspin3}), one can rewrite them 
in terms of the higher spin currents in the linear version.
Finally, using the explicit relations appearing in Appendix $B$, one
arrives at the final results in (\ref{tildehigherspin2}).
This was also realized in \cite{AK1411}. 

One can analyze the other higher spin currents.
The following relations can be obtained
%%%%%%%%%%%%%%%%%%%%%%%%%%%%%%%%%%%%%%%%%%%%%%%%%%%%%%%%%%%%%%%%%%%%%%%%
\bea
\widetilde{V}_{\frac{3}{2}}^{(1), 0}(z) & = & 
-2 i \sqrt{2} \left( {\bf W_{+,non}^{(\frac{5}{2})}} + 
{\bf W_{-,non}^{(\frac{5}{2})}} \right)(z)
-\frac{4 \sqrt{2}}{(5+k)} 
 \left[
\hat{A}_{-} \hat{G}_{11} 
+ \hat{A}_{-}  {\bf U_{non}^{(\frac{3}{2})}}
- 2 \hat{A}_{3}  \hat{G}_{21}- 2 \hat{A}_{3}  {\bf T_{+,non}^{(\frac{3}{2})}}
\right. \nonu \\
& + & \left.
 \hat{B}_{-}  \hat{G}_{22} -  \hat{B}_{-}  {\bf V_{non}^{(\frac{3}{2})}} 
+2 \hat{B}_3   {\bf T_{+,non}^{(\frac{3}{2})}} + 
\frac{2 i}{3}  \pa \hat{G}_{21}
\right](z)
 +\frac{4 \sqrt{2}}{(5+k)}
 \left[
\hat{A}_{+}  \hat{G}_{22} 
- \hat{A}_{+}  {\bf V_{non}^{(\frac{3}{2})}}
\right. \nonu \\
& - &  \left.
2 \hat{A}_{3}  \hat{G}_{12}+ 2 \hat{A}_{3}  {\bf T_{-,non}^{(\frac{3}{2})}}
+
 \hat{B}_{+}  \hat{G}_{11} + \hat{B}_{+}  {\bf U_{non}^{(\frac{3}{2})}} 
-2 \hat{B}_3   {\bf T_{-,non}^{(\frac{3}{2})}} - 
\frac{2 i}{3}  \pa \hat{G}_{12}
\right](z), 
\nonu \\
\widetilde{V}_{\frac{3}{2}}^{(1), 1}(z) & = & 
-2  \sqrt{2} \left( {\bf U_{non}^{(\frac{5}{2})}} - 
{\bf V_{non}^{(\frac{5}{2})}} \right)(z)  
\nonu \\
& + & \frac{4 i \sqrt{2}}{(5+k)}
 \left[
\hat{A}_{+}  \hat{G}_{21} 
+ \hat{A}_{+}  {\bf T_{+,non}^{(\frac{3}{2})}}
- \hat{B}_{-}  \hat{G}_{12}+ 2 \hat{B}_{-}  {\bf T_{-,non}^{(\frac{3}{2})}}
+2 \hat{B}_3  {\bf U_{non}^{(\frac{3}{2})}} - 
\frac{2 i}{3} \pa  {\bf U_{non}^{(\frac{3}{2})}} 
\right](z)
\nonu \\
&+& \frac{4 i \sqrt{2}}{(5+k)}
 \left[
\hat{A}_{-}  \hat{G}_{12} 
-2 \hat{A}_{-}  {\bf T_{-,non}^{(\frac{3}{2})}}
-2 \hat{A}_{3}  {\bf V_{non}^{(\frac{3}{2})}}- 
\hat{B}_{+}  {\bf T_{+,non}^{(\frac{3}{2})}}
 + \frac{2 i}{3}  \pa  {\bf V_{non}^{(\frac{3}{2})}} 
\right](z),
\nonu \\
\widetilde{V}_{\frac{3}{2}}^{(1), 2}(z) & = & 
-2 i \sqrt{2} \left( {\bf U_{non}^{(\frac{5}{2})}} + 
{\bf V_{non}^{(\frac{5}{2})}} \right)(z)  
\nonu \\
& - & \frac{4  \sqrt{2}}{(5+k)}
 \left[
\hat{A}_{+}  \hat{G}_{21} 
+ \hat{A}_{+}  {\bf T_{+,non}^{(\frac{3}{2})}}
- \hat{B}_{-}  \hat{G}_{12}+ 2 \hat{B}_{-}  {\bf T_{-,non}^{(\frac{3}{2})}}
+2 \hat{B}_3  {\bf U_{non}^{(\frac{3}{2})}} -
\frac{2 i}{3} \pa  {\bf U_{non}^{(\frac{3}{2})}} 
\right](z)
\nonu \\
&+& \frac{4  \sqrt{2}}{(5+k)}
 \left[
\hat{A}_{-}  \hat{G}_{12} 
-2 \hat{A}_{-}  {\bf T_{-,non}^{(\frac{3}{2})}}
-2 \hat{A}_{3}  {\bf V_{non}^{(\frac{3}{2})}}- 
\hat{B}_{+}  {\bf T_{+,non}^{(\frac{3}{2})}}
 + \frac{2 i}{3}  \pa  {\bf V_{non}^{(\frac{3}{2})}} 
\right](z),
\nonu \\
 \widetilde{V}_{\frac{3}{2}}^{(1), 3}(z) & = & 
2  \sqrt{2} \left( {\bf W_{+,non}^{(\frac{5}{2})}} - 
{\bf W_{-,non}^{(\frac{5}{2})}} \right)(z)  
-\frac{4 i \sqrt{2}}{(5+k)} 
 \left[
\hat{A}_{-} \hat{G}_{11} 
+ \hat{A}_{-}  {\bf U_{non}^{(\frac{3}{2})}}
- 2 \hat{A}_{3}  \hat{G}_{21}- 2 \hat{A}_{3}  {\bf T_{+,non}^{(\frac{3}{2})}}
\right. \nonu \\
& + & \left.
 \hat{B}_{-}  \hat{G}_{22} -  \hat{B}_{-}  {\bf V_{non}^{(\frac{3}{2})}} 
+2 \hat{B}_3   {\bf T_{+,non}^{(\frac{3}{2})}} + 
\frac{2 i}{3}  \pa \hat{G}_{21}
\right](z)
 -\frac{4 i \sqrt{2}}{(5+k)}
 \left[
\hat{A}_{+}  \hat{G}_{22} 
- \hat{A}_{+}  {\bf V_{non}^{(\frac{3}{2})}}
\right. \nonu \\
& - &  \left.
2 \hat{A}_{3}  \hat{G}_{12}+ 2 \hat{A}_{3}  {\bf T_{-,non}^{(\frac{3}{2})}}
+
 \hat{B}_{+}  \hat{G}_{11} + \hat{B}_{+}  {\bf U_{non}^{(\frac{3}{2})}} 
-2 \hat{B}_3   {\bf T_{-,non}^{(\frac{3}{2})}} - 
\frac{2 i}{3}  \pa \hat{G}_{12}
\right](z). 
\nonu
\eea
These observations are also appeared in \cite{AK1411}. 

Now the final higher spin-$3$ current has the following relation
%%%%%%%%%%%%%%%%%%%%%%%%%%%%%%%%%%%%%%%%%%%%%%%%%%%%%%%%%%%%%%%
\bea
\widetilde{V}_{2}^{(1)}(z) & = & 4 \left[
{\bf W_{non}^{(3)}} +\frac{8(-3+k)}{(19+23k)} {\bf T^{(1)}} \, \hat{T}
+\frac{2i}{(5+k)} \hat{A}_{-}  {\bf U_{-,non}^{(2)}}
+\frac{4i}{(5+k)} \hat{A}_3  {\bf T_{non}^{(2)}}
\right.
\nonu \\
& - & \frac{2i}{(5+k)} \hat{B}_{-}  {\bf V_{-,non}^{(2)}} 
-\frac{4i}{(5+k)} \hat{B}_3  {\bf T_{non}^{(2)}}
-\frac{1}{(5+k)}  \pa {\bf W_{non}^{(2)}}
+\frac{1}{2(5+k)} \hat{B}_{+}  {\bf T^{(1)}}  \hat{B}_{-}
\nonu \\
&-& \frac{i}{(5+k)} \pa {\bf T^{(1)}}  \hat{A}_3
+\frac{i}{(5+k)} \pa {\bf T^{(1)}}  \hat{B}_3
+\frac{i}{(5+k)} {\bf T^{(1)}}  \pa \hat{A}_3
-\frac{1}{2(5+k)} \hat{B}_{-}  {\bf T^{(1)}}  \hat{B}_{+}
\nonu \\
& - & \frac{2}{(5+k)^2} \hat{A}_{+} \pa \hat{A}_{-}
- \frac{8i(48+97k+29k^2)}{(5+k)(3+7k)(19+23k)} \hat{A}_3  \hat{T} 
 +
\frac{2}{(5+k)^2} \hat{A}_3 
 \pa \hat{A}_3
\nonu \\
&-& \frac{(2+k)}{(5+k)^2} \hat{B}_{+}  \pa \hat{B}_{-}
+\frac{k}{(5+k)^2} \hat{B}_{-} \pa \hat{B}_{+}
 - 
\frac{i}{(5+k)^2} \hat{B}_3 \hat{B}_{+}  \hat{B}_{-}
\nonu \\
&+& \frac{8i k(-27+k)}{(5+k)(3+7k)(19+23k)} \hat{B}_3  \hat{T}
 +\frac{2(-6+k)}{(5+k)^2}
\hat{B}_3  \pa \hat{A}_3
\nonu \\
& - & 
\frac{1}{(5+k)} \pa \hat{T} 
-\frac{6i}{(5+k)^2} \pa^2 \hat{A}_3 -\frac{i(3+8k)}{3(5+k)^2}  \pa^2 
\hat{B}_3
+  \frac{2}{(5+k)} \hat{G}_{12}  \hat{G}_{21}
\nonu \\
&-& \frac{2i}{(5+k)^2} \hat{A}_{-}  \hat{A}_{+}  \hat{A}_3
-\frac{6i}{(5+k)^2} \hat{A}_3  \hat{A}_{+}  \hat{A}_{-}
-\frac{8i}{(5+k)^2} \hat{A}_3  \hat{A}_3  \hat{A}_3
\nonu \\
&- &  \frac{i(-4+k)}{(5+k)^2} \hat{B}_{+}  \hat{A}_3  \hat{B}_{-}
+\frac{i(-4+k)}{(5+k)^2} \hat{B}_{-}  \hat{A}_3  \hat{B}_{+}
+\frac{i}{(5+k)^2} \hat{B}_{-}  \hat{B}_{+}  \hat{B}_3
\nonu \\
&+& \left.
\frac{8i}{(5+k)^2} \hat{B}_3  \hat{A}_{+}  \hat{A}_{-}
+\frac{16i}{(5+k)^2} \hat{B}_3  \hat{A}_3  \hat{A}_3
-\frac{8i}{(5+k)^2} \hat{B}_3  \hat{A}_3  \hat{B}_3 \right](z).
\nonu
\eea

It would be interesting to see the OPEs between the higher spin currents 
in the basis above and how they behave differently compared to 
the previous results in Part I and Part II.

%%%%%%%%%%%%%%%%%%%%%%%%%%%%%%%%%%%%%%%%%%%%%%%%%%%%%%%%%%%%%%%%%%%%%
%%%%%%%%%%%%%%%%%%%%%%%%%%%%%%%%%%%%%%%%%%%%%%%%%%%%%%%%%%%%%%%%%%%%%
\section{ 
Some details on the construction of 
the $16$ second lowest higher spin currents 
%\label{higher2}
}
%B%%%%%%%%%%%%%%%%%%%%%%%%%%%%%%%%%%%%%%%%%%%%%%%%%%%%%%%%%%%%%%%%%%%
%%%%%%%%%%%%%%%%%%%%%%%%%%%%%%%%%%%%%%%%%%%%%%%%%%%%%%%%%%%%%%%%%%%%

In section $5$, the higher spin currents of spin-$\frac{5}{2}$ and spin-$3$
are obtained explicitly. In this Appendix, the remaining 
higher spin currents are obtained. 
Due to the presence of four spin-$\frac{1}{2}$ currents and the spin-$1$
current, the structures of the OPEs are rather complicated. 
It is rather nontrivial to extract the correct primary field from the 
given singular term because there exist many quasi primary fields.

%%%%%%%%%%%%%%%%%%%%%%%%%%%%%%%%%%%%%%%%%%%%%%%%%%%%%%%%%%%%%%%%%%%%%
\subsection{ The remaining higher spin currents in the next $16$
higher spin currents }

Let us consider the OPE between the spin-$\frac{3}{2}$ current and 
the higher spin-$3$ current found in (\ref{g1221qr5half})
\bea
%%%%%%%%%%%%%%%%%%%%%%%%%%%%%%%%%%%%%%%%%%%%%%%%%%%%%%%%%%%%%%%%%%
G_{21}(z) \, {\bf Q_{-}^{(3)}}(w) & = & 
%%%%%%%%%%%%%%%%%%%%%%%%%%%%%%%%%%%%%%%%%%%%%%%%%%%%%%%%%%%%%%%
-\frac{1}{(z-4)^4}
\frac{6k(303+334k+47k^2)}{(5+k)^3(17+13k)} F_{11}(w)
\nonu \\
& + & \frac{1}{(z-w)^3} \left[-
\frac{4(21+137k+52k^2)}{(5+k)^2(17+13k)}  U F_{11}
-\frac{12}{(5+k)^2} F_{12} F_{21} F_{11} 
\right. \nonu \\
&-& \frac{16i(114+122k+81k^2+13k^3)}{(5+k)^3(17+13k)} F_{11} A_3
- \frac{4i(21+21k+4k^2)}{(5+k)^3} F_{21} A_{+}
\nonu \\
&-& \frac{4i(21+86k+13k^2)}{(5+k)^2(17+13k)} F_{12} B_{-}
+\frac{4i(-105-247k+73k^2+39k^3)}{(5+k)^3(17+13k)} F_{11} B_3
\nonu \\
& - & \left. 
%\frac{6k(303+334k+47k^2)}{(5+k)^3(17+13k)} \pa F_{11}
\pa (\mbox{pole-4}) 
-\frac{(2+k)(105+637k+104k^2)}{(5+k)^2(17+13k)} G_{11} 
\right](w)
\nonu \\
&+& \frac{1}{(z-w)^2} \left[ 
\frac{2(11+3k)}{(5+k)} {\bf Q^{(\frac{5}{2})}}
+\frac{2(11+3k)}{(5+k)} {\bf U^{(\frac{5}{2})}}
-\frac{i(3+5k)}{2(5+k)^2} A_3 G_{11}
\right. \nonu \\
&-& \frac{3}{(5+k)^2} A_3 A_3 F_{11}
+\frac{20(3+k)}{(5+k)^3} A_3 B_3 F_{11} 
+\frac{(39+11k)}{(5+k)^3} A_3 B_{-} F_{12}
\nonu \\
&+& \frac{2i(-795-397k-21k^2+13k^3)}{(5+k)^3(17+13k)}
\pa A_3 F_{11}
-\frac{(45+17k)}{(5+k)^3} B_3 B_3 F_{11}
\nonu \\
& - & \frac{2i(-651+322k+29k^2)}{3(5+k)^3(17+13k)} A_3 \pa F_{11}
- \frac{i(2079+1244k+181k^2)}{2(5+k)^2(17+13k)} A_{+} G_{21}
\nonu \\
&-& \frac{2(3+k)}{(5+k)^3} A_{+} A_3 F_{21}
-\frac{(9+k)}{(5+k)^3} A_{+} A_{-} F_{11}
+\frac{18(3+k)}{(5+k)^3} A_{+} B_3 F_{21}
\nonu \\
&-& \frac{3(11+3k)}{(5+k)^3} A_{+} B_{-} F_{22}
-\frac{4i(15+4k)}{3(5+k)^3} \pa A_{+} F_{21}
-\frac{4i(15+4k)}{3(5+k)^3} A_{+} \pa F_{21}
\nonu \\
&+& \frac{3i(3+k)}{2(5+k)^2} B_3 G_{11}
-  \frac{2(-1671-271k+223k^2+39k^3)}{3(5+k)^3(17+13k)}
\pa U F_{11}
\nonu \\
& + & 
\frac{4i(618+1537k+566k^2+39k^3)}{3(5+k)^3(17+13k)}
B_3 \pa F_{11}
-\frac{6i}{(5+k)^2} B_{-} G_{12}
\nonu \\
&-& \frac{3(7+3k)}{(5+k)^3} B_{-} B_3 F_{12}
+\frac{i(1518+715k+308k^2+39k^3)}{3(5+k)^3(17+13k)} 
\pa B_{-} F_{12}
\nonu \\
&- & \frac{i(741-784k+265k^2+78k^3)}{3(5+k)^3(17+13k)}
B_{-} \pa F_{12}
-\frac{8(3+k)}{(5+k)^3} B_{+} B_{-} F_{11}
\nonu \\
&+& \frac{(2367+2215k+521k^2+33k^3)}{(5+k)^3(17+13k)} \pa^2 F_{11}
-\frac{(9+k)(163+23k)}{(5+k)^3(17+13k)} F_{11} F_{12} G_{21}
\nonu \\
& +& \frac{2(-3+22k+9k^2)}{3(5+k)^4} \pa F_{11} F_{12} F_{21} 
-\frac{2(381+194k+21k^2)}{3(5+k)^4} F_{11} \pa F_{12} F_{21}
\nonu \\
&+& \frac{8(96+43k+3k^2)}{3(5+k)^4} F_{11} F_{12} \pa F_{21}
-\frac{8(3+k)}{(5+k)^3} F_{11} F_{21} G_{12}
\nonu \\
&-& \frac{(15+7k)}{(5+k)^3} F_{11} F_{22} G_{11}
+\frac{10(3+k)}{(5+k)^3} F_{12} F_{21} G_{11}
-\frac{i(39+11k)}{(5+k)^3} U B_{-} F_{12}
\nonu \\
& + & \frac{(1977+1234k+233k^2)}{6(5+k)^2(17+13k)} \pa G_{11}
-\frac{4(153+88k+11k^2)}{(5+k)^2(17+13k)} T F_{11}
\nonu \\
&- & \frac{(27+11k)}{2(5+k)^2} U G_{11} -\frac{16i(3+k)}{(5+k)^3}
U A_3 F_{11}
-\frac{18i(3+k)}{(5+k)^3} U A_{+} F_{21}
\nonu \\
&-& \frac{2i(9+k)}{(5+k)^3} U B_3 F_{11}
+  
\frac{(-1365+116k+391k^2+78k^3)}{3(5+k)^3(17+13k)} F_{11} G_{22} F_{11}
\nonu \\
& + & \frac{8(-303+71k+82k^2)}{3(5+k)^3(17+13k)} U \pa F_{11}
-\frac{(63+19k)}{(5+k)^3} U U F_{11}
\nonu \\
& -& \left. \frac{2(105+42k+k^2)}{(5+k)^4} 
F_{11} \pa F_{11} F_{22}
\right](w)
\nonu \\
& + & \frac{1}{(z-w)} \left[ 
\frac{1}{5} \pa \{ G_{21} \, {\bf Q_{-}^{(3)}} \}_{-2}
+ {\bf Q^{(\frac{7}{2})}}
\right. \nonu \\
&
-
&
\frac{16(21+137k+52k^2)}{(5+k)(17+13k)(67+39k)}
\left( T U F_{11} -\frac{3}{8} \pa^2 (U F_{11})  \right)
 \nonu \\
&- & \frac{48}{(5+k)(67+39k)}
\left( T  F_{12} F_{21} F_{11} -\frac{3}{8} \pa^2 ( F_{12} F_{21} F_{11})  \right)
\nonu \\
& -& \frac{64i(114+122k+81k^2+13k^3)}{(5+k)^2(17+13k)(67+39k)}
\left( T F_{11} A_3 -\frac{3}{8} \pa^2 ( F_{11} A_3)  \right)
\nonu \\
&- & 
 \frac{16i(21+21k+4k^2)}{(5+k)^2(67+39k)}
\left( T F_{21} A_{+} -\frac{3}{8} \pa^2 ( F_{21} A_{+})  \right)
\nonu \\
&-&  \frac{16i(21+86k+13k^2)}{(5+k)(17+13k)(67+39k)}
\left( T F_{12} B_{-} -\frac{3}{8} \pa^2 ( F_{12} B_{-})  \right)
\nonu \\
&+&  \frac{16i(-105-247k+73k^2+39k^3)}{(5+k)^2(17+13k)(67+39k)}
\left( T F_{11} B_3 -\frac{3}{8} \pa^2 ( F_{11} B_3)  \right)
\nonu \\
&+&  \frac{192k(303+334k+47k^2)}{5(5+k)^2(17+13k)(43+47k)}
\left( T \pa F_{11}-\frac{1}{4} \pa T F_{11} -\frac{1}{3} \pa^3 F_{11}  \right)
\nonu \\
&-& \left.
  \frac{4(2+k)(105+637k+104k^2)}{(5+k)(17+13k)(67+39k)}
\left( T G_{11} -\frac{3}{8} \pa^2 G_{11}  \right)
\right](w) + \cdots.
\nonu 
\eea
In the third order pole there is no higher spin current.
It is clear that there are no higher spin currents in the context of
footnote \ref{nonlineardef} which will appear in Appendix $F$. 
It is easy to check that 
the various terms in the first order pole are quasi primary fields.
Note that 
the second term from the last expression has different relative coefficients
$-\frac{1}{4}$ and $-\frac{1}{3}$ rather than $-\frac{3}{8}$. 
There are no nonlinear terms between the higher spin currents in the above. 
 
Let us consider the OPE between the spin-$\frac{3}{2}$ current and 
the higher spin-$3$ current found in (\ref{g2112qr})
\bea
%%%%%%%%%%%%%%%%%%%%%%%%%%%%%%%%%%%%%%%%%%%%%%%%%%%%%%%%%%%%%
G_{21}(z) \, {\bf R_{-}^{(3)}}(w) & = & 
%%%%%%%%%%%%%%%%%%%%%%%%%%%%%%%%%%%%%%%%%%%%%%%%%%%%%%%%%%%%%%
\frac{1}{(z-4)^4}
\frac{6k(237+314k+61k^2)}{(5+k)^3(17+13k)} F_{22}(w)
\nonu \\
& + & \frac{1}{(z-w)^3} \left[
 - 
%\frac{6k(237+314k+61k^2)}{(5+k)^3(17+13k)} \pa F_{22}
\pa (\mbox{pole-4}) 
-
\frac{8(1683+1713k+391k^2+17k^3)}{3(5+k)^3(17+13k)}  U F_{22}
\right. \nonu \\
& - & \frac{12}{(5+k)^2} F_{21} F_{12} F_{22} 
+ \frac{4i(-1683-1200k+197k^2+34k^3)}{3(5+k)^3(17+13k)} F_{22} A_3
\nonu \\
&+& \frac{4i(1071+1212k+431k^2+34k^3)}{3(5+k)^3(17+13k)} F_{12} A_{-}
\nonu \\
&+& \frac{4i(66+15k+k^2)}{(5+k)^3} F_{21} B_{+}
+\frac{4i(3366+3405k+556k^2+5k^3)}{3(5+k)^3(17+13k)} F_{22} B_3
\nonu \\
& - & \left.
\frac{(6732+6753k+1717k^2+68k^3)}{3(5+k)^2(17+13k)} G_{22} 
\right](w)
\nonu \\
&+& \frac{1}{(z-w)^2} \left[ 
\frac{4(7+k)}{(5+k)} {\bf R^{(\frac{5}{2})}}
-\frac{4(7+k)}{(5+k)} {\bf V^{(\frac{5}{2})}}
+ \frac{9i}{(5+k)^2} A_3 G_{22}
\right. \nonu \\
&+& \frac{(75+7k)}{(5+k)^3} A_3 A_3 F_{22}
-\frac{10(9+k)}{(5+k)^3} A_3 B_3 F_{22} 
-\frac{9(9+k)}{(5+k)^3} A_3 B_{+} F_{21}
\nonu \\
&-& \frac{i(3+k)(765+842k+61k^2)}{3(5+k)^3(17+13k)}
\pa A_3 F_{22}
+\frac{3}{(5+k)^2} B_3 B_3 F_{22}
\nonu \\
& + & \frac{2i(-6171-4954k-263k^2+24k^3)}{3(5+k)^3(17+13k)} A_3 \pa F_{22}
- \frac{3i(1+k)}{2(5+k)^2} A_{-} G_{12}
\nonu \\
&+& \frac{3(13+k)}{(5+k)^3} A_{-} A_3 F_{12}
+\frac{4(9+k)}{(5+k)^3} A_{-} A_{+} F_{22}
-\frac{(51+7k)}{(5+k)^3} A_{-} B_3 F_{12}
\nonu \\
&+& \frac{6(7+k)}{(5+k)^3} A_{-} B_{+} F_{11}
-\frac{i(255+878k+439k^2+24k^3)}{3(5+k)^3(17+13k)} \pa A_{-} F_{12}
\nonu \\
& + & \frac{i(-1785-458k+527k^2+48k^3)}{3(5+k)^3(17+13k)} A_{-} \pa F_{12}
+ \frac{i(-12+k)}{(5+k)^2} B_3 G_{22}
\nonu \\
& - & 
\frac{2i(-2244-1067k+430k^2+37k^3)}{3(5+k)^3(17+13k)}
B_3 \pa F_{22}
+ \frac{(9+k)}{(5+k)^3} B_{+} B_3 F_{21}
\nonu \\
& - & \frac{i(561+1144k+311k^2+24k^3)}{2(5+k)^2(17+13k)} B_{+} G_{21}
+\frac{i(1+k)(21+2k)}{3(5+k)^3} 
\pa B_{+} F_{21}
\nonu \\
&- & \frac{i(-183+13k+4k^2)}{3(5+k)^3}
B_{+} \pa F_{21}
+\frac{2(3+k)}{(5+k)^3} B_{-} B_{+} F_{22}
\nonu \\
&-& \frac{(21573+14907k+2023k^2+113k^3)}{6(5+k)^3(17+13k)} \pa^2 F_{22}
-\frac{4(9+k)}{(5+k)^3} F_{22} F_{21} G_{12}
\nonu \\
& -& \frac{2(-147-2k+k^2)}{3(5+k)^4} \pa F_{22} F_{21} F_{12} 
+\frac{4(177+88k+7k^2)}{3(5+k)^4} F_{22} \pa F_{21} F_{12}
\nonu \\
&-& \frac{2(501+178k+13k^2)}{3(5+k)^4} F_{22} F_{21} \pa F_{12}
- \frac{2(15+k)}{(5+k)^3} F_{22} F_{11} G_{22}
\nonu \\
& - & \frac{2(3+k)(-1+3k)(17+4k)}{(5+k)^3(17+13k)} F_{22} F_{12} G_{21}
-\frac{5(9+k)}{(5+k)^3} F_{21} F_{12} G_{22}
\nonu \\
& + & \frac{(867+1310k+259k^2+24k^3)}{6(5+k)^2(17+13k)} \pa G_{22}
+\frac{(663+338k+43k^2)}{(5+k)^2(17+13k)} T F_{22}
\nonu \\
&+ & \frac{2(12+k)}{(5+k)^2} U G_{22} -\frac{4i(3+k)}{(5+k)^3}
U A_3 F_{22}
-\frac{i(51+7k)}{(5+k)^3} U A_{-} F_{12}
\nonu \\
&-& \frac{8i(9+k)}{(5+k)^3} U B_3 F_{22}
-\frac{9i(9+k)}{(5+k)^3} U B_{+} F_{21}
-  \frac{(-21+52k+k^2)}{3(5+k)^3}
\pa U F_{22}
\nonu \\
& - & \frac{2(-663-538k-107k^2+24k^3)}{3(5+k)^3(17+13k)} U \pa F_{22}
+\frac{(87+11k)}{(5+k)^3} U U F_{22}
\nonu \\
& + & 
\frac{(510+1345k+664k^2+37k^3)}{3(5+k)^3(17+13k)} F_{22} G_{11} F_{22}
\nonu \\
& +& \left. \frac{8(15+12k+k^2)}{(5+k)^4} 
F_{22} \pa F_{22} F_{11}
\right](w)
\nonu \\
& + & \frac{1}{(z-w)} \left[ 
\frac{1}{5} \pa \{ G_{21} \, {\bf R_{-}^{(3)}} \}_{-2}
+ {\bf R^{(\frac{7}{2})}}
\right.
\nonu \\
& - & \frac{32(1683+1713k+391k^2+17k^3)}{3(5+k)^2(17+13k)(67+39k)}
\left( T U F_{22} -\frac{3}{8} \pa^2 (U F_{22})  \right)
 \nonu \\
&- & \frac{48}{(5+k)(67+39k)}
\left( T  F_{21} F_{12} F_{22} -\frac{3}{8} \pa^2 ( F_{21} F_{12} F_{22})  \right)
\nonu \\
& +& \frac{16i(-1683-1200k+197k^2+34k^3)}{3(5+k)^2(17+13k)(67+39k)}
\left( T F_{22} A_3 -\frac{3}{8} \pa^2 ( F_{22} A_3)  \right)
\nonu \\
&+ & 
 \frac{16i(1071+1212k+431k^2+34k^3)}{3(5+k)^2(17+13k)(67+39k)}
\left( T F_{12} A_{-} -\frac{3}{8} \pa^2 ( F_{12} A_{-})  \right)
\nonu \\
&+&  \frac{16i(66+15k+k^2)}{(5+k)^2(67+39k)}
\left( T F_{21} B_{+} -\frac{3}{8} \pa^2 ( F_{21} B_{+})  \right)
\nonu \\
&+&  \frac{16i(3366+3405k+556k^2+5k^3)}{3(5+k)^2(17+13k)(67+39k)}
\left( T F_{22} B_3 -\frac{3}{8} \pa^2 ( F_{22} B_3)  \right)
\nonu \\
&-&  \frac{192k(237+314k+61k^2)}{5(5+k)^2(17+13k)(43+47k)}
\left( T \pa F_{22}-\frac{1}{4} \pa T F_{22} -\frac{1}{3} \pa^3 F_{22}  \right)
\nonu \\
&-& \left.
  \frac{4(6732+6753k+1717k^2+68k^3)}{3(5+k)(17+13k)(67+39k)}
\left( T G_{22} -\frac{3}{8} \pa^2 G_{22}  \right)
\right](w) \nonu \\
& + & \cdots.
\nonu 
\eea
There is no boldface higher spin current in the third order pole.
As before, various nonlinear terms in Part II do not appear in this OPE.
In this case also, various quasi primary fields appear in the first order 
pole. 
Furthermore,
there are no nonlinear terms between the higher spin currents in the above.

Let us consider the OPE between the spin-$\frac{3}{2}$ current and 
the higher spin-$3$ current found in (\ref{g22q5half})
\bea
%%%%%%%%%%%%%%%%%%%%%%%%%%%%%%%%%%%%%%%%%%%%%%%%%%%%%%%%%%
G_{21}(z) \, {\bf S^{(3)}}(w) & = & 
%%%%%%%%%%%%%%%%%%%%%%%%%%%%%%%%%%%%%%%%%%%%%%%%%%%%%%%%%%%
\frac{1}{(z-4)^4}
\frac{12k(189+182k+25k^2)}{(5+k)^3(17+13k)} F_{21}(w)
\nonu \\
& + & \frac{1}{(z-w)^3} \left[
-   
%\frac{12k(189+182k+25k^2)}{(5+k)^3(17+13k)} \pa F_{21}
\pa (\mbox{pole-4}) 
-\frac{4(-3+k)(113+61k)}{3(5+k)(17+13k)} {\bf T_{+}^{(\frac{3}{2})}}
\right.
\nonu \\
& + & 
\frac{2(3987+5676k+1895k^2+190k^3)}{3(5+k)^3(17+13k)}  U F_{21}
 \nonu \\
& +& \frac{8(-15+k)}{(5+k)^3} F_{21} F_{11} F_{22} 
- \frac{2i(3375+4596k+1427k^2+190k^3)}{3(5+k)^3(17+13k)} F_{21} A_3
\nonu \\
&+& \frac{2i(3969+3744k+1325k^2+190k^3)}{3(5+k)^3(17+13k)} F_{11} A_{-}
\nonu \\
&+& \frac{2i(-3987-4788k-751k^2+34k^3)}{3(5+k)^3(17+13k)} F_{22} B_{-}
\nonu \\
& - & \frac{2i(-3987-3432k+433k^2+278k^3)}{3(5+k)^3(17+13k)} F_{21} B_3
\nonu \\
& + & \left.
\frac{(7377+6955k+1578k^2+68k^3)}{3(5+k)^2(17+13k)} G_{21} 
\right](w)
\nonu \\
&+& \frac{1}{(z-w)^2} \left[
- \frac{2(4239+5758k+1735k^2+40k^3)}{(5+k)^3(17+13k)}
F_{21} {\bf T_{-}^{(\frac{3}{2})}} F_{21}
\right.
\nonu \\
& - & \frac{(489-142k-223k^2+8k^3)}{4(5+k)(17+13k)}
(F_{21} {\bf P^{(2)}}-
{\bf P^{(2)}} F_{21})
\nonu \\
&-& \frac{(-3+k)}{(5+k)} {\bf P_{+}^{(\frac{5}{2})}}
-\frac{2(11+3k)}{(5+k)} {\bf W_{+}^{(\frac{5}{2})}}
 \nonu \\
&-& \frac{6{\bf (-3+k)}}{(17+13k)} G_{21} {\bf T^{(1)}} +
\frac{(369-126k-215k^2+8k^3)}{(5+k)(17+13k)} \pa {\bf T_{+}^{(\frac{3}{2})}}
\nonu \\
&+& 
\frac{2i(14883+8577k-847k^2-341k^3+24k^4)}{
3(5+k)^4(17+13k)} F_{21} ((F_{11} B_{+}) F_{21})
 \nonu \\
&+& \frac{8i(77+30k+k^2)}{(5+k)^4}
F_{21} ((F_{12} B_3) F_{21})
- \frac{10(3+k)}{(5+k)^3} B_{-} B_3 F_{22}
\nonu \\
&-& \frac{2(1+k)}{(5+k)^3} A_3 A_3 F_{21}
+\frac{2(29+9k)}{(5+k)^3} A_3 B_3 F_{21} 
-\frac{4(7+2k)}{(5+k)^3} A_3 B_{-} F_{22}
\nonu \\
&+& \frac{2i(119+53k+2k^2)}{(5+k)^3}
\pa A_3 F_{21}
- \frac{3i(389+237k+28k^2)}{(5+k)^2(17+13k)} A_3 G_{21}
\nonu \\
& - & \frac{4i(-17+5k)}{(5+k)^3} A_3 \pa F_{21}
+ \frac{i(9+2k)}{(5+k)^2} A_{-} G_{11}
-  \frac{2i(3+k)}{(5+k)^2} B_{-} G_{22}
\nonu \\
&+& \frac{2(3+k)}{(5+k)^3} A_{-} A_3 F_{11}
+\frac{4}{(5+k)^3} A_{+} A_{-} F_{21}
-\frac{4(16+5k)}{(5+k)^3} A_{-} B_3 F_{11}
\nonu \\
&-& \frac{2(17+5k)}{(5+k)^3} A_{-} B_{-} F_{12}
-\frac{4i(141-422k-262k^2+9k^3)}{(5+k)^3(17+13k)} \pa A_{-} F_{11}
\nonu \\
& - & \frac{i(3397+254k-879k^2+24k^3)}{(5+k)^3(17+13k)} A_{-} \pa F_{11}
- 
\frac{4(5+2k)}{(5+k)^3} F_{11} F_{21} G_{22}
\nonu \\
&-& \frac{3i(51+161k+54k^2+4k^3)}{(5+k)^2(17+13k)} B_3 G_{21}
+\frac{8(5+2k)}{(5+k)^3} B_3 B_3 F_{21}
\nonu \\
& - & 
\frac{8i(14+k)}{(5+k)^3}
B_3 \pa F_{21}
+ \frac{4(157+76k+7k^2)}{(5+k)^4} F_{11} F_{21} \pa F_{22}
\nonu \\
& + & \frac{i(1971+1491k-535k^2-527k^3+24k^4)}{3(5+k)^3(17+13k)} 
\pa B_{-} F_{22}
\nonu \\
&+ & \frac{i(1122+903k-2020k^2-881k^3+24k^4)}{3(5+k)^3(17+13k)}
B_{-} \pa F_{22}
\nonu \\
& + & \frac{2(5+3k)}{(5+k)^3} B_{+} B_{-} F_{21}
+  \frac{(-317+458k+195k^2+12k^3)}{(5+k)^3(17+13k)} F_{11} F_{22} G_{21}
\nonu \\
& -& 
\frac{2(8253+855k-2977k^2-419k^3+24k^4)}{3(5+k)^4(17+13k)} 
F_{11} \pa F_{21} F_{22}
\nonu \\
&-& \frac{3(389+286k+77k^2+4k^3)}{(5+k)^3(17+13k)} F_{12} F_{21} G_{21}
-\frac{2(7+k)}{(5+k)^3} F_{21} F_{22} G_{11}
\nonu \\
&-& \frac{(-5007+1968k+1205k^2-770k^3+24k^4)}{
3(5+k)^3(17+13k)} \pa^2 F_{21} 
\nonu \\
& + & \frac{(882-543k-1060k^2-187k^3+8k^4)}{2(5+k)^2(17+13k)} \pa G_{21}
-\frac{20i(3+k)}{(5+k)^3}
U A_3 F_{21}
\nonu \\
& + & \frac{2(17+2k)(11+7k)}{(5+k)^2(17+13k)} T F_{21}
+  \frac{2(153+109k+32k^2)}{(5+k)^2(17+13k)} U G_{21} 
\nonu \\
& + & \frac{18i(3+k)}{(5+k)^3} U A_{-} F_{11}
+ \frac{2i(9+k)}{(5+k)^3} U B_3 F_{21}
-\frac{12i(4+k)}{(5+k)^3} U B_{-} F_{22}
\nonu \\
& + & \left.
\frac{4(819+1417k+493k^2+7k^3)}{(5+k)^3(17+13k)}
\pa U F_{21}
+\frac{2(29+9k)}{(5+k)^3} U U F_{21}
\right](w)
\nonu \\
& + & \frac{1}{(z-w)} \left[ 
\frac{1}{5} \pa \{ G_{21} \, {\bf S^{(3)}} \}_{-2}
\right.
\nonu \\
 & - & \frac{16(-3+k)(113+61k)}{3(17+13k)(67+39k)}
\left( T {\bf T_{+}^{(\frac{3}{2})}} -\frac{3}{8} \pa^2  {\bf T_{+}^{(\frac{3}{2})}}  \right) + {\bf S_{+}^{(\frac{7}{2})}}
\nonu \\
& + & \frac{8(3987+5676k+1895k^2+190k^3)}{3(5+k)^2(17+13k)(67+39k)}
\left( T U F_{21} -\frac{3}{8} \pa^2 (U F_{21})  \right)
 \nonu \\
&+ & \frac{32(-15+k)}{(5+k)^2(67+39k)}
\left( T  F_{21} F_{11} F_{22} -\frac{3}{8} \pa^2 ( F_{21} F_{11} F_{22})  \right)
\nonu \\
& -& \frac{8i(3375+4596k+1427k^2+190k^3)}{3(5+k)^2(17+13k)(67+39k)}
\left( T F_{21} A_3 -\frac{3}{8} \pa^2 ( F_{21} A_3)  \right)
\nonu \\
&+ & 
 \frac{8i(3969+3744k+1325k^2+190k^3)}{3(5+k)^2(17+13k)(67+39k)}
\left( T F_{11} A_{-} -\frac{3}{8} \pa^2 ( F_{11} A_{-})  \right)
\nonu \\
&+&  \frac{8i(-3987-4788k-751k^2+34k^3)}{3(5+k)^2(17+13k)(67+39k)}
\left( T F_{22} B_{-} -\frac{3}{8} \pa^2 ( F_{22} B_{-})  \right)
\nonu \\
&-&  \frac{8i(-3987-3432k+433k^2+278k^3)}{3(5+k)^2(17+13k)(67+39k)}
\left( T F_{21} B_3 -\frac{3}{8} \pa^2 ( F_{21} B_3)  \right)
\nonu \\
&-&  \frac{384k(189+182k+25k^2)}{5(5+k)^2(17+13k)(43+47k)}
\left( T \pa F_{21}-\frac{1}{4} \pa T F_{21} -\frac{1}{3} \pa^3 F_{21}  \right)
\nonu \\
&+& \left.
  \frac{4(7377+6955k+1578k^2+68k^3)}{3(5+k)(17+13k)(67+39k)}
\left( T G_{21} -\frac{3}{8} \pa^2 G_{21}  \right)
\right](w) \nonu \\
& + & \cdots.
\nonu 
\eea
One can reexpress the first term in the second order pole 
by moving the higher spin current to the right hand side
and will be written in term of known other terms.  
Note that one can rewrite the two terms containing the higher 
spin-$2$ current ${\bf P^{(2)}}(w)$ 
in terms of the known other currents and then there will be no
${\bf P^{(2)}}(w)$ dependence eventually.
From the previous experience, one can consider the nonlinear term 
between the $G_{21}$ and the higher spin-$1$ current and this will be 
removed by redefining the spin-$3$ current with $T \, {\bf T^{(1)}}(w)$
into the left hand side. In the first order pole, one sees many 
quasi primary fields. 
One can also consider the OPE between the $G_{21}(z)$ and the 
quasi primary field appearing in (\ref{bcgspin3}). Then 
from the results in Appendix $C$, the terms having 
${\bf T_{+}^{(\frac{3}{2})}}(w)$ in the second term of the 
first order pole can be seen from this OPE.  
Then, those nonlinear terms can be removed in the redefined 
higher spin-$3$ current by adding the quasi primary field to the left
hand side of above OPE.

Similarly one has
\bea
%%%%%%%%%%%%%%%%%%%%%%%%%%%%%%%%%%%%%%%%%%%%%%%%%%%%%%%
G_{12}(z) \, {\bf S^{(3)}}(w) & = & 
%%%%%%%%%%%%%%%%%%%%%%%%%%%%%%%%%%%%%%%%%%%%%%%%%%%%%%
\frac{1}{(z-4)^4}
\frac{48k(39+43k+8k^2)}{(5+k)^3(17+13k)} F_{12}(w)
\nonu \\
& + & \frac{1}{(z-w)^3} \left[
 -   
%\frac{48k(39+43k+8k^2)}{(5+k)^3(17+13k)} \pa F_{12}
\pa (\mbox{pole-4}) 
-\frac{4(-3+k)(113+61k)}{3(5+k)(17+13k)} {\bf T_{-}^{(\frac{3}{2})}}
\right.
\nonu \\
& - & 
\frac{2(3375+5412k+2051k^2+190k^3)}{3(5+k)^3(17+13k)}  U F_{12}
 \nonu \\
& -& \frac{32k}{(5+k)^3} F_{12} F_{11} F_{22} 
- \frac{2i(-5193-2364k+1115k^2+190k^3)}{3(5+k)^3(17+13k)} F_{12} A_3
\nonu \\
&+& \frac{2i(909+1608k+1481k^2+190k^3)}{3(5+k)^3(17+13k)} F_{22} A_{+}
\nonu \\
&-& \frac{2i(3375+4728k+1267k^2+122k^3)}{3(5+k)^3(17+13k)} F_{11} B_{+}
\nonu \\
& + & \frac{2i(3375+4596k+1427k^2+190k^3)}{3(5+k)^3(17+13k)} F_{12} B_3
\nonu \\
& - & \left.
\frac{(6765+6691k+1734k^2+68k^3)}{3(5+k)^2(17+13k)} G_{12} 
\right](w)
\nonu \\
&+& \frac{1}{(z-w)^2} \left[ 
 -  \frac{(-523+65k+235k^2+31k^3)}{8(5+k)^2(17+13k)}
(F_{12} {\bf P^{(2)}}-
{\bf P^{(2)}} F_{12}) \right.
\nonu \\
&-& \frac{(-3+k)}{(5+k)} {\bf P_{-}^{(\frac{5}{2})}}
+\frac{2(11+3k)}{(5+k)} {\bf W_{-}^{(\frac{5}{2})}}
 \nonu \\
&-& \frac{6{\bf (-3+k)}}{(17+13k)} G_{12} {\bf T^{(1)}} -
\frac{(677+145k+123k^2+15k^3)}{2(5+k)^2(17+13k)} \pa {\bf T_{-}^{(\frac{3}{2})}}
\nonu \\
&+& \frac{2(35+3k)}{(5+k)^3} A_3 A_3 F_{12}
+\frac{2(41+5k)}{(5+k)^3} A_3 B_3 F_{12} 
-\frac{(91+11k)}{(5+k)^3} A_3 B_{+} F_{11}
\nonu \\
&+& \frac{i(77+24k+3k^2)}{(5+k)^3}
\pa A_3 F_{12}
- \frac{3i(693+392k+43k^2)}{2(5+k)^2(17+13k)} A_3 G_{12}
\nonu \\
& + & \frac{i(4143+4325k+421k^2+15k^3)}{2(5+k)^3(17+13k)} A_3 \pa F_{12}
- \frac{i(9+k)}{(5+k)^2} A_{+} G_{22}
\nonu \\
&-& \frac{5(9+k)}{(5+k)^3} A_{+} A_3 F_{22}
+\frac{(25+k)}{(5+k)^3} A_{+} A_{-} F_{12}
-\frac{(37+5k)}{(5+k)^3} A_{+} B_3 F_{22}
\nonu \\
&-& \frac{2(23+3k)}{(5+k)^3} A_{+} B_{+} F_{21}
-\frac{i(263+1269k+469k^2+39k^3)}{2(5+k)^3(17+13k)} \pa A_{+} F_{22}
\nonu \\
& - & \frac{i(-1347-1703k-181k^2+15k^3)}{2(5+k)^3(17+13k)} A_{+} \pa F_{22}
\nonu \\
&-& \frac{3i(187+404k+121k^2+8k^3)}{2(5+k)^2(17+13k)} B_3 G_{12}
-\frac{8}{(5+k)^3} B_3 B_3 F_{12}
\nonu \\
& - & 
\frac{i(-2861-759k+833k^2+171k^3)}{2(5+k)^3(17+13k)}
B_3 \pa F_{12}
+  \frac{i(12+k)}{(5+k)^2} B_{+} G_{11}
\nonu \\
& + &  \frac{(9+k)}{(5+k)^3} B_{+} B_3 F_{11}
-  \frac{i(2091+3441k+1361k^2+139k^3)}{6(5+k)^3(17+13k)} 
\pa B_{+} F_{11}
\nonu \\
&+ & \frac{i(-11337-2229k+3745k^2+557k^3)}{6(5+k)^3(17+13k)}
B_{+} \pa F_{11}
+  \frac{(1+k)}{(5+k)^3} B_{+} B_{-} F_{12}
\nonu \\
&+& 
\frac{(35+3k)}{(5+k)^3} F_{11} F_{12} G_{22}
+\frac{4(6165+6006k+1313k^2+64k^3)}{3(5+k)^4(17+13k)} 
F_{11} F_{12} \pa F_{22}
\nonu \\
& -& 
\frac{2(5979+9435k+3629k^2+301k^3)}{3(5+k)^4(17+13k)} 
\pa F_{11} F_{12} F_{22}
+\frac{9i(9+k)}{(5+k)^3} U B_{+} F_{11}
\nonu \\
&-& \frac{4(822+1581k+556k^2+53k^3)}{3(5+k)^4(17+13k)} F_{11} \pa F_{12}  F_{22}
- \frac{10i(9+k)}{(5+k)^3} U B_3 F_{12}
\nonu \\
& + & \frac{(-1507-690k+13k^2+12k^3)}{(5+k)^3(17+13k)} F_{11} F_{22} G_{12}
-  \frac{3i(19+3k)}{(5+k)^3} U A_{+} F_{22}
\nonu \\
&-& \frac{3(491+330k+51k^2+4k^3)}{(5+k)^3(17+13k)} F_{12} F_{21} G_{12}
+\frac{(11+3k)}{(5+k)^3} F_{12} F_{22} G_{11}
\nonu \\
&+& \frac{(-14013-9129k-587k^2+65k^3)}{
6(5+k)^3(17+13k)} \pa^2 F_{12} 
\nonu \\
& + & \frac{(551+529k+261k^2+27k^3)}{2(5+k)^2(17+13k)} \pa G_{12}
+\frac{4i(3+k)}{(5+k)^3}
U A_3 F_{12}
\nonu \\
& + & \frac{(425+268k+27k^2)}{(5+k)^2(17+13k)} T F_{12}
+  \frac{(969+590k+37k^2)}{2(5+k)^2(17+13k)} U G_{12} 
\nonu \\
& + & 
\frac{(-277+877k+369k^2+15k^3)}{2(5+k)^3(17+13k)}
 U \pa F_{12}
+\frac{2(41+5k)}{(5+k)^3} U U F_{12}
\nonu \\
&+ & \left. \frac{(25+12k+3k^2)}{2(5+k)^3} F_{12} G_{21} F_{12}
+\frac{4(47+34k+3k^2)}{(5+k)^4} F_{12} \pa F_{12} F_{21}
\right](w)
\nonu \\
& + & \frac{1}{(z-w)} \left[ 
\frac{1}{5} \pa \{ G_{12} \, {\bf S^{(3)}} \}_{-2}
\right.
\nonu \\
 & - & \frac{16(-3+k)(113+61k)}{3(17+13k)(67+39k)}
\left( T {\bf T_{-}^{(\frac{3}{2})}} -\frac{3}{8} \pa^2  {\bf T_{-}^{(\frac{3}{2})}}  \right) + {\bf S_{-}^{(\frac{7}{2})}}
\nonu \\
& - & \frac{8(3375+5412k+2051k^2+190k^3)}{3(5+k)^2(17+13k)(67+39k)}
\left( T U F_{12} -\frac{3}{8} \pa^2 (U F_{12})  \right)
 \nonu \\
&- & \frac{128k}{(5+k)^2(67+39k)}
\left( T  F_{12} F_{11} F_{22} -\frac{3}{8} \pa^2 ( F_{12} F_{11} F_{22})  \right)
\nonu \\
& -& \frac{8i(-5193-2364k+1115k^2+190k^3)}{3(5+k)^2(17+13k)(67+39k)}
\left( T F_{12} A_3 -\frac{3}{8} \pa^2 ( F_{12} A_3)  \right)
\nonu \\
&+ & 
 \frac{8i(909+1608k+1481k^2+190k^3)}{3(5+k)^2(17+13k)(67+39k)}
\left( T F_{22} A_{+} -\frac{3}{8} \pa^2 ( F_{22} A_{+})  \right)
\nonu \\
&-&  \frac{8i(3375+4728k+1267k^2+122k^3)}{3(5+k)^2(17+13k)(67+39k)}
\left( T F_{11} B_{+} -\frac{3}{8} \pa^2 ( F_{11} B_{+})  \right)
\nonu \\
&+&  \frac{8i(3375+4596k+1427k^2+190k^3)}{3(5+k)^2(17+13k)(67+39k)}
\left( T F_{12} B_3 -\frac{3}{8} \pa^2 ( F_{12} B_3)  \right)
\nonu \\
&-&  
\frac{1536k(39+43k+8k^2)}{5(5+k)^2(17+13k)(43+47k)}
\left( T \pa F_{12}-\frac{1}{4} \pa T F_{12} -\frac{1}{3} \pa^3 F_{12}  \right)
\nonu \\
&-& \left.
 \frac{4(6765+6691k+1734k^2+68k^3)}{3(5+k)(17+13k)(67+39k)} 
\left( T G_{12} -\frac{3}{8} \pa^2  G_{12}  \right)
\right](w) \nonu \\
& + & \cdots.
\nonu 
\eea
Note that one can rewrite the two terms containing the higher 
spin-$2$ current ${\bf P^{(2)}}(w)$ 
in terms of the known other currents and then there will be no
${\bf P^{(2)}}(w)$ dependence eventually.
The nonlinear term 
between the $G_{12}$ and the higher spin-$1$ current will be 
removed by redefining the spin-$3$ current with $T \, {\bf T^{(1)}}(w)$
into the left hand side.
The terms having 
${\bf T_{-}^{(\frac{3}{2})}}(w)$ in the second term of the 
first order pole can be removed in the redefined 
higher spin-$3$ current by adding the quasi primary field in (\ref{bcgspin3}) 
to the left
hand side of above OPE.

Let us consider the OPE between the spin-$\frac{3}{2}$ current and 
the higher spin-$\frac{7}{2}$ current found in previous OPE
\bea
%%%%%%%%%%%%%%%%%%%%%%%%%%%%%%%%%%%%%%%%%%%%%%%%%%%%%%%%%%%%%%%%%
G_{21}(z) \, {\bf S_{-}^{(\frac{7}{2})}}(w) & = & 
%%%%%%%%%%%%%%%%%%%%%%%%%%%%%%%%%%%%%%%%%%%%%%%%%%%%%%%%%%%%%%%%%
\frac{1}{(z-w)^4} \left[
-\frac{16 {\bf (k-3)} (23 k+19) (61 k+113)}{15 (k+5) (13 k+17) (39 k+67)}
{\bf T^{(1)}} \right. \nonu \\
& - & 
\frac{8i}{5 (k+5)^3 (13 k+17) (39 k+67) (47 k+43)}
 \nonu \\
&\times& 
(812455 k^5+6422062 k^4+16948458 k^3+20543408 k^2+11175575 k
\nonu \\
& + & 1957962) A_3  + \frac{8ik}{15 (k+5)^3 (13 k+17) (39 k+67) (47 k+43)}
\nonu \\
& \times & (
410780 k^5+10154773 k^4+64740756 k^3+155849878 k^2 \nonu \\
&+ & 159556664 k+58487229) B_3
\nonu \\
&-& \frac{24 k (6277 k^3+34435 k^2+48679 k+19689)}{
5 (k+5)^3 (13 k+17) (47 k+43)} U
\nonu \\
&-& \frac{16}{15 (k+5)^4 (13 k+17) (39 k+67)} 
(4370 k^5+94701 k^4+517032 k^3 \nonu \\
& + & 914974 k^2+592158 
k+68301) F_{11} F_{22}
\nonu \\
&+&  \frac{16}{15 (k+5)^4 (13 k+17) (39 k+67)} 
(4370 k^5+63819 k^4+274248 k^3\nonu \\
&+ &  \left. 520234 k^2+269646 k-
68301) F_{12} F_{21} 
\right](w)
\nonu \\
&+& \frac{1}{(z-w)^3} \left[ 
\frac{12 (k-3)}{5 (k+5)} {\bf P^{(2)}}
-\frac{48 (k-3)}{5 (k+5)} {\bf W^{(2)}}
\right. \nonu \\
& - & \frac{16 (k-3) 
(1543 k^3+225 k^2-10179 k-11101)}{15 (k+5)^2 (13 k+17) (39 k+67)}
{\bf T^{(2)}}
\nonu \\
& + & \frac{128 i {\bf (k-3)}}{5 (k+5) (13 k+17)}
A_3 {\bf T^{(1)}}
+ \frac{32 i {\bf (k-3)} (k+1)}{5 (k+5) (13 k+17)}
B_3 {\bf T^{(1)}}
\nonu \\
& + &  \frac{8 {\bf (k-3)} (23 k+19) (61 k+113)}{15 (k+5) (13 k+17) (39 k+67)}
\pa {\bf T^{(1)}}
\nonu \\
& - & 
\frac{16 (1613 k^4+18563 k^3+106625 k^2+284877 k+223074)}{
15 (k+5)^2 (13 k+17) (39 k+67)}
T
\nonu \\
&-& \frac{16 (6218 k^4+72179 k^3+335333 k^2+710217 k+
525069)}{15 (k+5)^3 (13 k+17) (39 k+67)}
A_3 A_3
\nonu \\
&-& \frac{16 (7922 k^4+90179 k^3+180017 k^2-14607 k-
169263)}{15 (k+5)^3 (13 k+17) (39 k+67)}
A_3 B_3
\nonu \\
&+& \frac{4 i}{15 (k+5)^3 (13 k+17) (39 k+67) (47 k+43)} 
\nonu \\
& \times & (1268381 k^5+5934390 k^4-9953250 k^3-74286536 k^2-
106361763 k\nonu \\
& - & 46310742) \pa A_3 
-  \frac{32 i (k^2-10 k-131)}{5 (k+5)^4} A_{+} F_{21} F_{22}
\nonu \\
& - & \frac{64 i 
(637 k^4-1209 k^3-34981 k^2-100635 k-79596)}{
5 (k+5)^4 (13 k+17) (39 k+67)} A_3 F_{11} F_{22}
\nonu \\
&+& \frac{48 i (27 k^3+261 k^2+937 k+1023)}{5 (k+5)^4 (13 k+17)}
A_3 F_{12} F_{21}
\nonu \\
&+& \frac{16 i (125 k^3+2287 k^2+5575 k+4757)}{
5 (k+5)^4 (39 k+67)} A_{-} F_{11} F_{12}
\nonu \\
&-& \frac{16 (6218 k^4+65225 k^3+263723 k^2+481683 k+303399)}{
15 (k+5)^3 (13 k+17) (39 k+67)}
A_{+} A_{-} \nonu \\
&-& \frac{32 (234 k^5-701 k^4+6760 k^3+88144 k^2+185490 k+111537)}{
15 (k+5)^3 (13 k+17) (39 k+67)}
B_3 B_3
\nonu \\
& -&\frac{4 i}{15 (k+5)^3 (13 k+17) (39 k+67) (47 k+43)} 
\nonu \\
& \times & (410780 k^6+10444857 k^5+73693068 k^4
\nonu \\
& + & 211858366 k^3+283781640 k^2+173849361 k+38368728) \pa B_3
\nonu \\
&-& \frac{32 i (78 k^5-243 k^4+6026 k^3+65908 k^2+161880 k+
113967)}{5 (k+5)^4 (13 k+17) (39 k+67)}
B_3 F_{11} F_{22}
\nonu \\
&+& \frac{16 i (4 k^4+107 k^3+813 k^2+2649 k+2259)}{
5 (k+5)^4 (13 k+17)} B_3 F_{12} F_{21}
\nonu \\
&+& \frac{32 i (191 k^3+423 k^2-1527 k-3551)}{
5 (k+5)^4 (39 k+67)} B_{-} F_{12} F_{22}
\nonu \\
&-& \frac{16 (1543 k^4+46207 k^3+255643 k^2+426885 k+
223074)}{15 (k+5)^3 (13 k+17) (39 k+67)}
B_{+} B_{-}
\nonu \\
&- & \frac{16 i (11 k^2+78 k+227)}{5 (k+5)^4}
B_{+} F_{11} F_{21}
\nonu \\
&+& \frac{8 (16687 k^4+122554 k^3+344764 k^2+530214 k+322677)}{15 
(k+5)^3 (13 k+17) (39 k+67) } F_{11} G_{22}
\nonu \\
&-& \frac{8}{15 (k+5)^4 (13 k+17) (39 k+67)} 
 (8066 k^5+12481 k^4 \nonu \\
& + & 374680 k^3+3383510 k^2+7061982 k+
4345953)
\pa F_{11} F_{22}
\nonu \\
&+& 
\frac{8}{15 (k+5)^4 (13 k+17) (39 k+67)} (16806 k^5+280975 k^4
\nonu \\
& + & 2000920 k^3+6245866 k^2+8394762 k+4004175) F_{11} \pa F_{22}
\nonu \\
&-& \frac{4}{5 (k+5)^3 (13 k+17) (39 k+67) (47 k+43)} 
(212589 k^5+2511712 k^4 \nonu \\
& + & 6213530 k^3+971776 k^2
-8108847 k-4883080)
F_{12} G_{21}
\nonu \\
&-& \frac{8}{5 (k+5)^4 (13 k+17) (39 k+67)} 
(5602 k^5+84477 k^4 
\nonu \\
&+ & 531736 k^3+1929966 k^2+3036574 k+1647789)
\pa F_{12} F_{21}
\nonu \\
& + & \frac{8 }{15 (k+5)^4 (13 k+17) (39 k+67)}
(8066 k^5+78369 k^4 \nonu \\
& + & 751944 k^3+3118438 k^2+5205102 k+2947761)
F_{12} \pa F_{21}
\nonu \\
&+& \frac{8 (165 k^3+1186 k^2+1351 k+34)}{5 (k+5)^3 (13 k+17)}
F_{21} G_{12}
\nonu \\
&-& 
\frac{16 (1387 k^4+35356 k^3+179710 k^2+272436 k+127143)}{
15 (k+5)^3 (13 k+17) (39 k+67)}
F_{22} G_{11}
\nonu \\
&+& \frac{48 i 
(13 k^3+4187 k^2+2431 k-4879)}{5 (k+5)^3 (13 k+17) (39 k+67)}
U A_3 
\nonu \\
&+& \frac{32 i k (1443 k^3+20601 k^2+53657 k+38083)}{
5 (k+5)^3 (13 k+17) (39 k+67)}
U B_3
\nonu \\
&+& \frac{12 k (6277 k^3+34435 k^2+48679 k+19689)}{
5 (k+5)^3 (13 k+17) (47 k+43)}
\pa U
\nonu \\
&-& \frac{32 
(3109 k^4+39658 k^3+154630 k^2+225810 k+111537)}{
15 (k+5)^3 (13 k+17) (39 k+67)}
U U 
\nonu \\
& + & \frac{64 (76 k^3+373 k^2+2126 k+2613)}{5 (k+5)^4 (39 k+67)}
U F_{11} F_{22}
\nonu \\
&-& \left. \frac{16 (13 k-7)}{5 (k+5)^3} U F_{12} F_{21}
\right](w) 
\nonu \\
&+& \frac{1}{(z-w)^2} \left[ 
\frac{6 (k-3)}{5 (k+5)} {\bf P^{(3)}}
+6 {\bf S^{(3)}}
-\frac{12 (3 k+11)}{5 (k+5)} {\bf W^{(3)}}
\right. \nonu \\
&+& \frac{96 i {\bf (k-3)}}{5 (k+5) (13 k+17)} \pa A_3 {\bf T^{(1)}}
-\frac{96 i {\bf (k-3)}}{5 (k+5) (13 k+17)} A_3 \pa {\bf T^{(1)}}
\nonu \\
& + & \frac{24 i {\bf (k-3)} (k+1)}{5 (k+5) (13 k+17)} \pa B_3 {\bf T^{(1)}}
- \frac{24 i {\bf (k-3)} (k+1)}{5 (k+5) (13 k+17)}
B_3 \pa {\bf T^{(1)}}
\nonu \\
&-& \frac{12 (k-3)}{5 (13 k+17)} G_{12} {\bf T_{+}^{(\frac{3}{2})}}
+ \frac{20 (k-3) (k+5)}{(13 k+17) (39 k+67)} G_{21} {\bf T_{-}^{(\frac{3}{2})}}
\nonu \\
&-& \frac{4 (k-3) (71 k+163)}{5 (13 k+17) (39 k+67)} \pa {\bf T^{(2)}}
+ \frac{8 {\bf (k-3)} (23 k+19)}{5 (13 k+17) (39 k+67)} \pa^2 {\bf T^{(1)}}
\nonu \\
&-& \frac{32 {\bf (k-3)} (23 k+19)}{15 (13 k+17) (39 k+67)} T {\bf T^{(1)}}
- \frac{4 i (11 k+79)}{5 (k+5)^3} A_3 F_{11} G_{22}
\nonu \\
&+& \frac{16 i}{5 (k+5)^2} A_3 A_3 A_3
-\frac{96 i (k+3)}{5 (k+5)^3} A_3 A_3 B_3
-\frac{16}{5 (k+5)^2} \pa A_3 A_3
\nonu \\
&- & \frac{4 (1638 k^4-20469 k^3-190079 k^2-392311 k-251283)}{
5 (k+5)^3 (13 k+17) (39 k+67)} \pa A_3 B_3
\nonu \\
&+& \frac{4 (1638 k^4-8301 k^3-116759 k^2-254527 k-169275)}{
5 (k+5)^3 (13 k+17) (39 k+67)} A_3 \pa B_3
\nonu \\
&+& \frac{2 i}{5 (k+5)^3 (13 k+17) (39 k+67) (47 k+43)} 
(221563 k^5+2925184 k^4 \nonu \\
& + & 12258818 k^3+11607476 k^2-
12780037 k-14213564)
\pa^2 A_3
\nonu \\
&+& \frac{2 i (1287 k^4+29908 k^3+207762 k^2+192324 k-118657)}
{5 (k+5)^4 (13 k+17) (39 k+67)} A_3 \pa F_{11} F_{22}
\nonu \\
&+& \frac{2 i (23595 k^4+304260 k^3+1360330 k^2+2098996 k+952003)}{
5 (k+5)^4 (13 k+17) (39 k+67)} A_3 F_{11} \pa F_{22}
\nonu \\
&-& 
\frac{2 i (3070 k^4+47971 k^3+191033 k^2+178769 k-963)}{
5 (k+5)^3 (13 k+17) (39 k+67)}
A_3 F_{12} G_{21}
\nonu \\
&+& \frac{8 i (115 k^3-485 k^2-4995 k-4619)}{5 (k+5)^4 (13 k+17)}
\pa A_3 F_{12} F_{21}
-\frac{96 i (k+3)}{5 (k+5)^3} A_{+} A_{-} B_3
\nonu \\
&+& \frac{2 i (307 k^3+5933 k^2+24657 k+19479)}{5 (k+5)^4 (13 k+17)}
A_3  \pa F_{12} F_{21}
+ \frac{16 i}{5 (k+5)^2} A_{+} A_{-} A_3
\nonu \\
&-& \frac{2 i (473 k^3+391 k^2-10701 k-10299)}{5 (k+5)^4 (13 k+17)}
A_3 F_{12} \pa F_{21}
+  \frac{4 i (31 k+131)}{5 (k+5)^3} B_3 F_{11} G_{22}
\nonu \\
&- & \frac{2 i (6 k^3-151 k^2-680 k-411)}{5 (k+5)^3 (13 k+17)}
A_3 F_{21} G_{12}
+ \frac{4 i (7 k+59)}{5 (k+5)^3} A_3 F_{22} G_{11}
\nonu \\
&+& \frac{2 i (6 k^3-221 k^2-1780 k-2241)}{5 (k+5)^3 (13 k+17)}
A_{-} F_{11} G_{12}
+ \frac{4 i (11 k+59)}{5 (k+5)^3} A_{-} F_{12} G_{11}
\nonu \\
&-& \frac{8 i (99 k^3+2695 k^2+14461 k+16281)}{5 (k+5)^4 (39 k+67)}
\pa A_{-} F_{11} F_{12}
-\frac{16 (5 k+13)}{5 (k+5)^3} \pa B_3 B_3
\nonu \\
&-& \frac{16 i (87 k^3-216 k^2-4617 k-6298)}{5 (k+5)^4 (39 k+67)}
A_{-} \pa F_{11} F_{12}
+  \frac{16 i (5 k+13)}{5 (k+5)^3} B_3 B_3 B_3
\nonu \\
&+& \frac{16 i (186 k^3+2479 k^2+9844 k+9983)}{5 (k+5)^4 (39 k+67)}
A_{-} F_{11} \pa F_{12}
+ \frac{4 i (11 k+59)}{5 (k+5)^3} B_{-} F_{12} G_{22}
\nonu \\
& -& \frac{4 (826 k^3-1759 k^2-19756 k-20675)}{
5 (k+5)^2 (13 k+17) (39 k+67)}
\pa A_{+} A_{-}
+  \frac{4 i (13 k-7)}{5 (k+5)^3} B_3 F_{22} G_{11}
\nonu \\
&+& \frac{4 (826 k^3-1759 k^2-19756 k-20675)}{
5 (k+5)^2 (13 k+17) (39 k+67)}
A_{+} \pa A_{-}
+ \frac{16 i (5 k+13)}{5 (k+5)^3} B_{+} B_{-} B_3
\nonu \\
&-& \frac{4 i (9 k+49)}{5 (k+5)^3} A_{+} F_{21} G_{22}
-\frac{16 i \left(k^2+30 k+129\right)}{5 (k+5)^4}
\pa A_{+} F_{21} F_{22}
\nonu \\
& -& \frac{32 i \left(k^2-k-32\right)}{5 (k+5)^4} A_{+} \pa F_{21} F_{22}
+ \frac{32 i \left(2 k^2+29 k+97\right)}{5 (k+5)^4} A_{+} F_{21} \pa F_{22}
\nonu \\
&+& \frac{2 i (3070 k^4+44065 k^3+181611 k^2+186907 k+32787)}{
5 (k+5)^3 (13 k+17) (39 k+67)}
A_{+} F_{22} G_{21}
 \nonu \\
&+& \frac{2}{5 (k+5)^3 (13 k+17) (39 k+67) (47 k+43)} 
i (23312 k^6+533811 k^5 \nonu 
\\
& + & 2300220 k^4 -  
431454 k^3-11134868 k^2-11008445 k-2079136)
\pa^2 B_3
\nonu \\
&-& \frac{12 i}{5 (k+5)^4 (13 k+17) (39 k+67)} 
(156 k^5+1295 k^4\nonu \\
& - & 3100 k^3-64246 k^2-105576 k-24321)
\pa B_3 F_{11} F_{22}
\nonu \\
&+& \frac{2 i}{5 (k+5)^4 (13 k+17) (39 k+67)} 
(936 k^5+11085 k^4 \nonu \\
& + & 232412 k^3+1112022 k^2+1962372 k+
1161445)
B_3 \pa F_{11} F_{22}
\nonu \\
&+& \frac{2 i}{5 (k+5)^4 (13 k+17) (39 k+67)} 
(936 k^5+29337 k^4 \nonu \\
& + & 64556 k^3-314882 k^2-937964 k-619951)
B_3 F_{11} \pa F_{22}
\nonu \\
&+& \frac{2 i (3070 k^4+47035 k^3+227457 k^2+492889 k+345429)}{
5 (k+5)^3 (13 k+17) (39 k+67)}
B_3 F_{12} G_{21}
\nonu \\
&+& \frac{4 i (12 k^4+87 k^3+2237 k^2+11541 k+11299)}{
5 (k+5)^4 (13 k+17)}
\pa B_3 F_{12} F_{21}
\nonu \\
&-& \frac{2 i (24 k^4+785 k^3+2959 k^2+8187 k+7077)}{
5 (k+5)^4 (13 k+17)}
B_3 \pa F_{12} F_{21}
\nonu \\
& - & \frac{2 i (24 k^4+317 k^3+7651 k^2+33375 k+30945)}{
5 (k+5)^4 (13 k+17)}
B_3 F_{12} \pa F_{21}
\nonu \\
&+& \frac{2 i (6 k^3+889 k^2+3800 k+3669)}{5 (k+5)^3 (13 k+17)}
B_3 F_{21} G_{12}
\nonu \\
&-& \frac{2 i (267 k^3+4789 k^2+22521 k+30351)}{5 (k+5)^4 (39 k+67)}
\pa B_{-} F_{12} F_{22}
\nonu \\
&+& \frac{2 i (345 k^3+4455 k^2+17427 k+22981)}{5 (k+5)^4 (39 k+67)}
B_{-} \pa F_{12} F_{22}
\nonu \\
&+& \frac{2 i (189 k^3+5123 k^2+27615 k+37721)}{5 (k+5)^4 (39 k+67)}
B_{-} F_{12} \pa F_{22}
\nonu \\
&+& \frac{2 i (6 k^3-221 k^2-1780 k-2241)}{5 (k+5)^3 (13 k+17)}
B_{-} F_{22} G_{12}
\nonu \\
&-& \frac{4 (1475 k^4+9959 k^3+31887 k^2+78277 k+64666)}{
5 (k+5)^3 (13 k+17) (39 k+67)}
\pa B_{+} B_{-}
\nonu \\
&+& \frac{4 (1475 k^4+9959 k^3+31887 k^2+78277 k+64666)}{
5 (k+5)^3 (13 k+17) (39 k+67)}
B_{+} \pa B_{-}
\nonu \\
&-& \frac{2 i (50 k^4-985 k^3+19149 k^2+172733 k+181413)}{
5 (k+5)^3 (13 k+17) (39 k+67)}
B_{+} F_{11} G_{21}
\nonu \\
&-& \frac{2 i (39 k^2+382 k+903)}{5 (k+5)^4} 
\pa B_{+} F_{11} F_{21}
+ \frac{2 i (37 k^2+410 k+1093)}{5 (k+5)^4} 
B_{+} \pa F_{11} F_{21}
\nonu \\
& + &  \frac{2 i (41 k^2+354 k+713)}{5 (k+5)^4}
B_{+} F_{11} \pa F_{21}
+ \frac{192 k}{(k+5)^2 (39 k+67)} F_{11} F_{12} F_{22} G_{21}
\nonu \\
&+& \frac{2}{5 (k+5)^4 (13 k+17) (39 k+67)} 
(1156 k^5+57927 k^4 \nonu \\
& + & 484204 k^3+1404610 k^2+
1377864 k+100911)
\pa^2 F_{11} F_{22}
\nonu \\
&-& \frac{4}{5 (k+5)^4 (13 k+17) (39 k+67)} 
(496 k^5+68487 k^4 \nonu \\
& + & 589748 k^3+1717458 k^2+2192244 k+1275327)
\pa F_{11} \pa F_{22}
\nonu \\
&+& \frac{2}{
5 (k+5)^4 (13 k+17) (39 k+67)
} (1156 k^5+57927 k^4 \nonu \\
& + & 484204 k^3+1404610 k^2+1377864 k+100911)
F_{11} \pa^2 F_{22}
\nonu \\
&+& \frac{(7321 k^4-33698 k^3-523364 k^2-1342878 k-943605)}{
5 (k+5)^3 (13 k+17) (39 k+67)}
\pa F_{11} G_{22}
\nonu \\
&-& \frac{(2251 k^4-18618 k^3-244984 k^2-676278 k-499395)}{
5 (k+5)^3 (13 k+17) (39 k+67)}
F_{11} \pa G_{22}
\nonu \\
&-& \frac{2}{5 (k+5)^4 (13 k+17) (39 k+67)} 
(1156 k^5+46107 k^4 \nonu \\
& + & 428708 k^3+1545554 k^2+2544784 k+1682763)
\pa^2 F_{12} F_{21}
\nonu \\
&+& \frac{4}{5 (k+5)^4 (13 k+17) (39 k+67)} 
(496 k^5+52191 k^4 \nonu \\
& + & 509484 k^3+1617994 k^2+2018524 k+686031)
\pa F_{12} \pa F_{21}
\nonu \\
&-& \frac{2}{5 (k+5)^4 (13 k+17) (39 k+67)} 
(1156 k^5+46107 k^4 \nonu \\
& + & 428708 k^3+1545554 k^2+2544784 k+
1682763)
F_{12} \pa^2 F_{21}
\nonu \\
&-& \frac{1}{5 (k+5)^3 (13 k+17) (39 k+67) (47 k+43)}
(946375 k^5+11001867 k^4 \nonu \\
&+ & 54265174 k^3+149710406 k^2+
178493139 k+70730399)
\pa F_{12} G_{21}
\nonu \\
&+& 
\frac{1}{5 (k+5)^3 (13 k+17) (39 k+67) (47 k+43)} (410085 k^5+4466445 k^4
\nonu \\
& + & 17966834 k^3+36726202 k^2+35009881 k+11610473)
F_{12} \pa G_{21}
\nonu \\
&+& \frac{2 (63 k^3+40 k^2-2381 k-2726)}{5 (k+5)^3 (13 k+17)}
\pa F_{21} G_{12} 
- \frac{4 i (9 k+49)}{5 (k+5)^3} B_{+} F_{21} G_{11}
\nonu \\
&-& \frac{2 (25 k^3+22 k^2-927 k-1292)}{5 (k+5)^3 (13 k+17)}
F_{21} \pa G_{12}
\nonu \\
&-& \frac{2 (6187 k^4+57202 k^3+194476 k^2+325326 k+195561)}{
5 (k+5)^3 (13 k+17) (39 k+67)}
\pa F_{22} G_{11}
\nonu \\
&+& \frac{2 
(1117 k^4+8400 k^3+33170 k^2+102888 k+86217)}{5 (k+5)^3 (13 k+17) (39 k+67)}
F_{22} \pa G_{11}
\nonu \\
&+& \frac{18 (k+3)}{5 (k+5)^2} G_{11} G_{22}
-\frac{2 (1744 k^3+23367 k^2+59474 k+42267)}{5 (k+5) (13 k+17) (39 k+67)}
G_{12} G_{21}
\nonu \\
&+& \frac{4 i (11369 k^4+728349 k^3+5556781 k^2+11678039 k+6652702)}{
5 (k+5)^2 (13 k+17) (39 k+67) (47 k+43)}
T A_3 
\nonu \\
&+& \frac{4 i}{15 (k+5)^2 (13 k+17) (39 k+67) (47 k+43)} 
(85352 k^5+1785237 k^4 \nonu \\
& + & 11949591 k^3+
28863259 k^2+26292429 k+7640412)
T B_3
\nonu \\
&+& \frac{4 (801 k^4+13102 k^3+73621 k^2+150124 k+102516)}{
5 (k+5)^2 (13 k+17) (39 k+67)}
\pa T
\nonu \\
&-& \frac{4 (5059 k^3+44554 k^2+120379 k+78948)}{
5 (k+5)^2 (13 k+17) (47 k+43)} T U
-\frac{16 (k-15)}{5 (k+5)^3} U A_3 B_3
\nonu \\
&-& \frac{16 (454 k^4+11473 k^3+59023 k^2+62739 k-39153)}{
15 (k+5)^3 (13 k+17) (39 k+67)}
T F_{11} F_{22}
\nonu \\
&+& \frac{32 (227 k^3+3820 k^2+12093 k+14508)}{
15 (k+5)^2 (13 k+17) (39 k+67)}
T F_{12} F_{21}
- \frac{16 (5 k+27)}{5 (k+5)^3} U A_3 A_3 
\nonu \\
& +& \frac{16 i (413 k^4+4082 k^3+9344 k^2+8238 k+4419)}
{5 (k+5)^3 (13 k+17) (39 k+67)}
\pa U A_3
\nonu \\
&-& \frac{16 i (413 k^4+6617 k^3+30703 k^2+55351 k+35172)}{
5 (k+5)^3 (13 k+17) (39 k+67)}
U \pa A_3
\nonu \\
&+& 
\frac{8 i (929 k^4-5539 k^3-66895 k^2-171009 k-123894)}{
5 (k+5)^3 (13 k+17) (39 k+67)}
\pa U B_3
\nonu \\
&-& \frac{8 i (929 k^4-469 k^3-24177 k^2-76783 k-62388)}{
5 (k+5)^3 (13 k+17) (39 k+67)}
U \pa B_3
\nonu \\
&-&  \frac{16 (5 k+27)}{5 (k+5)^3} U B_{+} B_{-}
+\frac{2 (5059 k^3+44554 k^2+120379 k+78948)}{
5 (k+5)^2 (13 k+17) (47 k+43)}
\pa^2 U
\nonu  \\
& - &  \frac{2 (903 k^3+28265 k^2+140557 k+158187)}{5 (k+5)^4 (39 k+67)}
U \pa F_{11} F_{22}
- \frac{16 (5 k+27)}{5 (k+5)^3} U A_{+} A_{-}
\nonu \\
&+& \frac{2 (2997 k^3+34907 k^2+141319 k+159393)}{
5 (k+5)^4 (39 k+67)}
U F_{11} \pa F_{22}
-\frac{16 (5 k+27)}{5 (k+5)^3} U B_3 B_3
\nonu \\
&-& \frac{2 (3070 k^4+55303 k^3+291205 k^2+593469 k+386433)}{
5 (k+5)^3 (13 k+17) (39 k+67)}
U F_{12} G_{21}
\nonu \\
&-& \frac{2 \left(83 k^2+710 k+1859\right)}{5 (k+5)^4} U \pa F_{12} F_{21}
+ \frac{2 \left(25 k^2+658 k+2281\right)}{5 (k+5)^4} U F_{12} \pa F_{21}
\nonu \\
&+& \frac{6 \left(2 k^3+123 k^2+520 k+543\right)}{5 (k+5)^3 (13 k+17)}
U F_{21} G_{12}
+ \frac{4 (11 k+31)}{5 (k+5)^3} U F_{11} G_{22}
\nonu \\
&-& \frac{4 (11 k+31)}{5 (k+5)^3} U F_{22} G_{11}
+ \frac{16 i}{5 (k+5)^2} U U A_3
+ \frac{16 i (5 k+13)}{5 (k+5)^3} U U B_3
\nonu \\
&-& \frac{16 (5 k+27)}{5 (k+5)^3} U U U
+ \frac{4 (167 k^3+1175 k^2+3401 k+1593)}{5 (k+5)^4 (13 k+17)}
F_{22} G_{11} F_{22} F_{11}
\nonu \\
&-& \frac{2 (585 k^4-53056 k^3-361318 k^2-565672 k-203211)}{
5 (k+5)^4 (13 k+17) (39 k+67)}
F_{11} G_{22} F_{11} F_{22}
\nonu \\
& + & \left. \frac{2 (29 k^2+26 k-211)}{5 (k+5)^4} 
F_{12} G_{21} F_{12} F_{21}
\right](w)
\nonu \\
&+& \frac{1}{(z-w)} \left[ 
\frac{1}{6} \pa \{ G_{21} \, {\bf S_{-}^{(\frac{7}{2})}} \}_{-2}
+ \frac{9 (k-3)}{(41 k+85)} \left( T {\bf P^{(2)}} -\frac{3}{10} 
\pa^2 {\bf P^{(2)}}\right)
\right. \nonu \\
&-& \frac{36 (k-3)}{(41 k+85)}
\left( T {\bf W^{(2)}} -\frac{3}{10} 
\pa^2 {\bf W^{(2)}}\right)
\nonu \\
&-& \frac{4 (k-3) 
(1543 k^3+225 k^2-10179 k-11101)}
{(k+5) (13 k+17) (39 k+67) (41 k+85)}
\left( T {\bf T^{(2)}} -\frac{3}{10} 
\pa^2 {\bf T^{(2)}}\right)
\nonu \\
&+& \frac{96 i {\bf (k-3)}}{(13 k+17) (41 k+85)} 
 \left( T A_3 {\bf T^{(1)}} -\frac{3}{10} 
\pa^2 (A_3 {\bf T^{(1)}}) \right)
\nonu \\
&+& \frac{24 i {\bf (k-3)} (k+1)}{(13 k+17) (41 k+85)}
 \left( T B_3 {\bf T^{(1)}} -\frac{3}{10} 
\pa^2 (B_3 {\bf T^{(1)}}) \right)
\nonu \\
& +& 
\frac{16 {\bf (k-3)} (61 k+113)}{9 (13 k+17) (39 k+67)}
\left( T  \pa {\bf T^{(1)}}  -\frac{1}{2} \pa T {\bf T^{(1)}} 
-\frac{1}{4} \pa^3 {\bf T^{(1)}} \right)
+ {\bf S^{(4)}}
\nonu \\
&-& 
\frac{2 (5k+13)^{-1}(7k+11)^{-1}(37+17k)^{-1}(73+53k)^{-1}(71 k+115)^{-1}}{
(k+5) (13 k+17) (23 k+19) (29 k+97) (39 k+67) (41 k+85)} 
\nonu \\
& \times & (18582783268720 k^{11}+468783625312793 k^{10}+5611734564211938 k^9
\nonu \\
& + & 41116614702193245 k^8 
+  200379415234559352 k^7
\nonu \\
& + & 671296144948119522 k^6
+  1562566286025217548 k^5
\nonu \\
& + & 2514238282804743954 k^4 + 
2733706958184300744 k^3
\nonu \\
& + & 1910595867995994437 k^2+771960599593492834 k
\nonu \\
& + & 
136491146486617425)
 \left( T T -\frac{3}{10} 
\pa^2 T \right)
\nonu \\
&- & \frac{4 (6218 k^4+72179 k^3+335333 k^2+710217 k+525069)}
{(k+5)^2 (13 k+17) (39 k+67) (41 k+85)}
 \nonu \\
& \times & \left( T A_3 A_3 -\frac{3}{10} 
\pa^2 (A_3 A_3) \right)
\nonu \\
& -&\frac{4 (7922 k^4+90179 k^3+180017 k^2-14607 k-169263)}
{(k+5)^2 (13 k+17) (39 k+67) (41 k+85)}
\nonu \\
& \times & \left( T A_3 B_3 -\frac{3}{10} 
\pa^2 (A_3 B_3) \right)
\nonu \\
&-& \frac{4 (6218 k^4+65225 k^3+263723 k^2+481683 k+303399)}{
5 (k+5)^2 (7 k+11) (13 k+17) (39 k+67)}
 \nonu \\
& \times & \left( T A_{+} A_{-} -\frac{1}{2} 
\pa^2 A_{+} A_{-} -\frac{1}{2} A_{+} \pa^2 A_{-} -\frac{i}{12} \pa^3 A_3 +
\frac{i}{2} \pa T A_3 \right)
\nonu \\
&-& \frac{8 (234 k^5-701 k^4+6760 k^3+88144 k^2+185490 k+111537)}{
(k+5)^2 (13 k+17) (39 k+67) (41 k+85)}
\nonu \\ 
& \times & \left( T B_3 B_3 -\frac{3}{10} 
\pa^2 (B_3 B_3) \right)
\nonu \\
&-& \frac{4 (1543 k^4+46207 k^3+255643 k^2+426885 k+223074)}{
5 (k+5)^2 (7 k+11) (13 k+17) (39 k+67)}
\nonu \\
& \times & \left( T B_{+} B_{-} 
 -\frac{1}{2} 
\pa^2 B_{+} B_{-} -\frac{1}{2} B_{+} \pa^2 B_{-} -\frac{i}{12} \pa^3 B_3 +
\frac{i}{2} \pa T B_3 
 \right)
\nonu \\
&-& \frac{48 i (637 k^4-1209 k^3-34981 k^2-100635 k-79596)}{
(k+5)^3 (13 k+17) (39 k+67) (41 k+85)}
\nonu \\
&\times &  \left( T A_3 F_{11} F_{22} -\frac{3}{10} 
\pa^2 (A_3 F_{11} F_{22}) \right)
\nonu \\
&+& \frac{36 i (27 k^3+261 k^2+937 k+1023)}{(k+5)^3 (13 k+17) (41 k+85)}
 \left( T A_3 F_{12} F_{21} -\frac{3}{10} 
\pa^2 (A_3 F_{12} F_{21}) \right)
\nonu \\
&+& \frac{24 i (125 k^3+2287 k^2+5575 k+4757)}{
5 (k+5)^3 (17 k+37) (39 k+67)}
\nonu \\
&\times &  
\left( T A_{-} F_{11} F_{12} 
 -\frac{1}{2} 
\pa^2 A_{-} F_{11} F_{12} - 3 A_{-} \pa F_{11} \pa F_{12}
\right. \nonu \\
& - &  
\frac{i}{4} \pa T F_{11} F_{22} +\frac{3i}{4} \pa^2 F_{11} \pa F_{22}
-\frac{i}{4} \pa T F_{12} F_{21} +\frac{3i}{4} \pa^2 F_{12} \pa F_{21} 
\nonu \\
&-& \left. \frac{i}{12}  \pa^3 F_{11}  F_{22} -\frac{i}{12} \pa^3 F_{12}
 F_{21} \right)
\nonu \\
&- & \frac{48 i (k^2-10 k-131)}{5 (k+5)^3 (17 k+37)}
\left( T A_{+} F_{21} F_{22} -\frac{1}{2} 
\pa^2 A_{+} F_{21} F_{22} - 3 A_{+} \pa F_{21} \pa F_{22}
\right. \nonu \\
& + &  
\frac{i}{4} \pa T F_{11} F_{22} +\frac{i}{12} F_{11} \pa^3 F_{22}
+\frac{i}{4} \pa T F_{12} F_{21} +\frac{i}{12} F_{12} \pa^3 F_{21} 
\nonu \\
&-& \left. \frac{3i}{4}  \pa F_{11} \pa^2 F_{22} -\frac{3i}{4} \pa F_{12}
\pa^2 F_{21} \right)
\nonu \\
&-& \frac{24 i (78 k^5-243 k^4+6026 k^3+65908 k^2+161880 k+113967)}{
(k+5)^3 (13 k+17) (39 k+67) (41 k+85)}
\nonu \\
& \times &
\left( T B_3 F_{11} F_{22} -\frac{3}{10} 
\pa^2 (B_3 F_{11} F_{22}) \right)
\nonu \\
&+& 
\frac{12 i (4 k^4+107 k^3+813 k^2+2649 k+2259)}{
(k+5)^3 (13 k+17) (41 k+85)}
\nonu \\
& \times & \left( T B_3 F_{12} F_{21} -\frac{3}{10} 
\pa^2 (B_3 F_{12} F_{21}) \right)
\nonu \\
&+& \frac{192 i (191 k^3+423 k^2-1527 k-3551)}{
5 (k+5)^3 (39 k+67) (53 k+73)}
\nonu \\
&\times & 
\left( T B_{-} F_{12} F_{22} 
 -\frac{3}{4} 
\pa^2 B_{-} F_{12} F_{22} - \frac{3}{4} B_{-}  F_{12} \pa^2 F_{22}
\right. \nonu \\
& + &  
\frac{3}{2} \pa B_{-} F_{12} \pa F_{22} -3  B_{-} \pa F_{12} \pa F_{22}
-\frac{i}{4} \pa T F_{11} F_{22} +\frac{i}{4} \pa T F_{12}  F_{21} 
\nonu \\
&+& \left. \frac{3i}{4}  \pa F_{11}  \pa^2 F_{22} -\frac{3i}{4} \pa F_{12}
 \pa^2 F_{21} \right)
\nonu \\
&-& \frac{6 i (11 k^2+78 k+227)}{5 (k+5)^3 (5 k+13)}
\left( T B_{+} F_{11} F_{21} 
-\frac{3}{2}  \pa B_{+} \pa F_{11} F_{21}  \right. \nonu \\
&- & \left. \frac{3}{2}  \pa B_{+} F_{11} \pa F_{21}   
+\frac{i}{4} \pa T F_{11} F_{22}
-\frac{i}{4} \pa T F_{12} F_{21}
\right)
\nonu \\
&+& \frac{48 (76 k^3+373 k^2+2126 k+2613)}{(k+5)^3 (39 k+67) (41 k+85)}
\left( T U F_{11} F_{22} -\frac{3}{10} 
\pa^2 ( U F_{11} F_{22}) \right)
\nonu \\
&-& \frac{12 (13 k-7)}{(k+5)^2 (41 k+85)} 
\left( T U F_{12} F_{21} -\frac{3}{10} 
\pa^2 ( U F_{12} F_{21}) \right)
\nonu \\
&+& \frac{2 (16687 k^4+122554 k^3+344764 k^2+530214 k+322677)}{
(k+5)^2 (13 k+17) (39 k+67) (41 k+85)}
\nonu \\
& \times & \left( T  F_{11} G_{22} -\frac{3}{10} 
\pa^2 (  F_{11} G_{22}) \right)
\nonu \\
&-& \frac{3}{(k+5)^2 (13 k+17) (39 k+67) (41 k+85) (47 k+43)} 
\nonu \\ 
& \times & (212589 k^5+2511712 k^4+6213530 k^3+971776 k^2
\nonu \\ 
& - & 8108847 k-4883080)  \left( T  F_{12} G_{21} -\frac{3}{10} 
\pa^2 (  F_{12} G_{21}) \right)
\nonu \\
&+& \frac{6 (165 k^3+1186 k^2+1351 k+34)}{
(k+5)^2 (13 k+17) (41 k+85)}
\left( T  F_{21} G_{12} -\frac{3}{10} 
\pa^2 (  F_{21} G_{12}) \right)
\nonu \\
&-& \frac{4 (1387 k^4+35356 k^3+179710 k^2+272436 k+127143)}{
(k+5)^2 (13 k+17) (39 k+67) (41 k+85)}
\nonu \\
& \times & \left( T  F_{22} G_{11} -\frac{3}{10} 
\pa^2 (  F_{22} G_{11}) \right)
\nonu \\
&+& \frac{36 i (13 k^3+4187 k^2+2431 k-4879)}{
(k+5)^2 (13 k+17) (39 k+67) (41 k+85)}
\left( T  U A_3 -\frac{3}{10} 
\pa^2 (  U A_3 ) \right)
\nonu \\
&+& \frac{24 i k (1443 k^3+20601 k^2+53657 k+38083)}{
(k+5)^2 (13 k+17) (39 k+67) (41 k+85)}
\left( T  U B_3 -\frac{3}{10} 
\pa^2 (  U B_3 ) \right)
\nonu \\
&-&  \frac{8 (3109 k^4+39658 k^3+154630 k^2+225810 k+111537)}{
(k+5)^2 (13 k+17) (39 k+67) (41 k+85)}
\nonu \\
& \times & \left( T  U U -\frac{3}{10} 
\pa^2 (  U U ) \right)
\nonu \\
&+& \frac{4 i}{15 (k+5)^2 (7 k+11) (13 k+17) (23 k+19) (39 k+67) (47 k+43)} 
\nonu \\
& \times & (36706876 k^6+292282887 k^5+654064523 k^4+
91424438 k^3 \nonu \\
& - & 1307825118 k^2-1527225197 k-528255129)
\nonu \\ 
& \times & \left( T  \pa A_3  -\frac{1}{2} \pa T A_3 
-\frac{1}{4} \pa^3 A_3 \right)
\nonu \\
&-& \frac{4 i}{45 (k+5)^2 (7 k+11) (13 k+17) (23 k+19) (39 k+67) (47 k+43)} 
\nonu \\
& \times & (28754600 k^7+771031757 k^6+6124561187 k^5+21312125212 k^4
\nonu \\
& + & 37135458430 k^3+32925947125 k^2+13350908007 k+1640263122)
\nonu \\
& \times & 
\left( T  \pa B_3  -\frac{1}{2} \pa T B_3 
-\frac{1}{4} \pa^3 B_3 \right)
\nonu \\
&+& \frac{8 k (6277 k^3+34435 k^2+48679 k+19689)}{
(k+5)^2 (13 k+17) (23 k+19) (47 k+43)}
\left( T  \pa U  -\frac{1}{2} \pa T U 
-\frac{1}{4} \pa^3 U \right)
\nonu \\
&-& 
\frac{ (39 k+67)^{-1} (53 k+73)^{-1}}{
90 (k+5)^3 (5 k+13) (13 k+17) (17 k+37) (23 k+19) (29 k+97) 
}
\nonu \\
&\times & (34502966080 k^9-393778019461 k^8-6294110274792 k^7
\nonu \\
& - & 13713341570588 k^6+98506625033288 k^5+629802093225522 k^4
\nonu \\
& + & 1546655751602536 k^3+1967790292232708 k^2 \nonu \\
& + & 1287742927169400 k
+343785031306731)
\nonu \\
& \times &
\left( T \pa F_{11} F_{22} -\pa T F_{11} F_{22} + 3 
T F_{11} \pa F_{22} - 3 \pa^2 F_{11} \pa F_{22} - F_{11} \pa^3 F_{22}\right)
\nonu \\
&-& 
\frac{(39 k+67)^{-1} (53 k+73)^{-1}}{
90 (k+5)^3 (5 k+13) (13 k+17) (17 k+37) (23 k+19) (29 k+97) 
}
\nonu \\
&\times & (151049118080 k^9+3422145311317 k^8+34193189819496 k^7
\nonu \\
& + & 202349378272988 k^6+763718572719544 k^5+1854666569799342 k^4
\nonu \\
& + & 2846500244497688 k^3+2637697241887420 k^2
\nonu \\
& + & 1332346363292616 k+278192359770597)
\nonu \\
& \times & 
\left( T \pa F_{12} F_{21} -\pa T F_{12} F_{21} + 3 
T F_{12} \pa F_{21} - 3 \pa^2 F_{12} \pa F_{21} - F_{12} \pa^3 F_{21}\right)
\nonu \\
&+& 
\frac{8 (39 k+67)^{-1} (53 k+73)^{-1}}{
45 (k+5)^3 (5 k+13) (13 k+17) (17 k+37) 
(23 k+19) (29 k+97)} 
\nonu \\
& \times & 
(20044114520 k^9+328144939453 k^8+3596066948700 k^7
\nonu \\
& + & 26963376212984 k^6+123301840570060 k^5+339182244047274 k^4
\nonu \\
& + & 563131131917300 k^3+550433364392848 k^2
\nonu \\
& + & 289951406263308 k+62921457327249)
\nonu \\
& \times &
\left( T  F_{11} \pa F_{22} - \frac{1}{4} \pa T F_{11} F_{22} - \frac{3}{4} 
\pa F_{11} \pa^2 F_{22}  - \frac{1}{4} F_{11} \pa^3 F_{22}\right)
\nonu \\
&+&
\frac{8(39 k+67)^{-1} (53 k+73)^{-1}}{
45 (k+5)^3 (5 k+13) (13 k+17) (17 k+37) (23 k+19) 
(29 k+97) } 
\nonu \\
& \times & 
(26343906520 k^9+540983572121 k^8+5334907265412 k^7
\nonu \\
& + & 32669384753056 k^6+128858905140884 k^5+325963171531530 k^4
\nonu \\
& + & 520830224450668 k^3+505789721434088 k^2
\nonu \\
& + & 271388045277732 k+61391736767061)
\nonu \\
& \times & \left.
 \left( T  F_{12} \pa F_{21} - \frac{1}{4} \pa T F_{12} F_{21} - \frac{3}{4} 
\pa F_{12} \pa^2 F_{21}  - \frac{1}{4} F_{12} \pa^3 F_{21}\right)
\right](w) +\cdots.
\nonu
\eea
In the fourth and fifth terms of the third order pole one sees 
the nonlinear terms. One can add the $G_{12} \, A_3  \, {\bf T^{(1)}}(w)$
or  $G_{12} \, B_3  \, {\bf T^{(1)}}(w)$
into the left hand side. Then the nonzero contribution 
 $A_3  \, {\bf T^{(1)}}(w)$ or 
 $B_3  \, {\bf T^{(1)}}(w)$ will appear in the third order pole in the 
right hand side because one has the central term between $G_{21}(z)$ 
and $G_{12}(w)$.
In the second order pole, the nonlinear terms containing ${\bf T^{(1)}}(w)$
can appear from the $\pa G_{12} \, {\bf T^{(1)}}(w)$ in the left hand side.
The nonlinear terms containing $ \pa {\bf T^{(1)}}(w)$
can appear from the $G_{12} \, \pa {\bf T^{(1)}}(w)$ in the left hand side
as before.
According to the result of (\ref{g11221221t1}),
if one introduces 
$G_{12} \, \pa {\bf T^{(1)}}(w)$ in the left hand side, 
the nonlinear term $G_{12} \, {\bf T_{+}^{(\frac{3}{2})}}(w)$
can appear in the second order pole of the right hand side.
Similarly, 
if one introduces 
$ \pa {\bf T^{(1)}} \, {\bf T_{-}^{(\frac{3}{2})}}(w)$ in the left hand side, 
the nonlinear term $G_{21} \, {\bf T_{-}^{(\frac{3}{2})}}(w)$
can appear in the second order pole of the right hand side.
Among the various quasi primary fields appearing in the first order pole,
the quasi primary field with 
relative coefficients $-\frac{1}{2}$ and $-\frac{1}{4}$ is special because
it is not $-\frac{3}{10}$ as other cases. There are various quasi primary 
fields which behave very differently and contain many terms.

%%%%%%%%%%%%%%%%%%%%%%%%%%%%%%%%%%%%%%%%%%%%%%%%%%%%%%%%%%%%%%%%%%%%%
\subsection{ The higher spin currents in the nonlinear version in terms
of those in the linear version  }
%C%%%%%%%%%%%%%%%%%%%%%%%%%%%%%%%%%%%%%%%%%%%%%%%%%%%%%%%%%%%%%%%%%%%

As done in Appendix $B$, one has the following relations
\bea
{\bf P_{+,non}^{(\frac{5}{2})}}(z) & = & \left[ 
 {\bf P_{+}^{(\frac{5}{2})}}
+\frac{6i}{(5+k)} A_3  {\bf T_{+}^{(\frac{3}{2})}}  
+ \frac{2i}{(5+k)} B_3  {\bf T_{+}^{(\frac{3}{2})}}
-\frac{2}{(5+k)} F_{11}  {\bf V_{+}^{(2)}} \right.
+  
\frac{8}{(5+k)^2} F_{12}  F_{21}  {\bf T_{+}^{(\frac{3}{2})}}
\nonu \\
& + &  \frac{8}{(5+k)^2} F_{11}  F_{22}  {\bf T_{+}^{(\frac{3}{2})}}
-\frac{2}{(5+k)} F_{21}  {\bf T^{(2)}}
-  \frac{8}{(5+k)^2} F_{21}  F_{22}  {\bf U^{(\frac{3}{2})}}
+\frac{2}{(5+k)} \pa F_{21}  {\bf T^{(1)}}
\nonu \\
& - &  \frac{1}{(5+k)} F_{21}  \pa {\bf T^{(1)}}
-  \frac{2}{(5+k)} F_{22} {\bf U_{+}^{(2)}}
-\frac{2}{(5+k)} U  {\bf T_{+}^{(\frac{3}{2})}}
-\frac{2 i}{(k+5)} A_{-}   {\bf U^{(\frac{3}{2})}} 
\nonu \\
& - & \frac{2 i}{(k+5)} B_{-}   {\bf V^{(\frac{3}{2})}} 
+ \frac{2}{3 (k+5)} F_{21}  {\bf P^{(2)}}
-\frac{2}{3 (k+5)} {\bf P^{(2)}} F_{21}
\nonu \\
& + & \frac{6i}{(5+k)} A_3  G_{21}
+\frac{6}{(5+k)^2} A_3  A_3  F_{21}
+ \frac{4}{(5+k)^2} A_3  B_{-}  F_{22}
+\frac{16i}{(5+k)^3} A_3  F_{11}  F_{21}  F_{22}
\nonu \\
& + & \frac{4i(k+3)}{3(5+k)^3} \pa A_3  F_{21}
- \frac{8i(k+6)}{3(5+k)^3} A_3  \pa F_{21}
-\frac{2i}{(5+k)} A_{-}  G_{11} -
\frac{4}{(5+k)^2} A_{-}  A_3  F_{11}
\nonu \\
&-& \frac{4}{(5+k)^2} A_{3}  B_3  F_{21}
-\frac{2i(5k+9)}{3(5+k)^3} \pa A_{-}  F_{11}
+\frac{8i(k+9)}{3(5+k)^3} A_{-} \pa F_{11}
+ \frac{8i}{(5+k)^3} A_{-}  F_{11}  F_{12}  F_{21}
\nonu \\
& + & \frac{2}{(5+k)^2} A_{+}  A_{-}  F_{21}
+\frac{2i}{(5+k)} B_3  G_{21}
-  \frac{2}{(5+k)^2} B_3  B_3  F_{21}
-  \frac{4i(k+3)}{3(5+k)^3} \pa B_3  F_{21}
\nonu \\
& + & \frac{8i(k+6)}{3(5+k)^2} B_3  \pa F_{21}
-  \frac{2i(5k+9)}{3(5+k)^3} \pa B_{-}  F_{22}
- \frac{16i(k+3)}{3(5+k)^3} B_{-}  \pa F_{22}
- \frac{2}{(5+k)^2} B_{+}  B_{-}  F_{21}
\nonu \\
&+ & \frac{4}{(5+k)^2} F_{11}  F_{21}  G_{22}
+\frac{8}{(5+k)^3} \pa F_{11}  F_{21}  F_{22}
+  \frac{8}{(5+k)^3} F_{11}  \pa F_{21}  F_{22}
+  \frac{4}{(5+k)^2} F_{11}  F_{22}  G_{21}
\nonu \\
& + &  \frac{8}{(5+k)^2} F_{12}  F_{21}  G_{21}
+ \frac{8 (k-3)}{3 (k+5)^3} \pa^2 F_{21}
-  \frac{4}{(5+k)^2} F_{21}  F_{22}  G_{11}
+  \frac{4(k+6)}{3(5+k)^2} \pa G_{21}
\nonu \\
& - & \frac{2}{(5+k)} U  G_{21}
+\frac{8i}{(5+k)^2} U  A_3  F_{21}
- \frac{4i}{(5+k)^2} U  A_{-}  F_{11}
-\frac{2}{(5+k)^2} U  U  F_{21}
\nonu \\
 & - & \left. \frac{8}{(5+k)^3} F_{21}  \pa F_{21}  F_{12} 
+ \frac{8(k+6)}{3(5+k)^3} \pa U  F_{21}
+\frac{8(k+6)}{3(5+k)^3} U  \pa F_{21}
\right](z),
\nonu \\
{\bf P_{-,non}^{(\frac{5}{2})}}(z) & = &
 \left[ {\bf P_{-}^{(\frac{5}{2})}}
-\frac{6i}{(5+k)} A_3  {\bf T_{-}^{(\frac{3}{2})}}
- \frac{2i}{(5+k)} B_3   {\bf T_{-}^{(\frac{3}{2})}}
-\frac{2}{(5+k)} F_{11}  {\bf V_{-}^{(2)}} 
-\frac{8}{(k+5)^2} F_{12}  F_{21}   {\bf T_{-}^{(\frac{3}{2})}}
\right.
\nonu \\
& + & \frac{8}{(5+k)^2} F_{11}  F_{12}  {\bf V^{(\frac{3}{2})}}
-  \frac{8}{(5+k)^2} F_{11}  F_{22}   {\bf T_{-}^{(\frac{3}{2})}}
+\frac{2}{(5+k)} F_{12}  {\bf T^{(2)}}
+ \frac{2 i}{(k+5)} B_{+}  {\bf U^{(\frac{3}{2})}}
\nonu \\
& + & 
\frac{2}{(5+k)} \pa F_{12}  {\bf T^{(1)}}
-  \frac{1}{(5+k)} F_{12}  \pa {\bf T^{(1)}}
+ \frac{2 i}{(k+5)} A_{+}  {\bf V^{(\frac{3}{2})}}
+ \frac{8}{3 (k+5)} \pa  {\bf T_{-}^{(\frac{3}{2})}}
\nonu \\
&  - & \frac{2}{(5+k)} F_{22}  {\bf U_{-}^{(2)}}
-\frac{2}{(5+k)} U   {\bf T_{-}^{(\frac{3}{2})}}
\nonu \\
& + & \frac{6i}{(5+k)} A_3  G_{12}
-\frac{6}{(5+k)^2} A_3  A_3  F_{12}
+ \frac{4}{(k+5)^2} A_3  B_3  F_{12}
-  \frac{4}{(5+k)^2} A_3  B_{+}  F_{11}
\nonu \\
& + & \frac{16i}{(5+k)^3} A_3  F_{11}  F_{12}  F_{22}
-\frac{8i}{3(5+k)^2} A_3  \pa F_{12}
-\frac{2i}{(5+k)} A_{+}  G_{22}
+\frac{4}{(5+k)^2} A_{+}  A_3 F_{22}
\nonu \\
&-& \frac{2}{(5+k)^2} A_{+}  A_{-}  F_{12}
-\frac{8i}{(5+k)^3} A_{+}  F_{12}  F_{21}  F_{22}
-  \frac{10i}{3(5+k)^2} \pa A_{+}  F_{22} +
\frac{8i}{3(5+k)^2} A_{+}  \pa F_{22}
\nonu \\
& + & \frac{2i}{(5+k)} B_3  G_{12}
+
\frac{2}{(5+k)^2} B_3  B_3  F_{12}
+ \frac{8i}{3(5+k)^2} B_3  \pa F_{12}
+ 
\frac{2}{(5+k)^2} B_{+}  B_{-}  F_{12}
\nonu \\
& - & \frac{2i}{3(5+k)^2} \pa B_{+}  F_{11}
-  \frac{8i}{3(5+k)^2} B_{+} \pa F_{11}
- \frac{4}{(5+k)^2} F_{11}  F_{12}  G_{22}
-\frac{16}{3(5+k)^3} \pa F_{11}  F_{12}  F_{22}
\nonu \\
&+& \frac{8}{3(5+k)^3} F_{11}  \pa F_{12}  F_{22}
+\frac{8}{3(5+k)^3} F_{11}  F_{12}  \pa F_{22}
+\frac{4}{(5+k)^2} F_{11}  F_{22}  G_{12}
\nonu \\
& + &  \frac{8}{(5+k)^2} F_{12}  F_{21}  G_{12}
+  \frac{4}{(5+k)^2} F_{12}  F_{22}  G_{11}
-\frac{8}{3(5+k)} \pa G_{12}
+ \frac{2}{(5+k)} U  G_{12} 
\nonu \\
& + & \frac{8i}{(5+k)^2} U  A_3  F_{12}
+ \frac{16 i}{(k+5)^3} F_{12}  B_3  F_{12}  F_{21}
-  \frac{4i}{(5+k)^2} U  A_{+}  F_{22}
+\frac{2}{(5+k)^2} U  U  F_{12}
\nonu \\
& + & \left. \frac{8}{3(5+k)^2} F_{12}  G_{21}  F_{12}
-  \frac{16}{3(5+k)^2} \pa U  F_{12} -\frac{8}{3(5+k)^2}
U  \pa F_{12} \right](z),
\nonu \\
{\bf Q_{non}^{(\frac{5}{2})}}(z) & = & 
\left[ {\bf Q^{(\frac{5}{2})}} 
+\frac{2i}{(5+k)} A_3  {\bf U^{(\frac{3}{2})}} 
+ \frac{2i}{(5+k)} B_3  {\bf U^{(\frac{3}{2})}}
- \frac{2i}{(5+k)} B_{-}  {\bf T_{-}^{(\frac{3}{2})}} 
- \frac{2}{(5+k)} U   {\bf U^{(\frac{3}{2})}}
\right.
\nonu \\
& - & \frac{2}{(5+k)} F_{11}  {\bf W^{(2)}}
 -
\frac{8}{(5+k)^2} F_{11}  F_{12}  {\bf T_{+}^{(\frac{3}{2})}}
- \frac{2i}{(5+k)} A_{+}  {\bf T_{+}^{(\frac{3}{2})}} 
-  \frac{1}{(5+k)} F_{11}  \pa {\bf T^{(1)}}
\nonu \\
&+ & \frac{8}{(5+k)^2} F_{11}  F_{21}  {\bf T_{-}^{(\frac{3}{2})}}
+\frac{2}{(5+k)} \pa F_{11}  {\bf T^{(1)}}
+  \frac{2}{(5+k)} F_{12}  {\bf U_{+}^{(2)}}
+  \frac{2}{(5+k)} F_{21}   {\bf U_{-}^{(2)}}
\nonu \\
&+ & \frac{2i}{(5+k)} A_3  G_{11}
-
\frac{4}{(5+k)^2} A_3  B_3  F_{11}
-\frac{2}{(k+5)^2} A_3  A_3  F_{11}
-  \frac{4i}{(5+k)^2} \pa A_3  F_{11} 
\nonu \\
& + & \frac{16i}{(5+k)^3} A_3  F_{11}  F_{12}  F_{21}
+\frac{2i}{(5+k)} B_3  G_{11}
-  \frac{2i}{(5+k)} A_{+}  G_{21}
-\frac{4}{(5+k)^2} A_{+}  A_3  F_{21}
\nonu \\
& + & \frac{2}{(5+k)^2} A_{+}  A_{-}  F_{11}
- 
\frac{4}{(5+k)^2} A_{+}  B_{-}  F_{22}
-\frac{8i}{(5+k)^3} A_{+}  F_{11} F_{21}  F_{22}
+  \frac{2i}{(5+k)^2} \pa A_{+}  F_{21}
\nonu \\
& - & \frac{2}{(5+k)^2} B_3  B_3  F_{11}
+\frac{16i}{(5+k)^3} B_3  F_{11} F_{12} F_{21}
-\frac{2}{(k+5)^2} U  U  F_{11}
- 
\frac{4}{(5+k)^2} B_{-}  B_3  F_{12}
\nonu \\
& + & \frac{8i}{(5+k)^3} B_{-}  F_{11}  F_{12}  F_{22}
-  \frac{2i}{(5+k)^2} \pa B_{-}  F_{12}
+\frac{2}{(5+k)^2} B_{+}  B_{-}  F_{11}
- \frac{4}{(5+k)^2} F_{11}  F_{12}  G_{21}
\nonu \\
&- & \frac{8}{(5+k)^3} F_{11} \pa F_{12} F_{21}
+\frac{8}{(5+k)^3} F_{11}  F_{12}  \pa F_{21}
-
\frac{4}{(5+k)^2} F_{11}  F_{21}  G_{12}
-  \frac{2}{(5+k)} U  G_{11}
\nonu \\
& - &  \frac{4i}{(5+k)^2} U  A_{+}  F_{21}
-\frac{4i}{(5+k)^2} U  B_{-}  F_{12}
+  \frac{4}{(5+k)^2} \pa U   F_{11}
-\frac{4}{(5+k)^2} F_{11}  G_{22}  F_{11}
\nonu \\
& + & \left.
\frac{4}{(5+k)^2} F_{12}  F_{21}  G_{11} \right](z),
\nonu \\
{\bf R_{non}^{(\frac{5}{2})}}(z) & = & 
\left[ {\bf R^{(\frac{5}{2})}} 
-\frac{2i}{(5+k)} A_3  {\bf V^{(\frac{3}{2})}} 
+\frac{2i}{(5+k)} A_{-}  {\bf T_{-}^{(\frac{3}{2})}}
- \frac{2i}{(5+k)} B_3  {\bf V^{(\frac{3}{2})}} 
+\frac{2}{(5+k)}  F_{21}  {\bf V_{-}^{(2)}}
\right.
\nonu \\
&+ & \frac{2}{(5+k)} F_{22}  {\bf W^{(2)}}
-\frac{8}{(5+k)^2} F_{22}  F_{21}  {\bf T_{-}^{(\frac{3}{2})}}
-\frac{2}{(5+k)} U  {\bf V^{(\frac{3}{2})}}
+\frac{2}{(5+k)} F_{12}  {\bf V_{+}^{(2)}}
\nonu \\
& + & 
\frac{8}{(5+k)^2} F_{22}  F_{12}   {\bf T_{+}^{(\frac{3}{2})}}
+ \frac{2 i}{(k+5)} B_{+}   {\bf T_{+}^{(\frac{3}{2})}}
+  \frac{2}{(5+k)} \pa F_{22}  {\bf T^{(1)}}
-\frac{1}{(5+k)} F_{22}  \pa {\bf T^{(1)}}
\nonu \\
& + & \frac{2i}{(5+k)} A_3  G_{22}
+
\frac{2}{(5+k)^2} A_3  A_3  F_{22}
+  \frac{4}{(5+k)^2}  A_3  B_3 F_{22} 
+  \frac{16i}{(5+k)^3} A_{3}  F_{22}  F_{21}   F_{12}
\nonu \\
& - & \frac{2i}{(5+k)} A_{-}   G_{12}
-\frac{4i}{3(5+k)^2} \pa A_3    F_{22}
+ \frac{4}{(5+k)^2} A_{-}  A_3  F_{12}
+\frac{4}{(5+k)^2} A_{-}  B_{+}  F_{11}
\nonu \\
& - & \frac{8i}{(5+k)^3} A_{-}  F_{22} F_{12}  F_{11}
+  \frac{2i}{(5+k)^2} \pa A_{-}  F_{12}
-\frac{2}{(5+k)^2} A_{-}  A_{+} F_{22}
+\frac{2i}{(5+k)} B_3  G_{22}
\nonu \\
& + & 
\frac{2}{(5+k)^2} B_3  B_3   F_{22}
+  \frac{16i}{(5+k)^3} B_3   F_{22}  F_{21}  F_{12}
+\frac{4}{(5+k)^2} B_{+}  B_3  F_{21}
\nonu \\
&+& \frac{8i}{(5+k)^3} B_{+}  F_{22}  F_{21}  F_{11}
-\frac{2i}{(5+k)^2} \pa B_{+}   F_{21}
-  \frac{2}{(5+k)^2}  B_{-}  B_{+}  F_{22}
+ \frac{4}{(5+k)^2} F_{22}  F_{21}  G_{12}
\nonu \\
& + & \frac{8}{(5+k)^3}  F_{22}  \pa F_{21}  F_{12}
-  \frac{8}{(5+k)^3} F_{22}  F_{21} \pa F_{12}
+    \frac{2}{(5+k)} U   G_{22}
-\frac{4i}{(5+k)^2} U  A_{-}  F_{12}
\nonu \\
&+&  \frac{4}{(5+k)^2} F_{22}   F_{12}  G_{21}
- \frac{4i}{(5+k)^2} U   B_{+}  F_{21}
-\frac{4}{(5+k)^2} \pa U   F_{22}
-  \frac{4}{(5+k)^2} F_{21}   F_{12}  G_{22}
\nonu \\
& + & \left. \frac{4}{(5+k)^2} F_{22}   G_{11}  F_{22}
+\frac{2}{(5+k)^2} U  U  F_{22} \right](z).
\label{higherspin5half}
\eea
One can also write down the above expressions in terms of the fields 
in the nonlinear version.

What about other higher spin current? For example, 
let us look at the equation $(2.7)$ of Part II.
One would like to determine the explicit relation between 
${\bf Q_{+,non}^{(3)}}(z)$ and its linear version ${\bf Q}_{+}^{(3)}(z)$.
The upper OPE in $(2.7)$ of Part II is given by 
the OPE between $\hat{G}_{21}(z)$ and ${\bf Q_{non}^{(\frac{5}{2})}}(w)$.
Then one can express each field in terms of its linear version plus 
other terms which are defined also in the linear version.  
Then one can express the equation $(2.7)$ of Part II in terms of
the OPE between  ${G}_{21}(z)$ and ${\bf Q^{(\frac{5}{2})}}(w)$ and other 
terms which can be calculated from the previous results.
The former is given by (\ref{g2112qr}) where one can see the first order 
pole containing ${\bf Q_{+}^{(3)}}(w)$. The latter 
can be calculated explicitly. For example, the  
next term in $\hat{G}_{21}(z)$ of (\ref{ghatinlinear}) is given by
$U F_{21}(z)$ and the next term in  ${\bf Q_{non}^{(\frac{5}{2})}}(w)$ of 
above OPE (\ref{higherspin5half}) is given by $A_3 \, 
{\bf U^{(\frac{3}{2})}}(w)$. Then it is straightforward to calculate the 
OPE between $U F_{21}(z)$ and  $A_3 \, 
{\bf U^{(\frac{3}{2})}}(w)$ and read off the first order pole 
in the linear version.
Eventually one can equate the first order pole of $(2.7)$ of Part II 
(where the other fields except ${\bf Q_{+,non}^{(3)}}(w)$
should be written in the linear version)
to the first order pole of (\ref{g2112qr}) plus first order pole from 
various OPEs from other contributions as explained above. 

%%%%%%%%%%%%%%%%%%%%%%%%%%%%%%%%%%%%%%%%%%%%%%%%%%%%%%%%%%%%%%%%%%%%%
\subsection{ The higher spin currents in different basis }
%C%%%%%%%%%%%%%%%%%%%%%%%%%%%%%%%%%%%%%%%%%%%%%%%%%%%%%%%%%%%%%%%%%%%

As done in (\ref{v3half}), one can calculate the OPEs between 
the spin-$\frac{3}{2}$
currents and the higher spin-$2$ current in (\ref{v02}),
use the property of the fourth equation of Appendix $D.1$
with $s=2$
and obtain the following results
%%%%%%%%%%%%%%%%%%%%%%%%%%%%%%%%%%%%%%%%%%%%%%%%%%%%%%%
\bea
V_{\frac{1}{2}}^{(2), 0}(z) & = & \left[ 
-2 i \sqrt{2} {\bf P_{+}^{(\frac{5}{2})}}
 -2 i \sqrt{2} {\bf W_{+}^{(\frac{5}{2})}}  
+ \frac{i \sqrt{2} \left(41 k^2+400 k+783\right)}{5 (k+5)^2}   
G_{21}  {\bf T^{(1)}} \right.
\nonu \\
& + & 
\frac{2 i \sqrt{2} \left(41 k^2+400 k+783\right)}{5 (k+5)^2} 
{\bf T^{(1)}}  {\bf T_{+}^{(\frac{3}{2})}}
- \frac{i \left(41 k^2+400 k+783\right)}{5 \sqrt{2} (k+5)^2}  
(F_{21}  {\bf P^{(2)}}
-
{\bf P^{(2)}} F_{21})
\nonu \\
& - & \frac{4 \sqrt{2}}{(k+5)^2}  A_3  G_{21}
+ \frac{8 i \sqrt{2}}{(k+5)^3}  A_3  A_3  F_{21} + 
\frac{4 i \sqrt{2} \left(3 k^2+30 k+59\right)}{5 (k+5)^3} A_3  B_3  F_{21}
\nonu \\
&- & \frac{4 i \sqrt{2} \left(3 k^2+30 k+59\right)}{5 (k+5)^3} 
A_3  B_{-}  F_{22}
+\frac{4 \sqrt{2} \left(9 k^2+80 k+127\right)}{5 (k+5)^4}  
A_3  F_{11}  F_{21}  F_{22}
\nonu \\
& + & \frac{2 \sqrt{2} \left(50 k^3+617 k^2+2150 k+2151\right)}{15 (k+5)^4} 
\pa A_3  F_{21}
+  \frac{4 \sqrt{2}}{(k+5)^2} A_{-}  G_{11}
\nonu \\
& + & 
\frac{2 \sqrt{2} \left(50 k^3+617 k^2+2270 k+2751\right)}{15 (k+5)^4}
A_3  \pa F_{21}
 -
 \frac{4 i \sqrt{2} \left(3 k^2+30 k+59\right)}{5 (k+5)^3}
A_{-}  B_3  F_{11}
\nonu \\
& - & \frac{4 i \sqrt{2} \left(3 k^2+30 k+59\right)}{5 (k+5)^3} 
A_{-}  B_{-}  F_{12}
-   \frac{4 \sqrt{2} \left(25 k^3+124 k^2-695 k-2298\right)}{15 (k+5)^4}
\pa A_{-}  F_{11}
\nonu \\
& - & \frac{4 \sqrt{2} \left(25 k^3+124 k^2-665 k-2148\right)}{15 (k+5)^4} 
A_{-} \pa F_{11}
-\frac{6 \sqrt{2} \left(k^2+10 k+23\right)}{5 (k+5)^2}  B_3  G_{21}
\nonu \\
&-& \frac{4 \sqrt{2} \left(9 k^2+80 k+127\right)}{5 (k+5)^4}  
A_{-}  F_{11}  F_{12}  F_{21}
+\frac{8 i \sqrt{2}}{(k+5)^3}  A_{+}  A_{-}  F_{21}
\nonu \\
& - & \frac{12 i \sqrt{2} \left(k^2+10 k+23\right)}{5 (k+5)^3} 
B_3  B_3  F_{21}
- \frac{4 \sqrt{2} \left(9 k^2+80 k+127\right)}{5 (k+5)^4}
B_3  F_{11}  F_{21}  F_{22}
\nonu \\
& - & \frac{2 \sqrt{2} (k+3) \left(14 k^2+35 k-463\right)}{15 (k+5)^4}  
\pa B_3  F_{21}
-\frac{2 \sqrt{2} \left(50 k^3+617 k^2+2270 k+2751\right)}{15 (k+5)^4}  
B_3  \pa F_{21}
\nonu \\
&+& \frac{6 \sqrt{2} \left(k^2+10 k+23\right)}{5 (k+5)^2} B_{-}  G_{22}
-\frac{4 \sqrt{2} \left(9 k^2+80 k+127\right)}{5 (k+5)^4} 
B_{-}  F_{12}  F_{21}  F_{22}
\nonu \\
&- & \frac{2 \sqrt{2} \left(91 k^3+853 k^2+1393 k-681\right)}{15 (k+5)^4} 
\pa B_{-}  F_{22}
\nonu \\
& - & \frac{2 \sqrt{2} \left(73 k^3+583 k^2+79 k-2751\right)}{15 (k+5)^4} 
B_{-}  \pa F_{22}
- \frac{12 i \sqrt{2} \left(k^2+10 k+23\right)}{5 (k+5)^3}  
B_{+}  B_{-}  F_{21}
\nonu \\
&+ &\frac{4 i \sqrt{2} \left(3 k^2+25 k+44\right)}{5 (k+5)^3}  
F_{11}  F_{21}  G_{22}
- \frac{4 i \sqrt{2} \left(23 k^2+280 k+729\right)}{15 (k+5)^4} 
\pa F_{11}  F_{21}  F_{22}
\nonu \\
& - & \frac{4 i \sqrt{2} \left(41 k^2+400 k+783\right)}{15 (k+5)^4}  
F_{11}  \pa F_{21}  F_{22}
- \frac{4 i \sqrt{2} \left(59 k^2+520 k+837\right)}{15 (k+5)^4}
F_{11}  F_{21}  \pa F_{22}
\nonu \\
& - &  \frac{2 i \sqrt{2} \left(3 k^2+20 k+9\right)}{5 (k+5)^3}
F_{11}  F_{22}  G_{21}
+ \frac{2 i \sqrt{2} \left(3 k^2+30 k+79\right)}{5 (k+5)^3}  
F_{12}  F_{21}  G_{21}
\nonu \\
& - & \frac{2 i \sqrt{2} (k-3) \left(41 k^2+400 k+783\right)}{5 (k+5)^4}  
\pa^2 F_{21}
- \frac{4 i \sqrt{2} (k+7)}{(k+5)^3}  F_{21}  F_{22}  G_{11}
\nonu \\
& + & \frac{i \sqrt{2} \left(50 k^3+617 k^2+2270 k+2751\right)}{15 (k+5)^3}  
\pa G_{21}
+ \frac{2 i \sqrt{2}}{k+5} U  G_{21}
+\frac{4 \sqrt{2} (k+3)}{(k+5)^3}  U  A_3  F_{21}
\nonu \\
&-& \frac{4 \sqrt{2} (k+3)}{(k+5)^3} U  A_{-}  F_{11}
- \frac{4 \sqrt{2} \left(3 k^2+35 k+94\right)}{5 (k+5)^3}  U  B_3  F_{21}
\nonu \\
& + &  \frac{4 \sqrt{2} \left(3 k^2+35 k+94\right)}{5 (k+5)^3}
U  B_{-}  F_{22}
+\frac{4 i \sqrt{2} \left(9 k^2+80 k+127\right)}{5 (k+5)^4}  
 U  F_{11}  F_{21}  F_{22} 
\nonu \\
&+&  \frac{2 i \sqrt{2} \left(50 k^3+617 k^2+2270 k+2751\right)}{15 (k+5)^4} 
\pa U  F_{21}
\nonu \\
& + & \frac{2 i \sqrt{2} \left(50 k^3+617 k^2+2270 k+2751\right)}{15 (k+5)^4} 
U  \pa F_{21}
\nonu \\
& + & \left.
\frac{4 i \sqrt{2}}{(k+5)^2} U  U  F_{21}
- \frac{8 i \sqrt{2} \left(3 k^2+20 k+9\right)}{5 (k+5)^4} F_{21} 
\pa F_{21}  F_{12}
\right](z)
\nonu \\
 & + &
 \left[ -2 i \sqrt{2} {\bf P_{-}^{(\frac{5}{2})}}
+ 2 i \sqrt{2}  {\bf W_{-}^{(\frac{5}{2})}}
- \frac{i \sqrt{2} (k+3) (5 k+34)}{3 (k+5)^2} ( F_{12}   {\bf P^{(2)}}
- {\bf P^{(2)}}  F_{12} ) \right.
\nonu \\
& + & \frac{i \sqrt{2} \left(41 k^2+400 k+783\right)}{5 (k+5)^2}  
G_{12}  {\bf T^{(1)}}
-  \frac{2 i \sqrt{2} \left(173 k^2+1690 k+3369\right)}{15 (k+5)^2}  
\pa   {\bf T_{-}^{(\frac{3}{2})}}
\nonu \\
& - & \frac{2 i \sqrt{2} \left(41 k^2+400 k+783\right)}{5 (k+5)^2}  
{\bf T^{(1)}}    {\bf T_{-}^{(\frac{3}{2})}}
\nonu \\
& - & \frac{4 \sqrt{2}}{(k+5)^2}  A_3  G_{12}
-  \frac{8 i \sqrt{2}}{(k+5)^3} A_3  A_3  F_{12}
- \frac{4 i \sqrt{2} \left(3 k^2+30 k+59\right)}{5 (k+5)^3} 
A_3  B_3  F_{12}
\nonu \\
& + & \frac{4 i \sqrt{2} \left(3 k^2+30 k+59\right)}{5 (k+5)^3} 
A_3  B_{+}  F_{11}
-\frac{4 \sqrt{2} \left(9 k^2+80 k+127\right)}{5 (k+5)^4}  
A_3  F_{11}  F_{12}  F_{22}
\nonu \\
&+& \frac{4 \sqrt{2}}{(k+5)^2}
 A_{+}  G_{22}
+ \frac{4 i \sqrt{2} \left(3 k^2+30 k+59\right)}{5 (k+5)^3}  
A_{+}  B_3 F_{22}
-\frac{8 i \sqrt{2}}{(k+5)^3}  A_{+}  A_{-}  F_{12}
\nonu \\
 & + & \frac{4 i \sqrt{2} \left(3 k^2+30 k+59\right)}{5 (k+5)^3}  
A_{+}  B_{+}  F_{21} 
 + \frac{4 \sqrt{2} \left(9 k^2+80 k+127\right)}{5 (k+5)^4} 
A_{+}  F_{12}  F_{21}  F_{22}
\nonu \\
& + & \frac{8 \sqrt{2}}{(k+5)^3}  \pa A_{+}  F_{22} 
- \frac{6 \sqrt{2} \left(k^2+10 k+23\right)}{5 (k+5)^2} B_3  G_{12}
+\frac{12 i \sqrt{2} \left(k^2+10 k+23\right)}{5 (k+5)^3}
 B_3  B_3  F_{12}
\nonu \\
& + & \frac{4 \sqrt{2} \left(9 k^2+80 k+127\right)}{5 (k+5)^4}  
B_3   F_{11}  F_{12}  F_{22}
+\frac{6 \sqrt{2} \left(k^2+10 k+23\right)}{5 (k+5)^2} B_{+}  G_{11}
\nonu \\
 & + & \frac{12 i \sqrt{2} \left(k^2+10 k+23\right)}{5 (k+5)^3}
 B_{+}  B_{-}  F_{12}
- \frac{12 \sqrt{2} \left(k^2+10 k+23\right)}{5 (k+5)^3} \pa B_{+}  F_{11}
\nonu \\
&+ & \frac{4 \sqrt{2} \left(9 k^2+80 k+127\right)}{5 (k+5)^4} 
B_{+}  F_{11}  F_{12}  F_{21}
-\frac{4 i \sqrt{2} (k+7)}{(k+5)^3}  F_{11}  F_{12}  G_{22}
\nonu \\
 & - & \frac{16 i \sqrt{2} \left(3 k^2+25 k+44\right)}{5 (k+5)^4}  
\pa F_{11}  F_{12}  F_{22}
- \frac{8 i \sqrt{2} \left(3 k^2+30 k+79\right)}{5 (k+5)^4} 
F_{11}  \pa F_{12}  F_{22}
\nonu \\
& - & \frac{16 i \sqrt{2} (k+7)}{(k+5)^4}  F_{11}  F_{12}  \pa F_{22}
- \frac{2 i \sqrt{2} \left(3 k^2+20 k+9\right)}{5 (k+5)^3}
F_{11}  F_{22}  G_{12}
\nonu \\
& + & \frac{2 i \sqrt{2} \left(3 k^2+30 k+79\right)}{5 (k+5)^3}
 F_{12}  F_{21}  G_{12}
+ \frac{4 i \sqrt{2} \left(3 k^2+25 k+44\right)}{5 (k+5)^3} 
F_{12}  F_{22}  G_{11}
\nonu \\
& + &  \frac{i \sqrt{2} \left(173 k^2+1690 k+3369\right)}{15 (k+5)^2} 
\pa G_{12}
- \frac{2 i \sqrt{2}}{k+5}  U  G_{12} 
+ \frac{4 \sqrt{2} (k+3)}{(k+5)^3} U  A_3  F_{12}
\nonu \\
&- & \frac{4 \sqrt{2} (k+3)}{(k+5)^3} U  A_{+}  F_{22}
-\frac{4 \sqrt{2} \left(3 k^2+35 k+94\right)}{5 (k+5)^3}  U  B_3  F_{12}
\nonu \\
& + &  
\frac{4 \sqrt{2} \left(3 k^2+35 k+94\right)}{5 (k+5)^3} U  B_{+}  F_{11}
+ \frac{4 i \sqrt{2} \left(9 k^2+80 k+127\right)}{5 (k+5)^4} 
U  F_{11}  F_{12}  F_{22}
\nonu \\
& - & \left. \frac{4 i \sqrt{2}}{(k+5)^2} U  U  F_{12} 
+ \frac{8 i \sqrt{2} \left(3 k^2+20 k+9\right)}{5 (k+5)^4}
F_{12}  \pa F_{12}  F_{21}\right](z),
\nonu \\
%%%%%%%%%%%%%%%%%%%%%%%%%%%%%%%%%%%%%%%%%%%%%%%%%%%%%%%%%%%%%%%%%%%
V_{\frac{1}{2}}^{(2),1} & = &
\left[-2 \sqrt{2} {\bf Q^{(\frac{5}{2})}} 
-2 \sqrt{2}  {\bf U^{(\frac{5}{2})}} 
+ \frac{\sqrt{2} \left(41 k^2+400 k+783\right)}{5 (k+5)^2} 
G_{11}  {\bf T^{(1)}}
\right. \nonu \\
& - &  \frac{2 \sqrt{2} \left(41 k^2+400 k+783\right)}{5 (k+5)^2} 
\pa  {\bf U^{(\frac{3}{2})}}
+ \frac{2 \sqrt{2} \left(41 k^2+400 k+783\right)}{5 (k+5)^2} 
 {\bf T^{(1)}}  {\bf U^{(\frac{3}{2})}}
\nonu \\
&- &  \frac{4 i \sqrt{2}}{(k+5)^2} A_3  G_{11}
- \frac{4 \sqrt{2} \left(3 k^2+30 k+59\right)}{5 (k+5)^3}
 A_3  B_3  F_{11}
+  \frac{8 \sqrt{2}}{(k+5)^3}  A_3  A_3  F_{11}
\nonu \\
& + & \frac{4 i \sqrt{2} (k+3) (5 k+34)}{3 (k+5)^3} \pa A_3  F_{11}
+ \frac{4 i \sqrt{2} (k+3) (5 k+34)}{3 (k+5)^3} A_3  \pa F_{11} 
\nonu \\
& - & \frac{4 \sqrt{2} \left(3 k^2+30 k+59\right)}{5 (k+5)^3}
A_3  B_{-}  F_{12}
+  \frac{4 i \sqrt{2} \left(9 k^2+80 k+127\right)}{5 (k+5)^4}  
A_3  F_{11}  F_{12}  F_{21}
\nonu \\
&- &  \frac{4 i \sqrt{2}}{(k+5)^2}  A_{+}  G_{21}
-  \frac{4 \sqrt{2} \left(3 k^2+30 k+59\right)}{5 (k+5)^3} 
A_{+}  B_3  F_{21}
+ \frac{8 \sqrt{2}}{(k+5)^3}  A_{+}  A_{-}  F_{11}
\nonu \\
&+& \frac{4 \sqrt{2} \left(3 k^2+30 k+59\right)}{5 (k+5)^3}
 A_{+}  B_{-}  F_{22}
+\frac{4 i \sqrt{2} \left(9 k^2+80 k+127\right)}{5 (k+5)^4}  
A_{+}  F_{11} F_{21}  F_{22}
\nonu \\
& + & \frac{4 i \sqrt{2} \left(5 k^2+49 k+96\right)}{3 (k+5)^3}  
\pa A_{+}  F_{21}
+ \frac{4 i \sqrt{2} (k+3) (5 k+34)}{3 (k+5)^3}  A_{+}  \pa F_{21}
\nonu \\
& + & \frac{6 i \sqrt{2} \left(k^2+10 k+23\right)}{5 (k+5)^2} B_3  G_{11}
 -  \frac{12 \sqrt{2} \left(k^2+10 k+23\right)}{5 (k+5)^3}
B_3  B_3  F_{11}
\nonu \\
& + & \frac{4 i \sqrt{2} \left(7 k^2+65 k+96\right)}{15 (k+5)^3} \pa B_3
 F_{11} +
\frac{4 i \sqrt{2} (k+3) (5 k+34)}{3 (k+5)^3} B_3  \pa F_{11}
\nonu \\
& + & \frac{4 i \sqrt{2} \left(9 k^2+80 k+127\right)}{5 (k+5)^4} 
B_3  F_{11} F_{12} F_{21}
+ \frac{6 i \sqrt{2} \left(k^2+10 k+23\right)}{5 (k+5)^2} B_{-}  G_{12}
\nonu \\
&-& \frac{4 i \sqrt{2} \left(9 k^2+80 k+127\right)}{5 (k+5)^4}
B_{-}  F_{11}  F_{12} F_{22}
+ \frac{4 i \sqrt{2} \left(16 k^2+155 k+303\right)}{15 (k+5)^3} \pa B_{-} 
F_{12}
\nonu \\
&+& \frac{4 i \sqrt{2} (k+3) (5 k+34)}{3 (k+5)^3} B_{-}  \pa F_{12}
-  \frac{12 \sqrt{2} \left(k^2+10 k+23\right)}{5 (k+5)^3}
B_{+}  B_{-}  F_{11}
\nonu \\
& - & \frac{4 \sqrt{2} (k+7)}{(k+5)^3}  F_{11}  F_{12}  G_{21}
+  \frac{8 \sqrt{2} \left(41 k^2+400 k+783\right)}{15 (k+5)^4}
\pa F_{11}  F_{12} F_{21}
\nonu \\
&- & \frac{32 \sqrt{2} \left(8 k^2+85 k+189\right)}{15 (k+5)^4} 
F_{11} \pa F_{12} F_{21}
+ \frac{16 \sqrt{2} \left(5 k^2+46 k+81\right)}{3 (k+5)^4} 
F_{11}  F_{12}  \pa F_{21}
\nonu \\
& - & \frac{4 \sqrt{2} \left(3 k^2+25 k+44\right)}{5 (k+5)^3}
 F_{11}  F_{21}  G_{12}
- \frac{2 \sqrt{2} \left(3 k^2+30 k+79\right)}{5 (k+5)^3} F_{11} F_{22} G_{11}
\nonu \\
&+& \frac{2 \sqrt{2} \left(3 k^2+20 k+9\right)}{5 (k+5)^3}  
F_{12} F_{21} G_{11} 
-\frac{\sqrt{2} \left(73 k^2+710 k+1329\right)}{15 (k+5)^2}
\pa G_{11}
\nonu \\
& + & \frac{2 \sqrt{2}}{k+5}   U  G_{11}
+ \frac{4 i \sqrt{2} (k+3)}{(k+5)^3}
U A_3 F_{11}
+ \frac{4 i \sqrt{2} (k+3)}{(k+5)^3} U  A_{+}  F_{21}
\nonu \\
& + & 
\frac{4 i \sqrt{2} \left(3 k^2+35 k+94\right)}{5 (k+5)^3}
U B_3 F_{11} +
  \frac{4 i \sqrt{2} \left(3 k^2+35 k+94\right)}{5 (k+5)^3} 
U  B_{-}  F_{12}
\nonu \\
& + &   \frac{4 \sqrt{2} (k+3) (5 k+34)}{3 (k+5)^3} \pa U   F_{11}
+ \frac{4 \sqrt{2} (k+3) (5 k+34)}{3 (k+5)^3} U \pa F_{11}
+ \frac{4 \sqrt{2}}{(k+5)^2} U U F_{11}
\nonu \\
& + & \left.  \frac{4 \sqrt{2} \left(9 k^2+80 k+127\right)}{5 (k+5)^4}  
U F_{11}  F_{12} F_{21}
- \frac{8 \sqrt{2} \left(3 k^2+20 k+9\right)}{5 (k+5)^4}  
F_{11}  \pa F_{11}  F_{22} \right](z)
\nonu \\
&+& 
\left[ 2 \sqrt{2} {\bf R^{(\frac{5}{2})}}
- 2 \sqrt{2}  {\bf V^{(\frac{5}{2})}}
+  \frac{4 \sqrt{2} \left(5 k^2+46 k+81\right)}{(k+5) (7 k+3)}
( F_{22}  {\bf P^{(2)}} -  {\bf P^{(2)}} F_{22}) 
\right. 
\nonu \\
&- & \frac{\sqrt{2} \left(41 k^2+400 k+783\right)}{5 (k+5)^2} 
G_{22}  {\bf T^{(1)}}
+ \frac{2 \sqrt{2} \left(1061 k^3+10009 k^2+17763 k-2673\right)}
{15 (k+5)^2 (7 k+3)}  \pa  {\bf V^{(\frac{3}{2})}}
\nonu \\
& + &  \frac{2 \sqrt{2} \left(41 k^2+400 k+783\right)}{5 (k+5)^2} 
 {\bf T^{(1)}}  {\bf V^{(\frac{3}{2})}}
\nonu \\
& + & \frac{4 i \sqrt{2}}{(k+5)^2} A_3  G_{22}
+\frac{8 \sqrt{2}}{(k+5)^3}
 A_3  A_3  F_{22}
-\frac{4 \sqrt{2} \left(3 k^2+30 k+59\right)}{5 (k+5)^3}  A_3  B_3 F_{22} 
\nonu \\
& - &  
\frac{4 \sqrt{2} \left(3 k^2+30 k+59\right)}{5 (k+5)^3} A_3 B_{+} F_{21}
+ \frac{8 i \sqrt{2} \left(29 k^3+91 k^2-1143 k-2637\right)}{5 (k+5)^3 (7 k+3)}
\pa A_3    F_{22}
\nonu \\
&+& \frac{4 i \sqrt{2} \left(5 k^3+24 k^2-109 k-264\right)}{(k+5)^3 (7 k+3)}
A_3 \pa F_{22}
- \frac{4 i \sqrt{2} \left(9 k^2+80 k+127\right)}{5 (k+5)^4} A_{3}  F_{22}  F_{21}   F_{12}
\nonu \\
& +  & \frac{4 i \sqrt{2}}{(k+5)^2}  A_{-}   G_{12}
+ \frac{8 \sqrt{2}}{(k+5)^3} A_{-} A_{+} F_{22}
- \frac{4 \sqrt{2} \left(3 k^2+30 k+59\right)}{5 (k+5)^3}   
A_{-}  B_3  F_{12}
\nonu \\
&+& \frac{4 \sqrt{2} \left(3 k^2+30 k+59\right)}{5 (k+5)^3}
 A_{-}  B_{+}  F_{11}
-\frac{4 i \sqrt{2} \left(9 k^2+80 k+127\right)}{5 (k+5)^4}  
A_{-}  F_{22} F_{12}  F_{11}
\nonu \\
& + & \frac{4 i \sqrt{2} \left(5 k^3+24 k^2-95 k-258\right)}{(k+5)^3 (7 k+3)}  
\pa A_{-}  F_{12}
+ \frac{4 i \sqrt{2} \left(5 k^3+24 k^2-109 k-264\right)}{(k+5)^3 (7 k+3)}   
A_{-}  \pa F_{12}
\nonu \\
& - & \frac{6 i \sqrt{2} \left(k^2+10 k+23\right)}{5 (k+5)^2}  
B_3  G_{22}
- \frac{12 \sqrt{2} \left(k^2+10 k+23\right)}{5 (k+5)^3}
 B_3  B_3   F_{22}
\nonu \\
&-& \frac{4 i \sqrt{2} \left(15 k^3+100 k^2-119 k-708\right)}{(k+5)^3 (7 k+3)}
B_3 \pa F_{22}
\nonu \\
& - &  \frac{4 i \sqrt{2} \left(9 k^2+80 k+127\right)}{5 (k+5)^4}  
B_3   F_{22}  F_{21}  F_{12}
-\frac{6 i \sqrt{2} \left(k^2+10 k+23\right)}{5 (k+5)^2} B_{+}   G_{21}
\nonu \\
&+& \frac{4 i \sqrt{2} \left(9 k^2+80 k+127\right)}{5 (k+5)^4} 
B_{+}  F_{22}  F_{21}  F_{11}
+ \frac{4 i \sqrt{2} \left(46 k^3+639 k^2+2788 k+3747\right)}
{5 (k+5)^3 (7 k+3)} \pa B_{+}   F_{21}
\nonu \\
& + & \frac{4 i \sqrt{2} (k+3) \left(5 k^2+69 k+236\right)}{(k+5)^3 (7 k+3)}  
 B_{+}   \pa F_{21}-
\frac{12 \sqrt{2} \left(k^2+10 k+23\right)}{5 (k+5)^3}  
B_{-}  B_{+}  F_{22}
\nonu \\
& + & 
\frac{8 \sqrt{2} (k-9) \left(5 k^2+46 k+81\right)}{(k+5)^3 (7 k+3)}
\pa^2 F_{22} - \frac{4 \sqrt{2} (k+7)}{(k+5)^3}
 F_{22}  F_{21}  G_{12}
\nonu \\
& - &  \frac{8 \sqrt{2} \left(3 k^2+20 k+9\right)}{5 (k+5)^4}  
\pa F_{22}   F_{21}  F_{12}
-\frac{16 \sqrt{2} \left(3 k^2+20 k+9\right)}{5 (k+5)^4}   
F_{22}  \pa F_{21}  F_{12}
\nonu \\
& - &  \frac{4 \sqrt{2} \left(3 k^2+25 k+44\right)}{5 (k+5)^3}
 F_{22}   F_{12}  G_{21}
-\frac{2 \sqrt{2} \left(3 k^2+30 k+79\right)}{5 (k+5)^3}
F_{22} F_{11} G_{22}
\nonu \\
&+& \frac{2 \sqrt{2} \left(3 k^2+20 k+9\right)}{5 (k+5)^3}
  F_{21}   F_{12}  G_{22}
-\frac{\sqrt{2} \left(1211 k^3+12529 k^2+31053 k+18567\right)}
{15 (k+5)^2 (7 k+3)}
\pa G_{22}
\nonu \\
&+& \frac{2 \sqrt{2}}{k+5} U  G_{22}
-\frac{4 i \sqrt{2} (k+3)}{(k+5)^3} U  A_3  F_{22}
-\frac{4 i \sqrt{2} (k+3)}{(k+5)^3} U  A_{-}  F_{12}
\nonu \\
& - & \frac{4 i \sqrt{2} \left(3 k^2+35 k+94\right)}{5 (k+5)^3} U 
B_3  F_{22}
-\frac{4 i \sqrt{2} \left(3 k^2+35 k+94\right)}{5 (k+5)^3}
U  B_{+}  F_{21}
\nonu \\
&+& \frac{8 \sqrt{2} \left(4 k^3-179 k^2-1978 k-3747\right)}
{5 (k+5)^3 (7 k+3)}
\pa U  F_{22}
-\frac{4 \sqrt{2} (k+3) \left(5 k^2+69 k+236\right)}{(k+5)^3 (7 k+3)}
U  \pa F_{22}
\nonu \\
&+& \frac{4 \sqrt{2} \left(9 k^2+80 k+127\right)}{5 (k+5)^4}
U  F_{22}  F_{21}  F_{12}
-\frac{4 \sqrt{2} \left(33 k^3+62 k^2-1741 k-3954\right)}{5 (k+5)^3 (7 k+3)}
F_{22}  G_{11}  F_{22}
\nonu \\
& - & \left.  \frac{8 \sqrt{2} \left(3 k^2+20 k+9\right)}{5 (k+5)^4}
F_{22}   \pa F_{22}  F_{11}
+ \frac{4 \sqrt{2}}{(k+5)^2} U  U  F_{22} \right](z),
\nonu \\
%%%%%%%%%%%%%%%%%%%%%%%%%%%%%%%%%%%%%%%%%%%%%%%%%%%%%%%%%%%%%
V_{\frac{1}{2}}^{(2),2} & = &
\left[-2 i \sqrt{2} {\bf Q^{(\frac{5}{2})}} 
-2 i \sqrt{2}  {\bf U^{(\frac{5}{2})}} 
+ \frac{i \sqrt{2} \left(41 k^2+400 k+783\right)}{5 (k+5)^2} 
G_{11}  {\bf T^{(1)}}
\right. \nonu \\
& - &  \frac{2 i \sqrt{2} \left(41 k^2+400 k+783\right)}{5 (k+5)^2} 
\pa  {\bf U^{(\frac{3}{2})}}
+ \frac{2 i \sqrt{2} \left(41 k^2+400 k+783\right)}{5 (k+5)^2} 
 {\bf T^{(1)}}  {\bf U^{(\frac{3}{2})}}
\nonu \\
&+ &  \frac{4  \sqrt{2}}{(k+5)^2} A_3  G_{11}
- \frac{4 i \sqrt{2} \left(3 k^2+30 k+59\right)}{5 (k+5)^3}
 A_3  B_3  F_{11}
-  \frac{8  \sqrt{2}}{(k+5)^3}  A_3  A_3  F_{11}
\nonu \\
& - & \frac{4  \sqrt{2} (k+3) (5 k+34)}{3 (k+5)^3} \pa A_3  F_{11}
- \frac{4  \sqrt{2} (k+3) (5 k+34)}{3 (k+5)^3} A_3  \pa F_{11} 
\nonu \\
& - & \frac{4 i \sqrt{2} \left(3 k^2+30 k+59\right)}{5 (k+5)^3}
A_3  B_{-}  F_{12}
-  \frac{4  \sqrt{2} \left(9 k^2+80 k+127\right)}{5 (k+5)^4}  
A_3  F_{11}  F_{12}  F_{21}
\nonu \\
&+ &  \frac{4  \sqrt{2}}{(k+5)^2}  A_{+}  G_{21}
-  \frac{4 i \sqrt{2} \left(3 k^2+30 k+59\right)}{5 (k+5)^3} 
A_{+}  B_3  F_{21}
+ \frac{8 i \sqrt{2}}{(k+5)^3}  A_{+}  A_{-}  F_{11}
\nonu \\
&+& \frac{4 i \sqrt{2} \left(3 k^2+30 k+59\right)}{5 (k+5)^3}
 A_{+}  B_{-}  F_{22}
-\frac{4  \sqrt{2} \left(9 k^2+80 k+127\right)}{5 (k+5)^4}  
A_{+}  F_{11} F_{21}  F_{22}
\nonu \\
& - & \frac{4  \sqrt{2} \left(5 k^2+49 k+96\right)}{3 (k+5)^3}  
\pa A_{+}  F_{21}
- \frac{4  \sqrt{2} (k+3) (5 k+34)}{3 (k+5)^3}  A_{+}  \pa F_{21}
\nonu \\
& - & \frac{6  \sqrt{2} \left(k^2+10 k+23\right)}{5 (k+5)^2} B_3  G_{11}
 -  \frac{12 i \sqrt{2} \left(k^2+10 k+23\right)}{5 (k+5)^3}
B_3  B_3  F_{11}
\nonu \\
& - & \frac{4  \sqrt{2} \left(7 k^2+65 k+96\right)}{15 (k+5)^3} \pa B_3
 F_{11} -
\frac{4  \sqrt{2} (k+3) (5 k+34)}{3 (k+5)^3} B_3  \pa F_{11}
\nonu \\
& - & \frac{4  \sqrt{2} \left(9 k^2+80 k+127\right)}{5 (k+5)^4} 
B_3  F_{11} F_{12} F_{21}
- \frac{6  \sqrt{2} \left(k^2+10 k+23\right)}{5 (k+5)^2} B_{-}  G_{12}
\nonu \\
&+& \frac{4  \sqrt{2} \left(9 k^2+80 k+127\right)}{5 (k+5)^4}
B_{-}  F_{11}  F_{12}  F_{22}
- \frac{4  \sqrt{2} \left(16 k^2+155 k+303\right)}{15 (k+5)^3} \pa B_{-} 
F_{12}
\nonu \\
&-& \frac{4  \sqrt{2} (k+3) (5 k+34)}{3 (k+5)^3} B_{-}  \pa F_{12}
-  \frac{12 i \sqrt{2} \left(k^2+10 k+23\right)}{5 (k+5)^3}
B_{+}  B_{-}  F_{11}
\nonu \\
& - & \frac{4 i \sqrt{2} (k+7)}{(k+5)^3}  F_{11}  F_{12}  G_{21}
+  \frac{8 i \sqrt{2} \left(41 k^2+400 k+783\right)}{15 (k+5)^4}
\pa F_{11}  F_{12} F_{21}
\nonu \\
&+ & \frac{32 i \sqrt{2} \left(8 k^2+85 k+189\right)}{15 (k+5)^4} 
F_{11} \pa F_{12} F_{21}
+ \frac{16 i \sqrt{2} \left(5 k^2+46 k+81\right)}{3 (k+5)^4} 
F_{11}  F_{12}  \pa F_{21}
\nonu \\
& - & \frac{4 i \sqrt{2} \left(3 k^2+25 k+44\right)}{5 (k+5)^3}
 F_{11}  F_{21}  G_{12}
- \frac{2 i \sqrt{2} \left(3 k^2+30 k+79\right)}{5 (k+5)^3} F_{11} F_{22} G_{11}
\nonu \\
&+& \frac{2 i \sqrt{2} \left(3 k^2+20 k+9\right)}{5 (k+5)^3}  
F_{12} F_{21} G_{11} 
-\frac{i \sqrt{2} \left(73 k^2+710 k+1329\right)}{15 (k+5)^2}
\pa G_{11}
\nonu \\
& + & \frac{2 i \sqrt{2}}{k+5}   U  G_{11}
- \frac{4  \sqrt{2} (k+3)}{(k+5)^3}
U A_3 F_{11}
- \frac{4  \sqrt{2} (k+3)}{(k+5)^3} U  A_{+}  F_{21}
\nonu \\
& - & 
\frac{4  \sqrt{2} \left(3 k^2+35 k+94\right)}{5 (k+5)^3}
U B_3 F_{11} -
  \frac{4  \sqrt{2} \left(3 k^2+35 k+94\right)}{5 (k+5)^3} 
U  B_{-}  F_{12}
\nonu \\
& + &   \frac{4 i \sqrt{2} (k+3) (5 k+34)}{3 (k+5)^3} \pa U   F_{11}
+ \frac{4 i \sqrt{2} (k+3) (5 k+34)}{3 (k+5)^3} U \pa F_{11}
+ \frac{4 i \sqrt{2}}{(k+5)^2} U U F_{11}
\nonu \\
& + & \left.  \frac{4 i \sqrt{2} \left(9 k^2+80 k+127\right)}{5 (k+5)^4}  
U F_{11}  F_{12}  F_{21}
- \frac{8 i \sqrt{2} \left(3 k^2+20 k+9\right)}{5 (k+5)^4}  
F_{11}  \pa F_{11}  F_{22} \right](z)
\nonu \\
&+& 
\left[ - 2 i \sqrt{2} {\bf R^{(\frac{5}{2})}}
+ 2 i \sqrt{2}  {\bf V^{(\frac{5}{2})}}
- \frac{4 i \sqrt{2} \left(5 k^2+46 k+81\right)}{(k+5) (7 k+3)}
( F_{22}  {\bf P^{(2)}} -  {\bf P^{(2)}} F_{22}) 
\right. 
\nonu \\
&+ & \frac{i \sqrt{2} \left(41 k^2+400 k+783\right)}{5 (k+5)^2} 
G_{22}  {\bf T^{(1)}}
- \frac{2 i \sqrt{2} \left(1061 k^3+10009 k^2+17763 k-2673\right)}
{15 (k+5)^2 (7 k+3)}  \pa  {\bf V^{(\frac{3}{2})}}
\nonu \\
& - &  \frac{2 i \sqrt{2} \left(41 k^2+400 k+783\right)}{5 (k+5)^2} 
 {\bf T^{(1)}}  {\bf V^{(\frac{3}{2})}}
\nonu \\
& + & \frac{4  \sqrt{2}}{(k+5)^2} A_3  G_{22}
-\frac{8 i \sqrt{2}}{(k+5)^3}
 A_3  A_3  F_{22}
+\frac{4 i \sqrt{2} \left(3 k^2+30 k+59\right)}{5 (k+5)^3}  A_3  B_3 F_{22} 
\nonu \\
& + &  
\frac{4 i \sqrt{2} \left(3 k^2+30 k+59\right)}{5 (k+5)^3} A_3 B_{+} F_{21}
+ \frac{8  \sqrt{2} \left(29 k^3+91 k^2-1143 k-2637\right)}{5 (k+5)^3 (7 k+3)}
\pa A_3    F_{22}
\nonu \\
&+& \frac{4  \sqrt{2} \left(5 k^3+24 k^2-109 k-264\right)}{(k+5)^3 (7 k+3)}
A_3 \pa F_{22}
- \frac{4  \sqrt{2} \left(9 k^2+80 k+127\right)}{5 (k+5)^4} A_{3}  F_{22}  F_{21}   F_{12}
\nonu \\
& +  & \frac{4  \sqrt{2}}{(k+5)^2}  A_{-}   G_{12}
- \frac{8 i \sqrt{2}}{(k+5)^3} A_{-} A_{+} F_{22}
+ \frac{4 i \sqrt{2} \left(3 k^2+30 k+59\right)}{5 (k+5)^3}   
A_{-}  B_3  F_{12}
\nonu \\
&-& \frac{4 i \sqrt{2} \left(3 k^2+30 k+59\right)}{5 (k+5)^3}
 A_{-}  B_{+}  F_{11}
-\frac{4  \sqrt{2} \left(9 k^2+80 k+127\right)}{5 (k+5)^4}  
A_{-}  F_{22} F_{12}  F_{11}
\nonu \\
& + & \frac{4  \sqrt{2} \left(5 k^3+24 k^2-95 k-258\right)}{(k+5)^3 (7 k+3)}  
\pa A_{-}  F_{12}
+ \frac{4  \sqrt{2} \left(5 k^3+24 k^2-109 k-264\right)}{(k+5)^3 (7 k+3)}   
A_{-}  \pa F_{12}
\nonu \\
& - & \frac{6  \sqrt{2} \left(k^2+10 k+23\right)}{5 (k+5)^2}  
B_3  G_{22}
+ \frac{12 i \sqrt{2} \left(k^2+10 k+23\right)}{5 (k+5)^3}
 B_3  B_3   F_{22}
\nonu \\
&-& \frac{4  \sqrt{2} \left(15 k^3+100 k^2-119 k-708\right)}{(k+5)^3 (7 k+3)}
B_3 \pa F_{22}
\nonu \\
& - &  \frac{4  \sqrt{2} \left(9 k^2+80 k+127\right)}{5 (k+5)^4}  
B_3   F_{22}  F_{21}  F_{12}
-\frac{6  \sqrt{2} \left(k^2+10 k+23\right)}{5 (k+5)^2} B_{+}   G_{21}
\nonu \\
&+& \frac{4  \sqrt{2} \left(9 k^2+80 k+127\right)}{5 (k+5)^4} 
B_{+}  F_{22}  F_{21}  F_{11}
+ \frac{4  \sqrt{2} \left(46 k^3+639 k^2+2788 k+3747\right)}
{5 (k+5)^3 (7 k+3)} \pa B_{+}   F_{21}
\nonu \\
& + & \frac{4  \sqrt{2} (k+3) \left(5 k^2+69 k+236\right)}{(k+5)^3 (7 k+3)}  
 B_{+}   \pa F_{21}+
\frac{12 i \sqrt{2} \left(k^2+10 k+23\right)}{5 (k+5)^3}  
B_{-}  B_{+}  F_{22}
\nonu \\
& - & 
\frac{8 i \sqrt{2} (k-9) \left(5 k^2+46 k+81\right)}{(k+5)^3 (7 k+3)}
\pa^2 F_{22} + \frac{4 i \sqrt{2} (k+7)}{(k+5)^3}
 F_{22}  F_{21}  G_{12}
\nonu \\
& + &  \frac{8 i \sqrt{2} \left(3 k^2+20 k+9\right)}{5 (k+5)^4}  
\pa F_{22}  F_{21}  F_{12}
+\frac{16 i \sqrt{2} \left(3 k^2+20 k+9\right)}{5 (k+5)^4}   
F_{22}  \pa F_{21}  F_{12}
\nonu \\
& + &  \frac{4 i \sqrt{2} \left(3 k^2+25 k+44\right)}{5 (k+5)^3}
 F_{22}  F_{12}  G_{21}
+\frac{2 i \sqrt{2} \left(3 k^2+30 k+79\right)}{5 (k+5)^3}
F_{22} F_{11} G_{22}
\nonu \\
&-& \frac{2 i \sqrt{2} \left(3 k^2+20 k+9\right)}{5 (k+5)^3}
  F_{21}   F_{12}  G_{22}
+\frac{ i \sqrt{2} \left(1211 k^3+12529 k^2+31053 k+18567\right)}
{15 (k+5)^2 (7 k+3)}
\pa G_{22}
\nonu \\
&-& \frac{2 i \sqrt{2}}{k+5} U  G_{22}
-\frac{4  \sqrt{2} (k+3)}{(k+5)^3} U  A_3  F_{22}
-\frac{4  \sqrt{2} (k+3)}{(k+5)^3} U  A_{-}  F_{12}
\nonu \\
& - & \frac{4  \sqrt{2} \left(3 k^2+35 k+94\right)}{5 (k+5)^3} U 
B_3  F_{22}
-\frac{4  \sqrt{2} \left(3 k^2+35 k+94\right)}{5 (k+5)^3}
U  B_{+}  F_{21}
\nonu \\
&-& \frac{8 i \sqrt{2} \left(4 k^3-179 k^2-1978 k-3747\right)}
{5 (k+5)^3 (7 k+3)}
\pa U  F_{22}
+\frac{4 i \sqrt{2} (k+3) \left(5 k^2+69 k+236\right)}{(k+5)^3 (7 k+3)}
U \pa F_{22}
\nonu \\
&-& \frac{4 i \sqrt{2} \left(9 k^2+80 k+127\right)}{5 (k+5)^4}
U  F_{22}  F_{21}  F_{12}
+\frac{4 i \sqrt{2} \left(33 k^3+62 k^2-1741 k-3954\right)}{5 (k+5)^3 (7 k+3)}
F_{22}  G_{11}  F_{22}
\nonu \\
& + & \left.  \frac{8 i \sqrt{2} \left(3 k^2+20 k+9\right)}{5 (k+5)^4}
F_{22}  \pa F_{22}  F_{11}
- \frac{4 i \sqrt{2}}{(k+5)^2} U  U  F_{22} \right](z),
\nonu \\
%%%%%%%%%%%%%%%%%%%%%%%%%%%%%%%%%%%%%%%%%%%%%%%%%%%%%%%%%%%%%%
V_{\frac{1}{2}}^{(2), 3}(z) & = & \left[ 
2  \sqrt{2} {\bf P_{+}^{(\frac{5}{2})}}
 + 2  \sqrt{2} {\bf W_{+}^{(\frac{5}{2})}}  
- \frac{ \sqrt{2} \left(41 k^2+400 k+783\right)}{5 (k+5)^2}   
G_{21}  {\bf T^{(1)}} \right.
\nonu \\
& - & 
\frac{2  \sqrt{2} \left(41 k^2+400 k+783\right)}{5 (k+5)^2} 
{\bf T^{(1)}}  {\bf T_{+}^{(\frac{3}{2})}}
+ \frac{ \left(41 k^2+400 k+783\right)}{5 \sqrt{2} (k+5)^2}  
(F_{21}  {\bf P^{(2)}}
-
{\bf P^{(2)}} F_{21})
\nonu \\
& - & \frac{4 i \sqrt{2}}{(k+5)^2}  A_3  G_{21}
- \frac{8  \sqrt{2}}{(k+5)^3}  A_3  A_3  F_{21} -
\frac{4  \sqrt{2} \left(3 k^2+30 k+59\right)}{5 (k+5)^3} A_3  B_3  F_{21}
\nonu \\
&+ & \frac{4  \sqrt{2} \left(3 k^2+30 k+59\right)}{5 (k+5)^3} 
A_3  B_{-}  F_{22}
+\frac{4 i \sqrt{2} \left(9 k^2+80 k+127\right)}{5 (k+5)^4}  
A_3  F_{11} F_{21}  F_{22}
\nonu \\
& + & \frac{2 i \sqrt{2} \left(50 k^3+617 k^2+2150 k+2151\right)}{15 (k+5)^4} 
\pa A_3  F_{21}
+  \frac{4 i \sqrt{2}}{(k+5)^2} A_{-}  G_{11}
\nonu \\
& + & 
\frac{2 i \sqrt{2} \left(50 k^3+617 k^2+2270 k+2751\right)}{15 (k+5)^4}
A_3  \pa F_{21}
 + 
 \frac{4  \sqrt{2} \left(3 k^2+30 k+59\right)}{5 (k+5)^3}
A_{-}  B_3  F_{11}
\nonu \\
& + & \frac{4  \sqrt{2} \left(3 k^2+30 k+59\right)}{5 (k+5)^3} 
A_{-}  B_{-}  F_{12}
-   \frac{4 i \sqrt{2} \left(25 k^3+124 k^2-695 k-2298\right)}{15 (k+5)^4}
\pa A_{-}  F_{11}
\nonu \\
& - & \frac{4 i \sqrt{2} \left(25 k^3+124 k^2-665 k-2148\right)}{15 (k+5)^4} 
A_{-} \pa F_{11}
-\frac{6 i \sqrt{2} \left(k^2+10 k+23\right)}{5 (k+5)^2}  B_3  G_{21}
\nonu \\
&-& \frac{4 i \sqrt{2} \left(9 k^2+80 k+127\right)}{5 (k+5)^4}  
A_{-}  F_{11}  F_{12}  F_{21}
-\frac{8  \sqrt{2}}{(k+5)^3}  A_{+}  A_{-}  F_{21}
\nonu \\
& + & \frac{12  \sqrt{2} \left(k^2+10 k+23\right)}{5 (k+5)^3} 
B_3  B_3  F_{21}
- \frac{4 i \sqrt{2} \left(9 k^2+80 k+127\right)}{5 (k+5)^4}
B_3  F_{11} F_{21}  F_{22}
\nonu \\
& - & \frac{2 i \sqrt{2} (k+3) \left(14 k^2+35 k-463\right)}{15 (k+5)^4}  
\pa B_3 F_{21}
-\frac{2 i \sqrt{2} \left(50 k^3+617 k^2+2270 k+2751\right)}{15 (k+5)^4}  
B_3  \pa F_{21}
\nonu \\
&+& \frac{6 i \sqrt{2} \left(k^2+10 k+23\right)}{5 (k+5)^2} B_{-}  G_{22}
-\frac{4 i \sqrt{2} \left(9 k^2+80 k+127\right)}{5 (k+5)^4} 
B_{-}  F_{12}  F_{21}  F_{22}
\nonu \\
&- & \frac{2 i \sqrt{2} \left(91 k^3+853 k^2+1393 k-681\right)}{15 (k+5)^4} 
\pa B_{-}  F_{22}
\nonu \\
& - & \frac{2 i \sqrt{2} \left(73 k^3+583 k^2+79 k-2751\right)}{15 (k+5)^4} 
B_{-}  \pa F_{22}
+ \frac{12  \sqrt{2} \left(k^2+10 k+23\right)}{5 (k+5)^3}  
B_{+}  B_{-}  F_{21}
\nonu \\
&- &\frac{4  \sqrt{2} \left(3 k^2+25 k+44\right)}{5 (k+5)^3}  
F_{11}  F_{21}  G_{22}
+ \frac{4  \sqrt{2} \left(23 k^2+280 k+729\right)}{15 (k+5)^4} 
\pa F_{11}  F_{21}  F_{22}
\nonu \\
& + & \frac{4  \sqrt{2} \left(41 k^2+400 k+783\right)}{15 (k+5)^4}  
F_{11} \pa F_{21}  F_{22}
+ \frac{4  \sqrt{2} \left(59 k^2+520 k+837\right)}{15 (k+5)^4}
F_{11}  F_{21}  \pa F_{22}
\nonu \\
& + &  \frac{2  \sqrt{2} \left(3 k^2+20 k+9\right)}{5 (k+5)^3}
F_{11}  F_{22}  G_{21}
- \frac{2  \sqrt{2} \left(3 k^2+30 k+79\right)}{5 (k+5)^3}  
F_{12}  F_{21}  G_{21}
\nonu \\
& + & \frac{2  \sqrt{2} (k-3) \left(41 k^2+400 k+783\right)}{5 (k+5)^4}  
\pa^2 F_{21}
+ \frac{4  \sqrt{2} (k+7)}{(k+5)^3}  F_{21}  F_{22}  G_{11}
\nonu \\
& - & \frac{ \sqrt{2} \left(50 k^3+617 k^2+2270 k+2751\right)}{15 (k+5)^3}  
\pa G_{21}
- \frac{2  \sqrt{2}}{k+5} U  G_{21}
+\frac{4 i \sqrt{2} (k+3)}{(k+5)^3}  U  A_3  F_{21}
\nonu \\
&-& \frac{4 i \sqrt{2} (k+3)}{(k+5)^3} U  A_{-}  F_{11}
- \frac{4 i \sqrt{2} \left(3 k^2+35 k+94\right)}{5 (k+5)^3}  U  B_3  F_{21}
\nonu \\
& + &  \frac{4 i \sqrt{2} \left(3 k^2+35 k+94\right)}{5 (k+5)^3}
U B_{-}  F_{22}
-\frac{4  \sqrt{2} \left(9 k^2+80 k+127\right)}{5 (k+5)^4}  
 U  F_{11}  F_{21}  F_{22} 
\nonu \\
&-&  
\frac{2  \sqrt{2} \left(50 k^3+617 k^2+2270 k+2751\right)}{15 (k+5)^4} 
\pa U  F_{21}
 - 
\frac{4  \sqrt{2}}{(k+5)^2} U  U  F_{21}
\nonu \\
& - & \left.
\frac{2  \sqrt{2} \left(50 k^3+617 k^2+2270 k+2751\right)}{15 (k+5)^4} 
U  \pa F_{21}
+ \frac{8  \sqrt{2} \left(3 k^2+20 k+9\right)}{5 (k+5)^4} F_{21} 
\pa F_{21}  F_{12}
\right](z)
\nonu \\
 & + &
 \left[ -2  \sqrt{2} {\bf P_{-}^{(\frac{5}{2})}}
+ 2  \sqrt{2}  {\bf W_{-}^{(\frac{5}{2})}}
- \frac{ \sqrt{2} (k+3) (5 k+34)}{3 (k+5)^2} ( F_{12}   {\bf P^{(2)}}
- {\bf P^{(2)}}  F_{12} ) \right.
\nonu \\
& + & \frac{ \sqrt{2} \left(41 k^2+400 k+783\right)}{5 (k+5)^2}  
G_{12} {\bf T^{(1)}}
-  \frac{2  \sqrt{2} \left(173 k^2+1690 k+3369\right)}{15 (k+5)^2}  
\pa   {\bf T_{-}^{(\frac{3}{2})}}
\nonu \\
& - & \frac{2  \sqrt{2} \left(41 k^2+400 k+783\right)}{5 (k+5)^2}  
{\bf T^{(1)}}   {\bf T_{-}^{(\frac{3}{2})}}
\nonu \\
& + & \frac{4 i \sqrt{2}}{(k+5)^2}  A_3  G_{12}
-  \frac{8  \sqrt{2}}{(k+5)^3} A_3  A_3  F_{12}
- \frac{4  \sqrt{2} \left(3 k^2+30 k+59\right)}{5 (k+5)^3} 
A_3  B_3 F_{12}
\nonu \\
& + & \frac{4  \sqrt{2} \left(3 k^2+30 k+59\right)}{5 (k+5)^3} 
A_3  B_{+}  F_{11}
+\frac{4 i \sqrt{2} \left(9 k^2+80 k+127\right)}{5 (k+5)^4}  
A_3  F_{11} F_{12}  F_{22}
\nonu \\
&-& \frac{4 i \sqrt{2}}{(k+5)^2}
 A_{+}  G_{22}
+ \frac{4  \sqrt{2} \left(3 k^2+30 k+59\right)}{5 (k+5)^3}  
A_{+}  B_3 F_{22}
-\frac{8  \sqrt{2}}{(k+5)^3}  A_{+}  A_{-}  F_{12}
\nonu \\
 & + & \frac{4  \sqrt{2} \left(3 k^2+30 k+59\right)}{5 (k+5)^3}  
A_{+}  B_{+}  F_{21} 
 - \frac{4 i \sqrt{2} \left(9 k^2+80 k+127\right)}{5 (k+5)^4} 
A_{+}  F_{12}  F_{21}  F_{22}
\nonu \\
& - & \frac{8 i \sqrt{2}}{(k+5)^3}  \pa A_{+}  F_{22} 
+ \frac{6 i \sqrt{2} \left(k^2+10 k+23\right)}{5 (k+5)^2} B_3  G_{12}
+\frac{12  \sqrt{2} \left(k^2+10 k+23\right)}{5 (k+5)^3}
 B_3  B_3  F_{12}
\nonu \\
& - & \frac{4 i \sqrt{2} \left(9 k^2+80 k+127\right)}{5 (k+5)^4}  
B_3   F_{11}  F_{12}  F_{22}
-\frac{6 i \sqrt{2} \left(k^2+10 k+23\right)}{5 (k+5)^2} B_{+}  G_{11}
\nonu \\
 & + & \frac{12  \sqrt{2} \left(k^2+10 k+23\right)}{5 (k+5)^3}
 B_{+}  B_{-}  F_{12}
+ \frac{12 i \sqrt{2} \left(k^2+10 k+23\right)}{5 (k+5)^3} \pa B_{+}  F_{11}
\nonu \\
&- & \frac{4 i \sqrt{2} \left(9 k^2+80 k+127\right)}{5 (k+5)^4} 
B_{+}  F_{11}  F_{12}  F_{21}
-\frac{4  \sqrt{2} (k+7)}{(k+5)^3}  F_{11}  F_{12}  G_{22}
\nonu \\
 & - & \frac{16  \sqrt{2} \left(3 k^2+25 k+44\right)}{5 (k+5)^4}  
\pa F_{11} F_{12}  F_{22}
- \frac{8  \sqrt{2} \left(3 k^2+30 k+79\right)}{5 (k+5)^4} 
F_{11}  \pa F_{12}  F_{22}
\nonu \\
& - & \frac{16  \sqrt{2} (k+7)}{(k+5)^4}  F_{11}  F_{12} \pa F_{22}
- \frac{2  \sqrt{2} \left(3 k^2+20 k+9\right)}{5 (k+5)^3}
F_{11}  F_{22}  G_{12}
\nonu \\
& + & \frac{2  \sqrt{2} \left(3 k^2+30 k+79\right)}{5 (k+5)^3}
 F_{12}  F_{21}  G_{12}
+ \frac{4  \sqrt{2} \left(3 k^2+25 k+44\right)}{5 (k+5)^3} 
F_{12}  F_{22}  G_{11}
\nonu \\
& + &  \frac{ \sqrt{2} \left(173 k^2+1690 k+3369\right)}{15 (k+5)^2} 
\pa G_{12}
- \frac{2  \sqrt{2}}{k+5}  U  G_{12} 
- \frac{4  i \sqrt{2} (k+3)}{(k+5)^3} U  A_3  F_{12}
\nonu \\
&+ & \frac{4 i \sqrt{2} (k+3)}{(k+5)^3} U  A_{+}  F_{22}
+\frac{4 i \sqrt{2} \left(3 k^2+35 k+94\right)}{5 (k+5)^3}  U  B_3  F_{12}
\nonu \\
& - &  
\frac{4 i \sqrt{2} \left(3 k^2+35 k+94\right)}{5 (k+5)^3} U  B_{+}  F_{11}
+ \frac{4  \sqrt{2} \left(9 k^2+80 k+127\right)}{5 (k+5)^4} 
U  F_{11}  F_{12} F_{22}
\nonu \\
& - & \left. \frac{4  \sqrt{2}}{(k+5)^2} U  U  F_{12} 
+ \frac{8  \sqrt{2} \left(3 k^2+20 k+9\right)}{5 (k+5)^4}
F_{12}  \pa F_{12}  F_{21}\right](z).
\nonu
\eea
Compared to the description in section $4$ (the equations (\ref{v3half}), 
(\ref{vspin2}), (\ref{v5half}) and (\ref{bcgspin3})), 
the above expressions are rather complicated because 
the higher spin-$2$ in (\ref{v02}) has many extra terms except 
${\bf P^{(2)}}(z)$. 
Furthermore, the higher spin currents in the nonlinear version
\cite{BCG} has the explicit relations with its linear version via
the equations $(4.22)$-$(4.25)$ of \cite{BCG}. Then one can write down 
$\widetilde{V}_0^{(2)}(z) \equiv V_0^{(2)}(z)$ and 
$\widetilde{V}_{\frac{1}{2}}^{(2),a}(z) 
\equiv V_{\frac{1}{2}}^{(2),a}(z)$. 
Once the other higher spin currents $V_1^{(2), \pm i}(z)$, 
$V_{\frac{3}{2}}^{(2),a}(z)$ and $V_2^{(2)}(z)$ are found explicitly, then 
the corresponding higher spin currents,
 $\widetilde{V}_1^{(2), \pm i}(z)$, 
$\widetilde{V}_{\frac{3}{2}}^{(2),a}(z)$ and $\widetilde{V}_2^{(2)}(z)$, 
can be written  in terms of the corresponding higher spin currents in the
linear versions plus four spin-$\frac{1}{2}$ currents and the spin-$1$ 
current.

%%%%%%%%%%%%%%%%%%%%%%%%%%%%%%%%%%%%%%%%%%%%%%%%%%%%%%%%%%%%%%%%%%%%%
%%%%%%%%%%%%%%%%%%%%%%%%%%%%%%%%%%%%%%%%%%%%%%%%%%%%%%%%%%%%%%%%%%%%%
\section{ The $136$ OPEs between the $16$ lowest higher spin currents }
%C%%%%%%%%%%%%%%%%%%%%%%%%%%%%%%%%%%%%%%%%%%%%%%%%%%%%%%%%%%%%%%%%%%%
%%%%%%%%%%%%%%%%%%%%%%%%%%%%%%%%%%%%%%%%%%%%%%%%%%%%%%%%%%%%%%%%%%%%

In this final Appendix, the complete OPEs between the 
$16$ lowest higher spin currents are presented.
The highest spin in the right hand side of these OPEs 
is given by $3$. The composite fields of spin-$\frac{7}{2}$,
spin-$4$, spin-$\frac{9}{2}$ and spin-$5$ are not present in this paper. 
They can be obtained from its nonlinear version of Part II as described 
in Appendix $E$ (just below the equation (\ref{higherspin5half})) 
in principle.
All the $16$ higher spin currents are known in terms of coset fields
described in section $3$ and one has all the singular terms of these
OPEs completely. The nontrivial things here are to 
write them in terms of known $16$ currents and $16$ higher spin currents
(and their derivatives). In this Appendix, the final OPEs are presented 
after determing the unknown strucuture constants.
Some simple comments on each OPE are added.  

%%%%%%%%%%%%%%%%%%%%%%%%%%%%%%%%%%%%%%%%%%%%%%%%%%%%%%%%%%%%%%%%%%%%%
\subsection{ The OPEs between the first ${\cal N}=2$ multiplet and itself 
(and the other three ${\cal N}=2$ multiplets) }
%C%%%%%%%%%%%%%%%%%%%%%%%%%%%%%%%%%%%%%%%%%%%%%%%%%%%%%%%%%%%%%%%%%%%

From the relation (\ref{newspinonelinear}),
one has the following OPE
%%%%%%%%%%%%%%%%%%%%%%%%%%%%%%%%%%%%%%%%%%%%%%%%%%%%%%%%%%%%%%
\bea
{\bf T^{(1)}}(z) \, {\bf T^{(1)}}(w)  & = & \frac{1}{(z-w)^2} \,
\left[\frac{6k}{(5+k)} \right] +\cdots.
\nonu 
\eea
This is the same as the OPE in the nonlinear version 
because one has the relation (\ref{newspinonelinear}).

From the expressions (\ref{newspinonelinear}) and (\ref{t3halfnon})
with the help of Part II, 
%%%%%%%%%%%%%%%%%%%%%%%%%%%%%%%%%%%%%%%%%%%%%%%%%%%%%%%%%%%%%%%
\bea
{\bf T^{(1)}}(z) \, {\bf T_{\pm}^{(\frac{3}{2})}}(w) & = &  \frac{1}{(z-w)}  \,
\left[ \pm {\bf T_{\pm}^{(\frac{3}{2})}} -\frac{1}{(5+k)}  U  
\left(
\begin{array}{c}
 F_{21} \nonu \\
 F_{12} 
 \end{array}
\right)
\pm \frac{4}{(5+k)^2}  
\left(
\begin{array}{c}
 F_{21} \nonu \\
 F_{12} 
 \end{array}
\right)
F_{11} F_{22} 
\right. \nonu \\
& \pm &   \frac{i}{(5+k)}
\left(
\begin{array}{c}
 F_{21} \nonu \\
 F_{12} 
 \end{array}
\right)
A_3 
\mp  \frac{i}{(5+k)} 
\left(
\begin{array}{c}
F_{11} A_{-} \nonu \\
 F_{22} A_{+}
\end{array}
\right)
\pm \frac{i}{(5+k)} 
\left(
\begin{array}{c}
F_{22} B_{-} \nonu \\
F_{11} B_{+}
\end{array}
\right)
\nonu \\
& \mp & \left.
\frac{i}{(5+k)} 
\left(
\begin{array}{c}
F_{21} \nonu \\
 F_{12} 
\end{array}
\right) B_3 \right](w)  +\cdots.
\nonu 
\eea
Due to the presence of spin-$\frac{1}{2}$ currents, 
there are many terms containing them in the above.

Together with (\ref{t2non}), one obtains the following OPE
%%%%%%%%%%%%%%%%%%%%%%%%%%%%%%%%%%%%%%%%%%%%%%%%%%%%%%%%%%%
\bea
{\bf T^{(1)}}(z) \, {\bf T^{(2)}}(w) & = & \frac{1}{(z-w)^2}  \,
\left[ -\frac{6i}{(5+k)} A_3 -\frac{2ik}{(5+k)} B_3 +
\frac{2(-3+k)}{(5+k)^2} F_{11} F_{22} - \frac{2(3+k)}{(5+k)^2}
F_{12} F_{21} \right](w)  \nonu \\
& + & \frac{1}{(z-w)} \, \left[ \frac{2i}{(5+k)^2} F_{11} F_{12} A_{-}-
\frac{2i}{(5+k)^2}  F_{21} F_{22} A_{+} -\frac{2i}{(5+k)^2}
F_{12} F_{22} B_{-} \right. \nonu \\
&+&  \frac{2i}{(5+k)^2}  F_{11} F_{21} B_{+} -\frac{1}{(5+k)}
F_{11} G_{22}   \nonu \\
&+ & \left. \frac{1}{(5+k)} F_{22} G_{11} -\frac{4}{(5+k)^2}  F_{11} F_{22} U 
\right](w) +\cdots.
\nonu
\eea
Compared to the OPE in the nonlinear version, the above OPE 
has nontrivial first order pole. The second order pole does not have 
the higher spin-$1$ current.

With (\ref{uv3halfnon1}), one calculates the following OPEs
%%%%%%%%%%%%%%%%%%%%%%%%%%%%%%%%%%%%%%%%%%%%%%%%%%%%%%%%%%%%%%%%%
\bea
{\bf T^{(1)}}(z) \, 
\left(
\begin{array}{c}
{\bf U^{(\frac{3}{2})}} \\
{\bf V^{(\frac{3}{2})}} \\
\end{array} \right) (w) 
& = & \frac{1}{(z-w)}  
\, \left[ \pm
\left(
\begin{array}{c}
{\bf U^{(\frac{3}{2})}} \\
{\bf V^{(\frac{3}{2})}} \\
\end{array} \right)  \mp \frac{4}{(5+k)^2}  F_{12} F_{21} 
\left(
\begin{array}{c}
F_{11} \nonu \\
F_{22}
\end{array}
\right)
\right. \nonu \\
& 
\mp &  \frac{i}{(5+k)} 
\left(
\begin{array}{c}
F_{11} \nonu \\
F_{22} 
\end{array}
\right)
 A_3 
\mp \frac{i}{(5+k)} \left(
\begin{array}{c} 
F_{21} \nonu \\
F_{12} 
\end{array}
\right) A_{\pm}
- \frac{1}{(5+k)} U \left(
\begin{array}{c} 
F_{21} \nonu \\
F_{12} 
\end{array}
\right)
\nonu \\
& 
\mp & \left. \frac{i}{(5+k)} 
\left(
\begin{array}{c}
F_{12} \nonu \\
F_{21} 
\end{array} 
\right)
B_{\mp}
 \mp \frac{i}{(5+k)} 
\left(
\begin{array}{c} 
F_{11} \nonu \\
F_{22} 
\end{array}
\right)
 B_3
\right](w) 
  +\cdots.
\nonu 
\eea
If one ignores the spin-$\frac{1}{2}$ current terms, then this OPE 
has the same form as the one in the nonlinear version.

From the previous expression for higher spin-$1$ current and the expression
(\ref{uvplusminusnon}), one can describe the following OPEs
%%%%%%%%%%%%%%%%%%%%%%%%%%%%%%%%%%%%%%%%%%%%%%%%%%%%%%%%%%%
\bea
{\bf T^{(1)}}(z) \, 
\left(
\begin{array}{c} 
{\bf U_{+}^{(2)}} \\
{\bf V_{-}^{(2)}} \\
\end{array} \right) (w) & = &  \frac{1}{(z-w)^2}\,
\left[ \pm \frac{2k}{(5+k)} i {B}_{\mp}  \mp \frac{4k}{(5+k)^2}
\left(
\begin{array}{c}
F_{11} F_{21} \nonu \\
F_{12} F_{22} 
\end{array}
\right) 
\right] (w)  
\nonu \\
&+& \frac{1}{(z-w)} \,
\left[ -\frac{2i}{(5+k)^2} 
F_{11} F_{22}  B_{\mp}
+\frac{2i}{(5+k)^2} F_{12} F_{21} B_{\mp}
\right. \nonu \\
& \pm &  \frac{4i}{(5+k)^2} 
\left(
\begin{array}{c}
F_{11} F_{21} \nonu \\
F_{22} F_{12} \end{array}
\right) B_3 
+  \frac{4}{(5+k)^2} \left(
\begin{array}{c}
F_{11} F_{21}  \nonu \\
F_{22} F_{12}  
\end{array}
\right) U
\nonu \\
 & + & \left. 
\frac{1}{(5+k)} 
\left(
\begin{array}{c} 
F_{11} G_{21} \nonu \\
F_{22} G_{12}
\end{array}
\right) -\frac{1}{(5+k)}
\left(
\begin{array}{c}
F_{21} G_{11} \nonu \\
F_{12} G_{22} 
\end{array}
\right) \right](w)  +\cdots.
\nonu 
\eea
Again, there are many spin-$\frac{1}{2}$ current terms 
which are new in the linear version.

Similarly, from the equation (\ref{uvminusplusnon}),
one calculates the following OPEs 
%%%%%%%%%%%%%%%%%%%%%%%%%%%%%%%%%%%%%%%%%%%%%%%%%%%%%%%%
\bea
{\bf T^{(1)}}(z) \, 
\left(
\begin{array}{c} 
{\bf U_{-}^{(2)}} \\
{\bf V_{+}^{(2)}} \\
\end{array} \right) (w) & = &  \frac{1}{(z-w)^2}\,
\left[ \mp \frac{6}{(5+k)} i {A}_{\pm} -\frac{12}{(5+k)^2} 
\left(
\begin{array}{c}
F_{11} F_{12}  \nonu \\
F_{22} F_{21} 
\end{array}
\right)
\right] (w) 
\nonu \\
&+& \frac{1}{(z-w)} \left[ \mp \frac{4i}{(5+k)^2} 
\left(
\begin{array}{c}
F_{11} F_{12} \nonu \\
F_{22} F_{21} 
\end{array}
\right)
A_3
+\frac{2i}{(5+k)^2} F_{11} F_{22} A_{\pm}
\right. \nonu \\
&+& \frac{2i}{(5+k)^2} F_{12} F_{21} A_{\pm}
-\frac{1}{(5+k)} 
\left(
\begin{array}{c}
F_{11} G_{12} \nonu \\
F_{22} G_{21} 
\end{array}
\right)
\nonu \\
&+ & \left. 
\frac{1}{(5+k)} 
\left(
\begin{array}{c}
F_{12} G_{11} \nonu \\
F_{21} G_{22} 
\end{array}
\right)
-\frac{4}{(5+k)^2} 
\left(
\begin{array}{c}
F_{11} F_{12} \nonu \\
 F_{22} F_{21}  
\end{array}
\right)
U
\right](w)  +\cdots.
\nonu 
\eea
Due to the spin-$\frac{1}{2}$ currents, this OPE has many 
new terms in the above.

From the relations (\ref{u5halfnon}) and (\ref{v5halfnon}), one 
obtains the following OPEs
%%%%%%%%%%%%%%%%%%%%%%%%%%%%%%%%%%%%%%%%%%%%%%%%%%%%%%%%%%%%%%%%%%
\bea
{\bf T^{(1)}}(z) \left(
\begin{array}{c}
{\bf U^{(\frac{5}{2})}} \nonu \\
{\bf V^{(\frac{5}{2})}}  \end{array}
\right)(w) 
& = & 
\mp \frac{1}{(z-w)^3} \frac{12k}{(5+k)^2}  
\left(
\begin{array}{c}
F_{11} \nonu \\
F_{22}  
 \end{array}
\right)(w) 
\nonu \\
& + & \frac{1}{(z-w)^2} \left[ 
%\frac{12k}{(5+k)^2} \pa 
%\left(
%\begin{array}{c}
%F_{11} \nonu \\
%-F_{22}  
% \end{array}
%\right) 
- \pa \, \mbox{(pole-3)} 
+ \frac{4i(6+k)}{3(5+k)^2}
\left(
\begin{array}{c}
F_{11} \nonu \\
F_{22}  
 \end{array}
\right) B_3
\right. \nonu \\
& \mp &  \frac{4(-3+k)}{3(5+k)}
 \left(
\begin{array}{c}
G_{11} \nonu \\
G_{22}  
 \end{array}
\right)
\mp \frac{8(-3+k)}{3(5+k)^2} U 
\left(
\begin{array}{c}
F_{11} \nonu \\
F_{22}  
 \end{array}
\right) \nonu \\
&+ &  \frac{4(-3+k)}{3(5+k)^3} F_{12} F_{21} 
\left(
\begin{array}{c}
F_{11} \nonu \\
F_{22}  
 \end{array}
\right) 
-\frac{4i(3+2k)}{3(5+k)^2} \left(
\begin{array}{c}
F_{11} \nonu \\
F_{22}  
 \end{array}
\right) A_3
\nonu \\
&- & \left. \frac{4i(3+2k)}{3(5+k)^2} 
\left(
\begin{array}{c}
F_{21} \nonu \\
F_{12}  
 \end{array}
\right) A_{\pm}
+\frac{4i(6+k)}{3(5+k)^2}
\left(
\begin{array}{c}
F_{12} \nonu \\
F_{21}  
 \end{array}
\right) B_{\mp} 
\right](w) \nonu \\
& + & \frac{1}{(z-w)} 
 \left[ \pm \left(
\begin{array}{c}
{\bf U^{(\frac{5}{2})}} \nonu \\
{\bf V^{(\frac{5}{2})}}  \end{array}
\right)  + 
 \left(
\begin{array}{c}
{\bf Q^{(\frac{5}{2})}} \nonu \\
{\bf R^{(\frac{5}{2})}}  \end{array}
\right)(w) 
\right](w)   +\cdots.
\nonu  
\eea
Compared to the OPE in the nonlinear version, 
the above OPE does not contain the higher spin-$\frac{3}{2}$ currents.

Together with (\ref{newspinonelinear}) and (\ref{w2non}), 
the following OPE can be obtained
%%%%%%%%%%%%%%%%%%%%%%%%%%%%%%%%%%%%%%%%%%%%%%%%%%%%%%%%%%%%%%%%%%%%%%
\bea
{\bf T^{(1)}}(z) \, {\bf W^{(2)}}(w) & = & \frac{1}{(z-w)^2} \,
\left[  \frac{6i}{(5+k)} {A}_3  -\frac{2ik}{(5+k)} 
{B}_3 +  \frac{2(3+k)}{(5+k)^2}  F_{11} F_{22} -\frac{2(-3+k)}{(5+k)^2}
F_{12} F_{21} \right](w)    
\nonu \\
& + & \frac{1}{(z-w)} \, \left[ -\frac{2i}{(5+k)^2} F_{11} F_{12} A_{-}+
\frac{2i}{(5+k)^2}  F_{21} F_{22} A_{+} -\frac{2i}{(5+k)^2}
F_{12} F_{22} B_{-} \right. \nonu \\
&+&  \frac{2i}{(5+k)^2}  F_{11} F_{21} B_{+} +\frac{1}{(5+k)}
F_{12} G_{21}   \nonu \\
&- & \left. \frac{1}{(5+k)} F_{21} G_{12} +\frac{4}{(5+k)^2}  F_{12} F_{21} U 
\right](w) +\cdots.
\nonu 
\eea
Note that in this OPE there exists nontrivial first order pole and 
the second order pole does not contain the higher spin-$1$ current.

Furthermore, the relations (\ref{w+5halfnon}) and (\ref{w-5halfnon})
determine the following OPEs 
\bea
%%%%%%%%%%%%%%%%%%%%%%%%%%%%%%%%%%%%%%%%%%%%%%%%%%
{\bf T^{(1)}}(z) \,
{\bf W_{\pm}^{(\frac{5}{2})}}(w) 
& = & 
%%%%%%%%%%%%%%%%%%%%%%%%%%%%%%%%%%%%%%%%%%%%%%%%%%%%%%%
\mp \frac{1}{(z-w)^3} \frac{12k}{(5+k)^2}  
\left(
\begin{array}{c}
F_{21} \nonu \\
F_{12}  
 \end{array}
\right)(w) 
\nonu \\
& + & \frac{1}{(z-w)^2} \left[ 
%\frac{12k}{(5+k)^2} \pa 
%\left(
%\begin{array}{c}
%F_{21} \nonu \\
%-F_{12}  
% \end{array}
%\right) 
- \pa \, \mbox{(pole-3)} 
+  \frac{4i(6+k)}{3(5+k)^2}
\left(
\begin{array}{c}
F_{21} \nonu \\
F_{12}  
 \end{array}
\right) B_3
\right. \nonu \\
&\mp &  \frac{4(-3+k)}{3(5+k)}
 \left(
\begin{array}{c}
G_{21} \nonu \\
G_{12}  
 \end{array}
\right)
\mp \frac{8(-3+k)}{3(5+k)^2} U 
\left(
\begin{array}{c}
F_{21} \nonu \\
F_{12}  
 \end{array}
\right) \nonu \\
&- &  \frac{4(-3+k)}{3(5+k)^3} 
\left(
\begin{array}{c}
F_{21} \nonu \\
F_{12}  
 \end{array}
\right) F_{11} F_{22}
+\frac{4i(3+2k)}{3(5+k)^2} \left(
\begin{array}{c}
F_{21} \nonu \\
F_{12}  
 \end{array}
\right) A_3
\nonu \\
&- & \left. \frac{4i(3+2k)}{3(5+k)^2} 
\left(
\begin{array}{c}
F_{11} \nonu \\
F_{22}  
 \end{array}
\right) A_{\mp}
-\frac{4i(6+k)}{3(5+k)^2}
\left(
\begin{array}{c}
F_{22} \nonu \\
F_{11}  
 \end{array}
\right) B_{\mp} 
\right](w) \nonu \\
& + &  \frac{1}{(z-w)} 
 \left[ 
\pm {\bf W_{\pm}^{(\frac{5}{2})}}   +
{\bf P_{\pm}^{(\frac{5}{2})}} 
\right](w)   +\cdots.
\nonu 
\eea
Note that there are no higher spin-$\frac{3}{2}$ currents 
which were present in the nonlinear version.

For the higher spin-$3$ current, one can use the relation (\ref{w3non})
and obtain the following OPE
%%%%%%%%%%%%%%%%%%%%%%%%%%%%%%%%%%%%%%%%%%%%%%%%%%%%%%%%%%
\bea
{\bf T^{(1)}}(z) \, {\bf W^{(3)}}(w) & = & 
\frac{1}{(z-w)^3} \, \frac{24k}{(5+k)^2} 
U(w) 
\nonu \\
& + & \frac{1}{(z-w)^2}  \, \left[-  {\bf P^{(2)}}
+ 4 
{\bf W^{(2)}} +\frac{4(-3+k)}{3(5+k)} {\bf T^{(2)}} 
- \frac{4 {\bf (-3+k)}}{(17+13k)} {\bf T^{(1)}}  {\bf T^{(1)}}
\right. 
\nonu \\
& + &  \frac{8(-3+k)(17+4k)}{3(5+k)(17+13k)} T 
+  \frac{4(3+2k)}{3(5+k)^2} {A}_3 {A}_3 -
   \frac{4(-3+k)}{3(5+k)^2} {A}_3 {B}_3+
 \frac{4i (3+2k)}{3(5+k)^2} \pa {A}_3  
\nonu \\
& + & 
 \frac{4(3+2k)}{3(5+k)^2} {A}_{+} {A}_{-} 
-   \frac{4(6+k)}{3(5+k)^2} {B}_3 {B}_3 -
   \frac{4i(6+k)}{3(5+k)^2} \pa {B}_3-
 \frac{4(6+k)}{3(5+k)^2} {B}_{+} {B}_{-} 
 \nonu \\
&+& \frac{4i(-3+k)}{3(5+k)^3}  A_3 F_{11} F_{22}
-\frac{8i(15+k)}{3(5+k)^3} A_{-} F_{11} F_{12}
+\frac{4i(9+k)}{(5+k)^3} A_{+} F_{21} F_{22}
\nonu \\
&-& \frac{4i(-3+k)}{3(5+k)^3} B_{3} F_{11} F_{22}
-\frac{4i(15+7k)}{3(5+k)^3} B_{-} F_{12} F_{22}
+\frac{8i(3+k)}{(5+k)^3} B_{+} F_{11} F_{21}
\nonu \\
&-& \frac{(15+7k)}{3(5+k)^2} F_{11} G_{22}
+\frac{4(-63-9k+2k^2)}{3(5+k)^3} \pa F_{11} F_{22}
-\frac{4(27+9k+2k^2)}{3(5+k)^3} F_{11} \pa F_{22}
 \nonu \\
&+& \frac{(9+k)}{(5+k)^2} F_{12} G_{21}
+\frac{8(-3+k)(6+k)}{3(5+k)^3} \pa F_{12} F_{21}
-\frac{8(-3+k)(2+k)}{3(5+k)^3} F_{12} \pa F_{21}
\nonu \\
&- & \frac{3}{(5+k)} F_{21} G_{12} -\frac{(21+5k)}{3(5+k)^2} F_{22} G_{11}
+
\frac{12i}{(5+k)^2} U A_3 +\frac{4ik}{(5+k)^2} U B_3  
\nonu \\
&-& 
\left. \frac{12k}{(5+k)^2} \pa U +\frac{8(-3+k)}{3(5+k)^2} U U 
-\frac{16(-3+k)}{3(5+k)^2} U F_{11} F_{22} +\frac{12}{(5+k)^2} U F_{12} F_{21} 
\right](w)
\nonu \\
& + & \frac{1}{(z-w)} \, \left[  
\frac{4 i}{(k+5)^2} A_3 \pa (F_{11} F_{22})
+\frac{4 i}{(k+5)^2} A_3  \pa (F_{12}  F_{21})
-\frac{4 i}{(k+5)^2} \pa A_{-} F_{11} F_{12}
\right. \nonu \\
&+& \frac{4 i}{(k+5)^2} A_{-} \pa (F_{11} F_{12})
-\frac{4 i}{(k+5)^2} \pa A_{+} F_{21} F_{22}
+\frac{4 i}{(k+5)^2} A_{+} \pa (F_{21} F_{22})
\nonu \\
&+& \frac{4 i}{(k+5)^2} B_3 \pa (F_{11} F_{22})
-\frac{4 i}{(k+5)^2} B_3 \pa (F_{12} F_{21})
-\frac{4 i}{(k+5)^2} \pa B_{-} F_{12} F_{22}
\nonu \\
&+& \frac{4 i}{(k+5)^2} B_{-} \pa (F_{12} F_{22})
-\frac{4 i}{(k+5)^2} \pa B_{+} F_{11} F_{21}
+\frac{4 i}{(k+5)^2} B_{+} \pa (F_{11} F_{21})
\nonu \\
&-& \frac{3}{(k+5)} \pa F_{11} G_{22}
+\frac{1}{(k+5)} F_{11} \pa G_{22}
-\frac{3}{(k+5)} \pa F_{12} G_{21}
+\frac{1}{(k+5)} F_{12} \pa G_{21}
\nonu \\
&-& \frac{3}{(k+5)} \pa F_{21} G_{12}
+ \frac{1}{(k+5)} F_{21} \pa G_{12}
-\frac{3}{(k+5)} \pa F_{22} G_{11}
+ 
\frac{1}{(k+5)} F_{22} \pa G_{11}
\nonu \\
&-& \frac{8}{(k+5)^2} U \pa F_{11} F_{22}
+ \frac{8}{(k+5)^2} U F_{11} \pa F_{22}
+ \frac{4}{(k+5)^2} \pa U F_{12} F_{21}
\nonu \\
&-& \frac{8}{(k+5)^2} U \pa F_{12} F_{21}
+ \frac{8}{(k+5)^2} U F_{12} \pa F_{21}
+ \frac{2}{(k+5)^2} F_{22} G_{11} F_{22} F_{11}
\nonu \\
&+ & \left.
 \frac{2}{(k+5)^2} F_{11} G_{22} F_{11} F_{22}
+ \frac{4}{(k+5)^2} F_{21} G_{12} F_{21} F_{12}
 \right](w)  + 
\cdots.
\nonu
\eea
The fields appearing in the 
second and third lines of the first order pole
can be seen in the corresponding OPE in the nonlinear 
version.  
Compared to the corresponding OPE in the nonlinear version, 
the above OPE does not contain any composite fields 
between the higher spin currents and the currents from the large 
${\cal N}=4$ linear superconformal algebra \footnote{
There are no composite fields written in terms of the field with a boldface 
and the field without a boldface. There are only composite fields with 
a boldface and only composite fields without a boldface. In this Appendix,
what we mean by 'the nonlinear terms in the nonlinear version' are the 
composite fields between any higher spin currents and any currents 
in the large ${\cal N}=4$ linear superconformal algebra.
\label{nonlineardef}
}.
Also note  the presence of the quadratic term (nonlinear term) 
of higher spin-$1$ current.
Using the following relations, some of the terms in the OPE 
can be written in terms of other terms
%%%%%%%%%%%%%%%%%%%%%%%%%%%%%%%%%%%%%%%%%%%%%%%%%%%%%
\bea
F_{22} G_{11} F_{22} F_{11} & = & ( i \pa A_3 + i \pa B_3  + \pa U ) F_{22} F_{11}, 
\nonu \\
F_{11} G_{22} F_{11} F_{22} & = & ( -i \pa A_3 - i \pa B_3  + \pa U ) F_{11} F_{22}, 
\nonu \\
F_{21} G_{12} F_{21} F_{12} & = & ( i \pa A_3 - i \pa B_3  + \pa U ) F_{21} F_{12}.
\nonu
\eea

Let us move to the OPEs between the next higher spin-$\frac{3}{2}$
current and other higher spin currents (except the previous higher 
spin-$1$ current).
From the description of (\ref{g1221t1}), the following OPEs 
can be obtained
%%%%%%%%%%%%%%%%%%%%%%%%%%%%%%%%%%%%%%%%%%%%%%%%%%%%%%%%%%%%%%%%%%%%%%%%
\bea
{\bf T_{\pm}^{(\frac{3}{2})}}(z) \, 
{\bf T_{\pm}^{(\frac{3}{2})}}(w)  & = & \frac{1}{(z-w)} \, \left[
-\frac{1}{(k+5)} A_{\mp} B_{\mp}   \mp \frac{2 i}{(k+5)^2} A_{\mp} 
\left(
\begin{array}{c}
F_{11} F_{21} \nonu \\
F_{22} F_{12} 
\end{array}
\right) \right. \nonu \\
& \pm & \left. \frac{2 i}{(k+5)^2} B_{\mp}
 \left(
\begin{array}{c}
F_{21} F_{22} \nonu \\
F_{12} F_{11} 
\end{array}
\right)  
-\frac{2}{(k+5)^2} 
\left(
\begin{array}{c}
F_{21} \pa F_{21} \nonu \\
F_{12} \pa F_{12} 
\end{array}
\right)  
\right](w) + \cdots.
\nonu 
\eea
These are new nontrivial OPEs. In the nonlinear version, 
the corresponding OPEs are not present.
Even the terms which did not appear in the corresponding OPEs 
in the nonlinear version appear in the above OPEs.

Again the relations (\ref{g1221t1}) determine 
the following OPE
%%%%%%%%%%%%%%%%%%%%%%%%%%%%%%%%%%%%%%%%%%%%%%%%%%%
\bea
{\bf T_{+}^{(\frac{3}{2})}}(z) \, 
{\bf T_{-}^{(\frac{3}{2})}}(w)  & = & -\frac{1}{(z-w)^3} \, \left[
\frac{(4+7k)}{(5+k)} \right]
\nonu \\
& + & \frac{1}{(z-w)^2} \, \left[ 
-  {\bf T^{(1)}} +
\frac{7i}{(5+k)} {A}_3 +\frac{i(1+2k)}{(5+k)}
{B}_3  -\frac{(k-3)}{(k+5)^2} F_{11} F_{22} \right. \nonu \\
& + & \left.  \frac{(k+3)}{(k+5)^2}
F_{12} F_{21}
 \right](w) \nonu \\
& + & \frac{1}{(z-w)} \, \left[
-   {\bf T^{(2)}} -\frac{1}{2} \pa {\bf T^{(1)}} 
\right. 
\nonu \\
& - &   T 
-  \frac{1}{2(5+k)} {A}_3 {A}_3 +
   \frac{1}{(5+k)} {A}_3 {B}_3+
 \frac{3i}{(5+k)} \pa {A}_3  
\nonu \\
& - & 
 \frac{1}{2(5+k)} {A}_{+} {A}_{-} 
-   \frac{1}{2(5+k)} {B}_3 {B}_3 +
   \frac{ik}{(5+k)} \pa {B}_3-
 \frac{1}{2(5+k)} {B}_{+} {B}_{-} 
 \nonu \\
&+& \frac{2i}{(5+k)^2}  A_3 F_{11} F_{22}
+\frac{i}{(5+k)^2} A_{-} F_{11} F_{12}
+\frac{i}{(5+k)^2} A_{+} F_{21} F_{22}
\nonu \\
&-& \frac{2i}{(5+k)^2} B_{3} F_{11} F_{22}
-\frac{i}{(5+k)^2} B_{-} F_{12} F_{22}
-\frac{i}{(5+k)^2} B_{+} F_{11} F_{21}
\nonu \\
&-& 
\frac{(3+k)}{(5+k)^2} \pa F_{11} F_{22}
+\frac{6}{(5+k)^2} F_{11} \pa F_{22}
-
\frac{2}{(5+k)^2} \pa F_{12} F_{21}
\nonu \\
& + & \left. \frac{1}{(5+k)} F_{12} \pa F_{21}- 
 \frac{1}{2(5+k)} U U   
  \right](w) +  \cdots.
\nonu 
\eea
This OPE is rather complicated due to the spin-$\frac{1}{2}$ currents 
and spin-$1$ current which are present in the linear version.

Together with the previous relations and (\ref{g1221t3half}), the following 
OPEs can be obtained
%%%%%%%%%%%%%%%%%%%%%%%%%%%%%%%%%%%%%%%%%%%%%%%%%%%%%%%%%%%%%%
\bea
{\bf T_{\pm}^{(\frac{3}{2})}}(z) \, 
{\bf T^{(2)}}(w) & = & \frac{1}{(z-w)^2} \, 
\left[ \frac{3}{2}
{\bf T_{\pm}^{(\frac{3}{2})}} 
 \mp \frac{(3 k+11)}{2 (k+5)^2} U 
\left(
\begin{array}{c}
 F_{21} \nonu \\
F_{12}  
\end{array}
\right)  
+\frac{2}{(k+5)^2}
 \left(
\begin{array}{c}
 F_{21} \nonu \\
F_{12}  
\end{array}
\right)  
F_{11} F_{22}
\right](w) 
\nonu \\
&+& \frac{i (3 k+11)}{2 (k+5)^2}
 \left(
\begin{array}{c}
 F_{21} \nonu \\
F_{12}  
\end{array}
\right) 
A_3 
-\frac{i (3 k-1)}{2 (k+5)^2}
 \left(
\begin{array}{c}
 F_{11} \nonu \\
F_{22}  
\end{array}
\right) A_{\mp}
-\frac{i (k-11)}{2 (k+5)^2}
 \left(
\begin{array}{c}
 F_{22} \nonu \\
F_{11}  
\end{array}
\right)
B_{\mp}
\nonu \\
&- & \left. \frac{i (3 k+11)}{2 (k+5)^2}
 \left(
\begin{array}{c}
 F_{21} \nonu \\
F_{12}  
\end{array}
\right) B_3 
 \pm \frac{1}{(k+5)}
 \left(
\begin{array}{c}
 G_{21} \nonu \\
G_{12}  
\end{array}
\right)
\right](w) \nonu \\
& + & 
\frac{1}{(z-w)} \,
\left[ 
 \frac{1}{2}
\pa {\bf T_{\pm}^{(\frac{3}{2})}} 
 \mp \frac{1}{(k+5)^2} A_3 B_{\mp} 
 \left(
\begin{array}{c}
 F_{22} \nonu \\
F_{11}  
\end{array}
\right)
+ \frac{i (k+1)}{2 (k+5)^2}  
 \pa \left(
\begin{array}{c}
 A_3 F_{21} \nonu \\
A_3 F_{12}  
\end{array}
\right)
\right. \nonu \\
&+ & \frac{i}{2 (k+5)} A_{\mp}
 \left(
\begin{array}{c}
 G_{11} \nonu \\
G_{22}  
\end{array}
\right)
 \mp \frac{1}{(k+5)^2} A_{\mp} A_3
 \left(
\begin{array}{c}
 F_{11} \nonu \\
F_{22}  
\end{array}
\right)
\pm
\frac{1}{(k+5)^2} A_{\mp} B_3 
 \left(
\begin{array}{c}
 F_{11} \nonu \\
F_{22}  
\end{array}
\right)
\nonu \\
& -& \frac{i (k-1)}{2 (k+5)^2} \pa A_{\mp} 
 \left(
\begin{array}{c}
 F_{11} \nonu \\
F_{22}  
\end{array}
\right)
-\frac{i (k+7)}{2 (k+5)^2} A_{\mp} \pa 
 \left(
\begin{array}{c}
 F_{11} \nonu \\
F_{22}  
\end{array}
\right)
\mp \frac{1}{(k+5)^2}
A_{\pm} A_{\mp} 
 \left(
\begin{array}{c}
 F_{21} \nonu \\
F_{12}  
\end{array}
\right)
\nonu \\
& - & \frac{i (k+1)}{2 (k+5)^2}
\pa 
 \left(
\begin{array}{c}
B_3  F_{21} \nonu \\
B_3 F_{12}  
\end{array}
\right)
+ \frac{i}{2 (k+5)} B_{\mp} 
\left(
\begin{array}{c}
 G_{22} \nonu \\
G_{11}  
\end{array}
\right)
\pm \frac{1}{ (k+5)^2} B_{\mp} B_3
\left(
\begin{array}{c}
 F_{22} \nonu \\
F_{11}  
\end{array}
\right)
\nonu \\
& - &  
\frac{i (k-5)}{2 (k+5)^2} \pa B_{\mp}
\left(
\begin{array}{c}
 F_{22} \nonu \\
F_{11}  
\end{array}
\right)
+ \frac{i (k+7)}{2 (k+5)^2} B_{\mp} \pa
\left(
\begin{array}{c}
 F_{22} \nonu \\
F_{11}  
\end{array}
\right)
\pm \frac{1}{(k+5)^2} B_{\pm} B_{\mp}
\left(
\begin{array}{c}
 F_{21} \nonu \\
F_{12}  
\end{array}
\right)
\nonu \\
& \mp & \frac{1}{(k+5)^2}
\left(
\begin{array}{c}
F_{11} F_{21} G_{22} \nonu \\
F_{22} F_{12} G_{11}  
\end{array}
\right)
\mp
\frac{8}{(k+5)^3} 
\left(
\begin{array}{c}
\pa F_{11} F_{21} F_{22} \nonu \\
\pa F_{22} F_{12} F_{11}  
\end{array}
\right)
\nonu \\
& \pm & \frac{(k-3)}{2 (k+5)^2} \pa^2
\left(
\begin{array}{c}
 F_{21} \nonu \\
F_{12}  
\end{array}
\right)
\pm  
\frac{1}{(k+5)^2} 
\left(
\begin{array}{c}
 F_{21} F_{22} G_{11} \nonu \\
 F_{12} F_{11} G_{22}  
\end{array}
\right)
\nonu \\
& + & \frac{i}{(k+5)^2} U A_{\mp}
\left(
\begin{array}{c}
 F_{11} \nonu \\
F_{22}  
\end{array}
\right)
+\frac{i}{(k+5)^2} U B_{\mp} 
\left(
\begin{array}{c}
 F_{22} \nonu \\
F_{11}  
\end{array}
\right)
\nonu \\
& \mp & \left. 
\frac{1}{2 (k+5)} \pa
\left(
\begin{array}{c}
U  F_{21} \nonu \\
 U F_{12}  
\end{array}
\right)
\mp
\frac{2 (k+1)}{(k+5)^3} 
\left(
\begin{array}{c}
F_{11} F_{21} \pa F_{22} \nonu \\
 F_{22} F_{12} \pa F_{11}  
\end{array}
\right)
 \right](w)  +  \cdots.
\nonu 
\eea
In this OPE, the second order pole  has 
spin-$\frac{3}{2}$ currents living in the large ${\cal N}=4$
linear superconformal algebra. There are many spin-$\frac{1}{2}$ current 
dependent terms.
Furthermore, some terms containing the above spin-$\frac{3}{2}$ currents 
in the first order pole do not appear in the 
corresponding OPEs in the nonlinear version. 

From the result of (\ref{g1122t1}), one can compute the following OPEs
%%%%%%%%%%%%%%%%%%%%%%%%%%%%%%%%%%%%%%%%%%%%%%%%%%%%%%%%%%%%%%%%%%%%%%%%%%%
\bea
{\bf T_{\pm}^{(\frac{3}{2})}}(z) \, 
\left(
\begin{array}{c}
{\bf U^{(\frac{3}{2})}} \\
{\bf V^{(\frac{3}{2})}} \\
\end{array} \right) (w) & = &  \frac{1}{(z-w)^2} \, 
\left[ \pm \frac{i}{(5+k)} 
{B}_{\mp} + \frac{2 k}{(k+5)^2} 
\left(
\begin{array}{c}
 F_{11} F_{21} \nonu \\
F_{22} F_{12}  
\end{array}
\right)
 \right] (w) \nonu \\
& + & 
\frac{1}{(z-w)} \, \left[ 
-\frac{1}{(k+5)} A_3 B_{\mp}
\mp \frac{2 i}{(k+5)^2} A_3 
\left(
\begin{array}{c}
 F_{11} F_{21} \nonu \\
F_{22} F_{12}  
\end{array}
\right)
\pm \frac{i}{2 (k+5)} \pa B_{\mp}
\right. \nonu \\
&+& 
\frac{i}{(k+5)^2} B_{\mp} 
 F_{11} F_{22} 
+ \frac{i}{(k+5)^2} B_{\mp} F_{12} F_{21}
+\frac{(k+1)}{(k+5)^2}
\left(
\begin{array}{c}
 \pa F_{11} F_{21} \nonu \\
\pa F_{22} F_{12}  
\end{array}
\right) \nonu \\
&+ & \left. \frac{(k-1)}{(k+5)^2}
\left(
\begin{array}{c}
 F_{11} \pa F_{21} \nonu \\
F_{22} \pa F_{12}  
\end{array}
\right)
\right](w) +\cdots.
\nonu 
\eea
These OPEs are new and the corresponding OPEs 
in the nonlinear version are trivial.

Similarly, the following OPEs can be calculated  
\bea
%%%%%%%%%%%%%%%%%%%%%%%%%%%%%%%%%%%%%%%%%%%%%%%%%%%%%%%%%%%%%%%
{\bf T_{\pm}^{(\frac{3}{2})}}(z) \, 
\left(
\begin{array}{c}
{\bf V^{(\frac{3}{2})}} \\
{\bf U^{(\frac{3}{2})}} \\
\end{array} \right) (w) & = &  \frac{1}{(z-w)^2} \, 
\left[\mp \frac{7 i}{(k+5)}  {A}_{\mp} 
\mp \frac{6}{(k+5)^2} 
\left(
\begin{array}{c}
 F_{21}  F_{22} \nonu \\
F_{11}  F_{12}  
\end{array}
\right)
\right] (w) \nonu \\
& + & 
\frac{1}{(z-w)} \, \left[ \mp \left(
\begin{array}{c} 
 {\bf V_{+}^{(2)}} \\
{\bf U_{-}^{(2)}} \\
\end{array} \right)   
-\frac{1}{(k+5)} A_{\mp} B_3
\mp \frac{7 i}{2 (k+5)} \pa A_{\mp}
\right. \nonu \\
&- & \frac{i}{(k+5)^2} A_{\mp} F_{11} F_{22}
+ \frac{i}{(k+5)^2} A_{\mp} F_{12} F_{21}
 \mp \frac{2 i}{(k+5)^2} B_3
\left(
\begin{array}{c}
 F_{22}  F_{21} \nonu \\
F_{11}  F_{12}  
\end{array}
\right)
\nonu \\
&+& \left.
\frac{4}{(k+5)^2} 
\left(
\begin{array}{c}
 \pa F_{22}  F_{21} \nonu \\
\pa F_{11}  F_{12}  
\end{array}
\right)
+ \frac{2}{(k+5)^2}
\left(
\begin{array}{c}
  F_{22}  \pa F_{21} \nonu \\
F_{11}  \pa F_{12}  
\end{array}
\right)
\right](w) +\cdots.
\nonu 
\eea
Compared to the corresponding OPEs in the nonlinear version,
the above OPEs have spin-$\frac{1}{2}$ current dependent terms.
The fields in the first line of the first order pole except the 
higher spin current do not appear in the 
corresponding OPEs in the nonlinear version.

The equations (\ref{g1221vu3half}) determine the following OPEs
\bea
%%%%%%%%%%%%%%%%%%%%%%%%%%%%%%%%%%%%%%%%%%%%%%%%%%%%%%%%%%%%%%%%%%%%
{\bf T_{\pm}^{(\frac{3}{2})}}(z) \, 
\left(
\begin{array}{c}
{\bf U_{+}^{(2)}} \\
{\bf V_{-}^{(2)}} \\
\end{array} \right)(w) & = & 
\frac{1}{(z-w)^2} \, \left[ \pm \frac{2 i k}{(k+5)^2} 
\left(
\begin{array}{c}
   F_{21} \nonu \\
  F_{12}  
\end{array}
\right)
B_{\mp}
\right](w)
\nonu \\
&+& \frac{1}{(z-w)} \,
\left[  \frac{1}{(k+5)^2}
A_3 B_{\mp} 
\left(
\begin{array}{c}
   F_{21} \nonu \\
  F_{12}  
\end{array}
\right)
-\frac{1}{(k+5)^2} A_{\mp} B_{\mp} 
\left(
\begin{array}{c}
   F_{11} \nonu \\
  F_{22}  
\end{array}
\right)
\right. \nonu \\
& \mp & \frac{i}{2 (k+5)} B_{\mp} 
\left(
\begin{array}{c}
   G_{21} \nonu \\
  G_{12}  
\end{array}
\right)
+ \frac{1}{(k+5)^2}
B_{\mp} B_3 
\left(
\begin{array}{c}
   F_{21} \nonu \\
 F_{12}  
\end{array}
\right)
\nonu \\
& -&
\frac{1}{(k+5)^2} B_{\mp} B_{\mp} 
\left(
\begin{array}{c}
   F_{22} \nonu \\
 F_{11}  
\end{array}
\right)
\pm \frac{i k}{(k+5)^2}
\pa B_{\mp}
\left(
\begin{array}{c}
   F_{21} \nonu \\
 F_{12}  
\end{array}
\right)
\nonu \\
& \mp & \frac{3 i}{(k+5)^2} B_{\mp}
\pa 
 \left(
\begin{array}{c}
   F_{21} \nonu \\
 F_{12}  
\end{array}
\right)
+ \frac{1}{(k+5)^2}
\left(
\begin{array}{c}
   F_{11} F_{21} G_{21} \nonu \\
 F_{22} F_{12} G_{12}  
\end{array}
\right)
\nonu \\
& \mp & \left. \frac{i}{(k+5)^2}
U B_{\mp} 
\left(
\begin{array}{c}
   F_{21}  \nonu \\
 F_{12}   
\end{array}
\right)
+
\frac{2 (k+1)}{(k+5)^3}
\left(
\begin{array}{c}
   F_{11} F_{21} \pa F_{21} \nonu \\
 F_{22} F_{12} \pa F_{12}  
\end{array}
\right)
\right] (w) +\cdots.
\nonu 
\eea
Note that the nonlinear term appearing in the nonlinear version
disappears in these OPEs. In other words, there are no composite fields
with a boldface and without a boldface.
The field containing the spin-$\frac{3}{2}$ current in the first order pole
does not appear in the corresponding OPEs in the nonlinear version. 

Furthermore, the relations (\ref{g1221uv3half}) allow us to
calculate the following OPEs
\bea
%%%%%%%%%%%%%%%%%%%%%%%%%%%%%%%%%%%%%%%%%%%%%%%%%%%%%%%%%%%%%%%%%%%
{\bf T_{\pm}^{(\frac{3}{2})}}(z) \,
\left(
\begin{array}{c}
{\bf U_{-}^{(2)}} \nonu \\
 {\bf V_{+}^{(2)}}
\end{array}
\right)
(w) & = & 
\frac{1}{(z-w)^3} \, \left[\frac{6 k}{(k+5)^2} 
\left(
\begin{array}{c}
F_{11} \nonu \\
F_{22}
\end{array}
\right) \right](w)
\nonu \\
& + & \frac{1}{(z-w)^2} \,
\left[  \mp \frac{2 (k+3)}{(k+5)}  
\left(
\begin{array}{c}
{\bf U^{(\frac{3}{2})}} \nonu \\
{\bf V^{(\frac{3}{2})}} 
\end{array}
\right) 
-\frac{1}{(k+5)} 
\left(
\begin{array}{c}
G_{11} \nonu \\
G_{22}
\end{array}
\right)
\right.
\nonu \\
& \pm & \frac{2 i (k-1)}{(k+5)^2} 
\left(
\begin{array}{c}
F_{11} \nonu \\
F_{22}
\end{array}
\right) A_3
\pm \frac{4 i}{(k+5)^2} 
\left(
\begin{array}{c}
F_{11} \nonu \\
F_{22}
\end{array}
\right) B_3
+ \frac{2 (k+2)}{(k+5)^2} U
\left(
\begin{array}{c}
F_{11} \nonu \\
F_{22}
\end{array}
\right)
\nonu \\
& \pm & 
\frac{2}{(k+5)^2}
F_{12} F_{21}
 \left(
\begin{array}{c}
F_{11} \nonu \\
F_{22}
\end{array}
\right)
\pm \frac{4 i}{(k+5)^2}
 \left(
\begin{array}{c}
F_{12} \nonu \\
F_{21}
\end{array}
\right) B_{\mp}
\nonu \\
& \pm & \left. \frac{2 i (k+2)}{(k+5)^2}
 \left(
\begin{array}{c}
F_{21} \nonu \\
F_{12}
\end{array}
\right) A_{\pm}
\right](w) 
\nonu \\
& + & \frac{1}{(z-w)} \, \left[  \mp \frac{1}{2} 
\left(
\begin{array}{c}
{\bf Q^{(\frac{5}{2})}} \nonu \\
   {\bf R^{(\frac{5}{2})}} 
\end{array}
\right)
-
\left(
\begin{array}{c}
{\bf U^{(\frac{5}{2})}} \nonu \\
{\bf V^{(\frac{5}{2})}}
\end{array}
\right) 
\mp \frac{2 (k+3)}{3 (k+5)} \pa
\left(
\begin{array}{c}
{\bf U^{(\frac{3}{2})}} \nonu \\
{\bf V^{(\frac{3}{2})}}
\end{array}
\right) 
\right. \nonu \\
& \mp & \frac{i}{2 (k+5)} A_3 
\left(
\begin{array}{c}
G_{11} \nonu \\
G_{22}
\end{array}
\right) 
\nonu \\
&+ &  \frac{2}{(k+5)^2} 
A_3 A_3 
\left(
\begin{array}{c}
F_{11} \nonu \\
F_{22}
\end{array}
\right)
-\frac{2}{(k+5)^2} 
A_3 B_3 
\left(
\begin{array}{c}
F_{11} \nonu \\
F_{22}
\end{array}
\right)
\nonu \\
& - & \frac{2}{(k+5)^2}
A_3 B_{\mp} 
\left(
\begin{array}{c}
F_{12} \nonu \\
F_{21}
\end{array}
\right)
\pm \frac{2 i (2 k+3)}{3 (k+5)^2} 
\pa A_3 
\left(
\begin{array}{c}
F_{11} \nonu \\
F_{22}
\end{array}
\right)
\nonu \\
& \pm & \frac{2 i (k+9)}{3 (k+5)^2} A_3  \pa
\left(
\begin{array}{c}
F_{11} \nonu \\
F_{22}
\end{array}
\right)
+ \frac{1}{(k+5)^2} A_{\pm} A_3 
\left(
\begin{array}{c}
F_{21} \nonu \\
F_{12}
\end{array}
\right)
\nonu \\
&+& \frac{1}{(k+5)^2} A_{\pm} A_{\mp} 
\left(
\begin{array}{c}
F_{11} \nonu \\
F_{22}
\end{array}
\right)
-\frac{1}{(k+5)^2} A_{\pm} B_3
\left(
\begin{array}{c}
F_{21} \nonu \\
F_{12}
\end{array}
\right)
\nonu \\
&+& \frac{1}{(k+5)^2} A_{\pm} B_{\mp} 
\left(
\begin{array}{c}
F_{22} \nonu \\
F_{11}
\end{array}
\right)
\pm
\frac{i (2 k+3)}{3 (k+5)^2} 
\pa A_{\pm} 
\left(
\begin{array}{c}
F_{21} \nonu \\
F_{12}
\end{array}
\right)
\nonu \\
& \pm & \frac{i (2 k+9)}{3 (k+5)^2} A_{\pm} \pa
\left(
\begin{array}{c}
F_{21} \nonu \\
F_{12}
\end{array}
\right)
\pm \frac{i}{2 (k+5)} B_3 
\left(
\begin{array}{c}
G_{11} \nonu \\
G_{22}
\end{array}
\right)
\nonu \\
& \pm & \frac{i (k+9)}{3 (k+5)^2} 
\pa B_3 
\left(
\begin{array}{c}
F_{11} \nonu \\
F_{22}
\end{array}
\right)
\pm \frac{2 i k}{3 (k+5)^2} B_3 
\pa \left(
\begin{array}{c}
F_{11} \nonu \\
F_{22}
\end{array}
\right)
\nonu \\
&\pm & \frac{i}{2 (k+5)} B_{\mp}
 \left(
\begin{array}{c}
G_{12} \nonu \\
G_{21}
\end{array}
\right)
\mp \frac{i (k-3)}{3 (k+5)^2} \pa B_{\mp} 
 \left(
\begin{array}{c}
F_{12} \nonu \\
F_{21}
\end{array}
\right)
\nonu \\
& \pm & \frac{2 i k}{3 (k+5)^2} B_{\mp} \pa
 \left(
\begin{array}{c}
F_{12} \nonu \\
F_{21}
\end{array}
\right)
-\frac{3 (k+3)}{2 (k+5)^2} \pa^2
\left(
\begin{array}{c}
F_{11} \nonu \\
F_{22}
\end{array}
\right)
\nonu \\
&-& \frac{1}{(k+5)^2}
 \left(
\begin{array}{c}
F_{11} F_{12} G_{21} \nonu \\
F_{22} F_{21} G_{12}
\end{array}
\right)
\pm \frac{2 (k+9)}{3 (k+5)^3} \pa
 \left(
\begin{array}{c}
F_{11}  \nonu \\
F_{22}
\end{array}
\right) F_{12} F_{21}
\nonu \\
& + & 
\frac{8 (k+3)}{3 (k+5)^3} 
\left(
\begin{array}{c}
F_{11} \pa F_{12} F_{21} \nonu \\
F_{22} \pa F_{21} F_{12}
\end{array}
\right)
-\frac{4 (k+3)}{3 (k+5)^3}
\left(
\begin{array}{c}
F_{11} F_{12} \pa F_{21} \nonu \\
F_{22} F_{21} \pa F_{12}
\end{array}
\right)
\nonu \\
& \mp & \frac{1}{(k+5)^2} 
F_{11} F_{22}
\left(
\begin{array}{c}
G_{11}  \nonu \\
G_{22} 
\end{array}
\right)
\mp \frac{1}{(k+5)^2} F_{12} F_{21}
\left(
\begin{array}{c}
G_{11}  \nonu \\
G_{22} 
\end{array}
\right)
\nonu \\
&+& \frac{1}{(k+5)} T
\left(
\begin{array}{c}
F_{11}  \nonu \\
F_{22} 
\end{array}
\right)
+ \frac{1}{2 (k+5)} U
\left(
\begin{array}{c}
G_{11}  \nonu \\
G_{22} 
\end{array}
\right)
\pm \frac{i}{(k+5)^2} U A_{\pm} 
\left(
\begin{array}{c}
F_{21}  \nonu \\
F_{12} 
\end{array}
\right)
\nonu \\
& \pm & \frac{2 i}{(k+5)^2} U B_3 
\left(
\begin{array}{c}
F_{11}  \nonu \\
F_{22} 
\end{array}
\right)
\pm \frac{2 i}{(k+5)^2} 
U B_{\mp} 
\left(
\begin{array}{c}
F_{12}  \nonu \\
F_{21} 
\end{array}
\right)
\nonu \\
&+& \frac{2 (k+3)}{3 (k+5)^2} U
\pa 
\left(
\begin{array}{c}
F_{11}  \nonu \\
F_{22} 
\end{array}
\right)
+\frac{2}{(k+5)^2} U U 
\left(
\begin{array}{c}
F_{11}  \nonu \\
F_{22} 
\end{array}
\right)
\nonu \\
&+ & \left. \frac{2 (k+3)}{3 (k+5)^2} 
\left(
\begin{array}{c}
F_{11} G_{22} F_{11} \nonu \\
F_{22} G_{11} F_{22}
\end{array}
\right) \mp
\frac{2 (k+1)}{(k+5)^3} 
F_{11} F_{22}
\pa
\left(
\begin{array}{c}
 F_{11} \nonu \\
 F_{22}
\end{array}
\right)
\right](w) 
+  \cdots.
\nonu 
\eea
The nonlinear terms (between the higher spin currents and the currents 
of large ${\cal N}=4$ linear superconformal algebra) 
appearing in the nonlinear version
disappear in this OPE.
The structure constants in the above two OPEs are related to each other
in very simple form. However, the corresponding OPEs in the nonlinear version
cannot be written in this way because there  are no simple relations between
the structure constants. 
Some terms containing the spin-$\frac{3}{2}$ current 
in the above OPEs do not appear in the corresponding 
OPEs in the nonlinear version.

Similarly, the following OPEs can be described 
\bea
%%%%%%%%%%%%%%%%%%%%%%%%%%%%%%%%%%%%%%%%%%%%%%%%%%%%%%%%%%%%%%%%%%%%%%%%
{\bf T_{\pm}^{(\frac{3}{2})}}(z) \, 
\left(
\begin{array}{c}
{\bf V_{+}^{(2)}} \\
{\bf U_{-}^{(2)}} \\
\end{array} \right) (w) & = & 
\frac{1}{(z-w)^2} \,
\left[ \mp \frac{6 i}{(k+5)^2} A_{\mp}
\left(
\begin{array}{c}
 F_{21} \nonu \\
 F_{12}
\end{array}
\right)
\right](w)
\nonu \\
&+ & \frac{1}{(z-w)} \,
\left[ 
\mp \frac{i}{2 (k+5)} A_{\mp} 
\left(
\begin{array}{c}
 G_{21} \nonu \\
 G_{12}
\end{array}
\right)
-\frac{1}{(k+5)^2} A_{\mp} A_3 
\left(
\begin{array}{c}
 F_{21} \nonu \\
 F_{12}
\end{array}
\right)
\right.
\nonu \\
&+& \frac{1}{(k+5)^2} A_{\mp} A_{\mp} 
\left(
\begin{array}{c}
 F_{11} \nonu \\
 F_{22}
\end{array}
\right)
-\frac{1}{(k+5)^2} A_{\mp} B_3 
\left(
\begin{array}{c}
 F_{21} \nonu \\
 F_{12}
\end{array}
\right)
\nonu \\
&+& \frac{1}{(k+5)^2} A_{\mp} B_{\mp}
\left(
\begin{array}{c}
 F_{22} \nonu \\
 F_{11}
\end{array}
\right)
\mp \frac{3 i}{(k+5)^2} 
\pa A_{\mp} 
\left(
\begin{array}{c}
 F_{21} \nonu \\
 F_{12}
\end{array}
\right)
\nonu \\
& \pm & \frac{3 i}{(k+5)^2} A_{\mp} \pa
\left(
\begin{array}{c}
 F_{21} \nonu \\
 F_{12}
\end{array}
\right)
-\frac{1}{(k+5)^2}
\left(
\begin{array}{c}
 F_{21} F_{22} G_{21} \nonu \\
 F_{12} F_{11} G_{12}
\end{array}
\right)
\nonu \\
& \mp &\left.
\frac{i}{(k+5)^2} U A_{\mp}
\left(
\begin{array}{c}
 F_{21}  \nonu \\
 F_{12}
\end{array}
\right)
-\frac{8}{(k+5)^3}
\left(
\begin{array}{c}
 F_{22} F_{21} \pa F_{21}  \nonu \\
 F_{11} F_{12} \pa F_{12}
\end{array}
\right)
\right](w)
+\cdots.
\nonu 
\eea
These OPEs are new in the sense that the corresponding OPEs in the 
nonlinear version are trivial.

Furthermore, one has the following OPEs
\bea
%%%%%%%%%%%%%%%%%%%%%%%%%%%%%%%%%%%%%%%%%%%%%%%%%%%%%%%%%%%%%%%%%%%%%%%%
{\bf T_{\pm}^{(\frac{3}{2})}}(z) \, 
\left(
\begin{array}{c}
{\bf V_{-}^{(2)}} \\
{\bf U_{+}^{(2)}} \\
\end{array} \right) (w) & = & 
\frac{1}{(z-w)^3} \,
\left[ -\frac{6 k}{(k+5)^2} 
\left(
\begin{array}{c}
 F_{22} \nonu \\
 F_{11}
\end{array}
\right)
\right](w)
\nonu \\
&+& 
\frac{1}{(z-w)^2} \,
\left[ \mp \frac{(k+9)}{(k+5)} 
\left(
\begin{array}{c}
 {\bf V^{(\frac{3}{2})}} \\
{\bf U^{(\frac{3}{2})}} \\
\end{array} \right)
 + 
\frac{(k+8)}{(k+5)}
\left(
\begin{array}{c}
 {G}_{22} \\
 {G}_{11} \\
\end{array} \right)
\right.
\nonu \\
&+ & \frac{(k+7)}{(k+5)^2} U
\left(
\begin{array}{c}
 {F}_{22} \\
 {F}_{11} \\
\end{array} \right)
\mp \frac{2}{(k+5)^2} 
F_{12} F_{21}
\left(
\begin{array}{c}
 {F}_{22} \\
 {F}_{11} \\
\end{array} \right)
\nonu \\
& \mp & \frac{i (k+1)}{(k+5)^2}
\left(
\begin{array}{c}
 {F}_{22} \\
{F}_{11} \\
\end{array} \right)
A_3  \mp \frac{i (k+1)}{(k+5)^2} 
\left(
\begin{array}{c}
 {F}_{12} \\
{F}_{21} \\
\end{array} \right)
A_{\mp}
\nonu \\
& \mp & \left.
\frac{i (k+7)}{(k+5)^2} 
\left(
\begin{array}{c}
 {F}_{21} \\
{F}_{12} \\
\end{array} \right)
B_{\pm} \pm \frac{i (k-7)}{(k+5)^2}
  \left(
\begin{array}{c}
 {F}_{22} \\
{F}_{11} \\
\end{array} \right) B_3
\right](w)
\nonu \\
& + & \frac{1}{(z-w)} \, \left[  \mp \frac{1}{2} \left(
\begin{array}{c} 
{\bf R^{(\frac{5}{2})}} \\
{\bf Q^{(\frac{5}{2})}} \\
\end{array} \right)
 \mp \frac{(k+9)}{3 (k+5)} \pa
\left(
\begin{array}{c} 
{\bf V^{(\frac{3}{2})}} \\
{\bf U^{(\frac{3}{2})}} \\
\end{array} \right)
\right. \nonu \\
& \pm & \frac{i}{2 (k+5)} A_3
\left(
\begin{array}{c}
 {G}_{22} \\
{G}_{11} \\
\end{array} \right)
+\frac{2}{(k+5)^2} 
A_3 B_3
\left(
\begin{array}{c}
 {F}_{22} \\
 {F}_{11} \\
\end{array} \right)
\nonu \\
&+& \frac{1}{(k+5)^2}
A_3 B_{\pm}
\left(
\begin{array}{c}
 {F}_{21} \\
 {F}_{12} \\
\end{array} \right)
\mp \frac{2 i (k+3)}{3 (k+5)^2} \pa A_3
\left(
\begin{array}{c}
 {F}_{22} \\
 {F}_{11} \\
\end{array} \right)
\nonu \\
& \mp & \frac{i (k+3)}{3 (k+5)^2} A_3
\p \left(
\begin{array}{c}
 {F}_{22} \\
 {F}_{11} \\
\end{array} \right)
\pm \frac{i}{2 (k+5)} A_{\mp} 
\left(
\begin{array}{c}
 {G}_{12} \\
 {G}_{21} \\
\end{array} \right)
\nonu \\
&+& \frac{2}{(k+5)^2} 
A_{\mp} B_3 
\left(
\begin{array}{c}
 {F}_{12} \\
 {F}_{21} \\
\end{array} \right)
-\frac{1}{(k+5)^2} A_{\mp} B_{\pm}
\left(
\begin{array}{c}
 {F}_{11} \\
 {F}_{22} \\
\end{array} \right)
\nonu \\
& \mp & \frac{i (k-3)}{3 (k+5)^2}
\pa A_{\mp} 
\left(
\begin{array}{c}
 {F}_{12} \\
 {F}_{21} \\
\end{array} \right)
\mp \frac{i (k+3)}{3 (k+5)^2} A_{\mp}
\pa
\left(
\begin{array}{c}
 {F}_{12} \\
 {F}_{21} \\
\end{array} \right)
\nonu \\
& \mp & \frac{i}{2 (k+5)} B_3 
\left(
\begin{array}{c}
 {G}_{22} \\
 {G}_{11} \\
\end{array} \right)
-\frac{2}{(k+5)^2} B_3 B_3 
\left(
\begin{array}{c}
 {F}_{22} \\
 {F}_{11} \\
\end{array} \right)
\nonu \\
& \pm & \frac{i (k-15)}{3 (k+5)^2} \pa B_3
\left(
\begin{array}{c}
 {F}_{22} \\
 {F}_{11} \\
\end{array} \right)
\mp \frac{i (k+21)}{3 (k+5)^2} B_3 \pa
\left(
\begin{array}{c}
 {F}_{22} \\
 {F}_{11} \\
\end{array} \right)
\nonu \\
&-& \frac{1}{(k+5)^2} B_{\pm} B_3 
\left(
\begin{array}{c}
 {F}_{21} \\
 {F}_{12} \\
\end{array} \right)
\mp \frac{i (k+6)}{3 (k+5)^2} \pa B_{\pm}
\left(
\begin{array}{c}
 {F}_{21} \\
 {F}_{12} \\
\end{array} \right)
\nonu \\
& \mp & \frac{i (k+12)}{3 (k+5)^2} B_{\pm} \pa
\left(
\begin{array}{c}
 {F}_{21} \\
 {F}_{12} \\
\end{array} \right)
-\frac{1}{(k+5)^2} B_{\mp} B_{\pm}
\left(
\begin{array}{c}
 {F}_{22} \\
 {F}_{11} \\
\end{array} \right)
\nonu \\
&+& \frac{(k+15)}{2 (k+5)^2} \pa^2
\left(
\begin{array}{c}
 {F}_{22} \\
 {F}_{11} \\
\end{array} \right)
\mp \frac{4 (k+3)}{3 (k+5)^3} F_{12} F_{21} \pa 
\left(
\begin{array}{c}
 {F}_{22} \\
 {F}_{11} \\
\end{array} \right)
\nonu \\
&- & \frac{2 (k+9)}{3 (k+5)^3} 
\left(
\begin{array}{c}
 {F}_{22} \pa F_{21} F_{12} \\
 {F}_{11} \pa F_{12} F_{21} \\
\end{array} \right)
+ \frac{4 (k+9)}{3 (k+5)^3}
\left(
\begin{array}{c}
 {F}_{22}  F_{21} \pa F_{12} \\
 {F}_{11}  F_{12} \pa F_{21} \\
\end{array} \right)
\nonu \\
& - & \frac{1}{(k+5)^2} 
\left(
\begin{array}{c}
 {F}_{22}  F_{12} G_{21} \\
 {F}_{11}  F_{21}  G_{12} \\
\end{array} \right)
\pm \frac{1}{(k+5)^2} F_{11} F_{22} 
\left(
\begin{array}{c}
 G_{22} \\
 G_{11} \\
\end{array} \right)
\nonu \\
& \mp & \frac{1}{(k+5)^2} F_{12} F_{21}
\left(
\begin{array}{c}
 G_{22} \\
 G_{11} \\
\end{array} \right)
+ \frac{(k+9)}{3 (k+5)} \pa 
\left(
\begin{array}{c}
 G_{22} \\
 G_{11} \\
\end{array} \right)
\nonu \\
&-& \frac{1}{(k+5)} T
\left(
\begin{array}{c}
 F_{22} \\
 F_{11} \\
\end{array} \right)
-\frac{1}{2 (k+5)} U
\left(
\begin{array}{c}
 G_{22} \\
 G_{11} \\
\end{array} \right)
\pm \frac{2 i}{(k+5)^2} U A_3 
\left(
\begin{array}{c}
 F_{22} \\
 F_{11} \\
\end{array} \right)
\nonu \\
&\pm & \frac{2 i}{(k+5)^2} 
U A_{\mp} 
\left(
\begin{array}{c}
 F_{12} \\
 F_{21} \\
\end{array} \right)
\pm \frac{i}{(k+5)^2} U B_{\pm} 
\left(
\begin{array}{c}
 F_{21} \\
 F_{12} \\
\end{array} \right)
\nonu \\
&+& \frac{(k+9)}{3 (k+5)^2} U \pa
\left(
\begin{array}{c}
 F_{22} \\
 F_{11} \\
\end{array} \right)
-\frac{2}{(k+5)^2} U U 
\left(
\begin{array}{c}
 F_{22} \\
 F_{11} \\
\end{array} \right)
\nonu \\
&+& \left. \frac{(k+9)}{3 (k+5)^2}
 \left(
\begin{array}{c}
 F_{22} G_{11} F_{22} \\
 F_{11}  G_{22} F_{11} \\
\end{array} \right)
-\frac{16 i}{(k+5)^3} 
\left(
\begin{array}{c}
 F_{22} \\
 F_{11} \\
\end{array} \right)
A_3 F_{11} F_{22}
\right](w)
\nonu \\
& + & \cdots.
\nonu 
\eea
Due to the spin-$\frac{1}{2}$ currents, these OPEs are rather 
complicated than the ones in the nonlinear version.
Some terms containing the spin-$\frac{3}{2}$ currents in the first order pole 
in the above 
OPEs do not appear in the corresponding OPEs in the nonlinear version. 

From the previous relation (\ref{g1221vu2}), the following OPEs can be
obtained
\bea
%%%%%%%%%%%%%%%%%%%%%%%%%%%%%%%%%%%%%%%%%%%%%%%%%%%%%%%%%%%%%%%%%%%%%%
{\bf T_{\pm}^{(\frac{3}{2})}}(z) \,
\left(
\begin{array}{c}
{\bf U^{(\frac{5}{2})}} \nonu \\
{\bf V^{(\frac{5}{2})}}
\end{array}
\right)
(w) & = & \frac{1}{(z-w)^3} \,
\left[ \frac{2 i k (4 k+21)}{3 (k+5)^2} {B}_{\mp} 
-\frac{16 k (k+3)}{3 (k+5)^3}
\left(
\begin{array}{c}
 F_{11} F_{21} \\
 F_{12} F_{22} \\
\end{array} \right)
\right](w) \nonu \\
& + & \frac{1}{(z-w)^2} \, \left[ 
\frac{8 (k+3)}{3 (k+5)} 
\left(
\begin{array}{c}
 {\bf U_{+}^{(2)}} \\
 {\bf  V_{-}^{(2)} } \\
\end{array} \right)
\mp \frac{2 (k-3)}{3 (k+5)^2}
A_3 B_{\mp}
\right.
\nonu \\
&+& \frac{2 i (k-3)}{3 (k+5)^3}
A_3 
\left(
\begin{array}{c}
 F_{11} F_{21} \\
 F_{22} F_{12} \\
\end{array} \right)
+ \frac{2 i}{(k+5)^2} B_3
\left(
\begin{array}{c}
 F_{11} F_{21} \\
 F_{22} F_{12} \\
\end{array} \right)
\nonu \\
&\pm & \frac{2 i (k+9)}{3 (k+5)^3} 
B_{\mp} F_{12} F_{21}
\pm  \frac{(k+1)}{(k+5)^2} 
\left(
\begin{array}{c}
 F_{11} G_{21} \\
 F_{22} G_{12} \\
\end{array} \right)
\nonu \\
& \pm & 
\frac{2 (k+21)}{3 (k+5)^3} 
\left(
\begin{array}{c}
 \pa F_{11} F_{21} \\
 \pa F_{22} F_{12} \\
\end{array} \right)
\pm \frac{2 (5 k+9)}{3 (k+5)^3}
\left(
\begin{array}{c}
 F_{11} \pa F_{21} \\
   F_{22} \pa F_{12} \\
\end{array} \right)
\nonu \\
& \mp & \frac{2 (k+2)}{(k+5)^2} 
\left(
\begin{array}{c}
 F_{21}  G_{11} \\
   F_{12}  G_{22} \\
\end{array} \right)
+ \frac{2 i k}{(k+5)^2} 
U B_{\mp}
\nonu \\
&\pm & \left.
\frac{2}{(k+5)^2} U
\left(
\begin{array}{c}
 F_{11}  F_{21} \\
 F_{22}  F_{12} \\
\end{array} \right)
\mp \frac{4 i(k+3)}{3(k+5)^3} B_{\mp} F_{11} F_{22}
\right](w) \nonu \\
&+ &\frac{1}{(z-w)}  \,  \left[ 
\frac{1}{2} 
\left(
\begin{array}{c}
{\bf Q_{+}^{(3)}}   \nonu \\
{\bf R_{-}^{(3)}}
\end{array}
\right)
+ \frac{1}{4} \pa (\mbox{pole-2})
%\frac{2 (k+3)}{3 (k+5)} \pa
%\left(
%\begin{array}{c}
%{\bf U_{+}^{(2)}}   \nonu \\
%{\bf V_{-}^{(2)}}
%\end{array}
%\right)
%\mp \frac{k-3}{6 (k+5)^2} \pa (A_3
%B_{\mp})
\right. \nonu \\
%&+ &
%\frac{i (k-3)}{6 (k+5)^3} \pa  
%\left(
%\begin{array}{c}
%A_3 F_{11} F_{21}   \nonu \\
%A_3 F_{22} F_{12}
%\end{array}
%\right)
%+ \frac{i}{2 (k+5)^2} \pa 
%\left(
%\begin{array}{c}
%B_3 F_{11}  F_{21}   \nonu \\
%B_3  F_{22}  F_{12}
%\end{array}
%\right)
%\nonu \\
%& - & \frac{i k (4 k+21)}{2 (k+5) (13 k+17)} \pa^2 
%B_{\mp} 
%\mp \frac{i (k+3)}{3 (k+5)^3} \pa (B_{\mp} F_{11} F_{22})
%\nonu \\
%&\pm & \frac{i (k+9)}{6 (k+5)^3} \pa (B_{\mp} F_{12} F_{21})
%\pm \frac{24 k^3+205 k^2+650 k+357}{6 (k+5)^3 (13 k+17)} 
%\left(
%\begin{array}{c}
%\pa^2 F_{11}  F_{21}   \nonu \\
%\pa^2  F_{22}  F_{12}
%\end{array}
%\right)
%\nonu \\
%&\pm &
% \frac{8 k^2+37 k+17}{(k+5)^2 (13 k+17)} 
%\left(
%\begin{array}{c}
%\pa F_{11}  \pa F_{21}   \nonu \\
%\pa  F_{22}  \pa F_{12}
%\end{array}
%\right)
%\nonu \\
%& \pm & 
%\frac{24 k^3+257 k^2+562 k+153}{6 (k+5)^3 (13 k+17)}
%\left(
%\begin{array}{c}
% F_{11}  \pa^2 F_{21}   \nonu \\
%  F_{22}  \pa^2 F_{12}
%\end{array}
%\right)
%\pm
%\frac{k+1}{4 (k+5)^2}
%\pa \left(
%\begin{array}{c}
% F_{11}   G_{21}   \nonu \\
%  F_{22}  G_{12}
%\end{array}
%\right)
%\nonu \\
%&\mp&  \frac{k+2}{2 (k+5)^2}
%\pa
% \left(
%\begin{array}{c}
% F_{21}   G_{11}   \nonu \\
%  F_{12}  G_{22}
%\end{array}
%\right)
& + &  \frac{i k (4 k+21)}{(k+5) (13 k+17)} \left( T B_{\mp}
-\frac{1}{2} \pa^2 B_{\mp} \right)
\nonu \\
&\mp& \left.
\frac{8 k (k+3)}{(k+5)^2 (13 k+17)}
\left( T
\left(
\begin{array}{c}
 F_{11}   F_{21}   \nonu \\
  F_{22}  F_{12}
\end{array}
\right) -\frac{1}{2} \pa^2 
\left(
\begin{array}{c}
 F_{11}   F_{21}   \nonu \\
  F_{22}  F_{12}
\end{array}
\right) \right)
%+ \frac{i k}{2 (k+5)^2} \pa (U B_{\mp})
%\nonu \\
%&\pm & \left.
%\frac{1}{2 (k+5)^2}
%\pa
%\left(
%\begin{array}{c}
% U F_{11}   F_{21}   \nonu \\
% U  F_{22}  F_{12}
%\end{array}
%\right)
\right](w)
+\cdots.
\nonu 
\eea
One can easily see that the nonlinear terms (between the higher spin currents
and the $16$ currents from the large ${\cal N}=4$ linear superconformal 
algebra or in the spirit of the footnote \ref{nonlineardef})
appearing in the corresponding 
OPEs in the nonlinear version  disappear in the above OPEs.
Even the above OPEs have simple relations between the structure constants
and so one can write down them together in one single equation as above.
The last two expressions in the 
first order pole contain the quasi primary fields.
The first order pole of the above OPEs is rather simple 
because the many composite nonderivative terms which were present in the 
corresponding OPEs in the nonlinear version disappear.

One also has the following OPEs
\bea
%%%%%%%%%%%%%%%%%%%%%%%%%%%%%%%%%%%%%%%%%%%%%%%%%%%%%%%%%%%%%%%%%%%%%%%%
{\bf T_{\pm}^{(\frac{3}{2})}}(z) \, 
\left(
\begin{array}{c}
{\bf V^{(\frac{5}{2})}} \\
{\bf U^{(\frac{5}{2})}} 
\end{array}
\right)
(w) & = & \frac{1}{(z-w)^3} \, 
\left[ \frac{2 i (5 k+18)}{(k+5)^2} {A}_{\mp}
+ \frac{8 (k+9)}{(k+5)^3} 
\left(
\begin{array}{c}
 F_{21}   F_{22}   \nonu \\
  F_{11}  F_{12}
\end{array}
\right)
 \right](w)
\nonu \\
& + & 
\frac{1}{(z-w)^2} \, \left[ \frac{4 (k+9)}{3 (k+5)}
\left(
\begin{array}{c}
 {\bf V_{+}^{(2)}} \\
{\bf U_{-}^{(2)}}
\end{array}
\right)
+
\frac{2 i}{(k+5)^2}
A_3
\left(
\begin{array}{c}
 F_{22}   F_{21}   \nonu \\
  F_{11}  F_{12}
\end{array}
\right)
\right. \nonu \\
&\pm & \frac{2 (k-3)}{3 (k+5)^2}
A_{\mp} B_3
\pm \frac{2 i (k+9)}{3 (k+5)^3} A_{\mp} F_{11} F_{22}
\pm \frac{4 i (k+3)}{3 (k+5)^3} A_{\mp} F_{12} F_{21}
\nonu \\
&-& \frac{2 i (k-3)}{3 (k+5)^3} B_3 
\left(
\begin{array}{c}
 F_{22}   F_{21}   \nonu \\
  F_{11}  F_{12}
\end{array}
\right)
\mp \frac{4}{(k+5)^2} 
\left(
\begin{array}{c}
 F_{22}   G_{21}   \nonu \\
  F_{11}  G_{12}
\end{array}
\right)
\nonu \\
& \pm & \frac{2 (5 k+9)}{3 (k+5)^3} 
\left(
\begin{array}{c}
 \pa F_{22}   F_{21}   \nonu \\
  \pa F_{11}  F_{12}
\end{array}
\right)
\pm \frac{(k+7)}{(k+5)^2}
\left(
\begin{array}{c}
 F_{21}   G_{22}   \nonu \\
  F_{12}  G_{11}
\end{array}
\right)
\nonu \\
&- & \left. \frac{6 i}{(k+5)^2} U A_{\mp}
\mp \frac{2}{(k+5)^2} U
  \left(
\begin{array}{c}
 F_{22}   F_{21}   \nonu \\
  F_{11}  F_{12}
\end{array}
\right)
\pm \frac{2(k+21)}{3(k+5)^3} 
\left(
\begin{array}{c}
 F_{22} \pa   F_{21}   \nonu \\
  F_{11} \pa  F_{12}
\end{array}
\right)
\right](w) \nonu \\
& + & \frac{1}{(z-w)} \, \left[ \frac{1}{2} 
\left(
\begin{array}{c}
{\bf R_{+}^{(3)}} \\
  {\bf Q_{-}^{(3)}}
\end{array}
\right) +\frac{1}{4} \pa (\mbox{pole-2}) 
\right. \nonu \\
&+& \frac{3 i (5 k+18)}{(k+5) (13 k+17)} \left( T A_{\mp} -\frac{1}{2} \pa^2 
A_{\mp}\right)
\nonu \\
&\pm & \left.
\frac{12 (k+9)}{(k+5)^2 (13 k+17)} \left(
 T
\left(
\begin{array}{c}
 F_{21}   F_{22}   \nonu \\
  F_{12}  F_{11}
\end{array}
\right) -\frac{1}{2} \pa^2 
\left(
\begin{array}{c}
 F_{21}   F_{22}   \nonu \\
  F_{12}  F_{11}
\end{array}
\right)
 \right)  
\right](w)
+\cdots.
\nonu 
\eea
There are no nonlinear terms (footnote \ref{nonlineardef}) which appeared in the nonlinear version. The structure constants of above  OPEs behave simply 
and so one can write them together. The last two expressions in the 
first order pole contain the quasi primary fields.
Note that for the spin-$\frac{1}{2}$ currents dependent terms, the numerical 
factor in the derivative term is given by the same value $\frac{1}{2}$. 

The relation (\ref{g1122vu3half}) gives the higher spin-$2$ current and
one obtains the following OPEs
%%%%%%%%%%%%%%%%%%%%%%%%%%%%%%%%%%%%%%%%%%%%%%%%%%%%%%%%%%%%%%%%%%%%%%%
\bea
{\bf T_{\pm}^{(\frac{3}{2})}}(z) \, 
{\bf W^{(2)}}(w) & = & 
\frac{1}{(z-w)^3} \,
\left[ \pm \frac{6 k}{(k+5)^2}
\left(
\begin{array}{c}
 F_{21}      \nonu \\
  F_{12} 
\end{array}
\right)
 \right](w) \nonu \\
&+& 
\frac{1}{(z-w)^2} \, 
\left[ -\frac{(k-3)}{2 (k+5)}
{\bf T_{\pm}^{(\frac{3}{2})}} 
\pm \frac{(k-3)}{2 (k+5)^2} U
\left(
\begin{array}{c}
 F_{21}      \nonu \\
  F_{12} 
\end{array}
\right)
\right. \nonu \\
&-& \frac{i (k+9)}{2 (k+5)^2} 
\left(
\begin{array}{c}
 F_{21}      \nonu \\
  F_{12} 
\end{array}
\right) A_3 + \frac{i (k-3)}{2 (k+5)^2} 
\left(
\begin{array}{c}
 F_{11}      \nonu \\
  F_{22} 
\end{array}
\right) A_{\mp}
\nonu \\
&- & \left. \frac{i (k-3)}{2 (k+5)^2} 
 \left(
\begin{array}{c}
 F_{22}      \nonu \\
  F_{11} 
\end{array}
\right) B_{\mp}
-\frac{3 i (k+1)}{2 (k+5)^2} 
 \left(
\begin{array}{c}
 F_{21}      \nonu \\
  F_{12} 
\end{array}
\right) B_3
\right] (w) \nonu \\
& + & 
\frac{1}{(z-w)} \, \left[  -\frac{1}{2} 
{\bf P_{\pm}^{(\frac{5}{2})}} \mp
{\bf W_{\pm}^{(\frac{5}{2})}}  
-\frac{(k-3)}{6 (k+5)}  \pa {\bf T_{\pm}^{(\frac{3}{2})}}
\right. \nonu \\
& -& \frac{i}{2 (k+5)}
A_3 
 \left(
\begin{array}{c}
 G_{21}      \nonu \\
  G_{12} 
\end{array}
\right)
\mp \frac{1}{(k+5)^2} A_3 B_{\mp}
 \left(
\begin{array}{c}
 F_{22}      \nonu \\
  F_{11} 
\end{array}
\right)
-\frac{i (k+9)}{6 (k+5)^2} \pa A_3 
 \left(
\begin{array}{c}
 F_{21}      \nonu \\
  F_{12} 
\end{array}
\right)
\nonu \\
& - & \frac{i (k-3)}{6 (k+5)^2} A_3 \pa
 \left(
\begin{array}{c}
 F_{21}      \nonu \\
  F_{12} 
\end{array}
\right)
\pm \frac{1}{(k+5)^2} A_{\mp} A_3
 \left(
\begin{array}{c}
 F_{11}      \nonu \\
  F_{22} 
\end{array}
\right)
\nonu \\
& \mp & \frac{1}{(k+5)^2} A_{\mp} B_3 
 \left(
\begin{array}{c}
 F_{11}      \nonu \\
  F_{22} 
\end{array}
\right) \mp \frac{2}{(k+5)^2} A_{\mp} B_{\mp} 
 \left(
\begin{array}{c}
 F_{12}      \nonu \\
  F_{21} 
\end{array}
\right)
\nonu \\
&+& \frac{i (k+3)}{6 (k+5)^2} \pa A_{\mp}
 \left(
\begin{array}{c}
 F_{11}      \nonu \\
  F_{22} 
\end{array}
\right) +
\frac{i (k+15)}{6 (k+5)^2}
A_{\mp} \pa
 \left(
\begin{array}{c}
 F_{11}      \nonu \\
  F_{22} 
\end{array}
\right)
\nonu \\
&\pm & \frac{1}{(k+5)^2} A_{\pm} A_{\mp} 
 \left(
\begin{array}{c}
 F_{21}      \nonu \\
  F_{12} 
\end{array}
\right) + \frac{i}{2 (k+5)} B_3
 \left(
\begin{array}{c}
 G_{21}      \nonu \\
  G_{12} 
\end{array}
\right)
\nonu \\
&-& \frac{i (5 k-3)}{6 (k+5)^2} \pa B_3 
\left(
\begin{array}{c}
 F_{21}      \nonu \\
  F_{12} 
\end{array}
\right) + \frac{i (k-3)}{6 (k+5)^2}
B_3 \pa 
\left(
\begin{array}{c}
 F_{21}      \nonu \\
  F_{12} 
\end{array}
\right)
\nonu \\
&\pm & \frac{1}{(k+5)^2} B_{\mp} B_3 
\left(
\begin{array}{c}
 F_{22}      \nonu \\
  F_{11} 
\end{array}
\right)
-\frac{i (k-9)}{6 (k+5)^2} \pa B_{\mp} 
\left(
\begin{array}{c}
 F_{22}      \nonu \\
  F_{11} 
\end{array}
\right)
\nonu \\
&- & 
\frac{i (k-21)}{6 (k+5)^2} B_{\mp} \pa 
\left(
\begin{array}{c}
 F_{22}      \nonu \\
  F_{11} 
\end{array}
\right) \pm \frac{1}{(k+5)^2} B_{\pm} B_{\mp} 
\left(
\begin{array}{c}
 F_{21}      \nonu \\
  F_{12} 
\end{array}
\right)
\nonu \\
&\mp&  \frac{1}{(k+5)^2} 
\left(
\begin{array}{c}
 F_{11} F_{21} G_{22}      \nonu \\
  F_{22} F_{12} G_{11} 
\end{array}
\right)
\mp \frac{4 (k+3)}{3 (k+5)^3} 
\left(
\begin{array}{c}
 \pa F_{11} F_{21} F_{22}      \nonu \\
  \pa F_{22} F_{12} F_{11} 
\end{array}
\right)
\nonu \\
&- & \frac{2(k-3)}{3 (k+5)^3}
\pa
\left(
\begin{array}{c}
  F_{21}       \nonu \\
  F_{12}  
\end{array}
\right) F_{11} F_{22}
\pm \frac{2 (k+9)}{3 (k+5)^3} 
\left(
\begin{array}{c}
  F_{11} F_{21} \pa F_{22}       \nonu \\
  F_{22} F_{12} \pa F_{11}  
\end{array}
\right)
\nonu \\
&\mp & \frac{(k+6)}{(k+5)^2} \pa^2
\left(
\begin{array}{c}
  F_{21}        \nonu \\
  F_{12}  
\end{array}
\right) \mp
\frac{1}{(k+5)^2} 
\left(
\begin{array}{c}
  F_{21} F_{22} G_{11}        \nonu \\
  F_{12}  F_{11} G_{22} 
\end{array}
\right)
\nonu \\
& \pm & \frac{1}{(k+5)} T 
\left(
\begin{array}{c}
  F_{21}         \nonu \\
  F_{12}  
\end{array}
\right) \pm 
\frac{1}{2 (k+5)} U 
\left(
\begin{array}{c}
  G_{21}         \nonu \\
  G_{12}  
\end{array}
\right)
-\frac{2 i}{(k+5)^2} U A_3 
\left(
\begin{array}{c}
  F_{21}         \nonu \\
  F_{12}  
\end{array}
\right)
\nonu \\
&+& \frac{i}{(k+5)^2} U A_{\mp} 
\left(
\begin{array}{c}
  F_{11}         \nonu \\
  F_{22}  
\end{array}
\right) 
+ \frac{2 i}{(k+5)^2} U B_3 
\left(
\begin{array}{c}
  F_{21}         \nonu \\
  F_{12}  
\end{array}
\right) 
\nonu \\
&-& \frac{i}{(k+5)^2} U B_{\mp}
\left(
\begin{array}{c}
  F_{22}         \nonu \\
  F_{11}  
\end{array}
\right) \pm  \frac{(k-3)}{6 (k+5)^2} \pa 
\left(
\begin{array}{c}
  U F_{21}         \nonu \\
  U F_{12}  
\end{array}
\right)
\nonu \\
&\pm & \left. \frac{2}{(k+5)^2} U U 
 \left(
\begin{array}{c}
 F_{21}         \nonu \\
 F_{12}  
\end{array}
\right) + 
\frac{2}{(k+5)^2} \pa
\left(
\begin{array}{c}
 F_{21}         \nonu \\
 F_{12}  
\end{array}
\right) F_{12} F_{21}
\right](w)
+  \cdots.
\nonu 
\eea
These OPEs are rather complicated due to the presence of 
spin-$\frac{1}{2}$ currents.
Some terms containing the spin-$\frac{3}{2}$ currents in the above
OPEs do not appear in the nonlinear version.

From the relation (\ref{g1221w2}), the following OPEs can be obtained
\bea
%%%%%%%%%%%%%%%%%%%%%%%%%%%%%%%%%%%%%%%%%%%%%%%%%%%%%%%%%%%%%%%%%%%%%%%
{\bf T_{\pm}^{(\frac{3}{2})}}(z) \, 
{\bf W_{\pm}^{(\frac{5}{2})}}(w) & = & 
\frac{1}{(z-w)^2} \, \left[ \mp  \frac{2 (k-3)}{3 (k+5)^2}
 {A}_{\mp} {B}_{\mp}  +\frac{2 i (k-3)}{3 (k+5)^3}
A_{\mp}
\left(
\begin{array}{c}
 F_{11} F_{21}         \nonu \\
 F_{22} F_{12}  
\end{array}
\right)
\right. \nonu \\
&-& \frac{2 i (k-3)}{3 (k+5)^3} B_{\mp}
\left(
\begin{array}{c}
 F_{21} F_{22}         \nonu \\
 F_{12} F_{11}  
\end{array}
\right) \mp \frac{(k+3)}{(k+5)^2} 
 \left(
\begin{array}{c}
 F_{21} G_{21}         \nonu \\
 F_{12} G_{12}  
\end{array}
\right)
\nonu \\
&\pm &\left. \frac{8 (k-3)}{3 (k+5)^3}
 \left(
\begin{array}{c}
 F_{21} \pa F_{21}         \nonu \\
 F_{12} \pa F_{12}  
\end{array}
\right)
\right](w)
\nonu \\
&+ &\frac{1}{(z-w)} \, \left[ 
 \pm \frac{(4 k+21)}{3 (k+5)^2}  \pa A_{\mp} B_{\mp} 
\mp \frac{(5 k+18)}{3 (k+5)^2} A_{\mp} \pa B_{\mp}
\right. \nonu \\
& - & \frac{i}{(k+5)^2} A_{\mp} 
 \left(
\begin{array}{c}
 F_{11} G_{21}         \nonu \\
 F_{22} G_{12}  
\end{array}
\right)
+ \frac{8 i (k+3)}{3 (k+5)^3} \pa A_{\mp}
 \left(
\begin{array}{c}
 F_{11} F_{21}         \nonu \\
 F_{22} F_{12}  
\end{array}
\right)
 \nonu \\
&-&  \frac{2 i (5 k+21)}{3 (k+5)^3} A_{\mp} 
 \left(
\begin{array}{c}
 \pa F_{11} F_{21}         \nonu \\
 \pa F_{22} F_{12}  
\end{array}
\right)
-\frac{4 i (k+3)}{3 (k+5)^3} A_{\mp}
 \left(
\begin{array}{c}
 F_{11} \pa F_{21}         \nonu \\
 F_{22} \pa F_{12}  
\end{array}
\right)
\nonu \\
&+& \frac{i}{(k+5)^2} A_{\mp} 
\left(
\begin{array}{c}
 F_{21} G_{11}         \nonu \\
 F_{12} G_{22}  
\end{array}
\right)
+ \frac{i}{(k+5)^2} B_{\mp} 
\left(
\begin{array}{c}
 F_{21} G_{22}         \nonu \\
 F_{12} G_{11}  
\end{array}
\right)
\nonu \\
&+& \frac{4 i (k+9)}{3 (k+5)^3} 
\pa B_{\mp}
\left(
\begin{array}{c}
 F_{21} F_{22}         \nonu \\
 F_{12} F_{11}  
\end{array}
\right)
-\frac{2 i (k+9)}{3 (k+5)^3} B_{\mp}
\left(
\begin{array}{c}
 \pa F_{21} F_{22}         \nonu \\
 \pa F_{12} F_{11}  
\end{array}
\right)
\nonu \\
&- & \frac{8 i (k+6)}{3 (k+5)^3} B_{\mp} 
\left(
\begin{array}{c}
  F_{21} \pa F_{22}         \nonu \\
  F_{12} \pa F_{11}  
\end{array}
\right)
-\frac{i}{(k+5)^2} B_{\mp} 
\left(
\begin{array}{c}
  F_{22} G_{21}         \nonu \\
  F_{11} G_{12}  
\end{array}
\right)
\nonu \\
&\mp & \frac{(k+3)}{(k+5)^2} 
\left(
\begin{array}{c}
  \pa F_{21} G_{21}         \nonu \\
  \pa F_{12} G_{12}  
\end{array}
\right)
\pm \frac{2}{(k+5)^2} U A_{\mp} B_{\mp}
\pm \frac{2 (k-3)}{3 (k+5)^3} 
\left(
\begin{array}{c}
  F_{21} \pa F_{21}         \nonu \\
  F_{12} \pa F_{12}  
\end{array}
\right)
\nonu \\
&\pm & \frac{2}{(k+5)^2} 
\left(
\begin{array}{c}
  F_{21} \pa F_{21}         \nonu \\
  F_{12} \pa F_{12}  
\end{array}
\right) U
-\frac{2 i}{(k+5)^2} 
\left(
\begin{array}{c}
  F_{21} \pa F_{21}         \nonu \\
  F_{12} \pa F_{12}  
\end{array}
\right) A_3
\nonu \\
& + & \left.
 \frac{2 i}{(k+5)^2} 
\left(
\begin{array}{c}
  F_{21} \pa F_{21}         \nonu \\
  F_{12} \pa F_{12}  
\end{array}
\right) B_3
\right](w) +\cdots.
\nonu 
\eea
The nonlinear terms which appeared in the nonlinear version 
disappear in these OPEs.

Moreover, the following OPE can be obtained
\bea
%%%%%%%%%%%%%%%%%%%%%%%%%%%%%%%%%%%%%%%%%%%%%%%%%%%%%%%%%%%%%%%%%%%%%%%%%%
{\bf T_{+}^{(\frac{3}{2})}}(z) \, 
{\bf W_{-}^{(\frac{5}{2})}}(w) & = & 
\frac{1}{(z-w)^3} \, \left[  
 -\frac{4 {\bf (k-3)}}{3 (k+5)} {\bf T^{(1)}}
-\frac{2 i (5 k+18)}{(k+5)^2} A_3 
+\frac{2 i k (4 k+21)}{3 (k+5)^2} B_3
\right. \nonu \\
&- & \left.
\frac{6 k}{(k+5)^2} U
-\frac{4 \left(2 k^2+9 k+27\right)}{3 (k+5)^3} F_{11} F_{22}
+ \frac{4 (k-3) (2 k+9)}{3 (k+5)^3} F_{12} F_{21}
\right](w)
\nonu \\
& + & \frac{1}{(z-w)^2} 
\, \left[ 
-\frac{4 (k-3)}{3 (k+5)} {\bf T^{(2)}} 
-4 {\bf W^{(2)}}
+ \frac{1}{2} {\bf P^{(2)}}
+ \frac{8 i (k+9)}{3 (k+5)^3} A_{-} F_{11} F_{12}
\right.
\nonu \\
& - & \frac{8 i (k+9)}{3 (k+5)^3} A_{+} F_{21} F_{22}   
+ \frac{16 i (k+3)}{3 (k+5)^3} B_{-} F_{12} F_{22}
-\frac{16 i (k+3)}{3 (k+5)^3} B_{+} F_{11} F_{21}
\nonu \\
&+& \frac{2 (k-3)}{3 (k+5)^2} F_{11} G_{22}
+ \frac{8}{(k+5)^2} \pa (F_{11} F_{22})
-\frac{2}{(k+5)} F_{12} G_{21}
\nonu \\
&-& \frac{8 (k-3)}{3 (k+5)^3} \pa (F_{12} F_{21})
+ \frac{2}{(k+5)} F_{21} G_{12}
-\frac{2 (k-3)}{3 (k+5)^2} F_{22} G_{11}
\nonu \\
& + & \left.  
\frac{8 (k-3)}{3 (k+5)^3} U F_{11} F_{22}
-\frac{8}{(k+5)^2} U F_{12} F_{21}
\right](w)
\nonu \\
& + &  
\frac{1}{(z-w)} \, \left[ 
 \frac{1}{2} {\bf P^{(3)}} +\frac{1}{8} 
{\bf \pa P^{(2)} } - \pa {\bf W^{(2)}}
-{\bf W^{(3)}} -\frac{(k-3)}{3 (k+5)} \pa {\bf T^{(2)}}
\right. \nonu \\
& - &
\frac{2 {\bf (k-3)}}{(13 k+17)} \left( T {\bf T^{(1)}} -\frac{1}{2} 
 \pa^2 {\bf T^{(1)}} \right)
\nonu \\
& + &  \frac{3 i (5 k+18)}{2 (k+5) (13 k+17)} \pa^2 A_3
-\frac{2 i}{(k+5)^2} A_3 \pa (F_{11} F_{22})
\nonu \\
& -& \frac{2 i}{(k+5)^2} A_3 \pa (F_{12} F_{21})
+ \frac{8 i (k+6)}{3 (k+5)^3} \pa A_{-} F_{11} F_{12}
-\frac{4 i (k+3)}{3 (k+5)^3} A_{-} \pa (F_{11} F_{12})
\nonu \\
&+& \frac{4 i (k+3)}{3 (k+5)^3} \pa A_{+} F_{21} F_{22}
-\frac{8 i (k+6)}{3 (k+5)^3} A_{+} \pa (F_{21} F_{22})
-\frac{i k (4 k+21)}{2 (k+5) (13 k+17)} \pa^2 B_3
\nonu \\
&-& \frac{2 i}{(k+5)^2} B_3 \pa (F_{11} F_{22})
+ \frac{2 i}{(k+5)^2} B_3 \pa (F_{12} F_{21})
+ \frac{2 i (5 k+21)}{3 (k+5)^3} \pa B_{-} F_{12} F_{22}
\nonu \\
&-& \frac{2 i (k+9)}{3 (k+5)^3} B_{-} \pa (F_{12} F_{22})
+ \frac{2 i (k+9)}{3 (k+5)^3} \pa B_{+} F_{11} F_{21}
-\frac{2 i (5 k+21)}{3 (k+5)^3} B_{+} \pa (F_{11} F_{21})
\nonu \\
&+  & \frac{(2 k^2+35 k+61)}{(k+5)^2 (13 k+17)} \pa^2 (F_{11} F_{22})
+ \frac{(5 k+21)}{3 (k+5)^2} \pa F_{11} G_{22}
-\frac{(k+9)}{3 (k+5)^2} F_{11} \pa G_{22}
\nonu \\
&-& \frac{(k-3) \left(6 k^2+83 k+169\right)}{3 (k+5)^3 (13 k+17)} 
\pa^2 (F_{12} F_{21}) + \frac{1}{(k+5)} \pa F_{12} G_{21}
-\frac{1}{(k+5)} F_{12} \pa G_{21} 
\nonu \\
&+& \frac{2}{(k+5)} \pa F_{21} G_{12}
+ \frac{4 (k+6)}{3 (k+5)^2} \pa F_{22} G_{11}
-\frac{2 (k+3)}{3 (k+5)^2} F_{22} \pa G_{11}
\nonu \\
&-& \frac{3 i (5 k+18)}{(k+5) (13 k+17)} T A_3
+ \frac{i k (4 k+21)}{(k+5) (13 k+17)} T B_3
-\frac{9 k}{(k+5) (13 k+17)} T U
\nonu \\
&-& \frac{2 \left(2 k^2+9 k+27\right)}{(k+5)^2 (13 k+17)}
T F_{11} F_{22}
+ \frac{2 (k-3) (2 k+9)}{(k+5)^2 (13 k+17)} T F_{12} F_{21}
\nonu \\
&+& \frac{9 k}{2 (k+5) (13 k+17)} \pa^2 U 
+ \frac{8 (k+3)}{3 (k+5)^3} \pa U F_{11} F_{22}
+ \frac{2 (7 k+27)}{3 (k+5)^3} U \pa F_{11} F_{22}
\nonu \\
&-& \frac{2 (5 k+33)}{3 (k+5)^3} U F_{11} \pa F_{22}
-\frac{4}{(k+5)^2} \pa U F_{12} F_{21}
+ \frac{2}{(k+5)^2} U \pa F_{12} F_{21}
\nonu \\
&- & \left.  \frac{6}{(k+5)^2} U F_{12} \pa F_{21}
 -\frac{2}{(k+5)^2} F_{11} G_{22} F_{11} F_{22}
-\frac{2}{(k+5)^2} F_{21} G_{12} F_{21} F_{12}
 \right](w) \nonu \\
& + & \cdots.
\nonu 
\eea
Some nonlinear terms in the second order pole in the nonlinear 
version disappear in the above OPE. 
Note that there exist the nonlinear terms with coefficient $(k-3)$ 
factor containing the higher spin-$1$
current in the first order pole.
This is the only nonlinear term between the higher spin currents 
and the currents from the large ${\cal N}=4$ linear superconformal algebra.
We will see this feature in other OPEs in this Appendix.

Similarly, one has
\bea
%%%%%%%%%%%%%%%%%%%%%%%%%%%%%%%%%%%%%%%%%%%%%%%%%%%%%%%%%%%%%%%%%%%%%%%%%%%%
{\bf T_{-}^{(\frac{3}{2})}}(z) \,
 {\bf W_{+}^{(\frac{5}{2})}}(w) & = &
\frac{1}{(z-w)^3} \, \left[ 
 -\frac{4 {\bf (k-3)}}{3 (k+5)} {\bf T^{(1)}}
-\frac{2 i (5 k+18)}{(k+5)^2} A_3 
+\frac{2 i k (4 k+21)}{3 (k+5)^2} B_3
\right. \nonu \\
&+ & \left.
\frac{6 k}{(k+5)^2} U
-\frac{4 \left(2 k^2+9 k+27\right)}{3 (k+5)^3} F_{11} F_{22}
+ \frac{4 (k-3) (2 k+9)}{3 (k+5)^3} F_{12} F_{21}
 \right](w)
\nonu \\
& + & \frac{1}{(z-w)^2}
\left[
\frac{4 (k-3)}{3 (k+5)} {\bf T^{(2)}} 
+4 {\bf W^{(2)}}
- \frac{1}{2} {\bf P^{(2)}}
-\frac{2 i (k+21)}{3 (k+5)^3} A_{-} F_{11} F_{12}
\right.
\nonu \\
& + &  \frac{2 i (k+21)}{3 (k+5)^3}  A_{+} F_{21} F_{22}   
-\frac{2 i (5 k+9)}{3 (k+5)^3} B_{-} F_{12} F_{22}
+ \frac{2 i (5 k+9)}{3 (k+5)^3}  B_{+} F_{11} F_{21}
\nonu \\
&+& \frac{(k-3)}{3 (k+5)^2}  F_{11} G_{22}
 -\frac{4}{(k+5)^2} \pa (F_{11} F_{22})
 + \frac{2}{(k+5)^2}F_{12} G_{21}
\nonu \\
&+& \frac{8 (k-3)}{3 (k+5)^3}  \pa (F_{12} F_{21})
 -\frac{2}{(k+5)^2} F_{21} G_{12}
-\frac{(k-3)}{3 (k+5)^2}  F_{22} G_{11}
\nonu \\
& - & \left.  
\frac{8 (k-3)}{3 (k+5)^3}  U F_{11} F_{22}
+ \frac{4}{(k+5)^2} U F_{12} F_{21}
+ \frac{12 i}{(k+5)^2} U A_3 + \frac{4 i k}{(k+5)^2} U B_3
\right](w) \nonu \\
& + & \frac{1}{(z-w)} \left[ \frac{1}{2} {\bf P^{(3)}} -\frac{1}{8} 
{\bf \pa P^{(2)} } + \pa {\bf W^{(2)}}
-{\bf W^{(3)}} +\frac{(k-3)}{3 (k+5)} \pa {\bf T^{(2)}}
\right. \nonu \\
& - & \frac{2 {\bf (k-3)}}{(13 k+17)} \left( T {\bf T^{(1)}} -
\frac{1}{2} \pa^2 {\bf T^{(1)}} \right) 
- \frac{(3 k+13)}{(k+5)^2} \pa A_3 B_3 + 
\frac{(3 k+13)}{(k+5)^2} A_3 \pa B_3
\nonu \\
& + &  \frac{3 i (5 k+18)}{2 (k+5) (13 k+17)} \pa^2 A_3
+ \frac{2 i}{(k+5)^2} A_3 F_{11} G_{22}
-\frac{2 i}{(k+5)^2} A_3 F_{22} G_{11}
\nonu \\
& - & \frac{2 i (k-3)}{(k+5)^3} \pa A_3 F_{11} F_{22}
+ \frac{2 i (k+1)}{(k+5)^3} A_3 \pa (F_{11} F_{22})
+ \frac{i}{(k+5)^2} A_{-} F_{11} G_{12}
\nonu \\
& -& \frac{3 i}{(k+5)^2} A_3 \pa (F_{12} F_{21})
+ \frac{ i (k+9)}{3 (k+5)^3} \pa A_{-} F_{11} F_{12}
-\frac{5 i (k+9)}{3 (k+5)^3} A_{-} \pa F_{11} F_{12}
\nonu \\
&+& \frac{i (k-15)}{3 (k+5)^3} A_{-} F_{11} \pa F_{12}
-\frac{i}{(k+5)^2} A_{-} F_{12} G_{11}
+ \frac{i}{(k+5)^2} A_{+} F_{21} G_{22}
\nonu \\
&+& \frac{2 i (k+15)}{3 (k+5)^3} \pa A_{+} F_{21} F_{22}
+\frac{2 i (k+3)}{3 (k+5)^3} A_{+} \pa F_{21} F_{22}
\nonu \\
&-& \frac{4 i (k+6)}{3 (k+5)^3} A_{+} F_{21} \pa F_{22}
-\frac{i}{(k+5)^2} A_{+} F_{22} G_{21}
-\frac{i k (4 k+21)}{2 (k+5) (13 k+17)} \pa^2 B_3
\nonu \\
&-& \frac{2 i}{(k+5)^2} B_3 F_{11} G_{22}
+ \frac{2 i}{(k+5)^2} B_3 F_{22} G_{11}
-\frac{i}{(k+5)^2} B_{-} F_{12} G_{22}
\nonu \\
&+& \frac{8 i}{(k+5)^2} B_3 \pa (F_{11} F_{22})
+ \frac{3 i}{(k+5)^2} B_3 \pa (F_{12} F_{21})
+ \frac{2 i (k+3)}{3 (k+5)^3} \pa B_{-} F_{12} F_{22}
\nonu \\
&-& \frac{4 i k}{3 (k+5)^3} B_{-} \pa F_{12} F_{22}
-\frac{10 i (k+3)}{3 (k+5)^3} B_{-} F_{12} \pa F_{22}
+ \frac{i}{(k+5)^2} B_{-} F_{22} G_{12}
\nonu \\
&- & 
\frac{i}{(k+5)^2} B_{+} F_{11} G_{21}
+ \frac{ i (7k+15)}{3 (k+5)^3} \pa B_{+} F_{11} F_{21}
\nonu \\
& - & \frac{ i (5 k+21)}{3 (k+5)^3} B_{+} \pa F_{11} F_{21}
+ \frac{i (k+9)}{3 (k+5)^3} B_{+} F_{11} \pa F_{21}
+\frac{i}{(k+5)^2} B_{+} F_{21} G_{11}
\nonu \\
&+  & \frac{2 \left(k^2-2 k+5\right)}{(k+5)^2 (13 k+17)} \pa^2 (F_{11} F_{22})
+ \frac{(5 k+3)}{6 (k+5)^2}  \pa F_{11} G_{22}
-\frac{(k+3)}{6 (k+5)^2} F_{11} \pa G_{22}
\nonu \\
&-& \frac{(k-3) \left(6 k^2+31 k+101\right)}{3 (k+5)^3 (13 k+17)}
\pa^2 (F_{12} F_{21}) - \frac{1}{(k+5)^2} \pa F_{12} G_{21}
+\frac{1}{(k+5)^2} F_{12} \pa G_{21} 
\nonu \\
&-& \frac{2}{(k+5)^2} \pa F_{21} G_{12}
+ \frac{ (2k+3)}{3 (k+5)^2} \pa F_{22} G_{11}
-\frac{k}{3 (k+5)^2} F_{22} \pa G_{11}
\nonu \\
&-& \frac{3 i (5 k+18)}{(k+5) (13 k+17)} T A_3
+ \frac{i k (4 k+21)}{(k+5) (13 k+17)} T B_3
+\frac{17(k+2)}{(k+5) (13 k+17)} T U
\nonu \\
&-& \frac{2 \left(2 k^2+9 k+27\right)}{(k+5)^2 (13 k+17)}
T F_{11} F_{22}
+ \frac{2 (k-3) (2 k+9)}{(k+5)^2 (13 k+17)} T F_{12} F_{21}
\nonu \\
&+& \frac{2}{(k+5)^2} U A_3 A_3 
-\frac{4}{(k+5)^2} U A_3 B_3
+ \frac{3 i}{(k+5)^2} \pa U A_3
+ \frac{5 i}{(k+5)^2} U \pa A_3
\nonu \\
&+& \frac{2}{(k+5)^2} U A_{+} A_{-}
+ \frac{2}{(k+5)^2} U B_3 B_3 
+ \frac{i k}{(k+5)^2} \pa U B_3
+ \frac{i (k+2)}{(k+5)^2} U \pa B_3
\nonu \\
&+& 
\frac{2}{(k+5)^2} U B_{+} B_{-}
- \frac{17 (k+2)}{2 (k+5) (13 k+17)} \pa^2 U 
+ \frac{2 (5 k+33)}{3 (k+5)^3} U \pa F_{11} F_{22}
\nonu \\
&-& \frac{2 (7 k+27)}{3 (k+5)^3} U F_{11} \pa F_{22}
+ \frac{3}{(k+5)^2} U \pa F_{12} F_{21}
\nonu \\
&- &   \frac{1}{(k+5)^2} U F_{12} \pa F_{21}
+  \frac{2}{(k+5)^2} U U U 
+ \frac{(k+9)}{3 (k+5)^3} F_{22} G_{11} F_{22} F_{11}
\nonu \\
& - &  \frac{(k-15)}{3(k+5)^3} F_{11} G_{22} F_{11} F_{22}
-\frac{2}{(k+5)^2} F_{21} G_{12} F_{21} F_{12}
\nonu \\
& - & \left. \frac{1}{(k+5)^2} F_{12} G_{21} F_{12} F_{21} 
\right](w) +  \cdots.
\nonu 
\eea
One cannot combine this OPE and previous OPE because  
some terms of this OPE do not appear in the previous OPE.
There exist the nonlinear terms with coefficient $(k-3)$ 
factor containing the higher spin-$1$
current in the first order pole.
Some nonlinear terms in the second order pole in the nonlinear 
version disappear in the above OPE. 

From the relation (\ref{g1221w5half}), one calculates 
the following OPEs 
\bea
%%%%%%%%%%%%%%%%%%%%%%%%%%%%%%%%%%%%%%%%%%%%%%%%%%%%%%%%%%%%%%%%%%%%%%%%%%%%
{\bf T_{\pm}^{(\frac{3}{2})}}(z) \, 
{\bf W^{(3)}}(w) & = & 
\frac{1}{(z-w)^4} \, \left[ -\frac{18 k}{(k+5)^2} 
\left(
\begin{array}{c}
F_{21} \\
F_{12}
\end{array}
\right)
\right](w)
\nonu \\ 
&+& \frac{1}{(z-w)^3} \, \left[  \pm 
\frac{2 (k-3) (61 k+113)}{3 (k+5) (13 k+17)}
 {\bf T_{\pm}^{(\frac{3}{2})}} 
-\frac{2 (k-3) (61 k+113)}{3 (k+5)^2 (13 k+17)} 
U 
\left(
\begin{array}{c}
F_{21} \\
F_{12}
\end{array}
\right)
\right.
\nonu \\
&\pm &  \frac{4 (k-3) (5 k+73)}{3 (k+5)^3 (13 k+17)}
\left(
\begin{array}{c}
F_{21} \\
F_{12}
\end{array}
\right) F_{11} F_{22} \pm
\frac{2 i \left(61 k^2+164 k-33\right)}{3 (k+5)^2 (13 k+17)} 
\left(
\begin{array}{c}
F_{21} \\
F_{12}
\end{array}
\right) A_3
\nonu \\
&\mp & \frac{2 i \left(61 k^2+164 k-33\right)}{3 (k+5)^2 (13 k+17)}
\left(
\begin{array}{c}
F_{11} \\
F_{22}
\end{array}
\right) A_{\mp}
\mp \frac{2 i \left(17 k^2+172 k+339\right)}{3 (k+5)^2 (13 k+17)}
\left(
\begin{array}{c}
F_{22} \\
F_{11}
\end{array}
\right) B_{\mp}
\nonu \\
&\pm & \left.
 \frac{2 i \left(17 k^2+172 k+339\right)}{3 (k+5)^2 (13 k+17)}
\left(
\begin{array}{c}
F_{21} \\
F_{12}
\end{array}
\right) B_3
+ \frac{18 k}{(k+5)^2} \pa 
\left(
\begin{array}{c}
F_{21} \\
F_{12}
\end{array}
\right)
\right](w)
\nonu \\
& + & \frac{1}{(z-w)^2} \, \left[ \pm \frac{5}{2} 
{\bf P_{\pm}^{(\frac{5}{2})}} + 5 {\bf W_{\pm}^{(\frac{5}{2})}} 
\pm \frac{2 (k-3)}{(13 k+17)} \pa {\bf T_{\pm}^{(\frac{3}{2})}} 
-\frac{6 {\bf (k-3)}}{(13 k+17)} {\bf T^{(1)}}  {\bf T_{\pm}^{(\frac{3}{2})}} 
\right. \nonu \\
&\mp& \frac{10}{(k+5)^2} 
\pa \left(
\begin{array}{c}
F_{21} \\
F_{12}
\end{array}
\right) F_{12} F_{21}
\pm \frac{3 i}{2 (k+5)} A_3 
\left(
\begin{array}{c}
G_{21} \\
G_{12}
\end{array}
\right)
\nonu \\
&+& \frac{2}{(k+5)^2} A_3 A_3 
\left(
\begin{array}{c}
F_{21} \\
F_{12}
\end{array}
\right)
-\frac{4}{(k+5)^2} A_3 B_3 
\left(
\begin{array}{c}
F_{21} \\
F_{12}
\end{array}
\right)
\nonu \\
&+& \frac{4}{(k+5)^2} A_3 B_{\mp} 
\left(
\begin{array}{c}
F_{22} \\
F_{11}
\end{array}
\right)
\pm \frac{2 i \left(k^2+93 k+104\right)}{(k+5)^2 (13 k+17)}
\pa A_3 
\left(
\begin{array}{c}
F_{21} \\
F_{12}
\end{array}
\right)
\nonu \\
& \pm & \frac{2 i \left(k^2-128 k-185\right)}{(k+5)^2 (13 k+17)}
A_3 \pa 
\left(
\begin{array}{c}
F_{21} \\
F_{12}
\end{array}
\right)
\mp \frac{3 i}{2 (k+5)} A_{\mp} 
\left(
\begin{array}{c}
G_{11} \\
G_{22}
\end{array}
\right)
\nonu \\
&+& \frac{4}{(k+5)^2} A_{\mp} B_3 
\left(
\begin{array}{c}
F_{11} \\
F_{22}
\end{array}
\right)
+ \frac{4}{(k+5)^2} A_{\mp} B_{\mp}
\left(
\begin{array}{c}
F_{12} \\
F_{21}
\end{array}
\right)
\nonu \\
&\mp & \frac{2 i \left(k^2+80 k+87\right)}{(k+5)^2 (13 k+17)} 
\pa A_{\mp} 
\left(
\begin{array}{c}
F_{11} \\
F_{22}
\end{array}
\right) \mp
\frac{2 i \left(k^2-128 k-185\right)}{(k+5)^2 (13 k+17)} 
A_{\mp} \pa 
\left(
\begin{array}{c}
F_{11} \\
F_{22}
\end{array}
\right)
\nonu \\
&+& \frac{2}{(k+5)^2} A_{\pm} A_{\mp} 
\left(
\begin{array}{c}
F_{21} \\
F_{12}
\end{array}
\right)
\mp \frac{3 i}{2 (k+5)} B_3 
\left(
\begin{array}{c}
G_{21} \\
G_{12}
\end{array}
\right)
\nonu \\
&+& \frac{2}{(k+5)^2} B_3 B_3 
\left(
\begin{array}{c}
F_{21} \\
F_{12}
\end{array}
\right)
\pm \frac{i \left(37 k^2+112 k+115\right)}{(k+5)^2 (13 k+17)} 
\pa B_3 
\left(
\begin{array}{c}
F_{21} \\
F_{12}
\end{array}
\right)
\nonu \\
&\mp & \frac{4 i \left(20 k^2+33 k+1\right)}{(k+5)^2 (13 k+17)} 
B_3 \pa 
\left(
\begin{array}{c}
F_{21} \\
F_{12}
\end{array}
\right) \pm
\frac{3 i}{2 (k+5)} B_{\mp} 
\left(
\begin{array}{c}
G_{22} \\
G_{11}
\end{array}
\right)
\nonu \\
&\mp & \frac{i \left(37 k^2+86 k+81\right)}{(k+5)^2 (13 k+17)} 
\pa B_{\mp} 
\left(
\begin{array}{c}
F_{22} \\
F_{11}
\end{array}
\right)
\pm
\frac{4 i \left(20 k^2+33 k+1\right)}{(k+5)^2 (13 k+17)} 
B_{\mp} 
\pa
\left(
\begin{array}{c}
F_{22} \\
F_{11}
\end{array}
\right)
\nonu \\
&+& \frac{2}{(k+5)^2} B_{\pm} B_{\mp} 
\left(
\begin{array}{c}
F_{21} \\
F_{12}
\end{array}
\right) +
\frac{2}{(k+5)^2} 
\left(
\begin{array}{c}
F_{11} F_{21} G_{22} \\
F_{22} F_{12} G_{11}
\end{array}
\right)
\nonu \\
&+& \frac{2 (61 k+97)}{(k+5)^2 (13 k+17)}
\left(
\begin{array}{c}
\pa F_{11} F_{21} F_{22} \\
\pa F_{22} F_{12} F_{11}
\end{array}
\right)
\pm \frac{8 (k-3)}{(k+5)^2 (13 k+17)} 
\left(
\begin{array}{c}
\pa F_{21}  \\
\pa F_{12} 
\end{array}
\right) F_{11} F_{22} 
\nonu \\
&-& \frac{2 (69 k+73)}{(k+5)^2 (13 k+17)} 
\left(
\begin{array}{c}
F_{11} F_{21} \pa F_{22}  \\
F_{22} F_{12} \pa F_{11} 
\end{array}
\right)
\mp \frac{2}{(k+5)^2} F_{11} F_{22}
 \left(
\begin{array}{c}
G_{21}   \\
G_{12}  
\end{array}
\right)
\nonu \\
&-& \frac{3 (k+3)}{2 (k+5)^2} \pa^2 
\left(
\begin{array}{c}
F_{21}   \\
F_{12}  
\end{array}
\right) +
\frac{2}{(k+5)^2}
\left(
\begin{array}{c}
F_{21} F_{22} G_{11}  \\
F_{12} F_{11} G_{22} 
\end{array}
\right)
\nonu \\
&-& \frac{1}{(k+5)} T
\left(
\begin{array}{c}
F_{21}   \\
F_{12}  
\end{array}
\right)
-\frac{3}{2 (k+5)} U
\left(
\begin{array}{c}
G_{21}   \\
G_{12}  
\end{array}
\right)
\pm \frac{6 i}{(k+5)^2} U A_3 
\left(
\begin{array}{c}
F_{21}   \\
F_{12}  
\end{array}
\right)
\nonu \\
&\mp & \frac{6 i}{(k+5)^2} U A_{\mp}
\left(
\begin{array}{c}
F_{11}   \\
F_{22}  
\end{array}
\right)
\mp \frac{6 i}{(k+5)^2} U B_3 
\left(
\begin{array}{c}
F_{21}   \\
F_{12}  
\end{array}
\right)
\nonu \\
&\pm & \frac{6 i}{(k+5)^2} U B_{\mp}
\left(
\begin{array}{c}
F_{22}   \\
F_{11}  
\end{array}
\right) -
\frac{2 (k-3)}{(k+5) (13 k+17)} \pa 
\left(
\begin{array}{c}
U F_{21}   \\
U F_{12}  
\end{array}
\right)
\nonu \\
&-& \left. \frac{4}{(k+5)^2} U U
\left(
\begin{array}{c}
F_{21}   \\
F_{12}  
\end{array}
\right)
\right](w) +  \frac{1}{(z-w)} \, 
\left[ + \cdots \right](w) 
+  \cdots.
\nonu 
\eea
In the third order pole there is no spin-$\frac{3}{2}$ current.
One can write the two OPEs together 
due to the simple property of the structure constants.
The nonlinear terms in the spirit of the footnote \ref{nonlineardef}
are not present but the nonlinear term between the higher spin currents
is present. One can also remove this term by adding the $T {\bf T^{(1)}}(w)$
into the left hand side in the higher spin-$3$ current
because the OPE ${\bf T_{\pm}^{(\frac{3}{2})}}(z) \,
T {\bf T^{(1)}}(w)$ contributes to the above ${\bf T^{(1)}} \, 
{\bf T^{(\frac{3}{2})}}(w)$ term where the factor $(k-3)$ exists.
In this paper, one considers the spin-$3$ fields in the right hand side of 
the OPE. In the first order pole of above OPE, there are many nontrivial 
terms which can be obtained from the previous results for the corresponding 
OPE in the nonlinear
version.

Now one can consider the OPEs between the higher spin-$2$ current 
and the $13$ higher spin currents
%%%%%%%%%%%%%%%%%%%%%%%%%%%%%%%%%%%%%%%%%%%%%%%%%%%%%%%%%%%%%%%%%%%%%
\bea
{\bf T^{(2)}}(z) \, {\bf T^{(2)}}(w) & = & 
\frac{1}{(z-w)^4} \, 
\left[ \frac{9k}{(5+k)} \right] \nonu \\
& + & 
\frac{1}{(z-w)^2} \, \left[ 
\frac{2 (k+4)}{(k+5)} T + 
\frac{2 (k+4)}{(k+5)^2} A_3 A_3
-\frac{4 (k+4)}{(k+5)^2} A_3 B_3
\right. \nonu \\
&+& \frac{2 i (k+1)}{(k+5)^2} \pa A_3 
-\frac{4 i}{(k+5)^2} A_3 F_{11} F_{22}
-\frac{2 i}{(k+5)^2} A_{-} F_{11} F_{12}
\nonu \\
&+& \frac{2 (k+1)}{(k+5)^2} A_{+} A_{-} -\frac{2 i}{(k+5)^2} A_{+} F_{21} F_{22}
+ \frac{2 (k+4)}{(k+5)^2} B_3 B_3
\nonu \\
&+& \frac{8 i}{(k+5)^2} \pa B_3
+ \frac{4 i}{(k+5)^2} B_3 F_{11} F_{22}
+\frac{2 i}{(k+5)^2} B_{-} F_{12} F_{22}
+ \frac{8}{(k+5)^2} B_{+} B_{-}
\nonu \\
&+& \frac{2 i}{(k+5)^2} B_{+} F_{11} F_{21}
-\frac{(k-3)}{(k+5)^2} F_{11} G_{22}
+ \frac{2 \left(k^2+4 k+27\right)}{(k+5)^3} \pa F_{11} F_{22}
\nonu \\
&-& \frac{2 \left(k^2+4 k+27\right)}{(k+5)^3} F_{11} \pa F_{22}
+ \frac{2}{(k+5)} \pa F_{12} F_{21}
-\frac{2}{(k+5)} F_{12} \pa F_{21}
\nonu \\
&- & \left. \frac{(k-3)}{(k+5)^2} F_{22} G_{11}
+ \frac{2 (k+4)}{(k+5)^2} U U
\right](w) +  
\frac{1}{(z-w)}  \, \frac{1}{2} \pa (\mbox{pole-2})(w) 
+\cdots.
\nonu
\eea
There is no higher spin-$2$ current in the above which appears 
in the corresponding OPE in the nonlinear version. 
Also one sees many other terms which do not appear in the corresponding
OPE in the nonlinear version.

From the relation (\ref{g1122t1}), the following OPEs can be obtained
%%%%%%%%%%%%%%%%%%%%%%%%%%%%%%%%%%%%%%%%%%%%%%%%%%%%%%%%%%%%%%%%%%%
\bea
{\bf T^{(2)}}(z) \, 
\left(
\begin{array}{c}
{\bf U^{(\frac{3}{2})}} \\
{\bf V^{(\frac{3}{2})}} 
\end{array}
\right)
(w) & = & 
\frac{1}{(z-w)^3} \, \left[ \mp \frac{6 k}{(k+5)^2}
\left(
\begin{array}{c}
F_{11} \\
F_{22} 
\end{array}
\right)
\right](w)
\nonu \\
& + & \frac{1}{(z-w)^2}\, 
\left[ -\frac{(k-3)}{2 (k+5)} 
\left(
\begin{array}{c}
{\bf U^{(\frac{3}{2})}} \\
{\bf V^{(\frac{3}{2})}} 
\end{array}
\right) \pm
 \frac{(k-3)}{2 (k+5)^2}  U
\left(
\begin{array}{c}
F_{11} \\
F_{22} 
\end{array}
\right)
\right. \nonu \\
&+& \frac{i (k+9)}{2 (k+5)^2} 
\left(
\begin{array}{c}
F_{11} \\
F_{22} 
\end{array}
\right) A_3 
+ \frac{i (k-3)}{2 (k+5)^2} 
\left(
\begin{array}{c}
F_{21} \\
F_{12} 
\end{array}
\right) A_{\pm}
\nonu \\
&+& \left.  \frac{i (k-3)}{2 (k+5)^2} 
\left(
\begin{array}{c}
F_{12} \\
F_{21} 
\end{array}
\right) B_{\mp}
-\frac{3 i (k+1)}{2 (k+5)^2} 
\left(
\begin{array}{c}
F_{11} \\
F_{22} 
\end{array}
\right) B_3 \mp \frac{6 k}{(k+5)^2}
\pa
\left(
\begin{array}{c}
F_{11} \\
F_{22} 
\end{array}
\right)
\right] (w) 
\nonu \\
& + & \frac{1}{(z-w)} \, \left[ 
\frac{1}{2} 
\left(
\begin{array}{c}
{\bf Q^{(\frac{5}{2})}} \\
 {\bf R^{(\frac{5}{2})}} \end{array}
\right) \pm
\left(
\begin{array}{c}
{\bf U^{(\frac{5}{2})} } \\
{\bf V^{(\frac{5}{2})}} 
 \end{array}
\right)
-\frac{(k-3)}{3 (k+5)} \pa 
\left(
\begin{array}{c}
{\bf U^{(\frac{3}{2})}} \\
{\bf V^{(\frac{3}{2})}} 
\end{array}
\right)
\right. \nonu \\
&- & \frac{i}{2 (k+5)} A_3 
\left(
\begin{array}{c}
G_{11} \\
G_{22} 
\end{array}
\right) \pm
\frac{1}{(k+5)^2} A_3 B_{\mp}
\left(
\begin{array}{c}
F_{12} \\
F_{21} 
\end{array}
\right)
\nonu \\
&+& \frac{i \left(5 k^2+66 k-27\right)}{3 (k+5)^2 (2 k+9)} 
\pa A_3 
\left(
\begin{array}{c}
F_{11} \\
F_{22} 
\end{array}
\right)
+ \frac{i (k+15)}{3 (k+5)^2} A_3 \pa 
\left(
\begin{array}{c}
F_{11} \\
F_{22} 
\end{array}
\right)
\nonu \\
&\pm & \frac{1}{(k+5)^2} A_{\pm} A_3 
\left(
\begin{array}{c}
F_{21} \\
F_{12} 
\end{array}
\right)
\mp \frac{1}{(k+5)^2} A_{\pm} A_{\mp} 
\left(
\begin{array}{c}
F_{11} \\
F_{22} 
\end{array}
\right)
\nonu \\
&\pm & \frac{1}{(k+5)^2} A_{\pm} B_3 
\left(
\begin{array}{c}
F_{21} \\
F_{12} 
\end{array}
\right) \mp
\frac{2}{(k+5)^2} A_{\pm} B_{\mp} 
\left(
\begin{array}{c}
F_{22} \\
F_{11} 
\end{array}
\right)
\nonu \\
& + & \frac{i (k-6)}{3 (k+5)^2} \pa A_{\pm}
\left(
\begin{array}{c}
F_{21} \\
F_{12} 
\end{array}
\right) + \frac{i (k-12)}{3 (k+5)^2} A_{\pm}
\pa
\left(
\begin{array}{c}
F_{21} \\
F_{12} 
\end{array}
\right) 
\nonu \\
&-& \frac{i}{2 (k+5)} B_3 
\left(
\begin{array}{c}
G_{11} \\
G_{22} 
\end{array}
\right) -
\frac{i (5 k+3)}{3 (k+5)^2} B_3 \pa
\left(
\begin{array}{c}
F_{11} \\
F_{22} 
\end{array}
\right)
\nonu \\
&-& \frac{i (k-12) (k-9)}{3 (k+5)^2 (2 k+9)} \pa B_3 
\left(
\begin{array}{c}
F_{11} \\
F_{22} 
\end{array}
\right) \pm
\frac{1}{(k+5)^2} B_{\mp} B_3 
\left(
\begin{array}{c}
F_{12} \\
F_{21} 
\end{array}
\right)
\nonu \\
&+& \frac{i k}{3 (k+5)^2} \pa B_{\mp}
\left(
\begin{array}{c}
F_{12} \\
F_{21} 
\end{array}
\right) +
\frac{i (k+6)}{3 (k+5)^2} B_{\mp} \pa 
\left(
\begin{array}{c}
F_{12} \\
F_{21} 
\end{array}
\right)
\nonu \\
& \mp & \frac{1}{(k+5)^2} B_{\pm} B_{\mp} 
\left(
\begin{array}{c}
F_{11} \\
F_{22} 
\end{array}
\right) \mp
\frac{3 (k-1)}{(k+5)^2} \pa^2
\left(
\begin{array}{c}
F_{11} \\
F_{22} 
\end{array}
\right)
\nonu \\
&\pm & \frac{1}{(k+5)^2} 
\left(
\begin{array}{c}
F_{11} F_{12} G_{21} \\
F_{22} F_{21} G_{12}
\end{array}
\right) -
\frac{2 (k-3)}{3 (k+5)^3} 
\pa
\left(
\begin{array}{c}
F_{11}  \\
F_{22} 
\end{array}
\right) F_{12} F_{21}
\nonu \\
& \mp & \frac{2 (k+9)}{3 (k+5)^3}
\left(
\begin{array}{c}
F_{11} \pa F_{12} F_{21} \\
F_{22} \pa F_{21} F_{12}
\end{array}
\right) \pm
\frac{4 (k+3)}{3 (k+5)^3} 
\left(
\begin{array}{c}
F_{11}  F_{12} \pa F_{21} \\
F_{22}  F_{21} \pa F_{12}
\end{array}
\right)
\nonu \\
&\mp & \frac{1}{(k+5)^2} 
\left(
\begin{array}{c}
F_{11}  F_{21} G_{12} \\
F_{22}  F_{12} G_{21}
\end{array}
\right) \mp
\frac{1}{(k+5)} T 
\left(
\begin{array}{c}
F_{11}   \\
F_{22} 
\end{array}
\right) \mp
\frac{1}{2 (k+5)} U
\left(
\begin{array}{c}
G_{11}   \\
G_{22} 
\end{array}
\right)
\nonu \\
&-& \frac{2 i}{(k+5)^2} U A_3 
\left(
\begin{array}{c}
F_{11}   \\
F_{22} 
\end{array}
\right)
-\frac{i}{(k+5)^2} U A_{\pm}
\left(
\begin{array}{c}
F_{21}   \\
F_{12} 
\end{array}
\right)
\nonu \\
&-& \frac{2 i}{(k+5)^2} U B_3
\left(
\begin{array}{c}
F_{11}   \\
F_{22} 
\end{array}
\right)
-\frac{i}{(k+5)^2} U B_{\mp} 
\left(
\begin{array}{c}
F_{12}   \\
F_{21} 
\end{array}
\right)
\nonu \\
& \mp & \frac{(k^2+48 k-27)}{3 (k+5)^2 (2 k+9)} 
\pa U 
\left(
\begin{array}{c}
F_{11}   \\
F_{22} 
\end{array}
\right) \pm
\frac{(k-3)}{3 (k+5)^2} U \pa
\left(
\begin{array}{c}
F_{11}   \\
F_{22} 
\end{array}
\right)
\nonu \\
& \mp & \frac{2}{(k+5)^2} U U
\left(
\begin{array}{c}
F_{11}   \\
F_{22} 
\end{array}
\right)  \pm
\frac{(k-1) (k+18)}{(k+5)^2 (2 k+9)} 
\left(
\begin{array}{c}
F_{11} G_{22} F_{11}  \\
F_{22} G_{11} F_{22}
\end{array}
\right) 
\nonu \\
&+ & \left. \frac{2}{(k+5)^2} 
\pa 
 \left(
\begin{array}{c}
F_{11}   \\
F_{22} 
\end{array}
\right) F_{11} F_{22} 
\right](w) +\cdots.
\nonu 
\eea
There are no nonlinear terms as in the footnote \ref{nonlineardef}.
Also the OPE is rather complicated due to the presence of 
spin-$\frac{1}{2}$ currents as before. 
One has combined OPEs due to the symmetry of structure constants.
Some terms containing the spin-$\frac{3}{2}$ currents in the above
OPEs were not present in the corresponding OPEs in the 
nonlinear version.

One uses the relation (\ref{g1221vu3half}) to obtain the following 
OPEs
%%%%%%%%%%%%%%%%%%%%%%%%%%%%%%%%%%%%%%%%%%%%%%%%%%%%%%%%%%%%%%%%%%%
%%%%%%%%%%%%%%%%%%%%%%%%%%%%%%%%%%%%%%%%%%%%%%%%%%%%%%%%%%%%%%%%%%%%%%
\bea
{\bf T^{(2)}}(z) \, 
\left(
\begin{array}{c}
{\bf U_{+}^{(2)}} \\
{\bf V_{-}^{(2)}}
\end{array}
\right)
(w) & = & \frac{1}{(z-w)^3}
\, \left[\frac{2 i k (k+8)}{(k+5)^2} {B}_{\mp}
\mp
\frac{4 k}{(k+5)^2} 
\left(
\begin{array}{c}
F_{11} F_{21}  \\
F_{22} F_{12}
\end{array}
\right) 
 \right](w) \nonu \\
& + & 
\frac{1}{(z-w)^2} \, \left[ \pm 
\frac{2 (k+4)}{(k+5)^2} A_{3} B_{\mp}  
+ \frac{2 i}{(k+5)^2} A_3 
\left(
\begin{array}{c}
F_{11} F_{21}  \\
F_{22} F_{12}
\end{array}
\right) 
\right. \nonu \\
&\mp & \frac{2 k}{(k+5)^2} B_{\mp} B_3 
+ \frac{i k (k+7)}{(k+5)^2} \pa B_{\mp}
\mp \frac{i}{(k+5)^2} B_{\mp} F_{11} F_{22}
\nonu \\
&\mp & \frac{i}{(k+5)^2} B_{\mp} F_{12} F_{21}
\pm \frac{(k-3)}{2 (k+5)^2} 
\left(
\begin{array}{c}
F_{11} G_{21}  \\
F_{22} G_{12}
\end{array}
\right) 
\mp \frac{2 (k+1)^2}{(k+5)^3} 
\left(
\begin{array}{c}
\pa F_{11} F_{21}  \\
\pa F_{22} F_{12}
\end{array}
\right) 
\nonu \\
&\mp & \left.
\frac{2 \left(k^2+8 k-1\right)}{(k+5)^3} 
\left(
\begin{array}{c}
F_{11} \pa F_{21}  \\
F_{22} \pa F_{12}
\end{array}
\right) \pm
\frac{(k-3)}{2 (k+5)^2} 
\left(
\begin{array}{c}
F_{21} G_{11}  \\
F_{12} G_{22}
\end{array}
\right)
\right](w) 
\nonu \\
& + & \frac{1}{(z-w)} \, \left[ 
\frac{1}{2} 
\left(
\begin{array}{c}
{\bf Q_{+}^{(3)}} \\
 {\bf R_{-}^{(3)}} 
\end{array}
\right)
-\frac{i}{(k+5)^2} A_3 A_3 B_{\mp}
\pm \frac{(k+7)}{2 (k+5)^2} \pa A_3 B_{\mp}
\right. \nonu \\
&\pm & \frac{(3 k+11)}{2 (k+5)^2} A_3 \pa B_{\mp}
+ \frac{i}{(k+5)^2} A_3 
\left(
\begin{array}{c}
F_{11} G_{21} \\
F_{22} G_{12} 
\end{array}
\right)
\nonu \\
&-& \frac{i (k-3)}{2 (k+5)^3} \pa A_3 
\left(
\begin{array}{c}
F_{11} F_{21} \\
F_{22} F_{12} 
\end{array}
\right)
+ \frac{i (3 k+7)}{2 (k+5)^3} A_3 
\left(
\begin{array}{c}
\pa F_{11} F_{21} \\
\pa F_{22} F_{12} 
\end{array}
\right)
\nonu \\
&+& \frac{i (7 k+27)}{2 (k+5)^3} A_3 
\left(
\begin{array}{c}
F_{11} \pa F_{21} \\
F_{22} \pa F_{12} 
\end{array}
\right)
-\frac{i}{(k+5)^2} A_3 
\left(
\begin{array}{c}
F_{21} G_{11} \\
F_{12} G_{22} 
\end{array}
\right)
\nonu \\
&+& \frac{i}{(k+5)^2} A_{\mp}
\left(
\begin{array}{c}
F_{11} G_{11} \\
F_{22} G_{22} 
\end{array}
\right)
-\frac{i}{(k+5)^2} A_{\pm} A_{\mp} B_{\mp}
\nonu \\
&+& \frac{i}{(k+5)^2} A_{\pm} 
\left(
\begin{array}{c}
F_{21} G_{21} \\
F_{12} G_{12} 
\end{array}
\right)
-\frac{i}{(k+5)^2} B_3
\left(
\begin{array}{c}
F_{11} G_{21} \\
F_{22} G_{12} 
\end{array}
\right)
\nonu \\
&+& \frac{i (k-11)}{2 (k+5)^3} B_3
\left(
\begin{array}{c}
\pa F_{11} F_{21} \\
\pa F_{22} F_{12} 
\end{array}
\right)
-\frac{i (3 k-1)}{2 (k+5)^3} B_3
\left(
\begin{array}{c}
F_{11} \pa F_{21} \\
F_{22} \pa F_{12} 
\end{array}
\right)
\nonu \\
&+& \frac{i}{2 (k+5)^2} \pa B_3 
\left(
\begin{array}{c}
F_{11} F_{21} \\
F_{22} F_{12} 
\end{array}
\right)
-\frac{i}{(k+5)^2} B_3
\left(
\begin{array}{c}
F_{21} G_{11} \\
F_{12} G_{22} 
\end{array}
\right)
\nonu \\
&+& \frac{i}{(k+5)^2} B_{\mp} B_3 B_3 
\mp \frac{(k+1)}{(k+5)^2} \pa B_{\mp} B_3
\mp \frac{(k+2)}{(k+5)^2} B_{\mp} \pa B_3
\nonu \\
&+& \frac{i \left(14 k^3+67 k^2-345 k-306\right)}{6 (k+5)^2 (13 k+17)} 
\pa^2 B_{\mp}
+ \frac{i}{(k+5)^2} B_{\mp} 
\left(
\begin{array}{c}
F_{11} G_{22} \\
F_{22} G_{11} 
\end{array}
\right)
\nonu \\
&-& \frac{i (k-9)}{(k+5)^3} B_{\mp} 
\left(
\begin{array}{c}
\pa F_{11} F_{22} \\
\pa F_{22} F_{11} 
\end{array}
\right)
\mp \frac{i (k+7)}{(k+5)^3} \pa B_{\mp} F_{11} F_{22}
\nonu \\
&+& \frac{i (k-5)}{(k+5)^3} B_{\mp}
\left(
\begin{array}{c}
F_{11} \pa F_{22} \\
F_{22} \pa F_{11} 
\end{array}
\right)
-\frac{i (k+9)}{2 (k+5)^3} B_{\mp}
\left(
\begin{array}{c}
F_{12} \pa F_{21} \\
F_{21} \pa F_{12} 
\end{array}
\right)
\nonu \\
&-& \frac{i (k-7)}{2 (k+5)^3} B_{\mp} 
\left(
\begin{array}{c}
\pa F_{12}  F_{21} \\
\pa F_{21}  F_{12} 
\end{array}
\right)
\mp \frac{i (k+9)}{2 (k+5)^3} \pa B_{\mp} 
F_{12} F_{21}
\nonu \\
&+& \frac{i}{(k+5)^2} B_{\mp}
\left(
\begin{array}{c}
F_{21}  G_{12} \\
F_{12}  G_{21} 
\end{array}
\right)
+ \frac{2 i}{(k+5)^2} B_{\mp} 
\left(
\begin{array}{c}
F_{22}  G_{11} \\
F_{11}  G_{22} 
\end{array}
\right)
\nonu \\
&+& \frac{i}{(k+5)^2} B_{\pm} B_{\mp} B_{\mp}
\pm \frac{8 (k+1)}{(k+5)^3} 
\left(
\begin{array}{c}
\pa^2 F_{11}  F_{21} \\
\pa^2 F_{22}  F_{12} 
\end{array}
\right)
\mp 
\frac{2}{(k+5)^2} 
\left(
\begin{array}{c}
\pa F_{11}  \pa F_{21} \\
\pa F_{22}  \pa F_{12} 
\end{array}
\right)
\nonu \\
&\mp & \frac{(28 k^3+227 k^2-1110 k-1581)}{6 (k+5)^3 (13 k+17)}
\left(
\begin{array}{c}
F_{11}  \pa^2 F_{21} \\
F_{22}  \pa^2 F_{12} 
\end{array}
\right)
\pm \frac{1}{2 (k+5)} 
\left(
\begin{array}{c}
G_{11}  G_{21} \\
G_{22}  G_{12} 
\end{array}
\right)
\nonu \\
&+& \frac{i \left(4 k^2+34 k+17\right)}{(k+5) (13 k+17)} 
T B_{\mp}
\mp \frac{2 \left(4 k^2+25 k+17\right)}{(k+5)^2 (13 k+17)} 
T
\left(
\begin{array}{c}
F_{11}  F_{21} \\
F_{22}  F_{12} 
\end{array}
\right)
\nonu \\
&\mp & \frac{2}{(k+5)^2} U A_3 B_{\mp}
+ \frac{i (k-2)}{2 (k+5)^2} \pa U B_{\mp}
-\frac{i (k-2)}{2 (k+5)^2} U \pa B_{\mp}
\nonu \\
&\pm & \frac{1}{(k+5)^2} U
\left(
\begin{array}{c}
F_{11}  G_{21} \\
F_{22}  G_{12} 
\end{array}
\right) \pm
\frac{(k-11)}{2 (k+5)^3}
U
\left(
\begin{array}{c}
\pa F_{11}  F_{21} \\
\pa F_{22}  F_{12} 
\end{array}
\right)
\nonu \\
&\mp & \frac{(3 k-1)}{2 (k+5)^3} \pa U
\left(
\begin{array}{c}
F_{11}  F_{21} \\
F_{22}  F_{12} 
\end{array}
\right)
\pm \frac{(5 k+9)}{2 (k+5)^3} U
\left(
\begin{array}{c}
F_{11}  \pa F_{21} \\
F_{22}  \pa F_{12} 
\end{array}
\right)
\nonu \\
& \mp & \frac{1}{(k+5)^2} U 
\left(
\begin{array}{c}
F_{21}  G_{11} \\
F_{12}  G_{22} 
\end{array}
\right) + \frac{i}{(k+5)^2}
U U B_{\mp}
-\frac{8 i}{(k+5)^3} 
\left(
\begin{array}{c}
F_{11}  A_{-} \pa F_{11} \\
F_{22}  A_{+} \pa F_{22} 
\end{array}
\right)
\nonu \\
&+& \frac{i \left(28 k^3+227 k^2+138 k+51\right)}{3 (k+5)^3 (13 k+17)}
\left(
\begin{array}{c}
\pa F_{21}  A_{+}  F_{21} \\
\pa F_{12}  A_{-} F_{12} 
\end{array}
\right)
\nonu \\
&+& \frac{i \left(28 k^3+305 k^2+318 k+153\right)}{3 (k+5)^3 (13 k+17)}
\left(
\begin{array}{c}
F_{21}  A_{+}  \pa F_{21} \\
F_{12}  A_{-} \pa F_{12} 
\end{array}
\right)
\nonu \\
&\pm & \frac{(k-19)}{4 (k+5)^2} 
\left(
\begin{array}{c}
\pa F_{11}  G_{21} \\
\pa F_{22}  G_{12} 
\end{array}
\right)
\pm \frac{(k-3)}{4 (k+5)^2} 
\left(
\begin{array}{c}
F_{11}  \pa G_{21} \\
F_{22}  \pa G_{12} 
\end{array}
\right)
\nonu \\
&\mp & 
\left.  
\frac{(k+8)}{2 (k+5)^2} 
\left(
\begin{array}{c}
\pa F_{21} G_{11} \\
\pa F_{12} G_{22} 
\end{array}
\right) \pm
\frac{(k-2)}{2 (k+5)^2} 
\left(
\begin{array}{c}
F_{21} \pa G_{11} \\
F_{12} \pa G_{22} 
\end{array}
\right)
\right](w) +\cdots.
\nonu 
\eea
The OPE does not contain any nonlinear terms between the higher spin currents
and the currents of large ${\cal N}=4$ linear superconformal algebra.
In the nonlinear version one cannot combine the above two OPEs.
There is no higher spin-$2$ current in the second order pole of above
OPEs.

By using the relation (\ref{g1221uv3half}), the following OPEs can be obtained
\bea
%%%%%%%%%%%%%%%%%%%%%%%%%%%%%%%%%%%%%%%%%%%%%%%%%%%%%%%%%%%%%%%%%%%%%%%%
{\bf T^{(2)}}(z) \, 
\left(
\begin{array}{c}
{\bf U_{-}^{(2)}} \\
{\bf V_{+}^{(2)}} 
\end{array}
\right)(w) & = & \frac{1}{(z-w)^3}
\, \left[ \frac{6 i (2 k+5)}{(k+5)^2} {A}_{\pm} 
\pm \frac{12}{(k+5)^2}
\left(
\begin{array}{c}
F_{11} F_{12} \\
F_{22} F_{21} 
\end{array}
\right)
\right](w) \nonu \\
& + & 
\frac{1}{(z-w)^2} \, \left[ \pm
\frac{6}{(k+5)^2} A_{\pm} A_3 
\mp \frac{2 (k+4)}{(k+5)^2} A_{\pm} B_3
+ \frac{6 i (k+2)}{(k+5)^2} \pa A_{\pm}
\right. \nonu \\
&\mp & \frac{i}{(k+5)^2} A_{\pm } F_{11} F_{22}
\pm \frac{i}{(k+5)^2} A_{\pm} F_{12} F_{21}
+\frac{2 i}{(k+5)^2} B_3 
\left(
\begin{array}{c}
F_{11} F_{12} \\
F_{22} F_{21} 
\end{array}
\right)
\nonu \\
&\mp & \frac{(k-3)}{2 (k+5)^2} 
\left(
\begin{array}{c}
F_{11} G_{12} \\
F_{22} G_{21} 
\end{array}
\right)
\pm \frac{32}{(k+5)^3} 
\left(
\begin{array}{c}
\pa F_{11} F_{12} \\
\pa F_{22} F_{21} 
\end{array}
\right)
\nonu \\
& \pm &\left.  \frac{4 (3 k+7)}{(k+5)^3} 
\left(
\begin{array}{c}
F_{11} \pa F_{12} \\
F_{22} \pa F_{21} 
\end{array}
\right) \mp \frac{(k-3)}{2 (k+5)^2} 
\left(
\begin{array}{c}
F_{12} G_{11} \\
F_{21} G_{22} 
\end{array}
\right)
\right](w) 
\nonu \\
& + & \frac{1}{(z-w)} \, \left[ 
\frac{1}{2} 
\left(
\begin{array}{c}
{\bf Q_{-}^{(3)} } \\
 {\bf R_{+}^{(3)} }
\end{array}
\right)
+ \frac{i}{(k+5)^2} A_3 
\left(
\begin{array}{c}
F_{11} G_{12} \\
F_{22} G_{21} 
\end{array}
\right)
\right. \nonu \\
&+& \frac{i}{2 (k+5)^2} \pa A_3 
\left(
\begin{array}{c}
F_{11} F_{12} \\
F_{22} F_{21} 
\end{array}
\right) -\frac{i (3 k-1)}{2 (k+5)^3}
A_3 
\left(
\begin{array}{c}
\pa F_{11} F_{12} \\
\pa F_{22} F_{21} 
\end{array}
\right)
\nonu \\
&+& \frac{i (k-11)}{2 (k+5)^3} A_3 
\left(
\begin{array}{c}
F_{11} \pa F_{12} \\
F_{22} \pa F_{21} 
\end{array}
\right) +
\frac{i}{(k+5)^2} A_3 
\left(
\begin{array}{c}
F_{12} G_{11} \\
F_{21} G_{22} 
\end{array}
\right)
\nonu \\
&+& \frac{i}{(k+5)^2} A_{\pm} A_3 A_3 
\pm \frac{4}{(k+5)^2} \pa A_{\pm} A_3 
\pm \frac{1}{(k+5)^2} A_{\pm} \pa A_3 
\nonu \\
&+& \frac{i}{(k+5)^2} A_{\pm} A_{\pm} A_{\mp}
-\frac{i}{(k+5)^2} A_{\pm} B_3 B_3
\mp \frac{(3 k+11)}{2 (k+5)^2} \pa A_{\pm} B_3
\nonu \\
&\mp & \frac{(k+3)}{2 (k+5)^2} A_{\pm} \pa B_3
-\frac{i}{(k+5)^2} A_{\pm} B_{\pm} B_{\mp}
+\frac{i \left(24 k^2-13 k-185\right)}{2 (k+5)^2 (13 k+17)} \pa^2 A_{\pm}
\nonu \\
&-& \frac{i}{(k+5)^2} A_{\pm }
\left(
\begin{array}{c}
F_{11} G_{22} \\
F_{22} G_{11} 
\end{array}
\right)
\mp \frac{i (3 k+11)}{2 (k+5)^3} \pa A_{\pm} F_{11} F_{22}
\nonu \\
&+& \frac{i (5 k-3)}{2 (k+5)^3} A_{\pm}
\left(
\begin{array}{c}
\pa F_{11} F_{22} \\
\pa F_{22} F_{11} 
\end{array}
\right)
-\frac{i (3 k-5)}{2 (k+5)^3} A_{\pm} 
\left(
\begin{array}{c}
F_{11} \pa F_{22} \\
F_{22} \pa F_{11} 
\end{array}
\right)
\nonu \\
&-& \frac{i}{(k+5)^2} A_{\pm} 
\left(
\begin{array}{c}
F_{12} G_{21} \\
F_{21} G_{12} 
\end{array}
\right)
\pm \frac{i (k+3)}{(k+5)^3} \pa A_{\pm } F_{12} F_{21}
\nonu \\
&+& \frac{i (k+3)}{(k+5)^3} A_{\pm} 
\left(
\begin{array}{c}
\pa F_{12} F_{21} \\
\pa F_{21} F_{12} 
\end{array}
\right) 
-\frac{i (k-1)}{(k+5)^3} A_{\pm}
\left(
\begin{array}{c}
F_{12} \pa F_{21} \\
F_{21} \pa F_{12} 
\end{array}
\right) 
\nonu \\
&- &  \frac{2 i}{(k+5)^2} A_{\pm} 
\left(
\begin{array}{c}
F_{22} G_{11} \\
F_{11} G_{22} 
\end{array}
\right) 
-\frac{i}{(k+5)^2} B_3 
\left(
\begin{array}{c}
F_{11} G_{12} \\
F_{22} G_{21} 
\end{array}
\right) 
\nonu \\
&+& \frac{i (k-3)}{2 (k+5)^3} \pa B_3 
\left(
\begin{array}{c}
F_{11} F_{12} \\
F_{22} F_{21} 
\end{array}
\right) +
\frac{i (k+13)}{2 (k+5)^3} B_3 
\left(
\begin{array}{c}
\pa F_{11} F_{12} \\
\pa F_{22} F_{21} 
\end{array}
\right)
\nonu \\
&+& \frac{i (5 k+33)}{2 (k+5)^3} B_3 
\left(
\begin{array}{c}
F_{11} \pa F_{12} \\
F_{22} \pa F_{21} 
\end{array}
\right)
+ \frac{i}{(k+5)^2} B_3 
\left(
\begin{array}{c}
F_{12} G_{11} \\
F_{21} G_{22} 
\end{array}
\right) 
\nonu \\
&-& \frac{i}{(k+5)^2} B_{\mp}
\left(
\begin{array}{c}
F_{12} G_{12} \\
F_{21} G_{21} 
\end{array}
\right) -\frac{i}{(k+5)^2}
B_{\pm}
\left(
\begin{array}{c}
F_{11} G_{11} \\
F_{22} G_{22} 
\end{array}
\right)
\nonu \\
&\mp & \frac{(103 k^2+326 k+591)}{2 (k+5)^3 (13 k+17)} 
\left(
\begin{array}{c}
\pa^2 F_{11} F_{12} \\
\pa^2 F_{22} F_{21} 
\end{array}
\right)
\nonu \\
&\pm & \frac{(131 k^2+318 k-149)}{(k+5)^3 (13 k+17)} 
\left(
\begin{array}{c}
\pa F_{11} \pa F_{12} \\
\pa F_{22} \pa F_{21} 
\end{array}
\right)
\nonu \\
&\pm & \frac{(k-91)}{2 (k+5)^2 (13 k+17)} 
\left(
\begin{array}{c}
F_{11} \pa^2 F_{12} \\
F_{22} \pa^2 F_{21} 
\end{array}
\right) \mp
\frac{(k+1)}{(k+5)^2}
\left(
\begin{array}{c}
\pa F_{11} G_{12} \\
\pa F_{22} G_{21} 
\end{array}
\right)
\nonu \\
&\mp & \frac{(k+19)}{4 (k+5)^2} 
\left(
\begin{array}{c}
\pa F_{12} G_{11} \\
\pa F_{21} G_{22} 
\end{array}
\right)
\mp \frac{(k-5)}{4 (k+5)^2} 
\left(
\begin{array}{c}
F_{12} \pa G_{11} \\
F_{21} \pa G_{22} 
\end{array}
\right)
\nonu \\
&\mp & \frac{1}{2 (k+5)} 
\left(
\begin{array}{c}
G_{11} G_{12} \\
G_{22} G_{21} 
\end{array}
\right)
+ \frac{i (28 k+71)}{(k+5) (13 k+17)} T A_{\pm}
\nonu \\
&\pm & \frac{2 (19 k+71)}{(k+5)^2 (13 k+17)}  T
\left(
\begin{array}{c}
F_{11} F_{12} \\
F_{22} F_{21} 
\end{array}
\right)
\mp \frac{2}{(k+5)^2} U A_{\pm} B_3
\nonu \\
&-& \frac{i}{2 (k+5)^2} \pa U A_{\pm}
+ \frac{i}{2 (k+5)^2} U \pa A_{\pm} 
\mp \frac{1}{(k+5)^2} U 
\left(
\begin{array}{c}
F_{11} G_{12} \\
F_{22} G_{21} 
\end{array}
\right)
\nonu \\
&\pm & \frac{(k-11)}{2 (k+5)^3} \pa U 
\left(
\begin{array}{c}
F_{11} F_{12} \\
F_{22} F_{21} 
\end{array}
\right)
\mp \frac{(3 k-1)}{2 (k+5)^3} U 
\left(
\begin{array}{c}
\pa F_{11} F_{12} \\
\pa F_{22} F_{21} 
\end{array}
\right)
\nonu \\
&\pm & \frac{(k+21)}{2 (k+5)^3} U
\left(
\begin{array}{c}
F_{11} \pa F_{12} \\
 F_{22} \pa F_{21} 
\end{array}
\right)
\pm \frac{1}{(k+5)^2} 
U
\left(
\begin{array}{c}
F_{12} G_{11} \\
 F_{21} G_{22} 
\end{array}
\right)
\nonu \\
&+& \frac{i}{(k+5)^2} U U A_{\pm}
-\frac{2 i (k+1)}{(k+5)^3} 
\left(
\begin{array}{c}
F_{11} B_{+} \pa F_{11} \\
 F_{22} B_{-} \pa F_{22} 
\end{array}
\right)
\nonu \\
& + &  \left.  \frac{8 i}{(k+5)^3} 
\left(
\begin{array}{c}
F_{12} B_{-} \pa F_{12} \\
 F_{21} B_{+} \pa F_{21} 
\end{array}
\right)
\right](w) +\cdots.
\nonu 
\eea
In this case, one sees the symmetry between the structure constants in the 
two OPEs. The nonlinear terms appearing in the corresponding OPEs 
in the nonlinear version do not appear in the above OPEs. 
There is no higher spin-$2$ current in the second order pole of above
OPEs. One sees that there are many (nonderivative) terms 
in the first order pole
which do not appear in the corresponding OPEs in the nonlinear version.

With the help of (\ref{g1221vu2}), one calculates the following OPEs 
\bea
%%%%%%%%%%%%%%%%%%%%%%%%%%%%%%%%%%%%%%%%%%%%%%%%%%%%%%%%%%%%%%%%%%%
{\bf T^{(2)}}(z) \,
\left(
\begin{array}{c} 
{\bf U^{(\frac{5}{2})}} \\
{\bf V^{(\frac{5}{2})}} 
\end{array}
\right)(w) & = & 
\frac{1}{(z-w)^4} \, \left[ \frac{6 (k-3) k}{(k+5)^3} 
\left(
\begin{array}{c}
F_{11} \\
 F_{22} 
\end{array}
\right)
\right](w)
\nonu \\
&+& \frac{1}{(z-w)^3}
\, \left[ 
-\frac{2 \left(4 k^2+39 k+66\right)}{3 (k+5)^2}
\left(
\begin{array}{c}
G_{11} \\
 G_{22} 
\end{array}
\right) \right. \nonu \\
&-&  \frac{8 \left(2 k^2+15 k+33\right)}{3 (k+5)^3} U
\left(
\begin{array}{c}
F_{11} \\
F_{22} 
\end{array}
\right)
\mp \frac{4}{(k+5)^2} 
\left(
\begin{array}{c}
F_{11} \\
F_{22} 
\end{array}
\right) F_{12} F_{21}
\nonu \\
&\mp & \frac{4 i \left(4 k^2+15 k-33\right)}{3 (k+5)^3}
\left(
\begin{array}{c}
F_{11} \\
F_{22} 
\end{array}
\right) A_3 
 \mp \frac{4 i \left(4 k^2+21 k+21\right)}{3 (k+5)^3} 
\left(
\begin{array}{c}
F_{21} \\
F_{12} 
\end{array}
\right) A_{\pm}
\nonu \\
&\mp& \left. \frac{4 i \left(k^2+15 k+66\right)}{3 (k+5)^3} 
\left(
\begin{array}{c}
F_{12} \\
F_{21} 
\end{array}
\right) B_{\mp} \pm
\frac{4 i \left(k^2-k-22\right)}{(k+5)^3} 
\left(
\begin{array}{c}
F_{11} \\
F_{22} 
\end{array}
\right) B_3 
\right](w)
\nonu \\
& + & \frac{1}{(z-w)^2} \, \left[
\mp \frac{(k-3)}{6 (k+5)}
\left( 
\begin{array}{c}
{\bf Q^{(\frac{5}{2})}} \\
{\bf  R^{(\frac{5}{2})} } 
\end{array}
\right)
-\frac{(k-3)}{6 (k+5)}
\left(\begin{array}{c} 
{\bf U^{(\frac{5}{2})}} \\
{\bf V^{(\frac{5}{2})}} 
\end{array}
\right)
\right. \nonu \\
&\pm& \frac{i (5 k+27)}{3 (k+5)^2} A_3
\left(
\begin{array}{c}
G_{11} \\
G_{22} 
\end{array}
\right)
-\frac{2}{(k+5)^2} A_3 A_3 
\left(
\begin{array}{c}
F_{11} \\
F_{22} 
\end{array}
\right)
\nonu \\
&+& \frac{2 (k+9)}{3 (k+5)^3} 
A_3
\left(
\begin{array}{c}
B_{-} F_{12} \\
B_{+} F_{21} 
\end{array}
\right)
 \mp \frac{2 i \left(8 k^2+27 k-81\right)}{9 (k+5)^3}
\pa A_3 
\left(
\begin{array}{c}
F_{11} \\
F_{22} 
\end{array}
\right)
\nonu \\
&\mp& \frac{8 i (k+3) (2 k+9)}{9 (k+5)^3} A_3
\pa \left(
\begin{array}{c}
F_{11} \\
F_{22} 
\end{array}
\right) \pm
\frac{i (k+1)}{(k+5)^2} A_{\pm}
\left(
\begin{array}{c}
G_{21} \\
G_{12} 
\end{array}
\right)
\nonu \\
&-& \frac{4 (k+3)}{3 (k+5)^3} A_{\pm} A_3
\left(
\begin{array}{c}
F_{21} \\
F_{12} 
\end{array}
\right)
-\frac{2 (k+9)}{3 (k+5)^3}
A_{\pm} A_{\mp}
\left(
\begin{array}{c}
F_{11} \\
F_{22} 
\end{array}
\right)
\nonu \\
&-& \frac{4 (k+3)}{3 (k+5)^3} A_{\pm} B_3 
\left(
\begin{array}{c}
F_{21} \\
F_{12} 
\end{array}
\right)
+ \frac{2 (k-3)}{3 (k+5)^3} A_{\pm} B_{\mp} 
\left(
\begin{array}{c}
F_{22} \\
F_{11} 
\end{array}
\right)
\nonu \\
&\mp & \frac{4 i (k+3) (4 k-3)}{9 (k+5)^3} \pa A_{\pm}
\left(
\begin{array}{c}
F_{21} \\
F_{12} 
\end{array}
\right)
\mp \frac{4 i (k+3) (4 k+27)}{9 (k+5)^3} A_{\pm} \pa
\left(
\begin{array}{c}
F_{21} \\
F_{12} 
\end{array}
\right)
\nonu \\
&\mp & \frac{7 i (k+3)}{3 (k+5)^2} B_3
\left(
\begin{array}{c}
G_{11} \\
G_{22} 
\end{array}
\right)
+ \frac{2}{(k+5)^2} B_3 B_3 
\left(
\begin{array}{c}
F_{11} \\
F_{22} 
\end{array}
\right)
\nonu \\
&\pm & \frac{2 i (k-3) (10 k+39)}{9 (k+5)^3}
\pa B_3 
\left(
\begin{array}{c}
F_{11} \\
F_{22} 
\end{array}
\right)
\mp \frac{4 i (k+9) (k+12)}{9 (k+5)^3} B_3 
\pa
\left(
\begin{array}{c}
F_{11} \\
F_{22} 
\end{array}
\right)
\nonu \\
&\mp& \frac{4 i}{(k+5)^2} B_{\mp}
\left(
\begin{array}{c}
G_{12} \\
G_{21} 
\end{array}
\right) +
\frac{2 (k+9)}{3 (k+5)^3} B_{\mp} B_3 
\left(
\begin{array}{c}
F_{12} \\
F_{21} 
\end{array}
\right)
\nonu \\
&\pm & \frac{2 i (k-12) (k+9)}{9 (k+5)^3} \pa B_{\mp} 
\left(
\begin{array}{c}
F_{12} \\
F_{21} 
\end{array}
\right) 
\mp \frac{2 i (k+9) (8 k+15)}{9 (k+5)^3} B_{\mp} \pa
\left(
\begin{array}{c}
F_{12} \\
F_{21} 
\end{array}
\right)
\nonu \\
&+& \frac{4 (k+3)}{3 (k+5)^3} B_{\pm} B_{\mp} 
\left(
\begin{array}{c}
F_{11} \\
F_{22} 
\end{array}
\right)
+ \frac{2 (k-3) (2 k+9)}{3 (k+5)^3} \pa^2
\left(
\begin{array}{c}
F_{11} \\
F_{22} 
\end{array}
\right) 
\nonu \\
&+& \frac{2 (k+9)}{3 (k+5)^3}
\left(
\begin{array}{c}
F_{11} F_{12} G_{21} \\
F_{22} F_{21} G_{12} 
\end{array}
\right) 
\mp \frac{16 (k+3) (k+9)}{9 (k+5)^4} 
\pa 
\left(
\begin{array}{c}
F_{11}  \\
F_{22} 
\end{array}
\right) F_{12} F_{21} 
\nonu \\
&-& \frac{16 (k+3)^2}{9 (k+5)^4} 
\left(
\begin{array}{c}
F_{11} \pa F_{12} F_{21} \\
F_{22} \pa F_{21} F_{12}
\end{array}
\right)
-\frac{4 (k+9)^2}{9 (k+5)^4} 
\left(
\begin{array}{c}
F_{11} F_{12} \pa F_{21} \\
F_{22} F_{21} \pa F_{12}
\end{array}
\right)
\nonu \\
&+& \frac{4 (k+3)}{3 (k+5)^3} 
\left(
\begin{array}{c}
F_{11} F_{21} G_{12} \\
F_{22} F_{12} G_{21}
\end{array}
\right)
\pm \frac{2}{(k+5)^2} F_{11} F_{22}
\left(
\begin{array}{c}
G_{11} \\
G_{22}
\end{array}
\right)
\nonu \\
&-& \frac{(8 k^2+87 k+171)}{9 (k+5)^2} 
\pa
\left(
\begin{array}{c}
G_{11} \\
G_{22}
\end{array}
\right)
+ \frac{4 (k-3)}{3 (k+5)^2} T
\left(
\begin{array}{c}
F_{11} \\
F_{22}
\end{array}
\right)
\nonu \\
&-& \frac{(k-3)}{3 (k+5)^2} U
\left(
\begin{array}{c}
G_{11} \\
G_{22}
\end{array}
\right)
\pm \frac{8 i (k+3)}{3 (k+5)^3} U A_3
\left(
\begin{array}{c}
F_{11} \\
F_{22}
\end{array}
\right)
\nonu \\
&\pm & \frac{4 i (k+3)}{3 (k+5)^3} U A_{\pm} 
\left(
\begin{array}{c}
F_{21} \\
F_{12}
\end{array}
\right)
\mp \frac{4 i (k+9)}{3 (k+5)^3} U B_3
\left(
\begin{array}{c}
F_{11} \\
F_{22}
\end{array}
\right)
\nonu \\
&\mp & \frac{2 i (k+9)}{3 (k+5)^3} U B_{\mp}
\left(
\begin{array}{c}
F_{12} \\
F_{21}
\end{array}
\right)
-\frac{2 \left(8 k^2+87 k+171\right)}{9 (k+5)^3}
\pa U 
\left(
\begin{array}{c}
F_{11} \\
F_{22}
\end{array}
\right)
\nonu \\
&-& \frac{16 \left(k^2+3 k+18\right)}{9 (k+5)^3} 
U \pa 
\left(
\begin{array}{c}
F_{11} \\
F_{22}
\end{array}
\right)
+ \frac{2 (k-3)}{3 (k+5)^3} U U
\left(
\begin{array}{c}
F_{11} \\
F_{22}
\end{array}
\right) 
\nonu \\
&- & \left. 
\frac{4 (k-3)}{3 (k+5)^3}
\left(
\begin{array}{c}
F_{11} \pa F_{11} F_{22} \\
F_{22} \pa F_{22} F_{11}
\end{array}
\right)
 \right](w) 
\nonu \\
& + & 
\frac{1}{(z-w)} \, \left[ +\cdots
  \right](w) + \cdots.
\nonu  
\eea
There is no higher spin-$\frac{3}{2}$ current in the 
third order pole.
Again, one can combine two OPEs in one expression as above due to the 
symmetry of structure constants. It is an open problem to fill out the 
first order pole of the above OPE and see whether there are nonlinear terms 
in the context of the footnote \ref{nonlineardef} 
from the corresponding OPEs in the 
nonlinear results.

One uses the relation (\ref{g1122vu3half}) to obtain the following 
OPE
%%%%%%%%%%%%%%%%%%%%%%%%%%%%%%%%%%%%%%%%%%%%%%%%%%%%%%%%%%
\bea
{\bf T^{(2)}}(z) \, {\bf W^{(2)}}(w) & = & 
\frac{1}{(z-w)^4} \left[ -\frac{3 (k-3) k }{(k+5)^2} \right] 
\nonu \\
& + & \frac{1}{(z-w)^2}
\left[
\frac{2 (k-3)}{3 (k+5)} {\bf T^{(2)}} 
+2 {\bf W^{(2)}}
- \frac{1}{2} {\bf P^{(2)}}
\right. \nonu \\
&-& \frac{2 (k-3)}{3 (k+5)} T
-\frac{2 (k+6)}{3 (k+5)^2} A_3 A_3
-\frac{2 (k-3)}{3 (k+5)^2} A_3 B_3
-\frac{2 i (k-3)}{3 (k+5)^2} \pa A_3
\nonu \\
&+  &  \frac{2 i (k-3)}{3 (k+5)^3} A_3 F_{11} F_{22}
-\frac{4 i (k+15)}{3 (k+5)^3}  A_{-} F_{11} F_{12}
-\frac{2 (k-3)}{3 (k+5)^2} A_{+} A_{-}
\nonu \\
& + & \frac{2 i (k+9)}{(k+5)^3}   A_{+} F_{21} F_{22}   
+ \frac{2 (2 k+3)}{3 (k+5)^2} B_3 B_3
-\frac{2 i (k-3)}{3 (k+5)^2} \pa B_3 
\nonu \\
& - & 
\frac{2 i (k-3)}{3 (k+5)^3} B_3 F_{11} F_{22}
-\frac{2 i (7 k+15)}{3 (k+5)^3} B_{-} F_{12} F_{22}
-\frac{2 (k-3)}{3 (k+5)^2} B_{+} B_{-}
\nonu \\
& + &  \frac{4 i (k+3)}{(k+5)^3}  B_{+} F_{11} F_{21}
 -\frac{(k-3)}{6 (k+5)^2} F_{11} G_{22}
 -\frac{2 (k+3) (k+9)}{3 (k+5)^3} \pa F_{11} F_{22}
 \nonu \\
& + & \frac{2 \left(k^2-6 k-63\right)}{3 (k+5)^3} 
F_{11} \pa F_{22}+
 \frac{3}{2 (k+5)} F_{12} G_{21} -\frac{2 (k-3) k}{3 (k+5)^3} 
\pa F_{12} F_{21}
\nonu \\
&+& \frac{2 (k-3) (k+8)}{3 (k+5)^3}  F_{12} \pa F_{21}
 -\frac{(k+9)}{2 (k+5)^2}  F_{21} G_{12}
+  \frac{(k-3)}{6 (k+5)^2}  F_{22} G_{11}
\nonu \\
& - & 
\frac{8 (k-3)}{3 (k+5)^3}  U F_{11} F_{22}
+ \frac{6}{(k+5)^2} U F_{12} F_{21}
+ \frac{6 i}{(k+5)^2} U A_3 + \frac{2 i k}{(k+5)^2} U B_3
\nonu \\
&- & \left. \frac{2 (k-3)}{3 (k+5)^2} U U
\right](w) \nonu \\
& + & \frac{1}{(z-w)} \, \left[  
-\frac{1}{4} {\bf \pa P^{(2)} } + \pa {\bf W^{(2)}}
 +\frac{(k-3)}{3 (k+5)} \pa {\bf T^{(2)}}
-\frac{2 (k+6)}{3 (k+5)^2} \pa A_3 A_3
\right. \nonu \\
& - &  \frac{(7 k+27)}{3 (k+5)^2}
 \pa A_3 B_3 + \frac{(5 k+33)}{3 (k+5)^2} A_3 \pa B_3
+ \frac{2 i (k-3)}{3 (k+5)^3}  \pa A_3 F_{11} F_{22}
\nonu \\
& + & \frac{4 i (k+3)}{3 (k+5)^3} A_3 \pa (F_{11} F_{22}) 
+  \frac{i}{(k+5)^2} \pa A_3 F_{12} F_{21}
- \frac{i}{(k+5)^2} A_3 \pa (F_{12} F_{21})
\nonu \\
&
+ & \frac{i}{(k+5)^2}
A_{-} F_{11} G_{12}-
  \frac{2 i (k+15)}{3 (k+5)^3} \pa A_{-} F_{11} F_{12}
- \frac{4 i (2 k+9)}{3 (k+5)^3}   A_{-} \pa F_{11} F_{12}
\nonu \\
&+& \frac{4 i(k-6)}{3(k+5)^3}  A_{-} F_{11} \pa F_{12}
+\frac{i}{(k+5)^2} A_{-} F_{12} G_{11}
-\frac{(k-3)}{3 (k+5)^2} \pa (A_{+} A_{-})
\nonu \\
&
- &  \frac{i}{(k+5)^2} A_{+} F_{21} G_{22}
+ \frac{i (k+9)}{(k+5)^3} \pa A_{+} F_{21} F_{22}
+ \frac{i (3 k+11)}{(k+5)^3} A_{+} \pa F_{21} F_{22}
\nonu \\
&-& \frac{i (k-7)}{(k+5)^3}  A_{+} F_{21} \pa F_{22}
-\frac{i}{(k+5)^2} A_{+} F_{22} G_{21}
-\frac{i (k-3)}{3 (k+5)^2}  \pa^2 B_3
\nonu \\
&+& \frac{2 (2 k+3)}{3 (k+5)^2}  \pa B_3 B_3
+ \frac{2 i (k+9)}{3 (k+5)^3}  B_3 \pa (F_{11} F_{22})
+ \frac{i}{(k+5)^2} B_{-} F_{12} G_{22}
\nonu \\
& - & \frac{i}{(k+5)^2} \pa B_3 F_{12} F_{21} + 
\frac{ i}{(k+5)^2} B_3 \pa (F_{12} F_{21})
-\frac{i (7 k+15)}{3 (k+5)^3} \pa B_{-} F_{12} F_{22}
\nonu \\
&-& \frac{i (7 k-9)}{3 (k+5)^3} B_{-} \pa F_{12} F_{22}
-\frac{i (7 k+39)}{3 (k+5)^3}  B_{-} F_{12} \pa F_{22}
+ \frac{i}{(k+5)^2} B_{-} F_{22} G_{12}
\nonu \\
&- & \frac{(k-3)}{3 (k+5)^2} \pa (B_{+} B_{-}) 
-
\frac{i}{(k+5)^2} B_{+} F_{11} G_{21}
+  \frac{2 i (k+3)}{(k+5)^3} \pa B_{+} F_{11} F_{21}
\nonu \\
& + & \frac{2 i (k-1)}{(k+5)^3}  B_{+} \pa F_{11} F_{21}
+ \frac{2 i (k+7)}{(k+5)^3} B_{+} F_{11} \pa F_{21}
-\frac{i}{(k+5)^2} B_{+} F_{21} G_{11}
\nonu \\
&+  & 
  \frac{(2 k+15)}{3 (k+5)^2}  \pa F_{11} G_{22}
- \frac{k}{3 (k+5)^2}  F_{11} \pa G_{22}
-   \frac{1}{(k+5)^2} \pa F_{12} G_{21}
\nonu \\
& + & \frac{(k+4)}{(k+5)^2} F_{12} \pa G_{21} 
- \frac{(k+7)}{(k+5)^2} \pa F_{21} G_{12}
- \frac{2}{(k+5)^2} F_{21} \pa G_{12}
\nonu \\
& + & \frac{(5 k+27)}{6 (k+5)^2}  \pa F_{22} G_{11}
- \frac{(k+3)}{6 (k+5)^2}  F_{22} \pa G_{11}
-\frac{(k-3)}{3 (k+5)} \pa T
\nonu \\
& + & 
 \frac{3 i}{(k+5)^2} \pa (U A_3)
+ \frac{i k}{(k+5)^2} \pa (U B_3)
- \frac{8 k}{3 (k+5)^3}  \pa U  F_{11} F_{22}
\nonu \\
&+& \frac{2 (k+21)}{3 (k+5)^3}  U  \pa F_{11} F_{22}
- \frac{2 (5 k+9)}{3 (k+5)^3} U F_{11} \pa F_{22}
+ \frac{3}{(k+5)^2} \pa U F_{12} F_{21}
\nonu \\
& + &  \frac{1}{(k+5)^2}  U \pa F_{12} F_{21}
+ \frac{5}{(k+5)^2}   U F_{12} \pa F_{21}
-  \frac{2 (k-3)}{3 (k+5)^2}   \pa U U 
\nonu \\
& + & \frac{4 (k+3)}{3 (k+5)^3}   F_{11} G_{22} F_{11} F_{22}
- \frac{i (k-3)}{3 (k+5)^2} \pa^2 A_3
- \frac{(k^2+15 k+18)}{3 (k+5)^3} \pa^2 F_{11} F_{22}
\nonu \\
& - &  \frac{6}{(k+5)^2} \pa F_{11} \pa F_{22}
 +\frac{(k^2-3 k-72)}{3 (k+5)^3}  F_{11} \pa^2 F_{22}
 - \frac{(k-3)^2}{3 (k+5)^3} \pa^2 F_{12} F_{21}
\nonu \\
& +  & \left.  \frac{8 (k-3)}{3 (k+5)^3} \pa F_{12} \pa F_{21}
+  \frac{(k-3)}{3 (k+5)^2}  F_{12} \pa^2 F_{21}  
\right](w) +\cdots.
\nonu 
\eea
The central term in the above OPE 
is present.
The nonlinear terms appearing in the corresponding OPE in the 
nonlinear version do not appear in this OPE.

The following OPEs can be obtained from (\ref{g1221w2})
\bea
%%%%%%%%%%%%%%%%%%%%%%%%%%%%%%%%%%%%%%%%%%%%%%%%%%%%%%%%%%%%%%%%%%%%%
{\bf T^{(2)}}(z) \, 
{\bf W_{\pm}^{(\frac{5}{2})}}(w) & = & 
\frac{1}{(z-w)^4} \, \left[-\frac{18 k}{(k+5)^2} 
\left(
\begin{array}{c}
F_{21}  \\
F_{12}
\end{array}
\right) \right](w)
\nonu \\
& + & 
\frac{1}{(z-w)^3}
\, \left[ 
-\frac{2 (k-3)}{3 (k+5)^2}
\left(
\begin{array}{c}
G_{21}  \\
G_{12}
\end{array}
\right)
 -\frac{4 (k-3)}{3 (k+5)^3} U
\left(
\begin{array}{c}
F_{21}  \\
F_{12}
\end{array}
\right) \right.
\nonu \\
&\mp &  \frac{4 (k-3)}{3 (k+5)^3}
\left(
\begin{array}{c}
F_{21}  \\
F_{12}
\end{array}
\right) F_{11} F_{22}
\pm
\frac{8 i (5 k+21)}{3 (k+5)^3} 
\left(
\begin{array}{c}
F_{21}  \\
F_{12}
\end{array}
\right)  A_3
\nonu \\
&\mp & \frac{16 i (k-3)}{3 (k+5)^3} 
\left(
\begin{array}{c}
F_{11}  \\
F_{22}
\end{array}
\right) A_{\mp}
\pm 
\frac{4 i (k-3) (k+1)}{3 (k+5)^3} 
\left(
\begin{array}{c}
F_{22}  \\
F_{11}
\end{array}
\right) B_{\mp}
\nonu \\
&\pm & \left. \frac{4 i \left(3 k^2+14 k+3\right)}{3 (k+5)^3} 
\left(
\begin{array}{c}
F_{21}  \\
F_{12}
\end{array}
\right) B_3
  \right](w)
\nonu \\
& + & \frac{1}{(z-w)^2} \, \left[ \pm \frac{5}{2} 
{\bf P_{\pm}^{(\frac{5}{2})}} 
+\frac{5}{2} {\bf W_{\pm}^{(\frac{5}{2})}}
\mp \frac{12}{(k+5)^2} 
 \pa \left(
\begin{array}{c}
F_{21}  \\
F_{12}
\end{array}
\right) F_{12} F_{21}
\right. 
\nonu \\
& \pm & \frac{i (3 k+13)}{(k+5)^2} A_3 
 \left(
\begin{array}{c}
G_{21}  \\
G_{12}
\end{array}
\right) +
\frac{2}{(k+5)^2} A_3 A_3 
 \left(
\begin{array}{c}
F_{21}  \\
F_{12}
\end{array}
\right)
\nonu \\
&-& \frac{4}{(k+5)^2} A_3 B_3
 \left(
\begin{array}{c}
F_{21}  \\
F_{12}
\end{array}
\right) +
\frac{2 (7 k+39)}{3 (k+5)^3} A_3 B_{\mp}
 \left(
\begin{array}{c}
F_{22}  \\
F_{11}
\end{array}
\right)
\nonu \\
&\pm & \frac{2 i (25 k+117)}{3 (k+5)^3} \pa A_3 
\left(
\begin{array}{c}
F_{21}  \\
F_{12}
\end{array}
\right) \mp
\frac{2 i (23 k+99)}{3 (k+5)^3} A_3
\pa
\left(
\begin{array}{c}
F_{21}  \\
F_{12}
\end{array}
\right)
\nonu \\
& \mp &
\frac{i (7 k+15)}{3 (k+5)^2} 
A_{\mp}
\left(
\begin{array}{c}
G_{11}  \\
G_{22}
\end{array}
\right)
-\frac{4 (k+3)}{3 (k+5)^3}
A_{\mp} A_3 
\left(
\begin{array}{c}
F_{11}  \\
F_{22}
\end{array}
\right)
\nonu \\
&+& \frac{8 (2 k+9)}{3 (k+5)^3} A_{\mp} B_3 
\left(
\begin{array}{c}
F_{11}  \\
F_{22}
\end{array}
\right)
+ \frac{6}{(k+5)^2} A_{\mp} B_{\mp} 
\left(
\begin{array}{c}
F_{12}  \\
F_{21}
\end{array}
\right)
\nonu \\
&\mp & \frac{52 i (k+3)}{3 (k+5)^3} \pa A_{\mp}
\left(
\begin{array}{c}
F_{11}  \\
F_{22}
\end{array}
\right) \pm
\frac{2 i (29 k+117)}{3 (k+5)^3} A_{\mp} \pa
\left(
\begin{array}{c}
F_{11}  \\
F_{22}
\end{array}
\right)
\nonu \\
&+& \frac{2 (k+9)}{3 (k+5)^3} A_{\pm} A_{\mp}
\left(
\begin{array}{c}
F_{21}  \\
F_{12}
\end{array}
\right) \mp
\frac{i (3 k+13)}{(k+5)^2} B_3
\left(
\begin{array}{c}
G_{21}  \\
G_{12}
\end{array}
\right)
\nonu \\
&+& \frac{2}{(k+5)^2} B_3 B_3 
\left(
\begin{array}{c}
F_{21}  \\
F_{12}
\end{array}
\right) \pm
\frac{2 i \left(6 k^2+35 k+33\right)}{3 (k+5)^3} 
\pa B_3 
\left(
\begin{array}{c}
F_{21}  \\
F_{12}
\end{array}
\right)
\nonu \\
&\mp & \frac{2 i \left(6 k^2+31 k+21\right)}{3 (k+5)^3}
B_3 \pa 
\left(
\begin{array}{c}
F_{21}  \\
F_{12}
\end{array}
\right) \pm
\frac{2 i (k+15)}{3 (k+5)^2} 
B_{\mp}
\left(
\begin{array}{c}
G_{22}  \\
G_{11}
\end{array}
\right)
\nonu \\
&- & \frac{2 (k+9)}{3 (k+5)^3} B_{\mp} B_3 
\left(
\begin{array}{c}
F_{22}  \\
F_{11}
\end{array}
\right) \mp
\frac{2 i (k+9) (3 k+4)}{3 (k+5)^3}
\pa B_{\mp} 
\left(
\begin{array}{c}
F_{22}  \\
F_{11}
\end{array}
\right)
\nonu \\
&\pm & \frac{4 i \left(4 k^2+23 k-3\right)}{3 (k+5)^3} 
B_{\mp} \pa
\left(
\begin{array}{c}
F_{22}  \\
F_{11}
\end{array}
\right) +
\frac{4 (k+3)}{3 (k+5)^3} 
B_{\pm} B_{\mp}
\left(
\begin{array}{c}
F_{21}  \\
F_{12}
\end{array}
\right)
\nonu \\
&+& \frac{8 (k+6)}{3 (k+5)^3} 
\left(
\begin{array}{c}
F_{11} F_{21} G_{22}  \\
F_{22} F_{12} G_{11}
\end{array}
\right) +
\frac{4 \left(9 k^2+82 k+201\right)}{3 (k+5)^4} 
\left(
\begin{array}{c}
\pa F_{11} F_{21} F_{22}  \\
\pa F_{22} F_{12} F_{11}
\end{array}
\right)
\nonu \\
&-& \frac{32 \left(k^2+10 k+27\right)}{3 (k+5)^4}
\left(
\begin{array}{c}
F_{11} F_{21} \pa F_{22}  \\
F_{22} F_{12} \pa F_{11}
\end{array}
\right)
\mp \frac{2}{(k+5)^2} F_{11} F_{22}
 \left(
\begin{array}{c}
G_{21}   \\
G_{12} 
\end{array}
\right)
\nonu \\
&+& \frac{4 (k+3) (k+9)}{3 (k+5)^3} \pa^2 
 \left(
\begin{array}{c}
F_{21}   \\
F_{12} 
\end{array}
\right) +
\frac{2 (5 k+21)}{3 (k+5)^3} 
 \left(
\begin{array}{c}
F_{21}  F_{22} G_{11}  \\
F_{12} F_{11} G_{22}
\end{array}
\right)
\nonu \\
&+& \frac{(k-3)}{3 (k+5)^2} \pa 
 \left(
\begin{array}{c}
G_{21}   \\
G_{12} 
\end{array}
\right) -\frac{4 (k+4)}{(k+5)^2}
T 
 \left(
\begin{array}{c}
F_{21}   \\
F_{12} 
\end{array}
\right) -\frac{(k+7)}{(k+5)^2}
U
 \left(
\begin{array}{c}
G_{21}   \\
G_{12} 
\end{array}
\right)
\nonu \\
&\pm & \frac{8 i}{(k+5)^2} U A_3 
 \left(
\begin{array}{c}
F_{21}   \\
F_{12} 
\end{array}
\right) \mp
\frac{4 i (5 k+27)}{3 (k+5)^3} U A_{\mp}
 \left(
\begin{array}{c}
F_{11}   \\
F_{22} 
\end{array}
\right)
\nonu \\
&\mp & \frac{8 i}{(k+5)^2} U B_3
\left(
\begin{array}{c}
F_{21}   \\
F_{12} 
\end{array}
\right) \pm
\frac{2 i (11 k+51)}{3 (k+5)^3}
U B_{\mp} 
\left(
\begin{array}{c}
F_{22}   \\
F_{11} 
\end{array}
\right)
\nonu \\
&+& \frac{2 (k-3)}{3 (k+5)^3} \pa U
\left(
\begin{array}{c}
F_{21}   \\
F_{12} 
\end{array}
\right)
-\frac{10 (k-3)}{3 (k+5)^3} U
\pa
\left(
\begin{array}{c}
F_{21}   \\
F_{12} 
\end{array}
\right)
\nonu \\
&-& \left.   
\frac{6}{(k+5)^2} U U 
\left(
\begin{array}{c}
F_{21}   \\
F_{12} 
\end{array}
\right) 
\right](w)
\nonu \\ 
& + & \frac{1}{(z-w)} \, \left[ 
+ \cdots \right](w) +\cdots.
\nonu 
\eea
The OPEs are described up to the second order pole.
One combines two OPEs in one single OPE using the symmetry between the 
structure constants.
One sees that there is no higher spin-$\frac{3}{2}$ current 
in the third order pole of the above OPEs.

The equation (\ref{g1221w5half}) allows us to calculate the following OPE
\bea
%%%%%%%%%%%%%%%%%%%%%%%%%%%%%%%%%%%%%%%%%%%%%%%%%%%%%%%%%%%%%%%%%%%%%
{\bf T^{(2)}}(z) \, {\bf W^{(3)}}(w) & = & \frac{1}{(z-w)^4}
\, \left[ \frac{36 i \left(18 k^2+63 k+41\right)}{(k+5)^2 (13 k+17)} {A}_3 -
\frac{12 i k \left(8 k^2+75 k+95\right)}{(k+5)^2 (13 k+17)}  {B}_3  \right. \nonu \\
&+& \left. \frac{12 \left(8 k^3+51 k^2+182 k+123\right)}{(k+5)^3 (13 k+17)} 
F_{11} F_{22} -
\frac{12 (k-3) \left(8 k^2+45 k+41\right)}{(k+5)^3 (13 k+17)} 
F_{12} F_{21}
  \right](w)
\nonu \\
& + & \frac{1}{(z-w)^3} \, \left[ 
 -\frac{8 i \left(7 k^2+44 k+141\right)}{3 (k+5)^3 (13 k+17)} 
 A_{-} F_{11} F_{12}
+  \frac{8 i \left(7 k^2+44 k+141\right)}{3 (k+5)^3 (13 k+17)}  
 A_{+} F_{21} F_{22}   
\right. 
\nonu \\
& - &  \frac{4 i \left(25 k^2+158 k-27\right)}{3 (k+5)^3 (13 k+17)}  
B_{-} F_{12} F_{22}
+  \frac{4 i \left(25 k^2+158 k-27\right)}{3 (k+5)^3 (13 k+17)}   
B_{+} F_{11} F_{21}
\nonu \\ 
& - & \frac{(k-3) (89 k+205)}{3 (k+5)^2 (13 k+17)}  F_{11} G_{22}
 - \frac{4}{(k+5)^2}  \pa (F_{11} F_{22})
 \nonu \\
& + & 
\frac{(3 k+11)}{(k+5)^2} F_{12} G_{21}+
\frac{4 (k-3) (11 k+103)}{3 (k+5)^3 (13 k+17)}
\pa (F_{12} F_{21})
\nonu \\
&-& 
 \frac{(3 k+11)}{(k+5)^2}  F_{21} G_{12}
+  \frac{(k-3) (89 k+205)}{3 (k+5)^2 (13 k+17)}  F_{22} G_{11}
\nonu \\
& - & 
\frac{4 (k-3) (11 k+103)}{3 (k+5)^3 (13 k+17)}  U F_{11} F_{22}
+ \frac{4}{(k+5)^2}  U F_{12} F_{21}
-\frac{24 i}{(k+5)^2}  U A_3  \nonu \\
& - & \left. \frac{8 i k}{(k+5)^2}  U B_3
\right](w) 
\nonu \\
& + & \frac{1}{(z-w)^2} \, \left[-3  {\bf P^{(3)}} + 3 
{\bf W^{(3)}} 
-\frac{8 {\bf (k-3)}}{(13 k+17)} {\bf T^{(1)}} {\bf T^{(2)}} 
\right. \nonu \\
&- & \frac{2 i}{(k+5)^2} A_3 A_3 A_3 + 
+ \frac{6 i}{(k+5)^2}
A_3 A_3 B_3 
+ \frac{2}{(k+5)^2}
 \pa A_3 A_3
 \nonu \\
& - & \frac{6 i}{(k+5)^2}  A_3 B_3 B_3 +
 \frac{3 (3 k+11)}{(k+5)^2} \pa A_3 B_3  -\frac{3 (3 k+11)}{(k+5)^2}
 A_3 \pa B_3
\nonu \\
& - &  \frac{6 i}{(k+5)^2} A_3 B_{+} B_{-} -\frac{8 i}{(k+5)^2}
A_3 F_{11} G_{22}
+ \frac{2 i (5 k+29)}{(k+5)^3}  A_3 \pa F_{11} F_{22} 
\nonu \\
& - &     \frac{10 i (k+1)}{(k+5)^3} A_3 F_{11} \pa F_{22}
+   \frac{4 i}{(k+5)^2}  A_3 F_{22} G_{11}
-\frac{3 i}{(k+5)^2}
A_{-} F_{11} G_{12}
\nonu \\
& - &  \frac{i \left(431 k^2+3082 k+3675\right)}{3 (k+5)^3 (13 k+17)}
   \pa A_{-} F_{11} F_{12}
+  \frac{i \left(739 k^2+4610 k+4383\right)}{3 (k+5)^3 (13 k+17)}   
A_{-} \pa F_{11} F_{12}
\nonu \\
&+&  \frac{i \left(115 k^2+1298 k+1119\right)}{3 (k+5)^3 (13 k+17)} 
A_{-} F_{11} \pa F_{12}
+ \frac{i}{(k+5)^2}
A_{-} F_{12} G_{11}
-\frac{2 i}{(k+5)^2}
 A_{+} A_{-} A_3 
\nonu \\
& + & \frac{6 i}{(k+5)^2} A_{+ } A_{-} B_3 
- \frac{14}{(k+5)^2} \pa A_{+} A_{-}
+ \frac{14}{(k+5)^2} A_{+} \pa A_{-}
\nonu \\
&
- & \frac{5 i}{(k+5)^2}  A_{+} F_{21} G_{22}
-\frac{2 i \left(77 k^2+772 k+687\right)}{3 (k+5)^3 (13 k+17)} 
\pa A_{+} F_{21} F_{22}
\nonu \\
& + & 
 \frac{4 i \left(59 k^2+463 k+600\right)}{3 (k+5)^3 (13 k+17)}
A_{+} \pa F_{21} F_{22}
+ \frac{4 i \left(20 k^2+373 k+549\right)}{3 (k+5)^3 (13 k+17)} 
A_{+} F_{21} \pa F_{22}
\nonu \\
& + & \frac{3 i}{(k+5)^2}  A_{+} F_{22} G_{21}
+ \frac{2 i}{(k+5)^2}  B_3 B_3 B_3
-\frac{2}{(k+5)^2} \pa B_3 B_3
\nonu \\
&+& \frac{i \left(8 k^3+63 k^2+203 k-136\right)}{(k+5)^2 (13 k+17)}
  \pa^2 B_3
+ \frac{4 i}{(k+5)^2} B_3 F_{11} G_{22}
\nonu \\
& - &  \frac{6 i (k-3)}{(k+5)^3}
\pa B_3 F_{11} F_{22} -
\frac{40 i}{(k+5)^3} B_3 \pa F_{11} F_{22}
+ \frac{4 i (3 k+13)}{(k+5)^3} B_3 F_{11} \pa F_{22}
\nonu \\
& - &  \frac{8 i}{(k+5)^2}
B_3 F_{22} G_{11} +
\frac{i}{(k+5)^2}
B_{-} F_{12} G_{22}
- \frac{4 i \left(136 k^2+767 k+675\right)}{3 (k+5)^3 (13 k+17)}
\pa B_{-} F_{12} F_{22} 
\nonu \\
& + & \frac{4 i \left(59 k^2+229 k+294\right)}{3 (k+5)^3 (13 k+17)}
B_{-} \pa F_{12} F_{22}
+ 
\frac{4 i \left(176 k^2+1123 k+1263\right)}{3 (k+5)^3 (13 k+17)}
 B_{-} F_{12} \pa F_{22}
\nonu \\
& - & \frac{3 i}{(k+5)^2}  B_{-} F_{22} G_{12}
+ \frac{2 i}{(k+5)^2}  B_{+} B_{-} B_3 
+ \frac{(3 k+5)}{(k+5)^2} \pa B_{+} B_{-} 
\nonu \\
& - & \frac{(3 k+5)}{(k+5)^2} B_{+} \pa B_{-}
+ \frac{3 i}{(k+5)^2}
B_{+} F_{11} G_{21}
 -\frac{i \left(275 k^2+1162 k+1431\right)}{3 (k+5)^3 (13 k+17)}  
\pa B_{+} F_{11} F_{21}
\nonu \\
& + & \frac{i \left(349 k^2+1214 k+609\right)}{3 (k+5)^3 (13 k+17)}  
B_{+} \pa F_{11} F_{21}
+ \frac{i \left(349 k^2+1838 k+1425\right)}{3 (k+5)^3 (13 k+17)} 
B_{+} F_{11} \pa F_{21}
\nonu \\
& - & 
\frac{5 i}{(k+5)^2} B_{+} F_{21} G_{11}
-\frac{(8 k^3+521 k^2+1810 k+1633)}{(k+5)^3 (13 k+17)}
\pa^2 F_{11} F_{22}
\nonu \\
& - & \frac{2 \left(8 k^3-649 k^2-1826 k-1121\right)}{(k+5)^3 (13 k+17)} 
\pa F_{11} \pa F_{22}
\nonu \\
& - & 
\frac{(8 k^3+521 k^2+1810 k+1633)}{(k+5)^3 (13 k+17)} F_{11} \pa^2 F_{22}
-  \frac{(467 k^2+1009 k+330)}{3 (k+5)^2 (13 k+17)} \pa F_{11} G_{22}
\nonu \\
& + & \frac{(118 k^2+341 k+435)}{3 (k+5)^2 (13 k+17)} F_{11} \pa G_{22}
+ \frac{8 (k-3) \left(3 k^2+40 k+113\right)}{3 (k+5)^3 (13 k+17)} 
\pa^2 F_{12} F_{21}
\nonu \\
&+& \frac{4 (k-3) \left(12 k^2+121 k+401\right)}{3 (k+5)^3 (13 k+17)} 
\pa F_{12} \pa F_{21}
\nonu \\
& + & \frac{8 (k-3) \left(3 k^2+40 k+113\right)}{3 (k+5)^3 (13 k+17)} 
F_{12} \pa^2 F_{21} - \frac{3 (3 k+13)}{2 (k+5)^2}
\pa F_{12} G_{21}
\nonu \\
& + & \frac{(5 k+23)}{2 (k+5)^2} F_{12} \pa G_{21} 
- \frac{3 (3 k+11)}{2 (k+5)^2} \pa F_{21} G_{12}
+ \frac{(k+1)}{2 (k+5)^2}  F_{21} \pa G_{12}
\nonu \\
& - & \frac{7 \left(17 k^2+460 k+627\right)}{6 (k+5)^2 (13 k+17)} 
\pa F_{22} G_{11}
+ \frac{(115 k^2+1064 k+813)}{6 (k+5)^2 (13 k+17)}  F_{22} \pa G_{11}
\nonu \\
& - &  \frac{3}{(k+5)} G_{11} G_{22}
+ \frac{2 i (44 k+109)}{(k+5) (13 k+17)} T A_3 
- \frac{2 i \left(8 k^2+62 k-17\right)}{(k+5) (13 k+17)} T B_3
\nonu \\
& + &  \frac{3}{(k+5)}  \pa T
-\frac{6 (4 k+17)}{(k+5) (13 k+17)}  T U + 
\frac{16 \left(k^2+13 k+20\right)}{(k+5)^2 (13 k+17)}
T F_{11} F_{22}
\nonu \\
& - &  \frac{4 (k-3) (4 k+21)}{(k+5)^2 (13 k+17)} T F_{12} F_{21}
- \frac{6}{(k+5)^2}
 U A_3 A_3 
+ \frac{12}{(k+5)^2} U A_3 B_3 \nonu \\
& + & \frac{9 i}{(k+5)^2}
 \pa U A_3 
- \frac{27 i}{(k+5)^2}  U \pa A_3
- \frac{6}{(k+5)^2}  U A_{+} A_{-}
-  \frac{6}{(k+5)^2}  U B_3 B_3 
\nonu \\
& + &  \frac{i (k+6)}{(k+5)^2} \pa U B_3 -
 \frac{i (5 k+12)}{(k+5)^2} U \pa B_3
- \frac{6}{(k+5)^2} U B_{+} B_{-}
\nonu \\
& + & \frac{3 (4 k+17)}{(k+5) (13 k+17)}  \pa^2 U 
 -\frac{4}{(k+5)^2} U F_{11} G_{22}
+ \frac{2 \left(41 k^2+310 k+717\right)}{3 (k+5)^3 (13 k+17)}  
\pa U  F_{11} F_{22}
\nonu \\
&-& \frac{64 \left(17 k^2+91 k+78\right)}{3 (k+5)^3 (13 k+17)}
U  \pa F_{11} F_{22}
+ 
\frac{16 \left(49 k^2+374 k+453\right)}{3 (k+5)^3 (13 k+17)}
U F_{11} \pa F_{22}
\nonu \\
& + & \frac{5}{(k+5)^2}  \pa U F_{12} F_{21}
-\frac{21}{(k+5)^2}  U \pa F_{12} F_{21}
+ \frac{15}{(k+5)^2} U F_{12} \pa F_{21}
\nonu \\
& +&  \frac{4}{(k+5)^2} U F_{22} G_{11} 
-\frac{2 i}{(k+5)^2}  U U A_3
+ \frac{2 i}{(k+5)^2} U U B_3
\nonu \\
& - & \frac{6}{(k+5)^2}  U U U 
+ \frac{24}{(k+5)^3}  F_{22} G_{11} F_{22} F_{11}
- \frac{i \left(44 k^2+173 k+341\right)}{(k+5)^2 (13 k+17)} 
\pa^2 A_3
\nonu \\
&- & \frac{15 i}{(k+5)^2}
\pa A_3 F_{12} F_{21}
+ \frac{15 i}{(k+5)^2} A_3 \pa (F_{12} F_{21})
\nonu \\
& + & \left.
\frac{15 i}{(k+5)^2} \pa B_3 F_{12} F_{21}
- \frac{15 i}{(k+5)^2} B_3 \pa (F_{12} F_{21})
\right](w) 
\nonu \\
& + &  \frac{1}{(z-w)} \, \left[ 
+\cdots \right](w)  +\cdots.
\nonu
\eea
There is no higher spin-$1$ current in the fourth order pole 
 in the above OPE.
Even there is no higher spin-$2$ current in the 
third order pole which was present in the nonlinear version.
One can also remove the term between the higher spin currents  
by adding the $T {\bf T^{(1)}}(w)$
into the left hand side in the higher spin-$3$ current
because the OPE ${\bf T^{(2)}}(z) \,
T {\bf T^{(1)}}(w)$ contributes to the above ${\bf T^{(1)}} \, 
{\bf T^{(2)}}(w)$ term where the factor $(k-3)$ exists.

%%%%%%%%%%%%%%%%%%%%%%%%%%%%%%%%%%%%%%%%%%%%%%%%%%%%%%%%%%%%%%%%%%%%%
\subsection{ The OPEs containing the second or third  ${\cal N}=2$ multiplets}
%C%%%%%%%%%%%%%%%%%%%%%%%%%%%%%%%%%%%%%%%%%%%%%%%%%%%%%%%%%%%%%%%%%%%

From the explicit result in (\ref{g1122t1}) for the higher spin-$\frac{3}{2}$
current one can obtain the following OPEs
%%%%%%%%%%%%%%%%%%%%%%%%%%%%%%%%%%%%%%%%%%%%%%%%%%%%%%%%%%%%
\bea
\left(
\begin{array}{c}
{\bf U^{(\frac{3}{2})}} \\
{\bf V^{(\frac{3}{2})}} 
\end{array}
\right)(z) \,
\left(
\begin{array}{c}
{\bf U^{(\frac{3}{2})}} \\
{\bf V^{(\frac{3}{2})}} 
\end{array}
\right)(w) & =& 
\frac{1}{(z-w)} \, \left[ \frac{1}{(k+5)} A_{\pm} B_{\mp} 
\pm \frac{2 i}{(k+5)^2} A_{\pm}
\left(
\begin{array}{c}
F_{11} F_{21} \\
F_{22} F_{12} 
\end{array}
\right)
\right. \nonu \\
& \mp & \left. \frac{2 i}{(k+5)^2} B_{\mp} 
 \left(
\begin{array}{c}
F_{11} F_{12} \\
F_{22} F_{21} 
\end{array}
\right)
-\frac{2}{(k+5)^2} 
 \left(
\begin{array}{c}
F_{11} \pa F_{11} \\
F_{22} \pa F_{22} 
\end{array}
\right)
\right](w) + \cdots.
\nonu 
\eea
These are new OPEs and the corresponding OPEs in the nonlinear version
are trivial.

Furthermore the result of (\ref{g1221vu3half}) implies the following OPEs
%%%%%%%%%%%%%%%%%%%%%%%%%%%%%%%%%%%%%%%%%%%%%%%%%%%%%%%%%%%%%%%%%
\bea
\left(
\begin{array}{c}
{\bf U^{(\frac{3}{2})}} \\
{\bf V^{(\frac{3}{2})}} 
\end{array}
\right)(z) \,
\left(
\begin{array}{c}
{\bf U_{+}^{(2)}} \\
{\bf V_{-}^{(2)}} 
\end{array}
\right)(w) & =& \frac{1}{(z-w)^2}
\,
\left[ \pm \frac{2 i k}{(k+5)^2} 
\left(
\begin{array}{c}
F_{11} \\
F_{22} 
\end{array}
\right) B_{\mp}
\right](w) 
\nonu \\
&+& \frac{1}{(z-w)} \, \left[ 
-\frac{1}{(k+5)^2} A_3 B_{\mp}
\left(
\begin{array}{c}
F_{11} \\
F_{22} 
\end{array}
\right)
\right. \nonu \\
&- & \frac{1}{(k+5)^2} A_{\pm} B_{\mp}
\left(
\begin{array}{c}
F_{21} \\
F_{12} 
\end{array}
\right)
 \mp \frac{i}{2 (k+5)} B_{\mp}
\left(
\begin{array}{c}
G_{11} \\
G_{22} 
\end{array}
\right)
\nonu \\
&+& \frac{1}{(k+5)^2} B_{\mp} B_3 
\left(
\begin{array}{c}
F_{11} \\
F_{22} 
\end{array}
\right) +
\frac{1}{(k+5)^2} B_{\mp} B_{\mp} 
\left(
\begin{array}{c}
F_{12} \\
F_{21} 
\end{array}
\right)
\nonu \\
& \mp & \frac{3 i}{(k+5)^2} B_{\mp} \pa
\left(
\begin{array}{c}
F_{11} \\
F_{22} 
\end{array}
\right) + 
\frac{1}{(k+5)^2} 
\left(
\begin{array}{c}
F_{11} F_{21} G_{11} \\
F_{22} F_{12} G_{22}
\end{array}
\right)
\nonu \\
&\mp & \frac{i}{(k+5)^2} U B_{\mp} 
\left(
\begin{array}{c}
F_{11}  \\
F_{22} 
\end{array}
\right) \pm
\frac{i k}{(k+5)^2} 
\left(
\begin{array}{c}
F_{11}  \\
F_{22} 
\end{array}
\right) \pa B_{\mp}
\nonu \\
&+ & \left. \frac{2 (k+1)}{(k+5)^3}
\left(
\begin{array}{c}
F_{11} F_{21} \pa F_{11} \\
F_{22} F_{12} \pa F_{22}
\end{array}
\right) 
\right](w)
+\cdots.
\nonu 
\eea
These are also 
new OPEs and the corresponding OPEs in the nonlinear version
are trivial.

Similarly, the following OPEs can be obtained from the previous 
result in (\ref{g1221uv3half})
%%%%%%%%%%%%%%%%%%%%%%%%%%%%%%%%%%%%%%%%%%%%%%%%%%%%%%%%%%%%%%%
\bea
\left(
\begin{array}{c}
{\bf U^{(\frac{3}{2})}} \\
{\bf V^{(\frac{3}{2})}} 
\end{array}
\right)(z) \,
\left(
\begin{array}{c}
{\bf U_{-}^{(2)}} \\
{\bf V_{+}^{(2)}} 
\end{array}
\right)(w) & =& 
\frac{1}{(z-w)^2} \,
\left[ \pm \frac{6 i}{(k+5)^2} 
\left(
\begin{array}{c}
F_{11}  \\
F_{22} 
\end{array}
\right) A_{\pm}
\right](w) 
\nonu \\
&+& \frac{1}{(z-w)} \,
\left[ 
\pm  \frac{i}{2 (k+5)} 
A_{\pm} 
\left(
\begin{array}{c}
G_{11} \\
G_{22} 
\end{array}
\right)
\right. \nonu \\
&- & \frac{1}{(k+5)^2}  A_{\pm} A_3
\left(
\begin{array}{c}
F_{11} \\
F_{22} 
\end{array}
\right)
- \frac{1}{(k+5)^2}
A_{\pm} A_{\pm}
\left(
\begin{array}{c}
F_{21} \\
F_{12} 
\end{array}
\right)
\nonu \\
&+& \frac{1}{(k+5)^2} A_{\pm} B_3 
\left(
\begin{array}{c}
F_{11} \\
F_{22} 
\end{array}
\right) +
\frac{1}{(k+5)^2}
 A_{\pm} B_{\mp} 
\left(
\begin{array}{c}
F_{12} \\
F_{21} 
\end{array}
\right)
\nonu \\
& + &  \frac{1}{(k+5)^2} 
\left(
\begin{array}{c}
F_{11} F_{12} G_{11} \\
F_{22} F_{21} G_{22}
\end{array}
\right) \pm 
\frac{i}{(k+5)^2} 
U A_{\pm}
\left(
\begin{array}{c}
F_{11}  \\
F_{22} 
\end{array}
\right)
\nonu \\
&\mp & \frac{3 i}{(k+5)^2} A_{\pm} \pa 
\left(
\begin{array}{c}
F_{11}  \\
F_{22} 
\end{array}
\right) \pm
\frac{3 i}{(k+5)^2}
\left(
\begin{array}{c}
F_{11}  \\
F_{22} 
\end{array}
\right) \pa A_{\pm}
\nonu \\
&- & \left. 
\frac{8}{(k+5)^3}
\left(
\begin{array}{c}
F_{11} F_{12} \pa F_{11} \\
F_{22} F_{21} \pa F_{22}
\end{array}
\right) 
\right](w) +\cdots.
\nonu 
\eea
These are new OPEs and the corresponding OPEs in the nonlinear version
are trivial.

From the information of the higher spin-$\frac{5}{2}$ currents in 
(\ref{g1221vu2}), the following OPEs can be obtained
%%%%%%%%%%%%%%%%%%%%%%%%%%%%%%%%%%%%%%%%%%%%%%%%%%%%%%%%%%%%%%%%%%%
\bea
\left(
\begin{array}{c}
{\bf U^{(\frac{3}{2})}} \\
{\bf V^{(\frac{3}{2})}} 
\end{array}
\right)(z) \,
\left(
\begin{array}{c}
{\bf U^{(\frac{5}{2})}} \\
{\bf V^{(\frac{5}{2})}} 
\end{array}
\right)(w) & =& 
\frac{1}{(z-w)^2} \, \left[ \pm 
\frac{2 (k-3)}{3 (k+5)^2} A_{\pm} B_{\mp} 
-\frac{2 i (k-3)}{3 (k+5)^3}
A_{\pm} 
\left(
\begin{array}{c}
F_{11} F_{21} \\
F_{22} F_{12} 
\end{array}
\right) \right. \nonu \\
&+& \frac{2 i (k-3)}{3 (k+5)^3} B_{\mp} 
\left(
\begin{array}{c}
F_{11} F_{12} \\
F_{22} F_{21} 
\end{array}
\right) \mp 
\frac{(k+3)}{(k+5)^2} 
\left(
\begin{array}{c}
F_{11} G_{11} \\
F_{22} G_{22} 
\end{array}
\right)
\nonu \\
& \pm & \left. \frac{8 (k-3)}{3 (k+5)^3} 
\left(
\begin{array}{c}
F_{11} \pa F_{11} \\
F_{22} \pa F_{22} 
\end{array}
\right)
\right](w)
\nonu \\
&+ & \frac{1}{(z-w)} \, \left[ 
\mp \frac{(4 k+21)}{3 (k+5)^2}
\pa A_{\pm} B_{\mp}
\pm \frac{(5 k+18)}{3 (k+5)^2} A_{\pm} \pa B_{\mp}
\right. \nonu \\
&+& \frac{i}{(k+5)^2} A_{\pm} 
\left(
\begin{array}{c}
F_{11} G_{21} \\
F_{22} G_{12} 
\end{array}
\right) -\frac{8 i (k+3)}{3 (k+5)^3} 
\pa A_{\pm} 
\left(
\begin{array}{c}
F_{11} F_{21} \\
F_{22} F_{12} 
\end{array}
\right)
\nonu \\
&+& \frac{4 i (k+3)}{3 (k+5)^3} 
A_{\pm}
\left(
\begin{array}{c}
\pa F_{11} F_{21} \\
\pa F_{22} F_{12} 
\end{array}
\right) +
\frac{2 i (5 k+21)}{3 (k+5)^3} 
A_{\pm} 
\left(
\begin{array}{c}
F_{11} \pa F_{21} \\
F_{22} \pa F_{12} 
\end{array}
\right)
\nonu \\
&-& \frac{i}{(k+5)^2} A_{\pm} 
\left(
\begin{array}{c}
F_{21} G_{11} \\
F_{12} G_{22} 
\end{array}
\right) -\frac{i}{(k+5)^2}
B_{\mp} 
\left(
\begin{array}{c}
F_{11} G_{12} \\
F_{22} G_{21} 
\end{array}
\right)
\nonu \\
&-& \frac{4 i (k+9)}{3 (k+5)^3} \pa B_{\mp} 
\left(
\begin{array}{c}
F_{11} F_{12} \\
F_{22} F_{21} 
\end{array}
\right)
+ \frac{2 i (k+9)}{3 (k+5)^3} B_{\mp} 
\left(
\begin{array}{c}
\pa F_{11} F_{12} \\
\pa F_{22} F_{21} 
\end{array}
\right)
\nonu \\
&+& \frac{8 i (k+6)}{3 (k+5)^3} 
B_{\mp} 
\left(
\begin{array}{c}
F_{11} \pa F_{12} \\
F_{22} \pa F_{21} 
\end{array}
\right) +
\frac{i}{(k+5)^2} B_{\mp}
\left(
\begin{array}{c}
F_{12} G_{11} \\
F_{21} G_{22} 
\end{array}
\right)
\nonu \\
&\mp & \frac{(k+3)}{(k+5)^2} 
\left(
\begin{array}{c}
\pa F_{11} G_{11} \\
\pa F_{22} G_{22} 
\end{array}
\right) \mp
\frac{2}{(k+5)^2} U A_{\pm} B_{\mp}
\nonu \\
&+& \frac{2 i}{(k+5)^2} 
\left(
\begin{array}{c}
F_{11} \pa F_{11} \\
F_{22} \pa F_{22} 
\end{array}
\right) A_3
+ \frac{2 i}{(k+5)^2} 
\left(
\begin{array}{c}
F_{11} \pa F_{11} \\
F_{22} \pa F_{22} 
\end{array}
\right) B_3
\nonu \\
& \pm & \left. \frac{2}{(k+5)^2} 
\left(
\begin{array}{c}
F_{11} \pa F_{11} \\
F_{22} \pa F_{22} 
\end{array}
\right) U
\pm \frac{2(k-3)}{3(k+5)^3}
\left(
\begin{array}{c}
F_{11} \pa^2 F_{11} \\
F_{22} \pa^2 F_{22} 
\end{array}
\right)
\right](w) + \cdots.
\nonu 
\eea
As before, one has the above combined OPEs 
due to the symmetry of structure constants of two OPEs 
and there are no nonlinear terms in the spirit of footnote 
\ref{nonlineardef}.

Now one can construct the following OPE 
%%%%%%%%%%%%%%%%%%%%%%%%%%%%%%%%%%%%%%%%%%%%%%%%%%%%%%%
\bea
{\bf U^{(\frac{3}{2})}}(z) \,
{\bf V^{(\frac{3}{2})}}(w) 
& = & \frac{1}{(z-w)^3} \, \left[-\frac{(7 k+4) }{(k+5)} \right]
\nonu \\
& +  & \frac{1}{(z-w)^2} \, 
\left[-\frac{7 i}{(k+5)} A_3 + \frac{i (2 k+1)}{(k+5)} B_3 
- {\bf T^{(1)}}  \right. \nonu \\
& - & \left. \frac{(k+3)}{(k+5)^2} F_{11} F_{22} +
\frac{(k-3)}{(k+5)^2} F_{12} F_{21}
\right](w) \nonu \\
&+& \frac{1}{(z-w)} \, 
\left[ -\frac{1}{2} \pa  {\bf T^{(1)}} 
-  {\bf W^{(2)}} - T  
\right. \nonu \\
&-& 
\frac{1}{2 (k+5)} A_3 A_3-
\frac{1}{(k+5)} A_3 B_3 
  -\frac{4 i}{(k+5)} \pa A_3 
 + \frac{2 i}{(k+5)^2} A_3 F_{12} F_{21}
\nonu \\
& + & \frac{i}{(k+5)^2}
A_{-} F_{11} F_{12} -
\frac{1}{2 (k+5)} A_{+} A_{-}
+ \frac{i}{(k+5)^2}
 A_{+} F_{21} F_{22}    
\nonu \\
&-& \frac{1}{2 (k+5)} B_3 B_3 + 
\frac{i k}{(k+5)} \pa B_3
+ \frac{2 i}{(k+5)^2} B_3 F_{12} F_{21}
\nonu \\
& - &  \frac{i}{(k+5)^2}
B_{-} F_{12} F_{22}
-\frac{1}{2 (k+5)} B_{+} B_{-}
    -\frac{i}{(k+5)^2}
B_{+} F_{11} F_{21}
\nonu \\ 
& - & 
 \frac{1}{(k+5)}  \pa F_{11} F_{22}
+ \frac{2}{(k+5)^2} F_{11} \pa F_{22}
 -\frac{6}{(k+5)^2}
\pa F_{12} F_{21}
  \nonu \\
& + & \left.  
\frac{(k+3)}{(k+5)^2} F_{12} \pa F_{21}
-\frac{1}{2(k+5)} U U
\right](w) +  
\cdots.
\nonu 
\eea
Compared to the nonlinear version of this OPE, 
the above OPE is rather complicated.

From the results in section $3$, one obtains the following OPE
\bea
%%%%%%%%%%%%%%%%%%%%%%%%%%%%%%%%%%%%%%%%%%%%%%%%%%%%%%%%%%%%%%%%
{\bf U^{(\frac{3}{2})}}(z) \, {\bf V_{+}^{(2)}}(w) & = & 
\frac{1}{(z-w)^3} \, \left[-\frac{6 k}{(k+5)^2} F_{21} \right](w)
\nonu \\
&+& 
\frac{1}{(z-w)^2} \,  \left[  \frac{2 (k+3)}{(k+5)}
{\bf T_{+}^{(\frac{3}{2})}} 
+ \frac{1}{(k+5)} G_{21}
-\frac{2 (k+2)}{(k+5)^2} U F_{21}
 \right. \nonu \\
&+& \frac{2}{(k+5)^2} F_{21} F_{11} F_{22}
+ \frac{2 i (k-1)}{(k+5)^2} F_{21} A_3
-\frac{2 i (k+2)}{(k+5)^2} F_{11} A_{-}
\nonu \\
&+ & \left. \frac{4 i}{(k+5)^2} F_{22} B_{-}
-\frac{4 i}{(k+5)^2} F_{21} B_3 
\right](w) 
\nonu \\
& + & \frac{1}{(z-w)} \, \left[   \frac{1}{2} 
{\bf P_{+}^{(\frac{5}{2})} } + {\bf W_{+}^{(\frac{5}{2})}} 
+ \frac{2 (k+3)}{3 (k+5)} \pa 
{\bf T_{+}^{(\frac{3}{2})}} 
\right. \nonu \\
& - & \frac{i}{2 (k+5)} A_3 G_{21} 
-\frac{2}{(k+5)^2} A_3 A_3 F_{21} 
-\frac{2}{(k+5)^2} A_3 B_3 F_{21}
\nonu \\
&+& \frac{2}{(k+5)^2} A_3 B_{-} F_{22}
+ \frac{2 i (k-3)}{3 (k+5)^2} \pa A_3 F_{21}
+ \frac{2 i (k+9)}{3 (k+5)^2} A_3 \pa F_{21}
\nonu \\
&+& \frac{1}{(k+5)^2} A_{-} A_3 F_{11}
+ \frac{1}{(k+5)^2} A_{-} B_3 F_{11}
+ \frac{1}{(k+5)^2} A_{-} B_{-} F_{12}
\nonu \\
&-& \frac{i (2 k+3)}{3 (k+5)^2} \pa A_{-} F_{11}
- \frac{i (2 k+9)}{3 (k+5)^2} A_{-} \pa F_{11}
-\frac{1}{(k+5)^2} A_{+} A_{-} F_{21}
\nonu \\
&-& \frac{i}{2 (k+5)} B_3 G_{21} + 
\frac{i (k-3)}{3 (k+5)^2}
\pa B_3 F_{21} 
-\frac{2 i k}{3 (k+5)^2}  B_3 \pa F_{21}
\nonu \\
&+& \frac{i}{2 (k+5)}  B_{-} G_{22}
-\frac{i (k-3)}{3 (k+5)^2} \pa B_{-} \pa F_{22}
+ \frac{2 i k}{3 (k+5)^2} B_{-} \pa F_{22}
\nonu \\
& + & \frac{4 (k+3)}{3 (k+5)^3} \pa F_{11} F_{21} F_{22}
-\frac{2 (k+9)}{3 (k+5)^3} F_{11} \pa F_{21} F_{22}
\nonu \\
& - & \frac{8 (k+3)}{3 (k+5)^3} F_{11} F_{21} \pa F_{22}
-\frac{1}{(k+5)^2} F_{11} F_{22} G_{21}
\nonu \\
& - & \frac{1}{(k+5)^2} F_{12} F_{21} G_{21}
+ \frac{3 (k+3)}{2 (k+5)^2} \pa^2 F_{21}
+ \frac{1}{(k+5)^2} F_{21} F_{22} G_{11}
\nonu \\ 
&-& \frac{1}{(k+5)} T F_{21}
-\frac{1}{2 (k+5)} U G_{21}
-\frac{i}{(k+5)^2} U A_{-} F_{11}
\nonu \\
&-& \frac{2 i}{(k+5)^2} U B_3 F_{21}
+ \frac{2 i}{(k+5)^2} U B_{-} F_{22}
-\frac{2 (k+3)}{3 (k+5)^2} \pa U F_{21}
\nonu \\
&-& \left. \frac{2 (k+3)}{3 (k+5)^2}
U \pa F_{21} -\frac{2}{(k+5)^2}
U U F_{21} -\frac{2 (k+1)}{(k+5)^3}
F_{21} \pa F_{21} F_{12}
\right](w) +\cdots.
\nonu 
\eea
Note the spin-$\frac{3}{2}$ current dependent terms in the 
above OPE which were not present in the corresponding OPE
in the nonlinear version.

Similarly one has the following OPE
\bea
%%%%%%%%%%%%%%%%%%%%%%%%%%%%%%%%%%%%%%%%%%%%%%%%%%%%%%%%%%%%%%%%%
{\bf U^{(\frac{3}{2})}}(z) \, {\bf V_{-}^{(2)}}(w) & = & 
\frac{1}{(z-w)^3} \, \left[ \frac{6 k}{(k+5)^2}  F_{12} \right](w)
\nonu \\
&+&
\frac{1}{(z-w)^2} \, \left[
\frac{(k+9)}{(k+5)}  {\bf T_{-}^{(\frac{3}{2})}}  
-\frac{(k+7)}{(k+5)^2} U F_{12}
-\frac{2}{(k+5)^2} F_{12} F_{11} F_{22}
\right. \nonu \\
&-& \frac{i (k+1)}{(k+5)^2} F_{12} A_3
+ \frac{i (k+1)}{(k+5)^2} F_{22} A_{+}
- \frac{i (k+7)}{(k+5)^2} F_{11} B_{+}
\nonu \\
&- & \left. \frac{i (k-7)}{(k+5)^2} F_{12} B_3 
-\frac{(k+8)}{(k+5)} G_{12}
\right](w) 
\nonu \\
& + & \frac{1}{(z-w)} \, \left[
\frac{1}{2} {\bf P_{-}^{(\frac{5}{2})} } 
+ \frac{(k+9)}{3 (k+5)} \pa {\bf T_{-}^{(\frac{3}{2})}}  
+ \frac{i}{2 (k+5)} A_3 G_{12}
\right. \nonu \\
&+& \frac{2}{(k+5)^2} A_3 B_3 F_{12}
-\frac{1}{(k+5)^2} A_3 B_{+} F_{11}
-\frac{i (k-3)}{3 (k+5)^2} \pa A_3 F_{12}
\nonu \\
&-& \frac{i (k+3)}{3 (k+5)^2} A_3 \pa F_{12}
-\frac{i}{2 (k+5)} A_{+} G_{22}
-\frac{2}{(k+5)^2} A_{+} B_3 F_{22}
\nonu \\
&-& \frac{1}{(k+5)^2} A_{+} B_{+} F_{21}
+ \frac{i (k-3)}{3 (k+5)^2} \pa A_{+} F_{22}
+ \frac{i (k+3)}{3 (k+5)^2} A_{+} \pa F_{22}
\nonu \\
&+& \frac{i}{2 (k+5)} B_3 G_{12}
+ \frac{2}{(k+5)^2} B_3 B_3 F_{12}
-\frac{2 i (k-6)}{3 (k+5)^2} \pa B_3 F_{12}
\nonu \\
&+& \frac{i (k+21)}{3 (k+5)^2} B_3 \pa F_{12}
-\frac{1}{(k+5)^2} B_{+} B_3 F_{11}
+ \frac{1}{(k+5)^2} B_{+} B_{-} F_{12}
\nonu \\
&- & \frac{i (k+6)}{3 (k+5)^2} \pa B_{+} F_{11}
-\frac{i (k+12)}{3 (k+5)^2} B_{+} \pa F_{11}
-\frac{2 (k+9)}{3 (k+5)^3} \pa F_{11} F_{12} F_{22}
\nonu \\
&+& \frac{4 (k+3)}{3 (k+5)^3} F_{11} \pa F_{12} F_{22}
+\frac{4 (k+9)}{3 (k+5)^3} F_{11} F_{12} \pa F_{22}
-\frac{1}{(k+5)^2} F_{11} F_{22} G_{12}
\nonu \\
&-&\frac{(k+15)}{2 (k+5)^2} \pa^2 F_{12}
-\frac{(k+9)}{3 (k+5)} \pa G_{12}
-\frac{(k+9)}{3 (k+5)^2} \pa U F_{12} 
\nonu \\
&-& \frac{(k+9)}{3 (k+5)^2} U \pa F_{12} 
+ \frac{1}{(k+5)^2} F_{12} F_{21} G_{12}
+ \frac{1}{(k+5)^2} F_{12} F_{22} G_{11}
\nonu \\
&+& \frac{1}{(k+5)} T F_{12}
+\frac{1}{2 (k+5)} U G_{12}
+\frac{2 i}{(k+5)^2} U A_3 F_{12}
\nonu \\
&-& \frac{2 i}{(k+5)^2} U A_{+} F_{22}
+  \frac{i}{(k+5)^2} U B_{+} F_{11}
+ \frac{2}{(k+5)^2} U U F_{12}
\nonu \\
&+ & \left. \frac{16 i}{(k+5)^3} F_{12} B_3 F_{12} F_{21}
\right](w) +\cdots.
\nonu 
\eea
In this case also the nonlinear terms appeared in the corresponding OPE 
in the nonlinear version disappear. 

One can also construct the following OPE from the 
previous results
%%%%%%%%%%%%%%%%%%%%%%%%%%%%%%%%%%%%%%%%%%%%%%%%%%%%%%%%%%%%%%%%%%%%
\bea
{\bf U^{(\frac{3}{2})}}(z) \, {\bf V^{(\frac{5}{2})}}(w) & = & 
\frac{1}{(z-w)^3} \, \left[  
\frac{2 i (5 k+18)}{(k+5)^2} {A}_3  + \frac{2 i k (4 k+21)}{3 (k+5)^2} {B}_3  -
\frac{4 (k-3)}{3 (k+5)} {\bf T^{(1)}} 
\right. \nonu \\
& - & \left. \frac{6 k}{(k+5)^2} U 
-\frac{4 (k-3) (2 k+9)}{3 (k+5)^3} F_{11} F_{22}
+ \frac{4 \left(2 k^2+9 k+27\right)}{3 (k+5)^3} F_{12} F_{21}
\right](w)
\nonu \\
& + & \frac{1}{(z-w)^2} \, \left[ \frac{1}{2} {\bf P^{(2)}} - 
\frac{8 (k+3)}{3 (k+5)}  ({\bf T^{(2)}} 
+  {\bf W^{(2)}})  
\right. \nonu \\
& + & \frac{4 (k-3)}{3 (k+5)^2}  A_3 B_3
-\frac{2 i (k-3)}{3 (k+5)^3} A_3 F_{11} F_{22}
+ \frac{2 i (k-3)}{3 (k+5)^3} A_3 F_{12} F_{21}
\nonu \\
&+& \frac{2 i (k+21)}{3 (k+5)^3} A_{-} F_{11} F_{12}
-\frac{2 i (k+21)}{3 (k+5)^3} A_{+} F_{21} F_{22}
+ \frac{2 i (k-3)}{3 (k+5)^3} B_3 F_{11} F_{22}
\nonu \\
&+& \frac{2 i (k-3)}{3 (k+5)^3} B_3 F_{12} F_{21}
+ \frac{16 i (k+3)}{3 (k+5)^3} B_{-} F_{12} F_{22}
- \frac{16 i (k+3)}{3 (k+5)^3} B_{+} F_{11} F_{21}
\nonu \\
&+& \frac{5 (k+3)}{3 (k+5)^2} F_{11} G_{22}
+ \frac{2 (11 k+39)}{3 (k+5)^3} \pa F_{11} F_{22}
+ \frac{2 (7 k+51)}{3 (k+5)^3} F_{11} \pa F_{22}
\nonu \\
&-& \frac{2 (k+3)}{(k+5)^2} F_{12} G_{21}
-\frac{2 (3 k-1)}{(k+5)^3} \pa F_{12} F_{21}
-\frac{2 (5 k+9)}{3 (k+5)^3} F_{12} \pa F_{21}
\nonu \\
&+& \frac{(k+3)}{(k+5)^2} F_{21} G_{12}
-\frac{2 (k+3)}{3 (k+5)^2} F_{22} G_{11}
-\frac{12 i}{(k+5)^2} U A_3 
\nonu \\
&+ & \left. \frac{2 (7 k+3)}{3 (k+5)^3} U F_{11} F_{22}
-\frac{6}{(k+5)^2} U F_{12} F_{21}
\right](w)
\nonu \\
& + & \frac{1}{(z-w)} \, \left[ \frac{1}{2} {\bf S^{(3)}} +\frac{1}{8} 
  \pa 
{\bf P^{(2)}}  -{\bf W^{(3)}}  
-\frac{2 (k+3)}{3 (k+5)} \pa ({\bf T^{(2)}} +{\bf W^{(2)}})
\right. \nonu \\
&-& 
\frac{2 {\bf (k-3)}}{(13 k+17)} \left( T {\bf T^{(1)}}
-\frac{1}{2} \pa^2 {\bf T^{(1)}} \right)
\nonu \\
& - & \frac{2 (4 k+21)}{3 (k+5)^2} \pa A_3 B_3
+ \frac{2 (5 k+18)}{3 (k+5)^2} A_3 \pa B_3
-\frac{i (19 k+87)}{6 (k+5)^3} \pa A_3 F_{11} F_{22}
\nonu \\
&+& \frac{i (17 k+93)}{6 (k+5)^3} A_3 \pa (F_{11} F_{22}) 
-\frac{2 i}{(k+5)^2} A_3 F_{12} G_{21}
\nonu \\
&+& \frac{2 i}{(k+5)^2} A_3 F_{21} G_{12}
-\frac{i}{(k+5)^2} A_{-} F_{11} G_{12}
-\frac{i (k+9)}{3 (k+5)^3} \pa A_{-} F_{11} F_{12}
\nonu \\
& - & \frac{i (k-15)}{3 (k+5)^3} A_{-} \pa F_{11} F_{12}
+ \frac{5 i (k+9)}{3 (k+5)^3} A_{-} F_{11} \pa F_{12}
+ \frac{i}{(k+5)^2} A_{-} F_{12} G_{11}
\nonu \\
&-& \frac{i}{(k+5)^2} A_{+} F_{21} G_{22}
-\frac{2 i (k+15)}{3 (k+5)^3} \pa A_{+} F_{21} F_{22}
+ \frac{4 i (k+6)}{3 (k+5)^3} A_{+} \pa F_{21} F_{22}
\nonu \\
&-& \frac{2 i (k+3)}{3 (k+5)^3} A_{+} F_{21} \pa F_{22}
+ \frac{i}{(k+5)^2} A_{+} F_{22} G_{21}
-\frac{i k (4 k+21)}{2 (k+5) (13 k+17)} \pa^2 B_3 
\nonu \\
&-& \frac{i (17 k+93)}{6 (k+5)^3} \pa B_3 F_{11} F_{22}
+ \frac{i (19 k+87)}{6 (k+5)^3} B_3 \pa (F_{11} F_{22})
\nonu \\
&-& \frac{2 i}{(k+5)^2} B_3 F_{12} G_{21}
+ \frac{i (k-51)}{6 (k+5)^3} \pa B_3 F_{12} F_{21}
+ \frac{i (k+45)}{6 (k+5)^3} B_3 \pa (F_{12} F_{21})
\nonu \\
&+& \frac{2 i}{(k+5)^2} B_3 F_{21} G_{12}
+ \frac{i}{(k+5)^2} B_{-} F_{12} G_{22}
-\frac{i (k-3)}{6 (k+5)^3} \pa B_{-} F_{12} F_{22}
\nonu \\
& + & \frac{i (23 k+75)}{6 (k+5)^3} B_{-} \pa F_{12} F_{22}
+ \frac{i (11 k+15)}{6 (k+5)^3} B_{-} F_{12} \pa F_{22}
-\frac{i}{(k+5)^2} B_{-} F_{22} G_{12}
\nonu \\
&+& \frac{i}{(k+5)^2} B_{+} F_{11} G_{21}
-\frac{i (17 k+45)}{6 (k+5)^3} \pa B_{+} F_{11} F_{21}
-\frac{i (5 k+33)}{6 (k+5)^3} B_{+} \pa F_{11} F_{21}
\nonu \\
&+& \frac{i (7 k+27)}{6 (k+5)^3} B_{+} F_{11} \pa F_{21}
-\frac{i}{(k+5)^2}  B_{+} F_{21} G_{11}
\nonu \\
& + &
\frac{(12 k^3+221 k^2+622 k-147)}{6 (k+5)^3 (13 k+17)}  \pa^2 F_{11} F_{22}
+ \frac{(4 k^2+45 k-3)}{(k+5)^2 (13 k+17)} \pa F_{11} \pa F_{22}
\nonu \\
&+& \frac{(12 k^3+169 k^2+710 k+57)}{6 (k+5)^3 (13 k+17)} F_{11} \pa^2 F_{22}
+ \frac{(5 k+33)}{12 (k+5)^2} \pa F_{11} G_{22}
\nonu \\
&+& \frac{(5 k+9)}{12 (k+5)^2} F_{11} \pa G_{22}
-\frac{(4 k^3+77 k^2+182 k+253)}{2 (k+5)^3 (13 k+17)} \pa^2 F_{12} F_{21}
\nonu \\
&-& \frac{(12 k^3+205 k^2+590 k+861)}{3 (k+5)^3 (13 k+17)} 
\pa F_{12} \pa F_{21}
-\frac{(5 k+9)}{4 (k+5)^2} \pa F_{12} G_{21}
\nonu \\
&-& \frac{(12 k^3+179 k^2+634 k+963)}{6 (k+5)^3 (13 k+17)} F_{12} \pa^2 F_{21}
-\frac{1}{4 (k+5)} F_{12} \pa G_{21}
\nonu \\
&-& \frac{k}{2 (k+5)^2} \pa F_{21} G_{12}
+ \frac{(k+2)}{2 (k+5)^2} F_{21} \pa G_{12}
-\frac{(k-6)}{6 (k+5)^2} \pa F_{22} G_{11}
\nonu \\
&-& \frac{(k+6)}{6 (k+5)^2}  F_{22} \pa G_{11}
+ \frac{3 i (5 k+18)}{(k+5) (13 k+17)}  T A_3
+ \frac{i k (4 k+21)}{(k+5) (13 k+17)} T B_3
\nonu \\
&-& \frac{17 (k+2)}{(k+5) (13 k+17)} T U
-\frac{2 (k-3) (2 k+9)}{(k+5)^2 (13 k+17)} T F_{11} F_{22}
\nonu \\
& + &  \frac{2 \left(2 k^2+9 k+27\right)}{(k+5)^2 (13 k+17)} T F_{12} F_{21}
-\frac{2}{(k+5)^2} U A_3 A_3
-\frac{4}{(k+5)^2} U A_3 B_3
\nonu \\
& -& \frac{3 i}{(k+5)^2} \pa U A_3
-\frac{5 i}{(k+5)^2} U \pa A_3
-\frac{2}{(k+5)^2} U A_{+} A_{-}
\nonu \\
&-& \frac{2}{(k+5)^2} U B_3 B_3
 -\frac{2 i}{(k+5)^2} U \pa B_3
 -\frac{2}{(k+5)^2} U B_{+} B_{-}
\nonu \\
&+& \frac{17 (k+2)}{2 (k+5) (13 k+17)} \pa^2 U
+ \frac{(7 k+3)}{6 (k+5)^3} \pa U F_{11} F_{22}
- \frac{(5 k+57)}{6 (k+5)^3} U \pa F_{11} F_{22}
\nonu \\
&+& \frac{(19 k+63)}{6 (k+5)^3} U F_{11} \pa F_{22}
-\frac{3}{2 (k+5)^2} \pa U F_{12} F_{21}
-\frac{11}{2 (k+5)^2}  U \pa F_{12} F_{21}
\nonu \\
&+& \frac{5}{2 (k+5)^2} U F_{12} \pa F_{21}
-\frac{2}{(k+5)^2}  U U U
-\frac{3 i (5 k+18)}{2 (k+5) (13 k+17)} \pa^2 A_3
\nonu \\
& + & \left.  
\frac{i (13 k+9)}{6 (k+5)^3} \pa A_3 F_{12} F_{21}
-\frac{i (11 k+15)}{6 (k+5)^3} A_3 \pa (F_{12} F_{21})
\right](w) +\cdots.
\nonu 
\eea
Note that there exists the 
nonlinear term containing the higher spin-$1$ current
in the first order pole with $(k-3)$ factor.

With the help of (\ref{g1122vu3half}), one obtains 
\bea
%%%%%%%%%%%%%%%%%%%%%%%%%%%%%%%%%%%%%%%%%%%%%%%%%%%%%%%%%%%%%%%%%%%%%%%%%%
\left(
\begin{array}{c}
{\bf U^{(\frac{3}{2})}} \\
{\bf V^{(\frac{3}{2})}}
\end{array}
\right) 
(z) \, {\bf W^{(2)}}(w) & = & 
\frac{1}{(z-w)^2} \, \left[ \frac{3}{2} 
\left(
\begin{array}{c}
{\bf U^{(\frac{3}{2})}} \\
{\bf V^{(\frac{3}{2})}}
\end{array}
\right) 
\pm  \frac{1}{(k+5)} 
\left(
\begin{array}{c}
G_{11} \\
G_{22}
\end{array}
\right) \right. \nonu \\
&\mp & \frac{(3 k+11)}{2 (k+5)^2} U
\left(
\begin{array}{c}
F_{11} \\
F_{22}
\end{array}
\right) - 
\frac{2}{(k+5)^2} F_{12} F_{21}
\left(
\begin{array}{c}
F_{11} \\
F_{22}
\end{array}
\right)
\nonu \\
& - & \frac{i (3 k+11)}{2 (k+5)^2} 
\left(
\begin{array}{c}
F_{11} \\
F_{22}
\end{array}
\right) A_3
-\frac{i (3 k-1)}{2 (k+5)^2} 
\left(
\begin{array}{c}
F_{21} \\
F_{12}
\end{array}
\right) A_{\pm}
\nonu \\
& + & \left.  \frac{i (k-11)}{2 (k+5)^2} 
\left(
\begin{array}{c}
F_{12} \\
F_{21}
\end{array}
\right) B_{\mp}
-\frac{i (3 k+11)}{2 (k+5)^2}
\left(
\begin{array}{c}
F_{11} \\
F_{22}
\end{array}
\right) B_3
\right](w)
\nonu \\
& + & \frac{1}{(z-w)} \, 
\left[ \frac{1}{2} \pa 
\left(
\begin{array}{c}
{\bf U^{(\frac{3}{2})}} \\
{\bf V^{(\frac{3}{2})}}
\end{array}
\right)
\mp 
\frac{1}{(k+5)^2} A_3 B_{\mp} 
\left(
\begin{array}{c}
F_{12} \\
F_{21}
\end{array}
\right)
\right. \nonu \\
&+& \frac{i \left(k^2-10 k-99\right)}{2 (k+5)^2 (2 k+9)} 
\pa A_3 
\left(
\begin{array}{c}
F_{11} \\
F_{22}
\end{array}
\right)
-\frac{i (k+1)}{2 (k+5)^2} 
A_3 \pa
\left(
\begin{array}{c}
F_{11} \\
F_{22}
\end{array}
\right)
\nonu \\
&+& \frac{i}{2 (k+5)} 
A_{\pm}
\left(
\begin{array}{c}
G_{21} \\
G_{12}
\end{array}
\right)
\pm \frac{1}{(k+5)^2} A_{\pm} A_3 
\left(
\begin{array}{c}
F_{21} \\
F_{12}
\end{array}
\right)
\nonu \\
&\mp & \frac{1}{(k+5)^2} A_{\pm} A_{\mp}
 \left(
\begin{array}{c}
F_{11} \\
F_{22}
\end{array}
\right)
\pm
\frac{1}{(k+5)^2} A_{\pm} B_3
 \left(
\begin{array}{c}
F_{21} \\
F_{12}
\end{array}
\right)
\nonu \\
&-& \frac{i (k-1)}{2 (k+5)^2} \pa A_{\pm} 
 \left(
\begin{array}{c}
F_{21} \\
F_{12}
\end{array}
\right)
-\frac{i (k+1)}{2 (k+5)^2} B_3 \pa
 \left(
\begin{array}{c}
F_{11} \\
F_{22}
\end{array}
\right)
\nonu \\
&-& \frac{i (k+7)}{2 (k+5)^2} A_{\pm} \pa
 \left(
\begin{array}{c}
F_{21} \\
F_{12}
\end{array}
\right)
+ \frac{i (k-9) (k+7)}{2 (k+5)^2 (2 k+9)} \pa B_3 
 \left(
\begin{array}{c}
F_{11} \\
F_{22}
\end{array}
\right)
\nonu \\
&-& \frac{i}{2 (k+5)} B_{\mp}
\left(
\begin{array}{c}
G_{12} \\
G_{21}
\end{array}
\right)
\mp \frac{1}{(k+5)^2} B_{\mp} B_3 
\left(
\begin{array}{c}
F_{12} \\
F_{21}
\end{array}
\right)
\nonu \\
&+& \frac{i (k-5)}{2 (k+5)^2} \pa B_{\mp} 
\left(
\begin{array}{c}
F_{12} \\
F_{21}
\end{array}
\right)
- \frac{i (k+7)}{2 (k+5)^2} B_{\mp} \pa 
\left(
\begin{array}{c}
F_{12} \\
F_{21}
\end{array}
\right)
\nonu \\
&\pm & \frac{1}{(k+5)^2} 
B_{\pm} B_{\mp} 
\left(
\begin{array}{c}
F_{11} \\
F_{22}
\end{array}
\right)
\pm \frac{(k-3)}{2 (k+5)^2} \pa^2 
\left(
\begin{array}{c}
F_{11} \\
F_{22}
\end{array}
\right)
\nonu \\
& \pm & \frac{1}{(k+5)^2} 
\left(
\begin{array}{c}
F_{11} F_{12} G_{21} \\
F_{22} F_{21} G_{12}
\end{array}
\right) \mp
\frac{2 (k+1)}{(k+5)^3} 
\left(
\begin{array}{c}
F_{11} \pa F_{12} F_{21} \\
F_{22} \pa F_{21} F_{12}
\end{array}
\right)
\nonu \\
&\mp & \frac{8}{(k+5)^3} 
\left(
\begin{array}{c}
F_{11} F_{12} \pa F_{21} \\
F_{22}  F_{21} \pa F_{12}
\end{array}
\right) \pm
\frac{1}{(k+5)^2} 
\left(
\begin{array}{c}
F_{11} F_{21} G_{12} \\
F_{22}  F_{12} G_{21}
\end{array}
\right)
\nonu \\
&+& \frac{i}{(k+5)^2} U A_{\pm}
\left(
\begin{array}{c}
F_{21}  \\
F_{12} 
\end{array}
\right)
-\frac{i}{(k+5)^2} U B_{\mp}
\left(
\begin{array}{c}
F_{12}  \\
F_{21} 
\end{array}
\right)
\nonu \\
&\mp & \frac{(5 k^2+28 k-9)}{2 (k+5)^2 (2 k+9)} 
\pa U 
\left(
\begin{array}{c}
F_{11}  \\
F_{22} 
\end{array}
\right)
\mp 
\frac{1}{2 (k+5)} U \pa 
\left(
\begin{array}{c}
F_{11}  \\
F_{22} 
\end{array}
\right)
\nonu \\
& \pm & \left.  \frac{3 (k-3) (k+6)}{2 (k+5)^2 (2 k+9)}
\left(
\begin{array}{c}
F_{11} G_{22} F_{11}  \\
F_{22} G_{11} F_{22}
\end{array}
\right)
\right](w)
+\cdots.
\nonu 
\eea
Note that there is a spin-$\frac{3}{2}$ current in the second and 
first order poles.
One also writes down the OPEs together under the symmetry of the structure
constants. 

Similarly one describes the following OPEs from the results in 
(\ref{g1221w2}) 
\bea
%%%%%%%%%%%%%%%%%%%%%%%%%%%%%%%%%%%%%%%%%%%%%%%%%%%%%%%%%%%%%%%%%%%%%%%%%%
\left(
\begin{array}{c}
{\bf U^{(\frac{3}{2})}} \\
{\bf V^{(\frac{3}{2})}} 
\end{array}
\right)
(z) \, {\bf W_{\pm}^{(\frac{5}{2})}}(w) & = & 
\frac{1}{(z-w)^3} 
\, \left[ -\frac{2 i k (4 k+21)}{3 (k+5)^2} {B}_{\mp} \pm
\frac{16 k (k+3)}{3 (k+5)^3}
\left(
\begin{array}{c}
F_{11} F_{21}  \\
F_{22} F_{12}
\end{array}
\right) 
\right] (w) 
\nonu \\
& + & \frac{1}{(z-w)^2} \, \left[  
-\frac{8 (k+3)}{3 (k+5)}
\left(
\begin{array}{c}
{\bf U_{+}^{(2)}} \\
{\bf V_{-}^{(2)}}
\end{array}
\right)  
\mp
\frac{2 (k-3)}{3 (k+5)^2} A_3 B_{\mp}
\right. \nonu \\
&+& \frac{2 i (k-3)}{3 (k+5)^3} A_3 
\left(
\begin{array}{c}
F_{11} F_{21}  \\
F_{22} F_{12}
\end{array}
\right)
 -\frac{2 i}{(k+5)^2} B_3 
\left(
\begin{array}{c}
F_{11} F_{21}  \\
F_{22} F_{12}
\end{array}
\right) 
\nonu \\
&\pm & \frac{2 i (k+9)}{3 (k+5)^3} B_{\mp} F_{11} F_{22}
\mp \frac{4 i (k+3)}{3 (k+5)^3} B_{\mp} F_{12} F_{21}
\nonu \\
& \mp & \frac{2 (k+2)}{(k+5)^2} 
 \left(
\begin{array}{c}
F_{11} G_{21}  \\
F_{22} G_{12}
\end{array}
\right) 
\mp \frac{2 (5 k+9)}{3 (k+5)^3} 
 \left(
\begin{array}{c}
\pa F_{11} F_{21}  \\
\pa F_{22} F_{12}
\end{array}
\right) 
\nonu \\
&\mp & \frac{2 (k+21)}{3 (k+5)^3} 
 \left(
\begin{array}{c}
F_{11} \pa F_{21}  \\
F_{22} \pa F_{12}
\end{array}
\right) \pm
  \frac{(k+1)}{(k+5)^2} 
 \left(
\begin{array}{c}
F_{21} G_{11}  \\
F_{12} G_{22}
\end{array}
\right)
\nonu \\
&- & \left.  \frac{2 i k}{(k+5)^2} U B_{\mp}
\mp \frac{2}{(k+5)^2} U 
\left(
\begin{array}{c}
F_{11} F_{21}  \\
F_{22} F_{12}
\end{array}
\right)
\right](w) 
\nonu \\
&+ & \frac{1}{(z-w)} \, \left[ 
-\frac{1}{2} 
\left(
\begin{array}{c}
{\bf Q_{+}^{(3)}} \\
 {\bf R_{-}^{(3)}} 
\end{array}
\right)
-\frac{2 (k+3)}{3 (k+5)} 
\pa \left(
\begin{array}{c}
{\bf U_{+}^{(2)}} \\
{\bf V_{-}^{(2)}}
\end{array}
\right)
\pm \frac{(17 k+81)}{6 (k+5)^2}  \pa A_3 B_{\mp}
\right. \nonu \\
&\mp & \frac{(19 k+75)}{6 (k+5)^2}  A_3 \pa B_{\mp}
- \frac{2 i}{(k+5)^2}  A_3
\left(
\begin{array}{c}
F_{11} G_{21} \\
F_{22} G_{12} 
\end{array}
\right)
\nonu \\
&+& \frac{i (31 k+99)}{6 (k+5)^3}  \pa A_3 
\left(
\begin{array}{c}
F_{11} F_{21} \\
F_{22} F_{12} 
\end{array}
\right)
- \frac{i (29 k+105)}{6 (k+5)^3} A_3 
\pa \left(
\begin{array}{c}
 F_{11} F_{21} \\
 F_{22} F_{12} 
\end{array}
\right)
\nonu \\
&+& 
\frac{2i}{(k+5)^2} A_3 
\left(
\begin{array}{c}
F_{21} G_{11} \\
F_{12} G_{22} 
\end{array}
\right)
- \frac{i}{2 (k+5)^2} \pa B_3 
\left(
\begin{array}{c}
F_{11} F_{21} \\
F_{22} F_{12} 
\end{array}
\right)
\nonu \\
&-& \frac{5i}{2 (k+5)^2} B_3
\left(
\begin{array}{c}
\pa F_{11} F_{21} \\
\pa F_{22} F_{12} 
\end{array}
\right)
+\frac{3i}{2 (k+5)^2} B_3
\left(
\begin{array}{c}
F_{11} \pa F_{21} \\
F_{22} \pa F_{12} 
\end{array}
\right)
\nonu \\
&+&  \frac{i k (4 k+21)}{2 (k+5) (13 k+17)}
\pa^2 B_{\mp}
+ \frac{i}{(k+5)^2} B_{\mp} 
\left(
\begin{array}{c}
F_{11} G_{22} \\
F_{22} G_{11} 
\end{array}
\right)
\nonu \\
&-& \frac{i (k+15)}{3 (k+5)^3}  B_{\mp} 
\left(
\begin{array}{c}
\pa F_{11} F_{22} \\
\pa F_{22} F_{11} 
\end{array}
\right)
\pm \frac{i (5 k+39)}{3 (k+5)^3} \pa B_{\mp} F_{11} F_{22}
\nonu \\
&-& \frac{i (7 k+45)}{3 (k+5)^3}  B_{\mp}
\left(
\begin{array}{c}
F_{11} \pa F_{22} \\
F_{22} \pa F_{11} 
\end{array}
\right)
+ \frac{i}{(k+5)^2} B_{\mp} 
\left(
\begin{array}{c}
F_{12} G_{21} \\
F_{21} G_{12} 
\end{array}
\right)
\nonu \\
& - & \frac{5 i (k+9)}{6 (k+5)^3} B_{\mp}
\left(
\begin{array}{c}
F_{12} \pa F_{21} \\
F_{21} \pa F_{12} 
\end{array}
\right)
\mp   \frac{(k-3)}{3 (k+5)^3}
\left(
\begin{array}{c}
\pa^2 F_{11}  F_{21} \\
\pa^2 F_{22}  F_{12} 
\end{array}
\right)
\nonu \\
&-& \frac{i (17 k+105)}{6 (k+5)^3}  B_{\mp} 
\left(
\begin{array}{c}
\pa F_{12}  F_{21} \\
\pa F_{21}  F_{12} 
\end{array}
\right)
\pm \frac{7 i (k+9)}{6 (k+5)^3}  \pa B_{\mp} 
F_{12} F_{21}
\nonu \\
&-& \frac{i}{(k+5)^2} B_{\mp}
\left(
\begin{array}{c}
F_{21}  G_{12} \\
F_{12}  G_{21} 
\end{array}
\right)
- \frac{ i}{(k+5)^2} B_{\mp} 
\left(
\begin{array}{c}
F_{22}  G_{11} \\
F_{11}  G_{22} 
\end{array}
\right)
\nonu \\
&\mp & \frac{(24 k^3+205 k^2+650 k+357)}{6 (k+5)^3 (13 k+17)}
\left(
\begin{array}{c}
F_{11}  \pa^2 F_{21} \\
F_{22}  \pa^2 F_{12} 
\end{array}
\right)
\nonu \\
&-& \frac{i k (4 k+21)}{(k+5) (13 k+17)}
T B_{\mp}
\pm  \frac{8 k (k+3)}{(k+5)^2 (13 k+17)}
T
\left(
\begin{array}{c}
F_{11}  F_{21} \\
F_{22}  F_{12} 
\end{array}
\right)
\nonu \\
&\pm & \frac{4}{(k+5)^2} U A_3 B_{\mp}
- \frac{i k}{2 (k+5)^2} \pa (U B_{\mp})
\mp 
\frac{5}{2 (k+5)^2}
U
\left(
\begin{array}{c}
\pa F_{11}  F_{21} \\
\pa F_{22}  F_{12} 
\end{array}
\right)
\nonu \\
&\mp & \frac{1}{2 (k+5)^2} \pa U
\left(
\begin{array}{c}
F_{11}  F_{21} \\
F_{22}  F_{12} 
\end{array}
\right)
\pm \frac{3}{2 (k+5)^2} U
\left(
\begin{array}{c}
F_{11}  \pa F_{21} \\
F_{22}  \pa F_{12} 
\end{array}
\right)
\nonu \\
& +  & \frac{2 i}{(k+5)^2}
\left(
\begin{array}{c}
F_{11}  A_{-} \pa F_{11} \\
F_{22}  A_{+} \pa F_{22} 
\end{array}
\right)
+ \frac{i \left(8 k^2+63 k+51\right)}{(k+5)^2 (13 k+17)}
\left(
\begin{array}{c}
F_{21}  A_{+}  \pa F_{21} \\
F_{12}  A_{-} \pa F_{12} 
\end{array}
\right)
\nonu \\
&+& \frac{i \left(8 k^2+37 k+17\right)}{(k+5)^2 (13 k+17)}
\left(
\begin{array}{c}
\pa F_{21}  A_{+}  F_{21} \\
\pa F_{12}  A_{-} F_{12} 
\end{array}
\right)
\nonu \\
&\mp & \frac{(5k+13)}{4 (k+5)^2} 
\left(
\begin{array}{c}
\pa F_{11}  G_{21} \\
\pa F_{22}  G_{12} 
\end{array}
\right)
\mp \frac{(k+1)}{4 (k+5)^2} 
\left(
\begin{array}{c}
F_{11}  \pa G_{21} \\
F_{22}  \pa G_{12} 
\end{array}
\right)
\nonu \\
&\mp & 
\left.  
\frac{(k+4)}{2 (k+5)^2} 
\left(
\begin{array}{c}
\pa F_{21} G_{11} \\
\pa F_{12} G_{22} 
\end{array}
\right) \pm
\frac{(k+2)}{2 (k+5)^2} 
\left(
\begin{array}{c}
F_{21} \pa G_{11} \\
F_{12} \pa G_{22} 
\end{array}
\right)
\right](w) +\cdots.
\nonu 
\eea
Due to the spin-$\frac{1}{2}$ currents the OPEs are rather complicated and 
the nonlinear terms in the corresponding OPEs in the nonlinear version 
disappear. 

Moreover, one can construct the following OPEs
\bea
%%%%%%%%%%%%%%%%%%%%%%%%%%%%%%%%%%%%%%%%%%%%%%%%%%%%%%%%%%%%%%%%%%%%%%
\left(
\begin{array}{c}
{\bf U^{(\frac{3}{2})}} \\
{\bf V^{(\frac{3}{2})}} 
\end{array}
\right)
(z) \, {\bf W_{\mp}^{(\frac{5}{2})}}(w) & = &  
 \frac{1}{(z-w)^3}
\, \left[  \frac{2 i (5 k+18)}{(k+5)^2} {A}_{\pm} 
\pm \frac{8(k+9)}{(k+5)^3}
\left(
\begin{array}{c}
F_{11} F_{12} \\
F_{22} F_{21} 
\end{array}
\right)
\right](w) \nonu \\
& + & 
\frac{1}{(z-w)^2} \, \left[ 
- \frac{4 (k+9)}{3 (k+5)} 
\left(
\begin{array}{c}
{\bf U_{-}^{(2)} } \\
 {\bf V_{+}^{(2)} }
\end{array}
\right)
+ \frac{2 i}{(k+5)^2} A_3 
\left(
\begin{array}{c}
F_{11} F_{12} \\
F_{22} F_{21}
\end{array}
\right)
\right. \nonu \\
& \pm & 
 \frac{2 (k-3)}{3(k+5)^2} A_{\pm} B_3
\mp  \frac{4i(k+3)}{3(k+5)^3} A_{\pm } F_{11} F_{22}
\nonu \\
& - &  \frac{2i(k+9)}{3(k+5)^3} A_{\pm} F_{12} F_{21}
+\frac{2 i(k-3)}{3(k+5)^3} B_3 
\left(
\begin{array}{c}
F_{11} F_{12} \\
F_{22} F_{21} 
\end{array}
\right)
\nonu \\
&\pm & \frac{(k+7)}{ (k+5)^2} 
\left(
\begin{array}{c}
F_{11} G_{12} \\
F_{22} G_{21} 
\end{array}
\right)
\mp \frac{2(k+21)}{3(k+5)^3} 
\left(
\begin{array}{c}
\pa F_{11} F_{12} \\
\pa F_{22} F_{21} 
\end{array}
\right)
\nonu \\
& \mp &  \frac{2 (5 k+9)}{3(k+5)^3} 
\left(
\begin{array}{c}
F_{11} \pa F_{12} \\
F_{22} \pa F_{21} 
\end{array}
\right) \mp \frac{4}{ (k+5)^2} 
\left(
\begin{array}{c}
F_{12} G_{11} \\
F_{21} G_{22} 
\end{array}
\right) \nonu \\
&- & \left.  \frac{6 i}{(k+5)^2}  U A_{\pm} 
\pm \frac{2}{(k+5)^2} U
\left(
\begin{array}{c}
F_{11} F_{12} \\
F_{22} F_{21} 
\end{array}
\right)
\right](w) 
\nonu \\
& + & \frac{1}{(z-w)} \, \left[ 
\frac{1}{2} 
\left(
\begin{array}{c}
{\bf Q_{-}^{(3)} } \\
 {\bf R_{+}^{(3)} }
\end{array}
\right)
-\frac{(k+9)}{3 (k+5)} \pa 
\left(
\begin{array}{c}
{\bf U_{-}^{(2)} } \\
 {\bf V_{+}^{(2)} }
\end{array}
\right)
\right. \nonu \\
&+& \frac{i}{2 (k+5)^2} \pa A_3 
\left(
\begin{array}{c}
F_{11} F_{12} \\
F_{22} F_{21} 
\end{array}
\right) + \frac{5i }{2 (k+5)^2}
A_3 
\left(
\begin{array}{c}
\pa F_{11} F_{12} \\
\pa F_{22} F_{21} 
\end{array}
\right)
\nonu \\
&-& \frac{3i }{2 (k+5)^2} A_3 
\left(
\begin{array}{c}
F_{11} \pa F_{12} \\
F_{22} \pa F_{21} 
\end{array}
\right) 
 -   \frac{3 i (5 k+18)}{2 (k+5) (13 k+17)}
 \pa^2 A_{\pm}
\nonu \\
& \mp & \frac{(17 k+81)}{6 (k+5)^2}  \pa A_{\pm} B_3
\pm \frac{(19 k+75)}{6 (k+5)^2}  A_{\pm} \pa B_3
\nonu \\
&+& \frac{i}{(k+5)^2} A_{\pm }
\left(
\begin{array}{c}
F_{11} G_{22} \\
F_{22} G_{11} 
\end{array}
\right)
\mp \frac{i (17 k+57)}{6 (k+5)^3} \pa A_{\pm} F_{11} F_{22}
\nonu \\
&+& \frac{i (7 k+15)}{6 (k+5)^3} A_{\pm}
\left(
\begin{array}{c}
\pa F_{11} F_{22} \\
\pa F_{22} F_{11} 
\end{array}
\right)
+\frac{i (19 k+75)}{6 (k+5)^3} A_{\pm} 
\left(
\begin{array}{c}
F_{11} \pa F_{22} \\
F_{22} \pa F_{11} 
\end{array}
\right)
\nonu \\
&-& \frac{i}{(k+5)^2} A_{\pm} 
\left(
\begin{array}{c}
F_{12} G_{21} \\
F_{21} G_{12} 
\end{array}
\right)
\mp \frac{7i (k+3)}{3(k+5)^3} \pa A_{\pm } F_{12} F_{21}
\nonu \\
&-& \frac{5i (k+3)}{3(k+5)^3} A_{\pm} 
\left(
\begin{array}{c}
\pa F_{12} F_{21} \\
\pa F_{21} F_{12} 
\end{array}
\right) 
-\frac{i (11k+45)}{3(k+5)^3} A_{\pm}
\left(
\begin{array}{c}
F_{12} \pa F_{21} \\
F_{21} \pa F_{12} 
\end{array}
\right) 
\nonu \\
&+& \frac{i}{(k+5)^2} A_{\pm}
\left(
\begin{array}{c}
F_{21} G_{12} \\
F_{12} G_{21} 
\end{array}
\right) 
+ 
 \frac{2i}{(k+5)^2} B_3 
\left(
\begin{array}{c}
F_{12} G_{11} \\
F_{21} G_{22} 
\end{array}
\right) 
\nonu \\
&- &  \frac{ i}{(k+5)^2} A_{\pm} 
\left(
\begin{array}{c}
F_{22} G_{11} \\
F_{11} G_{22} 
\end{array}
\right) 
-\frac{2i}{(k+5)^2} B_3 
\left(
\begin{array}{c}
F_{11} G_{12} \\
F_{22} G_{21} 
\end{array}
\right) 
\nonu \\
&-& \frac{i (17k+141)}{6 (k+5)^3} \pa B_3 
\left(
\begin{array}{c}
F_{11} F_{12} \\
F_{22} F_{21} 
\end{array}
\right) +
\frac{i (19k+135)}{6 (k+5)^3} B_3 
\pa \left(
\begin{array}{c}
 F_{11} F_{12} \\
 F_{22} F_{21} 
\end{array}
\right)
\nonu \\
&\mp & \frac{(49 k^2+794 k+1977)}{6 (k+5)^3 (13 k+17)}
\left(
\begin{array}{c}
\pa^2 F_{11} F_{12} \\
\pa^2 F_{22} F_{21} 
\end{array}
\right)
\nonu \\
&\mp & \frac{25}{(k+5) (13 k+17)}
\left(
\begin{array}{c}
\pa F_{11} \pa F_{12} \\
\pa F_{22} \pa F_{21} 
\end{array}
\right)
\nonu \\
&\mp &  \frac{(101 k^2+706 k+1773)}{6 (k+5)^3 (13 k+17)}
\left(
\begin{array}{c}
F_{11} \pa^2 F_{12} \\
F_{22} \pa^2 F_{21} 
\end{array}
\right) \pm
\frac{(k+4)}{(k+5)^2}
\left(
\begin{array}{c}
\pa F_{11} G_{12} \\
\pa F_{22} G_{21} 
\end{array}
\right)
\nonu \\
&\pm & \frac{(3k+5)}{4 (k+5)^2} 
\left(
\begin{array}{c}
\pa F_{12} G_{11} \\
\pa F_{21} G_{22} 
\end{array}
\right)
\mp \frac{(k+7)}{4 (k+5)^2} 
\left(
\begin{array}{c}
F_{12} \pa G_{11} \\
F_{21} \pa G_{22} 
\end{array}
\right)
\nonu \\
&\pm &
\frac{1}{(k+5)^2}
\left(
\begin{array}{c}
F_{11} \pa G_{12} \\
F_{22} \pa G_{21} 
\end{array}
\right)+
  \frac{3i (5 k+18)}{(k+5) (13 k+17)} T A_{\pm}
\nonu \\
&\pm & \frac{12(k+9)}{(k+5)^2 (13 k+17)}  T
\left(
\begin{array}{c}
F_{11} F_{12} \\
F_{22} F_{21} 
\end{array}
\right)
\mp \frac{4}{(k+5)^2} U A_{\pm} B_3
\nonu \\
&-& \frac{3i}{2 (k+5)^2} \pa (U A_{\pm})
\mp  \frac{3}{2 (k+5)^2} U
\left(
\begin{array}{c}
F_{11} \pa F_{12} \\
 F_{22} \pa F_{21} 
\end{array}
\right)
\nonu \\
&\pm & \frac{1}{2 (k+5)^2} \pa U 
\left(
\begin{array}{c}
F_{11} F_{12} \\
F_{22} F_{21} 
\end{array}
\right)
\pm \frac{5}{2 (k+5)^2} U 
\left(
\begin{array}{c}
\pa F_{11} F_{12} \\
\pa F_{22} F_{21} 
\end{array}
\right)
\nonu \\
&-& \left.
\frac{2 i}{(k+5)^2} 
\left(
\begin{array}{c}
F_{11} B_{+} \pa F_{11} \\
 F_{22} B_{-} \pa F_{22} 
\end{array}
\right)
-   \frac{2 i}{(k+5)^2} 
\left(
\begin{array}{c}
F_{12} B_{-} \pa F_{12} \\
 F_{21} B_{+} \pa F_{21} 
\end{array}
\right)
\right](w) +\cdots.
\nonu 
\eea
One has these OPEs by writing two OPEs together with the symmetry 
of the structure constants.

From the relation in (\ref{g1221w5half}), one obtains the following OPEs
\bea
%%%%%%%%%%%%%%%%%%%%%%%%%%%%%%%%%%%%%%%%%%%%%%%%%%%%%%%%%%%%%%%%%%%%%%%%%%
\left(
\begin{array}{c}
{\bf U^{(\frac{3}{2})}} \\
{\bf V^{(\frac{3}{2})}}
\end{array}
\right)(z) \, {\bf W^{(3)}}(w) & = & 
\frac{1}{(z-w)^4} \, \left[ - \frac{18 k}{(k+5)^2}
\left(
\begin{array}{c}
F_{11} \\
F_{22} 
\end{array}
\right)
\right](w)
\nonu \\
& + & \frac{1}{(z-w)^3}\, 
\left[ \pm \frac{2 (k-3) (61 k+113)}{3 (k+5) (13 k+17)}
\left(
\begin{array}{c}
{\bf U^{(\frac{3}{2})}} \\
{\bf V^{(\frac{3}{2})}} 
\end{array}
\right) \right. \nonu \\
& - & 
 \frac{2 (k-3) (61 k+113)}{3 (k+5)^2 (13 k+17)}   U
\left(
\begin{array}{c}
F_{11} \\
F_{22} 
\end{array}
\right)
\mp \frac{4 (k-3) (5 k+73)}{3 (k+5)^3 (13 k+17)} 
F_{12} F_{21}
\left(
\begin{array}{c}
F_{11} \\
F_{22} 
\end{array}
\right)
 \nonu \\
& \mp & \frac{2 i \left(61 k^2+164 k-33\right)}{3 (k+5)^2 (13 k+17)}
\left(
\begin{array}{c}
F_{11} \\
F_{22} 
\end{array}
\right) A_3 
\mp \frac{2 i \left(61 k^2+164 k-33\right)}{3 (k+5)^2 (13 k+17)}
\left(
\begin{array}{c}
F_{21} \\
F_{12} 
\end{array}
\right) A_{\pm}
\nonu \\
&\pm & 
\frac{2 i \left(17 k^2+172 k+339\right)}{3 (k+5)^2 (13 k+17)}   
\left(
\begin{array}{c}
F_{12} \\
F_{21} 
\end{array}
\right) B_{\mp}
\nonu \\
& + & \left.  \frac{18 k}{(k+5)^2}
\pa
\left(
\begin{array}{c}
F_{11} \\
F_{22} 
\end{array}
\right)
\pm \frac{2 i \left(17 k^2+172 k+339\right)}{3 (k+5)^2 (13 k+17)}
\left(
\begin{array}{c}
F_{11} \\
F_{22} 
\end{array}
\right) B_3 
\right] (w) 
\nonu \\
& + & \frac{1}{(z-w)^2} \, \left[ 
\pm \frac{5}{2} 
\left(
\begin{array}{c}
{\bf Q^{(\frac{5}{2})}} \\
 {\bf R^{(\frac{5}{2})}} \end{array}
\right) + 5
\left(
\begin{array}{c}
{\bf U^{(\frac{5}{2})} } \\
{\bf V^{(\frac{5}{2})}} 
 \end{array}
\right)
\pm  \frac{2 (k-3)}{(13 k+17)} \pa 
\left(
\begin{array}{c}
{\bf U^{(\frac{3}{2})}} \\
{\bf V^{(\frac{3}{2})}} 
\end{array}
\right)
\right. \nonu \\
&- &  \frac{6 {\bf (k-3)}}{(13 k+17)} {\bf T^{(1)}} 
\left(
\begin{array}{c}
{\bf U^{(\frac{3}{2})}} \\
{\bf V^{(\frac{3}{2})}} 
\end{array}
\right) +
\frac{2}{(k+5)^2} A_3 A_3 
\left(
\begin{array}{c}
F_{11} \\
F_{22} 
\end{array}
\right)
\nonu \\
&+& \frac{4}{(k+5)^2} A_3 B_3 
\left(
\begin{array}{c}
F_{11} \\
F_{22} 
\end{array}
\right)
\mp  \frac{3 i}{2 (k+5)} B_{\mp}
\left(
\begin{array}{c}
G_{12} \\
G_{21} 
\end{array}
\right)
\nonu \\
&\mp & \frac{3i}{2 (k+5)} A_3 
\left(
\begin{array}{c}
G_{11} \\
G_{22} 
\end{array}
\right) +
\frac{4}{(k+5)^2} A_3 B_{\mp}
\left(
\begin{array}{c}
F_{12} \\
F_{21} 
\end{array}
\right)
\nonu \\
&\mp&  \frac{2 i \left(k^2+67 k+70\right)}{(k+5)^2 (13 k+17)} 
\pa A_3 
\left(
\begin{array}{c}
F_{11} \\
F_{22} 
\end{array}
\right)
\mp \frac{2 i \left(k^2-128 k-185\right)}{(k+5)^2 (13 k+17)}  A_3 \pa 
\left(
\begin{array}{c}
F_{11} \\
F_{22} 
\end{array}
\right)
\nonu \\
&\mp & \frac{3 i}{2(k+5)} A_{\pm}  
\left(
\begin{array}{c}
G_{21} \\
G_{12} 
\end{array}
\right)
+ \frac{2}{(k+5)^2} A_{\pm} A_{\mp} 
\left(
\begin{array}{c}
F_{11} \\
F_{22} 
\end{array}
\right)
\nonu \\
&+ & \frac{4}{(k+5)^2} A_{\pm} B_3 
\left(
\begin{array}{c}
F_{21} \\
F_{12} 
\end{array}
\right) -
\frac{4}{(k+5)^2} A_{\pm} B_{\mp} 
\left(
\begin{array}{c}
F_{22} \\
F_{11} 
\end{array}
\right)
\nonu \\
& \mp &  \frac{2 i \left(k^2+80 k+87\right)}{(k+5)^2 (13 k+17)} \pa A_{\pm}
\left(
\begin{array}{c}
F_{21} \\
F_{12} 
\end{array}
\right) \mp \frac{2 i \left(k^2-128 k-185\right)}{(k+5)^2 (13 k+17)}   A_{\pm}
\pa
\left(
\begin{array}{c}
F_{21} \\
F_{12} 
\end{array}
\right) 
\nonu \\
& \mp & \frac{3i}{2 (k+5)} B_3 
\left(
\begin{array}{c}
G_{11} \\
G_{22} 
\end{array}
\right) \mp
\frac{4i (20k^2 +33 k+1)}{ (k+5)^2(13k+17)} B_3 \pa
\left(
\begin{array}{c}
F_{11} \\
F_{22} 
\end{array}
\right)
\nonu \\
&\pm & \frac{i (37k^2+112k+115) (k-9)}
{3 (k+5)^2 (13 k+17)} \pa B_3 
\left(
\begin{array}{c}
F_{11} \\
F_{22} 
\end{array}
\right) + 
\frac{2}{(k+5)^2} B_{3} B_3 
\left(
\begin{array}{c}
F_{11} \\
F_{22} 
\end{array}
\right)
\nonu \\
& \pm &  \frac{i \left(37 k^2+86 k+81\right)}{(k+5)^2 (13 k+17)}
\pa B_{\mp}
\left(
\begin{array}{c}
F_{12} \\
F_{21} 
\end{array}
\right) \mp
\frac{4 i \left(20 k^2+33 k+1\right)}{(k+5)^2 (13 k+17)} B_{\mp} \pa 
\left(
\begin{array}{c}
F_{12} \\
F_{21} 
\end{array}
\right)
\nonu \\
& + & \frac{2}{(k+5)^2} B_{\pm} B_{\mp} 
\left(
\begin{array}{c}
F_{11} \\
F_{22} 
\end{array}
\right) -
\frac{3 (k+3)}{2(k+5)^2} \pa^2
\left(
\begin{array}{c}
F_{11} \\
F_{22} 
\end{array}
\right)
\nonu \\
&+ & \frac{2}{(k+5)^2} 
\left(
\begin{array}{c}
F_{11} F_{12} G_{21} \\
F_{22} F_{21} G_{12}
\end{array}
\right) -
\frac{8 (k-3)}{ (k+5)^2 (13k+17)} 
\pa
\left(
\begin{array}{c}
F_{11}  \\
F_{22} 
\end{array}
\right) F_{12} F_{21}
\nonu \\
&- & \frac{2}{(k+5)^2} 
\left(
\begin{array}{c}
F_{11}  F_{21} G_{12} \\
F_{22}  F_{12} G_{21}
\end{array}
\right)
+  \frac{2}{(k+5)^2} 
\left(
\begin{array}{c}
F_{12}  F_{21} G_{11} \\
F_{21}  F_{12} G_{22}
\end{array}
\right)
\nonu \\
& - &
\frac{1}{(k+5)} T 
\left(
\begin{array}{c}
F_{11}   \\
F_{22} 
\end{array}
\right) -
\frac{3}{2 (k+5)} U
\left(
\begin{array}{c}
G_{11}   \\
G_{22} 
\end{array}
\right)
\nonu \\
&\mp & \frac{6 i}{(k+5)^2} U A_3 
\left(
\begin{array}{c}
F_{11}   \\
F_{22} 
\end{array}
\right)
\mp \frac{6i}{(k+5)^2} U A_{\pm}
\left(
\begin{array}{c}
F_{21}   \\
F_{12} 
\end{array}
\right)
\nonu \\
&\mp & \frac{6 i}{(k+5)^2} U B_3
\left(
\begin{array}{c}
F_{11}   \\
F_{22} 
\end{array}
\right)
\mp \frac{6i}{(k+5)^2} U B_{\mp} 
\left(
\begin{array}{c}
F_{12}   \\
F_{21} 
\end{array}
\right)
\nonu \\
& - & \frac{2 (k-3)}{(k+5) (13 k+17)}
\pa  
\left(
\begin{array}{c}
U F_{11}   \\
U F_{22} 
\end{array}
\right) 
+ \frac{2(61k+97)}{(k+5)^2(13k+17)} 
\left(
\begin{array}{c}
F_{11} F_{12} \pa F_{21}   \\
F_{22} F_{21} \pa F_{12} 
\end{array}
\right) 
\nonu \\
& - & \frac{4}{(k+5)^2} U U
\left(
\begin{array}{c}
F_{11}   \\
F_{22} 
\end{array}
\right)  
\pm \frac{10}{(k+5)^2}
\left(
\begin{array}{c}
\pa F_{11}  \\
\pa F_{22} 
\end{array}
\right) F_{11} F_{22} 
\nonu \\
&- & \left. \frac{2 (69 k+73)}{(k+5)^2 (13 k+17)} 
 \left(
\begin{array}{c}
F_{11}  \pa F_{12} F_{21}  \\
F_{22} \pa F_{21} F_{12} 
\end{array}
\right) 
\right](w) 
\nonu \\
& + & \frac{1}{(z-w)} \, \left[ + \cdots \right](w) + \cdots.
\nonu
\eea
There is no spin-$\frac{3}{2}$ current in the third order pole.
Again the nonlinear term containing the higher spin currents is present.
As discussed before, this can be removed by adding the $T \, {\bf T^{(1)}}(w)$ 
term into the left hand side. 
Other nonlinear terms in the context of the footnote \ref{nonlineardef}
disappear in this OPEs. It is an open problem to 
complete the first order pole by figuring out the possible 
composite spin-$\frac{7}{2}$ fields.

The relation (\ref{g1221vu3half}) determines the following OPEs
%%%%%%%%%%%%%%%%%%%%%%%%%%%%%%%%%%%%%%%%%%%%%%%%%%%%%%%%%%%%%%%%
\bea
\left(
\begin{array}{c}
{\bf U_{+}^{(2)}} \\
{\bf V_{-}^{(2)}}
\end{array}
\right)
(z) \, 
\left(
\begin{array}{c}
{\bf U_{+}^{(2)}} \\
{\bf V_{-}^{(2)}}
\end{array}
\right)
(w) & = & \frac{1}{(z-w)^2} \,
\left[ \frac{2k}{(5+k)^2}  {B}_{\mp} {B}_{\mp} \right](w)
\nonu \\
& + & \frac{1}{(z-w)} \, \left[  \frac{2k}{(5+k)^2}  {B}_{\mp} \pa B_{\mp}
\right](w)
+\cdots.
\nonu 
\eea
These are new OPEs where the corresponding OPEs in the nonlinear version 
are trivial.

By using the result in (\ref{g1221uv3half}), one has 
%%%%%%%%%%%%%%%%%%%%%%%%%%%%%%%%%%%%%%%%%%%%%%%%%%%%%%%%%%%%%%%%%%%%%%%
\bea
{\bf U_{+}^{(2)}}(z) \, {\bf U_{-}^{(2)}}(w) & = & \frac{1}{(z-w)^2} \,
\left[ \frac{2 (k+4)}{(k+5)^2}  {A}_{+} {B}_{-} 
+\frac{2 i}{(k+5)^2} A_{+} F_{11} F_{21}
\right. \nonu \\
&-& \left. \frac{2 i}{(k+5)^2} B_{-} F_{11} F_{12}
- \frac{(k-3)}{(k+5)^2} F_{11} G_{11}
+  \frac{4 (3 k-1)}{(k+5)^3} F_{11} \pa F_{11}
\right](w)
\nonu \\
& + & \frac{1}{(z-w)} \, \left[  
 \frac{2 i}{(k+5)^2}  A_3 F_{11} G_{11} -
 \frac{1}{(k+5)^2} \pa A_{+} B_{-}
\right. \nonu \\
&+ &  \frac{(2 k+9)}{(k+5)^2}  A_{+} \pa B_{-}
+ \frac{i}{(k+5)^2} A_{+} F_{11} G_{21}
+  \frac{8 i}{(k+5)^3} A_{+} \pa F_{11} F_{21}
\nonu \\
&+& \frac{4 i (k+3)}{(k+5)^3} A_{+} F_{11} \pa F_{21}
+ \frac{i}{(k+5)^2} A_{+} F_{21} G_{11}
- \frac{2 i}{(k+5)^2} B_3 F_{11} G_{11}
\nonu \\
& - & \frac{i}{(k+5)^2} B_{-} F_{11} G_{12}
- \frac{2 i}{(k+5)^2} \pa B_{-} F_{11} F_{12}
-\frac{8 i}{(k+5)^3} B_{-} \pa F_{11} F_{12}
\nonu \\
&+& \frac{8 i}{(k+5)^3} B_{-} F_{11} \pa F_{12}
-\frac{i}{(k+5)^2} B_{-} F_{12} G_{11}
- \frac{2 (k+4)}{(k+5)^2} \pa F_{11} G_{11}
\nonu \\
& + & \frac{2}{(k+5)^2} F_{11} \pa G_{11}
+ \frac{4 i (k+1)}{(k+5)^3} F_{11} A_3 \pa F_{11}
+ \frac{16 i}{(k+5)^3} F_{11} B_3 \pa F_{11}
\nonu \\
& +  & \left.   \frac{4}{(k+5)^2} F_{11} U \pa F_{11}
+ \frac{4 (k+1)}{(k+5)^3} F_{11} \pa^2 F_{11}
\right](w)
+\cdots.
\nonu 
\eea
As before, the nonlinear terms in the corresponding OPE in the 
nonlinear version disappear.

For the higher spin-$\frac{5}{2}$ currents, one has the result in 
(\ref{g1221vu2}) and can construct the following OPEs
\bea
%%%%%%%%%%%%%%%%%%%%%%%%%%%%%%%%%%%%%%%%%%%%%%%%%%%%%%%%%%%%%%%%%%%
\left(
\begin{array}{c}
{\bf U_{+}^{(2)}} \\
{\bf V_{-}^{(2)}} 
\end{array}
\right) (z) \, 
\left(
\begin{array}{c}
{\bf U^{(\frac{5}{2})}} \\
{\bf V^{(\frac{5}{2})}}
\end{array}
\right)
(w) & = & 
\frac{1}{(z-w)^3} \, 
\left[ -\frac{16 i k (k+3)}{3 (k+5)^3} B_{\mp}
\left(
\begin{array}{c}
F_{11} \\
F_{22} 
\end{array}
\right)
 \right](w) \nonu \\
&+& \frac{1}{(z-w)^2} \,
\left[ \pm \frac{2 (k+9)}{3 (k+5)^3} A_3 B_{\mp} 
 \left(
\begin{array}{c}
F_{11} \\
F_{22} 
\end{array}
\right)
\pm 
\frac{2 (k+9)}{3 (k+5)^3} A_{\pm} B_{\mp}
 \left(
\begin{array}{c}
F_{21} \\
F_{12} 
\end{array}
\right)
\right. \nonu \\
&+& \frac{i (7 k+9)}{3 (k+5)^2} B_{\mp}
 \left(
\begin{array}{c}
G_{11} \\
G_{22} 
\end{array}
\right) \mp
\frac{2 (k+9)}{3 (k+5)^3} B_{\mp} B_3 
 \left(
\begin{array}{c}
F_{11} \\
F_{22} 
\end{array}
\right)
\nonu \\
& \mp & \frac{2 (k+9)}{3 (k+5)^3} B_{\mp} B_{\mp}
 \left(
\begin{array}{c}
F_{12} \\
F_{21} 
\end{array}
\right)
-\frac{2 i k (k+1)}{(k+5)^3} \pa B_{\mp}
 \left(
\begin{array}{c}
F_{11} \\
F_{22} 
\end{array}
\right)
\nonu \\
&-& \frac{2 i (k+9) (2 k-3)}{3 (k+5)^3} B_{\mp} \pa 
 \left(
\begin{array}{c}
F_{11} \\
F_{22} 
\end{array}
\right)
\mp
\frac{4 (k+1) (k+9)}{3 (k+5)^4} 
 \left(
\begin{array}{c}
F_{11} F_{21} \pa F_{11} \\
F_{22}  F_{12} \pa F_{22}
\end{array}
\right)
\nonu \\
&\mp & \left. \frac{2 (k+9)}{3 (k+5)^3} 
\left(
\begin{array}{c}
F_{11} F_{21} G_{11} \\
F_{22}  F_{12} G_{22}
\end{array}
\right)
+ \frac{2 i (k+9)}{3 (k+5)^3} U B_{\mp}
 \left(
\begin{array}{c}
F_{11} \\
F_{22} 
\end{array}
\right)
 \right](w)
\nonu \\
&+ & \frac{1}{(z-w)} \, \left[ + \cdots
\right](w) + \cdots.
\nonu  
\eea
The nonlinear terms in the corresponding OPEs in the 
nonlinear version disappear.

As done before, one constructs 
\bea
%%%%%%%%%%%%%%%%%%%%%%%%%%%%%%%%%%%%%%%%%%%%%%%%%%%%%%%%%%%%%%%%%%%%%%
{\bf U_{+}^{(2)}}(z) \, {\bf V^{(\frac{3}{2})}}(w) 
& = & 
\frac{1}{(z-w)^3} \,
\left[ -\frac{6 k}{(k+5)^2} F_{21} \right](w) \nonu \\
&+& \frac{1}{(z-w)^2} \, \left[ 
- \frac{(k+9)}{(k+5)} {\bf T_{+}^{(\frac{3}{2})}} 
-\frac{(k+8)}{(k+5)} G_{21} 
-\frac{i (k+1)}{(k+5)^2} F_{11} A_{-}
\right. \nonu \\
&-& \frac{(k+7)}{(k+5)^2}  U F_{21}
+ \frac{2}{(k+5)^2} F_{21} F_{11} F_{22}
+ \frac{i (k+1)}{(k+5)^2} F_{21} A_3
\nonu \\
&+ & \left. \frac{i (k+7)}{(k+5)^2} F_{22} B_{-}
+ \frac{i (k-7)}{(k+5)^2} F_{21} B_3
-\frac{6 k}{(k+5)^2} \pa F_{21}
\right](w)
\nonu \\
& + & \frac{1}{(z-w)} \, \left[ 
\frac{1}{2} {\bf P_{+}^{(\frac{5}{2})}}  -\frac{2 (k+9)}{3 (k+5)} \pa
{\bf T_{+}^{(\frac{3}{2})}}
\right. \nonu \\
&+& \frac{i}{2 (k+5)} A_3 G_{21}
-\frac{2}{(k+5)^2} A_3 B_3 F_{21}
+ \frac{1}{(k+5)^2} A_3 B_{-} F_{22}
\nonu \\
&+& \frac{2 i (k+3)}{3 (k+5)^2} \pa A_3 F_{21}
+ \frac{2 i k}{3 (k+5)^2} A_3 \pa F_{21}
-\frac{i}{2 (k+5)} A_{-} G_{11}
\nonu \\
& + & \frac{2}{(k+5)^2} A_{-} B_3 F_{11}
+ \frac{1}{(k+5)^2} A_{-} B_{-} F_{12}
-\frac{2 i (k+3)}{3 (k+5)^2} \pa A_{-} F_{11}
\nonu \\
&-& \frac{2 i k}{3 (k+5)^2} A_{-} \pa F_{11}
+ \frac{i}{2 (k+5)} B_3 G_{21}
-\frac{2}{(k+5)^2} B_3 B_3 F_{21}
\nonu \\
&+& \frac{i (k-15)}{3 (k+5)^2} \pa B_3 F_{21}
+ \frac{4 i k}{3 (k+5)^2} B_3 \pa F_{21}
+ \frac{1}{(k+5)^2} B_{-} B_3 F_{22}
\nonu \\
&+& \frac{i (2 k+15)}{3 (k+5)^2} \pa B_{-} F_{22}
+ \frac{i (2 k+9)}{3 (k+5)^2} B_{-} \pa F_{22}
-\frac{1}{(k+5)^2} B_{+} B_{-} F_{21}
\nonu \\
&+& \frac{1}{(k+5)^2} F_{11} F_{21} G_{22}
-\frac{2 (k-3)}{3 (k+5)^3} \pa F_{11} F_{21} F_{22}
-\frac{2 (k+9)}{3 (k+5)^3} F_{11} \pa F_{21} F_{22}
\nonu \\
&-& \frac{8 (k+6)}{3 (k+5)^3} F_{11} F_{21} \pa F_{22}
-\frac{1}{(k+5)^2} F_{11} F_{22} G_{21}
+ \frac{1}{(k+5)^2} F_{12} F_{21} G_{21}
\nonu \\
&-& \frac{3 (k+1)}{2 (k+5)^2} \pa^2 F_{21}
 -\frac{(2 k+15)}{3 (k+5)} \pa G_{21}
-\frac{1}{(k+5)} T F_{21}
\nonu \\
&-& \frac{1}{2 (k+5)} U G_{21}
+ \frac{2 i}{(k+5)^2} U A_3 F_{21}
-\frac{2 i}{(k+5)^2} U A_{-} F_{11}
\nonu \\
&+& \frac{i}{(k+5)^2} U B_{-} F_{22}
-\frac{2 (k+6)}{3 (k+5)^2} \pa U F_{21}
-\frac{2 (k+6)}{3 (k+5)^2} U \pa F_{21}
\nonu \\
&- & \left.  
\frac{2}{(k+5)^2} U U F_{21} -
\frac{8}{(k+5)^3} F_{21} \pa F_{21} F_{12}
\right](w) +\cdots.
\nonu 
\eea
The nonderivative terms containing the spin-$\frac{3}{2}$ current
in the first order pole do not appear in the corresponding OPE in the 
nonlinear version.

Similarly one can obtain the following OPE
\bea
%%%%%%%%%%%%%%%%%%%%%%%%%%%%%%%%%%%%%%%%%%%%%%%%%%%%%%%%%%%%%%%%%%%%%%%%
{\bf U_{+}^{(2)}}(z) \, {\bf V_{+}^{(2)}}(w) 
& = & \frac{1}{(z-w)^2} \, \left[ -\frac{2 (k+4)}{(k+5)^2}
{A}_{-} {B}_{-} 
-\frac{2 i}{(k+5)^2} A_{-} F_{11} F_{21}
\right. \nonu \\
&+& \left. \frac{2 i}{(k+5)^2} B_{-} F_{21} F_{22}
-\frac{(k-3)}{(k+5)^2} F_{21} G_{21}
+  \frac{4 (3 k-1)}{(k+5)^3} F_{21} \pa F_{21}
\right] (w)
\nonu \\
& + & \frac{1}{(z-w)} \, 
\left[ 
 -\frac{2 i}{(k+5)^2} A_3 F_{21} G_{21}
+ \frac{1}{(k+5)^2} \pa A_{-} B_{-}
\right. \nonu \\
& - & \frac{(2 k+9)}{(k+5)^2} A_{-} \pa B_{-}
+ \frac{i}{(k+5)^2} A_{-} F_{11} G_{21}
-\frac{4 i (k+3)}{(k+5)^3} A_{-} \pa F_{11} F_{21}
 \nonu \\
&-&  \frac{8 i}{(k+5)^3} A_{-} F_{11} \pa F_{21}
+ \frac{i}{(k+5)^2} A_{-} F_{21} G_{11}
- \frac{2 i}{(k+5)^2} B_3 F_{21} G_{21}
\nonu \\
&+& \frac{i}{(k+5)^2} B_{-} F_{21} G_{22}
+ \frac{2 i}{(k+5)^2} \pa B_{-} F_{21} F_{22}
+ \frac{8 i}{(k+5)^3} B_{-} \pa F_{21} F_{22}
\nonu \\
&-& \frac{8 i}{(k+5)^3} B_{-} F_{21} \pa F_{22}
+ \frac{i}{(k+5)^2} B_{-} F_{22} G_{21}
-\frac{2 (k+4)}{(k+5)^2} \pa F_{21} G_{21}
\nonu \\
&+& \frac{2}{(k+5)^2} F_{21} \pa G_{21}
+ \frac{(5 k+1)}{(k+5)^3} F_{21} \pa^2 F_{21}
+ \frac{4}{(k+5)^2} F_{21} \pa F_{21} U
\nonu \\
&- &  \left.
 \frac{4 i (k+1)}{(k+5)^3} F_{21} \pa F_{21} A_3
+ \frac{16 i}{(k+5)^3} F_{21} \pa F_{21} B_3
\right](w) +\cdots.
\nonu 
\eea
One sees that there are no nonlinear terms in the context of 
the footnote \ref{nonlineardef}.

One also presents the other case
\bea
%%%%%%%%%%%%%%%%%%%%%%%%%%%%%%%%%%%%%%%%%%%%%%%%%%%%%%%%%%%%%%%%%%%%%%
{\bf U_{+}^{(2)}}(z) \, {\bf V_{-}^{(2)}}(w) 
&=& 
\frac{1}{(z-w)^4} \, \left[ -\frac{6 k (k+9)}{(k+5)^2}\right]
\nonu \\
&+& 
\frac{1}{(z-w)^3} \, \left[  
\frac{4 i k (k+8)}{(k+5)^2}  {B}_3   
- \frac{4 k}{(k+5)^2} F_{11} F_{22}
+ \frac{4 k}{(k+5)^2}  F_{12} F_{21}
\right](w)
\nonu \\
& + & \frac{1}{(z-w)^2} \, \left[ \frac{1}{2} {\bf P^{(2)}} - 
\frac{2 (k-3)}{3 (k+5)}  {\bf T^{(2)}} 
-2  {\bf W^{(2)}} 
-\frac{2 (2 k+15)}{3 (k+5)} T
\right. \nonu \\
& - & \frac{4 (k+3)}{3 (k+5)^2} A_3 A_3  +
\frac{2 (k-3)}{3 (k+5)^2}  A_3 B_3
-\frac{4 i (k+3)}{3 (k+5)^2} \pa A_3
\nonu \\
& + & \frac{4 i (k+9)}{3 (k+5)^3}
 A_3 F_{11} F_{22}
+  \frac{2 i}{(k+5)^2} A_3 F_{12} F_{21}
-\frac{4 (k+3)}{3 (k+5)^2} A_{+} A_{-}
\nonu \\
&+& \frac{10 i (k+9)}{3 (k+5)^3}  A_{-} F_{11} F_{12}
- \frac{8 i}{(k+5)^3} A_{+} F_{21} F_{22}
+ \frac{2 (k-15)}{3 (k+5)^2} B_3 B_3
\nonu \\
& + & \frac{2 i \left(3 k^2+22 k-15\right)}{3 (k+5)^2} \pa B_3
  -\frac{4 i (k+9)}{3 (k+5)^3}   B_3 F_{11} F_{22}
+  \frac{2 i}{(k+5)^2} B_3 F_{12} F_{21}
\nonu \\
& + & \frac{8 i k}{3 (k+5)^3}  B_{-} F_{12} F_{22}
-\frac{2 (2 k+15)}{3 (k+5)^2} B_{+} B_{-}
-  \frac{2 i (3 k+11)}{(k+5)^3} B_{+} F_{11} F_{21}
\nonu \\
&+& \frac{2 (k-3)}{3 (k+5)^2} F_{11} G_{22}
-\frac{2 \left(5 k^2+24 k+51\right)}{3 (k+5)^3}  \pa F_{11} F_{22}
\nonu \\
& - & \frac{2 \left(k^2-12 k-141\right)}{3 (k+5)^3}  F_{11} \pa F_{22}
- \frac{(k+9)}{(k+5)^2} F_{12} G_{21}
+\frac{2 \left(k^2-9 k-78\right)}{3 (k+5)^3} \pa F_{12} F_{21}
\nonu \\
& + & \frac{2 \left(5 k^2+31 k+102\right)}{3 (k+5)^3} F_{12} \pa F_{21}
+ \frac{(k+3)}{(k+5)^2}  F_{21} G_{12}
+ \frac{(k-3)}{3 (k+5)^2}  F_{22} G_{11}
\nonu \\
& - & \frac{6 i}{(k+5)^2} U A_3 
-\frac{2 i k}{(k+5)^2} U B_3 
-\frac{2 (2 k+15)}{3 (k+5)^2} U U
\nonu \\
&+ & \left. \frac{8 (k-3)}{3 (k+5)^3}  U F_{11} F_{22}
-\frac{6}{(k+5)^2}  U F_{12} F_{21}
\right](w)
\nonu \\
& + & \frac{1}{(z-w)} \, \left[ \frac{1}{2} {\bf S^{(3)}} 
+\frac{1}{2} {\bf P^{(3)}}
+\frac{1}{4} 
  \pa 
{\bf P^{(2)}}  -{\bf W^{(3)}}  
-\frac{(k-3)}{3 (k+5)} \pa {\bf T^{(2)}} 
\right. \nonu \\
& - &  \pa {\bf W^{(2)}} -\frac{2 i}{(k+5)^2} A_3 A_3 B_3
-\frac{4 (k+3)}{3 (k+5)^2} \pa A_3 A_3
 \nonu \\
& - & \frac{4 (2 k+9)}{3 (k+5)^2} \pa A_3 B_3
+ \frac{2 (5 k+18)}{3 (k+5)^2} A_3 \pa B_3
-\frac{2 i (k+3)}{3 (k+5)^2} \pa^2 A_3
\nonu \\
&+& \frac{i}{(k+5)^2} A_3 F_{11} G_{22}
-\frac{i (11 k+15)}{6 (k+5)^3} \pa A_3 F_{11} F_{22}
\nonu \\
&+& \frac{i (13 k+57)}{6 (k+5)^3} A_3 \pa F_{11} F_{22} 
+ \frac{i (25 k+117)}{6 (k+5)^3} A_3 F_{11} \pa F_{22}
-\frac{ i}{(k+5)^2} A_3 F_{12} G_{21}
\nonu \\
&+& \frac{i (7 k+27)}{2 (k+5)^3} \pa A_3 F_{12} F_{21}
-\frac{i (k-3)}{2 (k+5)^3} A_3 \pa F_{12} F_{21}
-\frac{i (5 k+17)}{2 (k+5)^3} A_3 F_{12} \pa F_{21}
\nonu \\
&+& \frac{ i}{(k+5)^2} A_3 F_{21} G_{12}
-\frac{i}{(k+5)^2} A_{3} F_{22} G_{11}
+\frac{5i (k+9)}{3 (k+5)^3} \pa A_{-} F_{11} F_{12}
\nonu \\
& - & \frac{i (k-15)}{3 (k+5)^3} A_{-} \pa F_{11} F_{12}
+ \frac{ i (11k+75)}{3 (k+5)^3} A_{-} F_{11} \pa F_{12}
- \frac{2i}{(k+5)^2} A_{+} A_{-} B_3
\nonu \\
&-& \frac{2(k+3)}{3(k+5)^2} \pa (A_{+} A_{-})
-\frac{4 i }{ (k+5)^3} \pa (A_{+} F_{21} F_{22})
+ \frac{2 i }{ (k+5)^2} B_3 B_3 B_3
\nonu \\
&+& \frac{2  (k-18)}{3 (k+5)^2}  \pa B_3 B_3
+  \frac{i \left(14 k^3+80 k^2-445 k-459\right)}{3 (k+5)^2 (13 k+17)} \pa^2 B_3 
\nonu \\
&+& 
\frac{i}{(k+5)^2} B_3 F_{11} G_{22}
- \frac{i (13 k+105)}{6 (k+5)^3}  \pa B_3 F_{11} F_{22}
\nonu \\
&-& \frac{i (k-99)}{6 (k+5)^3} B_3 \pa F_{11} F_{22}
+ \frac{11 i (k-3)}{6 (k+5)^3}  B_3  F_{11} \pa F_{22}
\nonu \\
&+& \frac{ i}{(k+5)^2} B_3 F_{12} G_{21}
- \frac{i (k+13)}{2 (k+5)^3} \pa B_3 F_{12} F_{21}
+ \frac{i (3k+55)}{2 (k+5)^3} B_3 \pa F_{12} F_{21}
\nonu \\
&+& \frac{i (7 k+11)}{2 (k+5)^3} B_3 F_{12} \pa F_{21}
-\frac{(28 k^3+149 k^2-354 k-459)}{3 (k+5)^3 (13 k+17)} \pa F_{11} \pa F_{22}
\nonu \\
&+& \frac{3 i}{(k+5)^2} B_3 F_{21} G_{12}
+ \frac{3i}{(k+5)^2} B_{3} F_{22} G_{11}
+\frac{i (11k+15)}{6 (k+5)^3} \pa B_{-} F_{12} F_{22}
\nonu \\
& + & \frac{i (11 k+15)}{6 (k+5)^3} B_{-} \pa F_{12} F_{22}
- \frac{i ( k+45)}{6 (k+5)^3} B_{-} F_{12} \pa F_{22}
+\frac{2i}{(k+5)^2} B_{+} B_{-} B_3
\nonu \\
&-& \frac{2(k+6)}{3(k+5)^2} \pa B_{+} B_{-}
-\frac{2 (k+9)}{3 (k+5)^2} B_{+} \pa B_{-}
-\frac{i (7 k+27)}{2 (k+5)^3} \pa B_{+} F_{11} F_{21}
\nonu \\
& - & \frac{i (7 k+27)}{2 (k+5)^3} B_{+} \pa F_{11} F_{21}
- \frac{i (3 k+7)}{2 (k+5)^3} B_{+} F_{11} \pa F_{21}
\nonu \\
& - & \frac{(80 k^3+217 k^2+426 k+561)}{6 (k+5)^3 (13 k+17)}
  \pa^2 F_{11} F_{22}
\nonu \\
&+& \frac{(8 k^3+285 k^2+1930 k+1989)}{2 (k+5)^3 (13 k+17)}  F_{11} \pa^2 F_{22}
+ \frac{(13 k-3)}{12 (k+5)^2} \pa F_{11} G_{22}
\nonu \\
&+& \frac{(k-15)}{12 (k+5)^2} F_{11} \pa G_{22}
- \frac{(8 k^2+219 k+255)}{2 (k+5)^2 (13 k+17)} \pa^2 F_{12} F_{21}
\nonu \\
&+&  \frac{(28 k^3+279 k^2+1298 k+1479)}{3 (k+5)^3 (13 k+17)}
\pa F_{12} \pa F_{21}
\nonu \\
& + & \frac{(80 k^3+399 k^2+1990 k+2295)}{6 (k+5)^3 (13 k+17)}
F_{12} \pa^2 F_{21}
-\frac{(5 k+21)}{4 (k+5)^2} \pa F_{12} G_{21}
\nonu \\
&-& \frac{(k+17)}{4 (k+5)^2}
 F_{12} \pa G_{21} + 
\frac{1}{2 (k+5)} G_{11} G_{22}
-\frac{1}{2 (k+5)} G_{12} G_{21}
\nonu \\
&+& \frac{k}{2 (k+5)^2} \pa F_{21} G_{12}
+ \frac{(k+4)}{2 (k+5)^2} F_{21} \pa G_{12}
+\frac{(k+6)}{6 (k+5)^2} \pa F_{22} G_{11}
\nonu \\
&+& \frac{(k-6)}{6 (k+5)^2}  F_{22} \pa G_{11}
+ \frac{2 i \left(4 k^2+34 k+17\right)}{(k+5) (13 k+17)}  T B_3
\nonu \\
&-& \frac{(2 k+15)}{3 (k+5)}  \pa T 
- \frac{2 \left(4 k^2+25 k+17\right)}{(k+5)^2 (13 k+17)} T F_{11} F_{22}
\nonu \\
& + &  \frac{2 \left(4 k^2+25 k+17\right)}{(k+5)^2 (13 k+17)} 
T F_{12} F_{21}
-\frac{4}{(k+5)^2} U A_3 B_3
\nonu \\
& -& \frac{3 i}{(k+5)^2} \pa (U A_3)
-\frac{2 i}{(k+5)^2} \pa U  B_3
-\frac{2i(k-1)}{(k+5)^2} U \pa B_3
\nonu \\
&+& \frac{1}{(k+5)^2} U F_{11} G_{22}
- \frac{(k+21)}{6 (k+5)^3} \pa U F_{11} F_{22}
+ \frac{(23 k+3)}{6 (k+5)^3} U \pa F_{11} F_{22}
\nonu \\
&+& \frac{(11 k-57)}{6 (k+5)^3} U F_{11} \pa F_{22}
-\frac{1}{(k+5)^2} U F_{12} G_{21}
-\frac{(3k+31)}{2 (k+5)^3} \pa U F_{12} F_{21}
\nonu \\
& - &   \frac{(11 k+39)}{2 (k+5)^3} U \pa F_{12} F_{21}
- \frac{(7k+19)}{2 (k+5)^3} U F_{12} \pa F_{21}
+\frac{1}{(k+5)^2}  U F_{21} G_{12}
\nonu \\
& - & \left.  
\frac{1}{(k+5)^2} U F_{22} G_{11}
+ \frac{2 i}{(k+5)^2}  U U B_3
-\frac{2 (2 k+15)}{3 (k+5)^2} \pa U U
\right](w) +\cdots.
\nonu 
\eea
The nontrivial central term of this OPE appears.
The nonlinear terms in the corresponding OPEs in the 
nonlinear version disappear in this OPE. 

For the other higher spin current, one can obtain
\bea
%%%%%%%%%%%%%%%%%%%%%%%%%%%%%%%%%%%%%%%%%%%%%%%%%%%%%%%%%%%%%%%%%%%%%%%%%
{\bf U_{+}^{(2)}}(z) \, 
{\bf V^{(\frac{5}{2})}}(w)  
& = & 
\frac{1}{(z-w)^4} \,
\left[ \frac{12 k(k+9)}{(k+5)^3} F_{21} \right](w) \nonu \\
&+& \frac{1}{(z-w)^3} \, \left[ 
 \frac{2 \left(4 k^2+40 k+63\right)}{3 (k+5)^2}  G_{21} 
+ \frac{4 i \left(4 k^2+25 k+9\right)}{3 (k+5)^3}  F_{11} A_{-}
\right. \nonu \\
&+& \frac{4 \left(4 k^2+31 k+63\right)}{3 (k+5)^3}   U F_{21}
- \frac{8(k+9)}{3(k+5)^3} F_{21} F_{11} F_{22}
\nonu \\
& - & \frac{4 i \left(4 k^2+25 k+9\right)}{3 (k+5)^3} F_{21} A_3
+   \frac{4 i \left(2 k^2-k-63\right)}{3 (k+5)^3} F_{22} B_{-}
\nonu \\
& + & \left. \frac{4 i \left(2 k^2+13 k+63\right)}{3 (k+5)^3} F_{21} B_3
\right](w)
\nonu \\
& + & \frac{1}{(z-w)^2} \, \left[ 
-\frac{(7 k+39)}{3 (k+5)}  ({\bf P_{+}^{(\frac{5}{2})}}  +
{\bf W_{+}^{(\frac{5}{2})}})
\right. \nonu \\
&-& \frac{4i(k+3)}{3 (k+5)^2} A_3 G_{21}
+\frac{16(k+6)}{3(k+5)^3} A_3 B_3 F_{21}
- \frac{2(7k+39)}{3(k+5)^3} A_3 B_{-} F_{22}
\nonu \\
&-& \frac{16 i (k+3) (k+9)}{9 (k+5)^3} \pa A_3 F_{21}
-\frac{2 i \left(8 k^2-9 k-189\right)}{9 (k+5)^3} A_3 \pa F_{21}
\nonu \\
& - & \frac{16(k+6)}{3(k+5)^3} A_{-} B_3 F_{11}
- \frac{2(7k+39)}{3(k+5)^3} A_{-} B_{-} F_{12}
+ \frac{16 i (k+3) (k+9)}{9 (k+5)^3}  \pa A_{-} F_{11}
\nonu \\
&+& \frac{2 i \left(8 k^2-9 k-189\right)}{9 (k+5)^3} A_{-} \pa F_{11}
+ \frac{2 i (k+21)}{3 (k+5)^2} B_3 G_{21}
- \frac{8 (k+3)}{3 (k+5)^3} B_3 B_3 F_{21}
\nonu \\
&-&\frac{4 i \left(5 k^2+51 k+18\right)}{9 (k+5)^3}  \pa B_3 F_{21}
+ \frac{2 i \left(32 k^2+225 k+117\right)}{9 (k+5)^3} B_3 \pa F_{21}
\nonu \\
& -  & \frac{i (3 k+17)}{(k+5)^2}  B_{-} G_{22}
-\frac{2 (k+9)}{3 (k+5)^3}
  B_{-} B_3 F_{22}
+  \frac{4 i (k+3)}{3 (k+5)^2}  A_{-} G_{11}
\nonu \\
&+& \frac{2 i \left(19 k^2+93 k-54\right)}{9 (k+5)^3}  \pa B_{-} F_{22}
- \frac{4 i \left(13 k^2+90 k+99\right)}{9 (k+5)^3} B_{-} \pa F_{22}
\nonu \\
& - & \frac{2 (5 k+21)}{3 (k+5)^3} B_{+} B_{-} F_{21}
+\frac{4 (k+9)^2}{9 (k+5)^4}  F_{11} \pa F_{21} F_{22}
+  \frac{4(k+3)}{3(k+5)^3} F_{12} F_{21} G_{21}
\nonu \\
&-& \frac{2(k+9)}{3(k+5)^3} F_{11} F_{21} G_{22}
-\frac{8 \left(13 k^2+114 k+261\right)}{9 (k+5)^4} \pa F_{11} F_{21} F_{22}
\nonu \\
&+&  \frac{4 \left(31 k^2+294 k+711\right)}{9 (k+5)^4}  F_{11} F_{21} \pa F_{22}
+\frac{8(k+6)}{3(k+5)^3} F_{11} F_{22} G_{21}
\nonu \\
&-& \frac{2 \left(4 k^2+27 k+27\right)}{3 (k+5)^3} \pa^2 F_{21}
-\frac{4}{(k+5)^2} F_{21} F_{22} G_{11}
\nonu \\
&  + &  \frac{4 (k+3) (2 k+15)}{9 (k+5)^2} \pa G_{21}
+\frac{4(2k+15)}{3(k+5)^2} T F_{21}
\nonu \\
&+& \frac{2(2k+9)}{3 (k+5)^2} U G_{21}
- \frac{16 i(k+6)}{3(k+5)^3} U A_3 F_{21}
+\frac{16 i(k+6)}{3(k+5)^3} U A_{-} F_{11}
\nonu \\
&-& \frac{2i(13k+69)}{3(k+5)^3} U B_{-} F_{22}
+\frac{8 (k+3)(2k+15)}{9 (k+5)^3} \pa U F_{21}
\nonu \\
& + & 
 \frac{2 \left(8 k^2+39 k+99\right)}{9 (k+5)^3}  U \pa F_{21}
+ \frac{8 i}{(k+5)^2} U B_3 F_{21}
\nonu \\
&+ & \left.  
\frac{16(k+6)}{3(k+5)^3} U U F_{21} +
\frac{4 \left(9 k^2+82 k+201\right)}{3 (k+5)^4}
F_{21} \pa F_{21} F_{12}
 \right](w) \nonu \\
& + &  \frac{1}{(z-w)} \, \left[  + \cdots \right](w) 
+ \cdots.
\nonu 
\eea
There is no higher spin current in the third order pole of this OPE.

The result (\ref{g1122vu3half}) determines the following OPEs
\bea
%%%%%%%%%%%%%%%%%%%%%%%%%%%%%%%%%%%%%%%%%%%%%%%%%%%%%%%%%%%%%%%%%%%%%%%%%%%
\left(
\begin{array}{c}
{\bf U_{+}^{(2)}} \\
{\bf V_{-}^{(2)}} 
\end{array}
\right)(z) \, {\bf W^{(2)}}(w) & = &
\frac{1}{(z-w)^3} 
\, \left[ -\frac{2 i k ( k+8)}{ (k+5)^2} {B}_{\mp} \pm
\frac{4 k }{ (k+5)^2}
\left(
\begin{array}{c}
F_{11} F_{21}  \\
F_{22} F_{12}
\end{array}
\right) 
\right] (w) 
\nonu \\
& + & \frac{1}{(z-w)^2} \, \left[  
\mp
\frac{2 (k+4)}{ (k+5)^2} A_3 B_{\mp}
-\frac{2 i }{ (k+5)^2} A_3 
\left(
\begin{array}{c}
F_{11} F_{21}  \\
F_{22} F_{12}
\end{array}
\right)
\right. \nonu \\
& \mp & \frac{2 k }{(k+5)^2} B_{\mp} B_3 
 -\frac{i k (k+9)}{(k+5)^2} \pa B_{\mp}
\nonu \\
& \pm & \frac{ i }{ (k+5)^2} B_{\mp} F_{11} F_{22}
\pm \frac{ i }{ (k+5)^2} B_{\mp} F_{12} F_{21}
\nonu \\
& \mp & \frac{ (k-3)}{2(k+5)^2} 
 \left(
\begin{array}{c}
F_{11} G_{21}  \\
F_{22} G_{12}
\end{array}
\right) 
\pm \frac{2 (k+1)^2}{(k+5)^3} 
 \left(
\begin{array}{c}
\pa F_{11} F_{21}  \\
\pa F_{22} F_{12}
\end{array}
\right) 
\nonu \\
&\pm & \left. \frac{2 \left(k^2+8 k-1\right)}{(k+5)^3}
 \left(
\begin{array}{c}
F_{11} \pa F_{21}  \\
F_{22} \pa F_{12}
\end{array}
\right) \mp
  \frac{(k-3)}{2(k+5)^2} 
 \left(
\begin{array}{c}
F_{21} G_{11}  \\
F_{12} G_{22}
\end{array}
\right)  
\right](w) 
\nonu \\
&+ & \frac{1}{(z-w)} \, \left[ 
-\frac{1}{2} 
\left(
\begin{array}{c}
{\bf Q_{+}^{(3)}} \\
 {\bf R_{-}^{(3)}} 
\end{array}
\right)
+ \frac{i}{(k+5)^2} A_3 A_3 B_{\mp}
\pm \frac{(3 k+13)}{2 (k+5)^2}  \pa A_3 B_{\mp}
\right. \nonu \\
&\mp & \frac{(7 k+31)}{2 (k+5)^2}  A_3 \pa B_{\mp}
- \frac{ i}{(k+5)^2}  A_3
\left(
\begin{array}{c}
F_{11} G_{21} \\
F_{22} G_{12} 
\end{array}
\right)
\nonu \\
&+& \frac{i (5 k+17)}{2 (k+5)^3}  \pa A_3 
\left(
\begin{array}{c}
F_{11} F_{21} \\
F_{22} F_{12} 
\end{array}
\right)
- \frac{i (7 k+27)}{2 (k+5)^3} A_3 
 \left(
\begin{array}{c}
 \pa F_{11} F_{21} \\
 \pa F_{22} F_{12} 
\end{array}
\right)
\nonu \\
& -& \frac{i (11 k+47)}{2 (k+5)^3}
A_3 \left(
\begin{array}{c}
 F_{11} \pa F_{21} \\
 F_{22} \pa F_{12} 
\end{array}
\right)
+ \frac{i}{(k+5)^2} A_{\mp} 
 \left(
\begin{array}{c}
 F_{11} G_{11} \\
 F_{22} G_{22} 
\end{array}
\right)
\nonu \\
&+& 
\frac{i}{(k+5)^2} A_3 
\left(
\begin{array}{c}
F_{21} G_{11} \\
F_{12} G_{22} 
\end{array}
\right)
+ \frac{i}{(k+5)^2} A_{\pm} A_{\mp} B_{\mp}
\nonu \\
&+ &  \frac{i}{(k+5)^2} A_{\pm}
\left(
\begin{array}{c}
F_{21} G_{21} \\
F_{12} G_{12} 
\end{array}
\right)
- \frac{i}{(k+5)^2} B_3 
\left(
\begin{array}{c}
F_{11} G_{21} \\
F_{22} G_{12} 
\end{array}
\right)
\nonu \\
& - &  \frac{i}{2 (k+5)^2} \pa B_3 
\left(
\begin{array}{c}
F_{11} F_{21} \\
F_{22} F_{12} 
\end{array}
\right)
 -  \frac{ i (k-15)}{2 (k+5)^3} B_{\mp}
\left(
\begin{array}{c}
F_{12} \pa F_{21} \\
F_{21} \pa F_{12} 
\end{array}
\right)
\nonu \\
&-& \frac{i(k+21)}{2 (k+5)^3} B_3
\left(
\begin{array}{c}
\pa F_{11} F_{21} \\
\pa F_{22} F_{12} 
\end{array}
\right)
+\frac{i(3k+31)}{2 (k+5)^3} B_3
\left(
\begin{array}{c}
F_{11} \pa F_{21} \\
F_{22} \pa F_{12} 
\end{array}
\right)
\nonu \\
&- & \frac{i}{(k+5)^2} B_3 
\left(
\begin{array}{c}
F_{21} G_{11} \\
F_{12} G_{22} 
\end{array}
\right)
- \frac{i}{(k+5)^2} B_{\mp} B_3 B_3
\nonu \\
&\mp  & \frac{(k-1)}{(k+5)^2} \pa B_{\mp} B_3
\mp \frac{(k-2)}{(k+5)^2} B_{\mp} \pa B_3
\nonu \\
&-& \frac{i (7 k+6) \left(2 k^2+19 k-51\right)}{6 (k+5)^2 (13 k+17)}
\pa^2 B_{\mp}
\nonu \\
&-& \frac{6i }{ (k+5)^3}  B_{\mp} 
\left(
\begin{array}{c}
\pa F_{11} F_{22} \\
\pa F_{22} F_{11} 
\end{array}
\right)
+ \frac{2i (k+6)}{3 (k+5)^3} \pa B_{\mp} F_{11} F_{22}
\nonu \\
&-& \frac{2i (k+4)}{ (k+5)^3}  B_{\mp}
\left(
\begin{array}{c}
F_{11} \pa F_{22} \\
F_{22} \pa F_{11} 
\end{array}
\right)
- \frac{i}{(k+5)^2} B_{\mp} 
\left(
\begin{array}{c}
F_{12} G_{21} \\
F_{21} G_{12} 
\end{array}
\right)
\nonu \\
&-& \frac{i (k+33)}{2 (k+5)^3}  B_{\mp} 
\left(
\begin{array}{c}
\pa F_{12}  F_{21} \\
\pa F_{21}  F_{12} 
\end{array}
\right)
\pm \frac{i (3k+19)}{2 (k+5)^3}  \pa B_{\mp} 
F_{12} F_{21}
\nonu \\
&-& \frac{2i}{(k+5)^2} B_{\mp}
\left(
\begin{array}{c}
F_{21}  G_{12} \\
F_{12}  G_{21} 
\end{array}
\right)
- \frac{ i}{(k+5)^2} B_{\mp} 
\left(
\begin{array}{c}
F_{22}  G_{11} \\
F_{11}  G_{22} 
\end{array}
\right)
\nonu \\
&\mp &  \frac{(7k+19)}{ (k+5)^3}
\left(
\begin{array}{c}
\pa^2 F_{11}  F_{21} \\
\pa^2 F_{22}  F_{12} 
\end{array}
\right)
-\frac{i}{(k+5)^2} B_{\pm} B_{\mp} B_{\mp}
\nonu \\
&\pm & \frac{(28 k^3+71 k^2-846 k-969)}{6 (k+5)^3 (13 k+17)} 
\left(
\begin{array}{c}
F_{11}  \pa^2 F_{21} \\
F_{22}  \pa^2 F_{12} 
\end{array}
\right)
\nonu \\
&\mp & \frac{(7 k+19)}{4 (k+5)^2} 
\left(
\begin{array}{c}
\pa F_{11} G_{21} \\
\pa F_{22}  G_{12} 
\end{array}
\right)
\pm   \frac{1}{4 (k+5)} 
\left(
\begin{array}{c}
F_{11} \pa G_{21} \\
F_{22}  \pa G_{12} 
\end{array}
\right)
\nonu \\
&\mp & \frac{(2 k+11)}{2 (k+5)^2} 
 \left(
\begin{array}{c}
\pa F_{21}  G_{11} \\
\pa F_{12}   G_{22} 
\end{array}
\right)
\pm
\frac{3}{2 (k+5)^2}
 \left(
\begin{array}{c}
F_{21}  \pa G_{11} \\
F_{12}   \pa G_{22} 
\end{array}
\right)
\nonu \\
&- & \frac{i \left(4 k^2+34 k+17\right)}{(k+5) (13 k+17)}
T B_{\mp}
\pm   \frac{2 \left(4 k^2+25 k+17\right)}{(k+5)^2 (13 k+17)}
T
\left(
\begin{array}{c}
F_{11}  F_{21} \\
F_{22}  F_{12} 
\end{array}
\right)
\nonu \\
&\pm & \frac{2}{(k+5)^2} U A_3 B_{\mp}
- \frac{i (k-2)}{2 (k+5)^2} \pa U B_{\mp}
+ \frac{i (k-2)}{2 (k+5)^2} U \pa B_{\mp}
\nonu \\
&\mp &
\frac{(9k+29)}{2 (k+5)^3}
U
\left(
\begin{array}{c}
\pa F_{11}  F_{21} \\
\pa F_{22}  F_{12} 
\end{array}
\right)
\mp \frac{1}{(k+5)^2} U
 \left(
\begin{array}{c}
F_{11}  G_{21} \\
 F_{22}  G_{12} 
\end{array}
\right)
\nonu \\
&\pm & \frac{(3k-1)}{2 (k+5)^3} \pa U
\left(
\begin{array}{c}
F_{11}  F_{21} \\
F_{22}  F_{12} 
\end{array}
\right)
\pm \frac{(3k+31)}{2 (k+5)^3} U
\left(
\begin{array}{c}
F_{11}  \pa F_{21} \\
F_{22}  \pa F_{12} 
\end{array}
\right)
\nonu \\
&\pm & \frac{1}{(k+5)^2}
U
\left(
\begin{array}{c}
F_{21}  G_{11} \\
F_{12}  G_{22} 
\end{array}
\right)
-\frac{i}{(k+5)^2} U U B_{\mp}
\nonu \\
& +  & \frac{4 i(k+3)}{(k+5)^3}
\left(
\begin{array}{c}
F_{11}  A_{-} \pa F_{11} \\
F_{22}  A_{+} \pa F_{22} 
\end{array}
\right)
\mp  \frac{1}{2 (k+5)} 
\left(
\begin{array}{c}
G_{11}  G_{21} \\
G_{22}  G_{12} 
\end{array}
\right)
\nonu \\
& -& 
\frac{i \left(28 k^3+227 k^2+450 k+459\right)}{3 (k+5)^3 (13 k+17)}
\left(
\begin{array}{c}
F_{21}  A_{+}  \pa F_{21} \\
F_{12}  A_{-} \pa F_{12} 
\end{array}
\right)
\nonu \\
&-& \left.
\frac{i \left(28 k^3+305 k^2+630 k+561\right)}{3 (k+5)^3 (13 k+17)}
\left(
\begin{array}{c}
\pa F_{21}  A_{+}  F_{21} \\
\pa F_{12}  A_{-} F_{12} 
\end{array}
\right)
\nonu 
\right](w) +\cdots.
\nonu 
\eea
There is no higher spin current in the second order pole.

The result (\ref{g1221w2}) determines the following OPEs
\bea
%%%%%%%%%%%%%%%%%%%%%%%%%%%%%%%%%%%%%%%%%%%%%%%%%%%%%%%%%
\left(
\begin{array}{c}
{\bf U_{+}^{(2)}} \\
{\bf V_{-}^{(2)}}
\end{array}
\right) (z) \, {\bf W_{\pm}^{(\frac{5}{2})}}(w) & = &
 \frac{1}{(z-w)^3}
\,
\left[ - \frac{16 i k(k+3)}{3(k+5)^3} 
\left(
\begin{array}{c}
F_{21} \\
F_{12} 
\end{array}
\right) B_{\mp}
\right](w) 
\nonu \\
&+& \frac{1}{(z-w)^2} \, \left[ 
\mp \frac{2(k+9)}{3(k+5)^3} A_3 B_{\mp}
\left(
\begin{array}{c}
F_{21} \\
F_{12} 
\end{array}
\right)
\right. \nonu \\
&\pm& \frac{2(k+9)}{3(k+5)^3} A_{\mp} B_{\mp}
\left(
\begin{array}{c}
F_{11} \\
F_{22} 
\end{array}
\right)
+\frac{i(7k+9)}{3(k+5)^2} B_{\mp}
\left(
\begin{array}{c}
G_{21} \\
G_{12} 
\end{array}
\right)
\nonu \\
&\mp & \frac{2(k+9)}{3(k+5)^3} B_{\mp} B_3 
\left(
\begin{array}{c}
F_{21} \\
F_{12} 
\end{array}
\right) \pm
\frac{2(k+9)}{3(k+5)^3} B_{\mp} B_{\mp} 
\left(
\begin{array}{c}
F_{22} \\
F_{11} 
\end{array}
\right)
\nonu \\
& - & \frac{2 i (k+9) (2 k-3)}{3 (k+5)^3}   B_{\mp} \pa
\left(
\begin{array}{c}
F_{21} \\
F_{12} 
\end{array}
\right) \mp
\frac{2(k+9)}{3(k+5)^3} 
\left(
\begin{array}{c}
F_{11} F_{21} G_{21} \\
F_{22} F_{12} G_{12}
\end{array}
\right)
\nonu \\
&+ & \frac{2i(k+9)}{3(k+5)^3} U B_{\mp} 
\left(
\begin{array}{c}
F_{21}  \\
F_{12} 
\end{array}
\right) -
\frac{2i k(k+1)}{(k+5)^3} 
\pa B_{\mp}
\left(
\begin{array}{c}
F_{21}  \\
F_{12} 
\end{array}
\right) 
\nonu \\
& \mp & \left. \frac{4 (k+1)(k+9)}{3(k+5)^4}
\left(
\begin{array}{c}
F_{11} F_{21} \pa F_{21} \\
F_{22} F_{12} \pa F_{12}
\end{array}
\right) 
\right](w)
\nonu \\
&+& \frac{1}{(z-w)} \, \left[ + \cdots \right](w) 
+\cdots.
\nonu 
\eea
It is an open problem to obtain the first order pole of these OPEs.
The nonlinear terms in the corresponding OPEs in the 
nonlinear version disappear.

Similarly the other combination provides the following OPEs
\bea
%%%%%%%%%%%%%%%%%%%%%%%%%%%%%%%%%%%%%%%%%%%%%%%%%%%%%%%%%%%%%%%%%%%%%%%%%
\left(
\begin{array}{c}
{\bf U_{+}^{(2)}} \\
{\bf V_{-}^{(2)}}
\end{array}
\right) (z) \, {\bf W_{\mp}^{(\frac{5}{2})}}(w) & = &
\frac{1}{(z-w)^4} \, \left[ \mp \frac{12 k(k+9)}{(k+5)^3}
\left(
\begin{array}{c}
F_{11} \\
F_{22} 
\end{array}
\right)
\right](w)
\nonu \\
& + & \frac{1}{(z-w)^3}\, 
\left[ \mp
\frac{2 \left(4 k^2+40 k+63\right)}{3 (k+5)^2}
\left(
\begin{array}{c}
G_{11} \\
G_{22} 
\end{array}
\right) \right. \nonu \\
& \mp & 
    \frac{4 \left(4 k^2+31 k+63\right)}{3 (k+5)^3}
U
\left(
\begin{array}{c}
F_{11} \\
F_{22} 
\end{array}
\right)
-  \frac{8 (k+9)}{3 (k+5)^3}
F_{12} F_{21}
\left(
\begin{array}{c}
F_{11} \\
F_{22} 
\end{array}
\right)
 \nonu \\
& - & 
\frac{4 i \left(4 k^2+25 k+9\right)}{3 (k+5)^3}
\left(
\begin{array}{c}
F_{11} \\
F_{22} 
\end{array}
\right) A_3 
-  \frac{4 i \left(4 k^2+25 k+9\right)}{3 (k+5)^3}
\left(
\begin{array}{c}
F_{21} \\
F_{12} 
\end{array}
\right) A_{\pm}
\nonu \\
&+ & \left.
\frac{4 i \left(2 k^2-k-63\right)}{3 (k+5)^3}
\left(
\begin{array}{c}
F_{12} \\
F_{21} 
\end{array}
\right) B_{\mp}
- \frac{4 i \left(2 k^2+13 k+63\right)}{3 (k+5)^3}
\left(
\begin{array}{c}
F_{11} \\
F_{22} 
\end{array}
\right) B_3 
\right] (w) 
\nonu \\
& + & \frac{1}{(z-w)^2} \, \left[ 
\frac{(7 k+39)}{3 (k+5)}
\left(
\begin{array}{c}
{\bf Q^{(\frac{5}{2})}} \\
 {\bf R^{(\frac{5}{2})}} \end{array}
\right) \pm
\frac{(7 k+39)}{3 (k+5)}
\left(
\begin{array}{c}
{\bf U^{(\frac{5}{2})} } \\
{\bf V^{(\frac{5}{2})}} 
 \end{array}
\right)
\right. \nonu \\
&- & \frac{4i(k+3)}{3 (k+5)^2} A_3 
\left(
\begin{array}{c}
G_{11} \\
G_{22} 
\end{array}
\right) 
\pm
\frac{16 (k+6)}{3 (k+5)^3} A_3 B_3 
\left(
\begin{array}{c}
F_{11} \\
F_{22} 
\end{array}
\right) 
\nonu \\
&\pm&
\frac{2(7k+39)}{3(k+5)^3} A_3 B_{\mp}
\left(
\begin{array}{c}
F_{12} \\
F_{21} 
\end{array}
\right)
\mp  
\frac{2(7k+39)}{3(k+5)^3} A_{\pm} B_{\mp} 
\left(
\begin{array}{c}
F_{22} \\
F_{11} 
\end{array}
\right)
\nonu \\
&-&   \frac{16 i (k+3) (k+9)}{9 (k+5)^3}
\pa A_3 
\left(
\begin{array}{c}
F_{11} \\
F_{22} 
\end{array}
\right)
- \frac{2 i \left(8 k^2-9 k-189\right)}{9 (k+5)^3}  A_3 \pa 
\left(
\begin{array}{c}
F_{11} \\
F_{22} 
\end{array}
\right)
\nonu \\
&- & \frac{4 i(k+3)}{3(k+5)^2} A_{\pm}  
\left(
\begin{array}{c}
G_{21} \\
G_{12} 
\end{array}
\right)
\pm \frac{16(k+6)}{3(k+5)^3} A_{\pm} B_3 
\left(
\begin{array}{c}
F_{21} \\
F_{12} 
\end{array}
\right)
\nonu \\
& - & \frac{16 i (k+3) (k+9)}{9 (k+5)^3}  \pa A_{\pm}
\left(
\begin{array}{c}
F_{21} \\
F_{12} 
\end{array}
\right) - \frac{2 i \left(8 k^2-9 k-189\right)}{9 (k+5)^3}  A_{\pm}
\pa
\left(
\begin{array}{c}
F_{21} \\
F_{12} 
\end{array}
\right) 
\nonu \\
& - & \frac{2i(k+21)}{3 (k+5)^2} B_3 
\left(
\begin{array}{c}
G_{11} \\
G_{22} 
\end{array}
\right) -
 \frac{2 i \left(32 k^2+225 k+117\right)}{9 (k+5)^3}
B_3 \pa
\left(
\begin{array}{c}
F_{11} \\
F_{22} 
\end{array}
\right)
\nonu \\
& - &  \frac{i (3 k+17)}{(k+5)^2} B_{\mp}
\left(
\begin{array}{c}
G_{12} \\
G_{21} 
\end{array}
\right)
\mp
\frac{2 (k+9)}{3 (k+5)^3} 
B_{\mp} B_3
\left(
\begin{array}{c}
F_{12} \\
F_{21} 
\end{array}
\right)
\nonu \\
&+& \frac{4 i \left(5 k^2+51 k+18\right)}{9 (k+5)^3} \pa B_3 
\left(
\begin{array}{c}
F_{11} \\
F_{22} 
\end{array}
\right) 
\pm  
\frac{8(k+3)}{3(k+5)^3} B_{3} B_3 
\left(
\begin{array}{c}
F_{11} \\
F_{22} 
\end{array}
\right)
\nonu \\
& +  & \frac{2 i \left(19 k^2+93 k-54\right)}{9 (k+5)^3}
\pa B_{\mp}
\left(
\begin{array}{c}
F_{12} \\
F_{21} 
\end{array}
\right) 
\nonu \\
& - & \frac{4 i \left(13 k^2+90 k+99\right)}{9 (k+5)^3}
 B_{\mp} \pa 
\left(
\begin{array}{c}
F_{12} \\
F_{21} 
\end{array}
\right)
\nonu \\
& \pm & \frac{2(5k+21)}{3(k+5)^3} B_{\pm} B_{\mp} 
\left(
\begin{array}{c}
F_{11} \\
F_{22} 
\end{array}
\right) \pm \frac{2 \left(4 k^2+27 k+27\right)}{3 (k+5)^3}
 \pa^2
\left(
\begin{array}{c}
F_{11} \\
F_{22} 
\end{array}
\right)
\nonu \\
&\pm & \frac{4}{(k+5)^2} 
\left(
\begin{array}{c}
F_{11} F_{12} G_{21} \\
F_{22} F_{21} G_{12}
\end{array}
\right) -
\frac{4 (k+9)^2}{ 9(k+5)^4} 
\pa
\left(
\begin{array}{c}
F_{11}  \\
F_{22} 
\end{array}
\right) F_{12} F_{21}
\nonu \\
&\mp &  \frac{4 \left(31 k^2+294 k+711\right)}{9 (k+5)^4}
\left(
\begin{array}{c}
F_{11}  \pa F_{12} F_{21} \\
F_{22}  \pa F_{21} F_{12}
\end{array}
\right)
\nonu \\
&\pm & \frac{8 \left(13 k^2+114 k+261\right)}{9 (k+5)^4}
\left(
\begin{array}{c}
F_{11}  F_{12} \pa F_{21} \\
F_{22}  F_{21} \pa F_{12}
\end{array}
\right)
\nonu \\
&\mp & \frac{2(k+9)}{3(k+5)^3} 
\left(
\begin{array}{c}
F_{11}  F_{21} G_{12} \\
F_{22}  F_{12} G_{21}
\end{array}
\right)
\pm \frac{4 (k+3)}{3 (k+5)^3} 
\left(
\begin{array}{c}
F_{11}  F_{22} G_{11} \\
F_{22}  F_{11} G_{22}
\end{array}
\right)
\nonu \\
& \pm &  \frac{8(k+6)}{3(k+5)^3} 
\left(
\begin{array}{c}
F_{12}  F_{21} G_{11} \\
F_{21}  F_{12} G_{22}
\end{array}
\right)
\mp
\frac{4 (k+3) (2 k+15)}{9 (k+5)^2}
\pa
\left(
\begin{array}{c}
G_{11} \\
G_{22}
\end{array}
\right)
\nonu \\
& \mp &
\frac{4(2k+15)}{3(k+5)^2} T 
\left(
\begin{array}{c}
F_{11}   \\
F_{22} 
\end{array}
\right) \mp
\frac{2(2k+9)}{3 (k+5)^2} U
\left(
\begin{array}{c}
G_{11}   \\
G_{22} 
\end{array}
\right)
\nonu \\
&- & \frac{16 i(k+6)}{3(k+5)^3} U A_3 
\left(
\begin{array}{c}
F_{11}   \\
F_{22} 
\end{array}
\right)
- \frac{16i(k+6)}{3(k+5)^3} U A_{\pm}
\left(
\begin{array}{c}
F_{21}   \\
F_{12} 
\end{array}
\right)
\nonu \\
&- & \frac{8 i}{(k+5)^2} U B_3
\left(
\begin{array}{c}
F_{11}   \\
F_{22} 
\end{array}
\right)
- \frac{2i(13k+69)}{3(k+5)^3} U B_{\mp} 
\left(
\begin{array}{c}
F_{12}   \\
F_{21} 
\end{array}
\right)
\nonu \\
& \mp & \frac{8 (k+3) (2 k+15)}{9 (k+5)^3}
\pa U  
\left(
\begin{array}{c}
 F_{11}   \\
 F_{22} 
\end{array}
\right)
\nonu \\
 &\mp &
\frac{2 \left(8 k^2+39 k+99\right)}{9 (k+5)^3} U
\pa \left(
\begin{array}{c}
 F_{11}   \\
 F_{22} 
\end{array}
\right)
\mp  \frac{16(k+6)}{3(k+5)^3} U U
\left(
\begin{array}{c}
F_{11}   \\
F_{22} 
\end{array}
\right)  
\nonu \\
& + & \left.
\frac{4 \left(9 k^2+82 k+201\right)}{3 (k+5)^4}
\left(
\begin{array}{c}
\pa F_{11}  \\
\pa F_{22} 
\end{array}
\right) F_{11} F_{22} \right](w) 
\nonu \\
& + & \frac{1}{(z-w)} \, \left[ + \cdots \right](w) + \cdots.
\nonu
\eea
There is no higher spin current in the third order pole.

One also determines the following OPEs 
from the information on the higher spin-$3$ current in (\ref{g1221w5half})
%%%%%%%%%%%%%%%%%%%%%%%%%%%%%%%%%%%%%%%%%%%%%%%%%%%%%%%%%%%%%%%%%%%%%%%%%%%%%%
\bea
\left(
\begin{array}{c}
{\bf U_{+}^{(2)}} \\
{\bf V_{-}^{(2)}} 
\end{array}
\right)(z) \, {\bf W^{(3)}}(w) & = &
\frac{1}{(z-w)^4} 
\, \left[ \pm  \frac{12 i k \left(8 k^2+75 k+95\right)}{(k+5)^2 (13 k+17)}
 {B}_{\mp} \right. \nonu \\
& - & \left.
\frac{96 k \left(2 k^2+9 k+11\right)}{(k+5)^3 (13 k+17)}
\left(
\begin{array}{c}
F_{11} F_{21}  \\
F_{22} F_{12}
\end{array}
\right) 
\right] (w) 
\nonu \\
& + & \frac{1}{(z-w)^3} \, \left[  
\pm \frac{8 i \left(25 k^2+158 k-27\right)}{3 (k+5)^3 (13 k+17)}  B_3 
\left(
\begin{array}{c}
F_{11} F_{21}  \\
F_{22} F_{12}
\end{array}
\right)
\right. \nonu \\
&- & \frac{4 i \left(25 k^2+158 k-27\right)}{3 (k+5)^3 (13 k+17)}  
B_{\mp} F_{11} F_{22}
+ \frac{4 i \left(25 k^2+158 k-27\right)}{3 (k+5)^3 (13 k+17)}  
B_{\mp} F_{12} F_{21}
\nonu \\
& + & \frac{2 \left(103 k^2+260 k-27\right)}{3 (k+5)^2 (13 k+17)}
 \left(
\begin{array}{c}
F_{11} G_{21}  \\
F_{22} G_{12}
\end{array}
\right) 
+ \frac{8 \left(25 k^2+158 k-27\right)}{3 (k+5)^3 (13 k+17)}
 \left(
\begin{array}{c}
\pa F_{11} F_{21}  \\
\pa F_{22} F_{12}
\end{array}
\right) 
\nonu \\
&+ & 
\frac{8 \left(25 k^2+158 k-27\right)}{3 (k+5)^3 (13 k+17)}
 \left(
\begin{array}{c}
F_{11} \pa F_{21}  \\
F_{22} \pa F_{12}
\end{array}
\right) -
\frac{2 \left(103 k^2+260 k-27\right)}{3 (k+5)^2 (13 k+17)}
 \left(
\begin{array}{c}
F_{21} G_{11}  \\
F_{12} G_{22}
\end{array}
\right)
\nonu \\
& \pm & \left.
  \frac{8 i k}{(k+5)^2} 
U B_{\mp}
+ \frac{8 \left(25 k^2+158 k-27\right)}{3 (k+5)^3 (13 k+17)} 
U
\left(
\begin{array}{c}
F_{11} F_{21}  \\
F_{22} F_{12}
\end{array}
\right)
\right](w) 
\nonu \\
&+ & \frac{1}{(z-w)^2} \, \left[ 
\pm 3
\left(
\begin{array}{c}
{\bf Q_{+}^{(3)}} \\
 {\bf R_{-}^{(3)}} 
\end{array}
\right)
-\frac{8 {\bf (k-3)}}{(13 k+17)} {\bf T^{(1)}} 
\left(
\begin{array}{c}
{\bf U_{+}^{(2)}} \\
{\bf V_{-}^{(2)}} 
\end{array}
\right)
\right. \nonu \\
&
\mp & \frac{6i}{(k+5)^2} A_3 A_3 B_{\mp}
- \frac{3(3 k+11)}{ (k+5)^2}  \pa A_3 B_{\mp}
 \nonu \\
&+ & \frac{3(3 k+13)}{ (k+5)^2}  A_3 \pa B_{\mp}
\pm \frac{8 i}{(k+5)^2}  A_3
\left(
\begin{array}{c}
F_{11} G_{21} \\
F_{22} G_{12} 
\end{array}
\right)
\nonu \\
&\mp& \frac{3i (5 k+17)}{ (k+5)^3}  \pa A_3 
\left(
\begin{array}{c}
F_{11} F_{21} \\
F_{22} F_{12} 
\end{array}
\right)
\pm \frac{i (5 k+17)}{ (k+5)^3} A_3 
 \left(
\begin{array}{c}
 \pa F_{11} F_{21} \\
 \pa F_{22} F_{12} 
\end{array}
\right)
\nonu \\
& \pm & \frac{5i (5 k+17)}{ (k+5)^3}
A_3 \left(
\begin{array}{c}
 F_{11} \pa F_{21} \\
 F_{22} \pa F_{12} 
\end{array}
\right)
\mp \frac{2i}{(k+5)^2} A_{\mp} 
 \left(
\begin{array}{c}
 F_{11} G_{11} \\
 F_{22} G_{22} 
\end{array}
\right)
\nonu \\
&\mp& 
\frac{4i}{(k+5)^2} A_3 
\left(
\begin{array}{c}
F_{21} G_{11} \\
F_{12} G_{22} 
\end{array}
\right)
\mp \frac{6i}{(k+5)^2} A_{\pm} A_{\mp} B_{\mp}
\nonu \\
&\pm &  \frac{2i}{(k+5)^2} A_{\pm}
\left(
\begin{array}{c}
F_{21} G_{21} \\
F_{12} G_{12} 
\end{array}
\right)
\pm \frac{2i}{(k+5)^2} B_3 
\left(
\begin{array}{c}
F_{11} G_{21} \\
F_{22} G_{12} 
\end{array}
\right)
\nonu \\
& \pm & \frac{i \left(269 k^2+1906 k+1269\right)}{3 (k+5)^3 (13 k+17)}   
\pa B_3 
\left(
\begin{array}{c}
F_{11} F_{21} \\
F_{22} F_{12} 
\end{array}
\right)
\nonu \\
& \pm & \frac{i \left(113 k^2+298 k-567\right)}{3 (k+5)^3 (13 k+17)}   B_3
\left(
\begin{array}{c}
\pa F_{11} F_{21} \\
\pa F_{22} F_{12} 
\end{array}
\right)
\nonu \\
& \mp & \frac{i (5 k+9) (71 k+403)}{3 (k+5)^3 (13 k+17)}  B_3
\left(
\begin{array}{c}
F_{11} \pa F_{21} \\
F_{22} \pa F_{12} 
\end{array}
\right)
\nonu \\
& \mp & \frac{2i}{(k+5)^2} B_3 
\left(
\begin{array}{c}
F_{21} G_{11} \\
F_{12} G_{22} 
\end{array}
\right)
\mp \frac{i}{(k+5)^2} B_{\mp} 
\left(
\begin{array}{c}
F_{12} G_{21} \\
F_{21} G_{12} 
\end{array}
\right)
\nonu \\
&-  & \frac{6(k+1)}{(k+5)^2} \pa B_{\mp} B_3
+ \frac{6(k+2)}{(k+5)^2} B_{\mp} \pa B_3
\nonu \\
&\mp&  \frac{i \left(8 k^3+63 k^2+177 k-170\right)}{(k+5)^2 (13 k+17)}
\pa^2 B_{\mp}
\mp \frac{3 i}{(k+5)^2} B_{\mp}
\left(
\begin{array}{c}
F_{11} G_{22} \\
 F_{22} G_{11} 
\end{array}
\right)
\nonu \\
& \pm & \frac{4 i \left(59 k^2+619 k+804\right)}{3 (k+5)^3 (13 k+17)}  
B_{\mp} 
\left(
\begin{array}{c}
\pa F_{11} F_{22} \\
\pa F_{22} F_{11} 
\end{array}
\right)
\nonu \\
& - & \frac{2 i \left(155 k^2+1264 k+1197\right)}{3 (k+5)^3 (13 k+17)}  
\pa B_{\mp} F_{11} F_{22}
\nonu \\
& \pm &  \frac{4 i \left(59 k^2+463 k+600\right)}{3 (k+5)^3 (13 k+17)}  
B_{\mp}
\left(
\begin{array}{c}
F_{11} \pa F_{22} \\
F_{22} \pa F_{11} 
\end{array}
\right)
\nonu \\
& \mp & \frac{i \left(119 k^2+802 k+1227\right)}{3 (k+5)^3 (13 k+17)}  
B_{\mp}
\left(
\begin{array}{c}
F_{12} \pa F_{21} \\
F_{21} \pa F_{12} 
\end{array}
\right)
\nonu \\
& \pm & \frac{i \left(349 k^2+2774 k+2649\right)}{3 (k+5)^3 (13 k+17)}  
B_{\mp} 
\left(
\begin{array}{c}
\pa F_{12}  F_{21} \\
\pa F_{21}  F_{12} 
\end{array}
\right)
\nonu \\
& - & \frac{i(41k^2+622k+1125) }{ 3(k+5)^3(13k+17)}  
\pa B_{\mp} 
F_{12} F_{21}
\mp \frac{2i}{(k+5)^2} B_{\mp} B_3 B_3
\nonu \\
& \pm & \frac{3i}{(k+5)^2} B_{\mp}
\left(
\begin{array}{c}
F_{21}  G_{12} \\
F_{12}  G_{21} 
\end{array}
\right)
\pm \frac{5 i}{(k+5)^2} B_{\mp} 
\left(
\begin{array}{c}
F_{22}  G_{11} \\
F_{11}  G_{22} 
\end{array}
\right)
\nonu \\
&+ &  \frac{2(41k+61)}{ (k+5)^3}
\left(
\begin{array}{c}
\pa^2 F_{11}  F_{21} \\
\pa^2 F_{22}  F_{12} 
\end{array}
\right)
\mp \frac{2i}{(k+5)^2} B_{\pm} B_{\mp} B_{\mp}
\nonu \\
&+ & \frac{(48 k^3+1811 k^2+5374 k+2187)}{3 (k+5)^3 (13 k+17)}
\left(
\begin{array}{c}
F_{11}  \pa^2 F_{21} \\
F_{22}  \pa^2 F_{12} 
\end{array}
\right)
\nonu \\
&+ & \frac{(583 k^2+38 k-1329)}{6 (k+5)^2 (13 k+17)}
\left(
\begin{array}{c}
\pa F_{11} G_{21} \\
\pa F_{22}  G_{12} 
\end{array}
\right)
-   \frac{(41 k^2-470 k-303)}{6 (k+5)^2 (13 k+17)}
\left(
\begin{array}{c}
F_{11} \pa G_{21} \\
F_{22}  \pa G_{12} 
\end{array}
\right)
\nonu \\
&- & \frac{(116 k^2-737 k-1353)}{3 (k+5)^2 (13 k+17)}
 \left(
\begin{array}{c}
\pa F_{21}  G_{11} \\
\pa F_{12}   G_{22} 
\end{array}
\right)
-
\frac{(38 k^2+487 k+381)}{3 (k+5)^2 (13 k+17)}
 \left(
\begin{array}{c}
F_{21}  \pa G_{11} \\
F_{12}   \pa G_{22} 
\end{array}
\right)
\nonu \\
&+ & \frac{3}{ (k+5)} 
\left(
\begin{array}{c}
G_{11}  G_{21} \\
G_{22}  G_{12} 
\end{array}
\right)
\mp   \frac{16 i (k+4)}{(k+5)^3}
\left(
\begin{array}{c}
F_{11}  A_{-} \pa F_{11} \\
F_{22}  A_{+} \pa F_{22} 
\end{array}
\right)
\nonu \\
&\pm & \frac{2 i \left(8 k^2+62 k-17\right)}{(k+5) (13 k+17)}
T B_{\mp} 
-\frac{4 \left(8 k^2+61 k+17\right)}{(k+5)^2 (13 k+17)} 
T
\left(
\begin{array}{c}
F_{11}  F_{21} \\
F_{22}  F_{12} 
\end{array}
\right)
\nonu \\
&- & \frac{12}{(k+5)^2} U A_3 B_{\mp}
\mp \frac{i (k+6)}{ (k+5)^2} \pa U B_{\mp}
\pm \frac{i (5k+6)}{ (k+5)^2} U \pa B_{\mp}
\nonu \\
&+ &
\frac{(269 k^2+658 k-363)}{3 (k+5)^3 (13 k+17)}
U
\left(
\begin{array}{c}
\pa F_{11}  F_{21} \\
\pa F_{22}  F_{12} 
\end{array}
\right)
+ \frac{4}{(k+5)^2} U
 \left(
\begin{array}{c}
F_{11}  G_{21} \\
 F_{22}  G_{12} 
\end{array}
\right)
\nonu \\
&+ & \frac{(113 k^2+1546 k+1065)}{3 (k+5)^3 (13 k+17)}  \pa U
\left(
\begin{array}{c}
F_{11}  F_{21} \\
F_{22}  F_{12} 
\end{array}
\right)
\nonu \\
&- & \frac{(199 k^2+2294 k+3423)}{3 (k+5)^3 (13 k+17)}  U
\left(
\begin{array}{c}
F_{11}  \pa F_{21} \\
F_{22}  \pa F_{12} 
\end{array}
\right)
\nonu \\
&- & \frac{4}{(k+5)^2}
U
\left(
\begin{array}{c}
F_{21}  G_{11} \\
F_{12}  G_{22} 
\end{array}
\right)
\mp \frac{2i}{(k+5)^2} U U B_{\mp}
\nonu \\
& \mp& 
\frac{2 i \left(48 k^3-1465 k^2-3746 k-4137\right)}{3 (k+5)^3 (13 k+17)}
\left(
\begin{array}{c}
F_{21}  A_{+}  \pa F_{21} \\
F_{12}  A_{-} \pa F_{12} 
\end{array}
\right)
\nonu \\
& \mp & \left.
\frac{2 i \left(48 k^3-1387 k^2-3566 k-4035\right)}{3 (k+5)^3 (13 k+17)}
\left(
\begin{array}{c}
\pa F_{21}  A_{+}  F_{21} \\
\pa F_{12}  A_{-} F_{12} 
\end{array}
\right)
\nonu 
\right](w) \nonu \\
& + & \frac{1}{(z-w)} \, \left[ + \cdots \right](w) + \cdots.
\nonu
\eea
There is no higher spin current in the third order pole.
In this case, the nonlinear term between the higher spin currents 
is present. This can be removed by adding the $T \, {\bf T^{(1)}}(w)$
into the left hand side for spin-$3$ current and then this will produce 
the above nonlinear term.

Again the previous result in (\ref{g1221uv3half}) gives the following 
OPEs
%%%%%%%%%%%%%%%%%%%%%%%%%%%%%%%%%%%%%%%%%%%%%%%%%%%%%%%%%%%%
\bea
\left(
\begin{array}{c}
{\bf U_{-}^{(2)}} \\
{\bf V_{+}^{(2)}}
\end{array}
\right)
(z) \, 
\left(
\begin{array}{c}
{\bf U_{-}^{(2)}} \\
{\bf V_{+}^{(2)}}
\end{array}
\right)
(w) & = & \frac{1}{(z-w)^2} \,
\left[ \frac{6}{(5+k)^2}  {A}_{\pm} {A}_{\pm} \right](w)
\nonu \\
& + & \frac{1}{(z-w)} \, \left[  \frac{6}{(5+k)^2}  {A}_{\pm} \pa A_{\pm}
\right](w)
+\cdots.
\nonu 
\eea
These are new nontrivial OPEs. The corresponding OPEs are trivial in the 
nonlinear version. 

From the result in (\ref{g1221vu2}), one obtains the following OPEs
%%%%%%%%%%%%%%%%%%%%%%%%%%%%%%%%%%%%%%%%%%%%%%%%%%%%%%%%%%%%%%%%%%%
\bea
\left(
\begin{array}{c}
{\bf U_{-}^{(2)}} \\
{\bf V_{+}^{(2)}} 
\end{array}
\right) (z) \, 
\left(
\begin{array}{c}
{\bf U^{(\frac{5}{2})}} \\
{\bf V^{(\frac{5}{2})}}
\end{array}
\right)
(w) & = & 
\frac{1}{(z-w)^3} \, 
\left[ \frac{8 i (k+9)}{ (k+5)^3} A_{\pm}
\left(
\begin{array}{c}
F_{11} \\
F_{22} 
\end{array}
\right)
 \right](w) \nonu \\
&+& \frac{1}{(z-w)^2} \,
\left[  \frac{2 i (k+12)}{3 (k+5)^2} A_{\pm}
 \left(
\begin{array}{c}
G_{11} \\
G_{22} 
\end{array}
\right)
\mp \frac{4 (k+3)}{3 (k+5)^3} 
 A_{\pm} A_3
 \left(
\begin{array}{c}
F_{11} \\
F_{22} 
\end{array}
\right)
\right. \nonu \\
& \mp & \frac{4 (k+3)}{3 (k+5)^3}  A_{\pm} A_{\pm}
 \left(
\begin{array}{c}
F_{21} \\
F_{12} 
\end{array}
\right) \pm
 \frac{4 (k+3)}{3 (k+5)^3}
A_{\pm} B_3 
 \left(
\begin{array}{c}
F_{11} \\
F_{22} 
\end{array}
\right)
\nonu \\
& \pm & \frac{4 (k+3)}{3 (k+5)^3}   A_{\pm} B_{\mp}
 \left(
\begin{array}{c}
F_{12} \\
F_{21} 
\end{array}
\right)
\pm \frac{4 (k+3)}{3 (k+5)^3} 
 \left(
\begin{array}{c}
F_{11} F_{12} G_{11} \\
F_{22} F_{21} G_{22} 
\end{array}
\right)
\nonu \\
&+ & \frac{4 i (k+3)}{3 (k+5)^3}  U A_{\pm} 
 \left(
\begin{array}{c}
F_{11} \\
F_{22} 
\end{array}
\right)
+\frac{4 i (k+3)}{(k+5)^3}
A_{\pm} \pa
 \left(
\begin{array}{c}
F_{11}  \\
F_{22} 
\end{array}
\right)
\nonu \\
&\mp & \left.  \frac{32 (k+3)}{3 (k+5)^4}
\left(
\begin{array}{c}
F_{11} F_{12} \pa F_{11} \\
F_{22}  F_{21} \pa F_{22}
\end{array}
\right)
\mp \frac{24}{(k+5)^3}
 \left(
\begin{array}{c}
F_{11} G_{12} F_{11} \\
F_{22} G_{21} F_{22}
\end{array}
\right)
 \right](w)
\nonu \\
&+ & \frac{1}{(z-w)} \, \left[
+ \cdots \right](w) + \cdots.
\nonu
\eea
The nonlinear terms in the corresponding OPEs in the nonlinear version
disappear. The first order pole is not determined in this paper. 

Similarly one has the following OPE
%%%%%%%%%%%%%%%%%%%%%%%%%%%%%%%%%%%%%%%%%%%%%%%%%%%%%%%%%%%%
\bea
{\bf U_{-}^{(2)}}(z) \, {\bf V^{(\frac{3}{2})}}(w) & = & 
\frac{1}{(z-w)^3} \, \left[ \frac{6 k}{(k+5)^2}  F_{12} \right](w)
\nonu \\
&+&
\frac{1}{(z-w)^2} \, \left[
-\frac{2(k+3)}{(k+5)}  {\bf T_{-}^{(\frac{3}{2})}}  
-\frac{2(k+2)}{(k+5)^2} U F_{12}
-\frac{2}{(k+5)^2} F_{12} F_{11} F_{22}
\right. \nonu \\
&-& \frac{2i (k-1)}{(k+5)^2} F_{12} A_3
+ \frac{2i (k+2)}{(k+5)^2} F_{22} A_{+}
- \frac{4i}{(k+5)^2} F_{11} B_{+}
\nonu \\
&+ & \left. \frac{4i}{(k+5)^2} F_{12} B_3 
+\frac{1}{(k+5)} G_{12} + \frac{6 k}{(k+5)^2} \pa F_{12}
\right](w) 
\nonu \\
& + & \frac{1}{(z-w)} \, \left[
\frac{1}{2} {\bf P_{-}^{(\frac{5}{2})} } 
-  {\bf W_{-}^{(\frac{5}{2})} } 
- \frac{4(k+3)}{3 (k+5)} \pa {\bf T_{-}^{(\frac{3}{2})}}  
- \frac{i}{2 (k+5)} A_3 G_{12}
\right. \nonu \\
&+ &  \frac{2}{(k+5)^2} A_3 A_3 F_{12}
-\frac{1}{(k+5)^2} A_{+} A_3 F_{22}
\nonu \\
&+& \frac{2}{(k+5)^2} A_3 B_3 F_{12}
-\frac{2}{(k+5)^2} A_3 B_{+} F_{11}
-\frac{i (8k+3)}{3 (k+5)^2} \pa A_3 F_{12}
\nonu \\
&-& \frac{4i (k-6)}{3 (k+5)^2} A_3 \pa F_{12}
+\frac{1}{(k+5)^2} A_{+} A_{-} F_{12}
-\frac{1}{(k+5)^2} A_{+} B_3 F_{22}
\nonu \\
&-& \frac{1}{(k+5)^2} A_{+} B_{+} F_{21}
+ \frac{i (4k+9)}{3 (k+5)^2} \pa A_{+} F_{22}
+ \frac{i (4k+3)}{3 (k+5)^2} A_{+} \pa F_{22}
\nonu \\
&-& \frac{i}{2 (k+5)} B_3 G_{12}
+\frac{ i (5k+18)}{3 (k+5)^2} \pa B_3 F_{12}
- \frac{2i (k-6)}{3 (k+5)^2} B_3 \pa F_{12}
\nonu \\
& + & \frac{i}{2(k+5)} B_{+} G_{11}
-  \frac{i (k+9)}{3 (k+5)^2} \pa B_{+} F_{11}
+ \frac{2i (k-6)}{3 (k+5)^2} B_{+} \pa F_{11}
\nonu \\
& - & \frac{2 (k-3)}{3 (k+5)^3} \pa F_{11} F_{12} F_{22}
+ \frac{1}{(k+5)^2} F_{11} F_{12} G_{22}
\nonu \\
&+& \frac{4 (k+3)}{3 (k+5)^3} F_{11} \pa F_{12} F_{22}
+\frac{2 (5k+21)}{3 (k+5)^3} F_{11} F_{12} \pa F_{22}
-\frac{1}{(k+5)^2} F_{11} F_{22} G_{12}
\nonu \\
&+&\frac{3(3k-5)}{2 (k+5)^2} \pa^2 F_{12}
- 
 \frac{1}{(k+5)^2} F_{12} F_{21} G_{12}
+ \frac{1}{(k+5)^2} \pa U F_{12}
\nonu \\
&+& \frac{1}{(k+5)} T F_{12}
+\frac{1}{2 (k+5)} U G_{12}
-\frac{2 i}{(k+5)^2} U B_3 F_{12}
-\frac{2 (2 k+3)}{3 (k+5)^2} U \pa F_{12}
\nonu \\
&-& \frac{ i}{(k+5)^2} U A_{+} F_{22}
+  \frac{2i}{(k+5)^2} U B_{+} F_{11}
+ \frac{2}{(k+5)^2} U U F_{12}
\nonu \\
&+ & \left. \frac{4 i(k+1)}{(k+5)^3} F_{12} B_3 F_{12} F_{21}
-\frac{(4 k+9)}{3 (k+5)^2} F_{12} G_{21} F_{12}
+ \frac{1}{(k+5)} \pa G_{12}
\right](w) +\cdots.
\nonu 
\eea
The spin-$\frac{3}{2}$ current in the second order pole 
was not present in the corresponding OPE in the nonlinear version.

%%%%%%%%%%%%%%%%%%%%%%%%%%%%%%%%%%%%%%%%

The other combination of the OPE can be obtained
\bea
%%%%%%%%%%%%%%%%%%%%%%%%%%%%%%%%%%%%%%%%%%%%%%%%%%%%%%%%%%%%%%%
{\bf U_{-}^{(2)}}(z) \, {\bf V_{+}^{(2)}}(w) 
&=& 
\frac{1}{(z-w)^4} \, \left[ -\frac{12 k (k+3)}{(k+5)^2}\right]
\nonu \\
&+& 
\frac{1}{(z-w)^3} \, \left[  
-\frac{12 i k (2k+5)}{(k+5)^2}  {A}_3   
- \frac{12}{(k+5)^2} F_{11} F_{22}
- \frac{12}{(k+5)^2}  F_{12} F_{21}
\right](w)
\nonu \\
& + & \frac{1}{(z-w)^2} \, \left[ -\frac{1}{2} {\bf P^{(2)}} +
\frac{2 (k-3)}{3 (k+5)}  {\bf T^{(2)}} 
+2  {\bf W^{(2)}} 
-\frac{2 (4 k+9)}{3 (k+5)} T
\right. \nonu \\
& - & \frac{8k}{3 (k+5)^2} A_3 A_3  -
\frac{2 (k-3)}{3 (k+5)^2}  A_3 B_3
-\frac{4 i (11k+27)}{3 (k+5)^2} \pa A_3
\nonu \\
& + & \frac{8 i (k+3)}{3 (k+5)^3}
 A_3 F_{11} F_{22}
+  \frac{2 i}{(k+5)^2} A_3 F_{12} F_{21}
-\frac{2 (4 k+9)}{3 (k+5)^2} A_{+} A_{-}
\nonu \\
&+& \frac{2 i (k-15)}{3 (k+5)^3}  A_{-} F_{11} F_{12}
+ \frac{4 i(k+7)}{(k+5)^3} A_{+} F_{21} F_{22}
- \frac{2 (k+9)}{3 (k+5)^2} B_3 B_3
\nonu \\
& - & \frac{2 i (k+9)}{3 (k+5)^2} \pa B_3
  -\frac{8 i (k+3)}{3 (k+5)^3}   B_3 F_{11} F_{22}
+  \frac{2 i}{(k+5)^2} B_3 F_{12} F_{21}
\nonu \\
& - & \frac{20 i (k+3)}{3 (k+5)^3}  B_{-} F_{12} F_{22}
-\frac{2 (k+9)}{3 (k+5)^2} B_{+} B_{-}
+  \frac{2 i (k+1)}{(k+5)^3} B_{+} F_{11} F_{21}
\nonu \\
&+& \frac{(k-3)}{3 (k+5)^2} F_{11} G_{22}
-\frac{4 \left(2 k^2+21 k+75\right)}{3 (k+5)^3}  \pa F_{11} F_{22}
\nonu \\
& + & \frac{4 \left(2k^2+3 k-15\right)}{3 (k+5)^3}  F_{11} \pa F_{22}
+ \frac{2(k+3)}{(k+5)^2} F_{12} G_{21}
-\frac{2 \left(4k^2+27 k+123\right)}{3 (k+5)^3} \pa F_{12} F_{21}
\nonu \\
& + & \frac{2 \left(4 k^2+17 k+9\right)}{3 (k+5)^3} F_{12} \pa F_{21}
- \frac{6}{(k+5)^2}  F_{21} G_{12}
+ \frac{2(k-3)}{3 (k+5)^2}  F_{22} G_{11}
\nonu \\
& + & \frac{6 i}{(k+5)^2} U A_3 
+\frac{2 i k}{(k+5)^2} U B_3 
-\frac{2 (4 k+9)}{3 (k+5)^2} U U
\nonu \\
&- & \left. \frac{8 (k-3)}{3 (k+5)^3}  U F_{11} F_{22}
+\frac{6}{(k+5)^2}  U F_{12} F_{21}
\right](w)
\nonu \\
& + & \frac{1}{(z-w)} \, \left[ -\frac{1}{2} {\bf S^{(3)}} 
+\frac{1}{2} {\bf P^{(3)}}
-\frac{1}{4} 
  \pa 
{\bf P^{(2)}}  + \pa {\bf W^{(2)}}  
+\frac{(k-3)}{3 (k+5)} \pa {\bf T^{(2)}} 
\right. \nonu \\
& - &  \frac{2 i}{(k+5)^2} A_3 A_3 A_3
-\frac{2(4k-3)}{3 (k+5)^2} \pa A_3 A_3
+ \frac{2 i}{(k+5)^2} A_3 B_3 B_3
\nonu \\
& - & \frac{(k-3)}{3 (k+5)^2} \pa A_3 B_3
- \frac{ (k+3)}{3 (k+5)^2} A_3 \pa B_3
- \frac{i \left(124 k^2+263 k-249\right)}{3 (k+5)^2 (13 k+17)} \pa^2 A_3
\nonu \\
&+& \frac{3i}{(k+5)^2} A_3 F_{11} G_{22}
+\frac{i (23 k+123)}{6 (k+5)^3} \pa A_3 F_{11} F_{22}
+ \frac{2 i}{(k+5)^2} A_3
 B_{+} B_{-} \nonu \\
&-& \frac{i (25 k+69)}{6 (k+5)^3} A_3 \pa F_{11} F_{22} 
+ \frac{i (11 k-81)}{6 (k+5)^3} A_3 F_{11} \pa F_{22}
+\frac{ 3i}{(k+5)^2} A_3 F_{12} G_{21}
\nonu \\
&+& \frac{i (7 k+43)}{2 (k+5)^3} \pa A_3 F_{12} F_{21}
-\frac{3i (3k+7)}{2 (k+5)^3} A_3 \pa F_{12} F_{21}
+\frac{i (3 k-25)}{2 (k+5)^3} A_3 F_{12} \pa F_{21}
\nonu \\
&+& \frac{ i}{(k+5)^2} A_3 F_{21} G_{12}
+\frac{i}{(k+5)^2} A_{3} F_{22} G_{11}
+ \frac{i}{(k+5)^2} A_{-} F_{11} G_{12}
\nonu \\
& + & \frac{2i (5k+21)}{3 (k+5)^3} \pa A_{-} F_{11} F_{12}
+ \frac{i}{(k+5)^2} A_{+} F_{21} G_{22}
\nonu \\
& - & \frac{8i (k+9)}{3 (k+5)^3} A_{-} \pa F_{11} F_{12}
- \frac{8 i (k+9)}{3 (k+5)^3} A_{-} F_{11} \pa F_{12}
-\frac{i}{(k+5)^2} A_{-} F_{12} G_{11}
\nonu \\
& - & \frac{2i}{(k+5)^2} A_{+} A_{-} A_3
- \frac{4(k+3)}{3(k+5)^2} \pa A_{+} A_{-}
-\frac{2 (2 k+3)}{3 (k+5)^2} A_{+} \pa A_{-}
\nonu \\
& + & \frac{4 i(k+7) }{ (k+5)^3} \pa A_{+} F_{21} F_{22}
- \frac{ i }{ (k+5)^2} A_{+} F_{22} G_{21}
\nonu \\
&-& \frac{i (k+9)}{3 (k+5)^2}  \pa^2 B_3 
- 
\frac{i}{(k+5)^2} B_3 F_{11} G_{22}
+ \frac{i (13 k+57)}{6 (k+5)^3}  \pa B_3 F_{11} F_{22}
\nonu \\
&-& \frac{i (23k+75)}{6 (k+5)^3} B_3 \pa F_{11} F_{22}
- \frac{5 i (7k+27)}{6 (k+5)^3}  B_3  F_{11} \pa F_{22}
\nonu \\
&+& \frac{ i}{(k+5)^2} B_3 F_{12} G_{21}
- \frac{i (5k+17)}{2 (k+5)^3} \pa B_3 F_{12} F_{21}
+ \frac{i (7k+27)}{2 (k+5)^3} B_3 \pa F_{12} F_{21}
\nonu \\
&+& \frac{i (11 k+47)}{2 (k+5)^3} B_3 F_{12} \pa F_{21}
-\frac{i}{(k+5)^2} B_{-} F_{12} G_{22}
\nonu \\
&-& \frac{ i}{(k+5)^2} B_3 F_{21} G_{12}
+ \frac{i}{(k+5)^2} B_{3} F_{22} G_{11}
+\frac{i (k+21)}{6 (k+5)^3} \pa B_{-} F_{12} F_{22}
\nonu \\
& - & \frac{i (35 k+111)}{6 (k+5)^3} B_{-} \pa F_{12} F_{22}
- \frac{i ( 47k+171)}{6 (k+5)^3} B_{-} F_{12} \pa F_{22}
+\frac{i}{(k+5)^2} B_{-} F_{22} G_{12}
\nonu \\
&-& \frac{(k+9)}{3(k+5)^2} \pa B_{+} B_{-}
-\frac{(k+9)}{3 (k+5)^2} B_{+} \pa B_{-}
+\frac{i (9 k+29)}{2 (k+5)^3} \pa B_{+} F_{11} F_{21}
\nonu \\
& - & \frac{3i}{2 (k+5)^2} B_{+} \pa F_{11} F_{21}
- \frac{7i}{2 (k+5)^2} B_{+} F_{11} \pa F_{21}
- \frac{i}{(k+5)^2} B_{+} F_{11} G_{21}
\nonu \\
& - & \frac{(104 k^3+841 k^2+3234 k+2001)}{6 (k+5)^3 (13 k+17)}
  \pa^2 F_{11} F_{22}
\nonu \\
&+& \frac{(104 k^3+679 k^2+1518 k+2079)}{6 (k+5)^3 (13 k+17)}  
F_{11} \pa^2 F_{22}
+ \frac{(29 k+57)}{12 (k+5)^2} \pa F_{11} G_{22}
\nonu \\
&-& \frac{(7k+27)}{12 (k+5)^2} F_{11} \pa G_{22}
- \frac{(104 k^3+451 k^2+2022 k+1083)}{6 (k+5)^3 (13 k+17)} \pa^2 F_{12} F_{21}
\nonu \\
&- &   \frac{(367 k^2+1622 k+471)}{3 (k+5)^3 (13 k+17)}
\pa F_{12} \pa F_{21}
- \frac{(235 k^2+974 k+531)}{(k+5)^3 (13 k+17)} \pa F_{11} \pa F_{22}
\nonu \\
& + & \frac{(104 k^2+445 k+681)}{6 (k+5)^2 (13 k+17)}
F_{12} \pa^2 F_{21}
+\frac{(13 k+33)}{4 (k+5)^2} \pa F_{12} G_{21}
\nonu \\
&+& \frac{1}{4 (k+5)}
 F_{12} \pa G_{21} + \frac{1}{2 (k+5)} G_{11} G_{22}
+ \frac{1}{2 (k+5)} G_{12} G_{21}
\nonu \\
&+& \frac{3(k+4)}{2 (k+5)^2} \pa F_{21} G_{12}
- \frac{(k+12)}{2 (k+5)^2} F_{21} \pa G_{12}
+\frac{(11k+48)}{6 (k+5)^2} \pa F_{22} G_{11}
\nonu \\
&-& \frac{(k+24)}{6 (k+5)^2}  F_{22} \pa G_{11}
- \frac{2 i (28 k+71)}{(k+5) (13 k+17)}  T A_3
+ \frac{2 (4 k+17)}{(k+5) (13 k+17)} T U
\nonu \\
&-& \frac{4(k+3)}{3 (k+5)}  \pa T 
- \frac{2 (19 k+71)}{(k+5)^2 (13 k+17)}  T F_{11} F_{22}
+ \frac{2}{(k+5)^2} U A_{+} A_{-}
\nonu \\
& - &  \frac{2 (19 k+71)}{(k+5)^2 (13 k+17)}
T F_{12} F_{21}
+\frac{2}{(k+5)^2} U A_3 A_3 +
\frac{2}{(k+5)^2} U B_3 B_3
\nonu \\
& +& \frac{4 i}{(k+5)^2} \pa (U A_3)
+\frac{i k}{(k+5)^2} \pa U  B_3
+\frac{i(k+2)}{(k+5)^2} U \pa B_3
\nonu \\
& + &  \frac{2}{(k+5)^2} U B_{+} B_{-}
-\frac{(4 k+17)}{(k+5) (13 k+17)} \pa^2 U
\nonu \\
&+& \frac{1}{(k+5)^2} U F_{11} G_{22}
+ \frac{(11k-57)}{6 (k+5)^3} \pa U F_{11} F_{22}
+ \frac{(37 k+201)}{6 (k+5)^3} U \pa F_{11} F_{22}
\nonu \\
&-& \frac{(47 k+219)}{6 (k+5)^3} U F_{11} \pa F_{22}
+\frac{1}{(k+5)^2} U F_{12} G_{21}
+\frac{(5k+41)}{2 (k+5)^3} \pa U F_{12} F_{21}
\nonu \\
& + &   \frac{(21 k+89)}{2 (k+5)^3} U \pa F_{12} F_{21}
- \frac{(7k+51)}{2 (k+5)^3} U F_{12} \pa F_{21}
-\frac{1}{(k+5)^2}  U F_{21} G_{12}
\nonu \\
& - & 
\frac{1}{(k+5)^2} U F_{22} G_{11}
- \frac{2 i}{(k+5)^2}  U U A_3
-\frac{2 (4 k+9)}{3 (k+5)^2} \pa U U
\nonu \\
&+& \left. 
\frac{2}{(k+5)^2} U U U
- \frac{2 (k+9)}{3 (k+5)^2} B_3 \pa B_3
\right](w) +\cdots. 
\nonu 
\eea
The nonlinear terms (with the footnote \ref{nonlineardef}) 
in this case do not appear in the above OPE. 

One continues to calculate the following OPE
\bea
%%%%%%%%%%%%%%%%%%%%%%%%%%%%%%%%%%%%%%%%%%%%%%%%%%%%%%%%%%%%%%%%%%%
{\bf U_{-}^{(2)}}(z) \, {\bf V_{-}^{(2)}}(w) & = & 
 \frac{1}{(z-w)^2} \, \left[ -\frac{2 (k+4)}{(k+5)^2}
{A}_{+} {B}_{+} 
+\frac{2 i}{(k+5)^2} A_{+} F_{22} F_{12}
\right. \nonu \\
&-& \left. \frac{2 i}{(k+5)^2} B_{+} F_{12} F_{11}
-\frac{(k-3)}{(k+5)^2} F_{12} G_{12}
+  \frac{4 (3 k-1)}{(k+5)^3} F_{12} \pa F_{12}
\right] (w)
\nonu \\
& + & \frac{1}{(z-w)} \, 
\left[ 
 -\frac{2 i}{(k+5)^2} A_3 F_{12} G_{12}
- \frac{(2k+9)}{(k+5)^2} \pa A_{+} B_{+}
+ \frac{2 i}{(k+5)^2} \pa A_{+} F_{22} F_{12}
\right. \nonu \\
& + & \frac{1}{(k+5)^2} A_{+} \pa B_{+}
+ \frac{i}{(k+5)^2} A_{+} F_{22} G_{12}
-\frac{2 i(k+1) }{(k+5)^3} A_{+} \pa F_{22} F_{12}
 \nonu \\
&+&  \frac{2 i(k+1)}{(k+5)^3} A_{+} F_{22} \pa F_{12}
+ \frac{i}{(k+5)^2} A_{+} F_{12} G_{22}
- \frac{2 i}{(k+5)^2} B_3 F_{12} G_{12}
\nonu \\
&+& \frac{i}{(k+5)^2} B_{+} F_{12} G_{11}
- \frac{2 i(k+1)}{(k+5)^3} B_{+} \pa F_{12} F_{11}
\nonu \\
&-& \frac{2 i(k+9)}{(k+5)^3} B_{+} F_{12} \pa F_{11}
+ \frac{i}{(k+5)^2} B_{+} F_{11} G_{12}
+\frac{ (k+11)}{(k+5)^2} \pa F_{12} G_{12}
\nonu \\
&-& \frac{(k-1)}{(k+5)^2} F_{12} \pa G_{12}
+ \frac{(7 k-5)}{(k+5)^3} F_{12} \pa^2 F_{12}
- \frac{4}{(k+5)^2} F_{12} \pa F_{12} U
\nonu \\
&- &  \left.
 \frac{4 i (k+1)}{(k+5)^3} F_{12} \pa F_{12} A_3
+ \frac{16 i}{(k+5)^3} F_{12} \pa F_{12} B_3
\right](w) +\cdots.
\nonu
\eea
It is obvious that there exist no boldface higher spin currents.

Furthermore, one has
\bea
%%%%%%%%%%%%%%%%%%%%%%%%%%%%%%%%%%%%%%%%%%%%%%%%%%%%%%%%%%%
{\bf U_{-}^{(2)}}(z) \, {\bf V^{(\frac{5}{2})}}(w) & = & 
\frac{1}{(z-w)^4} \, \left[ \frac{24 k(k+3)}{(k+5)^3}  F_{12} \right](w)
\nonu \\
&+&
\frac{1}{(z-w)^3} \, \left[
- \frac{4 \left(4 k^2+29 k+69\right)}{3 (k+5)^3}  U F_{12}
- \frac{16 (k+3)}{3 (k+5)^3}  F_{12} F_{11} F_{22}
\right. \nonu \\
&-& \frac{4 i \left(4 k^2+17 k+33\right)}{3 (k+5)^3}  F_{12} A_3
+\frac{4 i \left(4 k^2+11 k-21\right)}{3 (k+5)^3}  F_{22} A_{+}
\nonu \\
&+ & \left. \frac{4 i (17 k+69)}{3 (k+5)^3}  F_{12} B_3 
- \frac{2 \left(4 k^2+38 k+69\right)}{3 (k+5)^2} G_{12} 
- \frac{4 i (17 k+69)}{3 (k+5)^3} F_{11} B_{+}
\right](w) 
\nonu \\
& + & \frac{1}{(z-w)^2} \, \left[
\frac{4 (2 k+9)}{3 (k+5)} ( {\bf P_{-}^{(\frac{5}{2})} } 
-   {\bf W_{-}^{(\frac{5}{2})} }) 
+ \frac{2i(5k+9)}{3 (k+5)^2} A_3 G_{12}
\right. \nonu \\
&- &  \frac{4(k+9)}{3(k+5)^3} A_3 A_3 F_{12}
-\frac{2 i (2 k+7)}{(k+5)^2} A_{+} G_{22}
-\frac{4 (k+3)}{3 (k+5)^3} A_{+} A_3 F_{22}
\nonu \\
&+& \frac{4(5k+21)}{3(k+5)^3} A_3 B_3 F_{12}
-\frac{4(5k+21)}{3(k+5)^3} A_3 B_{+} F_{11}
- \frac{2 i \left(4 k^2-93 k-297\right)}{9 (k+5)^3} \pa A_3 F_{12}
\nonu \\
&-& \frac{2 i \left(8 k^2+165 k+513\right)}{9 (k+5)^3}  A_3 \pa F_{12}
-\frac{8(k+6)}{3(k+5)^3} A_{+} A_{-} F_{12}
-\frac{8(2k+9)}{3(k+5)^3} A_{+} B_3 F_{22}
\nonu \\
&-& \frac{8(2k+9)}{3(k+5)^3} A_{+} B_{+} F_{21}
+ \frac{8 i \left(2 k^2-12 k-81\right)}{9 (k+5)^3} \pa A_{+} F_{22}
+  \frac{2i(k+9)}{3(k+5)^2} B_{+} G_{11}
\nonu \\
& + & \frac{2 i \left(8 k^2+147 k+459\right)}{9 (k+5)^3} A_{+} \pa F_{22}
- \frac{2i(k+9)}{3 (k+5)^2} B_3 G_{12}
\nonu \\
& + & \frac{2 i (k+9) (4 k+9)}{9 (k+5)^3}  \pa B_3 F_{12}
- \frac{2 i \left(16 k^2+51 k-153\right)}{9 (k+5)^3} B_3 \pa F_{12}
\nonu \\
& - &
 \frac{16 i (k+3) (k+9)}{9 (k+5)^3}  \pa B_{+} F_{11}
+ \frac{2 i \left(16 k^2+51 k-153\right)}{9 (k+5)^3}  B_{+} \pa F_{11}
\nonu \\
& - & \frac{16 \left(5 k^2+54 k+153\right)}{9 (k+5)^4}  \pa F_{11} F_{12} F_{22}
+ \frac{4(k+3)}{3(k+5)^3} F_{11} F_{12} G_{22}
\nonu \\
&+& \frac{16 (k+3)^2}{9 (k+5)^4}   F_{11} \pa F_{12} F_{22}
+ \frac{16 \left(7 k^2+72 k+189\right)}{9 (k+5)^4}  F_{11} F_{12} \pa F_{22}
\nonu \\
& - & \frac{2(5k+21)}{3(k+5)^3} F_{11} F_{22} G_{12}
 -   \frac{2 \left(8 k^2+9 k+189\right)}{9 (k+5)^3}  U \pa F_{12}
\nonu \\
&-&\frac{2(7k+27)}{ (k+5)^3} \pa^2 F_{12}
+ \frac{2 (k+9)}{3 (k+5)^3}
  F_{12} F_{21} G_{12}
- \frac{2 (k+9) (4 k+11)}{3 (k+5)^3} \pa U F_{12}
\nonu \\
&+& \frac{4}{(k+5)^2} F_{12} F_{22} G_{11}
-\frac{2 (k+9) (4 k+9)}{9 (k+5)^2} \pa G_{12}
+ \frac{8 i}{(k+5)^2} U A_3 F_{12}
\nonu \\
&+& \frac{4(4k+9)}{3(k+5)^2} T F_{12}
+\frac{2(k+12)}{3 (k+5)^2} U G_{12}
-\frac{4 i(5k+21)}{3(k+5)^3} U B_3 F_{12}
\nonu \\
&-& \frac{ 4i(7k+33)}{3(k+5)^3} U A_{+} F_{22}
+  \frac{4i(5k+21)}{3(k+5)^3} U B_{+} F_{11}
+ \frac{4(5k+21)}{3(k+5)^3} U U F_{12}
\nonu \\
&+ & \left. \frac{64 i \left(k^2+10 k+27\right)}{3 (k+5)^4} 
F_{12} B_3 F_{12} F_{21}
+ \frac{2 (k+9) (4 k+15)}{9 (k+5)^3}  F_{12} G_{21} F_{12}
\right](w) \nonu \\
& + & 
\frac{1}{(z-w)} \, \left[ +\cdots \right](w) 
+ \cdots.
\nonu 
\eea
There is no higher spin current in the third order pole of the above OPE.
The structure of the first order pole is not determined.

The result in (\ref{g1122vu3half}) can be used
and one obtains the following OPEs
\bea
%%%%%%%%%%%%%%%%%%%%%%%%%%%%%%%%%%%%%%%%%%%%%%%%%%%%%%%%%%%%%%%%%%
\left(

\right)
\right](w) +\cdots.
\nonu 
\eea
There is no higher spin current in the second order pole.
One sees that the nonlinear terms appeared in the nonlinear version
do not appear in these OPEs. 

Again the result (\ref{g1221w2}) can be used and the following OPEs 
can be obtained
%%%%%%%%%%%%%%%%%%%%%%%%%%%%%%%%%%%%%%%%%%%%%%%%%%%%%%%%
\bea
\left(
%
\right) F_{11} F_{22} \right](w) 
\nonu \\
& + & \frac{1}{(z-w)} \, \left[  + \cdots \right](w) + \cdots.
\nonu
\eea
One does not see the higher spin currents in the third order pole 
of the above OPEs.
In principle, the first order pole can be obtained from the 
description in the section $7$.

Similarly the other combination leads to the following OPEs
%%%%%%%%%%%%%%%%%%%%%%%%%%%%%%%%%%%%%%%%%%%%%%%%%%%%%%%%
\bea
\left(
\begin{array}{c}
{\bf U_{-}^{(2)}} \\
{\bf V_{+}^{(2)}}
\end{array}
\right) (z) \, {\bf W_{\mp}^{(\frac{5}{2})}}(w) & = &
\frac{1}{(z-w)^3}\, 
\left[ - \frac{8 i (k+9)}{(k+5)^3}
A_{\pm}
\left(
\begin{array}{c}
F_{12} \\
F_{21} 
\end{array}
\right)  \right] (w) 
\nonu \\
& + & \frac{1}{(z-w)^2} \, \left[ 
- \frac{2i(k+12)}{3 (k+5)^2} A_{\pm} 
\left(
\begin{array}{c}
G_{12} \\
G_{21} 
\end{array}
\right) 
\pm
\frac{4 (k+3)}{3 (k+5)^3} A_{\pm} A_3 
\left(
\begin{array}{c}
F_{12} \\
F_{21} 
\end{array}
\right) 
\right.
\nonu \\
&\mp&
\frac{4(k+3)}{3(k+5)^3} A_{\pm} A_{\pm}
\left(
\begin{array}{c}
F_{22} \\
F_{11} 
\end{array}
\right)
\pm \frac{4 (k+3)}{3 (k+5)^3} A_{\pm} B_3 
\left(
\begin{array}{c}
F_{12} \\
F_{21} 
\end{array}
\right)
\nonu \\
&-&   \frac{24 i}{(k+5)^3} 
\pa A_{\pm} 
\left(
\begin{array}{c}
F_{12} \\
F_{21} 
\end{array}
\right)
-  \frac{4 i (k+3)}{(k+5)^3}  A_{\pm} \pa 
\left(
\begin{array}{c}
F_{12} \\
F_{21} 
\end{array}
\right)
\nonu \\
&\mp & \frac{4 (k+3)}{3(k+5)^3} A_{\pm}  B_{\pm}
\left(
\begin{array}{c}
F_{11} \\
F_{22} 
\end{array}
\right)
\pm \frac{4(k+3)}{3(k+5)^3} 
\left(
\begin{array}{c}
F_{12} F_{11} G_{12} \\
F_{21} F_{22} G_{21}
\end{array}
\right)
\nonu \\
&- & \left. 
 \frac{4i (k+3)}{3 (k+5)^3} U A_{\pm}  
\left(
\begin{array}{c}
F_{12} \\
F_{21} 
\end{array}
\right)
\pm
\frac{32 (k+3)}{3 (k+5)^4} 
\left(
\begin{array}{c}
F_{11} F_{12} \pa F_{12} \\
F_{22} F_{21} \pa F_{21} 
\end{array}
\right)
 \right](w) 
\nonu \\
& + & \frac{1}{(z-w)} \, \left[  + \cdots \right](w) + \cdots.
\nonu
\eea
It is clear that there are no boldface higher spin currents in the OPEs.

The result (\ref{g1221w5half}) can be used to calculate the following OPEs
\bea
%%%%%%%%%%%%%%%%%%%%%%%%%%%%%%%%%%%%%%%%%%%%%%%%%%%%%%%%%%%%%%%%%%
\left(
%
\right)
\right](w) \nonu \\
& + & \frac{1}{(z-w)} \, \left[ + \cdots \right](w) + \cdots.
\nonu 
\eea
There is no boldface higher spin current in the third order pole 
of the OPEs. As before, the nonlinear terms between the boldface 
higher spin currents can be removed by adding the 
$T \, {\bf T^{(1)}}(w)$ into the left hand side. 

As before, the result (\ref{g1221vu2}) determines the following OPEs
%%%%%%%%%%%%%%%%%%%%%%%%%%%%%%%%%%%%%%%%%%%%%%%%%%%%%%%%
\bea
\left(
\begin{array}{c}
{\bf U^{(\frac{5}{2})}} \\
{\bf V^{(\frac{5}{2})}} 
\end{array}
\right)(z) \,
\left(
\begin{array}{c}
{\bf U^{(\frac{5}{2})}} \\
{\bf V^{(\frac{5}{2})}} 
\end{array}
\right)(w) & =& 
\frac{1}{(z-w)^3} \, \left[ \frac{8 \left(8 k^2+51 k+135\right)}{9 (k+5)^3}  
A_{\pm} B_{\mp} 
\right. \nonu \\
& \pm &  \frac{32 i (k+3) (k+9)}{9 (k+5)^4}  A_{\pm}
\left(
\begin{array}{c}
F_{11} F_{21} \\
F_{22} F_{12} 
\end{array}
\right)
 \nonu \\
& \mp &  \frac{32 i (k+3) (k+9)}{9 (k+5)^4}  B_{\mp} 
 \left(
\begin{array}{c}
F_{11} F_{12} \\
F_{22} F_{21} 
\end{array}
\right)
\nonu \\
& + & \frac{16 \left(31 k^2+138 k+27\right)}{9 (k+5)^4} 
 \left(
\begin{array}{c}
F_{11} \pa F_{11} \\
F_{22} \pa F_{22} 
\end{array}
\right) \nonu \\
&- & \left. 
\frac{8 (k-3) (2 k+9)}{3 (k+5)^3}
 \left(
\begin{array}{c}
F_{11} G_{11} \\
F_{22} G_{22} 
\end{array}
\right)
\right](w) \nonu \\
& + & \frac{1}{(z-w)^2} \, \frac{1}{2} \pa (\mbox{pole-3})(w) + 
\frac{1}{(z-w)} \, \left[ + \cdots \right](w) + 
\cdots.
\nonu
\eea
It is an open problem to obtain the 
first order pole of the OPEs.

Also one can construct the following OPE
\bea
%%%%%%%%%%%%%%%%%%%%%%%%%%%%%%%%%%%%%%%%%%%%%%%%%%%%%%%%%%%%%%%
{\bf U^{(\frac{5}{2})}}(z) \, {\bf V^{(\frac{3}{2})}}(w) 
&=& 
\frac{1}{(z-w)^3} \, \left[  
-\frac{4 {\bf (k-3)}}{3 (k+5)} {\bf T^{(1)}}
+ \frac{2 i (5k+18)}{(k+5)^2}  {A}_3   
+ \frac{2 i k (4 k+21)}{3 (k+5)^2} B_3 
+ \frac{6 k}{(k+5)^2} U 
\right. \nonu \\
& 
- &  \left.  \frac{4 (k-3) (2 k+9)}{3 (k+5)^3}  F_{11} F_{22}
+ \frac{4 \left(2 k^2+9 k+27\right)}{3 (k+5)^3}  F_{12} F_{21}
\right](w)
\nonu \\
& + & \frac{1}{(z-w)^2} \, \left[ \frac{1}{2} {\bf P^{(2)}} 
-  \frac{8 (k+3)}{3 (k+5)} ( {\bf T^{(2)}} + {\bf W^{(2)}}) 
-\frac{4 {\bf (k-3)}}{3 (k+5)}  \pa {\bf T^{(1)}}
\right. \nonu \\
& + & 
\frac{4 (k-3)}{3 (k+5)^2}  A_3 B_3
+\frac{2 i (5k+18)}{ (k+5)^2} \pa A_3
-  
\frac{4 i k}{(k+5)^2} U B_3 
+\frac{6k}{ (k+5)^2} \pa U
\nonu \\
& - & \frac{2 i (k-3)}{3 (k+5)^3}
 A_3 F_{11} F_{22}
+  \frac{2 i(k-3)}{3(k+5)^3} A_3 F_{12} F_{21}
+  \frac{2 i (5k+9)}{3 (k+5)^3}  B_{-} F_{12} F_{22}
\nonu \\
&+& \frac{8 i (k+9)}{3 (k+5)^3}  A_{-} F_{11} F_{12}
- \frac{8 i(k+9)}{3(k+5)^3} A_{+} F_{21} F_{22}
-  \frac{2 i (5k+9)}{3(k+5)^3} B_{+} F_{11} F_{21}
\nonu \\
& + & \frac{2 i k (4k+21)}{3 (k+5)^2} \pa B_3
  +\frac{2 i (k-3)}{3 (k+5)^3}   B_3 F_{11} F_{22}
+  \frac{2 i(k-3)}{(k+5)^2} B_3 F_{12} F_{21}
\nonu \\
&-& \frac{(k+9)}{3 (k+5)^2} F_{11} G_{22}
- \frac{2 \left(4 k^2-5 k-93\right)}{3 (k+5)^3}  \pa F_{11} F_{22}
\nonu \\
& - & \frac{2 (4 k-21)}{3 (k+5)^2}   F_{11} \pa F_{22}
- \frac{(k+9)}{(k+5)^2} F_{12} G_{21}
+  \frac{2 \left(4 k^2+15 k+87\right)}{3 (k+5)^3} \pa F_{12} F_{21}
\nonu \\
& + & \frac{2 \left(4 k^2+19 k+75\right)}{3 (k+5)^3}  F_{12} \pa F_{21}
+ \frac{6}{(k+5)^2}  F_{21} G_{12}
+ \frac{2(2k+9)}{3 (k+5)^2}  F_{22} G_{11}
\nonu \\
&+ & \left. \frac{2 (k-27)}{3 (k+5)^3}  U F_{11} F_{22}
-\frac{6}{(k+5)^2}  U F_{12} F_{21}
\right](w)
\nonu \\
& + & \frac{1}{(z-w)} \, \left[ \frac{1}{2} {\bf S^{(3)}} 
+\frac{3}{8} 
  \pa 
{\bf P^{(2)}}  - {\bf W^{(3)}} - \frac{2 (k+3)}{(k+5)} \pa ({\bf T^{(2)}}  
+ {\bf W^{(2)}}) 
\right. \nonu \\
&- &  \frac{2 {\bf (k-3)}}{(13 k+17)} \left( T {\bf T^{(1)}} -\frac{1}{2} 
\pa^2 {\bf T^{(1)}} \right) -
\frac{2{\bf (k-3)}}{3 (k+5)} \pa^2 {\bf T^{(1)}}
\nonu \\
& + & \frac{(k-3)}{ (k+5)^2} \pa (A_3 B_3)
+ \frac{i (5 k+18) (23 k+19)}{2 (k+5)^2 (13 k+17)} \pa^2 A_3
\nonu \\
&-&
\frac{i (5 k+17)}{2 (k+5)^3} \pa A_3 F_{11} F_{22}
+ \frac{i (3 k+23)}{2 (k+5)^3} A_3 \pa (F_{11} F_{22}) 
\nonu \\
&-& \frac{i (3 k+23)}{2 (k+5)^3} \pa A_3 F_{12} F_{21}
+\frac{i (5k+17)}{2 (k+5)^3} A_3 \pa (F_{12} F_{21})
\nonu \\
& + & \frac{8i}{ (k+5)^3} \pa A_{-} F_{11} F_{12}
+  \frac{4i (k+7)}{ (k+5)^3} A_{-} \pa (F_{11} F_{12})
\nonu \\
& - & \frac{4 i(k+7) }{ (k+5)^3} \pa A_{+} F_{21} F_{22}
- \frac{ 8i }{ (k+5)^3} A_{+} \pa (F_{21} F_{22})
\nonu \\
&+&   \frac{i k (4 k+21) (23 k+19)}{6 (k+5)^2 (13 k+17)} \pa^2 B_3 
- 
 \frac{i (3 k+23)}{2 (k+5)^3}  \pa B_3 F_{11} F_{22}
\nonu \\
&+& \frac{i (5k+17)}{2 (k+5)^3} B_3 \pa (F_{11} F_{22})
+ \frac{(9 k+13)}{2 (k+5)^3} U F_{11} \pa F_{22}
\nonu \\
&+& 
 \frac{i (5k+17)}{2 (k+5)^3} \pa B_3 F_{12} F_{21}
- \frac{i (3k+23)}{2 (k+5)^3} B_3 \pa (F_{12} F_{21})
\nonu \\
&+& 
\frac{i (k-11)}{2 (k+5)^3} \pa B_{-} F_{12} F_{22}
+  \frac{i (9 k+29)}{2 (k+5)^3} B_{-} \pa (F_{12} F_{22})
\nonu \\
&-&
\frac{i (9 k+29)}{2 (k+5)^3} \pa B_{+} F_{11} F_{21}
 -  \frac{i(9k-11)}{2 (k+5)^2} B_{+} \pa (F_{11} F_{21})
\nonu \\
& - & \frac{(92 k^3-215 k^2-3210 k-3015)}{6 (k+5)^3 (13 k+17)}
  \pa^2 F_{11} F_{22}
\nonu \\
& - &
 \frac{(92 k^3-137 k^2-3342 k-3321)}{3 (k+5)^3 (13 k+17)} \pa F_{11} \pa F_{22}
-\frac{9}{2 (k+5)^2} \pa U F_{12} F_{21}
\nonu \\
&-& \frac{(92 k^3-59 k^2-3474 k-3627)}{6 (k+5)^3 (13 k+17)} 
F_{11} \pa^2 F_{22}
- \frac{(7 k+39)}{4 (k+5)^2} \pa F_{11} G_{22}
\nonu \\
&+& \frac{(k+1)}{4 (k+5)^2} F_{11} \pa G_{22}
+ \frac{(92 k^3+373 k^2+2718 k+2709)}{6 (k+5)^3 (13 k+17)} \pa^2 F_{12} F_{21}
\nonu \\
&+ &   \frac{(92 k^3+451 k^2+2586 k+2403)}{3 (k+5)^3 (13 k+17)}
\pa F_{12} \pa F_{21}
\nonu \\
& + & \frac{(92 k^3+529 k^2+2454 k+2097)}{6 (k+5)^3 (13 k+17)}
F_{12} \pa^2 F_{21}
-\frac{3(3 k+19)}{4 (k+5)^2} \pa F_{12} G_{21}
\nonu \\
&-& \frac{(k+17)}{4 (k+5)^2}
 F_{12} \pa G_{21} 
- \frac{3(k+2)}{2 (k+5)^2} \pa F_{21} G_{12}
+ \frac{(k+14)}{2 (k+5)^2} F_{21} \pa G_{12}
\nonu \\
& - & \frac{(k+6)}{2 (k+5)^2} \pa F_{22} G_{11}
+ \frac{(3k+14)}{2 (k+5)^2}  F_{22} \pa G_{11}
+ \frac{9 k}{(k+5) (13 k+17)} T U
\nonu \\
& + &  \frac{3 i (5 k+18)}{(k+5) (13 k+17)}  T A_3
+ \frac{i k (4 k+21)}{(k+5) (13 k+17)} T B_3
\nonu \\
&-& 
 \frac{2 (k-3) (2 k+9)}{(k+5)^2 (13 k+17)}  T F_{11} F_{22}
+ \frac{2 \left(2 k^2+9 k+27\right)}{(k+5)^2 (13 k+17)}
T F_{12} F_{21}
\nonu \\
& -& 
\frac{3 i k}{(k+5)^2} \pa (U  B_3)
+ \frac{3 k (23 k+19)}{2 (k+5)^2 (13 k+17)}  \pa^2 U
\nonu \\
&+& 
 \frac{(k-27)}{2 (k+5)^3} \pa U F_{11} F_{22}
- \frac{(7 k+67)}{2 (k+5)^3} U \pa F_{11} F_{22}
\nonu \\
& - & \left.    \frac{17}{2 (k+5)^2} U \pa F_{12} F_{21}
- \frac{1}{2 (k+5)^2} U F_{12} \pa F_{21}
\right](w) +\cdots.
\nonu 
\eea
Note the presence of a quasi primary field containing the higher 
spin-$1$ current in the first order pole. 

Furthermore the following OPE can be obtained
\bea
%%%%%%%%%%%%%%%%%%%%%%%%%%%%%%%%%%%%%%%%%%%%%%%%%%%%%%%%%%%%%%%%%%%%%%%%%
{\bf U^{(\frac{5}{2})}}(z) \, 
{\bf V_{+}^{(2)}}(w)  
& = & 
\frac{1}{(z-w)^4} \,
\left[ -\frac{24 k(k+3)}{(k+5)^3} F_{21} \right](w) \nonu \\
&+& \frac{1}{(z-w)^3} \, \left[ 
 -\frac{2 \left(4 k^2+38 k+69\right)}{3 (k+5)^2}  G_{21} 
 -\frac{4 i \left(4 k^2+11 k-21\right)}{3 (k+5)^3}  F_{11} A_{-}
\right. \nonu \\
&-&  \frac{4 \left(4 k^2+29 k+69\right)}{3 (k+5)^3}  U F_{21}
+ \frac{16 (k+3)}{3 (k+5)^3} F_{21} F_{11} F_{22}
 -\frac{4 i (17 k+69)}{3 (k+5)^3}  F_{21} B_3
\nonu \\
& + & \left.
\frac{4 i \left(4 k^2+17 k+33\right)}{3 (k+5)^3}  F_{21} A_3
-\frac{24 k (k+3)}{(k+5)^3}  \pa F_{21}
+  \frac{4 i \left(17k+69\right)}{3 (k+5)^3} F_{22} B_{-}
\right](w)
\nonu \\
& + & \frac{1}{(z-w)^2} \, \left[ 
\frac{4(2 k+9)}{3 (k+5)}  ({\bf P_{+}^{(\frac{5}{2})}}  +
{\bf W_{+}^{(\frac{5}{2})}})
\right. \nonu \\
&+& \frac{2i(5k+9)}{3 (k+5)^2} A_3 G_{21}
-\frac{4(5k+21)}{3(k+5)^3} A_3 B_3 F_{21}
+ \frac{4(5k+21)}{3(k+5)^3} A_3 B_{-} F_{22}
\nonu \\
&+& \frac{16 i \left(2 k^2+21 k+63\right)}{9 (k+5)^3}  \pa A_3 F_{21}
+ \frac{2 i \left(16 k^2-63 k-315\right)}{9 (k+5)^3}  A_3 \pa F_{21}
\nonu \\
& - & \frac{2 i (2k+7)}{ (k+5)^2}  A_{-} G_{11}
+ \frac{4 (k+9)}{3 (k+5)^3} A_3 A_3 F_{21}
+ \frac{4 (k+3)}{3 (k+5)^3} A_{-} A_3 F_{11}
\nonu \\
& + & \frac{8(2k+9)}{3(k+5)^3} A_{-} B_3 F_{11}
+ \frac{8(2k+9)}{3(k+5)^3} A_{-} B_{-} F_{12}
-\frac{4 i (k+3) (8 k+33)}{9 (k+5)^3}  \pa A_{-} F_{11}
\nonu \\
&-& \frac{2 i \left(16 k^2-81 k-585\right)}{9 (k+5)^3}  A_{-} \pa F_{11}
- \frac{2 i (k+9)}{3 (k+5)^2} B_3 G_{21}
+ \frac{8 (k+6)}{3 (k+5)^3} A_{+} A_{-} F_{21}
\nonu \\
&+ & \frac{4 i \left(4 k^2-3 k-99\right)}{9 (k+5)^3}  \pa B_3 F_{21}
- \frac{2 i \left(16 k^2+153 k+261\right)}{9 (k+5)^3} B_3 \pa F_{21}
\nonu \\
& +  & \frac{2i (k+9)}{3(k+5)^2}  B_{-} G_{22}
+ \frac{2(k+9)}{3(k+5)^3} F_{12} F_{21} G_{21}
-\frac{32 (k+3) (k+6)}{9 (k+5)^4} F_{11} \pa F_{21} F_{22}
\nonu \\
&-&  \frac{4 i \left(4 k^2-3 k-99\right)}{9 (k+5)^3} \pa B_{-} F_{22}
+  \frac{2 i \left(16 k^2+153 k+261\right)}{9 (k+5)^3}  B_{-} \pa F_{22}
\nonu \\
&+& \frac{4}{(k+5)^2} F_{11} F_{21} G_{22}
+ \frac{64 (k+6)^2}{9 (k+5)^4} \pa F_{11} F_{21} F_{22}
\nonu \\
&-& \frac{32 \left(4 k^2+39 k+99\right)}{9 (k+5)^4}    F_{11} F_{21} \pa F_{22}
-\frac{2(5k+21)}{3(k+5)^3} F_{11} F_{22} G_{21}
\nonu \\
&-& \frac{2 \left(2 k^2+15 k+45\right)}{(k+5)^3} \pa^2 F_{21}
+ \frac{4 (k+3)}{3 (k+5)^3} F_{21} F_{22} G_{11}
\nonu \\
&  - &  \frac{2 (k+6) (8 k+21)}{9 (k+5)^2} \pa G_{21}
-\frac{4(4k+9)}{3(k+5)^2} T F_{21}
\nonu \\
&-& \frac{2(k+12)}{3 (k+5)^2} U G_{21}
+ \frac{8 i}{(k+5)^2} U A_3 F_{21}
-\frac{4 i(7k+33)}{3(k+5)^3} U A_{-} F_{11}
\nonu \\
&+& \frac{4i(5k+21)}{3(k+5)^3} U B_{-} F_{22}
-\frac{8 \left(4 k^2+21 k+63\right)}{9 (k+5)^3} \pa U F_{21}
\nonu \\
& - & \frac{2 \left(16 k^2+165 k+225\right)}{9 (k+5)^3}
  U \pa F_{21}
- \frac{4 i(5¤¿+21)}{3(k+5)^3} U B_3 F_{21}
\nonu \\
&- & \left.  
\frac{4(5¤¿+21)}{3(k+5)^3} U U F_{21} -
\frac{32 \left(k^2+10 k+27\right)}{3 (k+5)^4}
F_{21} \pa F_{21} F_{12}
 \right](w) \nonu \\
& + & \frac{1}{(z-w)} \, \left[ + \cdots \right](w) 
+ \cdots.
\nonu 
\eea
One does not see the higher spin current in the third order pole.
The corresponding OPE in the nonlinear version
contains the higher spin current in that pole.

The following OPE can be obtained
\bea
%%%%%%%%%%%%%%%%%%%%%%%%%%%%%%%%%%%%%%%%%%%%%%%%%%%%%%%%%%%
{\bf U^{(\frac{5}{2})}}(z) \, {\bf V_{-}^{(2)}}(w) & = & 
\frac{1}{(z-w)^4} \, \left[ -\frac{12 k(k+9)}{(k+5)^3}  F_{12} \right](w)
\nonu \\
&+&
\frac{1}{(z-w)^3} \, \left[
\frac{4 \left(4 k^2+31 k+63\right)}{3 (k+5)^3}  U F_{12}
+ \frac{8 (k+9)}{3 (k+5)^3}   F_{12} F_{11} F_{22}
\right. \nonu \\
&+& \frac{4 i \left(4 k^2+25 k+9\right)}{3 (k+5)^3}   F_{12} A_3
-\frac{4 i \left(4 k^2+25 k+9\right)}{3 (k+5)^3}  F_{22} A_{+}
-\frac{4 i \left(2 k^2-k-63\right)}{3 (k+5)^3}  F_{11} B_{+}
\nonu \\
&- & \left. \frac{4 i \left(2 k^2+13 k+63\right)}{3 (k+5)^3}  F_{12} B_3 
+ \frac{2 \left(4 k^2+40 k+63\right)}{3 (k+5)^2} G_{12}
 -\frac{12 k (k+9)}{(k+5)^3} \pa F_{12}
\right](w) 
\nonu \\
& + & \frac{1}{(z-w)^2} \, \left[
-\frac{(7 k+39)}{3 (k+5)} ( {\bf P_{-}^{(\frac{5}{2})} } 
-   {\bf W_{-}^{(\frac{5}{2})} }) 
- \frac{4i(k+3)}{3 (k+5)^2} A_3 G_{12}
\right. \nonu \\
&-& \frac{16(k+6)}{3(k+5)^3} A_3 B_3 F_{12}
+\frac{2(7k+39)}{3(k+5)^3} A_3 B_{+} F_{11}
- \frac{4 i \left(k^2+16 k+75\right)}{3 (k+5)^3}  \pa A_3 F_{12}
\nonu \\
&+& \frac{2 i \left(16 k^2+159 k+243\right)}{9 (k+5)^3}   A_3 \pa F_{12}
+\frac{4i(k+3)}{3(k+5)^2} A_{+} G_{22}
+\frac{16(k+6)}{3(k+5)^3} A_{+} B_3 F_{22}
\nonu \\
&+& \frac{2(7k+39)}{3(k+5)^3} A_{+} B_{+} F_{21}
 -\frac{4 i \left(8 k^2+27 k-81\right)}{9 (k+5)^3} \pa A_{+} F_{22}
\nonu \\
& - & \frac{2 i \left(16 k^2+159 k+243\right)}{9 (k+5)^3}  A_{+} \pa F_{22}
+ \frac{2i(k+21)}{3 (k+5)^2} B_3 G_{12}
\nonu \\
& + & \frac{8 (k+3)}{3 (k+5)^3}   \pa B_3 F_{12}
+ \frac{2 i \left(20 k^2+147 k-261\right)}{9 (k+5)^3}  B_3 \pa F_{12}
\nonu \\
& - & \frac{i(3k+17)}{(k+5)^2} B_{+} G_{11}
+\frac{2 (k+9)}{3 (k+5)^3} B_{+} B_3 F_{11}
+ \frac{2 (5 k+21)}{3 (k+5)^3} B_{+} B_{-} F_{12}
\nonu \\
&+& 
\frac{2 i (k+9) (7 k+36)}{9 (k+5)^3}  \pa B_{+} F_{11}
-\frac{4 i \left(19 k^2+87 k-90\right)}{9 (k+5)^3}  B_{+} \pa F_{11}
\nonu \\
& + & \frac{4 (5 k+21)^2}{9 (k+5)^4}  \pa F_{11} F_{12} F_{22}
- \frac{4}{(k+5)^2} F_{11} F_{12} G_{22}
+ \frac{8 (k+3)}{3 (k+5)^3} B_3 B_3 F_{12}
\nonu \\
&- & \frac{4 (k+9) (5 k+21)}{9 (k+5)^4}   F_{11} \pa F_{12} F_{22}
-\frac{32 \left(4 k^2+39 k+99\right)}{9 (k+5)^4}   F_{11} F_{12} \pa F_{22}
\nonu \\
&  + & \frac{8 (k+6)}{3 (k+5)^3}  F_{11} F_{22} G_{12}
- \frac{4(2k+15)}{3(k+5)^2} T F_{12}
-\frac{2(2k+9)}{3 (k+5)^2} U G_{12}
\nonu \\
&-& \frac{2 \left(19 k^2+114 k-189\right)}{3 (k+5)^3} \pa^2 F_{12}
+ \frac{4 (k+3)}{3 (k+5)^3}
  F_{12} F_{21} G_{12}
\nonu \\
&-& \frac{2(k+9)}{3(k+5)^3} F_{12} F_{22} G_{11}
+ \frac{2 (2 k+3) (4 k+33)}{9 (k+5)^2} \pa G_{12}
- \frac{16 i(k+6)}{3(k+5)^3} U A_3 F_{12}
\nonu \\
& + & 
\frac{16 i (k+6)}{3 (k+5)^3} U A_{+} F_{22}+
\frac{8 i}{(k+5)^2} U B_3 F_{12}
+ \frac{4 \left(19 k^2+126 k+243\right)}{9 (k+5)^3} \pa U F_{12}
\nonu \\
&-&
  \frac{2i(13k+69)}{3(k+5)^3} U B_{+} F_{11}
- \frac{16(k+6)}{3(k+5)^3} U U F_{12}
+ \frac{2 \left(16 k^2+147 k+279\right)}{9 (k+5)^3} U \pa F_{12}
\nonu \\
&- & \left.   \frac{8 i \left(9 k^2+82 k+201\right)}{3 (k+5)^4}
F_{12} B_3 F_{12} F_{21}
-\frac{4 \left(11 k^2+75 k+144\right)}{9 (k+5)^3}  F_{12} G_{21} F_{12}
\right](w) \nonu \\
& + & \frac{1}{(z-w)} \, \left[ + \cdots \right](w)   + \cdots. 
\nonu 
\eea
There is no boldface higher spin current in the third order pole.

The other combination of OPE can be described as
\bea
%%%%%%%%%%%%%%%%%%%%%%%%%%%%%%%%%%%%%%%%%%%%%%%%%%%%%%%%%%%%%%%%%%%%%%
{\bf U^{(\frac{5}{2})}}(z) \, {\bf V^{(\frac{5}{2})}}(w) 
&=& 
\frac{1}{(z-w)^5} \, \left[ -\frac{32 k(k+3) (k+9)}{(k+5)^3}\right]
\nonu \\
&+& 
\frac{1}{(z-w)^4} \, \left[  
-\frac{8 i (k+9) (4 k+9)}{(k+5)^3} A_3 
+ \frac{16 i k (k+3) (2 k+15)}{3 (k+5)^3}  {B}_3   
 \right. \nonu \\
& - & \left.   \frac{8 \left(4 k^3+27 k^2+90 k+243\right)}{3 (k+5)^4}   
F_{11} F_{22}
+ \frac{8 (k-3) \left(4 k^2+33 k+81\right)}{3 (k+5)^4}  F_{12} F_{21}
\right](w)
\nonu \\
& + & \frac{1}{(z-w)^3} \, \left[ \frac{2 (k-3)}{3 (k+5)} {\bf P^{(2)}} - 
\frac{8 (k-3)^2}{9 (k+5)^2}  {\bf T^{(2)}} 
- \frac{8 (k-3)}{3 (k+5)}  {\bf W^{(2)}}
\right. \nonu \\ 
& - &  
\frac{4 \left(20 k^2+204 k+369\right)}{9 (k+5)^2} T 
 -  \frac{4 \left(2 k^2+87 k+369\right)}{9 (k+5)^3} B_3 B_3
\nonu \\
& - &  \frac{4 \left(20 k^2+123 k+99\right)}{9 (k+5)^3} A_3 A_3  +
 \frac{8 \left(k^2+5 k+16\right)}{(k+5)^3}  A_3 B_3
\nonu \\
& - & \frac{16 i \left(14 k^2+132 k+207\right)}{9 (k+5)^3} \pa A_3
 -   
\frac{8 i \left(13 k^2+66 k+117\right)}{9 (k+5)^4}  B_{+} F_{11} F_{21}
\nonu \\
& + & \frac{64 i (k+3) (k+9)}{9 (k+5)^4}
 A_3 F_{11} F_{22}
+ \frac{8 i \left(5 k^2+42 k+117\right)}{9 (k+5)^4}  A_3 F_{12} F_{21}
\nonu \\
& - &  \frac{4 \left(20 k^2+123 k+99\right)}{9 (k+5)^3} A_{+} A_{-}
+  \frac{4 (k-3) (10 k+51)}{9 (k+5)^3}  F_{22} G_{11}
\nonu \\
&+& \frac{8 i (k+1) (k+9)}{(k+5)^4}  A_{-} F_{11} F_{12}
+ \frac{32 i \left(k^2+12 k+63\right)}{9 (k+5)^4} A_{+} F_{21} F_{22}
\nonu \\
& + &\frac{4 i \left(12 k^3+124 k^2+183 k-369\right)}{9 (k+5)^3}  \pa B_3
  -\frac{64 i (k+3) (k+9)}{9 (k+5)^4}    B_3 F_{11} F_{22}
\nonu \\
& + &
 \frac{8 i \left(5 k^2+42 k+117\right)}{9 (k+5)^4}  B_3 F_{12} F_{21}
- \frac{16 \left(2 k^3+45 k^2+270 k+747\right)}{9 (k+5)^4} \pa F_{12} F_{21}
\nonu \\
& - & \frac{64 i (k+3)}{(k+5)^4}  B_{-} F_{12} F_{22}
- \frac{4 \left(2 k^2+87 k+369\right)}{9 (k+5)^3} B_{+} B_{-}
\nonu \\
&+& \frac{4 (k-3) (11 k+48)}{9 (k+5)^3} F_{11} G_{22}
- \frac{32 \left(4 k^3+29 k^2+132 k+387\right)}{9 (k+5)^4} \pa F_{11} F_{22}
\nonu \\
& + &  \frac{8 \left(4 k^3+53 k^2+294 k+549\right)}{9 (k+5)^4}  F_{11} \pa F_{22}
\nonu \\
& + & \frac{8 \left(16 k^3+145 k^2+534 k+693\right)}{9 (k+5)^4} F_{12} \pa F_{21}
+ \frac{8 (k-3) (k+6)}{3 (k+5)^3} F_{21} G_{12}
\nonu \\
& - & \frac{8 i (k-3)}{(k+5)^3}  U A_3 
- \frac{8 i (k-3) k}{3 (k+5)^3} U B_3 
- \frac{4 \left(20 k^2+177 k+369\right)}{9 (k+5)^3} U U
\nonu \\
&+ & \left. \frac{32 (k-3)^2}{9 (k+5)^4}   U F_{11} F_{22}
- \frac{8 (k-3)}{(k+5)^3}  U F_{12} F_{21}
\right](w)
\nonu \\
& + & \frac{1}{(z-w)^2} \, \left[ \frac{(k-3)}{3 (k+5)} {\bf S^{(3)}} 
+3 {\bf P^{(3)}}
+ \frac{(k-3)}{3 (k+5)}
  \pa 
{\bf P^{(2)}}  - \frac{2 (5 k+21)}{3 (k+5)}  {\bf W^{(3)}}  
\right. \nonu \\
& - &  \frac{4 (k-3)^2}{9 (k+5)^2}  \pa {\bf T^{(2)}} 
-  \frac{4 (k-3)}{3 (k+5)}  \pa {\bf W^{(2)}} 
\nonu \\
& - & \frac{8 i}{(k+5)^2} A_3 A_3 B_3
- \frac{4 \left(20 k^2+123 k+99\right)}{9 (k+5)^3} \pa A_3 A_3
+  \frac{8 i}{(k+5)^2} A_3 B_3 B_3
\nonu \\
& - & \frac{2 \left(9 k^2+86 k+117\right)}{3 (k+5)^3}  \pa A_3 B_3
+  \frac{2 \left(21 k^2+146 k+309\right)}{3 (k+5)^3}  A_3 \pa B_3
\nonu \\
 & + & 
\frac{8 i}{(k+5)^2} A_3 B_{+} B_{-}
- \frac{2 i \left(392 k^3+3697 k^2+5646 k-5211\right)}{9 (k+5)^3 (13 k+17)} 
\pa^2 A_3
\nonu \\
&+& \frac{10 i}{(k+5)^2} A_3 F_{11} G_{22}
+ \frac{i \left(17 k^2+570 k+2169\right)}{9 (k+5)^4} \pa A_3 F_{11} F_{22}
\nonu \\
&-& \frac{i \left(79 k^2+774 k+2151\right)}{9 (k+5)^4}   A_3 \pa F_{11} F_{22} 
+ \frac{i \left(173 k^2+1170 k+1269\right)}{9 (k+5)^4}  A_3 F_{11} \pa F_{22}
\nonu \\
& + & \frac{ 4i(k+9)}{3(k+5)^3} A_3 F_{12} G_{21}
 -   \frac{i \left(97 k^2+1122 k+3249\right)}{9 (k+5)^4}   A_3 F_{12} \pa F_{21}
\nonu \\
&+& \frac{i \left(155 k^2+1494 k+3915\right)}{9 (k+5)^4}  \pa A_3 F_{12} F_{21}
- \frac{i \left(133 k^2+1194 k+2709\right)}{9 (k+5)^4}  A_3 \pa F_{12} F_{21}
\nonu \\
&+& \frac{ 8i(k+3)}{3(k+5)^3} A_3 F_{21} G_{12}
-\frac{2i}{(k+5)^2} A_{3} F_{22} G_{11}
+ \frac{8 i (k+6)}{3 (k+5)^3} A_{-} F_{11} G_{12}
\nonu \\
& + & \frac{4 i \left(11 k^2+120 k+285\right)}{3 (k+5)^4} 
\pa A_{-} F_{11} F_{12}
 -\frac{8i}{(k+5)^2} A_{+} A_{-} B_3
\nonu \\
& - &  \frac{8 i \left(6 k^2+61 k+183\right)}{3 (k+5)^4}  A_{-} \pa F_{11} F_{12}
+  \frac{8 i \left(k^2+k-48\right)}{3 (k+5)^4}  A_{-} F_{11} \pa F_{12}
\nonu \\
&-& \frac{4 \left(10 k^2+39 k-63\right)}{9 (k+5)^3}   \pa A_{+} A_{-}
- \frac{8 (k+3) (5 k+27)}{9 (k+5)^3} A_{+} \pa A_{-}
\nonu \\
&+& \frac{2 i (7 k+39)}{3 (k+5)^3} A_{+} F_{21} G_{22}
+ \frac{8 i \left(5 k^2+78 k+333\right)}{9 (k+5)^4} 
\pa A_{+} F_{21} F_{22}
\nonu \\
&- & \frac{16 i (k+3) (2 k+21)}{9 (k+5)^4} A_{+} \pa F_{21} F_{22} 
+ \frac{16 i (k-6) (k+3)}{9 (k+5)^4}
A_{+} F_{21} \pa F_{22}
\nonu \\
& - & \frac{8 i(k+6) }{3 (k+5)^3} A_{+} F_{22} G_{21}
-  \frac{4 \left(2 k^2+87 k+369\right)}{9 (k+5)^3} \pa B_3 B_3
\nonu \\
& + &  \frac{2 i \left(44 k^4+491 k^3-649 k^2-7743 k-4743\right)}
{9 (k+5)^3 (13 k+17)}
 \pa^2 B_3 
\nonu \\
&-& 
\frac{2i}{(k+5)^2} B_3 F_{11} G_{22}
+ \frac{i \left(7 k^2-90 k-369\right)}{9 (k+5)^4}  \pa B_3 F_{11} F_{22}
\nonu \\
&-& \frac{i \left(17 k^2-150 k-1431\right)}{9 (k+5)^4}   B_3 \pa F_{11} F_{22}
- \frac{i \left(125 k^2+1506 k+4149\right)}{9 (k+5)^4}  B_3  F_{11} \pa F_{22}
\nonu \\
&+& \frac{ 4i(k+9)}{3(k+5)^3} B_3 F_{12} G_{21}
- \frac{i \left(109 k^2+1194 k+2925\right)}{9 (k+5)^4}  \pa B_3 F_{12} F_{21}
\nonu \\
& + & 
\frac{i \left(131 k^2+1494 k+4131\right)}{9 (k+5)^4}  B_3 \pa F_{12} F_{21}
-  
\frac{2 i (k+9)}{3 (k+5)^3} A_{-} F_{12} G_{11}
\nonu \\
&+&  \frac{i \left(167 k^2+1566 k+3591\right)}{9 (k+5)^4}  B_3 F_{12} \pa F_{21}
+ \frac{(5 k+23)}{(k+5)^2} G_{11} G_{22} 
\nonu \\
&+& \frac{8 i(k+3)}{3(k+5)^3} B_3 F_{21} G_{12}
+ \frac{10i}{(k+5)^2} B_{3} F_{22} G_{11}
-\frac{2 i (k+9)}{3 (k+5)^3} B_{-} F_{12} G_{22}
\nonu \\
& + &  \frac{i \left(37 k^2+282 k+645\right)}{3 (k+5)^4} \pa B_{-} F_{12} F_{22}
-\frac{(k-3)}{(k+5)^2} G_{12} G_{21}
\nonu \\
& - & \frac{i \left(15 k^2+238 k+591\right)}{3 (k+5)^4}  B_{-} \pa F_{12} F_{22}
- \frac{i \left(59 k^2+710 k+1851\right)}{3 (k+5)^4} B_{-} F_{12} \pa F_{22}
\nonu \\
& + & \frac{8i(k+6)}{3(k+5)^3} B_{-} F_{22} G_{12}
- \frac{2 (k+9) (11 k+51)}{9 (k+5)^3} \pa B_{+} B_{-}
\nonu \\
 & + & \frac{2 \left(7 k^2-24 k-279\right)}{9 (k+5)^3}   B_{+} \pa B_{-}
-\frac{8 i (k+6)}{3 (k+5)^3}
B_{+} F_{11} G_{21}
\nonu \\
&+& \frac{i \left(5 k^2+258 k+621\right)}{9 (k+5)^4}  \pa B_{+} F_{11} F_{21}
-  \frac{i \left(103 k^2+630 k+927\right)}{9 (k+5)^4}  B_{+} \pa F_{11} F_{21}
\nonu \\
 & - & \frac{i \left(115 k^2+942 k+2187\right)}{9 (k+5)^4}  
B_{+} F_{11} \pa F_{21}
+ \frac{2 i (7 k+39)}{3 (k+5)^3} B_{+} F_{21} G_{11}
\nonu \\
& - & \frac{(608 k^4+657 k^3-10193 k^2+219 k+1053)}{9 (k+5)^4 (13 k+17)}
  \pa^2 F_{11} F_{22} \nonu \\
& - &
\frac{2 \left(88 k^4+7049 k^3+56775 k^2+137475 k+81837\right)}{9 (k+5)^4 (13 k+17)}  \pa F_{11} \pa F_{22}
\nonu \\
&+& \frac{(432 k^4+9491 k^3+64429 k^2+164721 k+141543)}{9 (k+5)^4 (13 k+17)}  
F_{11} \pa^2 F_{22}
\nonu \\
 & + & \frac{(251 k^2+1266 k+423)}{18 (k+5)^3} \pa F_{11} G_{22}
-\frac{(25 k^2+342 k+909)}{18 (k+5)^3}   F_{11} \pa G_{22}
\nonu \\
& - & \frac{(48 k^4+803 k^3+4117 k^2+10577 k+5999)}{(k+5)^4 (13 k+17)}  
\pa^2 F_{12} F_{21}
\nonu \\
&+&  \frac{2 (k-3) \left(88 k^2+89 k-1419\right)}{9 (k+5)^3 (13 k+17)}
\pa F_{12} \pa F_{21}
-\frac{6}{(k+5)^2} U F_{22} G_{11}
\nonu \\
& + & \frac{(608 k^4+6353 k^3+34775 k^2+91563 k+94725)}{9 (k+5)^4 (13 k+17)}
F_{12} \pa^2 F_{21}
\nonu \\
& + & \frac{(9 k^2+114 k+233)}{2 (k+5)^3}   \pa F_{12} G_{21}
- \frac{(9 k^2+122 k+273)}{6 (k+5)^3}
 F_{12} \pa G_{21}  
\nonu \\
&+&  \frac{(13 k^2+81 k+168)}{3 (k+5)^3}  \pa F_{21} G_{12}
+ \frac{(k^2-7 k-132)}{3 (k+5)^3}  F_{21} \pa G_{12}
\nonu \\
& + & \frac{(29 k^2+483 k+1530)}{9 (k+5)^3}  \pa F_{22} G_{11}
+ \frac{(17 k^2-105 k-918)}{9 (k+5)^3}   F_{22} \pa G_{11}
\nonu \\
& - & 
 \frac{4 i \left(20 k^2+231 k+483\right)}{(k+5)^2 (13 k+17)}  T A_3
+ \frac{16 i k \left(5 k^2+57 k+123\right)}{3 (k+5)^2 (13 k+17)}  T B_3
\nonu \\
&-& \frac{4 (k+3) (10 k+81)}{9 (k+5)^2}  \pa T
 + \frac{16 (k+6) (4 k+17)}{3 (k+5)^2 (13 k+17)}  T U
\nonu \\ 
& - &  \frac{8 \left(10 k^3+123 k^2+552 k+903\right)}{3 (k+5)^3 (13 k+17)}   
T F_{11} F_{22}
+ \frac{16 (k+6)}{3 (k+5)^3} U A_{+} A_{-}
\nonu \\
& + & \frac{16 (k-3) \left(5 k^2+45 k+108\right)}{3 (k+5)^3 (13 k+17)} 
T F_{12} F_{21}  + \frac{16 (k+6)}{3 (k+5)^3} U A_3 A_3
\nonu \\
& - & \frac{8(5k+21)}{3(k+5)^3} U A_3 B_3
 - \frac{8 i(3k+8)}{(k+5)^3} \pa U A_3
+ \frac{8 i (8 k+45)}{3 (k+5)^3} U \pa A_3
\nonu \\
& - & \frac{2 i \left(5 k^2+33 k+60\right)}{3 (k+5)^3}   \pa U  B_3
+ \frac{2 i \left(k^2+53 k+108\right)}{3 (k+5)^3}  U \pa B_3
\nonu \\
&+& \frac{16 (k+6)}{3 (k+5)^3} U B_{+} B_{-}
-\frac{8 (k+6) (4 k+17)}{3 (k+5)^2 (13 k+17)} \pa^2 U
\nonu \\
&+& \frac{6}{(k+5)^2} U F_{11} G_{22}
-\frac{(k-3) (29 k+273)}{9 (k+5)^4}  \pa U F_{11} F_{22}
\nonu \\
& + & \frac{(259 k^2+2118 k+5139)}{9 (k+5)^4}  U \pa F_{11} F_{22}
+ \frac{16 (k+6)}{3 (k+5)^3} U B_3 B_3
\nonu \\
&-& \frac{(137 k^2+2130 k+6201)}{9 (k+5)^4}  U F_{11} \pa F_{22}
-\frac{(7k+3)}{ (k+5)^3} \pa U F_{12} F_{21}
\nonu \\
& + &  \frac{(43 k+375)}{3 (k+5)^3}  U \pa F_{12} F_{21}
-  \frac{(49 k+213)}{3 (k+5)^3}  U F_{12} \pa F_{21}
\nonu \\
& + & \left.  
\frac{16 (k+6)}{3(k+5)^3}  U U U
- \frac{4 \left(20 k^2+177 k+369\right)}{9 (k+5)^3}  \pa U U
\right](w) \nonu \\
& + & \frac{1}{(z-w)} \, \left[ + \cdots  \right](w) +\cdots.
\nonu 
\eea
There is no boldface higher spin current in the fourth order pole.

One can see the following OPEs from (\ref{g1122vu3half})
%%%%%%%%%%%%%%%%%%%%%%%%%%%%%%%%%%%%%%%%%%%%%%%%%%%%%%%%
\bea
\left(

\right) F_{11} F_{22} \right](w) 
\nonu \\
& + & \frac{1}{(z-w)} \, \left[ + \cdots \right](w) + \cdots.
\nonu
\eea
There is no boldface higher spin current in the third order pole.

Again the relation (\ref{g1221w2}) determines the following OPEs
%%%%%%%%%%%%%%%%%%%%%%%%%%%%%%%%%%%%%%%%%%%%%%%%%%%%%%%%%%%%%%%%%%%%%%%%%%%%%%
\bea
\left(
%
\right)
\nonu 
\right](w) \nonu \\
& + & \frac{1}{(z-w)} \, \left[ + \cdots \right](w) + \cdots.
\nonu
\eea
There is no boldface higher spin current in the third order pole.

The above same relation gives the other combination of the OPEs
\bea
%%%%%%%%%%%%%%%%%%%%%%%%%%%%%%%%%%%%%%%%%%%%%%%%%%%%%%%%%%%%%%%%%%
\left(
%
\right)
\right](w) \nonu \\
& + &  \frac{1}{(z-w)} \, \left[ + \cdots \right](w) + \cdots.
\nonu 
\eea
There is no boldface higher spin current in the third order pole.

Similarly the following OPEs can be obtained from (\ref{g1221w5half})
%%%%%%%%%%%%%%%%%%%%%%%%%%%%%%%%%%%%%%%%%%%%%%%%%%%%%%%%
\bea
\left(
%
\right) F_{11} F_{22} \right](w) 
\nonu \\
& + & \frac{1}{(z-w)^2} \, \left[ + \cdots \right](w) +
\frac{1}{(z-w)} \, \left[ + \cdots \right](w) + \cdots.
\nonu
\eea
There is no boldface higher spin current in the fourth order pole.

Now one can calculate the following OPE
%%%%%%%%%%%%%%%%%%%%%%%%%%%%%%%%%%%%%%%%%%%%%%%%%%%%%%%%%%%%%%%%%%%%%%%
\bea
{\bf V_{+}^{(2)}}(z) \, {\bf V_{-}^{(2)}}(w) & = & \frac{1}{(z-w)^2} \,
\left[ \frac{2 (k+4)}{(k+5)^2}  {A}_{-} {B}_{+} 
-\frac{2 i}{(k+5)^2} A_{-} F_{22} F_{12}
\right. \nonu \\
&+& \left. \frac{2 i}{(k+5)^2} B_{+} F_{22} F_{21}
- \frac{(k-3)}{(k+5)^2} F_{22} G_{22}
+  \frac{4 (3 k-1)}{(k+5)^3} F_{22} \pa F_{22}
\right](w)
\nonu \\
& + & \frac{1}{(z-w)} \, \left[  
 \frac{2 i}{(k+5)^2}  A_3 F_{22} G_{22} +
 \frac{(2k+9)}{(k+5)^2} \pa A_{-} B_{+}
\right. \nonu \\
&- &  \frac{1}{(k+5)^2}  A_{-} \pa B_{+}
+ \frac{i}{(k+5)^2} A_{-} F_{22} G_{12}
-  \frac{2 i(k+1)}{(k+5)^3} A_{-} \pa F_{22} F_{12}
\nonu \\
&+& \frac{2 i (k+1)}{(k+5)^3} A_{-} F_{22} \pa F_{12}
+ \frac{i}{(k+5)^2} A_{-} F_{12} G_{22}
- \frac{2 i}{(k+5)^2} B_3 F_{22} G_{22}
\nonu \\
& - & \frac{i}{(k+5)^2} B_{+} F_{22} G_{21}
- \frac{2 i}{(k+5)^2} \pa A_{-} F_{22} F_{12}
+\frac{2 i(k+1)}{(k+5)^3} B_{+} \pa F_{22} F_{21}
\nonu \\
&+& \frac{2 i(k+9)}{(k+5)^3} B_{+} F_{22} \pa F_{21}
-\frac{i}{(k+5)^2} B_{+} F_{21} G_{22}
+ \frac{ (k+11)}{(k+5)^2} \pa F_{22} G_{22}
\nonu \\
& - & \frac{(k-1)}{(k+5)^2} F_{22} \pa G_{22}
+ \frac{4 i (k+1)}{(k+5)^3} F_{22} A_3 \pa F_{22}
+ \frac{16 i}{(k+5)^3} F_{22} B_3 \pa F_{22}
\nonu \\
& -  & \left.   \frac{4}{(k+5)^2} F_{22} U \pa F_{22}
+ \frac{8 (k-1)}{(k+5)^3} F_{22} \pa^2 F_{22}
\right](w)
+\cdots.
\nonu 
\eea
The nonlinear terms disappear in the OPE.
The corresponding OPE in the nonlinear version contains those
nonlinear terms.

%%%%%%%%%%%%%%%%%%%%%%%%%%%%%%%%%%%%%%%%%%%%%%%%%%%%%%%%%%%%%%%%%%%%%
\subsection{ The OPEs between the last ${\cal N}=2$ multiplet and itself }
%C%%%%%%%%%%%%%%%%%%%%%%%%%%%%%%%%%%%%%%%%%%%%%%%%%%%%%%%%%%%%%%%%%%%

The result (\ref{g1122vu3half}) in section $3$ determines 
the following OPE
\bea
%%%%%%%%%%%%%%%%%%%%%%%%%%%%%%%%%%%%%%%%%%%%%%%%%%%%%%%%%%%%%%%%%%%%%%
{\bf W^{(2)}}(z) \, {\bf W^{(2)}}(w) 
&=& 
\frac{1}{(z-w)^4} \, \left[ \frac{9 k}{(k+5)}\right]
\nonu \\
& + & \frac{1}{(z-w)^2} \, \left[ 
\frac{2(k+4)}{ (k+5)} T 
+   \frac{2(k+4)}{(k+5)^2} A_3 A_3  +
 \frac{4(k+4)}{(k+5)^2}  A_3 B_3
+ \frac{2 i (k+1)}{(k+5)^2} \pa A_3
\right. \nonu \\
& - & 
\frac{4 i}{ (k+5)^2}  A_3 F_{12} F_{21}
- \frac{2 i}{(k+5)^2}  A_{-} F_{11} F_{12}
+ \frac{2 (k+1)}{(k+5)^2} A_{+} A_{-}
\nonu \\
& - & \frac{2 i}{(k+5)^2} A_{+} F_{21} F_{22}
+  \frac{2 (k+4)}{(k+5)^2} B_3 B_3
+ \frac{8 i}{(k+5)^2}  \pa B_3
\nonu \\
 &  - &  \frac{4 i }{ (k+5)^2}  B_3 F_{12} F_{21}
+  \frac{2 i }{(k+5)^2}  B_{-} F_{12} F_{22}
+ \frac{8}{ (k+5)^2} B_{+} B_{-}
\nonu \\
&+& 
\frac{2}{ (k+5)} \pa F_{11} F_{22}
- \frac{2}{ (k+5)}  F_{11} \pa F_{22}
+ \frac{2 \left(k^2+4 k+27\right)}{(k+5)^3} \pa F_{12} F_{21}
\nonu \\
& - & \frac{2 \left(k^2+4 k+27\right)}{(k+5)^3}  F_{12} \pa F_{21}
- \frac{(k-3)}{ (k+5)^2} F_{21} G_{12}
- \frac{ (k-3) }{ (k+5)^2}  F_{12} G_{21}
\nonu \\
& + & \left.
\frac{2(k+4)}{ (k+5)^2} U U
+ \frac{2 i }{ (k+5)^2}  B_{+} F_{11} F_{21}
\right](w)
+  \frac{1}{(z-w)} \, \frac{1}{2} \pa (\mbox{pole-2})(w) 
+  \cdots.
\nonu 
\eea
There are no boldface higher spin currents in the OPE.

With the results in  (\ref{g1122vu3half})
and (\ref{g1221w2}), the following OPEs can be obtained
\bea
%%%%%%%%%%%%%%%%%%%%%%%%%%%%%%%%%%%%%%%%%%%%%%%%%%%%%%%%%%%%%%%%%%%%%%%%%
{\bf W^{(2)}}(z) \, 
{\bf W_{\pm}^{(\frac{5}{2})}}(w)  
 & = & 
\frac{1}{(z-w)^4} \, \left[\frac{6 k(k-3)}{(k+5)^3} 
\left(
\begin{array}{c}
F_{21}  \\
F_{12}
\end{array}
\right) \right](w)
\nonu \\
& + & 
\frac{1}{(z-w)^3}
\, \left[ 
- \frac{2 \left(4 k^2+39 k+66\right)}{3 (k+5)^2}
\left(
\begin{array}{c}
G_{21}  \\
G_{12}
\end{array}
\right)
 \right.
\nonu \\
&\pm  &  \frac{4}{(k+5)^2}
\left(
\begin{array}{c}
F_{21}  \\
F_{12}
\end{array}
\right) F_{11} F_{22}
\pm
\frac{4 i \left(4 k^2+15 k-33\right)}{3 (k+5)^3}
\left(
\begin{array}{c}
F_{21}  \\
F_{12}
\end{array}
\right)  A_3
\nonu \\
&\mp & \frac{4 i \left(4 k^2+21 k+21\right)}{3 (k+5)^3}
\left(
\begin{array}{c}
F_{11}  \\
F_{22}
\end{array}
\right) A_{\mp}
\pm 
\frac{4 i \left(k^2+15 k+66\right)}{3 (k+5)^3}
\left(
\begin{array}{c}
F_{22}  \\
F_{11}
\end{array}
\right) B_{\mp}
\nonu \\
&\pm & \left. \frac{4 i \left(k^2-k-22\right)}{(k+5)^3}
\left(
\begin{array}{c}
F_{21}  \\
F_{12}
\end{array}
\right) B_3
 -\frac{8 \left(2 k^2+15 k+33\right)}{3 (k+5)^3}   U
\left(
\begin{array}{c}
F_{21}  \\
F_{12}
\end{array}
\right)
  \right](w)
\nonu \\
& + & \frac{1}{(z-w)^2} \, \left[ \mp \frac{(k-3)}{6 (k+5)}
{\bf P_{\pm}^{(\frac{5}{2})}} 
- \frac{(k-3)}{6 (k+5)} {\bf W_{\pm}^{(\frac{5}{2})}}
\right. 
\nonu \\
& \mp & \frac{i (5 k+27)}{3(k+5)^2} A_3 
 \left(
\begin{array}{c}
G_{21}  \\
G_{12}
\end{array}
\right) -
\frac{2}{(k+5)^2} A_3 A_3 
 \left(
\begin{array}{c}
F_{21}  \\
F_{12}
\end{array}
\right)
\nonu \\
&+& 
\frac{2 ( k+9)}{3 (k+5)^3} A_3 B_{\mp}
 \left(
\begin{array}{c}
F_{22}  \\
F_{11}
\end{array}
\right)
-\frac{(k-3)}{3(k+5)^2}
U
 \left(
\begin{array}{c}
G_{21}   \\
G_{12} 
\end{array}
\right)
\nonu \\
&\pm &  \frac{2 i (k-3) (8 k+45)}{9 (k+5)^3} \pa A_3 
\left(
\begin{array}{c}
F_{21}  \\
F_{12}
\end{array}
\right) \pm \frac{8 i (k+3) (2 k+9)}{9 (k+5)^3}
 A_3
\pa
\left(
\begin{array}{c}
F_{21}  \\
F_{12}
\end{array}
\right)
\nonu \\
& \pm &
\frac{i (k+1)}{ (k+5)^2} 
A_{\mp}
\left(
\begin{array}{c}
G_{11}  \\
G_{22}
\end{array}
\right)
+\frac{4 (k+3)}{3 (k+5)^3}
A_{\mp} A_3 
\left(
\begin{array}{c}
F_{11}  \\
F_{22}
\end{array}
\right)
\nonu \\
&-& \frac{4 (k+3)}{3 (k+5)^3} A_{\mp} B_3 
\left(
\begin{array}{c}
F_{11}  \\
F_{22}
\end{array}
\right)
- \frac{2(k-3)}{3(k+5)^3} A_{\mp} B_{\mp} 
\left(
\begin{array}{c}
F_{12}  \\
F_{21}
\end{array}
\right)
\nonu \\
&\mp & \frac{4 i (k+3) (4 k-3)}{9 (k+5)^3}  \pa A_{\mp}
\left(
\begin{array}{c}
F_{11}  \\
F_{22}
\end{array}
\right) \mp
 \frac{4 i (k+3) (4 k+27)}{9 (k+5)^3}  A_{\mp} \pa
\left(
\begin{array}{c}
F_{11}  \\
F_{22}
\end{array}
\right)
\nonu \\
&-& \frac{2 (k+9)}{3 (k+5)^3} A_{\pm} A_{\mp}
\left(
\begin{array}{c}
F_{21}  \\
F_{12}
\end{array}
\right) \mp
\frac{7i (k+3)}{3(k+5)^2} B_3
\left(
\begin{array}{c}
G_{21}  \\
G_{12}
\end{array}
\right)
\nonu \\
&+& \frac{2}{(k+5)^2} B_3 B_3 
\left(
\begin{array}{c}
F_{21}  \\
F_{12}
\end{array}
\right) \pm
 \frac{2 i (k-3) (10 k+39)}{9 (k+5)^3}
\pa B_3 
\left(
\begin{array}{c}
F_{21}  \\
F_{12}
\end{array}
\right)
\nonu \\
&\mp & \frac{4 i (k+9) (k+12)}{9 (k+5)^3} 
B_3 \pa 
\left(
\begin{array}{c}
F_{21}  \\
F_{12}
\end{array}
\right) \pm
\frac{4 i }{ (k+5)^2} 
B_{\mp}
\left(
\begin{array}{c}
G_{22}  \\
G_{11}
\end{array}
\right)
\nonu \\
&- & \frac{2 (k+9)}{3 (k+5)^3} B_{\mp} B_3 
\left(
\begin{array}{c}
F_{22}  \\
F_{11}
\end{array}
\right) \mp
\frac{2 i (k+9) (k-12)}{9 (k+5)^3}
\pa B_{\mp} 
\left(
\begin{array}{c}
F_{22}  \\
F_{11}
\end{array}
\right)
\nonu \\
&\pm & \frac{2 i (k+9) (8 k+15)}{9 (k+5)^3} 
B_{\mp} \pa
\left(
\begin{array}{c}
F_{22}  \\
F_{11}
\end{array}
\right) +
\frac{4 (k+3)}{3 (k+5)^3} 
B_{\pm} B_{\mp}
\left(
\begin{array}{c}
F_{21}  \\
F_{12}
\end{array}
\right)
\nonu \\
&-& \frac{4 (k+3)}{3 (k+5)^3} 
\left(
\begin{array}{c}
F_{11} F_{21} G_{22}  \\
F_{22} F_{12} G_{11}
\end{array}
\right) -
\frac{4 (k+9)^2}{9 (k+5)^4}
\left(
\begin{array}{c}
\pa F_{11} F_{21} F_{22}  \\
\pa F_{22} F_{12} F_{11}
\end{array}
\right)
\nonu \\
&-& \frac{16 (k+3) (k+9)}{9 (k+5)^4} 
\left(
\begin{array}{c}
F_{11} \pa F_{21}  F_{22}  \\
F_{22} \pa F_{12}  F_{11}
\end{array}
\right)
\nonu \\
&-& \frac{16 (k+3)^2}{9 (k+5)^4}
\left(
\begin{array}{c}
F_{11} F_{21} \pa F_{22}  \\
F_{22} F_{12} \pa F_{11}
\end{array}
\right)
\mp \frac{2}{(k+5)^2} F_{12} F_{21}
 \left(
\begin{array}{c}
G_{21}   \\
G_{12} 
\end{array}
\right)
\nonu \\
&+& \frac{2 (k-3) (2 k+9)}{3 (k+5)^3} \pa^2 
 \left(
\begin{array}{c}
F_{21}   \\
F_{12} 
\end{array}
\right) +
\frac{2 (k+9)}{3 (k+5)^3} 
 \left(
\begin{array}{c}
F_{21}  F_{22} G_{11}  \\
F_{12} F_{11} G_{22}
\end{array}
\right)
\nonu \\
&-& \frac{(8 k^2+87 k+171)}{9 (k+5)^2}  \pa  
 \left(
\begin{array}{c}
G_{21}   \\
G_{12} 
\end{array}
\right) +\frac{4 (k-3)}{3(k+5)^2}
T 
 \left(
\begin{array}{c}
F_{21}   \\
F_{12} 
\end{array}
\right) 
\nonu \\
&\mp & \frac{8 i(k+3)}{3(k+5)^3} U A_3 
 \left(
\begin{array}{c}
F_{21}   \\
F_{12} 
\end{array}
\right) \pm
\frac{4 i (k+3)}{3 (k+5)^3} U A_{\mp}
 \left(
\begin{array}{c}
F_{11}   \\
F_{22} 
\end{array}
\right)
\nonu \\
&\mp & \frac{4 i(k+9)}{3(k+5)^3} U B_3
\left(
\begin{array}{c}
F_{21}   \\
F_{12} 
\end{array}
\right) \pm
\frac{2 i (k+9)}{3 (k+5)^3}
U B_{\mp} 
\left(
\begin{array}{c}
F_{22}   \\
F_{11} 
\end{array}
\right)
\nonu \\
&-& \frac{2 \left(8 k^2+87 k+171\right)}{9 (k+5)^3}    \pa U
\left(
\begin{array}{c}
F_{21}   \\
F_{12} 
\end{array}
\right)
- \frac{16 \left(k^2+3 k+18\right)}{9 (k+5)^3}  U
\pa
\left(
\begin{array}{c}
F_{21}   \\
F_{12} 
\end{array}
\right)
\nonu \\
&+& \left.   
\frac{2(k-3)}{3(k+5)^3} U U 
\left(
\begin{array}{c}
F_{21}   \\
F_{12} 
\end{array}
\right) 
\mp 
\frac{4 (k-3)}{3 (k+5)^3} \pa
\left(
\begin{array}{c}
F_{21}   \\
F_{12} 
\end{array}
\right)  F_{12} F_{21}
\right](w)
\nonu \\ 
& + & \frac{1}{(z-w)} \, \left[ 
+ \cdots \right](w) +\cdots.
\nonu
\eea
There is no boldface higher spin current in the third order pole 
of the OPEs.

With the results in  (\ref{g1122vu3half}) and (\ref{g1221w5half}),
 one obtains the following OPE
\bea
%%%%%%%%%%%%%%%%%%%%%%%%%%%%%%%%%%%%%%%%%%%%%%%%%%%%%%%%%%%%%%%%%%%%%%
{\bf W^{(2)}}(z) \, {\bf W^{(3)}}(w) 
&=& 
\frac{1}{(z-w)^4} \, \left[  
- \frac{36 i \left(18 k^2+63 k+41\right)}{(k+5)^2 (13 k+17)}  A_3 
- \frac{12 i k \left(8 k^2+75 k+95\right)}{(k+5)^2 (13 k+17)}  {B}_3   
 \right. \nonu \\
& + & \left. \frac{12 (k-3) \left(8 k^2+45 k+41\right)}{(k+5)^3 (13 k+17)}      
F_{11} F_{22}
- \frac{12 \left(8 k^3+51 k^2+182 k+123\right)}{(k+5)^3 (13 k+17)}  F_{12} F_{21}
\right](w)
\nonu \\
& + & \frac{1}{(z-w)^3} \, \left[
 \frac{8 i \left(7 k^2+44 k+141\right)}{3 (k+5)^3 (13 k+17)}  
A_{-} F_{11} F_{12}
\right. \nonu \\
& - & \frac{4 i \left(25 k^2+158 k-27\right)}{3 (k+5)^3 (13 k+17)}  
B_{-} F_{12} F_{22}
+ \frac{4 i \left(25 k^2+158 k-27\right)}{3 (k+5)^3 (13 k+17)}  
B_{+} F_{11} F_{21}
\nonu \\
&-& \frac{ (3k+11)}{ (k+5)^2} F_{11} G_{22}
- \frac{4 (k-3) (11 k+103)}{3 (k+5)^3 (13 k+17)} \pa (F_{11} F_{22})
\nonu \\
& + &  \frac{4}{(k+5)^2}  \pa (F_{12} F_{21})
- \frac{8 i \left(7 k^2+44 k+141\right)}{3 (k+5)^3 (13 k+17)}  A_{+} F_{21} F_{22}
\nonu \\
& + & 
 \frac{(k-3) (89 k+205)}{3 (k+5)^2 (13 k+17)}  F_{12} G_{21}
-\frac{(k-3) (89 k+205)}{3 (k+5)^2 (13 k+17)}  F_{21} G_{12}
\nonu \\
& + & \frac{24 i }{(k+5)^2}  U A_3 
- \frac{8 i  k}{ (k+5)^2} U B_3 
+ \frac{(3 k+11)}{(k+5)^2}  F_{22} G_{11}
\nonu \\
&- & \left. \frac{4}{ (k+5)^2}   U F_{11} F_{22}
+ \frac{4 (k-3) (11 k+103)}{3 (k+5)^3 (13 k+17)}  U F_{12} F_{21}
\right](w)
\nonu \\
& + & \frac{1}{(z-w)^2} \, \left[ -3  {\bf S^{(3)}} 
 + 3  {\bf W^{(3)}}  
 -\frac{8 {\bf (k-3)}}{(13 k+17)} {\bf T^{(1)}} {\bf W^{(2)}} 
+ \frac{2 i}{(k+5)^2} A_3 A_3 A_3
\right. \nonu \\
& + & \frac{6 i}{(k+5)^2} A_3 A_3 B_3
- \frac{2 }{(k+5)^2} \pa A_3 A_3
+  \frac{6 i}{(k+5)^2} A_3 B_3 B_3
\nonu \\
& + & \frac{3(3k+11)}{ (k+5)^2}  \pa A_3 B_3
 -\frac{9}{(k+5)}  A_3 \pa B_3
 +  \frac{ 8i}{(k+5)^2} A_3 F_{12} G_{21}
\nonu \\
 & + & 
\frac{6 i}{(k+5)^2} A_3 B_{+} B_{-}
+ \frac{i \left(44 k^2+173 k+341\right)}{(k+5)^2 (13 k+17)}
\pa^2 A_3
\nonu \\
&+& 
\frac{15i }{ (k+5)^2} \pa A_3 F_{11} F_{22}
- \frac{15 i}{(k+5)^2}   A_3 \pa (F_{11} F_{22}) 
\nonu \\
&+& \frac{24 i}{(k+5)^3}  \pa A_3 F_{12} F_{21}
- \frac{2 i (5 k+29)}{(k+5)^3}  A_3 \pa F_{12} F_{21}
+ \frac{10 i (k+1)}{(k+5)^3}   A_3 F_{12} \pa F_{21}
\nonu \\
&-& \frac{ 4i}{(k+5)^2} A_3 F_{21} G_{12}
+ \frac{ i }{ (k+5)^2} A_{-} F_{11} G_{12}
+ \frac{14 }{ (k+5)^2}   \pa A_{+} A_{-}
\nonu \\
& + & \frac{i \left(431 k^2+3082 k+3675\right)}{3 (k+5)^3 (13 k+17)}
\pa A_{-} F_{11} F_{12}
- \frac{14}{ (k+5)^2} A_{+} \pa A_{-}
\nonu \\
& - &  \frac{i \left(115 k^2+1298 k+1119\right)}{3 (k+5)^3 (13 k+17)}  
A_{-} \pa F_{11} F_{12}
- \frac{i \left(739 k^2+4610 k+4383\right)}{3 (k+5)^3 (13 k+17)}  
A_{-} F_{11} \pa F_{12}
\nonu \\
& - & 
\frac{3 i }{ (k+5)^2} A_{-} F_{12} G_{11}
 +\frac{2i}{(k+5)^2} A_{+} A_{-} A_3
+\frac{6i}{(k+5)^2} A_{+} A_{-} B_3
\nonu \\
&+& \frac{3 i }{ (k+5)^2} A_{+} F_{21} G_{22}
+ \frac{2 i \left(77 k^2+772 k+687\right)}{3 (k+5)^3 (13 k+17)}
\pa A_{+} F_{21} F_{22}
\nonu \\
&- & 
\frac{4 i \left(59 k^2+463 k+600\right)}{3 (k+5)^3 (13 k+17)}
A_{+} F_{21} \pa F_{22}
-  \frac{5 i }{ (k+5)^2} A_{+} F_{22} G_{21}
\nonu \\
& - &  \frac{2}{ (k+5)^2} \pa B_3 B_3
+   \frac{i \left(8 k^3+63 k^2+203 k-136\right)}{(k+5)^2 (13 k+17)}
 \pa^2 B_3 
\nonu \\
&+& 
\frac{2i}{(k+5)^2} B_3 B_{3} B_{3}
+ \frac{15i}{(k+5)^2}  \pa B_3 F_{11} F_{22}
- \frac{15i }{ (k+5)^2}   B_3 \pa (F_{11} F_{22})
\nonu \\
&+& \frac{ 4i}{(k+5)^2} B_3 F_{12} G_{21}
- \frac{6i (k+1)}{ (k+5)^3}  \pa B_3 F_{12} F_{21}
-  
\frac{40 i }{ (k+5)^3}  B_3 \pa F_{12} F_{21}
\nonu \\
&+&  \frac{4i (3k+13 )}{ (k+5)^3}  B_3 F_{12} \pa F_{21}
- \frac{8 i}{(k+5)^2} B_3 F_{21} G_{12}
-\frac{3 i }{ (k+5)^2} B_{-} F_{12} G_{22}
\nonu \\
& + &  \frac{i \left(275 k^2+1162 k+1431\right)}{3 (k+5)^3 (13 k+17)} 
\pa B_{-} F_{12} F_{22}
  -  \frac{(3 k+5)}{(k+5)^2}    B_{+} \pa B_{-}
\nonu \\
& - & \frac{i \left(349 k^2+1214 k+609\right)}{3 (k+5)^3 (13 k+17)}  
B_{-} \pa F_{12} F_{22}
-\frac{ i }{ (k+5)^2}
B_{+} F_{11} G_{21}
\nonu \\
& + & \frac{5i}{(k+5)^2} B_{-} F_{22} G_{12}
+ \frac{2 i}{(k+5)^2} B_{+} B_{-} B_3 
+ \frac{ (3k+5)}{ (k+5)^2} \pa B_{+} B_{-}
\nonu \\
&+& \frac{4 i \left(136 k^2+767 k+675\right)}{3 (k+5)^3 (13 k+17)}  
\pa B_{+} F_{11} F_{21}
-  \frac{4 i \left(59 k^2+229 k+294\right)}{3 (k+5)^3 (13 k+17)}  
B_{+} \pa F_{11} F_{21}
\nonu \\
 & + &
 \frac{3 i }{ (k+5)^2} B_{+} F_{21} G_{11}
- \frac{i \left(349 k^2+1838 k+1425\right)}{3 (k+5)^3 (13 k+17)} 
B_{-} F_{12} \pa F_{22}
\nonu \\
& - & \frac{2 \left(4 k^3+48 k^2+115k-269\right)}{ (k+5)^3 (13 k+17)}
  \pa^2 F_{11} F_{22}
  +   \frac{3 (3 k+13)}{2 (k+5)^2}  \pa F_{11} G_{22}
 \nonu \\
&-& \frac{2 \left(4 k^3+100 k^2+365k-31\right)}{ (k+5)^3 (13 k+17)}    
F_{11} \pa^2 F_{22}
-  \frac{(5 k+23)}{2 (k+5)^2}  F_{11} \pa G_{22}
\nonu \\
& + & \frac{(8 k^3+521 k^2+1810 k+1633)}{(k+5)^3 (13 k+17)}
\pa^2 F_{12} F_{21}
 +  \frac{3 (3 k+11)}{2 (k+5)^2}  \pa F_{22} G_{11}
\nonu \\
&+&  \frac{2 \left(8 k^3-545 k^2-1326 k-645\right)}{(k+5)^3 (13 k+17)}
\pa F_{12} \pa F_{21}
-\frac{(k+1)}{2 (k+5)^2}   F_{22} \pa G_{11}
\nonu \\
& + & \frac{(8 k^3+521 k^2+1810 k+1633)}{(k+5)^3 (13 k+17)}
F_{12} \pa^2 F_{21}
\nonu \\
& + & \frac{(467 k^2+1009 k+330)}{3 (k+5)^2 (13 k+17)}   \pa F_{12} G_{21}
- \frac{(118 k^2+341 k+435)}{3 (k+5)^2 (13 k+17)}
 F_{12} \pa G_{21}  
\nonu \\
&+&  \frac{7 \left(17 k^2+460 k+627\right)}{6 (k+5)^2 (13 k+17)}  
\pa F_{21} G_{12}
- \frac{(115 k^2+1064 k+813)}{6 (k+5)^2 (13 k+17)}    F_{21} \pa G_{12}
\nonu \\
&+& 
\frac{3}{(k+5)} G_{12} G_{21}
+\frac{32(49k^2+374k+453)}{3(k+5)^4(13k+17)} F_{11} F_{12} \pa F_{21} F_{22}
\nonu \\
& - & 
\frac{2 i (44 k+109)}{(k+5) (13 k+17)}   T A_3
-  \frac{2 i \left(8 k^2+62 k-17\right)}{(k+5) (13 k+17)} T B_3
\nonu \\
&-& \frac{3}{ (k+5)}  \pa T
 + \frac{6 (4 k+17)}{(k+5) (13 k+17)}  T U
+  \frac{4 (k-3) (4 k+21)}{(k+5)^2 (13 k+17)}
T F_{11} F_{22}
\nonu \\
& - &  \frac{16 \left(k^2+13 k+20\right)}{(k+5)^2 (13 k+17)}
T F_{12} F_{21}  + \frac{6}{ (k+5)^2} U A_3 A_3
\nonu \\
& + & \frac{12}{(k+5)^2} U A_3 B_3
 - \frac{9 i}{(k+5)^2} \pa U A_3
+ \frac{27 i}{ (k+5)^2} U \pa A_3
\nonu \\
&+& \frac{6}{ (k+5)^2} U A_{+} A_{-}
+ \frac{6}{ (k+5)^2} U B_3 B_3
 +  \frac{ i (k+6)}{ (k+5)^2}   \pa U  B_3
\nonu \\
& - & \frac{5 i k}{ (k+5)^2}  U \pa B_3
+ \frac{6}{ (k+5)^2} U B_{+} B_{-}
- \frac{3 (4 k+17)}{(k+5) (13 k+17)} \pa^2 U
\nonu \\
&-& 
\frac{5}{ (k+5)^2}  \pa U F_{11} F_{22}
+ \frac{21}{ (k+5)^2}  U \pa F_{11} F_{22}
+ \frac{6}{(k+5)^2} U U U
\nonu \\
&-& \frac{15}{ (k+5)^2}  U F_{11} \pa F_{22}
+ \frac{4}{(k+5)^2} U F_{12} G_{21}
-  \frac{2 (k-3) (41 k-35)}{3 (k+5)^3 (13 k+17)} \pa U F_{12} F_{21}
\nonu \\
& + &  \frac{64 \left(17 k^2+91 k+78\right)}{3 (k+5)^3 (13 k+17)}  
U \pa F_{12} F_{21}
-  \frac{16 \left(49 k^2+374 k+453\right)}{3 (k+5)^3 (13 k+17)}  
U F_{12} \pa F_{21}
\nonu \\
& - & \left.  
\frac{4}{(k+5)^2} U F_{21} G_{12}
+ \frac{2i}{(k+5)^2}  U U A_3
+ \frac{2 i}{(k+5)^2}  U U B_3
\right](w) \nonu \\
& + & \frac{1}{(z-w)} \, \left[ + \cdots \right](w) +\cdots.
\nonu 
\eea
There are no boldface higher spin currents in the fourth and third 
order poles in the OPE.
The nonlinear term containing the boldface higher spin currents 
in the second order pole can be removed by redefining the higher 
spin-$3$ current as a quasi primary field.

Moreover, the following OPEs can be obtained as before
%%%%%%%%%%%%%%%%%%%%%%%%%%%%%%%%%%%%%%%%%%%%%%%%%%%%%%%%%%%%%%%%%%
\bea
{\bf W_{\pm}^{(\frac{5}{2})}}(z) \, 
{\bf W_{\pm}^{(\frac{5}{2})}}(w)  & = & \frac{1}{(z-w)^3} \, \left[
-\frac{8 \left(8 k^2+51 k+135\right)}{9 (k+5)^3}  A_{\mp} B_{\mp}  
 \mp \frac{32 i (k+3) (k+9)}{9 (k+5)^4} A_{\mp} 
\left(
\begin{array}{c}
F_{11} F_{21} \nonu \\
F_{22} F_{12} 
\end{array}
\right) \right. \nonu \\
& \pm & \frac{32 i (k+3) (k+9)}{9 (k+5)^4}  B_{\mp}
 \left(
\begin{array}{c}
F_{21} F_{22} \nonu \\
F_{12} F_{11} 
\end{array}
\right)  
+ \frac{16 \left(31 k^2+138 k+27\right)}{9 (k+5)^4} 
\left(
\begin{array}{c}
F_{21} \pa F_{21} \nonu \\
F_{12} \pa F_{12} 
\end{array}
\right) \nonu \\
&-& \left.   \frac{8 (k-3) (2 k+9)}{3 (k+5)^3}
\left(
\begin{array}{c}
F_{21} G_{21} \nonu \\
F_{12} G_{12} 
\end{array}
\right) 
\right](w) \nonu \\
& + & \frac{1}{(z-w)^2} \, \frac{1}{2} \pa (\mbox{pole-3})(w) 
+ \frac{1}{(z-w)} \, \left[ + \cdots \right](w) + \cdots.
\nonu 
\eea
The first order pole can be determined by using the general procedure in section
$7$.

Again the different combination of OPE can be obtained 
\bea
%%%%%%%%%%%%%%%%%%%%%%%%%%%%%%%%%%%%%%%%%%%%%%%%%%%%%%%%%%%%%%%%%%%%%%
{\bf W_{+}^{(\frac{5}{2})}}(z) \, {\bf W_{-}^{(\frac{5}{2})}}(w) 
&=& 
\frac{1}{(z-w)^5} \, \left[ -\frac{32 k(k+3) (k+9)}{(k+5)^3}\right]
\nonu \\
&+& 
\frac{1}{(z-w)^4} \, \left[  \frac{8 i (k+9) (4 k+9)}{(k+5)^3}
 A_3 
+ \frac{16 i k (k+3) (2 k+15)}{3 (k+5)^3}  {B}_3   
 \right. \nonu \\
& - & \left.  
 \frac{8 (k-3) \left(4 k^2+33 k+81\right)}{3 (k+5)^4} F_{11} F_{22}
+ \frac{8 \left(4 k^3+27 k^2+90 k+243\right)}{3 (k+5)^4}  F_{12} F_{21}
\right](w)
\nonu \\
& + & \frac{1}{(z-w)^3} \, \left[ \frac{2 (k-3)}{3 (k+5)} {\bf P^{(2)}} - 
\frac{8 (k-3)^2}{9 (k+5)^2}  {\bf T^{(2)}} 
- \frac{8 (k-3)}{3 (k+5)}  {\bf W^{(2)}}
\right. \nonu \\ 
& - &  
\frac{4 \left(20 k^2+204 k+369\right)}{9 (k+5)^2} T
- \frac{4 \left(20 k^2+123 k+99\right)}{9 (k+5)^3}  A_{+} A_{-}
 \nonu \\
& - &  \frac{4 \left(20 k^2+123 k+99\right)}{9 (k+5)^3} A_3 A_3  -
\frac{8 \left(7 k^2+57 k+126\right)}{9 (k+5)^3}  A_3 B_3
\nonu \\
&+&  \frac{8 i (k+15) (8 k+21)}{9 (k+5)^3} \pa A_3
+   \frac{16 (k-3) (k+6)}{9 (k+5)^3}  F_{22} G_{11}
\nonu \\
& + & \frac{32 i (k+3) (k+9)}{9 (k+5)^4}
 A_3 F_{11} F_{22}
+ \frac{8 i }{ (k+5)^2}  A_3 F_{12} F_{21}
\nonu \\
&+& \frac{8 i (k+1) (k+9)}{(k+5)^4}  A_{-} F_{11} F_{12}
+ \frac{32 i \left(k^2+12 k+63\right)}{9 (k+5)^4} A_{+} F_{21} F_{22}
\nonu \\
& - & \frac{4 \left(2 k^2+87 k+369\right)}{9 (k+5)^3} B_3 B_3
+ \frac{8 i}{ (k+5)^2}  B_3 F_{12} F_{21}
\nonu \\
& + &\frac{4 i \left(12 k^3+124 k^2+183 k-369\right)}{9 (k+5)^3}  \pa B_3
  -\frac{32 i (k+3) (k+9)}{9 (k+5)^4}    B_3 F_{11} F_{22}
\nonu \\
& - & \frac{64 i (k+3)}{(k+5)^4}  B_{-} F_{12} F_{22}
- \frac{4 \left(2 k^2+87 k+369\right)}{9 (k+5)^3} B_{+} B_{-}
\nonu \\
& - & 
\frac{8 i \left(13 k^2+66 k+117\right)}{9 (k+5)^4}  B_{+} F_{11} F_{21}
\nonu \\
&+& \frac{4 (k-3) (5 k+21)}{9 (k+5)^3} F_{11} G_{22}
- \frac{16 \left(8 k^3+69 k^2+252 k+423\right)}{9 (k+5)^4} \pa F_{11} F_{22}
\nonu \\
& + &  \frac{8 \left(4 k^3+93 k^2+594 k+1305\right)}{9 (k+5)^4}  F_{11} \pa F_{22}
+ \frac{4 (k-3) (2 k+9)}{3 (k+5)^3} F_{12} G_{21}
\nonu \\
& - &  \frac{16 \left(2 k^3+25 k^2+120 k+369\right)}{9 (k+5)^4} \pa F_{12} F_{21}
\nonu \\
& + & \frac{8 \left(16 k^3+123 k^2+558 k+1395\right)}{9 (k+5)^4}  
F_{12} \pa F_{21}
+ \frac{4 (k-3) (4k+21)}{3 (k+5)^3} F_{21} G_{12}
\nonu \\
& - & \frac{8 i (k-3)}{(k+5)^3}  U A_3 
- \frac{8 i (k-3) k}{3 (k+5)^3} U B_3 
- \frac{4 \left(20 k^2+177 k+369\right)}{9 (k+5)^3} U U
\nonu \\
&+ & \left. \frac{32 (k-3)^2}{9 (k+5)^4}   U F_{11} F_{22}
- \frac{8 (k-3)}{(k+5)^3}  U F_{12} F_{21}
\right](w)
\nonu \\
& + & \frac{1}{(z-w)^2} \, \left[ 3 {\bf S^{(3)}} 
+ \frac{(k-3)}{3 (k+5)} {\bf P^{(3)}}
+ \frac{(k-3)}{3 (k+5)}
  \pa 
{\bf P^{(2)}}  - \frac{2 (5 k+21)}{3 (k+5)}  {\bf W^{(3)}}  
\right. \nonu \\
& - &  \frac{4 (k-3)^2}{9 (k+5)^2}  \pa {\bf T^{(2)}} 
-  \frac{4 (k-3)}{3 (k+5)}  \pa {\bf W^{(2)}} 
\nonu \\
& - & \frac{8 i}{(k+5)^2} A_3 A_3 B_3
- \frac{4 \left(20 k^2+123 k+99\right)}{9 (k+5)^3} \pa A_3 A_3
-  \frac{8 i}{(k+5)^2} A_3 B_3 B_3
\nonu \\
& - & \frac{2 \left(59 k^2+462 k+891\right)}{9 (k+5)^3}  \pa A_3 B_3
+  \frac{2 \left(31 k^2+306 k+747\right)}{9 (k+5)^3}  A_3 \pa B_3
\nonu \\
 & - & 
\frac{8 i}{(k+5)^2} A_3 B_{+} B_{-}
- \frac{2 i \left(128 k^3+181 k^2+1110 k+8577\right)}{9 (k+5)^3 (13 k+17)}
\pa^2 A_3
\nonu \\
&-& \frac{4 i(k+9)}{3(k+5)^3} A_3 F_{11} G_{22}
- \frac{i \left(119 k^2+1134 k+3015\right)}{9 (k+5)^4} \pa A_3 F_{11} F_{22}
\nonu \\
&+& \frac{i \left(169 k^2+1554 k+3609\right)}{9 (k+5)^4}   
A_3 \pa F_{11} F_{22} 
\nonu \\
& - & \frac{ 10 i}{(k+5)^2} A_3 F_{12} G_{21}
+ \frac{i \left(133 k^2+1482 k+4149\right)}{9 (k+5)^4}  A_3 F_{11} \pa F_{22}
\nonu \\
&+& \frac{i (17 k-27)}{3 (k+5)^3}  \pa A_3 F_{12} F_{21}
+  \frac{i (49 k+261)}{3 (k+5)^3}  A_3 \pa F_{12} F_{21}
\nonu \\
& - &  \frac{i (35 k-33)}{3 (k+5)^3}   A_3 F_{12} \pa F_{21}
-   \frac{4 i \left(5 k^2+60 k+231\right)}{3 (k+5)^4}
\pa A_{-} F_{11} F_{12}
\nonu \\
&+& \frac{ 2 i}{(k+5)^2} A_3 F_{21} G_{12}
-\frac{8i(k+3)}{3(k+5)^3} A_{3} F_{22} G_{11}
- \frac{2 i (k+9)}{3 (k+5)^3} A_{-} F_{11} G_{12}
\nonu \\
& + & \frac{8 i \left(2 k^2+29 k+75\right)}{3 (k+5)^4}   
A_{-} \pa F_{11} F_{12}
+ \frac{8 i \left(9 k^2+91 k+210\right)}{3 (k+5)^4}  A_{-} F_{11} \pa F_{12}
\nonu \\
& + & 
\frac{8 i (k+6)}{3 (k+5)^3} A_{-} F_{12} G_{11}
 -\frac{8i}{(k+5)^2} A_{+} A_{-} B_3
\nonu \\
&-&  \frac{8 (k+3) (5 k+27)}{9 (k+5)^3}  \pa A_{+} A_{-}
- \frac{4 \left(10 k^2+39 k-63\right)}{9 (k+5)^3} A_{+} \pa A_{-}
\nonu \\
&-& \frac{8 i (k+6)}{3 (k+5)^3} A_{+} F_{21} G_{22}
- \frac{8 i (k+3) (k+27)}{9 (k+5)^4}
\pa A_{+} F_{21} F_{22}
\nonu \\
&+ & \frac{16 i \left(k^2+27 k+144\right)}{9 (k+5)^4}  A_{+} \pa F_{21} F_{22} 
+ \frac{16 i \left(4 k^2+51 k+189\right)}{9 (k+5)^4}   A_{+} F_{21} \pa F_{22}
\nonu \\
& + & \frac{2 i(7k+39) }{3 (k+5)^3} A_{+} F_{22} G_{21}
-  \frac{4 \left(2 k^2+87 k+369\right)}{9 (k+5)^3} \pa B_3 B_3
\nonu \\
& + &  \frac{2 i \left(44 k^4+491 k^3-649 k^2-7743 k-4743\right)}
{9 (k+5)^3 (13 k+17)}
 \pa^2 B_3 
\nonu \\
&+& 
\frac{4i(k+9)}{3(k+5)^3} B_3 F_{11} G_{22}
-\frac{i \left(145 k^2+1554 k+3825\right)}{9 (k+5)^4}   \pa B_3 F_{11} F_{22}
\nonu \\
&+&  \frac{i \left(95 k^2+1134 k+3231\right)}{9 (k+5)^4}   B_3 \pa F_{11} F_{22}
+ \frac{i \left(131 k^2+1206 k+2691\right)}{9 (k+5)^4}  B_3  F_{11} \pa F_{22}
\nonu \\
&-& \frac{ 2i}{(k+5)^2} B_3 F_{12} G_{21}
+\frac{i (25 k+93)}{3 (k+5)^3}   \pa B_3 F_{12} F_{21}
\nonu \\
& + & \frac{i (17 k+213)}{3 (k+5)^3}
  B_3 \pa F_{12} F_{21}
-   \frac{i (19 k+159)}{3 (k+5)^3}    B_3 F_{12} \pa F_{21}
\nonu \\
&+& \frac{10 i}{(k+5)^2} B_3 F_{21} G_{12}
+ \frac{8i(k+3)}{3(k+5)^3} B_{3} F_{22} G_{11}
+\frac{8 i (k+6)}{3 (k+5)^3} B_{-} F_{12} G_{22}
\nonu \\
& - &  \frac{i \left(19 k^2+270 k+651\right)}{3 (k+5)^4}  \pa B_{-} F_{12} F_{22}
\nonu \\
& + &  \frac{i \left(17 k^2+26 k-135\right)}{3 (k+5)^4}  B_{-} \pa F_{12} F_{22}
+ \frac{i \left(21 k^2+130 k+285\right)}{3 (k+5)^4}   B_{-} F_{12} \pa F_{22}
\nonu \\
& - & \frac{2i(7k+39)}{3(k+5)^3} B_{-} F_{22} G_{12}
- \frac{2 (k+9) (11 k+51)}{9 (k+5)^3} \pa B_{+} B_{-}
\nonu \\
 & + & \frac{2 \left(7 k^2-24 k-279\right)}{9 (k+5)^3}   B_{+} \pa B_{-}
+\frac{2 i (k+9)}{3 (k+5)^3}
B_{+} F_{11} G_{21}
\nonu \\
&- & \frac{i \left(163 k^2+1398 k+3267\right)}{9 (k+5)^4}  
\pa B_{+} F_{11} F_{21}
- \frac{i \left(7 k^2-162 k-441\right)}{9 (k+5)^4}   B_{+} \pa F_{11} F_{21}
\nonu \\
 & + &  \frac{i \left(125 k^2+1578 k+4221\right)}{9 (k+5)^4}  
B_{+} F_{11} \pa F_{21}
- \frac{8 i (k+6)}{3 (k+5)^3} B_{+} F_{21} G_{11}
\nonu \\
& - &  \frac{(608 k^4+5989 k^3+32739 k^2+97479 k+105129)}{9 (k+5)^4 (13 k+17)}
  \pa^2 F_{11} F_{22} \nonu \\
& - & \frac{2 (k-3) \left(88 k^3+269 k^2-4278 k-10971\right)}
{9 (k+5)^4 (13 k+17)}
  \pa F_{11} \pa F_{22}
\nonu \\
&+&  \frac{(144 k^4+2461 k^3+13355 k^2+29679 k+13713)}{3 (k+5)^4 (13 k+17)}
F_{11} \pa^2 F_{22}
\nonu \\
 & - & \frac{(61 k^2+1002 k+2349)}{18 (k+5)^3} \pa F_{11} G_{22}
+ \frac{(47 k^2+390 k+567)}{18 (k+5)^3}   F_{11} \pa G_{22}
\nonu \\
& - & \frac{(432 k^4+9335 k^3+61417 k^2+170877 k+154395)}{9 (k+5)^4 (13 k+17)}
\pa^2 F_{12} F_{21}
\nonu \\
&+&  \frac{2 \left(88 k^4+7309 k^3+59299 k^2+131439 k+70209\right)}
{9 (k+5)^4 (13 k+17)}
\pa F_{12} \pa F_{21}
\nonu \\
& + & \frac{(608 k^4+1021 k^3-8157 k^2-5697 k-9351)}{9 (k+5)^4 (13 k+17)}
F_{12} \pa^2 F_{21}
\nonu \\
& - &  \frac{(61 k^2+390 k+441)}{6 (k+5)^3}   \pa F_{12} G_{21}
+ \frac{(31 k^2+146 k+3)}{6 (k+5)^3}
 F_{12} \pa G_{21}  
\nonu \\
&+& \frac{(5 k^2-129 k-738)}{3 (k+5)^3}   \pa F_{21} G_{12}
+ \frac{(k+6) (9 k+13)}{3 (k+5)^3}  F_{21} \pa G_{12}
\nonu \\
& - & \frac{(19 k^2+183 k+864)}{9 (k+5)^3}  \pa F_{22} G_{11}
+ \frac{(17 k^2+81 k+36)}{9 (k+5)^3}   F_{22} \pa G_{11}
\nonu \\
& + & 
 \frac{4 i \left(20 k^2+231 k+483\right)}{(k+5)^2 (13 k+17)}  T A_3
+ \frac{16 i k \left(5 k^2+57 k+123\right)}{3 (k+5)^2 (13 k+17)}  T B_3
\nonu \\
&-& \frac{4 \left(10 k^2+93 k+126\right)}{9 (k+5)^2}  \pa T
 - \frac{16 (k+6) (4 k+17)}{3 (k+5)^2 (13 k+17)}  T U
\nonu \\ 
& - &   \frac{16 (k-3) \left(5 k^2+45 k+108\right)}{3 (k+5)^3 (13 k+17)}  
T F_{11} F_{22}
-\frac{(5k+23)}{(k+5)^2} G_{12} G_{21}
\nonu \\
& + & \frac{8 \left(10 k^3+123 k^2+552 k+903\right)}{3 (k+5)^3 (13 k+17)}
T F_{12} F_{21}  - \frac{16 (k+6)}{3 (k+5)^3} U A_3 A_3
\nonu \\
& - & \frac{8(5k+21)}{3(k+5)^3} U A_3 B_3
 + \frac{8 i(2k+11)}{(k+5)^3} \pa U A_3
- \frac{8 i (11 k+36)}{3 (k+5)^3} U \pa A_3
\nonu \\
&-& \frac{16 (k+6)}{3 (k+5)^3} U A_{+} A_{-}
- \frac{16 (k+6)}{3 (k+5)^3} U B_3 B_3
+ \frac{ (k-3)}{(k+5)^2} G_{11} G_{22} 
\nonu \\
& - & \frac{2 i \left(5 k^2+33 k+60\right)}{3 (k+5)^3}   \pa U  B_3
+ \frac{2 i \left(k^2+37 k+12\right)}{3 (k+5)^3}  U \pa B_3
\nonu \\
&-& \frac{16 (k+6)}{3 (k+5)^3} U B_{+} B_{-}
+\frac{8 (k+6) (4 k+17)}{3 (k+5)^2 (13 k+17)} \pa^2 U
-\frac{6}{(k+5)^2} U F_{12} G_{21}
\nonu \\
&+& \frac{(43 k^2+174 k+819)}{9 (k+5)^4}
  \pa U F_{11} F_{22}
- \frac{(149 k^2+1938 k+4941)}{9 (k+5)^4}  U \pa F_{11} F_{22}
\nonu \\ \
&+& \frac{(127 k^2+1206 k+3879)}{9 (k+5)^4}   U F_{11} \pa F_{22}
+\frac{(k-3)}{ (k+5)^3} \pa U F_{12} F_{21}
\nonu \\
& - &  \frac{(31 k+99)}{ (k+5)^3}  U \pa F_{12} F_{21}
+  \frac{(13 k+153)}{ (k+5)^3}  U F_{12} \pa F_{21}
\nonu \\
& + & \left.  
\frac{6}{(k+5)^2} U F_{21} G_{12}
- \frac{16 (k+6)}{3(k+5)^3}  U U U
- \frac{4 \left(20 k^2+177 k+369\right)}{9 (k+5)^3}  \pa U U
\right](w) \nonu \\
& + & \frac{1}{(z-w)} \, \left[ + \cdots \right](w) +\cdots.
\nonu 
\eea
One does not see any higher spin current in the fourth order pole.

One describes the following OPEs
\bea
%%%%%%%%%%%%%%%%%%%%%%%%%%%%%%%%%%%%%%%%%%%%%%%%%%%%%%%%%%%%%%%%%%%%%%%%%
{\bf W_{\pm}^{(\frac{5}{2})}}(z)
\, {\bf W^{(3)}}(w)    
 & = & 
\frac{1}{(z-w)^5} \, \left[ \pm \frac{24 (k-3) k}{(k+5)^2 (13 k+17)} 
\left(
\begin{array}{c}
F_{21}  \\
F_{12}
\end{array}
\right) \right](w)
\nonu \\
& + & 
\frac{1}{(z-w)^4}
\, \left[ \mp \frac{2 \left(248 k^3+2770 k^2+7275 k+4509\right)}
{3 (k+5)^2 (13 k+17)}  
\left(
\begin{array}{c}
G_{21}  \\
G_{12}
\end{array}
\right)
\right. \nonu \\
& \mp & 
\frac{4 \left(248 k^3+1717 k^2+5898 k+4509\right)}{3 (k+5)^3 (13 k+17)}
U
\left(
\begin{array}{c}
F_{21}  \\
F_{12}
\end{array}
\right) 
\nonu \\
& +  &  \frac{8 (5 k+9) \left(11 k^2+118 k+219\right)}{3 (k+5)^4 (13 k+17)}
\left(
\begin{array}{c}
F_{21}  \\
F_{12}
\end{array}
\right) F_{11} F_{22}
\nonu \\
& + & 
\frac{4 i \left(248 k^3+1195 k^2+966 k-189\right)}{3 (k+5)^3 (13 k+17)}
\left(
\begin{array}{c}
F_{21}  \\
F_{12}
\end{array}
\right)  A_3
\nonu \\
&- & \frac{4 i \left(248 k^3+1195 k^2+966 k-189\right)}{3 (k+5)^3 (13 k+17)}
\left(
\begin{array}{c}
F_{11}  \\
F_{22}
\end{array}
\right) A_{\mp}
\nonu \\
&-&  
\frac{4 i \left(46 k^3-409 k^2-4404 k-4509\right)}{3 (k+5)^3 (13 k+17)}
\left(
\begin{array}{c}
F_{22}  \\
F_{11}
\end{array}
\right) B_{\mp}
\nonu \\
& + & \left. \frac{4 i \left(46 k^3-409 k^2-4404 k-4509\right)}
{3 (k+5)^3 (13 k+17)}
\left(
\begin{array}{c}
F_{21}  \\
F_{12}
\end{array}
\right) B_3
  \right](w)
\nonu \\
& + & \frac{1}{(z-w)^3} \, \left[ \frac{2 (k-3) (41 k+109)}{3 (k+5) (13 k+17)}
({\bf P_{\pm}^{(\frac{5}{2})}} 
\pm
 {\bf W_{\pm}^{(\frac{5}{2})}})
\right. 
\nonu \\
& - & \frac{4i (k+4)}{(k+5)^2} A_3 
 \left(
\begin{array}{c}
G_{21}  \\
G_{12}
\end{array}
\right) \mp
\frac{8(k+9)}{3(k+5)^3} A_3 A_3 
 \left(
\begin{array}{c}
F_{21}  \\
F_{12}
\end{array}
\right)
\nonu \\
&\pm & 
\frac{8 ( k-3)}{3 (k+5)^3} A_3 B_{\mp}
 \left(
\begin{array}{c}
F_{22}  \\
F_{11}
\end{array}
\right)
\mp \frac{8 (k-3)}{3 (k+5)^3} A_3 B_3 
\left(
\begin{array}{c}
F_{21}  \\
F_{12}
\end{array}
\right)
\nonu \\
&+ & \frac{4 i \left(248 k^3+883 k^2-2250 k-3861\right)}{9 (k+5)^3 (13 k+17)} 
\pa A_3 
\left(
\begin{array}{c}
F_{21}  \\
F_{12}
\end{array}
\right) 
\nonu \\
& + &  \frac{8 i \left(124 k^3+929 k^2+3198 k+2889\right)}{9 (k+5)^3 (13 k+17)}
 A_3
\pa
\left(
\begin{array}{c}
F_{21}  \\
F_{12}
\end{array}
\right)
\nonu \\
& + &
\frac{4i (k+4)}{ (k+5)^2} 
A_{\mp}
\left(
\begin{array}{c}
G_{11}  \\
G_{22}
\end{array}
\right)
\pm \frac{8 (k-3)}{3 (k+5)^3}
A_{\mp} B_3 
\left(
\begin{array}{c}
F_{11}  \\
F_{22}
\end{array}
\right)
\nonu \\
&\pm & 
\frac{8(k-3)}{3(k+5)^3} A_{\mp} B_{\mp} 
\left(
\begin{array}{c}
F_{12}  \\
F_{21}
\end{array}
\right)
\pm
 \frac{16(k+3)}{3(k+5)^3} B_3 B_3 
\left(
\begin{array}{c}
F_{21}  \\
F_{12}
\end{array}
\right) 
\nonu \\
& -  & \frac{4 i \left(248 k^3+961 k^2-1446 k-2943\right)}
{9 (k+5)^3 (13 k+17)}  \pa A_{\mp}
\left(
\begin{array}{c}
F_{11}  \\
F_{22}
\end{array}
\right) \nonu \\
& - & \frac{8 i \left(124 k^3+929 k^2+3198 k+2889\right)}
{9 (k+5)^3 (13 k+17)}
  A_{\mp} \pa
\left(
\begin{array}{c}
F_{11}  \\
F_{22}
\end{array}
\right)
\nonu \\
&\mp& \frac{8 (k+9)}{3 (k+5)^3} A_{\pm} A_{\mp}
\left(
\begin{array}{c}
F_{21}  \\
F_{12}
\end{array}
\right) -
\frac{4i (k+4)}{(k+5)^2} B_3
\left(
\begin{array}{c}
G_{21}  \\
G_{12}
\end{array}
\right)
\nonu \\
& + &
  \frac{4 i \left(202 k^3+419 k^2-3120 k-3897\right)}{9 (k+5)^3 (13 k+17)}
\pa B_3 
\left(
\begin{array}{c}
F_{21}  \\
F_{12}
\end{array}
\right)
\nonu \\
&- & \frac{8 i \left(133 k^3+896 k^2+3249 k+2790\right)}{9 (k+5)^3 (13 k+17)}
B_3 \pa 
\left(
\begin{array}{c}
F_{21}  \\
F_{12}
\end{array}
\right) +
\frac{4 i(k+4) }{ (k+5)^2} 
B_{\mp}
\left(
\begin{array}{c}
G_{22}  \\
G_{11}
\end{array}
\right)
\nonu \\
&- &  
\frac{4 i \left(202 k^3+263 k^2-3792 k-4509\right)}{9 (k+5)^3 (13 k+17)}
\pa B_{\mp} 
\left(
\begin{array}{c}
F_{22}  \\
F_{11}
\end{array}
\right)
\nonu \\
&+ &  \frac{8 i \left(133 k^3+896 k^2+3249 k+2790\right)}{9 (k+5)^3 (13 k+17)}
B_{\mp} \pa
\left(
\begin{array}{c}
F_{22}  \\
F_{11}
\end{array}
\right) \nonu \\
& \pm & 
\frac{16 (k+3)}{3 (k+5)^3} 
B_{\pm} B_{\mp}
\left(
\begin{array}{c}
F_{21}  \\
F_{12}
\end{array}
\right)
\mp \frac{8 (k+9)}{3 (k+5)^3} 
\left(
\begin{array}{c}
F_{11} F_{21} G_{22}  \\
F_{22} F_{12} G_{11}
\end{array}
\right) \nonu \\
& \pm &
\frac{8 \left(101 k^3-173 k^2-4089 k-5031\right)}{9 (k+5)^4 (13 k+17)}
\left(
\begin{array}{c}
\pa F_{11} F_{21} F_{22}  \\
\pa F_{22} F_{12} F_{11}
\end{array}
\right)
\nonu \\
& \mp&
 \frac{8 (5 k+9) \left(11 k^2+118 k+219\right)}{9 (k+5)^4 (13 k+17)}
\left(
\begin{array}{c}
F_{11} \pa F_{21}  F_{22}  \\
F_{22} \pa F_{12}  F_{11}
\end{array}
\right)
\nonu \\
& \mp & \frac{8 \left(211 k^3+1205 k^2+225 k-1089\right)}{9 (k+5)^4 (13 k+17)} 
\left(
\begin{array}{c}
F_{11} F_{21} \pa F_{22}  \\
F_{22} F_{12} \pa F_{11}
\end{array}
\right)
\nonu \\
& - &  \frac{4}{(k+5)^2} F_{12} F_{21}
 \left(
\begin{array}{c}
G_{21}   \\
G_{12} 
\end{array}
\right) - \frac{4 (k-3)}{3 (k+5)^3}
F_{11} F_{22} 
  \left(
\begin{array}{c}
G_{21}   \\
G_{12} 
\end{array}
\right)
\nonu \\
& \pm&  \frac{16 (k-3) \left(8 k^2+43 k+27\right)}{(k+5)^3 (13 k+17)}  \pa^2 
 \left(
\begin{array}{c}
F_{21}   \\
F_{12} 
\end{array}
\right) \pm
\frac{16 (k+3)}{3 (k+5)^3} 
 \left(
\begin{array}{c}
F_{21}  F_{22} G_{11}  \\
F_{12} F_{11} G_{22}
\end{array}
\right)
\nonu \\
& \mp&  
\frac{2 \left(248 k^3+3004 k^2+8517 k+5733\right)}{9 (k+5)^2 (13 k+17)}
 \pa  
 \left(
\begin{array}{c}
G_{21}   \\
G_{12} 
\end{array}
\right) 
\nonu \\
&  \pm & \frac{16 (k-3) (4 k+17)}{3 (k+5)^2 (13 k+17)} 
T 
 \left(
\begin{array}{c}
F_{21}   \\
F_{12} 
\end{array}
\right)  \mp \frac{4(k-3)}{(k+5)^2}
U
 \left(
\begin{array}{c}
G_{21}   \\
G_{12} 
\end{array}
\right)
\nonu \\
&- & \frac{32 i}{(k+5)^3} U A_3 
 \left(
\begin{array}{c}
F_{21}   \\
F_{12} 
\end{array}
\right) +
\frac{32 i }{ (k+5)^3} U A_{\mp}
 \left(
\begin{array}{c}
F_{11}   \\
F_{22} 
\end{array}
\right)
\nonu \\
&- & \frac{8 i(k+1)}{(k+5)^3} U B_3
\left(
\begin{array}{c}
F_{21}   \\
F_{12} 
\end{array}
\right) +
\frac{8 i (k+1)}{ (k+5)^3}
U B_{\mp} 
\left(
\begin{array}{c}
F_{22}   \\
F_{11} 
\end{array}
\right)
\nonu \\
& \mp&  \frac{4 \left(248 k^3+2653 k^2+8058 k+5733\right)}
{9 (k+5)^3 (13 k+17)}     \pa U
\left(
\begin{array}{c}
F_{21}   \\
F_{12} 
\end{array}
\right)
\nonu \\
& \mp & \frac{16 \left(62 k^3+49 k^2+948 k+1089\right)}{9 (k+5)^3 (13 k+17)}  U
\pa
\left(
\begin{array}{c}
F_{21}   \\
F_{12} 
\end{array}
\right)
\nonu \\
& \mp & \left.   
\frac{8(k-3)}{3(k+5)^3} U U 
\left(
\begin{array}{c}
F_{21}   \\
F_{12} 
\end{array}
\right) 
- 
\frac{32 (k-3)}{3 (k+5)^3} \pa
\left(
\begin{array}{c}
F_{21}   \\
F_{12} 
\end{array}
\right)  F_{12} F_{21}
\right](w)
\nonu \\ 
& + &  \frac{1}{(z-w)^2} \, \left[ + \cdots
\right](w)+ \frac{1}{(z-w)} \, \left[ + \cdots
\right](w) +\cdots.
\nonu
\eea
One does not see any higher spin current in the fourth order pole.

Finally, the final OPE can be described as
\bea
%%%%%%%%%%%%%%%%%%%%%%%%%%%%%%%%%%%%%%%%%%%%%%%%%%%%%%%%%%%%%%%%%%%%%%
{\bf W^{(3)}}(z) \, {\bf W^{(3)}}(w) 
&=& 
\frac{1}{(z-w)^6} \, \left[ \frac{64k(31k^3+434k^2+1419k+1080)}
{(k+5)^3(13k+17)}\right]
\nonu \\
& + & \frac{1}{(z-w)^4} \, \left[ -\frac{2 (k-3) (5 k-23)}{(k+5) (13 k+17)}  
{\bf P^{(2)}} +
\frac{8 (k-3)^2 (5 k-23)}{3 (k+5)^2 (13 k+17)}
  {\bf T^{(2)}} 
\right. \nonu \\
& + &  \frac{8 (k-3) (5 k-23)}{(k+5) (13 k+17)}      
{\bf W^{(2)}}
-\frac{16 {\bf (k-3)^2} (11 k+7)}{(k+5) (13 k+17)^2} {\bf T^{(1)}} {\bf T^{(1)}}
 \nonu \\ 
& +  & \frac{8 \left(3244 k^4+41659 k^3+151375 k^2+203133 k+83997\right)}
{3 (k+5)^2 (13 k+17)^2}   T \nonu \\
& + & \frac{32 \left(61 k^3+383 k^2+573 k+243\right)}{3 (k+5)^3 (13 k+17)}   
A_3 A_3  -
\frac{8 (k-3)^2 (5 k-23)}{3 (k+5)^3 (13 k+17)}  A_3 B_3
\nonu \\
&+&  \frac{32 i \left(61 k^3+383 k^2+573 k+243\right)}{3 (k+5)^3 (13 k+17)}  
\pa A_3
\nonu \\
& - &  \frac{128 i \left(7 k^3+86 k^2+267 k+252\right)}{3 (k+5)^4 (13 k+17)}
 A_3 F_{11} F_{22}
 -\frac{24 i}{(k+5)^2}  A_3 F_{12} F_{21}
\nonu \\
& + &  \frac{32 \left(61 k^3+383 k^2+573 k+243\right)}{3 (k+5)^3 (13 k+17)}   
A_{+} A_{-}
\nonu \\
&- & \frac{8 i \left(127 k^3+1397 k^2+3453 k+5895\right)}
{3 (k+5)^4 (13 k+17)}   A_{-} F_{11} F_{12}
\nonu \\
& - & \frac{16 i \left(17 k^3+217 k^2+879 k+327\right)}{(k+5)^4 (13 k+17)}  
A_{+} F_{21} F_{22}
\nonu \\
& - & \frac{8 \left(5 k^3-827 k^2-5325 k-4941\right)}{3 (k+5)^3 (13 k+17)}   
B_3 B_3
- \frac{24 i}{ (k+5)^2}  B_3 F_{12} F_{21}
\nonu \\
& - & \frac{8 i \left(5 k^3-827 k^2-5325 k-4941\right)}{3 (k+5)^3 (13 k+17)}   
\pa B_3
\nonu \\
&+&  \frac{128 i \left(7 k^3+86 k^2+267 k+252\right)}{3 (k+5)^4 (13 k+17)}   
B_3 F_{11} F_{22}
\nonu \\
& + & \frac{16 i \left(41 k^3+757 k^2+2271 k+1395\right)}
{3 (k+5)^4 (13 k+17)}   B_{-} F_{12} F_{22}
\nonu \\
&- &  \frac{8 \left(5 k^3-827 k^2-5325 k-4941\right)}{3 (k+5)^3 (13 k+17)}  
B_{+} B_{-}
\nonu \\
&+& \frac{8 i \left(49 k^3+395 k^2+1395 k+1689\right)}{(k+5)^4 (13 k+17)}
    B_{+} F_{11} F_{21}
\nonu \\
&- & \frac{4 (k-3) \left(193 k^2+1064 k+975\right)}{3 (k+5)^3 (13 k+17)}  
F_{11} G_{22}
\nonu \\
& + & \frac{8 \left(244 k^4+2291 k^3+12045 k^2+35937 k+24723\right)}
{3 (k+5)^4 (13 k+17)}  \pa F_{11} F_{22}
\nonu \\
& - &  \frac{8 \left(244 k^4+2381 k^3+11811 k^2+33759 k+30933\right)}
{3 (k+5)^4 (13 k+17)}   F_{11} \pa F_{22}
\nonu \\
& - & \frac{8 (k-3) \left(28 k^2+179 k+243\right)}{(k+5)^3 (13 k+17)}  
F_{12} G_{21}
\nonu \\
& + &  \frac{16 \left(122 k^4+1183 k^3+5805 k^2+17973 k+13293\right)}
{3 (k+5)^4 (13 k+17)}  \pa F_{12} F_{21}
\nonu \\
& - &  \frac{16 \left(122 k^4+1163 k^3+6017 k^2+17241 k+14121\right)}
{3 (k+5)^4 (13 k+17)}
F_{12} \pa F_{21}
\nonu \\
& - & \frac{24 (k-3) (11 k+7)}{(k+5)^2 (13 k+17)}  F_{21} G_{12}
-  \frac{8 (k-3) \left(94 k^2+551 k+453\right)}{3 (k+5)^3 (13 k+17)}  
F_{22} G_{11}
\nonu \\
& + & \frac{24 i (k-3) (5 k-23)}{(k+5)^3 (13 k+17)}  U A_3 
+ \frac{8 i (k-3) k (5 k-23)}{(k+5)^3 (13 k+17)}  U B_3 
\nonu \\
&+& \frac{8 \left(244 k^3+2189 k^2+7062 k+4941\right)}{3 (k+5)^3 (13 k+17)}   
U U
\nonu \\
&- & \left. \frac{32 (k-3)^2 (5 k-23)}{3 (k+5)^4 (13 k+17)}     
U F_{11} F_{22}
+ \frac{24 (k-3) (5 k-23)}{(k+5)^3 (13 k+17)}   U F_{12} F_{21}
\right](w)
\nonu \\
& + & \frac{1}{(z-w)^3} \, \frac{1}{2} \pa
(\mbox{pole-4})(w)
+  \frac{1}{(z-w)^2} \, \left[ + \cdots \right](w)
+  \frac{1}{(z-w)} \, \left[ + \cdots \right](w)
 +\cdots.
\nonu 
\eea
Note the presence of the nonlinear term containing the higher spin-$1$ current
in the fourth order pole.
In the basis of (\ref{bcgspin3}), one realizes that this term disappears 
eventually. 

It would be interesting to write the above $136$ OPEs
in the simplified notations as in Appendix $A$ or Appendix $D$.

%%%%%%%%%%%%%%%%%%%%%%%%%%%%%%%%%%%%%%%%%%%%%%%%%%%%%%%%%%%%%%%%%%%%%%%%%%%
%%%%%%%%%%%%%%%%%%%%%%%%%%%%%%%%%%%%%%%%%%%%%%%%%%%%%%%%%%%%%%%%%%%%%%%%%%

\end{document}